# Searching for long-lived particles beyond the Standard Model at the Large Hadron Collider

March 6, 2019

Particles beyond the Standard Model (SM) can generically have lifetimes that are long compared to SM particles at the weak scale. When produced at experiments such as the Large Hadron Collider (LHC) at CERN, these long-lived particles (LLPs) can decay far from the interaction vertex of the primary proton-proton collision. Such LLP signatures are distinct from those of promptly decaying particles that are targeted by the majority of searches for new physics at the LHC, often requiring customized techniques to identify, for example, significantly displaced decay vertices, tracks with atypical properties, and short track segments. Given their non-standard nature, a comprehensive overview of LLP signatures at the LHC is beneficial to ensure that possible avenues of the discovery of new physics are not overlooked. Here we report on the joint work of a community of theorists and experimentalists with the ATLAS, CMS, and LHCb experiments — as well as those working on dedicated experiments such as MoEDAL, milliQan, MATHUSLA, CODEX-b, and FASER — to survey the current state of LLP searches at the LHC, and to chart a path for the development of LLP searches into the future, both in the upcoming Run 3 and at the High-Luminosity LHC. The work is organized around the current and future potential capabilities of LHC experiments to generally discover new LLPs, and takes a signature-based approach to surveying classes of models that give rise to LLPs rather than emphasizing any particular theory motivation. We develop a set of simplified models; assess the coverage of current searches; document known, often unexpected backgrounds; explore the capabilities of proposed detector upgrades; provide recommendations for the presentation of search results; and look towards the newest frontiers, namely high-multiplicity "dark showers", highlighting opportunities for expanding the LHC reach for these signals.

**Editors:**
Juliette Alimena[1] (Experimental Coverage, Backgrounds, Upgrades), James Beacham[2] (Document Editor, Simplified Models), Martino Borsato[3] (Backgrounds, Upgrades), Yangyang Cheng[4] (Upgrades), Xabier Cid Vidal[5] (Experimental Coverage), Giovanna Cottin[6] (Simplified Models, Reinterpretations), Albert De Roeck[7] (Experimental Coverage), Nishita Desai[8] (Reinterpretations), David Curtin[9] (Simplified Models), Jared A. Evans[10] (Simplified Models, Experimental Coverage), Simon Knapen[11] (Dark Showers), Sabine Kraml[12] (Reinterpretations), Andre Lessa[13] (Reinterpretations), Zhen Liu[14] (Simplified Models, Backgrounds, Reinterpretations), Sascha Mehlhase[15] (Backgrounds), Michael J. Ramsey-Musolf[16,126] (Simplified Models), Heather Russell[17] (Experimental Coverage), Jessie Shelton[18] (Simplified Models, Dark Showers), Brian Shuve[19,20] (Document Editor, Simplified Models, Simplified Models Library), Monica Verducci[21] (Upgrades), Jose Zurita[22,23] (Experimental Coverage)

**Contributors & Endorsers:**

Todd Adams(24), Michael Adersberger(25), Cristiano Alpigiani(26), Artur Apresyan(83), Robert John Bainbridge(27), Varvara Batozskaya(28), Hugues Beauchesne(29), Lisa Benato(30), S. Berlendis(31), Eshwen Bhal(32), Freya Blekman(33), Christina Borovilou(34), Jamie Boyd(7), Benjamin P. Brau(35), Lene Bryngemark(36), Oliver Buchmueller(27), Malte Buschmann(37), William Buttinger(7), Mario Campanelli(38), Cari Cesarotti(39), Chunhui Chen(40), Hsin-Chia Cheng(41), Sanha Cheong(42,43), Matthew Citron(44), Andrea Coccaro(45), V. Coco(7), Eric Conte(46), Félix Cormier(47), Louie D. Corpe(38), Nathaniel Craig(44), Yanou Cui(20), Elena Dall'Occo(48), C. Dallapiccola(35), M.R. Darwish(49), Alessandro Davoli(50,52), Annapaola de Cosa(51), Andrea De Simone(50,52), Luigi Delle Rose(53,54), Frank F. Deppisch(38), Biplab Dey(55), Miriam D. Diamond(9), Keith R. Dienes(31,56), Sven Dildick(57), Babette Döbrich(7), Marco Drewes(58), Melanie Eich(30), M. ElSawy(59,60), Alberto Escalante del Valle(61), Gabriel Facini(38), Marco Farina(62), Jonathan L. Feng(63), Oliver Fischer(22), H.U. Flaecher(32), Patrick Foldenauer(64), Marat Freytsis(65,11,66), Benjamin Fuks(67,68), Iftah Galon(69), Yuri Gershtein(70), Stefano Giagu(71), Andrea Giammanco(58), Vladimir V. Gligorov(72), Tobias Golling(73), Sergio Grancagnolo(74), Giuliano Gustavino(75), Andrew Haas(76), Kristian Hahn(77), Jan Hajer(58), Ahmed Hammad(78), Lukas Heinrich(7), Jan Heisig(58), J.C. Helo(79), Gavin Hesketh(38), Christopher S. Hill(1), Martin Hirsch(80), M. Hohlmann(81), W. Hulsbergen(48), John Huth(39), Philip Ilten(82), Thomas Jacques(50), Bodhitha Jayatilaka(83), Geng-Yuan Jeng(56), K.A. Johns(31), Toshiaki Kaji(84), Gregor Kasieczka(30), Yevgeny Kats(29), Malgorzata Kazana(85), Henning Keller(86), Maxim Yu. Khlopov(87,88), Felix Kling(63), Ted R. Kolberg(24), Igor Kostiuk(48,127), Emma Sian Kuwertz(7), Audrey Kvam(26), Greg Landsberg(89), Gaia Lanfranchi(90,7), Iñaki Lara(91), Alexander Ledovskoy(92), Dylan Linthorne(93), Jia Liu(94), Iacopo Longarini(95), Steven Lowette(33), Henry Lubatti(26), Margaret Lutz(35), Jingyu Luo(96), Judita Mamužić(80), Matthieu Marinangeli(97), Alberto Mariotti(33), Daniel Marlow(96), Matthew McCullough(7), Kevin McDermott(4), P. Mermod(73), David Milstead(98), Vasiliki A. Mitsou(80), Javier Montejo Berlingen(7), Filip Moortgat(7,99), Alessandro Morandini(50,52), Alice Polyxeni Morris(38), David Michael Morse(100), Stephen Mrenna(83), Benjamin Nachman(101), Miha Nemevšek(102), Fabrizio Nesti(103), Christian Ohm(104,105), Silvia Pascoli(106), Kevin Pedro(83), Cristián Peña(83,107), Karla Josefina Pena Rodriguez(30), Jónatan Piedra(108), James L. Pinfold(109), Antonio Policicchio(71), Goran Popara(110), Jessica Prisciandaro(58), Mason Proffitt(26), Giorgia Rauco(51), Federico Redi(97), Matthew Reece(39), Allison Reinsvold Hall(83), H. Rejeb Sfar(49), Sophie Renner(112), Amber Roepe(75), Manfredi Ronzani(76), Ennio Salvioni(113), Arka Santra(80), Ryu Sawada(114), Jakub Scholtz(106), Philip Schuster(42), Pedro Schwaller(112), Cristiano Sebastiani(71), Sezen Sekmen(115), Michele Selvaggi(7), Weinan Si(20), Livia Soffi(71), Daniel Stolarski(93), David Stuart(44), John Stupak III(75), Kevin Sung(77), Wendy Taylor(116), Sebastian Templ(61), Brooks Thomas(117), Emma Torró-Pastor(26), Daniele Trocino(99), Sebastian Trojanowski(118), Marco Trovato(119), Yuhsin Tsai(14), C.G. Tully(96), Tamás Álmos Vámi(120), Juan Carlos Vasquez(121), Carlos Vázquez Sierra(48), K. Vellidis(34), Basile Vermassen(99), Martina Vit(99), Devin G.E. Walker(122), Xiao-Ping Wang(119), Gordon Watts(26), Si Xie(107), Melissa Yexley(123), Charles Young(42), Jiang-Hao Yu(124,16), Piotr Zalewski(85), Yongchao Zhang(125)

(1) *Department of Physics, The Ohio State University, 191 W. Woodruff Ave., Columbus, OH 43210-1117, USA*
(2) *Department of Physics, Duke University, 120 Science Drive, Durham, NC 27710, USA*
(3) *Physikalisches Institut, Ruprecht-Karls-Universität Heidelberg, Im Neuenheimer Feld 226 69120, Heidelberg, Germany*
(4) *Cornell University, 245 East Avenue, Ithaca, NY 14853, USA*
(5) *Instituto Galego de Física de Altas Enerxías, U. Santiago de Compostela, IGFAE, Rúa de Xoaquín Díaz de Rábago, s/n, 15782 Santiago de*

Compostela, Spain
(6) Department of Physics, National Taiwan University, Taipei 10617, Taiwan
(7) CERN, Esplanade des particules 1, Geneva, Switzerland
(8) Department of Theoretical Physics, Tata Institute of Fundamental Physics, Homi Bhabha Road, Mumbai 400005, India
(9) Department of Physics, University of Toronto, 60 St. George Street,Toronto, ON M5S 1A7, Canada
(10) Department of Physics, University of Cincinnati, 400 Geology/Physics Bldg., Cincinnati, OH 45221, USA
(11) School of Natural Sciences, Institute for Advanced Study, 1 Einstein Drive, Princeton, NJ 08540, USA
(12) Laboratoire de Physique Subatomique et de Cosmologie, Université Grenoble-Alpes, CNRS/IN2P3, 53 Avenue des Martyrs, F-38026 Grenoble, France
(13) Universidade Federal do ABC, Av. dos Estados, 5001, Santo Andre, 09210-580 SP, Brazil
(14) Maryland Center for Fundamental Physics, Department of Physics, University of Maryland, College Park, MD 20742, USA
(15) Faculty of Physics, Ludwig-Maximilians-Universität München, Schellingstraße 4, 80799 Munich, Germany
(16) Amherst Center for Fundamental Interactions, Department of Physics, University of Massachusetts Amherst, Lederle Graduate Research Center 416, Amherst, MA 01003 USA
(17) Department of Physics, McGill University, Montreal, QC H3A 2T8, Canada
(18) University of Illinois at Urbana-Champaign, 1110 W. Green St., Urbana, IL 61801, USA
(19) Harvey Mudd College, 301 Platt Blvd., Claremont, CA 91711, USA
(20) University of California, Riverside, 900 University Avenue, Riverside, CA 92521, USA
(21) Universita' di Roma Tre and INFN, via della Vasca Navale 84 00146 Roma Italy
(22) Institute for Nuclear Physics (IKP), Karlsruhe Institute of Technology, Hermann-von-Helmholtz-Platz 1, D-76344 Eggenstein-Leopoldshafen, Germany
(23) Institute for Theoretical Particle Physics (TTP), Karlsruhe Institute of Technology, Engesserstraße 7, D-76128 Karlsruhe, Germany
(24) Florida State University, 77 Chieftan Way, Tallahassee, FL 32306, USA
(25) Faculty of Physics, Ludwig-Maximilians Universität München, Schellingstrasse 4, 80799 Munich, Germany
(26) University of Washington, Seattle, 1410 NE Campus Parkway Seattle, WA 98195, USA
(27) High Energy Physics Group, Blackett Laboratory, Imperial College, Prince Consort Road, London SW7 2AZ, UK
(28) National Centre for Nuclear Research (NCBJ), Hoża 69, 00-681 Warsaw, Poland
(29) Department of Physics, Ben-Gurion University, 1 Ben-Gurion Boulevard Beer-Sheva, Beer-Sheva 8410501, Israel
(30) Institut für Experimentalphysik, Universität Hamburg, Luruper Chaussee 149, 22761 Hamburg, Germany
(31) University of Arizona, 118 E. 4th St., Tucson AZ 85721, USA
(32) University of Bristol, HH Wills Physics Laboratory, Tyndall Avenue, Bristol, BS8 1TL, UK
(33) IIHE, Vrije Universiteit Brussel, Pleinlaan 2, 1050 Brussels, Belgium
(34) Physics Department, National and Kapodistrian University of Athens, Panepistimioupoli, Zografou, 15784, Greece
(35) University of Massachusetts, Amherst, 1126 LGRT, University of Massachusetts, Amherst, MA 01003-9337, USA
(36) Department of Physics, Division of Particle Physics, Lund University, Box 118, 221 00 Lund, Sweden
(37) Leinweber Center for Theoretical Physics, University of Michigan, 450 Church Street, Ann Arbor MI-48109, USA
(38) University College London, Gower Street WC1E 6BT London, UK
(39) Department of Physics, Harvard University, 17 Oxford Street, Cambridge, MA, 02138
(40) Department of Physics and Astronomy, Iowa State University, 2323 Osborn Drive, Physics 0012, Ames, IA 50011-3160, USA
(41) Department of Physics, University of California, Davis, One Shields Avenue, Davis, CA 95616, USA
(42) SLAC National Accelerator Laboratory, 2575 Sand Hill Rd, Menlo Park, CA 94025, USA
(43) Physics Department, Stanford University, 450 Serra Mall, Stanford, CA 94305, USA
(44) Department of Physics, Broida Hall, University of California, Santa Barbara, CA 93106, USA
(45) INFN Sezione di Genova, Via Dodecaneso, 33 - 16146 - Genova, Italy
(46) Institut Pluridisciplinaire Hubert Curien (IPHC), Département Recherches Subatomiques, Université de Strasbourg/CNRS-IN2P3, 23 Rue du Loess, F-67037 Strasbourg, France
(47) Department of Physics & Astronomy, The University of British Columbia, 2329 West Mall, Vancouver, BC V6T 1Z4, Canada
(48) Nikhef National Institute for Subatomic Physics, Science Park 105, 1098 XG Amsterdam, The Netherlands
(49) Antwerp University, Prinsstraat 13, 2000 Antwerpen, Belgium
(50) SISSA, Via Bonomea 265, Trieste, 34136, Italy
(51) University of Zürich, Winterthurerstrasse 190, 8057 Zurich, Switzerland
(52) INFN Sezione di Trieste, via Bonomea 265, 34136 Trieste, Italy
(53) Department of Physics and Astronomy, University of Florence, Via G. Sansone 1, 50019 Sesto Fiorentino, Italy
(54) University of Southampton, Highfield, Southampton SO17 1BJ, UK
(55) Institute of Particle Physics, Central China Normal University, Wuhan, Hubei, China
(56) Department of Physics, University of Maryland, 4296 Stadium Drive, College Park, MD 20742, USA
(57) Texas A&M University, 4242 TAMU College Station, TX 77843-4242, USA
(58) Centre for Cosmology, Particle Physics and Phenomenology (CP3), Université catholique de Louvain, Chemin du Cyclotron 2, B-1348, Louvain-la-Neuve, Belgium
(59) Basic Science department, Faculty of Engineering, The British University in Egypt, El Sherouk City, Misr-Ismalia Road, Postal No 11837, PO Box 43, Cairo, Egypt
(60) Physics department, Faculty of Science, Beni Suef University, Qism Bani Sweif, Bani Sweif, Beni Suef Governorate, Egypt
(61) Institute of High Energy Physics, Austrian Academy of Sciences, Nikolsdorfer Gasse 18, 1050 Vienna, Austria
(62) Stony Brook University, Stony Brook, NY 11794-3840, USA
(63) Department of Physics and Astronomy, University of California, Irvine, CA 92697-4575, USA
(64) Institut für Theoretische Physik, Universität Heidelberg, Philosophenweg 16, D-69120 Heidelberg, Germany
(65) Raymond and Beverly Sackler School of Physics and Astronomy, Tel Aviv University, Tel-Aviv 69978, Israel
(66) Institute of Theoretical Science, University of Oregon, Eugene, OR 97403, USA
(67) Laboratoire de Physique Théorique et Hautes Energies (LPTHE), UMR 7589, Sorbonne Université et CNRS, 4 place Jussieu, 75252 Paris

Cedex 05, France
[68] Institut Universitaire de France, 103 boulevard Saint-Michel, 75005 Paris, France
[69] New High Energy Theory Center, Rutgers, The State University of New Jersey, Piscataway, NJ 08854-8019, USA
[70] Rutgers University, 136 Frelinghuysen Rd, Piscataway, NJ 08854 USA
[71] Sapienza Università di Roma and INFN Roma1, P.le A. Moro 5, Roma, 00185, Italy
[72] LPNHE, Sorbonne Université, Paris Diderot Sorbonne Paris Cité, CNRS/IN2P3, Barre 12-22, 4 Place Jussieu, 75252 Paris CEDEX 05, France
[73] Département de Physique Nucléaire et Corpusculaire, Université de Genève, 24, quai Ernest-Ansermet CH-1211 Genève 4, Switzerland
[74] Humboldt-Universität, Newtonstraße 15 12489 Berlin, Germany
[75] The University of Oklahoma, 440 W. Brooks St. Norman, OK 73019, USA
[76] New York University, 726 Broadway, New York, NY 10003, USA
[77] Northwestern University, 2145 Sheridan Rd, Evanston, IL 60208, USA
[78] Departement of Physics, University of Basel, Klingelbergstrasse 82, 4056, Basel, Switzerland
[79] Departamento de Física y Astronomía, Facultad de Ciencias, Universidad de La Serena, Avenida Cisternas 1200, La Serena, Chile
[80] Instituto de Física Corpuscular / Consejo Superior de Investigaciones Científicas - University of Valencia (IFIC / CSIC - UV), Carrer del Catedrátic José Beltrán Martinez, 2, 46980 Paterna, València
[81] Dept. of Aerospace, Physics & Space Sciences, Florida Institute of Technology, 150 W. University Blvd., Melbourne, FL 32901, USA
[82] University of Birmingham, Edgbaston, Birmingham B15 2TT, UK
[83] Fermi National Accelerator Laboratory, PO Box 500, Batavia, IL 60510
[84] Waseda University, Ookubo 3-4-1, Shinjuku-ku, Tokyo 169-8555, Japan
[85] National Centre for Nuclear Research (NCBJ), Andrzeja Sołtana 7, 05-400 Otwock-Świerk, Poland
[86] RWTH Aachen University, Otto-Blumenthal-Straße, 52074 Aachen, Germany
[87] National Research Nuclear University MEPHI, Kashirskoe chaussee 31, Moscow 115409, Russia
[88] APC Laboratory, 10, rue Alice Domon et Léonie Duquet 75205, Paris Cedex 13, France
[89] Brown University, 182 Hope St, Providence, RI 02912, USA
[90] Laboratori Nazionali di Frascati, INFN, via E. Fermi 40 00044 Frascati RM, Italy
[91] Instituto de Física Teórica (UAM-CSIC), C/Nicolás Cabrera 13-15 Campus de Cantoblanco 28049 Madrid, Spain
[92] Physics Department, University of Virginia, PO Box 400714, 382 McCormick Rd , Charlottesville, VA, 22904-4714, USA
[93] Carleton University, 1125 Colonel By Dr, Ottawa, ON K1S 5B6, Canada
[94] Enrico Fermi Institute, University of Chicago, 5640 S. Ellis Avenue, RI-183, Chicago, IL 60637, USA
[95] Università degli studi di Roma "La Sapienza" and INFN Roma, Dipartimento di Fisica G. Marconi. Piazzale Aldo Moro 5, 00185, Roma, Italy
[96] Physics Department, Washington Road, Princeton University, Princeton, NJ 08544, USA
[97] École Polytechnique Fédérale de Lausanne, Route Cantonale, 1015 Lausanne, Switzerland
[98] Fysikum, Stockholms Universitet, Roslagstullsbacken 21, 114 21 Stockholm, Sweden
[99] Department of Physics and Astronomy, University of Ghent, Proeftuinstraat 86, B-9000 Ghent, Belgium
[100] Northeastern University, 360 Huntington Ave., Boston, Massachusetts 02115, USA
[101] Physics Division, Lawrence Berkeley National Laboratory, Berkeley, 1 Cyclotron Road, CA 94720, USA
[102] Jožef Stefan Institute, Jamova 39, Ljubljana 1000, Slovenia
[103] Università dell'Aquila, via Vetoio, L'Aquila, 67100, Italy
[104] KTH Royal Institute of Technology, AlbaNova Universitetscentrum, KTH, Physics Department, SE-106 91 Stockholm, Sweden
[105] Oskar Klein Centre for Cosmoparticle Physics, Stockholm University, SE-106 91 Stockholm, Sweden
[106] IPPP, Department of Physics, Durham University, South Road, Durham, DH1 3LE, UK
[107] California Institute of Technology, 1200 E California Blvd, MC 256-48, Pasadena, CA, 91125, USA
[108] IFCA (CSIC - Universidad de Cantabria), Avenida de los Castros, 39005 Santander, Cantabria, Spain
[109] Physics Department, University of Alberta, Edmonton, AB T6G 2E1, Canada
[110] Ruder Bošković Institute, Bijenička cesta 54, 10000 Zagreb, Croatia
[111] Université catholique de Louvain, Place de l'université 1, 1348 Ottignies-Louvain-la-Neuve, Belgium
[112] PRISMA Cluster of Excellence & Mainz Institute for Theoretical Physics, Johannes Gutenberg University, Staudingerweg 7, 55128 Mainz, Germany
[113] Technical University of Munich, Physics Department, James-Franck-Strasse 1, 85748 Garching, Germany
[114] International Center for Elementary Particle Physics, The University of Tokyo, Hongo 7-3-1, Bunkyo-ku, Tokyo 113-0033, Japan
[115] Department of Physics, Kyungpook National University, 80 Daehakro, Bukgu, Daegu 41566, South Korea
[116] York University, 4700 Keele St, Toronto, ON M3J 1P3, Canada
[117] Lafayette College, 730 High St. Easton, PA 18042, USA
[118] Consortium for Fundamental Physics, School of Mathematics and Statistics, University of Sheffield, Hounsfield Road, Sheffield S3 7RH, UK
[119] Argonne National Laboratory, 9700 Cass Avenue, Lemont, IL 60439, USA
[120] Wigner Research Centre for Physics, Konkoly-Thege Miklós út 29-33, Budapest 1121, Hungary
[121] Universidad Técnica Federico Santa María, Avenida España 1680, Valparaiso, Chile, 2340000
[122] Department of Physics and Astronomy, Dartmouth College, Hanover, NH 03755 USA
[123] Lancaster University, Lancaster LA1 4YW, UK
[124] CAS Key Laboratory of Theoretical Physics, Institute of Theoretical Physics, Chinese Academy of Sciences, Beijing 100190, China
[125] Department of Physics and McDonnell Center for the Space Sciences, Washington University, St. Louis, MO 63130, USA
[126] T.-D. Lee Institute and School of Physics and Astronomy, Shanghai Jiao Tong University, 800 Dongchuan Road, Shanghai, 200240, China
[127] Institute for Nuclear Research of the National Academy of Sciences (KINR), 47 Nauky Avenue, 03680, Kyiv, Ukraine

Contact editors: lhc-llp-admin@cern.ch

# Contents

# 1

# *Introduction*

**Document editors:** James Beacham, Brian Shuve

Particles in the Standard Model (SM) have lifetimes spanning an enormous range of magnitudes, from the $Z$ boson ($\tau \sim 2 \times 10^{-25}$ s) through to the proton ($\tau \gtrsim 10^{34}$ years) and electron (stable).

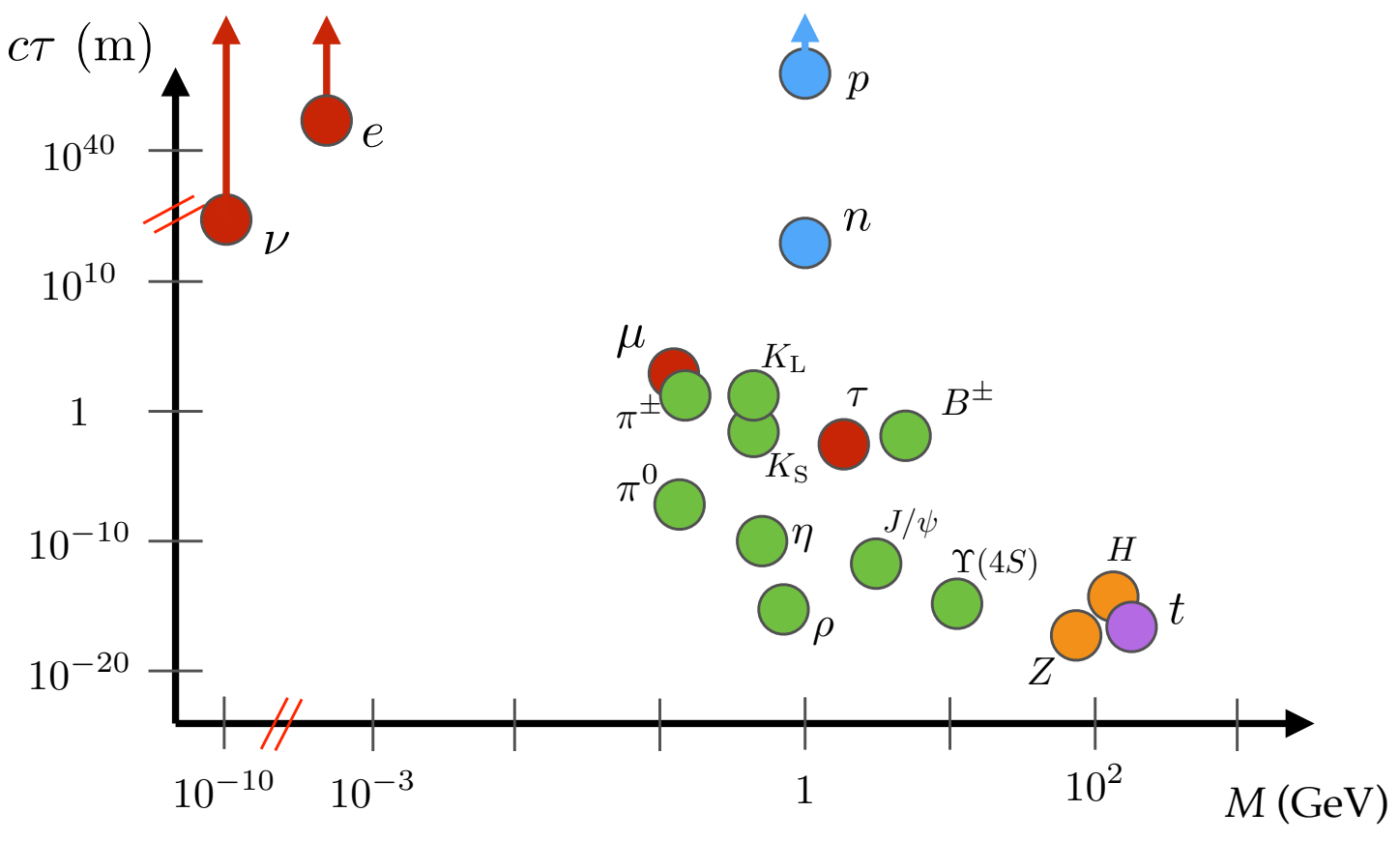

Figure 1.1: Particle lifetime $c\tau$, expressed in meters, as a function of particle mass, expressed in GeV, for a variety of particles in the Standard Model [1].

Similarly, models beyond the SM (BSM) typically predict new particles with a variety of lifetimes. In particular, new weak-scale particles can easily have long lifetimes for several reasons, including approximate symmetries that stabilize the long-lived particle (LLP), small couplings between the LLP and lighter states, and suppressed phase space available for decays. For particles moving close to the speed of light, this can lead to macroscopic, detectable displacements between the production and decay points of an unstable particle for $c\tau \gtrsim 10$ $\mu$m. [1]

[1] Recently, a comprehensive collection of the vast array of theoretical frameworks within which LLPs naturally arise has been assembled as part of the physics case document for the proposed MATHUSLA experiment [2]. Because the focus of the current document is on the experimental signatures of LLPs and explicitly not the theories that predict them, the combination of the MATHUSLA physics case document (and the large number of references therein) and the present document can be considered, together, a comprehensive view of the present status of theoretical motivation and experimental possibilities for the potential discovery of LLPs produced at the interaction points of the Large Hadron Collider.

The experimental signatures of LLPs at the LHC are varied and, by nature, are often very different from signals of SM processes. For example, LLP signatures can include tracks with unusual ionization and propagation properties; small, localized deposits of energy inside of the calorimeters without associated tracks; stopped particles that decay out of time with collisions; displaced vertices in the inner detector or muon spectrometer; and disappearing, appearing, and kinked tracks.

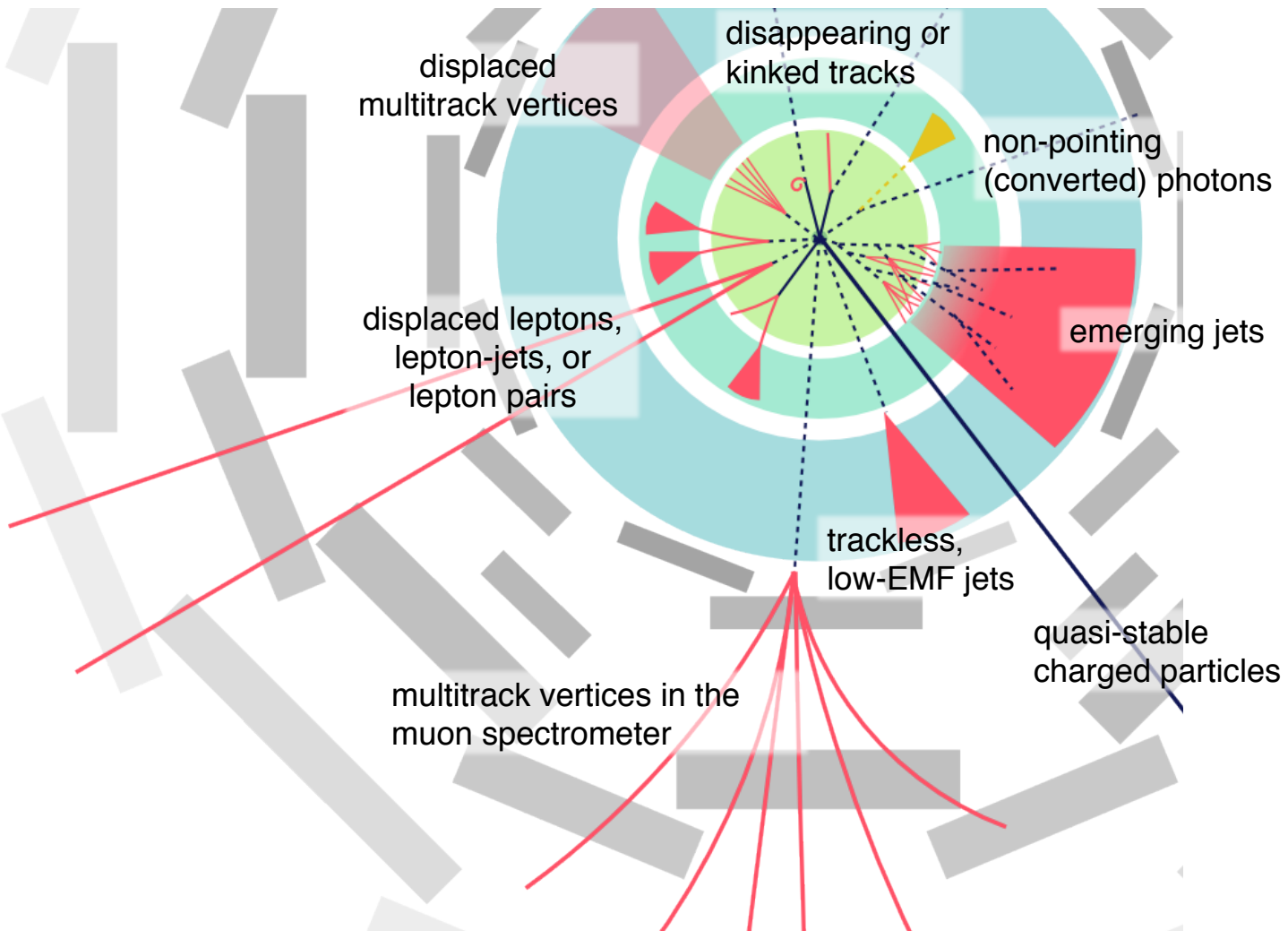

Figure 1.2: Schematic of the variety of challenging, atypical experimental signatures that can result from BSM LLPs in the detectors at the LHC. Shown is a cross-sectional plane in azimuthal angle, $\phi$, of a general purpose detector such as ATLAS or CMS. From Ref. [3].

Because the long-lived particles of the SM have masses $\lesssim 5$ GeV and have well-understood experimental signatures, the unusual signatures of BSM LLPs offer excellent prospects for the discovery of new physics at particle colliders. At the same time, standard reconstruction algorithms may reject events or objects containing LLPs precisely because of their unusual nature, and dedicated searches are needed to uncover LLP signals. These atypical signatures can also resemble noise, pile-up, or mis-reconstructed objects in the detector; due to the rarity of such mis-reconstructions, Monte Carlo (MC) simulations may not accurately model backgrounds for LLP searches, and dedicated methods are needed to do so.

Although small compared to the large number of searches for prompt decays of new particles, many searches for LLPs at the ATLAS, CMS, and LHCb experiments at the Large Hadron Collider (LHC) have already been performed; we refer the reader to Chapter 3 for descriptions of and references to these searches. Existing LLP searches have necessitated the development of novel methods for identifying signals of LLPs, and measuring and suppressing the relevant backgrounds. Indeed, in several scenarios searches for LLPs have sensitivities that greatly exceed the search for similar, promptly decaying new particles (as is true, for example, for directly produced staus in supersymmetry [4]). The excellent sensitivity of these searches, together with the lack of a definitive signal in any prompt channels at the LHC, have focused attention on other types of LLP signatures that are not currently covered. These include low-mass LLPs that do not pass trigger or selection thresholds of current searches, high multiplicities of LLPs produced in dark-sector showers, or unusual LLP production and decay modes that are not covered by current methods. Given the excellent sensitivity of LHC detectors to LLPs, along with the potentially large production cross sections of LLPs and the enormous amount of data expected to be collected when the LHC switches to high-luminosity running in the 2020s, it is imperative that the space of LLP signatures be explored as thoroughly as possible to ensure that no signals are missed. This is particularly important now, with the recent conclusion of LHC Run 2, as new triggering strategies for LLP signatures in the upcoming Run 3 can be investigated with urgency. Moreover, decisions are currently being made about detector upgrades for Phase 2 of the LHC, and design choices should be made to ensure that sensitivity to LLPs is retained or possibly improved through high-luminosity running, as may indeed be the case for many of the plans under consideration by the main experiments.

The increased interest in LLP signatures at the LHC is naturally complementary to the recognition that there are several BSM scenarios that give rise to particles difficult to optimally detect at the LHC — either promptly decaying particles or those with naturally long lifetimes — and that are searched for or planned to be searched for in fixed-target experiments, B-factories, and beam dump experiments [5–7].

The growing theoretical and experimental interest in LLPs has been mirrored by an increased activity in proposals for searches for LLPs produced at the main LHC interaction points — either within the existing ATLAS, CMS, and LHCb collaborations or with new, dedicated detectors — new experimental analyses, and meetings to communicate results and discuss new ideas. Workshops focused on LLPs at the University of Massachusetts, Amherst [2], in November of 2015; Fermilab; and KITP (UCSB) [3] in May of 2016, among others, highlighted the need for a community-wide effort to map the current space of both theoretical models for LLPs and the atypical experimental signatures that could be evidence of LLPs, assess the coverage of current experimental methods to these models, and identify areas where new searches are required. Additionally, the work presented in these meetings underscored the importance of presenting the results of experimental searches in a manner that allows for their application to different models, and generated new ideas for designing analyses with the goal of minimizing model dependence. Such largely model-independent presentation makes current searches more powerful by increasing their applicability to new scenarios, while reducing redundancies in searches and ensuring that gaps in coverage are identified and addressed. This task extends beyond the purview of any particular theoretical model or experiment, and requires an effort across collaborations to address the needs of the LLP community and illuminate a path forward.

This flurry of activity eventually coalesced in the establishment of a more central and regular platform — the LHC LLP Community

[2] "LHC Searches for Long-Lived BSM Particles: Theory Meets Experiment", `https://www.physics.umass.edu/acfi/seminars-and-workshops/lhc-searches-for-long-lived-bsm%2Dparticles-theory-meets-experiment`

[3] "Experimental Challenges for the LHC Run II", `http://online.kitp.ucsb.edu/online/experlhc16/`

— for experimentalists at the LHC and those in the theoretical and phenomenological communities to exchange ideas about LLP searches to ensure the full discovery potential of the LHC. This began with a mini-workshop at CERN in May of 2016 [4] and has continued with workshops in April of 2017 at CERN [5], October of 2017 at ICTP Trieste [6], May of 2018 again at CERN [7], and at Nikhef, in Amsterdam, in October of 2018 [8].

This is the work undertaken by the LHC LLP Community and presented in this document. Based on the most pressing needs identified by the community, we organize the work of this initiative into a few key realms:

- **Simplified models:** We seek to identify a minimal (but expandable) set of simplified models that capture, with a very limited number of free parameters, the most important LLP signatures motivated by theory and accessible at the LHC. The simplified models approach has been successfully applied to models such as supersymmetry (SUSY) and dark matter, and proposals exist for LLP simplified models in particular contexts. We aim to provide a basis of models that serves as a focal point for the other studies performed by the community, as well as a library that can be used in simulating LLP signal events, to allow for a common grammar to better understand how current and future searches cover LLP signature space.

- **Experimental coverage:** In spite of the many successful LLP searches undertaken by the ATLAS, CMS, and LHCb experiments, there remains a need for a systematic study of the complementary coverage of LLP searches to the parameter spaces of LLP models. Having developed a simplified model basis, we provide a comprehensive overview of the sensitivity of existing searches, highlighting gaps in coverage and high-priority searches to be undertaken in the future.

- **Backgrounds to LLP searches:** We provide a summary and analysis of backgrounds for LLP signals at the LHC, sources of which can be rare, unexpected, and largely irrelevant for searches for prompt BSM particles, and thus not fully well understood. We assemble the collected knowledge and experience of backgrounds to prior searches with the intention of providing insight into the opportunities and challenges of searching for LLP signatures.

- **Upgrades and triggering strategies:** We discuss the prospects for LLP searches with upgraded detectors for Phase 2 of LHC running, with a focus on how upgrades can offer new sensitivity to LLPs as well as mitigate the effects of pile-up. New opportunities for improving sensitivity of triggers and searches to LLPs are additionally presented for the upgrades planned for Run 3. This is tied to the crucial question of triggers for LLPs; we discuss the performance of current triggers for LLPs, as well as the effects of future upgrades to the trigger system. Most importantly, we identify a few concrete upgrade studies that should be performed by the experiments that are of prime interest to the community.

[4] "LHC Long-Lived Particle Mini-Workshop", https://indico.cern.ch/e/LHC_LLP_2016

[5] "Searches for long-lived particles at the LHC: First workshop of the LHC LLP Community", https://indico.cern.ch/e/LHC_LLP_April_2017

[6] "Searches for long-lived particles at the LHC: Second workshop of the LHC LLP Community", https://indico.cern.ch/e/LHC_LLP_October_2017

[7] "Searching for long-lived particles at the LHC: Third workshop of the LHC LLP Community", https://indico.cern.ch/e/LHC_LLP_May_2018

[8] "Searches for long-lived particles at the LHC: Fourth workshop of the LHC LLP Community", https://indico.cern.ch/e/LHC_LLP_October_2018

- **Reinterpretation of LLP searches:** Due to the non-standard nature of the objects used in analyses, LLP searches are notoriously hard to reinterpret for models beyond those considered by the experimental collaborations. Designing searches and presenting search results in a way that is broadly applicable to current and yet-to-be-developed LLP models is crucial to the impact and legacy of the LLP search program. We discuss the reinterpretation of the LLP searches by means of concrete examples to illustrate specific challenges and, based on the lessons learned from this procedure, we provide recommendations on the presentation of LLP experimental search results.

- **Dark showers:** Current LLP search strategies have limited sensitivity to models where the LLPs are very soft, highly collimated, and come in large multiplicities, as can occur in models of dark-sector showers. We report on recent progress in theoretically parameterizing the space of dark-shower models and signatures, as well as experimental searches to uncover these signals.

Finally, we provide information about current and proposed experiments to search for LLPs at the LHC via dedicated detectors. These include the MoEDAL monopole search, the milliQan millicharged particle experiment, the MATHUSLA surface detector for ultra-LLPs, the CODEX-b proposal for a new detector near LHCb, and the FASER proposal for a long, narrow detector located in the forward direction well downstream one of the collision points.

This is the first report of the LHC LLP Community initiative, and is expected to be followed by future reports as our collective understanding of these signatures as a means of discovering new physics at the LHC evolves.

# 2
# Simplified Models Yielding Long-Lived Particles

## Contents

**Chapter editors:** James Beacham, Giovanna Cottin, David Curtin, Jared Evans, Zhen Liu, Michael Ramsey-Musolf, Jessie Shelton, Brian Shuve

**Contributors:** Oliver Buchmueller, Alessandro Davoli, Andrea De Simone, Kristian Hahn, Jan Heisig, Thomas Jacques, Matthew McCullough, Stephen Mrenna, Marco Trovato, Jiang-Hao Yu

Long-lived particles (LLPs) arise in many well-motivated theories of physics beyond the SM, ranging from heavily studied scenarios such as the minimal supersymmetric SM (MSSM) [8–12] to newer theoretical frameworks such as neutral naturalness [13–15] and hidden sector dark matter [16–21]. Macroscopic decay lengths of new particles naturally arise from the presence and breaking of symmetries, which can be motivated by cosmology (such as dark matter and baryogenesis) [22–35], neutrino masses [36–49], as well as solutions to the hierarchy problem [13–15, 50–55]; indeed, LLPs are generically a prediction of new hidden sectors at and below the weak scale [56–64]. An extensive and encyclopedic compilation of theoretical motivations for LLPs has already been performed for the physics case of the proposed MATHUSLA experiment [2], and we refer the reader to this document and the references therein for an in-depth discussion of theoretical motivations for LLPs. Given the large number of theories predicting LLPs, however, it is clear that a comprehensive search program for LLPs is critical to fully leverage the LHC's immense capability to illuminate the physics of the weak scale and beyond.

The simplified model framework has proven to be a highly successful approach to characterizing signals of beyond the SM (BSM) physics. Simplified models have driven the development of searches for new signatures at the LHC and allowed existing searches to be reinterpreted for many models beyond the one(s) initially targeted in the analysis. Comprehensive simplified model programs exist for scenarios featuring prompt decays of new particles [65–71] or dark matter produced at colliders [72–83]. Simplified models are so successful because the majority of search sensitivity is driven by only a few broad aspects of a given BSM signature, such as the production process, overall production rate, and decay topology. Meanwhile, the sensitivity of searches is typically insensitive to other properties such as the spin of the particles involved [84–87].

To extend the simplified model approach to LLP signatures in a systematic way, we develop a proposal for a set of simplified models which aims to ensure that experimental results can be characterized as follows: (i) *powerful*, covering as much territory in model space as possible; (ii) *efficient*, reducing unnecessary redundancy among searches; (iii) *flexible*, so that they are broadly applicable to different types of models; and (iv) *durable*, providing a common framework for Monte Carlo (MC) simulation of signals and facilitating the communication of results of LLP searches so that they may be applied to new models for years to come. We elaborate on these goals in Section 2.1. This framework helps illuminate gaps in coverage and highlight areas where new searches are needed, and we undertake such a study in Chapter 3. Our efforts build on earlier work proposing simplified model programs for LLPs motivated by particular considerations such as SUSY or dark matter (DM) [88–93].

In our work, we concentrate on establishing an initial basis of simplified models representative of theories giving rise to final states with one or two LLPs [1]. The simplified model approach is very powerful for LLP signatures: the typically lower backgrounds for displaced signatures allow searches to be highly inclusive with respect to other objects in the event or the identification of objects originating from the decay of an LLP. This enables a single analysis to have sensitivity to a wide variety of models for LLP production and decay.

[1] Some models predict moderately higher LLP multiplicities, but the coverage of such signatures from 1-2 LLP searches is good provided the LLPs do not overlap in the detector. Our proposed simplified models are not, however, representative of high-multiplicity signatures such as dark showers (see Section 2.6 and Chapter 7).

We organize our simplified models in terms of **LLP channels** characterized by a combination of a particular LLP production mode with a particular decay mode. Because the production and decay positions of LLPs are physically distinct [2], it is often possible to factorize and consider separately their production and decay [3]. For each LLP channel, the lifetime of the LLP is taken to be a free parameter. We emphasize that the LLP channel defined here is *not* the same as an experimental signature that manifests in the detector: a single channel can give rise to many different signatures depending on where (or whether) [4] the LLP decays occur inside the detector, while a single experimental search for a particular signature could potentially cover many simplified model channels. In this chapter, we focus on the construction and simulation of a concrete basis of LLP simplified model channels; a partial mapping of existing searches into our basis of simplified models is discussed in Chapter 3, along with the highest-priority gaps in current coverage and proposals for new searches.

[2] Indeed, the decay position may be so far from the collision point that external detectors can also be used to search for ultra-long-lived neutral or milli-charged particles [94–98].

[3] In addition to production and decay, a third consideration is the propagation of particles through the detector. While neutral LLPs undergo straightforward propagation, states with electric or color charge (*e.g.*, SUSY $R$-hadrons), or particles with exotic charges such as magnetic monopoles or quirks, typically engage in a more complicated and often very uncertain traverse through the detector. This spoils the factorization of LLP production and decay. The subtleties related to LLPs with electric or color charge is discussed more in Section 2.4.3. A trickier question is how to best simulate such states: since LLPs with electric or color charge interact with the detector material, there must be an interface between the detector simulation software and the program implementing decay. This is discussed further in Section 2.5.2.

[4] The case of detector-stable particles is understood to be included in the simplified models by setting $c\tau \to \infty$. In this case there is manifestly no dependence on the decay mode. See Section 2.3.2 for further details.

As discussed in the existing simplified model literature, simplified models have their own limited range of applicability [71, 79–81, 99]. For example, the presentation of search results in terms of simplified models often assume 100% branching fractions into particular final states. In a UV model where the LLP decays in a very large number of ways, none of the individual simplified model searches may be sufficient to constrain it. Similarly, if the LLP is produced in a UV model with other associated objects that spoil the signal efficiency (for example, the production of energetic, prompt objects collimated with the LLP such that the signal fails isolation or displacement criteria; this is particularly important for high-multiplicity or dark-shower scenarios, as discussed in Chapter 7), then the simplified model result does not apply and a more targeted analysis is required to cover the model. Nevertheless, the simplified models framework allows us to organize possible production modes and signatures in a systematic way and identify if there are any interesting signals or parts of parameter space that are missed by current searches.Therefore, we present a proposal for simplified models here with the understanding that there exist scenarios where UV models remain important for developing searches and presenting results.

The basis of simplified models presented here is a starting point, rather than a final statement. The present goal is to provide a set of simplified models that covers the majority of the best-motivated and simplest UV models predicting LLPs, which we outline in Section 2.2. Many of these contain singly and doubly produced LLPs (or in some cases, three-to-four relatively isolated LLPs, which are typically covered well by searches for $1-2$ LLPs) and so we restrict our simplified model proposal to cover these multiplicities. By design, simplified models do not include all of the specific details and subtle features that may be found in a given complete model. Therefore, the provided list is meant to be expanded to cover new or more refined models as the LLP-search program develops. For instance, extending the simplified model framework to separately treat final states with heavy-flavor particles is of great interest (in analogy with the prompt case [100–102]); see Section 2.6 for a discussion of this and other limitations of the current framework along with future opportunities for expansion. High-multiplicity signatures such as dark showers or emerging jets present different experimental and theoretical issues, which are discussed in Chapter 7. Finally, a broader set of simplified models may be needed to present the results of experimental searches and to allow ready application of experimental results to UV models of interest (see Chapter 6).

## 2.1 *Goals of the Present Simplified Model Framework*

The purpose of the simplified model framework is to provide a simple, common language that experimentalists and theorists can use to describe theories of LLPs and the corresponding mapping between models and experimental signatures. We therefore want our simplified model space to:

1. Use a minimal but sufficient set of models to cover a wide range of the best-motivated theories of LLPs;

2. Furnish a simple map between models and signatures to enable a clear assessment of existing search coverage and possible gaps;

3. Expand flexibly when needed to incorporate theories and signatures not yet proposed;

4. Provide a concrete MC signal event generation framework for signals;

5. Facilitate the reinterpretation of searches by supplying a sufficiently varied set of standard benchmark models for which experimental efficiencies can be provided for validation purposes.

Note that points #1 and #5 are somewhat in tension with one another: we wish to have a compact set of models that can be the subject of systematic study in terms of experimental signatures, but expressing experimental results in terms of only this set of simplified models may make it challenging to reinterpret experimental searches for UV models that are not precisely described by one of the simplified models. In this section, we prioritize having a minimal set of simplified models for the purpose of studying experimental coverages and generating new search ideas, while we defer a discussion of simplified models in the presentation and reinterpretation of search results to Chapter 6. [5]

In the remainder of this chapter, we construct a proposal for a minimal basis of simplified models for events with one or two LLPs. We begin with a discussion of the well-motivated UV theories that predict the existence of LLPs, and identify a set of umbrella models that yield LLPs in Section 2.2. We next identify the relevant (simplified) production and decay modes for LLPs in Section 2.3, emphasizing that each channel for production and decay has a characteristic set of predictions for the number and nature of *prompt* accompanying objects (AOs) producing along with the LLP. In Section 2.4, we combine these production and decay modes into our simplified model basis set and highlight how different umbrella models naturally populate the various LLP channels. Section 2.5 and Appendix A present a framework and instructions for how the best-motivated simplified model channels can be simulated in Monte Carlo (MC) using a new model library provided in Appendix A. Finally, limitations of the existing framework, along with opportunities for its further development are outlined in Section 2.6.

[5] We note that, in general, more benchmark models may be needed for enabling reliable reinterpretation than the minimal set discussed here. An example where an extended set of simplified models is used can be seen in the heavy stable charged particle (HSCP) reinterpretation in Section 6.3.2 (Table 6.1).

## 2.2 *Existing Well-Motivated Theories for LLPs*

Here we provide a brief distillation of many of the best-motivated theories with LLPs into five over-arching categories, focusing in particular on those that give rise to single and double production of LLPs at colliders. We emphasize that each of these categories is a broad umbrella containing many different individual models containing LLPs; in many cases, the motivations and model details among theories within a particular category may be very different, but tend to predict similar types of LLPs. Additionally, the categories are not mutually exclusive, with several examples of UV models falling into one or more category. In all cases, long lifetimes typically arise from some combination of hierarchies of scales in interactions that mediate decays; small couplings; and phase space considerations (such as small mass splittings between particles or large multiplicities of final-state particles in a decay). Many of the broad theoretical motivations for LLPs have recently been summarized in the literature [2].

The UV umbrella models we consider are:

- **Supersymmetry-like theories (SUSY).** This category contains models with multiple new particles carrying SM gauge charges and a variety of allowed cascade decays. Here LLPs can arise as a result of approximate symmetries (such as $R$-parity [52, 103, 104] or indeed SUSY itself in the case of gauge mediation [105]) or through a hierarchy of mass scales (such as highly off-shell intermediaries in split SUSY [106], or nearly-degenerate multiplets [56, 57, 107], as in anomaly-mediated SUSY breaking [58]). Finally, models of SUSY hidden sectors such as Stealth SUSY [51] generically lead to LLPs. Our terminology classifies any non-SUSY models with new SM gauge-charged particles, such as composite Higgs or extra-dimensional models, under the SUSY-like umbrella because of the prediction of new particles above the weak scale with SM gauge charges. In this category, LLP production is typically dominated by SM gauge interactions, whether of the LLP itself or of a heavy parent particle that de-cays to LLPs.

- **Higgs-portal theories (Higgs).** In this category, LLPs couple predominantly to the SM-like Higgs boson. This possibility is well motivated because the SM Higgs field provides one of the leading renormalizable portals for new gauge-singlet particles to couple to the SM, and the experimental characterization of the Higgs boson leaves much scope for couplings of the Higgs to BSM physics [108, 109]. The most striking signatures here are exotic Higgs decays to low-mass particles [110] (as in many Hidden Valley scenarios [59, 60]), which can arise in models of neutral naturalness [13, 14, 111] and DM [112], as well as in more exotic scenarios such as relaxion models [113]. The Higgs is also special in that it comes with a rich set of associated production modes in addition to the dominant gluon-fusion process, with vector-boson fusion (VBF) and Higgs-strahlung (VH) production modes allowing novel opportunities for triggering on and suppressing backgrounds to Higgs-portal LLP signatures. Indeed, in many scenarios where LLPs are produced in exotic Higgs decays, associated-production modes can be the only way of triggering on the event.

- **Gauge-portal theories (ZP).** This category contains scenarios where new vector mediators can produce LLPs. These are similar to Higgs models, although here the vector mediator is predominantly produced from $q\bar{q}$ initial states without other associated objects except for gluon initial-state radiation (ISR). Examples include models where both SM fermions and LLPs carry a charge associated with a new $Z'$ (for a review, see Ref. [114]), as well as either Abelian or non-Abelian dark photon or dark $Z$ models [115] in which the couplings of new vector bosons to the SM are mediated by kinetic mixing. Scenarios with LLPs coupled to new gauge bosons are well motivated by theories of DM, particularly models with significant self-interactions [116–118] and/or sub-weak mass scales [17, 18, 20, 119, 120].

- **Dark-matter theories (DM):** non-SUSY and hidden-sector DM scenarios are collected in this category, which encompasses models where the cosmological DM is produced as a final state in the collider process. Examples of multi-component DM theories include models of new electroweak multiplets [57, 121–123], strongly interacting massive particles (SIMPs) [124], inelastic dark matter [125–128], models with DM coannihilation partners [91, 129–134] (including scenarios where the coannihilation partners are out of chemical equilibrium, giving distinctly predictions for the relic abundance [135–138]), and non-thermal "freeze-in" scenarios [139–146]. In many of these models, the collider phenomenology and LLP lifetime can be tied to the DM relic abundance [128, 139, 147]. For LLPs decaying inside the detector, an important feature distinguishing this category from the Higgs and gauge scenarios above is that an explicit detector-level signature of a dark matter candidate, *i.e.*, missing energy ($\not{E}_T$), is a necessary and irreducible component [30, 59, 60, 93, 126, 128, 148–150].

- **Heavy neutrino theories (RH$\nu$):** the see-saw mechanism of SM neutrino mass generation predicts new right-handed neutrino (RHN) states [151–155]. If the RHNs have masses in the GeV to TeV range, they typically have a long lifetime and can be probed at the LHC [38, 43–45, 47, 48, 156–167]. Examples of well-motivated, UV-complete models with RHNs include the neutrino minimal SM ($\nu$MSM) [168, 169] and the left−right symmetric model [170–173]. Characteristic features of models in this category are LLPs produced singly via SM neutral- and charged-current interactions, and lepton-rich signatures in terms of prompt and displaced objects (often in association with quarks). For example, in extended scenarios like left−right symmetric models, production through new right-handed $W$ and $Z$ bosons can result in between one and four LLPs, and cascade decays between RHNs can lead to phenomena such as doubly displaced decays. Additionally, RHNs can be produced via Higgs decays [40, 41, 44, 163, 165, 174].

It is possible for a given model to fit into two or more of the umbrella UV model categories. For example, a SUSY theory with a stable lightest SUSY particle (LSP) could have the LSP serve as a dark matter candidate, while alternatively DM could be a new electroweak multiplet, giving rise to SUSY-like signatures [57, 121–123]. In other models featuring particles charged under a confining gauge group (such as "quirks" [175]), there can exist many production possibilities for the LLPs, including via the Higgs portal and the annihilation of new TeV-scale states (see, for example, Ref. [176]). Thus, the umbrella models should not be considered as exclusive categories, but rather as over-arching scenarios that motivate particular classes of signatures (such as new SM gauge-charged particles in the SUSY-like category, or presence of $\not{E}_T$ in DM models).

In developing our simplified model framework below, we construct maps between these UV model categories and the simplified model channels to illuminate some of the best-motivated combinations of production and decay modes for LLPs. This allows us to focus on the most interesting channels and assess their coverage in Chapter 3.

## 2.3 *The Simplified Model Building Blocks*

As discussed above, production and decay can largely be factorized in LLP searches [6]. This allows us to specify the relevant production and decay modes for LLP models separately; we then put them together and map the space of models into the umbrella categories of motivated theories.

[6] Once again, we comment that non-factorization of production and decay due to LLP interaction with the detector material and non-trivial propagation effects arise in models with LLPs with electric or color charge, and we discuss these subtleties further in Section 2.5.2

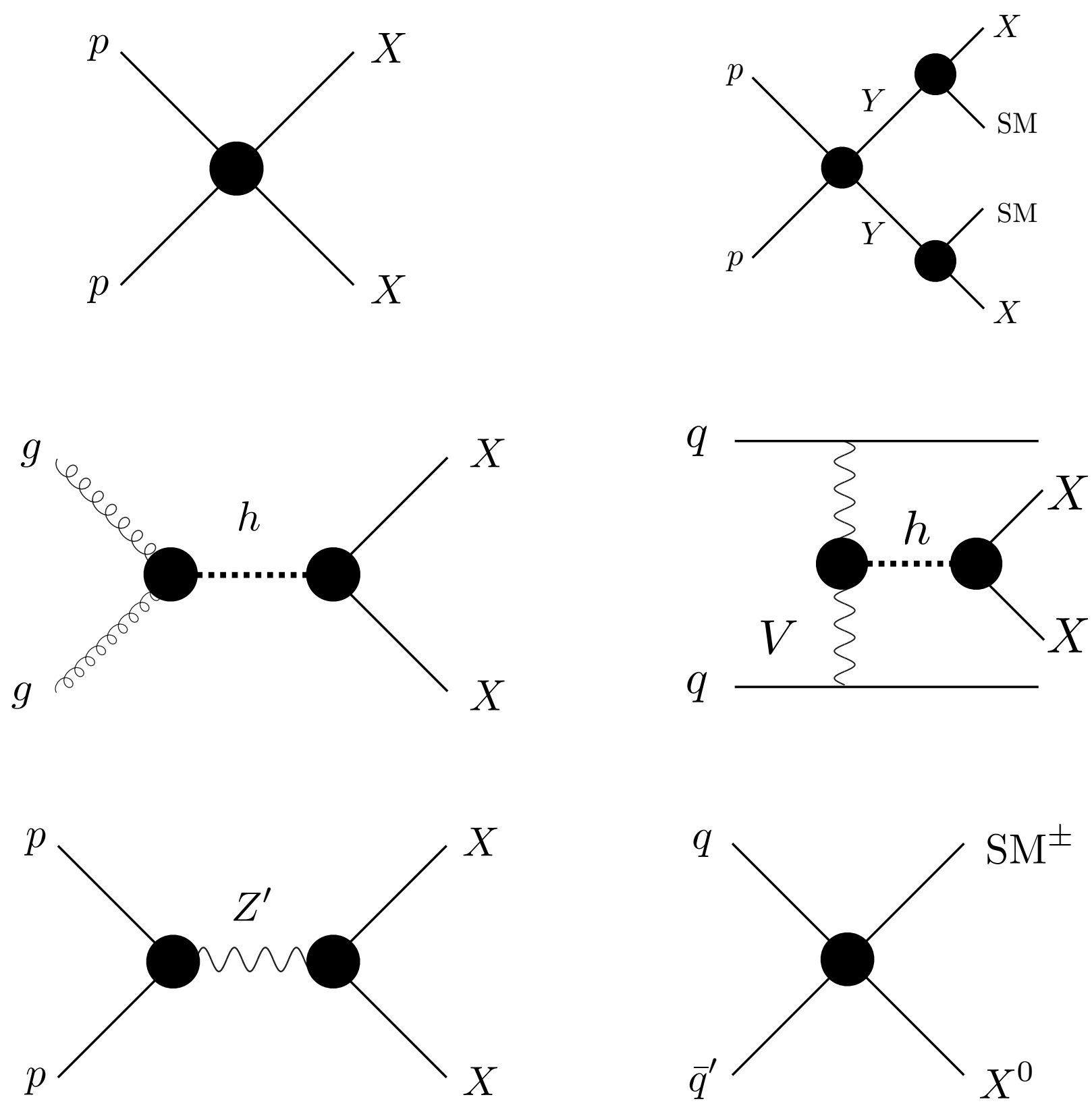

Figure 2.1: Schematic illustrations of LLP production modes in our simplified model framework. From top to bottom and left to right: direct pair production (DPP); heavy parent (HP); Higgs modes (HIG), including gluon fusion and VBF production (not shown here is $VH$ production); heavy resonance (RES); charged current (CC).

### 2.3.1 *Production Modes*

Motivated by our over-arching UV frameworks, we can identify a minimal set of interesting production modes for LLPs. Schematic diagrams for each production mode are shown in Figure 2.1. These production modes determine LLP signal rates both by relating the LLP production cross section to meaningful theory parameters such as gauge charges or Higgs couplings, and by determining the kinematic distribution of the LLP. Additionally, a given production mechanism also makes clear predictions for the number and type of *prompt* objects accompanying the LLP(s). These prompt AOs can be important for both triggering on events with LLPs and for background rejection, particularly when the LLP has a low mass or decays purely hadronically, and they can be either SM states (leptons, $\not{E}_T$, tagging jets) or BSM objects such as $Z'$ or dark photons [150, 177, 178].

- **Direct-Pair Production (DPP):** here the LLP is dominantly pair-produced non-resonantly from SM initial states. This is most straightforwardly obtained when the LLP is charged under a SM gauge interaction. In this case, an irreducible production cross section is then specified by the LLP gauge charge and mass. Such continuum DPP can also occur in the presence of a (heavy, virtual) mediator (*e.g.*, an initial quark—antiquark pair may exchange a virtual squark to pair produce bino-like neutralinos); in this case the production cross section is essentially a free parameter, as it is determined by the unknown heavy mediator masses and couplings.

- **Heavy Parent (HP):** the LLP is produced in the decays of on-shell heavy-parent particles that are themselves pair produced from the $pp$ initial state. The production cross section is essentially a free parameter, and is indirectly specified by the gauge charges and masses of the heavy parent particles. Heavy-parent production gives very different kinematics for the LLP than DPP, and often produces additional prompt AOs in the rapid cascade decays of the parents.

- **Higgs (HIG):** here the LLP is produced through its couplings to the SM-like Higgs boson. This case has an interesting interplay of possible production modes. The dominant production is via gluon fusion, which features no AOs beyond gluon ISR. Owing to its role in electroweak symmetry breaking, however, the Higgs has associated production modes (VBF, $VH$), each with its own characteristic features. The best prospects for discovery are for LLP masses below $m_h/2$, in which case the LLPs can be in decays of the on-shell SM-like Higgs boson. Higher-mass LLPs can still be produced via an off-shell Higgs, albeit at substantially lower rates [26, 179]. The LLP can be pair produced or singly produced through the Higgs portal depending on the model; an LLP $X$ can also be produced in association with $\not{E}_T$ via $h \to XX + \not{E}_T$ or $h \to X + \not{E}_T$. The cross section (or, equivalently, the Higgs branching fraction into the LLP) is a free parameter of the model. The Higgs mass can also be taken as a free parameter: there exist many theories that predict new exotic scalar states (such as the singlet-scalar extension to the SM [112]), and these new scalars can be produced in the same manner as the SM Higgs.

- **Heavy Resonance (RES):** here the LLP is produced in the decay of an on-shell resonance, such as a heavy $Z'$ gauge boson initiated by $q\bar{q}$ initial state. Note that production via an off-shell resonance is kinematically similar to the DPP mode. As with HIG, the LLP can be pair produced or singly produced (potentially in association with $\not{E}_T$). In RES models, ISR is the dominant source of prompt AOs. Models with new heavy scalars could conceivably fall into either RES or HIG; the main determining factor according to our organizational scheme is whether the scalar possesses Higgs-like production modes such as VBF and $VH$. Note that heavy resonance decays to SM particles also occur in these models, and searches for such resonances [180–186] may complement the sensitivity for decays to LLPs.

- **Charged Current (CC):** in models with weak-scale right-handed neutrinos, the LLP can be produced in the leptonic decays of $W/W'$. Single production is favored. Prompt charged leptons from the charged-current interaction are typical prompt AOs.

It is important to note that each of the above production mechanisms has its own "natural" set of triggers to record the signal. For example, HIG production can be accompanied by forward jets or leptons that are characteristic of VBF or $VH$ production. Similarly, CC production often results in prompt charged leptons, while HP production comes with AOs from the heavy-parent decay. However, the reader should be cautioned that this does not necessarily mean that the "natural" trigger is *optimal* for a particular signal. For example, the HIG modes suggest the use of VBF- or $VH$-based triggers, but if the LLP decays leptonically, it might be more efficient to trigger on the lepton from the LLP decay. Thus, the final word on which trigger is most effective for a given simplified model depends on the production mode as well as the nature and kinematics of the LLP decay. The prompt AOs of each production mode could still, however, be used to extend sensitivity to the model (see Section 6.3).

We also comment that some models may span several production modes. For example, a charged LLP that is part of an electroweak multiplet and nearly degenerate with a stable, neutral component [56–58, 92, 121–123, 187, 188] gives both DPP signatures (via $pp \to \chi^+\chi^-$) and CC production (via associated production $pp \to \chi^\pm\chi^0$). Comprehensive coverage of each of the above production modes will allow for a conservative determination of sensitivity for models that span many production modes.

### 2.3.2 *Decay Modes*

We now list a characteristic set of LLP decay modes. As we attempt to construct a minimal, manageable set of decay-mode building blocks, it is important to bear in mind that a given experimental search for LLPs can frequently be sensitive to a variety of possible LLP decay modes. As a result, it is not always necessary to perform separate searches for each possible decay mode as might otherwise be needed for prompt signatures.

The fact that LLP searches can be sensitive to many LLP decay modes is, in part, because LLPs that decay far from the collision point offer fewer avenues for particle identification. For example, for an LLP decaying inside of the calorimeter, most decay products are reconstructed as missing energy, or an energy deposition in the calorimeter. Consequently, particle identification criteria are typically relaxed in comparison to requirements on searches without displaced objects. Indeed, these "loose" collider objects can differ significantly from the corresponding "tight", prompt objects. This leads to more inclusive analyses that can cover a wider range of signatures with a single search.

Additionally, backgrounds for LLP searches are often small; for a comprehensive discussion of backgrounds to LLP searches, see Chapter 4. As a result, tight identification and/or reconstruction criteria typically found in exclusive prompt analyses are no longer needed to suppress backgrounds. For example, ATLAS has a displaced vertex search sensitive to di-lepton and multi-track vertices that is relatively inclusive with respect to other objects originating from near the displaced vertex [189]. Similarly, CMS has an analysis sensitive to events with one each of a high-impact-parameter muon and electron without reconstructing a vertex or any other objects [190]. For these examples, the backgrounds are sufficiently low that other requirements may be relaxed and the specific decay mode of the LLP may not be too important so long as certain objects (such as muons) are present or the decay occurs in a specific location. An even more extreme example in this regard is the search for highly-ionizing tracks sensitive to electrically and color-charged LLPs. While the searches are primarily targeted to detector-stable particles (heavy stable charged particles or $R$-hadrons) they can also be used to probe intermediate lifetimes for which only a certain fraction of LLPs traverse the tracker before decaying (see *e.g.* [135]). Both because of low backgrounds as well as modified particle identification criteria compared to prompt searches, LLP searches can often be inclusive and therefore covered by a more limited range of simplified models.

In some cases, however, the topology of a decay does matter. One potentially important factor that influences the sensitivity of a search to a particular model is whether the LLP decays into two SM objects vs. three, because the kinematics of multi-body decay are distinct from two-body decay and this may affect the acceptance of particular search strategies. An additional simplified model featuring a three-body decay of the LLP may consequently be needed to span the space of signatures.

Below, we describe an irreducible set of decay modes that can be used to characterize LLP signatures for various LLP charges (including neutral, electrically charged, and color charged). For each, we also provide an explicit example for how the decay would appear in a particular UV model. **We emphasize that the following decay modes are loosely defined with the understanding that their signatures are also representative of similar, related decay modes; for example $2j$ or $2j + \not{E}_T$ can also be proxies for $3j$ because searches for multi-body hadronic LLP decays can be sensitive to both and typically do not require reconstruction of a third jet.** It should also be noted that we are not recommending searches to be optimized to the exact, exclusive decay mode because that could suppress sensitivity to related but slightly more complicated LLP decays.

- **Di-photon decays:** the LLP can decay resonantly to $\gamma\gamma$ (like in Higgs-portal models or left−right symmetric models [191]) or to $\gamma\gamma$ + invisible (in DM models). This latter mode stands as a proxy for other $\gamma\gamma + X$ decays where the third object is not explicitly reconstructed, although whether $X$ is truly invisible can influence the triggers used. *Example: a singlino decaying to a singlet (which decays to $\gamma\gamma$) and a gravitino in Stealth SUSY [51].*

- **Single-photon decays:** the LLP decays to $\gamma$ + invisible (like in SUSY models). The SUSY model mandates a near-massless invisible particle, while other models (such as DM theories [127, 149]) allow for a heavy invisible particle. *Example: a bino decaying to photon plus gravitino in gauge-mediated models of SUSY breaking [192].*

- **Hadronic decays:** the LLP can decay into two jets ($jj$) (like in Higgs and gauge-portal models, or RPV SUSY), $jj$ + invisible (SUSY, dark matter, or neutrino models), or $j$ + invisible (SUSY). Here, "jet" ($j$) means either a light-quark parton, gluon, or $b$-quark. This category also encompasses decays directly into hadrons (for example, LLP decay into $\pi^+$ plus an invisible particle [56–58]). *Example: a scalar LLP decaying to $b\bar{b}$ due to mixing with the SM Higgs boson, as in models of neutral naturalness [13, 14, 111].*

- **Semi-leptonic decays:** the LLP can decay into a lepton + 1 jet (such as in leptoquark models) or 2 jets (like in SUSY or neutrino models). *Example: a right-handed neutrino decaying to a left-handed lepton and an on- or off-shell hadronically decaying $W$ boson (or $W'$ boson in a left−right symmetric model) [156].*

- **Leptonic decays:** the LLP can decay into $\ell^+\ell^-$(+invisible), or $\ell^\pm$ + invisible (as in Higgs-portal, gauge-portal, SUSY, or neutrino models). Here the symbol $\ell$ may be any flavor of charged lepton, but the decays are lepton flavor-universal and (for $\ell^+\ell^-$ decays) flavor-conserving. *Example: a wino decaying to a neutralino and an on- or off-shell leptonic $Z$ boson in SUSY [52].*

- **Flavored leptonic decays:** the LLP can decay into $\ell_\alpha$ + invisible, $\ell_\alpha^+\ell_\beta^-$ or $\ell_\alpha^+\ell_\beta^-$ + invisible where flavors $\alpha \neq \beta$ (as in SUSY or neutrino models). *Example: a neutralino decaying to two leptons and a neutrino in R-parity-violating SUSY [52]; or a right-handed neutrino decaying to two leptons and a neutrino [193].*

In all cases, both the LLP mass and proper lifetime are free parameters. Therefore, the case of detector-stable particles is automatically included by taking any of the above decay modes and taking the lifetime to infinity [7]. We emphasize that, depending on the location of the LLP within the detector, these decay modes may or may not be individually distinguishable: a displaced di-jet decay will look very different from a displaced di-photon decay in the tracker, but nearly identical if the decay occurs in the calorimeter. The goal

[7] As mentioned earlier, in the $c\tau \to \infty$ limit the decay mode becomes irrelevant. However, an exception is the search for particles that are stopped inside the detector material and decay out of time, which are discussed in Section 3.5.3.

here is to identify promising channels (as distinct from detector signatures).

As an example of how the above-listed decay modes cover the most important experimental signatures, we consider a scenario of an LLP decaying to top quarks. This scenario is very well motivated (for instance, with long-lived stops in SUSY) and might appear to merit its own decay category of an LLP decaying to one or more top quarks. However, the top quark immediately decays to final states that *are* covered in the above list, giving an effective semileptonic decay mode ($t \to b\ell^+\nu$) and a hadronic decay mode ($t \to bjj$) of the LLP. Similarly, LLP decays to four or more final states are typically covered by the above inclusive definitions of decay modes; this provides motivation not to over-optimize experimental searches to the specific, exclusive features of a particular decay mode.

While it would be ideal to have separate experimental searches for each of the above decay modes (when distinguishable), it is rare for specific models to allow the LLP to decay in only one manner; as in the example of an LLP decaying to a top quark, a number of decay modes typically occur with specific predictions for the branching fractions. As another example, if the LLP couples to the SM via mixing with the SM Higgs boson, then the LLP decays via mass-proportional couplings giving rise to $b$- and $\tau$-rich signatures. If, instead, the LLP decays through a kinetic mixing as in the case of dark photons or $Z$ bosons, then the LLP can decay to any particle charged under the weak interactions, giving rise to a relatively large leptonic branching fraction in addition to hadronic decay modes. This allows some level of prioritization of decay modes based on motivated UV-complete models; for example, the Higgs-portal model prioritizes searches for heavy-flavor quarks and leptons in LLP decay, while the gauge-portal model prioritizes searches for electrons and muons in LLP decay. Ultimately, however, it is desirable to retain independent sensitivity to each individual decay mode as much as possible. Indeed, for each decay mode listed above, models exist for which the given decay mode would be the main discovery channel.

**Invisible Final-State Particles:** where invisible particles appear as products of LLP decays, additional model dependence arises from the unknown mass of the invisible particle. The invisible particle could be a SM neutrino, DM, an LSP in SUSY, or another BSM particle. The phenomenology depends strongly on the mass splitting, $\Delta \equiv M_{\text{LLP}} - M_{\text{invisible}}$. If $\Delta \ll M_{\text{LLP}}$ (*i.e.*, $M_{\text{LLP}} \sim M_{\text{invisible}}$), the spectrum is compressed and the visible decay products of the LLP are soft. This could, for instance, lead to signatures such as disappearing tracks or necessitate the use of ISR jets to trigger on the LLP signature. If the mass splitting is large, $M_{\text{invisible}} \ll M_{\text{LLP}}$, then the signatures lose their dependence on the invisible particle mass.

We suggest three possible benchmarks: a compressed spectrum with $\Delta \ll M_{\text{LLP}}$ (example: a nearly degenerate chargino-neutralino pair, giving rise to soft leptons or disappearing tracks [56–58, 92, 121–123, 187, 188]); a massless invisible state, $\Delta = M_{\text{LLP}}$ (example: a next-to-lightest SUSY particle (NLSP) decaying to SM particles and a massless gravitino in gauge-mediated SUSY breaking [105, 194–199]); and an intermediate splitting corresponding to a democratic mass hierarchy, $\Delta \approx M_{\text{LLP}}/2$ (example: NLSPs in mini-split SUSY [54, 55, 89]).

## 2.4 *A Simplified Model Proposal*

In this section, we present a compact set of simplified model channels that, broadly speaking, covers the space of theoretical models in order to motivate new experimental searches. Such a minimal, compact set may not be optimal for reinterpretation of results (where variations on our listed production and decay modes may influence signal efficiencies and cross section sensitivities), but rather provides a convenient characterization of possible signals to ensure that no major discovery mode is missed. These models may therefore serve as a starting point for systematically understanding experimental coverage of LLP signatures and devising new searches, but may need to be extended in future for the purposes of facilitating reinterpretation. We undertake an in-depth discussion of these topics in Section 6.

We classify LLPs according to their SM gauge charges, as these dictate the dominant or allowed LLP production and decay modes, and can give rise to different signatures (for example, disappearing tracks and hadronized LLPs). We separately consider LLPs that are: (a) neutral; (b) electrically charged but color neutral; and (c) color charged. In the latter case, it is important to distinguish between the long-lived parton (which carries a charge under quantum chromodynamics, QCD) that hadronizes prior to decay, and the physical LLP, which is a color-singlet "$R$-hadron" (using the standard nomenclature inspired by SUSY). The decays of the $R$-hadron are still dominated by the parton-level processes.

All of the following models have the LLP mass and lifetime as free parameters. For heavy-parent (HP) production, the parent mass is an additional parameter, while for invisible decays, several different benchmarks for mass splittings between LLP and invisible final state may have to be separately considered as described in Section 2.3.2. The cross section may have a theoretically well-motivated target value depending on UV-model parameters, but phenomenologically can generally be taken as a free parameter.

We emphasize that in spite of the many simplified model channels proposed below, a small number of experimental LLP searches can have excellent coverage over a wide range of channels (at least for certain lifetime ranges). The list is intended to be comprehensive in order to identify whether there are new searches that could have a similarly high impact on the space of simplified models, and identify where the gaps in coverage are.

### 2.4.1 Neutral LLPs

The simplified model channels for neutral LLPs are shown in Table 2.1, where $X$ indicates the LLP.

In our initial proposal, which is the first iteration of the simplified model framework, it is sufficient to consider as "jets" all of the following: $j = u, d, s, c, b, g$. It is worth commenting that $b$-quarks pose unique challenges and opportunities. Since $b$-quarks are themselves LLPs, they appear with an additional displacement relative to the LLP decay location. They also often give rise to soft muons in their decays, which could in principle lead to additional trigger or selection possibilities. However, these subtleties can be addressed in further refinements of the simplified models; we discuss this further in Section 2.6. Similarly, we consider $e$, $\mu$, and $\tau$ to be included in the broad category of "leptons", with the proviso that searches should be designed where possible with sensitivity to each.

When multiple production modes are specified in one row of the table, this means that multiple especially well-motivated production channels give rise to similar signatures. Typically only one of these simplified model production modes will actually need to be included when developing and assessing sensitivity of an experimental search, but we sometimes include multiple different production modes as individuals may variously prefer one over the other.

In each entry of the table, we indicate which umbrella category of well-motivated UV models (Section 2.2) can predict a particular (production) $\times$ (decay) mode. An asterisk (*) on the umbrella model indicates that $\not{E}_T$ is required in the decay. A dagger ($^\dagger$) indicates that this particle production $\times$ decay scenario is not present in the *simplest and most minimal* implementations or spectra of the umbrella model, but could be present in extensions of the minimal models. While the HIG production signatures are best-motivated for the SM-like 125 GeV Higgs, exotic Higgses of other masses can still have the same production modes and so $m_H$ can be taken as a free parameter.

We remind the reader that the production modes listed in Table 2.1 encompass also the associated production of characteristic prompt objects. For example, the Higgs production modes not only proceed through gluon fusion, but also through VBF and $VH$ production, each of which results in associated prompt objects such as forward jets in VBF, and leptons or $\not{E}_T$ in $VH$. All of the production modes listed in Table 2.1 could be accompanied by ISR jets that aid in triggering or identifying signal events. It is therefore important that searches are designed to exploit such prompt AOs whenever they can improve signal sensitivity, especially with regard to triggering.

To demonstrate how to map full models onto the list of simplified models (and vice-versa), we consider a few concrete cases. For instance, if we consider a model of neutral naturalness where $X$ is a long-lived scalar that decays via Higgs mixing (for instance, $X$ could be the lightest quasi-stable glueball), then the process where the SM Higgs $h$ decays via $h \to XX$, $X \to b\bar{b}$ would be covered with the HIG production mechanism and a di-jet decay. Entirely unrelated models, such as the case where $X$ is a bino-like neutralino with RPV decays $h \to XX$, $X \to jjj$ could be covered with the same simplified model because most hadronic LLP searches do not have exclusive requirements on jet multiplicity. Similarly, a hidden-sector model with a dark photon, $A'$, produced in $h \to A'A'$, $A' \to f\bar{f}$ would also give rise to the di-jet signature when $f$ is a quark, whereas it would populate the $\ell^+\ell^-$ column if $f$ is a lepton. Finally, a scenario with multiple hidden-sector states $X_1$ and $X_2$, in which $X_2$ is an LLP and $X_1$ is a stable, invisible particle, could give rise to signatures like $h \to X_2X_2$, $X_2 \to X_1 jj$ that would be covered by the same HIG production, hadronic-decay simplified model; however, we see how $\not{E}_T$ can easily appear in the final state, and that the LLP decay products may not be entirely hadronic. Therefore, the simplified models in Table 2.1 can cover an incredibly broad range of signatures, but only if searches are not overly optimized to particular features such as $\not{E}_T$ and LLPs decaying entirely visibly (which would allow reconstruction of the LLP mass) [8].

| Production \ Decay | $\gamma\gamma(+\text{inv.})$ | $\gamma + \text{inv.}$ | $jj(+\text{inv.})$ | $jj\ell$ | $\ell^+\ell^-(+\text{inv.})$ | $\ell^+_\alpha \ell^-_{\beta\neq\alpha}(+\text{inv.})$ |
|---|---|---|---|---|---|---|
| DPP: sneutrino pair or neutralino pair | † | SUSY | SUSY | SUSY | SUSY | SUSY |
| HP: squark pair, $\tilde{q} \to jX$ or gluino pair $\tilde{g} \to jjX$ | † | SUSY | SUSY | SUSY | SUSY | SUSY |
| HP: slepton pair, $\tilde{\ell} \to \ell X$ or chargino pair, $\tilde{\chi} \to WX$ | † | SUSY | SUSY | SUSY | SUSY | SUSY |
| HIG: $h \to XX$ or $\to XX + \text{inv.}$ | Higgs, DM* | † | Higgs, DM* | RH$\nu$ | Higgs, DM* RH$\nu$* | RH$\nu$* |
| HIG: $h \to X + \text{inv.}$ | DM*, RH$\nu$ | † | DM* | RH$\nu$ | DM* | † |
| RES: $Z(Z') \to XX$ or $\to XX + \text{inv.}$ | $Z'$, DM* | † | $Z'$, DM* | RH$\nu$ | $Z'$, DM* | † |
| RES: $Z(Z') \to X + \text{inv.}$ | DM | † | DM | RH$\nu$ | DM | † |
| CC: $W(W') \to \ell X$ | † | † | RH$\nu$* | RH$\nu$ | RH$\nu$* | RH$\nu$* |

Table 2.1: **Simplified model channels for neutral LLPs.** The LLP is indicated by $X$. Each row shows a separate production mode and each column shows a separate possible decay mode, and therefore every cell in the table corresponds to a different simplified model channel of (production)$\times$(decay). We have cross-referenced the UV models from Section 2.2 with cells in the table to show how the most common signatures of complete models populate the simplified model space. The asterisk (*) shows that the model definitively predicts missing energy in the LLP decay. A dagger (†) indicates that this particle production $\times$ decay scenario is not present in the *simplest and most minimal* implementations or spectra of the umbrella model, but could be present in extensions of the minimal models. When two production modes are provided (with an "or"), either simplified model can be used to simulate the same simplified model channel.

[8] This should not, of course, be interpreted as saying that searches shouldn't be done that exploit these features. Instead, our position is that experiments should bear in mind the range of topologies and models covered by each cell in Table 2.1 when designing searches, and that some more inclusive signal regions should be established where possible.

### 2.4.2 *Electrically Charged LLPs:* $|Q| = 1$

For an electrically charged LLP, we need to consider far fewer production modes because of the irreducible gauge production associated with the electric charge. We still consider the additional possibility of a HP scenario where the parent has a QCD charge, as this could potentially dominate the production cross section, see *e.g.*, Ref. [88]. We summarize our proposals in Table 2.2.

Note that we group all resonant production into the $Z'$ simplified model. The reason is that the SM Higgs cannot decay into two on-shell charged particles due to the model-independent limits from LEP on charged particle masses, $M \gtrsim 75-90$ GeV (see, for example, Ref. [200]); because of this lower limit on the LLP mass, it is less important to use AOs for triggering and reconstructing charged LLP signatures than for neutral LLPs. Additionally, there are fewer allowed decay modes because of the requirement of charge conservation.

For concreteness, we recommend using $|Q| = 1$ as a benchmark for charged LLPs for the purpose of determining allowed decay modes. Although other values of $Q$ are possible, these often result in cosmologically stable charged relics or necessitate different decay modes than those listed here. Additionally, LLPs with $|Q| = 1$ are motivated within SUSY [56–58, 201–203] and within Type-III seesaw models of neutrino masses [204–207]. We note that there exist already dedicated searches for heavy quasi-stable charged particles with non-standard charges [208, 209]. Because such searches are by construction not intended to be sensitive to the decays of the LLP, the existing models are sufficient for characterizing these signatures and they do not need to be additionally included in our framework.

For massive particles with $|Q| = 1$ with intermediate or large lifetimes such that the LLP traverses a significant part (or all) of the tracker, the highly ionizing track of the LLP provides a prominent signature. This can be exploited for an efficient suppression of backgrounds while keeping identification and/or reconstruction criteria as loose and, hence, as inclusive as possible. In particular, for decay-lengths of the order of or larger than the detector size, the signature of highly ionizing tracks and anomalous time of flight (*i.e.*, searches for heavy stable charged particles; see Sections 3.5 and 6.4.1) constitute an important search strategy covering a large range of lifetimes present in the parameter space of theoretically motivated models. While the searches for heavy stable charged particles are largely inclusive with respect to additional objects in the event, they depend strongly on the velocity of the LLP. For $\beta \to$

| Production \ Decay | $\ell+$inv. | $jj(+$inv.$)$ | $jj\ell$ | $\ell\gamma$ |
|---|---|---|---|---|
| DPP: chargino pair or slepton pair | SUSY DM* | SUSY DM* | SUSY | † |
| HP: $\tilde{q} \to jX$ | SUSY DM* | SUSY DM* | SUSY | † |
| RES: $Z' \to XX$ | Z′, DM* | Z′, DM* | Z′ | † |
| CC: $W' \to X+$inv. | DM* | DM* | RH$\nu$ | † |

Table 2.2: **Simplified model channels for electrically charged LLPs** such that $|Q| = 1$. The LLP is indicated by $X$. Each row shows a separate production mode and each column shows a separate possible decay mode, and therefore every cell in the table corresponds to a different simplified model channel of (production)×(decay). We have cross-referenced the UV models from Section 2.2 with cells in the table to show how the most common signatures of complete models populate the simplified model space. The asterisk (*) shows that the model definitively predicts missing energy in the LLP decay. A dagger (†) indicates that this particle production × decay scenario is not present in the *simplest and most minimal* implementations or spectra of the umbrella model, but could be present in extensions of the minimal models. When two production modes are provided (with an "or"), both production simplified models can be used to cover the same experimental signatures.

1 one loses the discriminating power against minimally ionizing particles, while for small velocities, $\beta \lesssim 0.5$, the reconstruction becomes increasingly difficult due to timing issues. It is therefore important to include the heavy parent production scenario which covers a much larger kinematic range than direct production alone and which may feature a much wider range of signal efficiencies than the DPP scenario [90].

While the signatures in Table 2.2 form a minimal set, they also encompass some scenarios that merit special comment. One of these is the disappearing track signature [56–58, 92, 121–123, 187, 188], in which a charged LLP decays to a nearly degenerate neutral particle. The lifetime is long in this scenario due to the tiny mass splitting between the two states. Formally, these are included in the chargino or slepton DPP modes in Table 2.2 with decays to $\ell+$ inv. or $q\bar{q}'+$ inv. taken in the limit where the splitting between the charged LLP and the invisible final state is of $\mathcal{O}(200\text{ MeV})$. In the case of a hadronic decay, $X$ decays to a soft pion that is very challenging to reconstruct and so the track simply disappears. This is an important scenario that is already the topic of existing searches [210, 211]. As the degeneracy between the charged LLP and the neutral state is relaxed, other signatures are possible; this parameter range is well motivated both by SUSY and DM models with coannihilation [91, 129, 130].

| Production \ Decay | $j+\text{inv.}$ | $jj(+\text{inv.})$ | $j\ell$ | $j\gamma$ |
|---|---|---|---|---|
| DPP: squark pair or gluino pair | SUSY | SUSY | SUSY | † |

Table 2.3: **Simplified model channels for LLPs with color charge.** The LLP is indicated by $X$. Each row shows a separate production mode and each column shows a separate possible decay mode, and therefore every cell in the table corresponds to a different simplified model channel of (production)×(decay). We have cross-referenced the UV models from Section 2.2 with cells in the table to show how the most common signatures of complete models populate the simplified model space. A dagger (†) indicates that this particle production × decay scenario is not present in the *simplest and most minimal* implementations or spectra of the umbrella model, but could be present in extensions of the minimal models. When two production modes are provided (with an "or"), both production simplified models can be used to cover the same experimental signatures.

Finally, we comment on the challenges of simulating the charged LLP simplified models. Because the LLP bends and interacts with detector material prior to its decay, the simulation of the LLP propagation is important in correctly modeling the experimental signature. The subsequent decay of the LLP must either be hard-coded into the detector simulation, or allow for an interface with programs such as Pythia 8 to implement the decays. We discuss the challenges of simulating signals for LLPs with electric or color charge in Section 2.5.2.

### *2.4.3 LLPs with Color Charge*

An LLP charged under QCD is more constrained than even electrically charged LLPs. Because of the non-Abelian nature of the strong interactions, the gauge pair-production cross section of the LLP is specified by the LLP mass and its representation under the color group, $SU(3)_C$. We do not consider LLP production via a heavy parent particle because that cross section is unlikely to dominate the total production rate at the LHC relative to DPP. The simplified model channels are provided in Table 2.3.

A complication of the QCD-charged LLP is that the LLP hadronizes prior to its decay, forming an $R$-hadron bound state. The modeling of hadronization and subsequent propagation is directly related to many properties of the long-lived parton, such as electric charge, flavor, and spin. Event generators such as Pythia 8 have routines [212, 213] to simulate LLP hadronization, although it is unclear how precise these predictions are. For a point of comparison, using the default settings of Pythia 8 yields an estimate of the neutral $R$-hadron fraction from a gluino (color-octet fermion, $\tilde{g}$) of approximately 54%, while the neutral $R$-hadron fraction for a stop (scalar top partner) is estimated to be 44% [89]. After hadronization, the charge of the $R$-hadron may change as it passes through the detector. For instance, some estimates [214, 215] suggest that heavy, color-octet gluinos $\tilde{g}$ would predominantly form mesons (*e.g.*, $(u\tilde{g}\bar{d})$) at first. They eventually drop to the lower-energy neutral singlet baryon $\tilde{\Lambda} = (\tilde{g}uds)$ state when interacting with the protons and neutrons within the calorimeters.

The modeling of LLP hadronization and propagation is crucial to designing searches for color-charged LLPs and assessing their sensitivity. For example, only the charged $R$-hadrons can be found in heavy stable charged particle search; if the LLP charge changes as it passes through the detector, heavy stable charged particle searches may have limited sensitivity. To take this into account, the experimental searches include both tracker-only or tracker+calorimeter signal regions [4, 216], which enhances sensitivity to the scenario in which $R$-hadrons lose their charge by the time they reach the calorimeters.

Because no $R$-hadrons have been discovered to date and hence their properties cannot be directly measured, $R$-hadron modeling in detector simulations is challenging. We discuss the challenges of simulating the propagation and decays of LLPs with color charge in Section 2.5.2.

## 2.5 *Proposal for a Simplified Model Library*

The simplified models outlined in the above sections provide a common language for theorists and experimentalists to study the sensitivity of existing searches, propose new search ideas, and interpret results in terms of UV models. Each of these activities demands a simple framework for the simulation of signal events that can be used to evaluate signal efficiencies of different search strategies and map these back onto model parameters. Requiring individual users to create their own MC models for each simplified model is impractical, redundant, and invites the introduction of errors into the analysis process.

In this section, we propose and provide a draft version of a *simplified model library* consisting of model files and MC generator cards that can be used to generate events for various simplified models in a straightforward fashion. Because each experiment uses slightly different MC generators and settings, this allows each collaboration (as well as theorists) to generate events for each simplified model based on the provided files. Depending on how the LLP program expands and develops over the next few years, it may become expedient to expand the simplified model library to include sets of events in a standard format (such as the Les Houches format [217]) that can be directly fed into event-generator and detector-simulation programs. Given the factorization of produc-tion and decay of LLPs that is valid for neutral LLPs, this could involve two mini-libraries: a set of production events for LLPs and a set of decay events for LLPs, along with a protocol for "stitching" the events together.

The current version of the library is available at the LHC LLP Community website [9], hosted at CERN. In Appendix A, we also provide tables that list how to simulate each LLP simplified model channel with one of the specified base models. These proposals are based on the models outlined in Section 2.5.1 and often match the best-motivated simplified models from Section 2.4, and also building on the DM-inspired LLP simplified models proposed and detailed in Ref. [93]. The library currently focuses on models of neutral LLPs; simulating the propagation of charged LLPs along with the full range of decays listed in Sections 2.4.2—2.4.3 requires more careful collaboration with detector simulation and other MC programs to ensure that they can practically be used in experimental studies.

[9] http://cern.ch/longlivedparticles

We provide model files in the popular Universal Feynrules Output (UFO) format [218], which is designed to interface easily with parton-level simulation programs such as MadGraph5_aMC@NLO [219]. The goal is to cover as many of the simplified models of Section 2.4 with as few UFO models as possible; this limits the amount of upkeep needed to maintain the library and develops familiarity with the few UFO models needed to simulate the LLP simplified models. We provide specific instructions for how to simulate each simplified LLP channel along with the UFO models.

### 2.5.1 *Base Models for Library*

In order to reproduce the simplified model channels of Section 2.3, we need a collection of models that:

- Includes additional gauge bosons and scalars to allow vector- and scalar-portal production of LLPs (RES and HIG);
- Includes new gauge-charged fermions and scalars to cover direct and simple cascade production modes of LLPs (DPP and HP);
- Includes a RHN-like state with couplings to SM neutrinos and leptons (CC);
- *Recommended, but optional:* allows for the decays of the LLP particle through all of the decay modes listed in Section 2.3, either through renormalizable or higher-dimensional couplings. If couplings that allow LLP decay are included in the UFO model, then the decays can be performed directly at the matrix-element level in programs such as MadGraph5_aMC@NLO [219] and accompanying packages such as MadSpin [220]. Alternatively, it is possible for neutral LLPs to simulate the production and decay as a single process; in such cases, numerical instabilities sometimes arise, for which dedicated event generators are needed [48]. If the couplings needed for LLP are not in the UFO model, then LLPs can be left stable at the matrix-element level and decays implemented via Pythia 8 [212, 213], which allows for the straightforward implementation of decays according to a phase-space model, but does not correctly model the angular distribution of decay products. Instructions for implementing decays in Pythia are included with the model library files.

Fortunately, an extensive set of UFO models is already available for simulating the production of BSM particles. We note that extensions or generalizations of only three already-available UFO models are needed at the present time; the SUSY models in particular can cover many of the simplified models since they contain an enormous collection of new fermions and scalars. We also provide an optional fourth model, the Hidden Abelian Higgs Model, that can be helpful to simulate HIG and ZP theories.

1. **The Minimally Supersymmetric SM (MSSM):** the use of this model is motivated by and allows for the simulation of SUSY-like theories. The model contains a whole host of new particles with various gauge charges and spins. Therefore, an MSSM-based model allows for the simulation of many of the simplified model channels. In particular, we note that existing UFO variants of the MSSM that include gauge-mediated supersymmetry breaking (GMSB) couplings (including decays to light gravitinos), $R$-parity violation (making unstable the otherwise stable LSPs [52, 103, 104]), and the phenomenological MSSM (pMSSM) [221, 222] already cover most of the SUSY-motivated LLP scenarios. In some cases, the model is modified to give direct couplings between the Higgs states and gluons/photons.

2. **The Left−Right Symmetric Model (LRSM):** this UFO model is best for simulating UV theories with right-handed neutrinos (RH$\nu$). The UFO model supplements the SM by an additional $SU(2)_R$ symmetry, which gives additional charged and neutral gauge bosons. The model is available in the simplified models library and contains a right-handed neutrino which is the typical LLP candidate. The LLP can be produced via SM $W$, $Z$, or via the new gauge and Higgs bosons (both charged and neutral) present in the theory [10]. The LRSM therefore contains many of the charged and neutral current LLP production processes outlined in Section 2.4.1.

3. **Dark-Matter Simplified Models (DMSM)**: these UFO models are best for simulating UV theories in the DM class. These UFO models have been created by the LHC DM working group [79]. They typically consist of a new BSM mediator particle (such as a scalar of a $Z'$) coupled to invisible DM particles. The UFO models can either be modified to include an unstable LLP, or else the otherwise stable "DM" particle can be decayed via Pythia. The utility and applicability of the DM simplified model framework to LLPs has already been demonstrated with a detailed proposal and study of classes of DM simplified models for LLPs [93]. These models are particularly good for simulating LLP production via a heavy resonance (RES), and can also simulate continuum production of LLPs in the limit where the mediator is taken to be light and off-shell (DPP).

[10] Additional LRSM tools are available at https://sites.google.com/site/leftrighthep/.

4. **(optional) The Hidden Abelian Higgs Model (HAHM):** this UFO model contains new scalars and gauge bosons and so can be used to simulate both Higgs-portal and gauge-portal (ZP) theories. The model consists of the SM supplemented by a "hidden sector" consisting of a new $U(1)$ gauge boson and a corresponding Higgs field. The physical gauge and Higgs bosons couple to the SM via kinetic and mass mixing, respectively. The HAHM allows for straightforward simulation of Higgs-portal production of LLPs, as well as $Z'$ models and many hidden sector scenarios. The UFO implementation is from Ref. [223].

If additional decay modes are needed beyond those in the specified simplified models, then the library can be updated to include the new couplings mediating the decay. Alternatively, the LLPs can be left stable at parton level and decayed in event generators such as Pythia.

A detailed list of processes that can be used to simulate each simplified model channel is provided in Appendix A. The primary purpose of the library is to be used to simulate events for determining acceptances, and, as a result, the signal cross section is not important. Thus, for example, SM gauge interactions can be used as proxies for much weaker exotic interactions. Similarly, the spins of the particles are generally of subdominant importance: replacing the direct production of a fermion with the direct production of a scalar will not fundamentally alter the signature. As long as results are expressed in terms of sensitivity to cross sections and not couplings, the results can be qualitatively (and in many cases, quantitatively) applied to any similar production mode regardless of spin. However, we caution the reader that changing the spin of the LLP (or its parent) can change the angular distribution, and since in some cases LLP searches are typically more sensitive to aspects of event geometry than prompt searches, the second-order effects of spin could have more of an effect than for prompt simplified models.

### 2.5.2 *LLP Propagation and Interaction with Detector Material*

Long-lived particles with electric or QCD charges interact with the detector material prior to decay, and their propagation through the detector must be correctly modeled. The propagation of both LLPs with color charge (in the form of $R$-hadrons) and electrically charged LLPs can be implemented in the Geant4 (G4) toolkit [224]. For example, routines exist to simulate the propagation of color-charged LLPs [225, 226]. G4 also includes routines that can implement $N$-body decays of LLPs using a phase-space model. This works fine for decays of LLPs to leptons, photons, invisible particles such as neutrinos, as well as exclusive hadronic decays.

However, G4 cannot implement decays to partons that subsequently shower and hadronize. One solution to this limitation is employed by CMS [227, 228] and ATLAS [229] in their searches for stopped LLPs. In these analyses, the signal simulation proceeds in two stages. During the first stage, the production of the LLP and its subsequent interactions with the detector are simulated. Once the stopping point of the LLP is determined, a new event is simulated including the LLP decay; the LLP decay products are then manually moved to the stopping point from the first stage. G4 is then run a second time to determine the efficiency for reconstructing the LLP decay signal.

It would be preferable to fully automate the simulation of decays of charged LLPs after propagation in G4. There exists in G4 a class called G4ExtDecayer, which can be used to implement decays by interfacing with an external generator. This class has been used to interface G4 with Pythia 6 [11]. The interface with Pythia 6 has been used most recently to model LLP gluino propagation and decay in a search for displaced vertices and missing energy in ATLAS [230]. Work is ongoing to extend this functionality to Pythia 8 and to simplify the interface.

[11] See http://geant4-userdoc.web.cern.ch/geant4-userdoc/Doxygen/examples_doc/html/Exampledecayer6.html

An additional challenge of simulating LLP decays is that, if the LLP undergoes a multi-body decay, generators such as Pythia use a phase-space model to implement the decays. If more accuracy is required, it may be preferable to use the full matrix element via generators such as MadGraph5 [219, 231]. If the matrix element is important for computing the decay of the LLP, then either an interface with MadGraph is needed to implement the decay prior to passing the vertex back to Pythia 8 for showering and hadronization, or matrix-element-based methods within the event generator itself must be used.

Because of the need to interface with G4 in simulating the decays of LLPs with electric or color charges, we do not at this point include such decay modes in our simplified model library. The decays of such LLPs will be most easily simulated via an interface with Pythia 8 once it is finalized.

Finally, we comment that LLPs can have even stranger propagation properties than LLPs with electric or color charges. For example, quirks are LLPs that are charged under a hidden-sector gauge interaction that confines at macroscopic scales [175]. Because the confinement scale can be just about any distance, quirks can have very unusual properties; as a specific example, if electrically charged quirk-antiquirk pairs are bound on the millimeter or centimeter level, they behave as an electric dipole and therefore do not leave conventional tracks that bend in the magnetic field. Other confinement scales give rise to different behaviors, such as meta-stable heavy charged particles and non-helical tracks [232, 233]. In scenarios where the quirks carry color charge, the quirks hadronize and can undergo charge-flipping interactions as they move through the detector. These quirk scenarios can be challenging to model, and no public code exists that allows for the propagation and interaction of quirks with the detector material; we encourage the collaborations to validate and release any internal software they may have to study the propagation of quirks (for more discussion, see the discussion of quirks in Section 3.5) [12].

[12] Ideally, this software would be well-documented to facilitate sharing between experiments. A successful example of readily shareable software between experiments is the G4 package for $R$-hadrons and other particles' interaction with matter, found at http://r-hadrons.web.cern.ch/r-hadrons/

## 2.6 *Limitations of Simplified Models & Future Opportunities*

We conclude our discussion of simplified models with a more extensive discussion of the limitations of the current simplified model proposal in its application to models of various types, along with opportunities for future development. The presented framework is only the first step of a simplified model program that is comprehensive in terms of generating LHC signatures and allowing straightforward reinterpretation of experimental results for UV models. The framework we have developed with separate, modular components for LLP production and decay is amenable to expansion, and we encourage members of the theory and experimental communities to continue to do so over the coming years to ensure maximal utility of the simplified models framework.

One significant simplification we have undertaken in our framework is to define a "jet" as any of $j = u, c, d, s, b, g$. In reality, different partons give rise to different signatures, especially when one of the "jets" is a heavy-flavor quark. Jets initiated by $b$ and $c$ quarks have some useful distinguishing features, such as the fact that the underlying heavy-flavor meson decays at a distance slightly displaced from the proton interaction vertex and that there are often associated soft leptons resulting from meson decays. In particular, it is possible that the soft muons associated with $B$-meson decays could be used to enhance trigger and reconstruction prospects for LLPs decaying to $b$-jets [234]. However, heavy quarks also constitute an important backgrounds for LLP searches, and so LLPs decaying to $b$- and $c$-jets may necessitate dedicated treatment in future. Similarly, LLP decays to $\tau$ leptons may merit further specialized studies.

Another property of the current framework is that it is restricted to LLP signatures of low multiplicity. By "low multiplicity", we mean collider signatures with one or two LLPs. Searches inspired by these models are also suitable for many scenarios with three or four LLPs per event (which include models with dark-Higgs decays into lepton-jets [148], or left−right symmetric models [162]), since the LLP signatures are generally extremely rare and so only one or two typically need to be identified in a given event to greatly suppress backgrounds. Thus, as long as the search is inclusive with respect to possible additional displaced objects, the signature can be covered with low-multiplicity strategies. As the LLP multiplicity grows, however, the simplified model space we have presented requires modification. This is both because the individual LLPs grow softer, making them harder to reconstruct on an individual level, and they become less separated in the detector, which makes isolation and identification of signal a challenge. On the other hand, the high LLP multiplicity may allow for new handles for further rejecting backgrounds, and the kinematics can vary widely based on the model (for example, in some "quirky" scenarios, LLPs can be produced in a variety of ways with different kinematic distributions [113]). In extreme cases, signals can even mimic pile-up [235]. High-multiplicity signatures therefore require dedicated modeling, and we defer the study of these signatures to Chapter 7.

Finally, we conclude by noting that simplified models are intended to provide a general framework to cover a broad swath of models. Any simplified model set-up, however, cannot cover every single UV model without becoming as complex as the UV model space itself. As with the case of promptly decaying new particles, care must also be taken in the interpretation of simplified models [71, 79–81, 99]: for example, constraints on simplified models assuming 100% branching fractions of LLPs to a particular final state may not accurately represent the constraint on a full model due to the large multiplicity of possible decay modes. There will additionally always be very well-motivated models that predict specific signatures that are challenging to incorporate into the simplified model framework outlined here. Experimental searches for these signatures should still be done where possible, but we encourage theorists and experimentalists alike to think carefully about how to design such searches so as to retain maximal sensitivity to simplified models that may give rise to similar signatures.

# 3
# Experimental Coverage of Long-Lived Particle Signatures

Contents

**Chapter editors:** Juliette Alimena, Xabier Cid Vidal, Albert de Roeck, Jared Evans, Heather Russell, Jose Zurita

**Contributors:** David Curtin, Alberto Escalante del Valle, Philippe Mermod, Antonio Policicchio, Brian Shuve

A critical component of any discussion of long-lived particle searches at the LHC is the comprehensive review of the existing searches from ATLAS, CMS, and LHCb, and an assessment of their coverage and any gaps therein. This is an inherently challenging task, given the varied and atypical objects often defined and utilized in LLP analyses and the differences among the experiments. As such, the following discussion assumes little-to-no background on LLP search strategies and includes a high level of detail regarding the current analyses. The focus of the discussion is on the existing studies, while acknowledging that the landscape for new physics models and LLP signatures can be broader than the ones described here.

Backgrounds to most of these studies are typically small, as most LLP signatures are not naturally mimicked by any irreducible SM processeses. Backgrounds for LLP searches typically include peripheral or machine effects, those rarely important for searches for prompt physics, including cosmic muons, beam halo, detector noise, and cavern backgrounds. Such backgrounds are discussed in detail in Chapter 4. As rare as these backgrounds typically are, their rates are not completely negligible, and particular, model-dependent selection requirements (based on, for instance, the LLP mass range or specific decay modes) must be made to reduce backgrounds as much as possible and, in some cases, make the searches "background-free". Additionally, many default object reconstruction algorithms are not designed to detect particles originating from decays of LLPs, and so dedicated reconstruction of tracks, jets, leptons, or other objects may be required for LLP searches. Taken altogether, these factors make LLP searches very different from searches for prompt objects, and the following discussion additionally aims to collate and summarize the current techniques for LLP reconstruction at the LHC.

A particular challenge for many LLP signatures is the trigger. With the exception of certain dedicated ATLAS triggers in the calorimeters or muon spectrometer, there are no Level-1 (L1) triggers that directly exploit the displaced nature of LLP decays, and L1 trigger thresholds must be surpassed by standard objects (such as leptons or high-energy jets) for the event to be recorded. [1] Throughout this chapter, we highlight the role and limitations of the trigger(s) employed in current searches, and the design of customized LLP triggers is to be encouraged to probe new and otherwise inaccessible regions of parameter space.

[1] It is true that, depending on the signature, some of these caveats can be circumvented by a sensible use of existing prompt triggers. For example, photon triggers will collect displaced electrons, calorimeter/jet triggers will record displaced hadronic vertices, etc.

A detailed review of all existing searches is presented in Sections 3.1 through 3.5. This survey of the current experimental coverage aims to highlight the highest-priority searches still yet to be performed, which we summarize in Section 3.6. In all cases, we focus on the latest version of each analysis. Notably we will typically present searches based on data taken at a center-of-mass energy $\sqrt{s} = 13$ TeV, and discuss searches using Run 1 data only when the newer version is not yet available, or when there are conceptual differences between two versions of the same analysis.

Because long-lived particles travel macroscopic distances in the detectors, many of the search strategies rely on the identification of displaced objects, namely SM particles (charged leptons, photons, hadrons, jets) that are produced at a location away from the primary vertex (PV) where the hard $pp$ collision takes place. The secondary vertex at which the decay of the LLP occurs is referred to as a displaced vertex (DV). As far as possible, our classification of searches is linked to the parton-level objects produced in LLP decays, which allows a relatively straightforward linkage to LLP models (as well as simplified models; see Chapter 2). Borrowing the terminology from prompt searches, we consider the following categories for the analogous displaced objects produced in LLP decays: all-hadronic (jets), leptonic, semi-leptonic, and photonic. However, we caution the reader that these "jets" or "photons" may not be of the standard type, and so other objects may pass the selections of these analyses. The remaining searches fall in the "other long-lived exotics" category, mostly consisting of non-standard tracks (disappearing tracks, heavy stable charged particles, quirks, etc.), but also including some trackless signals, such as stopped particles and Strongly Interacting Massive Particles (SIMPs). These categories are not to be interpreted as exclusive; many models and searches could fit into several categories. [2] For example, Refs. [236–238] show how searches for different signatures and LLP lifetimes can be combined to cover large parts of the parameter spaces of particular UV models.

[2] Another important ingredient of the LLP searches is the possibility to reinterpret their results to a large variety of models, namely be able to *recast* them. While we refer the interested reader to Chapter 6 it is worth mentioning here that many existing searches publicly provide useful recasting information, such as efficiency maps or model-independent bounds on production cross sections.

## 3.1 *All-Hadronic Decays*

ATLAS has several searches for displaced decays with hadronic objects, including searches for two objects decaying in the hadronic calorimeter (HCAL) [239, 240]; decays within the muon system (MS) or inner detector (ID) [241]; ID decays in association with large $\not{E}_T$ [230]; and ID decays in association with large $\not{E}_T$, jets, or leptons [189]. CMS has inclusive searches for displaced jets using 13 (8) TeV data [242, 243] ([244]). Moreover, the CMS displaced jets searches are relatively inclusive and so also cover LLPs with semi-leptonic decays despite having no specific lepton requirements. CMS also has a search for displaced vertices in multijet events [245]. LHCb has searches for both one [246] and two [247] all-hadronic DVs in their detector. Here the discussion is restricted to summarizing the hadronic channels, while those studies including leptons [189, 242] will be revisited in Sections 3.2 and 3.3 for the fully-leptonic and semi-leptonic cases, respectively.

### 3.1.1 ATLAS Searches

The reconstruction of displaced tracks in the ATLAS ID [248] follows a two-step procedure. In the first iteration, the default track identification algorithm is applied, which uses hits in the pixel system, Semiconductor Tracker (SCT), and Transition Radiation Tracker (TRT) to reconstruct tracks with a small impact parameter. The hits not associated to a track during the first pass are used in a second run of the track finder, with loose requirements on the transverse and longitudinal impact parameters ($d_0$ and $z_0$) and the number of silicon hits that are shared (or not shared) with another track. This two-step procedure is referred to as the *large radius tracking* (LRT) algorithm by the ATLAS collaboration. Applying the LRT procedure is CPU-intensive, and thus it is only run once per data-processing campaign, on a subset of specially-requested events [248].

In searches where the LLPs decay exclusively in the ID, standard triggers are used to select events with high-$p_T$ jets, $\not{E}_T$, or high-$p_T$ leptons [189, 230]. An ATLAS 13 TeV search [230] uses a standard $\not{E}_T$ trigger and an offline requirement of $\not{E}_T > 250$ GeV. The 8 TeV search [189] covers a larger range of topologies, and the event must have either $\not{E}_T >$ 180 GeV or contain four, five, or six jets with $p_T > 90$, 65, or 55 GeV to pass the trigger. In both searches, the ID vertex is required to have at least 5 tracks and the invariant mass of the displaced vertex tracks to fullfil $m_{DV} > 10$ GeV. These searches are interpreted in the context of various SUSY scenarios involving gluinos or squarks decaying into leptons, jets and missing energy, namely R-Parity-Violating (RPV), General Gauge Mediation (GGM), and split SUSY. In the latter case R-hadrons [3] are considered. The particular LLP decay topology determines which trigger and analysis mode (specified by jet and lepton multiplicity, small/large $\not{E}_T$, etc.) has the best sensitivity. The LLPs covered by these searches are typically high mass ($\gtrsim 100$ GeV), and correspond to the direct-pair-production and heavy-parent production modes with hadronic decays (in the language of the simplified models presented in Section 2). However, these searches do not have sensitivity to low-mass LLPs, especially those resulting from the Higgs, $Z'$, or charged-current production portals and then decaying hadronically.

[3] R-hadrons form when BSM colored particles hadronize due to a lifetime larger than the hadronization scale. In split SUSY the R-hadrons are typically long-lived due to their decays being mediated by heavy squarks.

For LLPs decaying in the HCAL or MS, dedicated *CalRatio* and *MuonRoI* triggers are employed [239–241, 249], allowing the searches to place limited requirements on the non-displaced portion of the event. We describe these triggers in more detail shortly. The efficiency of these triggers is 50% – 70% for decays within the relevant geometric detector region, and negligible outside of them (see Figure 3 of Ref. [241]). The results of these analyses are interpreted in terms of a $\Phi \to ss$ model, where $\Phi$ is a heavy scalar boson with 100 GeV $< m_\Phi <$ 1000 GeV and $s$ is a long-lived, neutral scalar decaying to hadrons with branching fractions dictated by the Yukawa coupling. This can map to Higgs or $Z'$ production modes and hadronic decay mode in the simplified models.

The CalRatio trigger selects events with at least one trackless jet that has a very low fraction of energy deposited in the ECAL [4]. These CalRatio jets are characteristic of an LLP that decays within or just before the HCAL. The 13 TeV analysis [239] requires two CalRatio jets, where the exact CalRatio criteria are determined using a series of machine learning techniques to optimally discriminate the displaced decay signature from QCD jets and beam-induced background. Using the simplified $\Phi \to ss$ model with 125 GeV $< m_\Phi <$ 1000 GeV and 5 GeV $< m_s <$ 400 GeV, good sensitivity is observed for $c\tau$ between 0.05 and 35 m, depending on the $\Phi$ and LLP masses. Notably, SM-like Higgs boson decays to LLP pairs are constrained below 10% branching ratio in the most sensitive lifetime ranges, with exact limits dependent on the LLP mass [239]. The 8 TeV result also requires two CalRatio jets, and shows sensitivity for 100 GeV $< m_\Phi <$ 900 GeV and 10 GeV $< m_s <$ 150 GeV [240]

[4] The variable used to discriminate between CalRatio jets and standard jets is $\log_{10}(E_{\mathrm{HAD}}/E_{\mathrm{EM}})$, where $E_{\mathrm{HAD}}$ and $E_{\mathrm{EM}}$ are the fractions of the measured energies of the jets appearing in the HCAL and ECAL, respectively. The trigger selects trackless jets with $\log_{10}(E_{\mathrm{HAD}}/E_{\mathrm{EM}}) > 1.2$, which corresponds to an electromagnetic fraction of 0.067.

The MuonRoI trigger selects events with clusters of L1 Regions of Interest (RoIs) in the MS that are isolated from activity in the ID and calorimeters. It is efficient for LLPs that decay between 3 – 7 m transversely or 5 – 13 m longitudinally from the PV, for LLP masses greater than 10 GeV. After trigger selection, the ATLAS analysis in question requires either two reconstructed DVs in the MS [250] or one ID vertex and one MS vertex [241]. This ID–MS combination provides increased sensitivity to shorter lifetimes than an analysis only considering MS vertices, and shows good sensitivity to 100 GeV $< m_\Phi <$ 900 GeV and 10 GeV $< m_s <$ 150 GeV. Decays of a SM-like Higgs boson to LLP pairs are constrained below 1% in the most sensitive $c\tau$ regions (with cross section limits as low as 50 fb). The efficiency degrades for benchmarks with higher LLP boosts or very low mass LLPs, as fewer tracks are reconstructed. Another ATLAS search includes signal regions with only 1 DV in the MS, with sensitivity to SM-like Higgs decays to LLPs extending down to branching fractions of 0.1% [251]; this search also presents constraints on a wide range of models that helps facilitate reinterpretation for other BSM scenarios. In addition, a combination of the results from this search with the results from the 13 TeV CalRatio search was performed in Ref. [239] for the models common to both, and provides a summary of the ATLAS results for pair-produced neutral LLPs.

Recently, ATLAS presented a new study for hadronically decaying LLPs produced in association with a leptonically decaying $Z$ boson [252]. In this analysis, the LLP decays inside of the HCAL. The use of lepton triggers on the associated $Z$ decay products gives sensitivity to production of a *single* low-mass LLP, whereas other searches typically require 2 DVs; it is therefore an excellent example of the utility of prompt associated objects in obtaining sensitivity to low-mass LLPs. The model constraints are expressed in terms of a Higgs portal model where the SM-like Higgs decays to a bosonic

LLP.

With the exceptions of Refs. [189, 230, 252], which require prompt activity in addition to the DV and have comparatively high trigger thresholds, the ATLAS all-hadronic analyses require two DVs, and thus are insensitive to models that produce a single DV inside the detector [5].

[5] This may be the result of a signal that produces two DVs, but the lifetime is sufficiently long that only one DV appears inside the detector.

### *3.1.2 CMS Searches*

The CMS analyses [242–244] are based on a dedicated offline *displaced jet tagging* algorithm using tracker information to identify pairs of displaced jets. The triggers used here are based on large values of $H_T = \sum |p_{\mathrm{T,j}}|$ = 350 (500) GeV for 8 (13) TeV, where the $H_T$ sum runs over all jets with $p_{\mathrm{T,j}} > 40$ GeV and $|\eta_j| < 3.0$. The trigger for the 13 TeV analysis based on 2015 data additionally requires either two jets with $p_{\mathrm{T}} > 40$ GeV and no more than two associated prompt tracks ($d_0 < 1$ mm) and the $H_T$ threshold is lowered to 350 GeV if the two jets each have at least one track that originates far from the PV [6]. Only events with two or more displaced jets are kept in the analysis, while those with only one are used as a control sample to estimate the prompt jet misidentification rate. For $c\tau <$ 3 mm [7], the algorithm is inefficient as more than two tracks tend to have impact parameters less than 1 mm; for $c\tau > 1$ m the search is inefficient as most decays occur too far from the PV to form reconstructable tracks. A key difference among these searches is that the 8 TeV [244] and 13 TeV (2016 data) [243] analyses explicitly reconstruct the DV, while the 13 TeV (2015 data) [242] analysis does not.

[6] In this case this is defined by requiring that the transverse impact parameter significance $|d_0|/\sigma_{d_0}$ have a value greater than 5.

[7] Note that in principle the low $c\tau$ regime can be covered with standard triggers, however they require higher $H_T$ thresholds.

CMS interprets the signal in several benchmark models that can be mapped to the direct pair production simplified model production mode, including a neutral LLP decaying hadronically and a color-charged LLP decaying into a jet plus a lepton. For neutral LLP pair production decaying democratically into light jets, the trigger efficiencies for $c\tau = 30$ mm are reported to be 2, 41, 81, and 92% for 50, 100, 300, 1000 GeV masses, respectively. It is evident that the requirements on $H_T$ and on $p_{\mathrm{T,j}}$ make the search inefficient for low LLP masses. Indeed, a phenomenological recast of the 8 TeV analysis [244] in terms of rare decays of a SM-like Higgs boson with a mass of 125 GeV Higgs sets very mild bounds for LLP masses below $m_h/2$ [253]. Thus, the CMS search has limited sensitivity to low-mass, hadronically decaying LLPs through the Higgs, $Z'$, or charged-current simplified production modes.

As mentioned above, CMS also has a search for displaced vertices in multijet events [245], which was released near the time of the final editing of this manuscript.

### *3.1.3 LHCb Search*

The LHCb searches [246, 247] trigger directly on DVs with a transverse distance of $L_{xy} > 4$ mm) with four or more tracks, vetoing dense material regions in which hadronic interactions with the detector can mimic LLP decays. The trigger thresholds are, however, low. For example, the invariant mass of particles associated with the vertex must exceed 2 GeV and the scalar sum $p_T$ of tracks at the vertex must exceed 3 GeV. Jet reconstruction is then performed offline with standard algorithms. The benchmark model used by these searches is a scalar particle decaying to two neutral LLPs, $\pi_v$ (dark or "valley" pions), which corresponds to the Higgs simplified model with hadronic decay modes. The parent particle can be either a SM-like 125 GeV Higgs [246] or a Higgs-like scalar with mass in the 80–140 GeV range [247]. The search is performed for $\pi_v$ masses between 25 and 50 GeV and decay lenghts between 0.6 and 15 mm. It is expected that LHCb will extend their coverage to shorter lifetimes by improving the understanding of the material and to lower masses by using fat-jets and jet-substructure to access larger boosts [254]. In principle, the search is also sensitive to direct pair production of LLPs.

Because of the low thresholds, the LHCb search focuses on low-mass LLPs with short lifetimes, for which it has excellent sensitivity. However, its sensitivity for other signatures is limited by the geometry of the detector and the LHCb luminosity compared to ATLAS and CMS. A model-dependent direct comparison among the LHCb, ATLAS and CMS reaches for the Higgs production mode decaying into dark pion LLPs can be seen in Figure 3.1.

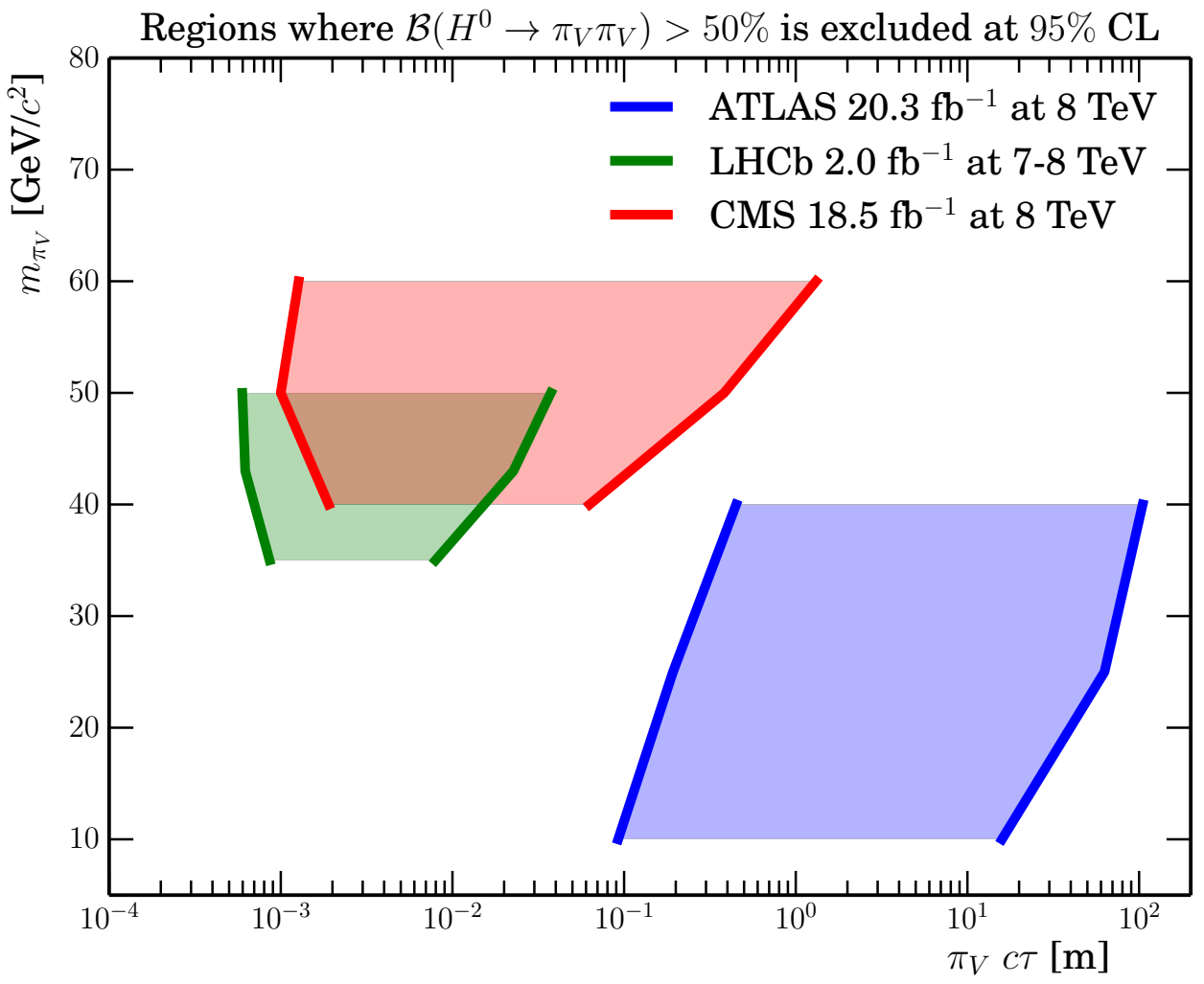

Figure 3.1: Comparison of the ATLAS [189], CMS [244] and LHCb [246] reaches for dark pions $\pi_V$ decaying into jets. The CMS result is taken from the recast done in reference [253] of the 8 TeV analysis [244]. In the shaded regions $B(H \to \pi_v\pi_v)$ is constrained to be below 50%. Note that the ATLAS reach extends to higher masses as well; the plot was produced using the benchmark scenarios presented in [189], hence the meaningful bound is on the lifetimes. Taken from Ref. [246].

### 3.1.4 Summary

Searches in hadronic final states do not currently cover LLP parent masses below $\sim$ 100 GeV in a comprehensive way. This is typically due to the large $p_{\mathrm{T,j}}$ requirements at the trigger level, with an exception being the DV reconstruction at LHCb. Additionally, the powerful ATLAS searches for LLPs decaying in the HCAL or MS require two LLP decays in the detector, meaning that as of this writing there is no sensitivity to singly produced long-lifetime LLPs with hadronic decays [8]. While the existing searches are typically sensitive to both direct pair production and heavy parent production of LLPs, not all of the searches provide benchmarks with a variety of LLP production kinematics and boost.

[8] Highly-inclusive searches for single LLPs decaying in the ATLAS MS have been proposed [255], finding that backgrounds are appreciable and need to be controlled using data-driven methods.

A potential way to extend the sensitivity of current analyses is to use other existing triggers exploiting such things as VBF production modes, leptons, $\not{E}_{\mathrm{T}}$, etc., to trigger on associated *prompt* objects and perform the hadronic DV reconstruction offline. The ATLAS 8 TeV study [189] does employ multiple triggers (such as lepton triggers), but in each case the triggered object must be associated with the DV (for a lepton trigger, the lepton must originate from the DV). If, instead, a prompt lepton or VBF trigger were used with the offline reconstruction of a separate displaced object, sensitivity could be recovered to low-mass hadronic DVs in a variety of simplified models, including Higgs production (via VBF or VH associated production modes) [253, 256] or charged-current production (in association with a prompt lepton) [257]. In particular, triggering on associated prompt objects would improve the efficiency of reconstructing low-mass hadronic LLPs produced in the decays of a SM-like 125 GeV Higgs. As there is no theoretical lower limit on the masses of light neutral LLPs, it is imperative to lower the LLP mass coverage as much as possible. If at all possible, a dedicated *online* reconstruction of DVs would allow for a further reduction on the $p_{\mathrm{T}}$ threshold, giving sensitivity to light LLP masses.

## 3.2 Leptonic Decays

All three experiments have searches for a pair of leptons coming from a DV [189, 258–262]. CMS also has a search requiring exactly one isolated muon and one isolated electron (i.e., events with additional isolated leptons are discarded) with large transverse impact parameters (0.2 cm $< |d_0| <$ 10 cm), but without any other additional requirement including, for example, that the reconstructed tracks do not need to point to a common vertex [190]. This loose selection makes the search sensitive to a variety of new physics scenarios. Light and boosted LLPs can decay into collimated light leptons, dubbed *lepton-jets* [120], which are searched for at both CMS [263, 264] and ATLAS. ATLAS has searches for both displaced [265, 266] and prompt lepton-jets [267]. The LHCb collaboration also looks for light, neutral LLPs decaying into $\mu^+\mu^-$ pairs by studying $B$-meson decays to kaons, for exclusive decay channels for both neutral [260] and charged [261] $B$-mesons, as well as dark photons that decay to muon pairs [262].

### 3.2.1 CMS Searches

The CMS searches trigger on leptons reconstructed using information from either the tracker [258] or the muon chambers [259], where the latter search uses only muons. In the tracker-based analysis, the LLP is reconstructed by forming pairs of charged leptons (where muons are required to have opposite signs), with $p_T$ cuts of 26 GeV for muons, and 36 (21) GeV for the leading (subleading) electron. This yields slightly larger efficiencies in the muon channel. The transverse impact parameter $|d_0|$ needs to be 12 times larger than its uncertainty $\sigma_d$ (approximately corresponding to a distance $\gtrsim$ 200 $\mu$m) to reject prompt backgrounds. In the MS-based analysis, muon candidates are reconstructed using hits in the muon chambers, and no information from the silicon tracker is used. In order to avoid biases from a loose beamspot constraint in the seeding step, these muons undergo an additional refit step. These candidates are referred to as *re-fitted stand-alone* (RSA) muons, and they need to fulfill $p_T > 26$ GeV, $|\eta| < 2$, and to be separated by $\Delta R > 0.2$. More importantly, these candidates are rejected if they can be matched to a $p_T > 10$ GeV track in the inner tracker, which efficiently excludes prompt muons and also renders this study fully complementary to the tracker-based one. Both these searches are interpreted in terms of decays of an SM-like Higgs $H$ ($H \to XX, X \to l^+l^-$) and RPV squarks, covering proper lifetimes of 0.01–$10^5$ cm for the Higgs scenario, and 0.1–$10^4$ cm for the SUSY case. The difference in the lower reach of $c\tau$ is due to the larger boost factor of the Higgs. These benchmarks map to the direct pair production, heavy parent and Higgs production simplified models, with flavor-conserving leptonic decays of the LLP. There is good sensitivity down to relatively low masses (LLPs of masses $\gtrsim$ 20 GeV produced in Higgs decays) due to the low lepton trigger thresholds.

Additionally, CMS has a search for one electron and one muon, each with large transverse impact parameter (200 $\mu$m $< |d_0| <$ 10 cm) [190]. Events are selected using a dedicated trigger for $e\mu$ pairs that applies a $p_T$ cut on the leptons (42 GeV for electrons, 40 GeV for muons) but, unlike standard triggers, places no restriction on the maximum $d_0$ or distance from the PV. Events with exactly one muon and exactly one electron are kept, and then separated into “prompt", “displaced control" and “signal" regions, defined as $|d_0| < 100$ $\mu$m, 100 $\mu$m $< |d_0| < 200$ $\mu$m, and $|d_0| > 200$ $\mu$m, respectively. This selection makes the signal region almost free of leptons coming from SM processes, with rare tau-leptons, $B$-mesons or $D$-mesons as the largest remaining background.

Although in the original search the results are interpreted in the context of long-lived RPV stops (excluding masses below 870 GeV for $c\tau = 2$ cm) [9], this search has been shown to be sensitive to many scenarios, including long-lived staus in gauge mediated SUSY breaking [198] and right-handed neutrinos [45]. Indeed, this search has sensitivity to LLPs produced via any of the simplified model production modes and (semi)-leptonic decays that give exactly one electron and one muon. On the other hand, models where long-lived particles decay only to *either* muons *or* electrons (*e.g.*, $\tilde{\mu} \to \mu \tilde{G}$) are unconstrained by this search. Furthermore, same-sign lepton signatures and signatures with additional leptons are not constrained by the current search but could be covered by extensions of the search [45, 198]. Due to the generality of tau-specific models, searches for hadronic tau channels is also desirable. This search has sensitivity to relatively low-mass LLPs; however, the 8 TeV analysis [268] has lower thresholds ($p_T > 22$ GeV on both leptons) albeit with a requirement for shorter decay distances ($|d_0| < 2$ cm), and so has superior sensitivity to very low-$p_T$ displaced signals. Maintaining low trigger thresholds is necessary to obtain sensitivity to the lowest-mass leptonic LLP signals.

[9] We note that a CMS prompt search for leptoquarks has been recasted using the same model, finding stringent constraints for lifetimes below a few millimeters. This reinterpretation is discussed in detail in Section 6.7.

### 3.2.2 *ATLAS Searches*

The primary ATLAS search for displaced leptons [189] triggers on muons without an ID track, electrons, or photons [10]. The trigger and offline $p_T$ criteria are relatively high, requiring one of the following: one muon of at least 50 GeV; one electron of at least 110 GeV; one photon of at least 130 GeV; or two electrons, photons, or an electron and a photon with minimum $p_T$ requirements for both objects in the 38–48 GeV range. The DV is formed from opposite-sign leptons, irrespective of flavor, and needs to be located more than 4 mm away from the PV in the transverse plane. DVs in regions with dense detector material are vetoed to suppression backgrounds from converted photons (e.g., $\gamma p \to e^+e^- p$). This search is in principle sensitive to events with a reconstructed DV mass $m_{DV} > 10$ GeV, but the high $p_T$ requirements for the leptons restrict the sensitivity to low-mass LLPs.

[10] Electrons with large transverse impact parameters $d_0$ tend to be missing a track at trigger level and are reconstructed as photons.

ATLAS has also recently released a search for pairs of muons that correspond to a displaced vertex [269]. This search is sensitive to LLP decays that occur sufficiently far from the interaction point that the muons are reconstructed only in the MS. The analysis has four separate trigger pathways: $\not{E}_T > 110$ GeV; one muon with $p_T > 60$ GeV and $|\eta| < 1.05$; two muons with $p_T \gtrsim 15$ GeV and $\Delta R_{\mu\mu} < 0.5$; or three muons with $p_T > 6$ GeV. Thus, the search has sensitivity to final states with high and low masses (down to $m_{\mu\mu} = 15$ GeV), as well as with various lepton multiplicities. Offline selections require the muons to have $p_T > 10$ GeV and opposite charge, and the search is efficient at reconstructing muons for transverse impact parameters up to $|d_0| = 200$ cm. Muons are also required to satisfy isolation requirements from jets as well as from nearby tracks. The search puts constraints on a SUSY model and a model of dark photon production in Higgs decays; this can be used to determine sensitivity to all of the simplified production modes from Chapter 2 in the $\mu\mu$ decay mode (or $\mu\mu$ in association with other objects). It also demonstrates how combining many trigger pathways and loose selection requirements can enhance sensitivity to a wide range of LLP models.

### 3.2.3 *LHCb Searches*

LHCb has a search that looks for the direct production of both promptly decaying and long-lived dark photons [262]. As a result of the direct production, dark photons do not tend to be highly boosted in the transverse direction. Events [11] are required to have a single muon with $p_T > 1.8$ GeV, or two muons with a product of transverse momenta $\gtrsim (1.5\text{ GeV})^2$. The displaced search constrains previously uncovered dark photon parameter space around masses of $\sim 300$ MeV.

[11] For the prompt dark photon search, events are reconstructed at trigger level so that all online reconstructed particles are recorded, while the rest of event information is discarded [270]. The prompt search constrains entirely new territory above 10 GeV.

The LHCb searches for displaced leptons in rare $B$ meson decays [260, 261] rely on standard techniques to identify the $B^\pm$ decay vertex and the kaons and pions in the event, and the dimuon invariant mass $m(\mu^+\mu^-)$ variable is scanned for excesses. The $X \rightarrow \mu^+\mu^-$ vertex is not required to be displaced from the $B^\pm$ vertex, and thus the constraints apply to both prompt and long-lived particles. The analysis probes LLP masses of 214 (250) MeV $< m_X <$ 4350 (4700) MeV for the $B^0 \rightarrow K^*\mu^+\mu^-$ ($B^+ \rightarrow K^+X, X \rightarrow \mu^+\mu^-$) process, with the mass range being limited by kinematics.

### 3.2.4 *Lepton-Jet Searches*

Searches for lepton-jets are focused on $\mathcal{O}(\text{GeV})$ LLP masses and distinctly boosted signatures, and thus we treat them separately.

The ATLAS 8 TeV search [265] considers three types of lepton-jets: those containing only muons, only electrons/pions, or a mixture of the two. The muon and electron/pion lepton-jets can contain either two or four leptons, while the mixed lepton-jet must contain two muons and a jet consistent with a displaced electron/pion pair. As these signatures contain relatively soft leptons, the ATLAS 8 TeV analysis uses a trigger that requires three muon tracks in the MS with $p_T > 6$ GeV. There is a built-in limitation to this trigger, which is that the L1 requirement of three separate muon RoIs makes it only sensitive to topologies with two lepton-jets in which one lepton-jet has a wide enough opening angle between two muons to create two level-one RoIs. For the electron lepton-jets, when the electrons are produced in the HCAL they are indistinguishable from a hadronic decay and thus the CalRatio trigger is used.

In the 13 TeV ATLAS analysis [266], a *narrow-scan* muon trigger is additionally used. This trigger starts off by selecting events with one muon with $p_T > 20$ GeV, then requires a second muon with $p_T > 6$ GeV within $\Delta R = 0.5$ of the leading muon.

Both the 8 and 13 TeV ATLAS searches are interpreted for Higgs-like scalar particles (with masses of 125 and 800 GeV) that decay effectively into either two or four lepton pairs, with each lepton pair assumed to come from a low-mass "dark" photon, $\gamma_D$. The ATLAS result excludes exotic Higgs branching ratios below 10% for dark photon lifetimes $2 < c\tau < 100$ mm. Note that here $\gamma_D$ is also allowed to decay to pions and so the results can also be interpreted for hadronically and semi-leptonically decaying LLPs. This corresponds to the Higgs production mode in the simplified models proposal with an admixture of flavor-conserving leptonic and hadronic LLP decays.

The CMS lepton-jet search focuses on fully muonic lepton-jets and has been performed with both the 8 TeV dataset [263] and part of the 13 TeV dataset [264]. The 13 TeV search is sensitive to di-muon parent particle masses up to 8.5 GeV. Events are selected with a di-muon trigger with standard isolation requirements. Further selection requires at least four muons, forming a minimum of two opposite-charged pairs. CMS uses a benchmark model with scalars decaying into either lighter scalars or dark photons, with varying scalar and dark photon mass. For the case of a 125 GeV Higgs they can exclude an exotic branching ratio of below 0.1% for some parameter points. The CMS search can be compared with the ATLAS results, as can be seen in Figure 3.2. We note that this study includes sensitivity to both prompt and displaced muonic lepton jets.

### 3.2.5 *Summary*

To summarize, the lepton searches rely on fairly standard lepton identification, with vertex reconstruction being performed mostly offline. Searches for leptonically decaying LLPs typically enjoy low trigger thresholds and good sensitivity to LLP production rates. Extending the success of the leptonic LLP program to future LHC running will necessitate maintaining low-threshold triggers for displaced leptons in a high-luminosity environment; this is a major challenge but one that must be overcome. Another outstanding challenge is coverage of LLP decays to $\tau$ leptons, which lie at the interface between hadronic and leptonic searches. Such decays are very well motivated from the theoretical point of view, as a Higgs-like scalar can typically decay about 300 times more often into $\tau^+\tau^-$ if kinematically allowed, and also one could have large rates into mixed decay modes such as $\tau^+\mu^-$. A displaced hadronic $\tau$ is a striking object, and most likely will have few backgrounds. Hence, limits on exclusive displaced $\tau$s would be of utmost importance [12].

As the leptonic searches explicitly require opposite-sign leptons, the same-sign lepton signature (motivated from Majorana neutrinos; see the LHCb search in Section 3.3, or heavy, doubly-charged

[12] We note that if the $\tau$ originates from outside the tracker, the hadronically decaying taus are indistinguishable from other displaced hadrons. For instance, in the ATLAS search utilizing muon RoIs [241] the results are interpreted for a model with a scalar particle with Higgs-like couplings to SM fermions, which includes a branching fraction into $\tau^+\tau^-$. However, if the $\tau$ originates from the ID, the low number of tracks associated to it (one to three) will not fulfill the requirements of the ATLAS study of five or more tracks associated to a DV.

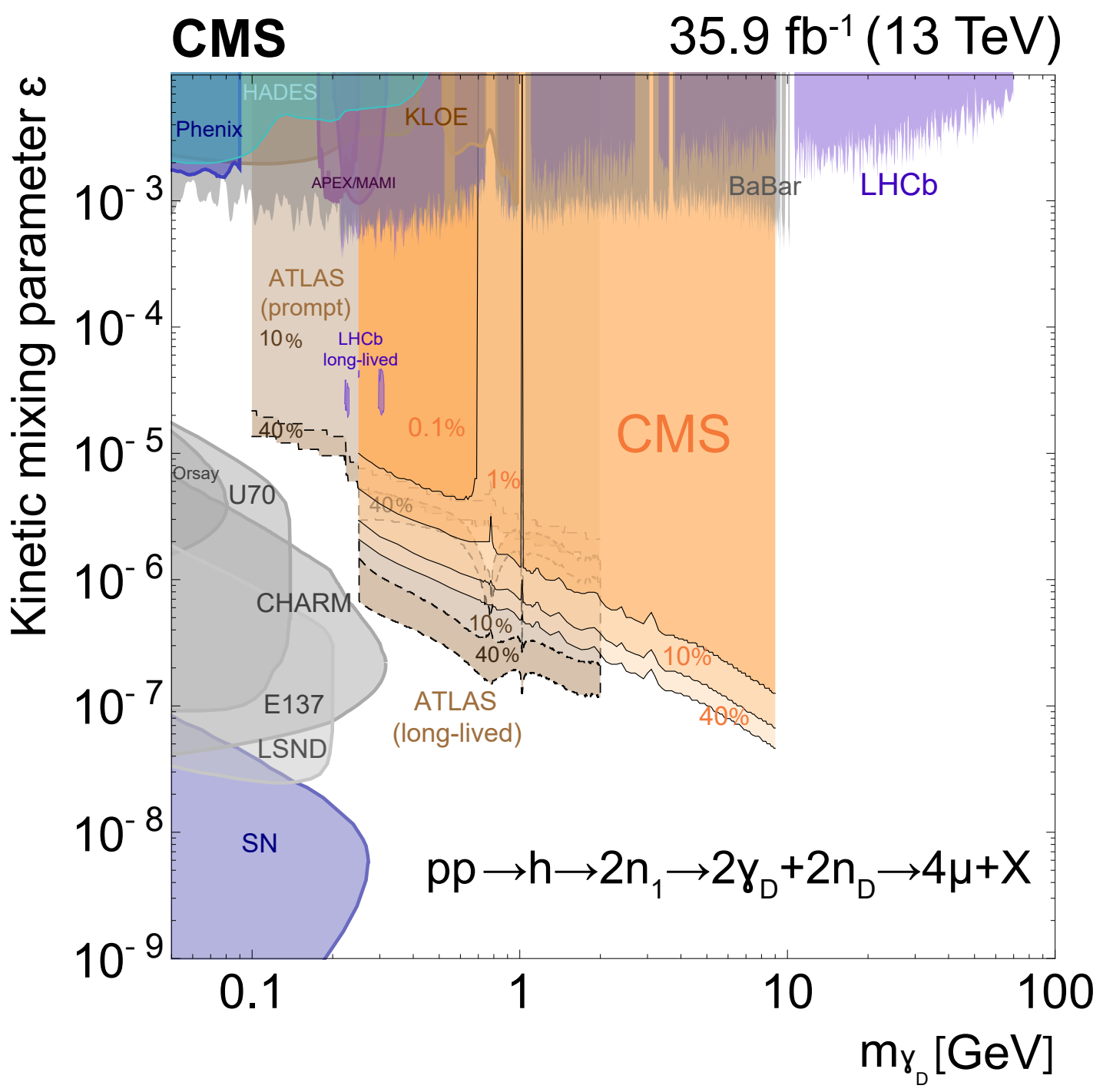

Figure 3.2: Comparison of the lepton-jet searches at ATLAS [265] and CMS [264] with respect to a dark photon scenario [148] vis-a-vis dark photon limits coming from low-energy experiments. Figure taken from Ref. [264].

LLPs) is currently neglected. Furthermore, the sensitivity of the CMS search for two high-$|d_0|$ leptons is currently only sensitive to opposite sign $e\mu$ pairs, with a veto on additional leptons. Relaxing these requirements would greatly enhance sensitivity to certain scenarios, especially the simplified models with flavor-conserving leptonic LLP decays.

Lepton-jet searches currently cover only final states with at least two LLPs and some muons in the final state [13], and the same statement currently applies to both the LHCb searches for dark photons [262, 270] and for LLPs produced in $B$-meson decays. The existing ATLAS lepton-jet studies express their results in terms of specific dark photon models [14], which makes it complicated to apply the results to other models. We refer the reader to Section 6 for a further discussion of this topic. In extending lepton-jet searches, it would be beneficial to have additional searches with a single lepton-jet or low-mass, leptonically-decaying LLP (which are motivated in models with hidden sectors and Majorana neutrinos, for example in Refs. [43, 128]). In addition, the status of coverage in the intermediate mass-transition region between “standard" displaced lepton pairs and lepton-jets is unclear, and may potentially harbor a gap; adopting the simplified model approach for leptonic LLP decays with masses varying between the GeV and weak scale would ensure comprehensive coverage of low-mass leptonically decaying LLPs.

[13] The ATLAS 8 TeV search [265] included a search channel with two electron-only lepton-jets, but the performance was poor and it was excluded from the final result.

[14] Recall that the lepton-jet studies also consider the $\gamma_D \to \pi^+\pi^-$ decay mode.

Finally, we comment on gaps in sensitivity to very low-mass leptonically decaying LLPs. A benchmark LLP model is the heavy neutral lepton (HNL), which corresponds to the charged-current production mode with (semi-)leptonic LLP decays in our simplified model framework. HNLs constitute an important physics case that leads to multi-lepton displaced signatures from $W$ decays, with nice prospects at ATLAS and CMS (see, e.g., Refs. [43, 47, 48]). While previous searches were not sensitive to this scenario due to either high-$p_T$ requirements or the requirement of two DVs in the same event, the presence of a prompt lepton from the $W$ allows the relaxation of these requirements in a dedicated analysis. Moreover, the prospects of triggering on a prompt lepton in such searches was studied recently in Ref. [47] and demonstrated in a prompt search in Ref. [271] [15]. The identification of two leptons from the vertex is a powerful discriminant against backgrounds from metastable particle decays and hadronic interactions in material. This permits a potentially cleaner exploration of the lower HNL mass range (3–6 GeV) than in the semi-leptonic channel (see Section 3.3) despite the lower branching ratio. It should be noted that HNL models can predict LLP decays to all three lepton flavors (either democratically or hierarchically), necessitating the capability to reconstruct displaced leptons of all flavors, including taus.

[15] Note that the displaced large transverse impact parameter $e\mu$ CMS search [190] fails to cover this scenario due to the aforementioned lepton veto, which eliminates sensitivity to the tri-lepton signals discussed in Ref. [43], as well as relatively high lepton $p_T$ trigger thresholds compared to the kinematics of 4-body $W$ decay.

## 3.3 *Semi-Leptonic Decays*

As semi-leptonic signatures include aspects of both hadronic and leptonic LLP decays, many of the previously discussed searches can partially cover these cases, and some do so explicitly. For instance, the ATLAS search for electrons and muons accompanied by tracks [189], the inclusive CMS search for DVs [242] (which contains a specific model interpretation called "$B$-lepton" addressing precisely this channel), and the search for a large impact parameter $e\mu$ pair by CMS [190] are all inclusive with respect to other hadrons produced in the LLP decay, provided the leptons are sufficiently isolated [16]. In addition, LHCb has dedicated searches for semi-leptonically decaying LLPs [272] and semi-leptonic decays of long-lived Majorana neutrinos coming from $B^-$ mesons [273]. The two CMS searches [190, 242] need no further explanation for how they cover semi-leptonic LLPs because of their very inclusive nature (see Section 3.2.1), but we now describe the rest of these searches in some detail.

[16] Note that the lepton isolation affects most of the semi-leptonic searches.

### 3.3.1 *LHCb Searches*

The LHCb search for semi-leptonic LLP decays selects events with a muon trigger, then reconstructs a DV offline [272]. The results are interpreted in terms of four distinctive topologies: single LLP production in association with a new particle (in this case a gluino), double LLP pair production via direct pair production, Higgs de-cay, or via squark pair production (heavy parent). The LLP then decays to two quarks and a muon (which maps to the $jj\ell$ simplified model decay). Material regions are vetoed for the DV, which results in the dominant background arising from heavy flavor production either directly or from $W/Z$ decays. The signal discrimination is obtained from a multivariate analysis based on the muon $p_T$ and impact parameter, and subsequently the search is optimized based on the LLP reconstructed mass and the muon isolation. This study is sensitive to low-mass LLPs with lifetimes between 1.5 and 30 mm, as can be seen in Figure 3.3.

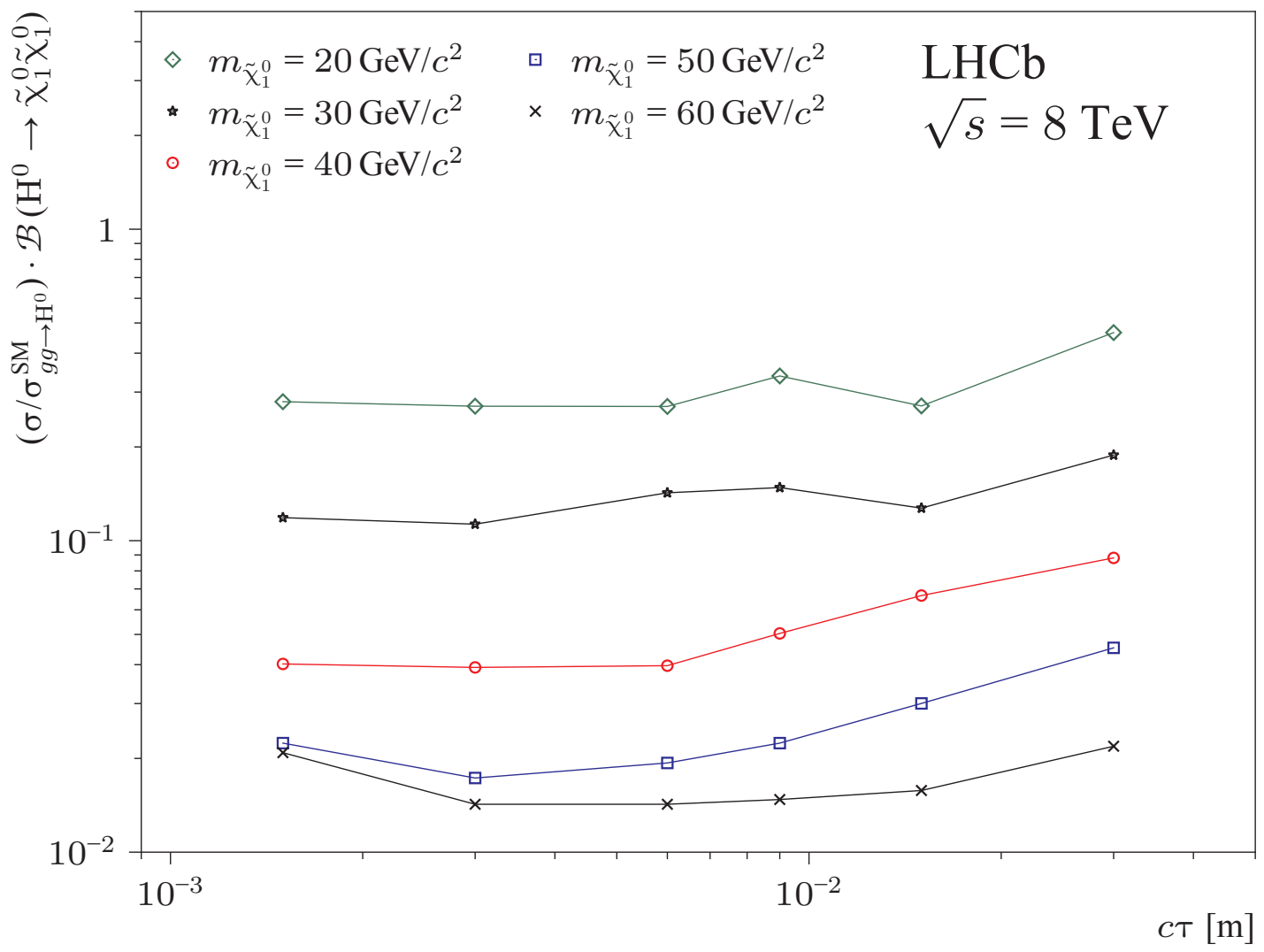

Figure 3.3: LHCb reach for displaced semi-leptonic decays. Taken from Ref. [272].

The LHCb search for Majorana neutrinos [273], $N$, probes Majorana neutrinos produced in leptonic $B$ decays, $B^\pm \to \mu^\pm N$. The Majorana neutrino subsequently decays exclusively to $N \to \mu^\pm \pi^\mp$; both prompt and displaced decays are considered [17]. A same-sign muon requirement, along with the reconstruction of the $N$ and $B$ meson masses, greatly reduces backgrounds to the search. The sensitivity of the search is limited by the restriction to muons in the final state (so models that predominantly decay to $e$ or $\tau$ are not constrained), and the same-sign lepton requirement gives sensitivity only to lepton-number-violating processes. More inclusive searches looking for additional $N$ production modes [275], decays with opposite-sign leptons, or searches targeting decays of heavier mesons like $B_c$ [276] could also improve the sensitivity to semi-leptonically decaying LLPs.

[17] Some care is required in interpreting the results of the search on a model with a Majorana neutrino, as the original theory interpretation is problematic [274].

### *3.3.2 ATLAS Search*

The ATLAS search for semi-leptonic LLP decays [189] looks for a vertex with a lepton accompanied by tracks. This search uses the same trigger as the dilepton vertex search described in Section 3.2.

The DV is required to have a lepton as well as at least four additional associated displaced tracks, and the invariant mass of the tracks must exceed 10 GeV. Thus, the search in principle can have sensitivity down to masses $\gtrsim$ 10 GeV, although the high $p_{\mathrm{T}}$ threshold for the displaced electron/muon typically limits sensitivity to low-mass LLPs; the vertex must contain a muon with $p_{\mathrm{T}} > 55$ GeV or an electron with $p_{\mathrm{T}} > 125$ GeV.

### 3.3.3 Summary

When considering the application of inclusive hadronic or leptonic searches to semi-leptonic LLP decays, it is important to understand how the simultaneous presence of leptons and jets in the signal can degrade the sensitivity. For instance, prompt jet searches explicitly can remove non-standard jets through jet cleaning cuts. Lepton isolation criteria can severely reduce the signal acceptance for a highly-boosted LLP decaying into a lepton and a jet, and they might also veto extra tracks in the events. Thus, boosted semi-leptonic decays (as might be found in the displaced decay of a low-mass, right-handed neutrino produced via $W$ decay) may not be covered by existing searches.

One of the major gaps in semi-leptonic LLP searches is at the smallest LLP masses. In this case, it can be challenging for the leptons from LLP decays to pass trigger thresholds and/or isolation criteria; backgrounds from heavy-flavor and other processes are also higher for semi-leptonic processes than fully leptonic ones. However, there are very good motivations for low-mass semi-leptonic LLPs from the HNL benchmark model [18] introduced in Section 3.2.5, which predicts LLPs for HNLs of masses below 30 GeV. The signature of HNLs from $W$ decays with displaced semi-leptonic HNL decays is an important item on the search agenda of ATLAS, CMS and LHCb [38, 43, 47, 48, 277, 278]. Recently, there has also been a proposal to search for low-mass HNLs in heavy ion collisions [279]. The semi-leptonic channel has the highest branching ratio (about 50% in the relevant mass range [280]) and can therefore offer the best discovery prospects at LHC experiments for HNL masses up to 30 GeV as long as a DV mass cut of around 6 GeV is made to mitigate backgrounds from B-mesons, $m_B \sim 5$ GeV, and backgrounds from random-crossings can be suppressed. The lower end of the 6–30 GeV mass range corresponds to a non-perturbative regime for the hadronization of the HNL decay products. As the number of charged hadrons significantly affects the DV reconstruction efficiency, the validation of event generator outputs for this process is an important issue currently being addressed by the community (see e.g., Ref. [47]). The ability of LHCb to trigger directly on the HNL decay products and better reconstruct displaced tracks can in some cases compensate for its lower acceptance and luminosity, as exemplified by a recent search for DVs composed of one muon and several tracks [272, 278]; similar arguments can be made in the case of heavy ion collisions [279]. This possibly enables LHCb to better probe the more challenging tau channel. Figure 3.4 shows the overall expected reach of LHC searches in the HNL coupling strength (for the muon channel) versus mass plane, using assumptions detailed in Ref. [277], similar to those in Refs. [38, 43].

[18] In the language of simplified models, this corresponds to the charged-current production mode, where the HNL LLP is produced in association with a prompt lepton. The HNL then decays semi-leptonically via the $jj\ell$ channel.

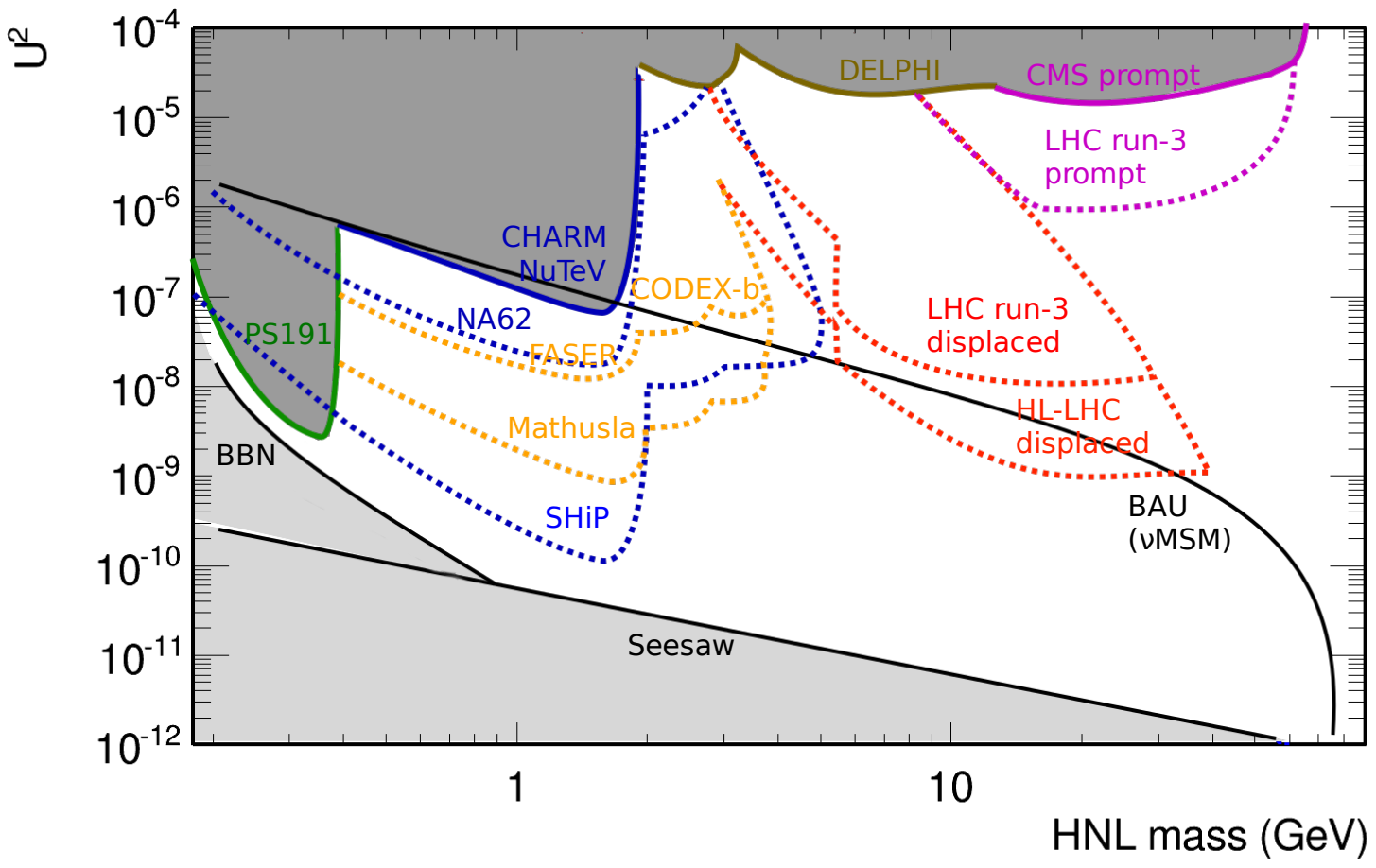

Figure 3.4: Summary of projected experimental sensitivities to HNLs in various experiments, in the coupling strength ($U^2$ for dominant mixing to $\nu_\mu$) vs. mass plane. The projections labelled "LHC Run 3" and "HL-LHC" are for HNLs in $W$ decays in general-purpose experiments, and the one labelled "Mathusla" assumes the full HL-LHC MATHUSLA dataset. We also show the recent CMS result for the prompt tri-lepton signature [271]. Prospects at proton beam-dump experiments are also shown for an already existing experiment, NA62 [281], and for the planned experiment SHiP [282]. Existing constraints from direct searches are indicated as coloured solid lines [283–287]. The lines labelled "Seesaw" and "BBN" show lower theoretical constraints from the observed neutrino masses (assuming a normal hierarchy) and primordial nucleosynthesis, respectively [288]. The line labelled "BAU" is an upper theoretical constraint in the $\nu$MSM model for accounting for baryon asymmetry in the universe while the lightest HNL is a dark-matter candidate [288].

The sensitivity of LHC experiments to HNLs is complementary to that of fixed-target experiments which can probe lower couplings thanks to high-intensity beams albeit at lower mass ranges (i.e., targeting HNLs from $c$ and $b$ decays). The CERN SPS provides great opportunities with the running NA62 experiment [289] and the planned SHiP experiment [282], which comprise a vacuum decay vessel and spectrometer tracker downstream of the target to reconstruct vertices of long-lived neutral particles [19]. These provide the best sensitivity to HNL masses up to 2 GeV, where they probe a region of the parameter space favored in models which simultaneously explain neutrino masses, matter-antimatter asymmetry and dark matter [288, 291–293] (see Figure 3.4).

[19] The proposed detector FASER would also have the capability to reconstruct such vertices [290].

## 3.4 Photonic Decays

There are two ways in which photons coming from LLP decays do not resemble standard photons. First, they cannot be traced back to the PV, thus giving rise to *non-pointing* photons. Second, they can arrive at the ECAL at a time slightly later than expected because the LLP moves slower than the speed of light; these are referred to as *delayed* photons. We note that both kinds of unusual photons can be vetoed in searches for photons that originate from the PV, and thus prompt searches typically provide weaker bounds on LLP scenarios than for promptly decaying signals. ATLAS has a search for non-pointing and delayed photons [294] using the full 8 TeV dataset, which supersedes the 7 TeV analysis [295]. In CMS there are studies for delayed photons in the ECAL [296] and for non-pointing photons detected via their conversion to $e^+e^-$ pairs [297]. The underlying topology in all these models is the neutralino decay into a gravitino and a photon ($\chi_1^0 \to \gamma\tilde{G}$), ubiquitous in gauge-mediated supersymmetry breaking (GMSB) scenarios [50, 298]. Hence all these studies require large $\not{E}_T$ in the final state. This corresponds to heavy parent production of LLPs with decays to a single photon and $\not{E}_T$ in the simplified model framework; all searches described below use the Snowmass Slopes Point 8 benchmark, which is not straightforward to map to a physical spectrum of heavy parent masses.

### 3.4.1 ATLAS Search

The ATLAS study [294] benefits from the capability of the liquid-argon electromagnetic calorimeters to measure the flight direction and the photons' time of flight. Resolutions on $\Delta z_\gamma$, the separation between the PV of the event and the extrapolated origin of the photon, and $|\Delta t_\gamma|$, difference of the arrival time of the photon with respect to the prompt case, are as low as 15 mm and 0.25 ns, respectively. The trigger demands two photons within $|\eta| < 2.5$, with transverse energies $E_T$ of 35 and 25 GeV. In addition, to guarantee the event comes from a proton–proton collision, a PV with five or more tracks with $p_T > 0.4$ GeV is required. The offline selection requires two photons with $E_T > 50$ GeV and $|\eta| < 2.3$, not in the barrel-endcap transition region ($1.37 < |\eta| < 1.52$), at least one of them in the barrel ($|\eta| < 1.37$) and with less than 4 GeV of energy deposited in the calorimeter in a cone of $\Delta R = 0.4$ around them (consituting the *isolation* criterion). In addition, the events are binned in $\not{E}_T$: the $\not{E}_T < 20$ GeV bin contains the prompt backgrounds, the 25 GeV $< \not{E}_T < 75$ GeV bin is used as the control region, and finally the signal analysis is performed in the $\not{E}_T >$ 75 GeV bin. This study covers lifetimes from 0.25 to 100 ns in the GMSB framework, the lower limit being a hard cut-off imposed experimentally, as the similarity between background and signal samples in that region makes discrimination rather difficult. The excluded signal rates in this range of lifetimes vary between 0.8 and 150 fb, with the best-constrained value obtained for $\tau \sim 2$ ns.

### 3.4.2 CMS Searches

The CMS study of delayed photons [296] follows a similar approach to ATLAS. The main difference is that it demands only one photon with $p_T > 80$ GeV, but in addition two jets are required. Furthermore, the vector sum of $\not{E}_T$ and $E_T^\gamma$, which is denoted $\not{E}_{T\ \text{nofl}}$, is additionally used for background discrimination. Collisional backgrounds have small $\not{E}_T$ and large $\not{E}_{T\ \text{no}\gamma}$, while the non-collisional backgrounds are characterised by large $\not{E}_T$ and small $\not{E}_{T\ \text{no}\gamma}$. For the signal events the two variables are large, hence they are both requested to be larger than 60 GeV. The time resolution is 0.372 ns, slightly worse than in the optimal scenario of the ATLAS search. Their reach in lifetimes lies in the 2–30 ns range, excluding signal rates of 10–30 fb.

The CMS study of non-pointing photons [297] relies on the photon converting to $e^+e^-$ pairs. It requires two photons, two additional jets, and $\not{E}_T > 60$ GeV. The photon trajectory is obtained from the conversion vertex as the line segment along the momenta of the $e^+e^-$ track pair, and the impact parameter, $|d_{XY}|$, is defined as the closest distance between the photon and the beam axis, which can be determined within approximately 1 mm. A comparison of the reach of these 8 TeV studies, as well as those using the 7 TeV dataset, can be found in Figure 3.5.

### 3.4.3 Summary

The gaps in these studies are straightforward to identify. The requirement of large $\not{E}_T$ is due to the fact that all of these studies have an underlying theoretical picture of neutralinos decaying into gravitinos and photons, motivated from GMSB scenarios. Hence, these searches do not cover cases without the presence of missing energy, including LLPs that decay to $\gamma\gamma$, $l\gamma$ or $j\gamma$. It may be possible to extract a constraint on such LLP decay modes due to mismeasurement of jets or the photon decay geometry which could mimic large missing energy; however, this would be sub-optimal compared to a dedicated search. With the exception of the CMS study [296] which requires two additional hard jets, all of these analyses require two displaced photons. A single displaced photon signature can occur in motivated models: it can easily arise, for example, from a very slightly mixed electroweak triplet and singlet as in SUSY theories (see the UV models in Section 2). Furthermore, as discussed in Section 3.1, a single LLP in the detector can also arise for very large lifetimes of neutral LLPs, which limits the reach of current searches at longer lifetimes. In such scenarios, it is possible that the photons from LLP decay can be quite soft, and obtaining sensitivity to models with single photons from LLP decay and/or low momenta may require triggering on associated prompt objects, similar to the recommendations in Section 3.1.4.

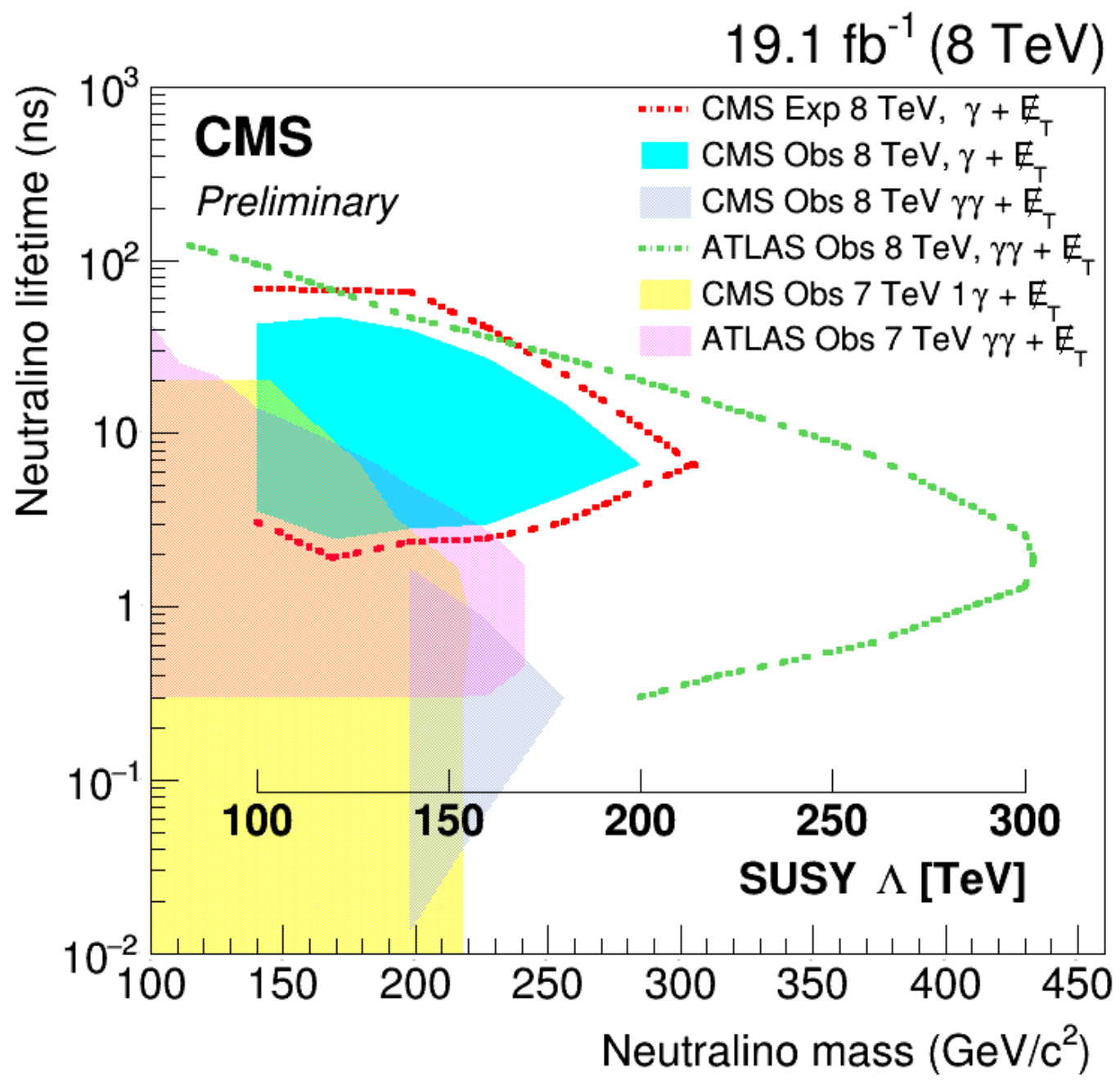

Figure 3.5: Summary of the $\gamma + \not{E}_T$ searches from ATLAS and CMS, displayed assuming the same GMSB model. Taken from Ref. [296].

## 3.5 *Other Exotic Long-Lived Signatures*

In the preceding sections, we presented analyses sensitive to LLPs decaying into objects such as jets, leptons, and photons. In many cases, however, LLPs give rise to signatures that are completely distinct from more conventional prompt signatures. In this section, we present analyses that exploit properties of other exotic long-lived signatures, such as non-standard tracks. We summarize in detail the existing searches for heavy, stable charged particles (HSCP); disappearing tracks (DT); stopped particles (SP) and monopoles, and describe existing ideas on how to look for quirks and SIMPs (Strongly Interacting Massive Particles).

Note that the terminology employed in some of these searches can be confusing, with occasional conflation of signatures with the underlying model. We provide a detailed explanation of each search, and we attempt a classification here based strictly on signature. We distinguish between three classes of signals: *tracks with anomalous ionization*, *tracks with anomalous geometry*, and finally *out-of-time* decays.

### 3.5.1 Tracks with Anomalous Ionization

In this category, we collect all searches for charged-particle tracks with anomalous ionization, i.e., those that are inconsistent with a charge $|Q| = e$. Here we include i) the so-called heavy, stable charged particles (HSCPs), which apply to stable, electrically charged particles but also charged particles that decay in the calorimeters and/or MS; and ii) magnetic monopoles.

*Heavy Stable Charged Particles (HSCPs)*

The searches for HSCPs at CMS [4, 299] and ATLAS [216, 300, 301] rely on two key properties. First, particles that are massive and/or electrically charged with $|Q| \neq e$ have a characteristic ionization loss ($dE/dx$) distinctively different than SM particles. This property can be measured in the tracker. Second, HSCPs are typically heavy and move with a speed less than the speed of light, $\beta = v/c < 1$. Thus, compared to a particle with $\beta \approx 1$, they require a longer time of flight (TOF) to reach the outermost components of the detector (calorimeters and muon chambers). As decays or interactions of the HSCP with the material in the detector can change the electric charge of the HSCP, both CMS and ATLAS perform separate *tracker-only* and *tracker + TOF* studies in the language of CMS [20]. The event selection relies on standard single-muon or large-missing-energy triggers. The offline selection relies on identifying the signal events from quality requirements on the tracks using discriminator variables built from track observables.

[20] ATLAS measures $\beta\gamma$ from $dE/dX$ and $\beta$ from time of flight and extracts an independent mass, $m_\beta$ and $m_{\beta\gamma}$, from each measurement.

The HSCP search conducted by LHCb [302] is slightly different. Instead of exploiting $dE/dx$ and time of flight, they use the lack of radiation in the ring imaging Cherenkov detector (RICH). Events are required to pass a high-$p_T$ single muon trigger ($p_T(\mu) >$ 15 GeV). Two opposite sign "muons" are then required, each with $p_T$ above 50 GeV and an invariant di-muon mass above 100 GeV, to suppress muons coming from DY production, the main background for this search. In addition, particles must have $\beta > 0.8$, set by the efficiency of the muon chambers to reconstruct slow particles. As electrons and hadrons interact more with the calorimeter than an HSCP, a deposit in the calorimeter of less than 1% of the momentum of the particle is required.

The theoretical interpretation of a signal or limit depends on whether the HSCP carries both color and electroweak charges. If it carries a color charge, the default benchmarks correspond to $R$-hadrons, namely HSCPs that hadronize with SM particles via the strong force, e.g., gluino-gluon or quark-squark states. In the absence of a color charge, the signal is exemplified by long-lived sleptons in the context of gauge-mediated SUSY. Both ATLAS [300] and CMS [4] studies employ these two scenarios, while LHCb [302] uses a stau benchmark model [21]. Finally, CMS also looks for HSCPs coupling only to hypercharge (and hence possessing only couplings to $\gamma$ and $Z$), while ATLAS has a search inspired by electroweakinos in SUSY: it considers the associated production of a neutral and an electrically charged LLP (chargino-neutralino), and thus only one HSCP plus missing energy are required. These scenarios correspond, in our simplified model framework, to the direct production of LLPs with electric or color charges and that do not decay, or decay at very large distances compared to the tracking volume.

[21] The ATLAS R-hadron searches using the 13 TeV dataset have recently been presented in Ref. [216].

To summarize, these searches present no obvious weak points. Standard triggers and tracking algorithms are used, and the analysis methods are well-understood and have been extensively validated against data. HSCP signatures do not suffer from the low-mass gap of many neutral LLPs due to constraints from LEP and other low-energy experiments. However, milli-charged particles are not covered by these searches and require dedicated detectors (see Section 5.3.3). While the HSCP search strategies are generally robust, we encourage the experimental collaborations to continue pursuing improvements for these searches. The small number of signal events that would be produced for HSCPs above current limits render the sensitivity highly dependent on the understanding of the background and control of the systematics.

*Magnetic Monopoles*

ATLAS [208] has a dedicated search for highly ionizing particles (HIPs), which encompass a variety of new physics scenarios, such as magnetic monopoles, dyons (particles with both magnetic and electric charge), stable microscopic black holes, etc. For the sake of concreteness, we focus on magnetic monopoles but the interpretation in terms of other models is straightforward. [22]

[22] At the time of writing there was no public result on a monopole search from CMS.

The main phenomenological feature is that magnetic charge is quantized in units of $g_D = 2\pi/e \approx 68.5$. Hence, a magnetic monopole behaves as a particle with at least 68.5 electron charges, leading to an unusually large ionization power in matter, so that they would quickly stop in the detector because HIPs lose energy at spectacular rates. Because of the large QED coupling of magnetic monopoles, a perturbative calculation of the cross section is invalid and there is no accurate determination of the production rate, but a naïve Drell-Yan production cross section is provided for the purposes of comparison. The specifics of the detector restrict the sensitivity of this search to magnetic charges $g < 2g_D$ because a large fraction of the monopoles stop in the material upstream of the electromagnetic calorimeter, as the latter is used for the L1 trigger [303]. We note that larger magnetic charges can be tested by the MoEDAL experiment [304], which is described in detail in Section 5.3.2. Additionally, theories with monopoles can also be tested in heavy ion collisions [305].

ATLAS [208] has a dedicated trigger for HIPs based on identifying relevant RoIs in the ECAL and subsequently counting the number of hits in the transition radiation tracker (TRT). As well, the fraction of TRT hits that are high-threshold (HT), meaning that they have an ionization larger than $\sim 3$ times that expected from a SM particle, is used as a discriminant. The analysis selects events based on the fraction of TRT-HT hits matched to an EM cluster deposit, and how the energy deposits are distributed in the different layers of the ECAL. It is important to note that due to the lack of a consistent theory, the signal simulation is performed by re-scaling Drell-Yan production at leading order and assuming no coupling to the $Z$ boson. The HIPs are assumed to have either spin-0 or 1/2. The spin does not affect the interaction with the material, but the angular distributions are different according to angular momentum conservation (keeping in mind that there is, of course, no perturbative theoretical prediction for the angular distribution). Cross section limits for $0.5 < |g|/g_D < 2$ are set for masses in the 890–1050 (1180–1340) GeV range for the spin-0 (1/2) case.

The coverage in LHC Run 2 of magnetic monopoles in the $g_D - \sigma$ plane is displayed in Figure 3.6.

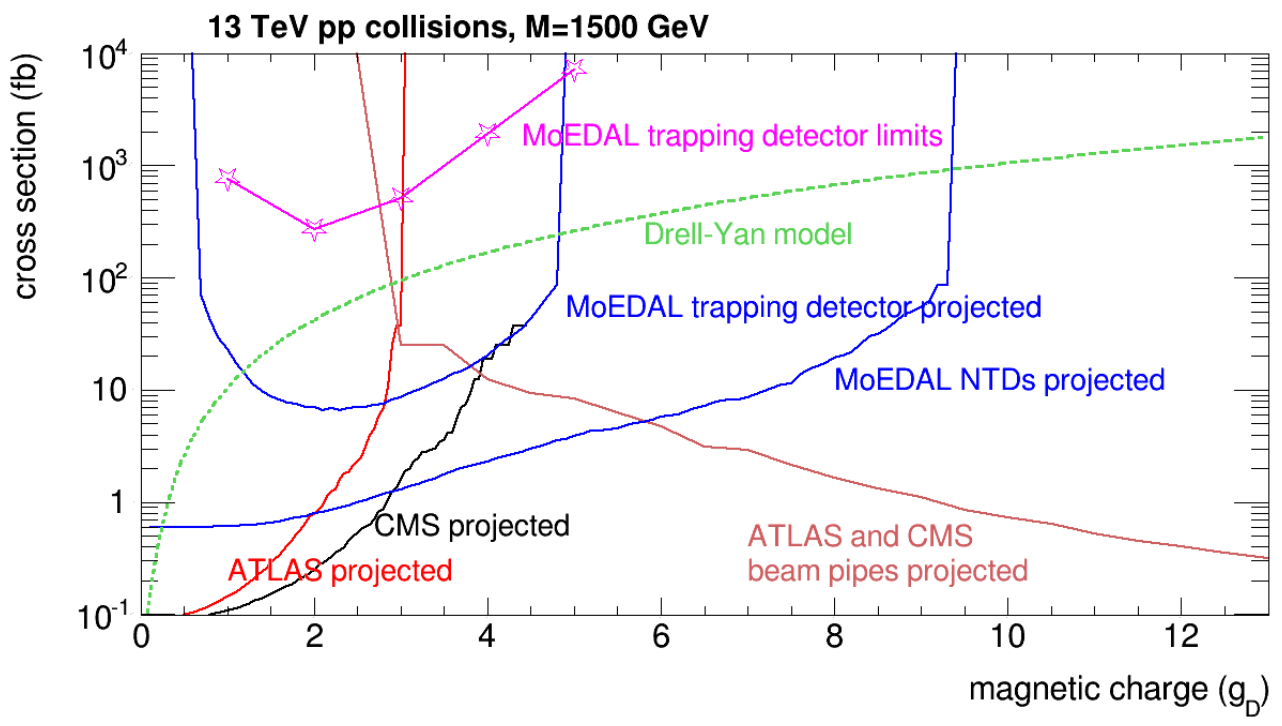

Figure 3.6: Comparison of the MoEDAL, ATLAS and CMS reaches for magnetic monopoles. The curves assume three signal events and a total integrated luminosity of 50 fb$^{-1}$ for 13 TeV collisions [306]. Updated version of existing figure in Ref. [303]. Note that the CMS curve relies on the expected performance of their detector.

### 3.5.2 *Tracks with Anomalous Geometry*

In this category we group searches where the tracks have an anomalous geometry, namely they disappear (track $\to$ nothing), or follow non-standard trajectories (such as quirks). There are additional anomalous track signatures that we do not cover such as kinked tracks, where a charged LLP decays to another charged particle that travels at a non-zero angle with respect to the LLP direction [105, 307–309]; however, there are currently no dedicated searches for such signatures.

Within this category we also have the emerging jets signature that has been recently studied by CMS [310]. Since emerging jets arise from dark sector radiation, they are described in detail in Chapter 7 in conjunction with the theoretical and phenomenological aspects of dark showers.

*Disappearing Tracks (DT)*

Massive charged particles traveling in the detector can decay to a lighter, almost mass-degenerate neutral state, emitting a soft SM charged particle (typically a pion or a muon). While at first glance a small mass gap naïvely seems like a hallmark of tuning, near degeneracies often occur naturally as a result of a symmetry. In fact, electroweak symmetry generically leads to small mass splittings between components of a single electroweak multiplet. For example, $\mathcal{O}$ (100 MeV) splittings arise between the different components of an electroweak multiplet [57, 121] due to EW gauge boson loops [23]. If the SM particle is sufficiently soft it cannot be reconstructed, and then a charged track seems to vanish: this is thus referred to as a disappearing track [24]. The actual lifetime of the charged particle is highly sensitive to the precise value of the mass splitting. For instance, the well studied cases of a fermionic doublet with $Y = 1/2$ and a fermionic triplet with $Y = 0$, reminiscent of a Higgsino and Wino in supersymmetry, respectively, have mass splittings of $\Delta$ = 355 and 166 MeV, up to small corrections, but the corresponding $c\tau$ values differ by almost an order of magnitude: 6.6 mm versus 5.5 cm [25]. This is because the lifetime, $c\tau$, depends on the third power of the mass splitting in these scenarios when the charged LLP decays to a charged pion [57, 121].

Before 2017, both ATLAS [311] and CMS [210] required a track to travel about 30 cm in order to be reconstructed, giving good coverage of the Wino scenario. This 30 cm value corresponds to four hits at ATLAS, three in the pixel layers plus one in the silicon tracker, and to seven hits in the pixel and trackers of CMS. The search employs a trigger requiring an ISR jet against which the charged particle recoils, along with the presence of large $\not{E}_T$. The disappearing track is reconstructed offline and needs to fulfill quality criteria (isolation, $p_T$ threshold, etc.). A phenomenological study [92] has shown that reducing the distance from 30 to 10 cm would give coverage to the elusive Higgsino scenario, moving the expected reach up to 400 GeV, surpassing the expected mono-jet reach of 250 GeV [312–314]. Later, ATLAS presented a study [211, 315] using 13 TeV data and exploiting the presence of a new innermost pixel layer (IBL). This addition allows for all four hits to be in the pixel, with the outermost pixel layer now at 12.25 cm, enhancing sensitivity to lower values of $c\tau$. The summary for disappearing tracks at ATLAS for the Wino case can be seen in the left panel of Figure 3.7, while in the right panel we show the constraints for Higgsinos from Ref. [92]. CMS also has a disappearing tracks search using 2015 and 2016 data at a center-of-mass energy of 13 TeV [316].

At LHCb the prospects for a disappearing track analysis with the present detector are poor. Currently, the momentum of the track can only be measured if the particle passes through the tracking station (TT), which is about 3 m away from the IP. Particles decaying in the VELO or RICH1 system will not leave a fully-measurable track and will be swamped in a background of SM processes such as kaon decays, which would give a similar signature in the detector components before the magnet. Detector improvements (additional magnets, better PID at low momentum, additional layers) might lead to some sensitivity for $\mathcal{O}$(cm) lived tracks, a golden opportunity for potential LLP discoveries at LHCb in the HL-LHC era.

[23] For a single fermion multiplet, the splitting can only be altered by higher-dimensional operators, and thus it is harder to vary $\Delta$ from the 1-loop EW value. For other cases, such as mixing with additional particles, the actual splitting can differ more substantially from this 1-loop EW value.

[24] If the charged particle could be reconstructed this case is often referred to as a *kinked track*. However, as the kinked portion has a very large impact parameter, without a serious attempt to capture the kink these tracks, too, simply disappear.

[25] While these values set a concrete physics target, we stress again that the mass splitting can be arbitrary in other corners of the BSM parameter space (even within SUSY). For instance, $\tilde{\tau} \to \tau\tilde{\nu}$ (where the stau and sneutrino masses are free independent parameters) or for scalar particles (e.g., $H^+ \to \mu^+ H^0$), where the mass splitting and the overall mass scale are set by arbitrary quartic couplings.

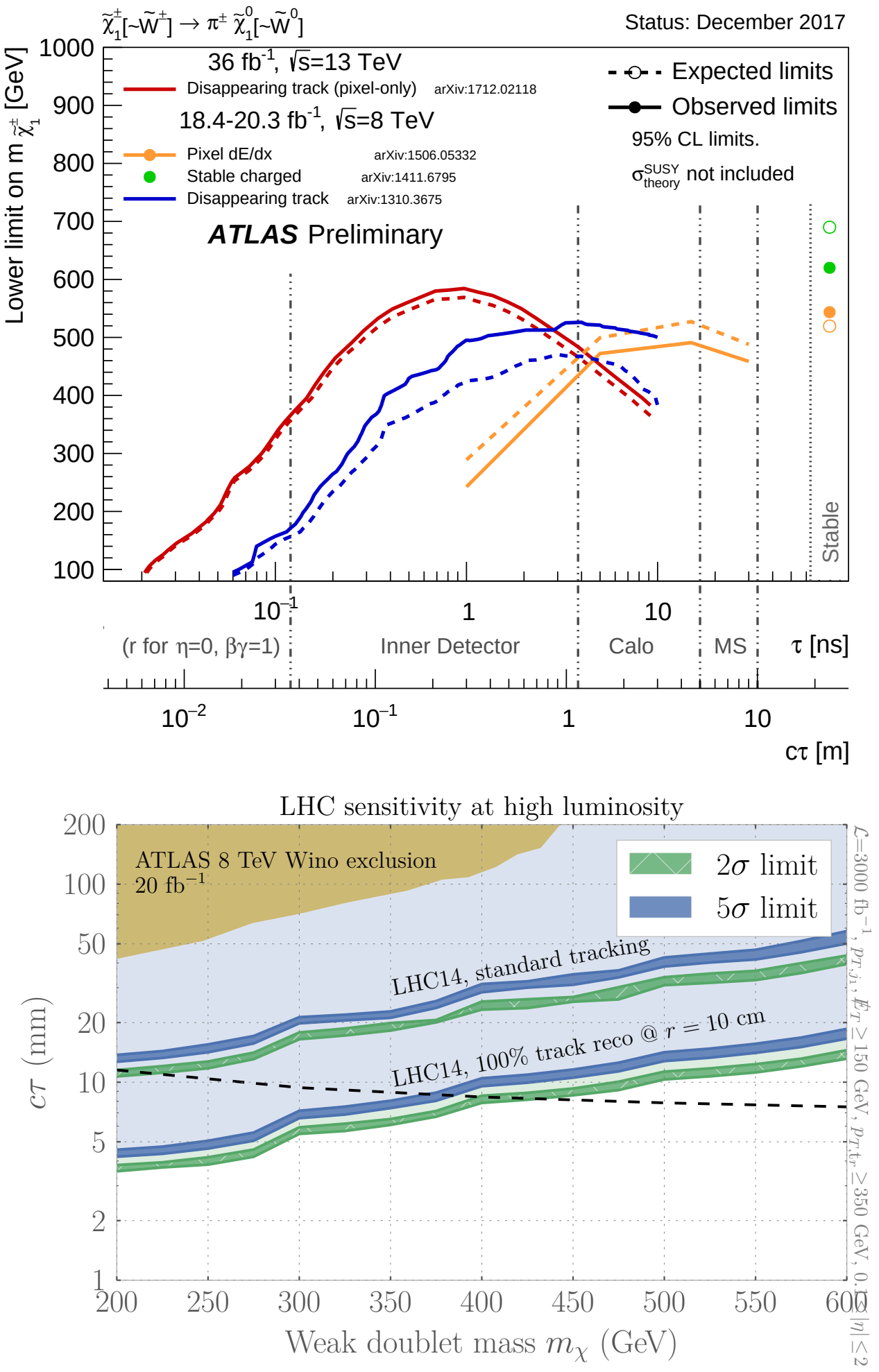

Figure 3.7: **Top:** Summary of ATLAS disappearing-track searches as applied to a Wino (electroweak triplet) benchmark scenario [317]. **Bottom:** HL-LHC projected constraints on the Higgsino scenario [92].

To summarize, the search for disappearing tracks presents a few challenges. Using an ISR jet trigger means a price is paid in terms of signal efficiency. For example, Ref. [92] has shown that significantly lowering the $p_T$ threshold of the jet or directly triggering on the momentum of the disappearing track [26] would lead to a factor of two increase in the number of signal events. It is also clear that better access to lower lifetimes is needed; this may only be possible, for instance, by adding new tracking layers as close as possible to the beampipe (and/or having double layers instead of single ones).

[26] While currently there are some proposals to trigger on tracks [318], those predominantly apply to standard tracks. In particular, the new fast track reconstruction (FTK) system at ATLAS requires 10 hits in inner detector silicon, which corresponds to a decay radius of less than 9 cm. However, pattern banks for hits including high-$|d_0|$ tracks are currently being considered out to $d_0 \sim 2$ cm [319, 320].

*Strongly Interacting Massive Particles (SIMPs)*

Strongly interacting massive particles (SIMPs) can be motivated by astrophysical observations of dark matter that do not fully agree with the WIMP paradigm (e.g., missing satellites, the core vs. cusp problem; see, e.g., Refs. [321–324] for further discussion). These particles are assumed to interact strongly with baryons. Consequently, the experimental signature is little to no signal in the tracker and the ECAL, and large energy deposits in the HCAL. Such a final state with trackless jets also arises in the context of emerging jets [325], and ATLAS has a trigger for trackless jets in association with a soft muon (where the muon is required to fire L1 of the trigger) [249]. Additionally, the CalRatio trigger and associated search for displaced hadronic decays (addressed in Section 3.1) is designed to be sensitive to a similar signature and could provide some coverage of this signature as well. Strictly speaking, SIMPs are not a track-based signature, but we include them here because the interactions of the SIMPs with the tracker are different from usual hadrons in jets, while the calorimeter signatures are similar.

An LHC phenomenological study of SIMPs was carried out in Ref. [326]. We summarize the main points of the study here. In their setup, SIMPs interact with the SM via an attractive potential (either scalar or vector mediator) coupling SIMP pairs to $q\bar{q}$ pairs. The proposed analysis selects events with high-$p_T$, back-to-back jets within the tracker, exploiting the charged energy fraction within a jet to discriminate signal from background. The astrophysical experimental constraints on this scenario are compared with the expected reach of this search and that of mono-jets in Figure 3.8. Currently there is an ongoing analysis in CMS pursuing this strategy.

*Quirks*

Quirks are particles charged under both the SM and a new confining gauge group [175], referred to here as "infracolor" (IC). The defining property of quirks is that the tree-level quirk masses $m_Q$ are above the confinement scale $\Lambda_{IC}$ (and thus similar to QCD but with no light-flavored quarks), so that there is never enough local energy density to create new quirks out of the vacuum. A pair consisting of a quirk and an anti-quirk can live in a quantum-mechanical bound state where the constituents are separated by macroscopic distances $\ell \sim \frac{m_Q}{\Lambda_{IC}^2}$, remaining connected by an infracolor flux tube. The infracolor flux tube exerts a force on each quirk that causes its trajectory to differ from the expected helicoidal ones for SM particles.

The collider phenomenology depends greatly on the size of $\ell$. If $\ell$ is much less than an Å, the rapid emission of infracolor glueballs results in the quirks annihilating before ever reaching the beampipe. For large enough confinement scales, the infracolor glueballs can decay back into SM particles on sufficiently long

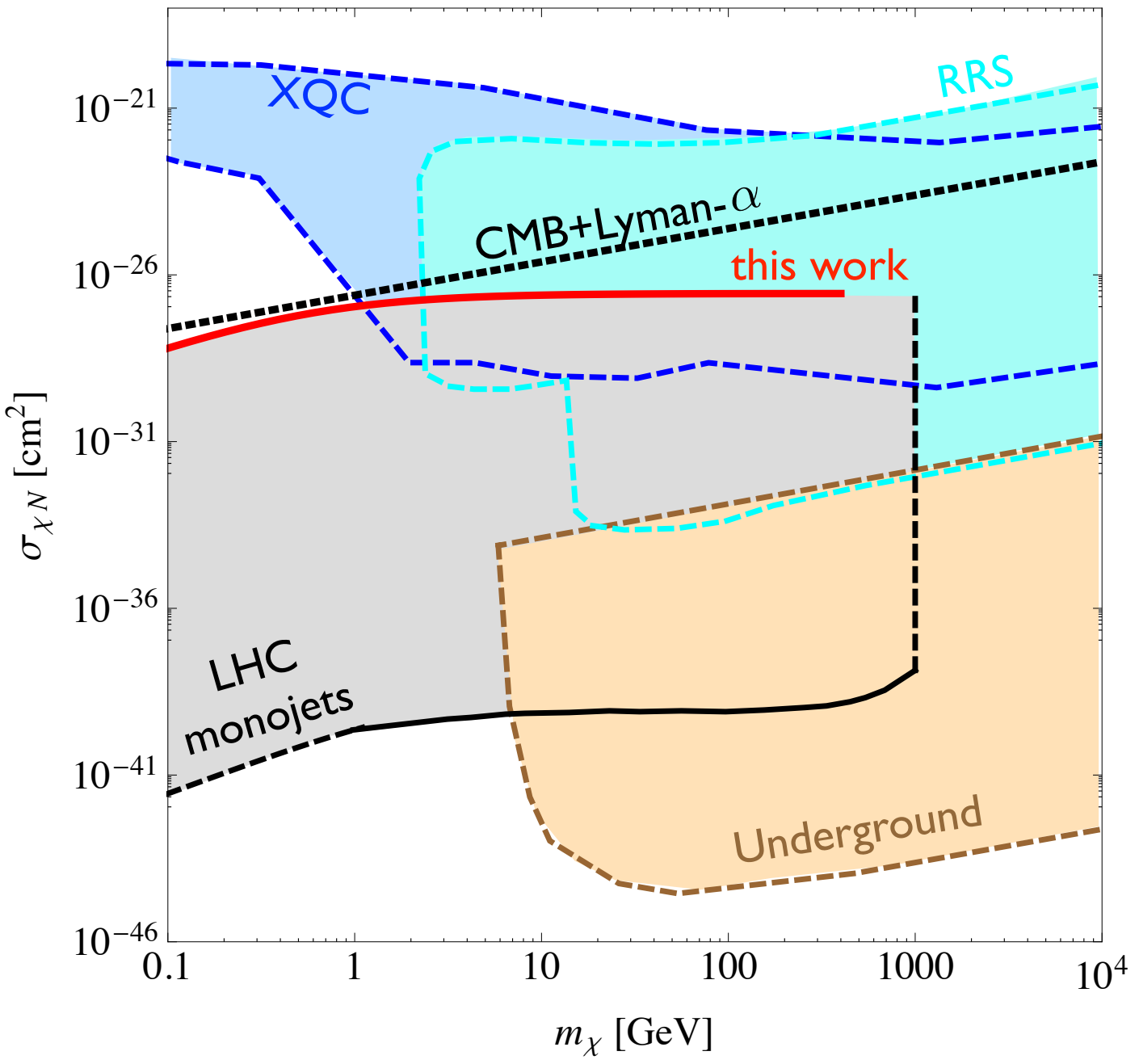

Figure 3.8: Astrophysical and collider constraints on a simple SIMP setup. Note that the relevance of the astrophysical constraints depends on the contribution of the SIMPs to the relic density. Taken from Ref. [326].

.

time scales that they can be distinguished from prompt signatures. While in this specific case, quirks produce hidden valley [59] or emerging jet [325] signatures due to these long-lived infracolor glueballs (see Chapter 7), we stress that elsewhere in parameter space quirks exhibit their own distinct phenomenology and are not merely a subset of hidden valleys, contrary to popular lore.

If $\ell$ is larger than an Åbut below the mm scale, the individual quirks are not distinguished from one another. However, the pair (which is overall neutral, and therefore does not bend in the magnetic field of the tracker) appears as a single, highly ionizing straight track with missing energy aligned with it, the latter arising from mis-measurement of the track momentum. The $D0$ collaboration has searched for precisely this signature [327], requiring an additional jet for trigger purposes. This search obtained lower bounds on quirk masses of 107, 119 and 133 GeV for SU(2), SU(3) and SU(5) gauge groups, respectively. However, no extensions of this search to higher mass have been performed at the LHC, and the existing HSCP searches require low uncertainties on the track momentum that a straight quirk track would not satisfy.

Conversely, if $\ell$ is very large then the existence of the confining force has no effect on the quirk motions, and HSCP searches apply directly to the quirk case, with quirks charged under QCD yielding $R$-hadrons and uncolored quirks leading to slepton-like signatures.

We refer the reader to the Section 3.5.1 for more information.

For intermediate values of $\ell$, there are interesting phenomenological prospects at the LHC that have been recently studied theoretically [232, 233, 328], but for which there are no current public searches by the LHC collaborations. The first study [232] has recast mono-jet and HSCP searches, finding bounds up to 1.7 TeV for colored quirk masses. In addition, it has also proposed using the CMS dataset taken with zero magnetic field. In this dataset, all SM particles are expected to follow a straight trajectory, but the quirks would still bend due to the string tension. The second study [233] has proposed a new algorithm to search for quirks, exploiting the fact that the quirk and anti-quirk pair should lie in the same plane with the highest sensitivity in the $\ell \sim$ 1-10 mm range. This avoids the necessity of fitting non-helicoidal trajectories, and has the potential to extend the sensitivity to quirks well beyond the current mono-jet and HSCP limits. The third study [328] considers quirks that lose energy through material interactions with the detector. A charged or colored quirk pair can come to a stop within the detector, and annihilate out-of-time with active $pp$ collisions, allowing for sensitivity from stopped particle searches across a wide range of characteristic length scales, $\ell \sim$ Å - km range. We refer the reader to the Section 3.5.3 for more information on stopped particle searches.

Because of the non-standard nature of the tracks, quirk phenomenology poses substantial challenges in their experimental reconstruction, and the lack of constraints on quirks have already attracted the attention of the ATLAS, CMS and LHCb experiments. It would be desirable to test how the phenomenological proposals in Refs. [232, 233, 328], among others, can perform in a realistic detector simulation of one of the LHC experiments.

### 3.5.3 *Out-of-Time Decays of Stopped Particles*

This category is unique because LLP decays occur out-of-time with the collision. Indeed, decays can even occur when the LHC is not running! The only member of this class is the search for stopped particles (SP), which we describe below.

If an HSCP is produced with very low kinetic energy, it can come to rest in the detector due to interactions with the detector materials. This most likely occurs in the calorimeters or the steel yoke in the muon system as a result of their high material densities. The HSCP can then decay at a later time when no collision is taking place (known as an out-of-time decay). This experimental restriction reduces the types of background processes affecting the search, with the dominant backgrounds coming from cosmic rays, beam halo, and instrumental noise.

In the Run 1 analyses from ATLAS [229] and CMS [227], events are selected with a dedicated trigger selecting bunch crossings which are empty and have no bunches of protons nearby. The anal-yses require a jet with $p_T$ ($E$) above 30 (50) GeV at ATLAS (CMS). ATLAS further supplements the hardware trigger by requiring $p_T(j) > 50$ GeV, $|\eta| < 1.3$ and $\not{E}_T > 50$ GeV, rejecting instrumental noise. In addition, CMS has updated the jet search in Run 2 and also provided a search that triggers on out-of-time muons, both of which use the 13 TeV dataset [228]. The latter also employs the displaced stand-alone (DSA) muon reconstruction algorithm [329].

An offline selection procedure is aimed at reducing the main backgrounds. Muons coming from cosmic rays can be identified due to their distinctive topology. The "beam halo" background is the result of protons interacting with residual gas in the beampipe, the beampipe itself, or collimators upstream from the interaction point. Most particles will not travel far before being absorbed by various structures, but muons will travel parallel to the beam and can leave calorimeter deposits out of time with a proton–proton collision. However, these deposits will often be accompanied by corresponding horizontal tracks in the muon systems and can thus be efficiently vetoed. Instrumental noise is rejected in CMS by exploiting the anomalous response in the HCAL.

Stopped particle searches provide an alternative way of probing charged particles besides more conventional HSCP searches. HSCP searches are typically more sensitive to any signature with a charged particle, so stopped particle searches are not often expected to be a discovery mode for most simple new physics scenarios with charged LLPs [27]. The typical added value of SP searches is that they can help identify and characterise positive signals in HSCPs, for instance by providing a cleaner extraction of the lifetime and also to properly identify the decay products. However, in some cases (such as for quirks, where HSCP searches are insensitive in much of parameter space), it has been argued that an SP search could actually be a discovery mode if modifications are made to the search strategy to improve sensitivity to quirks [328]. These modifications that would increase quirk acceptance or lower backgrounds include expanding the $\eta$ range, implementing higher energy thresholds, using the timing information, and considering shower evolution in the new CMS endcap calorimeter [330].

[27] The reason why the SP searches are less efficient than the HSCP ones is twofold. On one hand, only a fraction of LLPs stop in the detector, while the HSCP search only requires that the LLP crosses the detector. On the other hand, the SP is only looked for in a specific time-window that might fail to catch a large fraction of them.

## 3.6 *Discovery Opportunities: Overview of Gaps in Coverage*

In the preceding sections (3.1–3.5.3), we have examined the so-called "coverage" of existing searches for LLPs at the LHC with the explicit and express purpose of identifying uncovered realms and places where discoveries could be hiding. Here, we summarize these gaps and potential opportunities for LLP discovery in bullet form, as a to-do list for the experimental community.

1. All-hadronic LLP decays

    - Associated-object triggers (especially motivated by Higgs-like VBF and VH production modes) need to be more comprehen-

sively used to improve sensitivity to low-$p_T$ objects.

- Improvements are needed in sensitivity at lower masses & lifetimes (e.g., for LLPs produced in Higgs decays).
- Single hadronic DVs need to be looked for in searches that currently use two (such as decays in ATLAS HCAL and MS).
- Possibilities need to be explored for ATLAS and CMS for online reconstruction of hadronic displaced objects, as the inclusive $H_T$ triggers used by the two collaborations miss these objects unless they have a large $p_T$. (By constrast, LHCb can trigger on a displaced hadronic vertex [246, 247].)
- Low-mass hadronically decaying LLPs can look somewhat like tau leptons, so the question remains as to whether there is any possibility of using, for example, L1 tau triggers to seed displaced jet triggers at HLT and improve trigger efficiency; studies need to be performed by the experimental collaborations.
- The prospects for dedicated searches for displaced hadronic taus need to be investigated, since no dedicated searches currently exist.
- The potential for flavor-tagging displaced jets (b-displaced jets, c-displaced jets, etc) needs to be explored.

2. Leptonic

- Coverage needs to be provided for the intermediate region between boosted, low-mass LLPs (lepton jets) and high-mass, resolved LLPs (resolved ATLAS/CMS searches).
- Improvements need to be made to extend coverage to lower masses and to lower $p_T$ thresholds. Currently no prescription or plan for this exists, and so dedicated studies need to be done.
- Searches need to be done for different combinations of charge and flavor of displaced leptons (e.g., same-sign vs. opposite-sign, opposite-flavor vs. same-flavor).
- Searches need to be done for tau leptons in LLP decays, in particular if they come from the ID; an unanswered question remains as to whether displaced-jet triggers can be used for this purpose.

3. Semi-Leptonic

- Searches do not exist and need to be done for LLP masses below about 30 GeV; this mass range is theoretically well motivated by Majorana neutrinos.
- Searches need to be performed for all flavor combinations (for example, one CMS search only covers $e^{\pm}\mu^{\mp}$), as well as same-sign vs. opposite-sign leptons.

- Currently unknown improvements need to be made to relax or modify isolation criteria wherever possible to recover sensitivity to boosted semi-leptonic decays.
- Searches need to be done that better exploit triggering on associated objects for improved sensitivity to low-mass objects, or to employ high-multiplicity lepton triggers if there are multiple LLPs.

4. Photonic
   - There is currently no coverage for LLPs decaying into $l\gamma$, $j\gamma$, or without $\not{E}_{\mathrm{T}}$, and searches urgently need to be performed for this decay topology.
   - There is currently poor coverage (i.e., there exists no dedicated search) for single-$\gamma$ topologies. The only searches with sensitivity require two jets to be present in addition to $\not{E}_{\mathrm{T}}$ [296]. Studies are needed to assess the sensitivity of this search to signals with only one delayed photon and different jet multiplicities.
   - There is currently no coverage for softer non-pointing or delayed photons, and searches need to be performed for these kinematic realms.
   - Studies need to be performed to determine if triggers on associated objects may improve sensitivity to signals with a single photon, without $\not{E}_{\mathrm{T}}$, or for lower-$p_{\mathrm{T}}$ photons

5. Other exotic long-lived signatures
   - Disappearing tracks with $c\tau \sim$ mm are very hard to probe, and new ideas and detector components are needed to extend sensitivity to this potential discovery regime. It's unclear if the ATLAS insertable B-layer will be present in HL-LHC run and how sensitivity to the disappearing track topology will improve with the replacement of the current inner detector with the new ITk (Inner Tracker), or whether new tracking layers very close to the beam line can be added. It's an open question as to what is the lowest distance at which new layers (or double layers) can be inserted. Another open question that needs to be answered is whether there are any prospects for disappearing tracks at LHCb with an upgraded detector.
   - No dedicated searches for quirks exist at the LHC, a huge, open discovery possibility for ATLAS, CMS, and LHCb. Some LHC constraints exist by reinterpreting heavy stable charged particle searches, but dedicated searches need to be performed. There are significant challenges in modeling the propagation and interaction of quirks with the detector, as well as in fitting tracks to their trajectories, but new ideas have been proposed that need to be explored by the experimental collaborations that might allow improved sensitivity to quirks with less ambitious analysis methods.

# 4
# Common Sources of Backgrounds for LLP Searches

**Contents**

**Chapter editors:** Juliette Alimena, Martino Borsato, Zhen Liu, Sascha Mehlhase

## 4.1 Introduction

For many searches for LLPs, the main backgrounds do not stem from irreducible SM processes, but arise instead from *external sources*. Indeed, there can even be backgrounds of instrumental and/or algorithmic nature. Often, LLP searches are designed to have a very small number of background events, sometimes even zero events, that pass the full selection criteria. This chapter gives an overview of common LLP search backgrounds and the means to estimate or control them.

## 4.2 Long-Lived Particles in the Standard Model

Weak decays of SM particles can naturally give rise to displaced vertices at the boosts typically encountered at the LHC. Searches for LLP signatures at sufficiently low LLP mass and lifetime suffer from large backgrounds due to displaced SM decays. One simple example is found in the search for long-lived dark photons decaying to $\mu^+\mu^-$ at LHCb [262], which drastically loses sensitivity when the dark photon mass gets too close to the $K_S \to \pi^+\pi^-$ invariant mass, despite the very low $\pi \to \mu$ misidentification rate.

Moreover, $b$-hadrons can decay at displacements of a few mm and can be challenging to distinguish from LLPs with masses of a few tens of GeV that decay to a pair of jets. Requiring a large track multiplicity of the displaced vertex and performing a mass fit to the dijet invariant mass can help to significantly reduce the effect of this background (see, for example, Ref. [244, 246]). Backgrounds from heavy flavors are typically more abundant in the forward region, as arises, for example, if the signature under study is an LLP from the decay of a SM-like Higgs boson. However, the LHCb forward detector, which was designed to study these SM decays, is, in most cases, capable of rejecting heavy flavor backgrounds more effectively than can be done in ATLAS or CMS. Furthermore, displaced tracks from $b$-mesons, which usually have impact parameters ($d_0$) of less than 2 mm, can be rejected by using a larger criterion for the minimum track $d_0$.

## 4.3 *Real Particles Produced via Interactions with the Detector*

Particles produced in the $pp$ collision can interact with nuclei of the detector material, giving rise to displaced vertices, and can mimic LLP signals. Vertices from these interactions will be positioned in regions of the detector containing high densities of detector material and are therefore effectively vetoed by using detailed material maps.

The LHC detectors have developed tools internal to the collaborations to define a material volume to be vetoed. As the detector configurations changed slightly from Run 1 to Run 2, material maps have been determined separately for each data-taking period for both the ATLAS and CMS collaborations, using collision data. Maps can be found for ATLAS for Run 1 in Ref. [331], CMS for Run 1 in Ref. [332], ATLAS for Run 2 in Ref. [230], and CMS for Run 2 in Ref. [333]. Additionally, the Run 2 maps for both are shown here in Figures 4.1 and 4.2 for ATLAS and CMS, respectively.

LHCb recently developed a precise material map of the VErtex LOcator (VELO) using beam-gas collisions [334], shown in Figure 4.3. Beam-gas collisions can be distinguished from long-lived heavy flavor backgrounds and their utilization allows the map to cover precisely the whole VELO geometry, not only the region close to the interaction point. This map was used to veto photon conversions to di-muons, which is the main background affecting displaced dark-photon searches at low mass [262]. In analyses, this material map, together with properties of a reconstructed secondary vertex and its constituent tracks, is used to construct a $p$-value that is assigned to the hypothesis that the secondary vertex originates from a material interaction. As a rule of thumb, LHCb material interaction background is dominant for vertices at a distance from the beam axis larger than 6 mm (where the VELO material begins). Below 6 mm the background is dominated by heavy flavor decays [335].

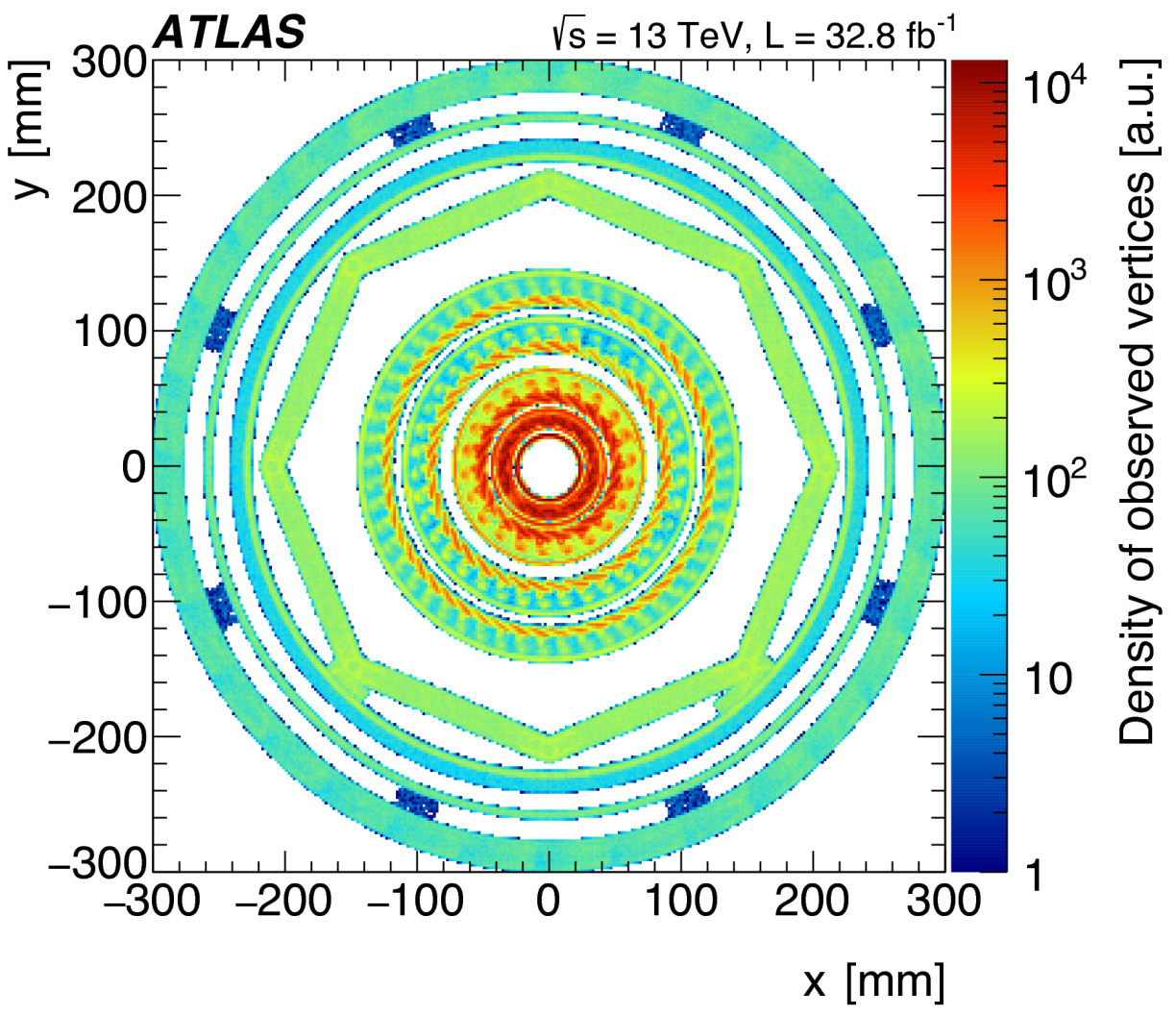

Figure 4.1: An example of a material map from ATLAS [230].

Because accurate material maps are essential to performing fully reliable sensitivity studies to signatures with displaced vertices, making them publicly available to the broader LLP community is of the highest priority. The availability of such tools in fast parametric simulations such as Delphes [336] would be very useful to reinterpret LLP search results. In Section 6.4.6.2, an example of a reinterpretation of an LLP search is presented, where a rough material veto was performed because these material maps were not publicly available, highlighting the shortcomings of such approaches in the absence of accurate material maps and emphasizing the benefits of making them available.

High-energy collision muons originating mainly from $W$ decays and creating secondary interactions in the tracker, calorimeters or muon system can be an additional source for displaced vertices mimicking LLP signals. This mostly minor background arises as vetoing these displaced vertices is based on a not 100% efficient detection of the high-$p_T$ track in the tracker.

Another important background, mainly for analyses targeting the reconstruction of decay vertices of long-lived particles reaching the muon systems, is hadronic or electromagnetic showers not contained in the calorimeter volume, so-called punch-through jets [241]. These punch-through jets occur especially in regions of reduced total interaction length in the calorimeters (e.g., transition regions between the barrel and the end-caps) and can be suppressed by either rejecting these $|\eta|$ regions or requiring a mini-mal number of hits in the muon system and isolating the displaced vertex from calorimeter jets as well as high-energy tracks and significant track activity in the inner tracking system. In order to not reject true vertices from displaced decays that occur near the end of the calorimeters, the calorimeter-jets veto should only consider jets with a minimum total energy deposit and, e.g., a minimum electromagnetic fraction of the total energy. The track isolation requirement aims at regions with a poor calorimeter measurement (again, transition regions in the calorimeters), where a single high-energy track or the sum of the track activity in a small cone around the displaced vertex could indicate a (punch-through) jet. On the other hand, punch-through jets, given their similarity to signal signatures, can also be used to evaluate systematic uncertainties due to imperfect modeling in the muon-system simulation.

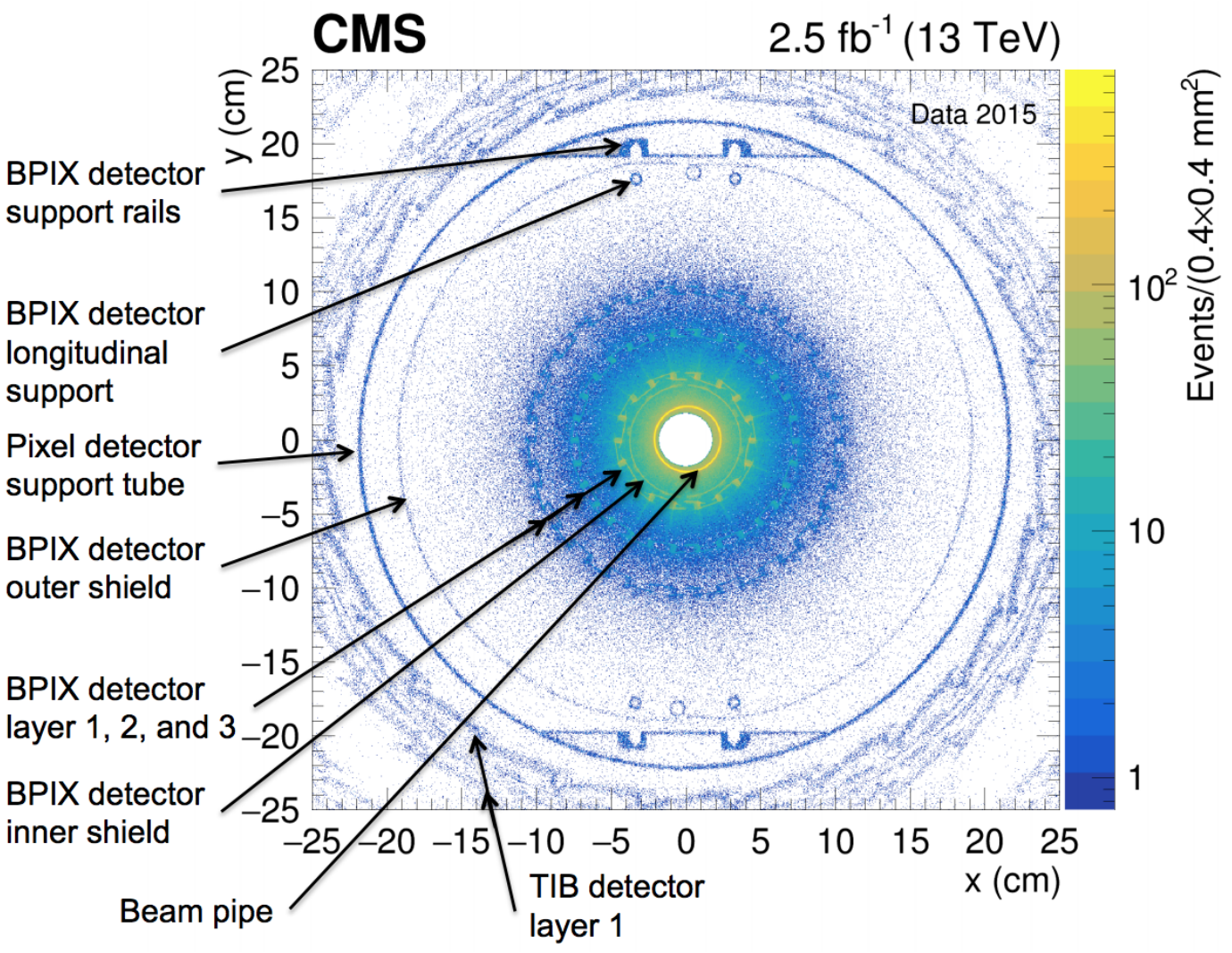

Figure 4.2: An example of a material map from CMS [333].

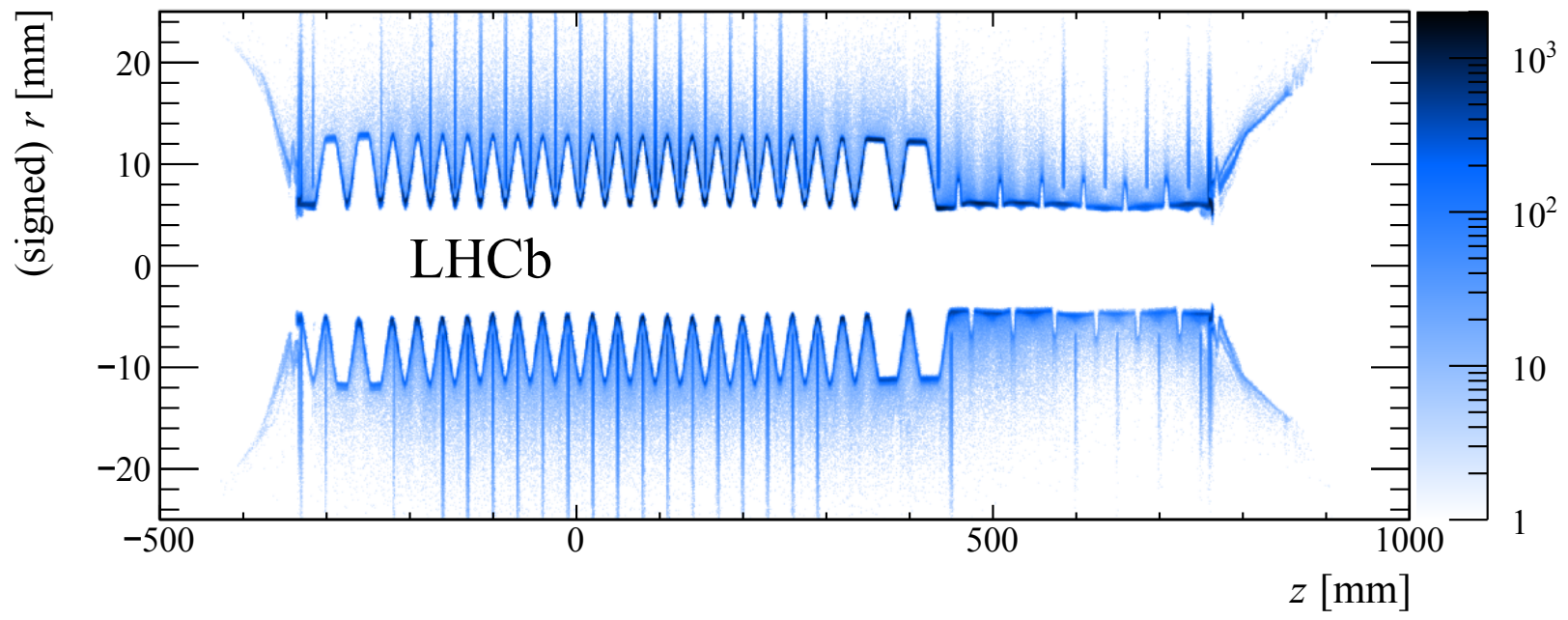

Figure 4.3: Reconstructed secondary vertices in the LHCb VELO from beam-gas collisions in the $zr$ plane integrated over $\phi$. These vertices are used to build the material map [334] to veto backgrounds from material interactions.

## 4.4 Real Particles Originating from Outside the Detector

There are several types of real particles generated outside the detector that could be sources of background in an LLP search.

### 4.4.1 Cosmic Muons

Cosmic rays from the atmosphere can enter the detector as cosmic-ray muons. These cosmic-ray muons can be reconstructed as displaced muons in the muon system or as displaced jets in the calorimeters. If cosmic-ray muons are reconstructed in the muon systems, they will typically appear as two back-to-back muons with $\phi$ values near $\pm\pi/2$. The rate of cosmic muons in the detector is about 500 Hz at L1, but depending on the HLT path and the offline selection used, the rate of cosmic-ray muons entering a given LLP analysis is generally much less.

Cosmic-ray muons are typically an important background source to consider for displaced signatures, especially those with large displacements [227, 229, 337–339]. Cosmic-ray muons are generally only an issue for LLP analyses in CMS and ATLAS since LHCb has coverage only in the forward direction.

For many analyses, cosmic-ray muons can be rejected with a simple veto on back-to-back dimuons. However, in some LLP analyses, this veto is not optimal for the signal acceptance or it is insufficient to suppress cosmic-ray backgrounds. Another often-used way to minimize the contribution from cosmic-ray muons is requiring high-momentum muons and/or high-energy jets, since cosmic-ray muons have a rapidly falling $p_T$ spectrum. In addition, if a search primarily looks for inner-tracker or calorimeter objects, cosmic-ray-muon events can be rejected by requiring little muon system activity [227, 337, 338].

If the cosmic-ray muon background is significant for an analysis, it can be estimated using data from dedicated cosmic data-taking runs or from empty bunches in $pp$ collision runs [227, 337, 338]. Cosmic ray muon simulations can be made, but in many LLP analyses, a data-driven approach is favoured if the simulation modeling is found to be insuffcent. Timing information in the calorimeters or the muon systems can be used to discriminate the signal from cosmic-ray muons, sometimes in conjunction with impact parameter variables.

### 4.4.2 Beam Halo

Another type of real particle generated outside the detector that could be a significant source of background for LLP searches is beam halo. Beam halo is produced when protons from the LHC beam scatter off the LHC collimators and produce debris, which can appear in the detector. Beam halo can create energy deposits in the calorimeters or hits in the muon system, both of which would be largely in the beam direction. These energy deposits or muon system hits would appear earlier than if they had been made from particles coming from the collision (see Figure 4.4). Beam halo is usually not modelled in MC simulation, since it is highly dependent on the beam parameters.

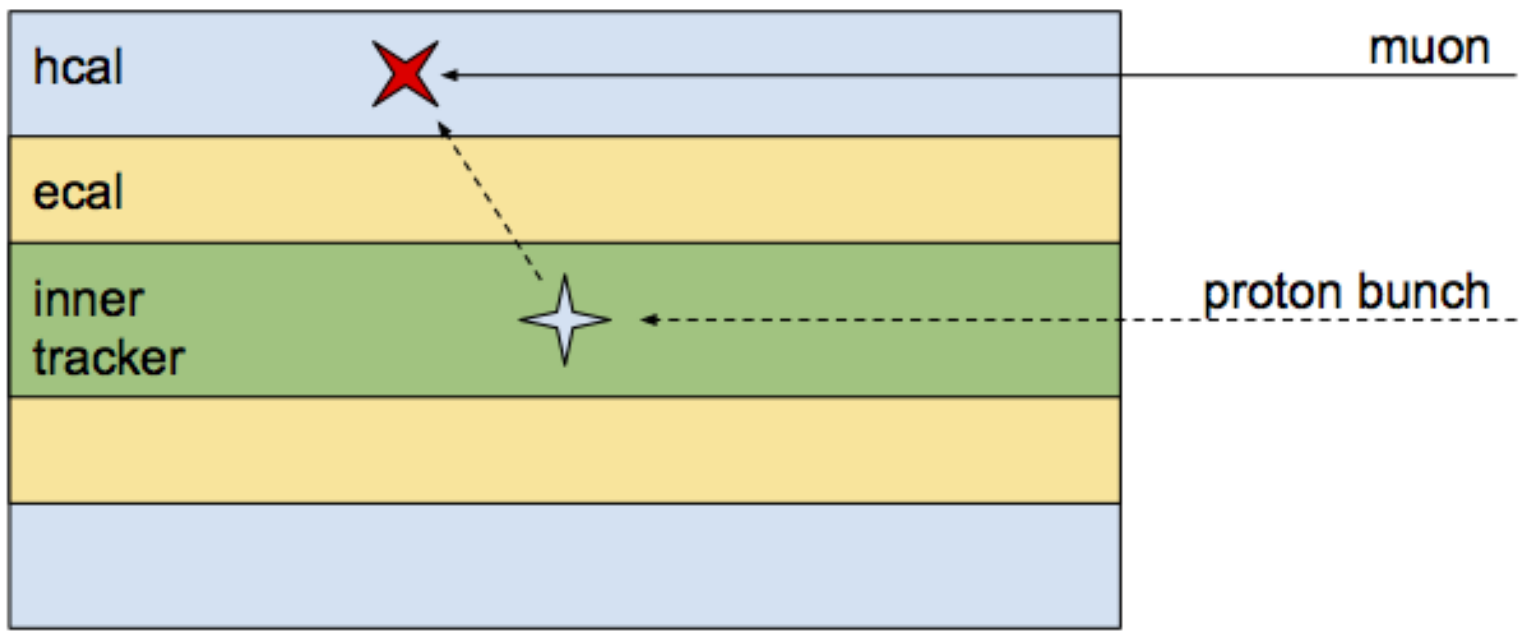

Figure 4.4: A sketch illustrating the timing differences due to the shorter, more direct path to the calorimeter cells between a beam-halo muon and a particle originating from the collision. The beam-halo muon is detected earlier than the particle from the collision.

Beam halo is most relevant for searches for displaced signatures without tracks in the inner tracker and for searches for decays in non-collision bunches (e.g., from stopped particles) [227, 337, 338], which are described in Section 3.5.

The contribution from beam halo can often be reduced by requiring high-momentum or high-energy reconstructed objects. One can also decrease the number of beam halo events by requiring central objects or vetoing forward muon system activity, since beam halo is usually in the very forward direction [227, 337, 338]. For inner tracker-based signatures, events from beam halo are rejected by requiring a minimum number of early hits; in this way beam halo is rejected due to its anomalous timing.

Beam halo background can be estimated using data control regions near $\phi = 0$ and $\pi$. One could also identify cells with a low number of (or zero) tracks that are assigned an early time.

### 4.4.3 *Cavern Radiation*

Diffuse backgrounds can also arise from proton–proton collisions filling the LHC caverns, consisting mostly of neutral, low-energy, and long-lived SM particles (i.e., neutrons and photons), leading to an overall increase in occupancy, especially in the muon systems. This so-called "cavern background" or "cavern radiation" can constitute a significant background in a LLP search. As simulations are resource-intensive, it is usually not at all modelled in MC simulation samples.

Cavern radiation is most relevant for searches looking in non-collision data, that is, stopped-particle searches, and for searches using muon system information to form tracks and vertices. It can be estimated from data by collecting events triggered by random triggers when there are no collisions, as was done in Ref. [229]. [1] Cavern radiation can also be estimated by overlaying a cavern radiation simulation with minimum-bias events from data.

[1] Note that these triggers are unlike those used to collect the search data for stopped-particle searches, which instead select events with physics objects during empty bunch crossings.

## 4.5 Fake-Particle Signatures

Another type of background for LLP searches is that from signatures that mimic real particles in the detector, but are in fact fake. Fake particles can originate from spurious detector noise. Noise appears differently for each detector, but in general, it is characterized by a single and concentrated energy deposit or hit that does not correspond in time or space to any other energy deposits or hits in the detector. Noise is usually difficult to model with MC simulation.

Calorimeter detector noise is most relevant for searches looking in non-collision bunches and low-energy collisions [227, 337, 338]. Muon system noise is most relevant for searches that are also highly affected by cosmic-ray muons.

Calorimeter noise can be rejected by vetoing single and concentrated energy deposits [227, 337, 338]. Muon system noise can be rejected by requiring high-quality muon tracks.

Noise in both the calorimeters and the muon systems could be estimated by looking at dedicated cosmic data-taking runs and then applying some selection criteria to reject cosmic-ray muons. The remaining events would most likely be noise.

## 4.6 Algorithmically Induced Fakes

For searches that aim to reconstruct the decay vertex of an LLP, and especially for long-lived particles decaying in the proximity of the interaction region, algorithmically induced fakes and/or instrumental backgrounds can be of importance. Algorithmic fakes can still be a significant background to LLP searches, even if a given detector is noise-free.

### 4.6.1 Random/Merged Vertices

This type of background, illustrated in Figure 4.5, is especially important in the environment close to the interaction region that experiences a high track density, and arises from two main sources. First, two or more individual tracks can cross each other and can be reconstructed as a displaced vertex. Second, two close-by, low-mass vertices can be reconstructed as one high-mass displaced vertex; such a final merging/cleaning step is often part of vertexing algorithms to reduce fakes in standard vertexing.

The former source is mostly suppressed by requirements originally targeting the removal of meta-stable SM particles: a minimum transverse impact parameter, $|d_0|$, for tracks and a minimum distance between the primary vertex and a given displaced vertex.

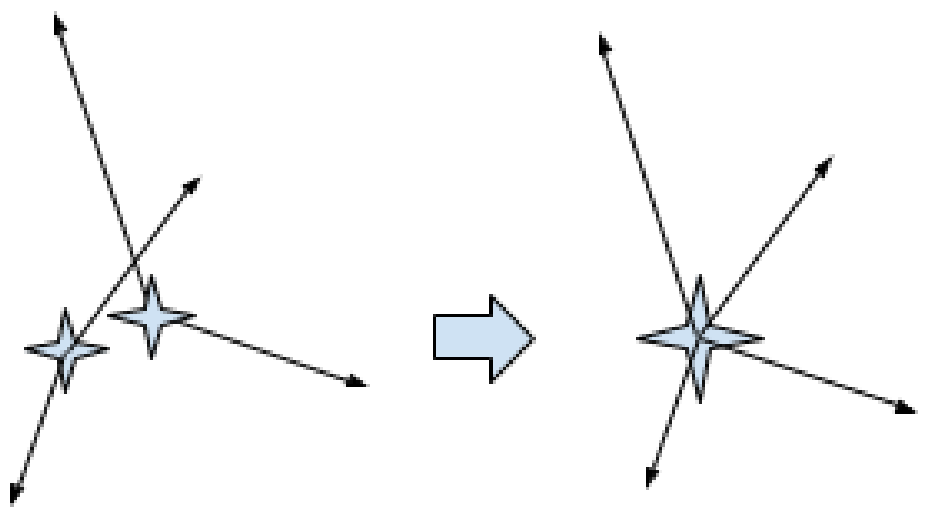

Figure 4.5: Illustration of two close-by, low-mass vertices being reconstructed as one high-mass vertex.

The latter source is harder to suppress, though can be estimated by randomly merging vertices from distinct events. By studying the number of reconstructed "merged" high-mass vertices as a function of distance between the two low-multiplicity low-mass vertices that were "merged" — both with vertices from the same event, as well as from different events and scaling them accordingly — an estimate for this background can be derived. This method has been successfully used in the ATLAS search for displaced vertices [230] and the ATLAS multitrack analysis [189].

### 4.6.2 *Randomly Crossing Tracks*

A background that is typically more relevant than merged vertices is the background stemming from low-mass displaced vertices crossed by unrelated tracks, resulting in the reconstruction of a high-mass vertex, as illustrated in Figure 4.6. The mass of the reconstructed displaced vertex is especially increased when the random track crosses the vertex in a direction that is perpendicular to the distance vector pointing from the primary vertex to the displaced one.

As demonstrated in detail in Ref. [189, 230], this background can be estimated by constructing vertices ($n$-track) from lower-multiplicity ones ($n-1$-track) by adding pseudo-tracks, drawn randomly from data-driven track templates derived for various radial detector regions. The normalization of the prediction is performed by comparing the $n-1$-track-based constructed vertices with the actual n-track vertices in all radial detector regions. One potential method for suppressing such backgrounds is to veto vertices where removing one track substantially decreases the mass of tracks associated with the vertex.

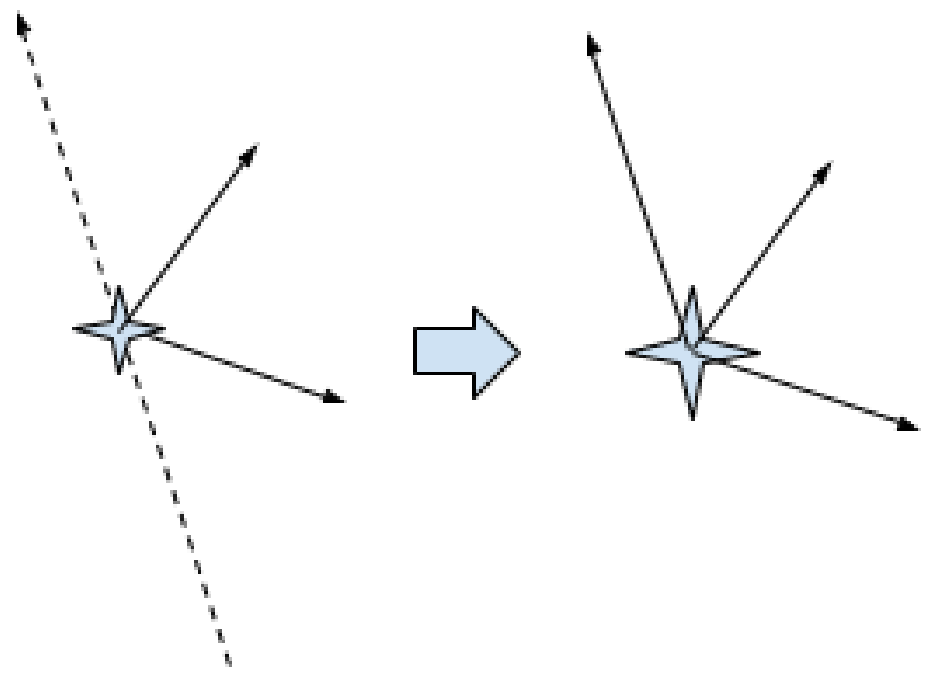

Figure 4.6: Illustration of a low-mass vertex crossed by an unrelated track and being reconstructed as a high-mass vertex instead.

## 4.7 *Summary*

LLP searches often have very low backgrounds, as opposed to searches for prompt particles. This makes LLP searches highly sensitive to signals of new physics.

There are, however, a few common sources of background that arise in different LLP searches: other, known, long-lived particles such as $b$-hadrons; real particles produced in the detector, such as particles produced in collisions that interact with the nuclei of the detector material; real particles produced outside the detector, such as cosmic muons or beam halo; fake particles, such as detector noise; and algorithmically induced fakes, such as two tracks that cross and are reconstructed as a displaced vertex, as described above. These backgrounds are generally atypical, difficult to model in simulation, and challenging to estimate. Thus, the possible appearance of unexpected background sources should be taken into account in any new LLP search, and the development of novel techniques and methods to estimate them is encouraged.

# 5
# Detector Upgrades

**Contents**

**Chapter editors:** Juliette Alimena, Martino Borsato, Yangyang Cheng, Monica Verducci

**Contributors:** Cristiano Alpigiani, Xabier Cid Vidal, David Curtin, Elena Dall'Occo, Sven Dildick, Jonathan L. Feng, Iftah Galon, Christopher Hill, Henning Keller, Felix Kling, Simon Knapen, Zhen Liu, Henry Lubatti, Philippe Mermod, Vasiliki A. Mitsou, James L. Pinfold, Jessica Prisciandaro, Livia Soffi, Sebastian Trojanowski, Carlos Vázquez Sierra, Si Xie, Charlie Young

The experimental searches for LLPs outlined in Chapter 3 are limited by the abilities of the ATLAS, CMS, and LHCb experiments to trigger on and reconstruct the objects that are associated with each signature. In the Phase 2 high-luminosity upgrade of the LHC (HL-LHC), the extremely high pile-up conditions necessitate the upgrade of all three detectors to maintain triggers at thresholds needed for sensitivity to electroweak, Higgs, and BSM physics (see Refs. [340, 341] for an overview); to reject particles originating from pile-up vertices; and to maintain object reconstruction in the high-luminosity environment. These upgrades include the addition of tracking layers to the forward regions of ATLAS and CMS, improvements to timing reconstruction in events, and the inclusion of tracking information at earlier stages of the trigger (or, in the case of LHCb, the full online reconstruction of every event).

While these upgrades are crucial to the success of the HL-LHC physics goals for conventional searches (such as for electroweak, Higgs, or SUSY signatures), they have the opportunity to be transformative for searches for LLPs. Signatures involving LLPs are often subject to small-to-negligible irreducible backgrounds, and improvements to the reconstruction, timing, and vertexing of displaced objects can typically suppress any instrumental or fake backgrounds. (See Chapter 4 for a discussion of sources of backgrounds for LLP searches.) At the same time, the introduction of tracking information to earlier stages of the trigger could be used to trigger on events that may contain hadronically decaying LLPs that must otherwise pass jet triggers, leading to an improvement in sensitivity to low-mass LLPs. Indeed, many of the gaps in coverage from current searches identified in Section 3.6 can be closed or reduced using the new technology from detector upgrades. Even more uniquely, LLP signatures may themselves motivate the introduction of new detector elements that are dedicated to exploring new lifetime frontiers in particle physics.

This chapter summarizes current and proposed plans for detector upgrades for the HL-LHC, paying special attention to those features of the detector upgrades that are most relevant for LLP searches. Where available, we show the results of projections for the sensitivity to various LLP scenarios of different improvements to the detector. We also highlight LLP studies for the Phase 2 upgrades that are not yet publicly available that should be done in order to assess (and, where possible, improve) the sensitivity of planned upgrades to LLP signatures. Finally, we include contributions from a number of existing and proposed experimental collaborations whose primary purposes are to search for LLPs produced at LHC interaction points using additional, dedicated detectors. These detectors complement the capabilities of ATLAS, CMS, and LHCb, allowing sensitivity to signatures that are otherwise not possible to reconstruct at the main detectors.

In Section 5.1, we combine the discussions of the planned ATLAS and CMS upgrades, facilitating for each detector component a direct comparison between the features of the upgraded detectors of both ATLAS and CMS. Since LHCb has a very different geometry and physics program from ATLAS and CMS, we have a separate discussion of planned LHCb upgrades in Section 5.2. Finally, we present the contributions of the dedicated LLP experiments in Sec-tion 5.3.

## 5.1 The ATLAS and CMS experiments

The planned upgrades to the ATLAS and CMS experiments for the HL-LHC will give the detectors increased coverage in the forward regions, better spatial and timing resolutions, and other new features including track triggers. The improved hardware capabilities, combined with software developments, give rise to exciting new prospects for future LLP searches. This section gives an overview of the upgrade scope (Section 5.1.1), discusses their physics potential (Sections 5.1.2-5.1.3), and presents new ideas for detector upgrade and LLP searches (Section 5.1.4).

Unless specified otherwise, the subsequent CMS experimental results are from its Technical Design Reports for the different sub-detector upgrades at HL-LHC, namely tracker [342], barrel calorimeter [343], endcap calorimeter [330], muon detectors [344], timing detector (Technical Proposal, Ref. [345]), and Level-1 Trigger (Interim Technical Design Report, Ref. [346]). ATLAS results are from the Technical Design Reports for the inner tracker pixel detector [347], the TDAQ system [348], the tile calorimeter [349], LAr calorimeter [350], muon spectrometer [351], and inner tracker strip detector [352].

### 5.1.1 Detector and Trigger Upgrades for High-Luminosity LHC

The High Luminosity LHC (HL-LHC) will begin with the third long shutdown (LS3) of the LHC in the coming decade (as of this writing estimated to begin at the end of 2023), where the machine and detectors will be upgraded to allow for pp running at a luminosity of $5 \times 10^{34}\,\text{cm}^{-2}\,\text{s}^{-1}$ in the nominal scenario, or potentially $7.5 \times 10^{34}\,\text{cm}^{-2}\,\text{s}^{-1}$ in the ultimate performance scenario. This will allow the ATLAS and CMS experiments to collect integrated luminosities ten times that of the current operations, which amounts to around 300 $\text{fb}^{-1}$ per year and 3000 $\text{fb}^{-1}$ during the projected HL-LHC lifetime of ten years (up to 4000 $\text{fb}^{-1}$if the ultimate instantaneous luminosity can be achieved).

The HL-LHC conditions create unique challenges in terms of high pile-up levels and high radiation dosage. About 140 pile-up events per bunch crossing, on average, are expected in the nominal scenario, and up to 200 pile-up events in the ultimate luminosity scenario. The radiation levels will be unprecedented: for the design integrated luminosity of 3000 $\text{fb}^{-1}$, a 1 MeV neutron equivalent fluence of $2.3 \times 10^{16}\ n_{eq}/\text{cm}^2$ and a total ionizing dose (TID) of 12MGy (1.2 Grad) is expected at the centre of the detectors, where the innermost silicon pixel tracking layers will be installed.

To meet the challenges of the HL-LHC operating conditions, and to fully profit from its physics capabilities, comprehensive upgrade programmes are planned for both the ATLAS and CMS experiments. This section summarizes the main detector and trigger upgrade plans for each sub-detector component of both experiments.

#### 5.1.1.1 Tracker

By the start of the HL-LHC, the inner trackers of both experiments must be replaced due to the significant radiation damage and performance degradation they have suffered. To maintain tracking performance in the high-density environment, and to cope with the increase of approximately a factor of ten in the integrated radiation dose, both the ATLAS and CMS experiments will entirely replace their inner tracking detectors.

*CMS Upgrade* The CMS tracker is composed of the inner pixel detector and the outer tracker. At the HL-LHC, the CMS inner pixel detector will include four cylindrical barrel layers covering the region of $|z| < 200$ mm, and forward extensions of up to twelve endcap disks on both sides (compared to the current configuration with three disks), which will extend its $|\eta|$ coverage from the current value of 2.4 to approximately 4. To maintain radiation hardness and reasonable occupancy, as well as to improve resolution, small, thin pixels will be used. For the studies in the CMS tracker TDR [342], pixels with a thickness of 150 $\mu$m and $25 \times 100$ $\mu m^2$ in size are used in the simulation [1]. The first layer of the barrel inner pixel detector will be positioned at a radius of 28 mm.

[1] An alternative option being considered is that of $50 \times 50$ $\mu m^2$. Larger pixel sizes of $50 \times 200$ $\mu m^2$ or $100 \times 100$ $\mu m^2$ are being considered in outer barrel layers and outer rings of the endcap as a potential option to reduce power consumption.

The CMS outer tracker is composed of six cylindrical barrel layers in the central region, covering the region of $|z| < 1200$ mm, complemented on each side by five endcap double-disks, in the region of $1200 < |z| < 2700$ mm. Modules are installed between $r \sim 21$ cm and $r \sim 112$ cm. Three sub-detectors are distinguished: the Tracker Barrel with pixel-strip modules (TBPS), the Tracker Barrel with strip-strip modules (TB2S), and the Tracker Endcap Double-Disks (TEDD). The inner rings of the TEDD disks use pixel-strip modules up to $r \sim 60$ cm, and the rest use strip-strip modules. The outer tracker modules, called $p_T$ modules, are composed of two single-sided, closely-spaced (1 to 4 mm separation) small pitch sensors read out by a set of front-end ASICs that correlate the signals in the two sensors and select the hit pairs (referred to as "stubs") compatible with particles above the chosen $p_T$ threshold. A $p_T$ threshold of 2 GeV corresponds to a data volume reduction of roughly one order of magnitude, which is sufficient to enable transmission of the stubs at 40 MHz. The "stubs" are used as input to the hardware trigger at Level-1 (L1), which enables track-finding at L1 for all tracks with a $p_T$ of 2 GeV or above. To improve the "stub"-finding efficiency and also to reduce material, the inner three outer tracker barrel layers, the TBPS, are made with flat modules in the center and tilted modules in the regions with larger $z$.

*ATLAS Upgrade* The ATLAS collaboration will replace its inner detector with a new, all-silicon tracker to maintain tracking performance in HL-LHC conditions.

The new ATLAS Inner Tracker (ITk) will consist of a greatly enlarged pixel system extending to roughly twice the radius and four times the length of the current pixel array, coupled with a much more finely-segmented strip detector. In total, the coverage of the full radius of the inner solenoid requires over three times the silicon area of the current detector.

The new system will consist of silicon barrel layers and disks (strips) or rings (pixels) with the possibility of inclined pixel modules to better cover the transition from the barrel to the end-cap regions. In detail, the strip detector has four barrel layers and six end-cap petal-design disks, both having double modules each with a small stereo angle. The strip detector, covering $|\eta| < 2.7$, is complemented by a five-layer pixel detector extending the coverage to $|\eta| < 4$. The combined strip-plus-pixel detectors provide a total of 13 hits for $|\eta| < 2.6$, with the exception of the barrel/end-cap transition of the Strip Detector, where the hit count is 11.

#### 5.1.1.2 Calorimetry

Both the ATLAS and CMS calorimetry consist of electromagnetic calorimeters and hadronic calorimeters. Different materials and designs are used for the two experiments.

*CMS Upgrade* For the CMS detector, the existing scintillating crystals in its electromagnetic calorimeter (ECAL) barrel (EB) will be kept for the duration of LHC. On the other hand, both front-end and back-end electronics will be replaced [343], which allows for higher transfer rates and more precise timing. The target timing resolution for the upgraded ECAL electronics is $\sim$ 30 ps for particles with $p_T \gtrsim 30$ GeV, which is the fundamental limit allowed by hardware and an order of magnitude smaller than the current limit. Current studies on the CMS hadronic calorimeter (HCAL) barrel radiation damage suggest there is no need for replacement at HL-LHC.

The CMS endcap calorimeter, including both the electromagnetic (EE) and the hadronic sections, will be replaced with a high-granularity, silicon-based calorimeter (HGCAL). The HGCAL, with fine granularity in both the lateral and longitudinal directions, enables 3D imaging in reconstructing energy clusters. The intrinsic high-precision timing capabilities of the silicon sensors will add an extra dimension to event reconstruction. The HGCAL is expected to provide a timing resolution of $\sim$ 10s of ps for high-energy particles with $p_T$ of tens of GeV.

*ATLAS Upgrade* The ATLAS Liquid Argon (LAr) calorimeter will be improved in Phase 2 with an electronics upgrade that will provide optimized super cells and full-granularity data to the trigger system by means of a new pre-processor. A similar upgrade of the ATLAS Tile calorimeter readout will use on-detector digitization and a new back end pre-processor. Both the LAr and Tile calorimeters expect to implement a 40 MHz readout system for Phase 2. The transmission of high-granularity calorimeter data (all cells with a transverse energy of two times the noise threshold) drives the bandwidth requirement for the upgraded trigger and data acquisition (TDAQ) system.

In addition, the outermost Tile calorimeter layer can be used to identify muons in the range $|\eta| < 1.3$ by better identifying particle energy depositions above the Minimum Ionizing Particle (MIP) threshold.

#### 5.1.1.3 *Muon System*

The muon system will be upgraded at both experiments to meet HL-LHC conditions, extend geometric coverage, and improve detector performance and trigger capabilities.

*CMS Upgrade* For the CMS detector, its current muon system consists of three different types of muon detectors. In the barrel region, drift tubes (DTs) are installed along with resistive plate chambers (RPCs). In the endcaps, there are cathode strip chambers (CSCs) together with RPCs. At the HL-LHC, the existing muon detectors will be improved with upgraded electronics to enable 40 MHz readout, as well as improve the timing resolution from the current 12.5 ns to 1 ns [344]. New muon detectors, namely gas electron multipliers (GEMs) and a new version of RPCs, will be added to the endcaps, covering the regions of $1.6 < |\eta| < 2.4$. Additional muon chambers, labeled ME0, will cover the very forward regions of $2.4 < |\eta| < 2.8$, a region also covered by the upgraded inner tracker. This will be used for the muon trigger at L1. The additional muon detectors are essential to achieve a high trigger efficiency with an acceptable rate, especially in the forward region. The additional hits in the new endcap muon stations, combined with improved algorithms, permit efficient triggering on displaced muon tracks even in the harsh environment of the HL-LHC.

*ATLAS Upgrade* Most of the front-end and detector readout of the ATLAS muon system, including the trigger electronics for the Resistive Plate Chambers (RPC), Thin Gap Chambers (TGC), and Monitored Drift Tube (MDT) chambers, will be replaced to face the higher trigger rates and longer latencies necessary for the new Level-0 (L0) trigger required by the HL-LHC conditions. The MDT chambers will be integrated into the L0 trigger in order to sharpen the momentum threshold. Some of the MDT chambers in the inner barrel layer will be replaced with new, small-diameter MDTs. Additional RPC chambers will be installed in the inner barrel layer to increase the acceptance and robustness of the trigger, and some chambers in high-rate regions will be refurbished. New TGC triplet chambers in the barrel–endcap transition region will replace the current TGC doublets to suppress the high trigger rate from random coincidences in this region. The electronics of all the subdetectors will need to be replaced due to obsolescence, aging, and radiation damage. During the Phase 1 upgrade (to take place from 2019 to 2021) the New Small Wheel (NSW) chambers will replace the Cathode Strip Chambers (CSC) and the MDT chambers of the innermost endcap wheel by Micromegas and small-strip TGCs. The replacement of the MDT front-end readout will address the trigger rate and latency requirements of the TDAQ system in Phase 2 and allow the use of MDT hit information to improve the muon $p_{\mathrm{T}}$ resolution in the L0 trigger. Additionally, in the upgraded detector all data from the barrel and endcap detectors will be transmitted to FPGAs at L0, which can be used to implement more advanced and flexible algorithms for muon reconstruction, including the use of neural networks and/or dedicated tracking for non-pointing muons [348].

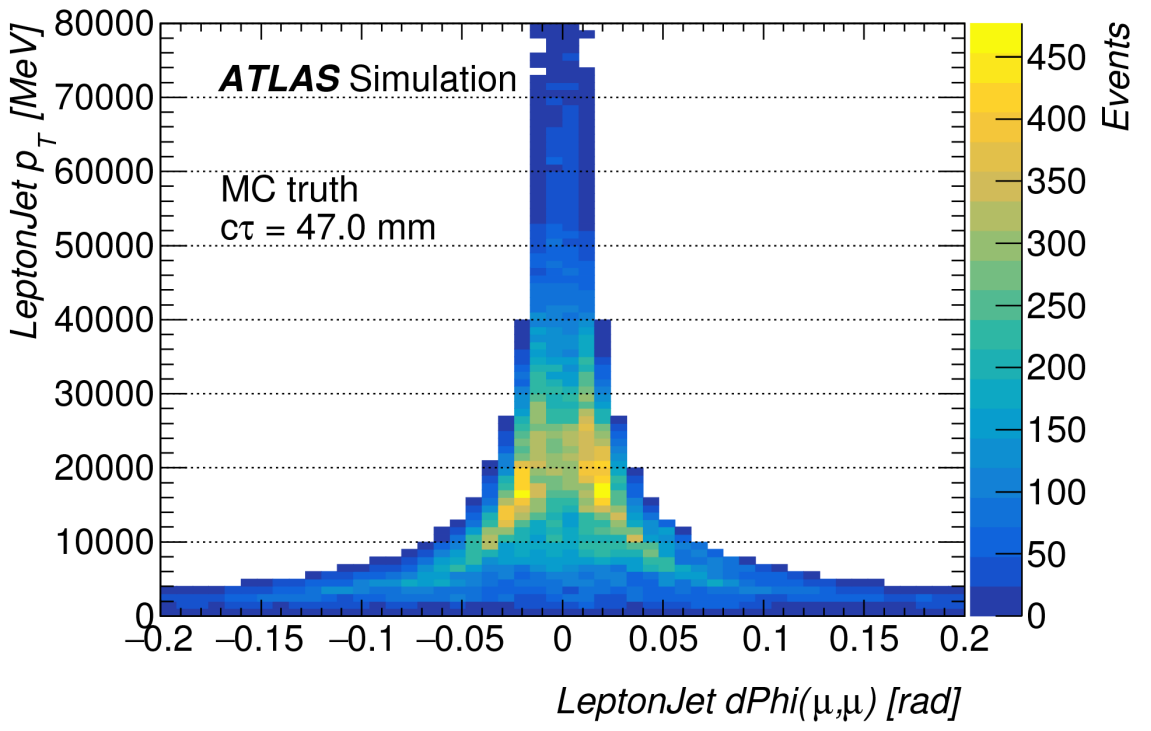

Figure 5.1: Opening angle between muons in a Hidden Valley model, where a sub-GeV-mass particle decays to $\mu^+\mu^-$. The opening angle is well below the resolution of the current system.

Some LLP scenarios (e.g., Hidden Valley models [59]) predict muon-jet final states, which result in collimated muons that are not identified with high efficiency by the current triggers (see Section 3.2). In Figure 5.1, the opening angle between muons in a Hidden Valley model is shown, for a particular combination of particle masses and parameters. The di-muon separation is much smaller than the current system can resolve (approximately 0.2 in $\Delta\phi(\mu,\mu)$). In the no-upgrade scenario, these can only be recorded by the single muon trigger. In the upgraded scenario, a dedicated trigger is under development for a dimuon trigger with a $p_{\mathrm{T}}$ threshold of $\approx$ 10 GeV.

#### 5.1.1.4 *Timing Detector*

Precision timing can be provided by the aforementioned calorimetry upgrades. However, the tens of ps timing resolution in the up-graded calorimeters is only achievable for particles with energy above tens of GeV. Moreover, timing information for delayed objects from calorimetry alone will be affected by the beamspot smearing, which corresponds to about 180 ps of uncertainty. Therefore, global event timing with the ability to reconstruct the vertex time and exploit time information in charged particle reconstruction requires a dedicated fast timing detector.

*CMS Upgrade* For the CMS experiment, the proposed MIP timing detector (MTD) will comprise a barrel and an endcap region made up of a single layer device between the tracker and calorimeters, and cover $|\eta|$ up to $\sim 3$. In the barrel, the proposal is to adapt the present Tracker Support Tube (TST) design by instrumenting the current location of the thermal screen with a thin, actively-cooled, stand-alone detector, based on lutetium-yttrium orthosilicate crystals activated with cerium (LYSO:Ce) and read-out with silicon photomultipliers. The endcap region can be instrumented with a hermetic, single layer of MIP-sensitive silicon devices with high timing resolution, with a pseudorapidity acceptance from about $|\eta| = 1.6$ to $|\eta| = 2.9$. The MTD is designed to provide timing resolution of a few tens of ps for charged tracks throughout the detector lifetime. The performance projection in Section 5.1.3 is evaluated with a 30 ps resolution for a $p_T$ threshold of 0.7 GeV in the barrel and a $p$ threshold of 0.7 GeV in the endcap, and covering the expected MTD fiducial region of $|\eta| < 3$.

*ATLAS Upgrade* The High-Granularity Timing Detector (HGTD) is intended to distinguish between collisions occurring very close in space but well-separated in time. Currently there is not yet a TDR for this project. The current proposed detector design is based on low-gain avalanche detector technology that will cover the $|\eta|$ region between 2.4 and 4, with a timing resolution of 30 ps for MIPs. High-precision timing will improve the track-to-vertex association in the forward region, impacting jet and lepton reconstruction, as well as offering unique capabilities for online and offline luminosity determination.

#### *5.1.1.5 Trigger*

The ATLAS and CMS experiments adopt a two-level trigger system: the hardware-based Level-1 trigger (L1) and the software-based high-level trigger (HLT).

*CMS Upgrade* For the CMS experiment, the L1 trigger currently only uses calorimeter and muon information. At the HL-LHC, with the aforementioned outer tracker upgrade of $p_T$ modules and stub-finding capabilities, tracking information will be included at L1 [346]. The L1 track trigger uses parallel processing and pattern recognition on stub information to achieve track finding at an output rate of 750 kHz.

The L1 tracking capability will be further complemented by the calorimeter and muon upgrades, which provide more precise position and momentum resolution, calorimeter shower shape, and more muon hits in the forward region. In the L1 trigger, the electron and photon trigger algorithms for HL-LHC will use information from the electromagnetic calorimeter as well as from the outer tracking detectors. The algorithm should preserve the ability to reconstruct electromagnetic clusters with $p_T$ above a few GeV with high efficiency (95% or greater above 10 GeV) as well as achieve high spatial resolution which should be as close as possible to the offline reconstruction. Following the upgrade of both on-detector and off-detector electronics for the barrel calorimeters at the HL-LHC, the EB will provide energy measurements with a granularity of (0.0174, 0.0174) in $(\eta, \phi)$, as opposed to the current input to the L1 trigger consisting of trigger towers with a granularity of (0.087, 0.087). The much finer granularity and resulting improvement in position resolution of the electromagnetic trigger algorithms is critical in improving electron/photon trigger efficiency and suppressing background at high pile-up.

The L1 Global Trigger (GT) will be upgraded with more sophisticated and effective global trigger calculations based on topology, plus an additional intermediate Correlator Trigger (CT) to fully exploit the increased information in the trigger objects, such as matching tracking info with finely-grained calorimeter information, or a combination of muon and track information. The upgraded detector readout and DAQ systems will allow 12.5 $\mu$s latency and a L1 rate of 750 KHz; the latter may be substantially reduced by adding L1 tracking information matched to improved L1 trigger objects from the calorimeters and muon system. At the high-level trigger (HLT), the processing power is expected to scale up by pile-up and L1 rate, with an output rate of 7.5 kHz and up to 10 kHz.

*ATLAS Upgrade* The ATLAS trigger and the data acquisition system are being planned with the intention of fully exploiting the physics potential of the HL-LHC. A baseline architecture has been proposed and documented in Ref. [348].

The hardware-based Level-0 Trigger system will be composed of separate calorimeter and muon triggers, as well as a Global Trigger and Central Trigger syb-systems. The result of the Level-0 trigger decision is transmittd to the data acquisition system at 1 MHz, and is followed by an upgraded Event Filter system to achieve a maximum stored event rate of 10 kHz.

The upgraded trigger system will take advantage of increased granularity provided by the calorimeters, will improve efficiency for muon-based triggers and perform hardware-based tracking profiting from the extended coverage of the planned silicon Inner Tracker (ITk). Options exist to further develop a hardware-based track trigger for quicker and less CPU-intensive rejection of the expected large increase in pile-up at the HL-LHC, to take full advantage of extended coverage provided by the ITk. Such a hardware track trigger would be an evolution of the current ATLAS Fast TracKer (FTK) reconstruction system.

### 5.1.2 *Object Performance: Tracking and Vertexing*

The ability of the detectors to reconstruct tracks and find vertices with high precision and efficiency in a high-density environment underlies the experimental reach for displaced objects. This section reviews the ATLAS and CMS experiments' projected tracking performance at the HL-LHC, highlighting improvements and new features with the upgrades.

#### 5.1.2.1 *CMS Performance*

*L1 Tracking* With the aforementioned tracker and L1 track trigger upgrades, the CMS experiment will be able to do track finding at L1 as well as offline at HL-LHC. Both L1 and offline tracking performance are discussed here.

All L1 tracking studies have been performed assuming 3 GeV stub $p_{\mathrm{T}}$ thresholds. In Figure 5.2, the L1 tracking efficiency for prompt muons and electrons for $t\bar{t}$ events in a scenario with 200 pile-up interactions per bunch crossing, on average, is presented. The tracking efficiency for muons exhibits a sharp turn-on at the 3 GeV stub $p_{\mathrm{T}}$ threshold, and saturates at approximately 98%. The tracking efficiency for electrons turns on more slowly and flattens out at 90%, mostly due to interaction with the detector material.

In Figure 5.3, the L1 tracking resolutions of the $p_{\mathrm{T}}$ and $z_0$ parameters of muons with $p_{\mathrm{T}} > 10$ GeV in $t\bar{t}$ events is shown for various average pile-up scenarios. The resolutions are defined in terms of an interval centered on the residual distribution that contains 68% or 90% of the tracks. Loss in tracking efficiency due to truncation effects (where there is insufficient time to transfer all the stub data) is determined from hardware and emulation to be at the level of $10^{-3}$ when considering $t\bar{t}$ samples with a pile-up rate of 200. As expected, resolutions degrade at forward pseudorapidity due to a corresponding increase in multiple scattering. In general, L1 parameter resolutions are excellent, which will provide for robust trigger object matching and charged particle reconstruction in the L1 trigger.

*Offline Tracking* Preliminary results on the offline tracking performance over the full acceptance of the CMS tracker are excellent, with further improvements expected as the detector design and simulation algorithms are optimized. In Figure 5.4, the resolution of the transverse momentum and the transverse impact parameter for single muons with $p_{\mathrm{T}} = 10$ GeV as a function of the pseudorapidity, both with the current detector and after the implementation of the HL-LHC upgrades, is shown. The better hit resolution of the HL-LHC tracker and the reduction of the material budget result in a significantly improved $p_T$ resolution. The transverse impact parameter resolution is also improved with respect to the current detector, ranging from below 10 $\mu$m in the central region to about 20 $\mu$m at the edge of the acceptance.

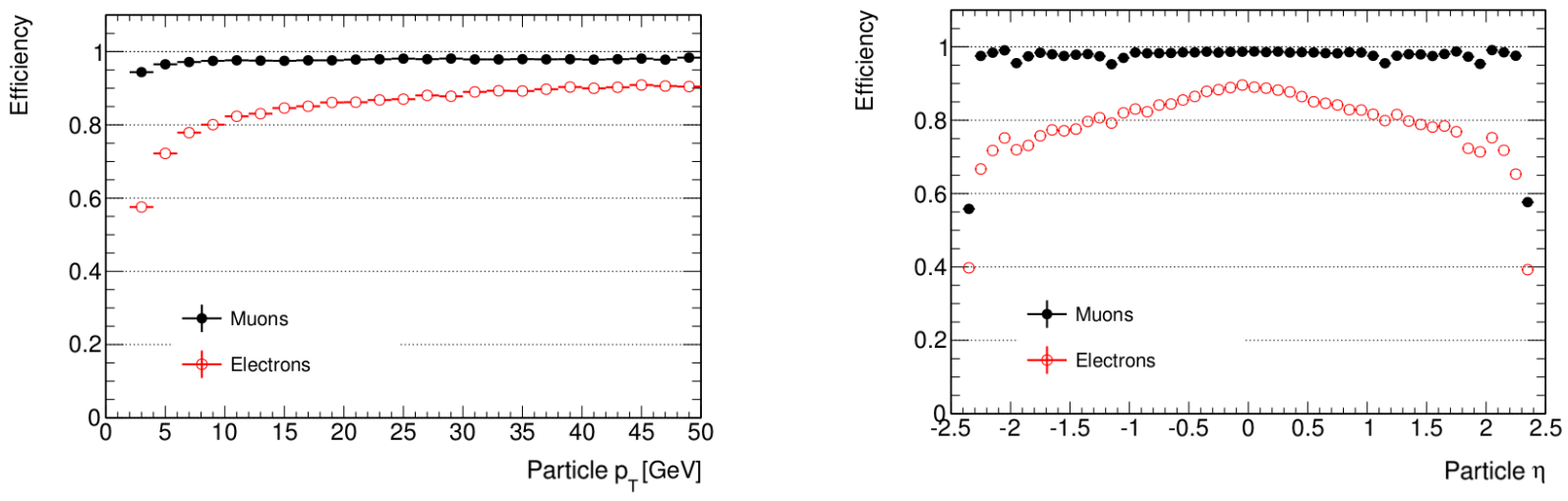

Figure 5.2: **Left:** L1 tracking efficiency versus generated particle $p_T$ for $|\eta| < 2.4$. **Right:** L1 tracking efficiency versus $\eta$ for $p_T > 3$ GeV. Results for muons (electrons) are shown as filled black (open red) circles, and are produced with $t\bar{t}$ events in a scenario with 200 pile-up events per bunch crossing, on average [342].

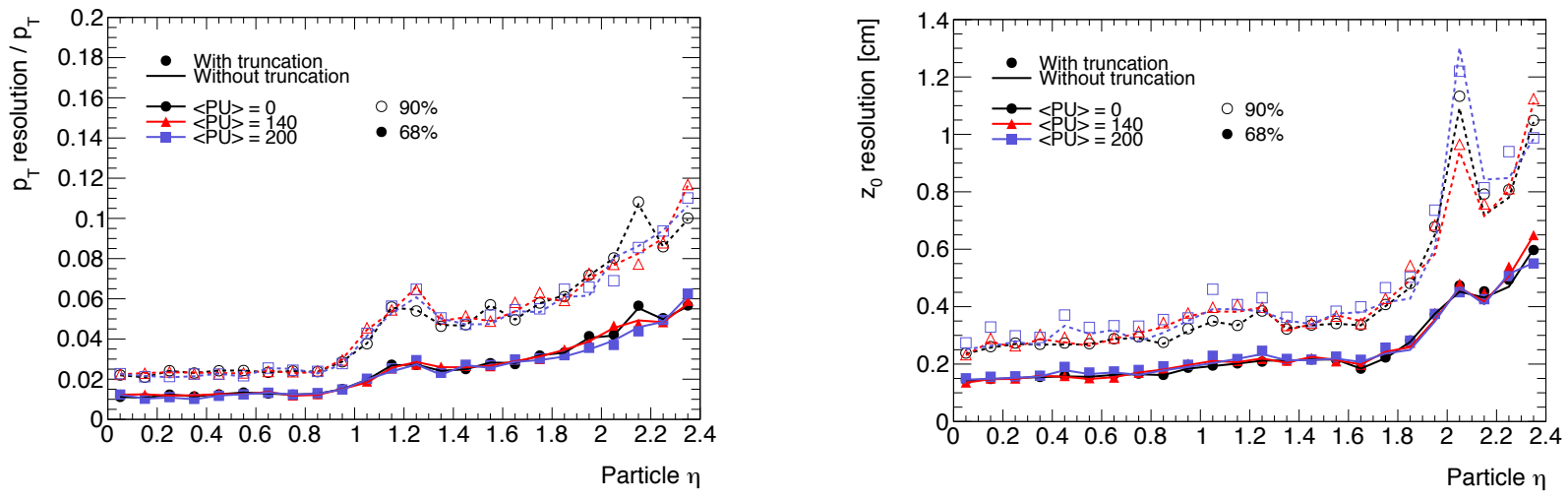

Figure 5.3: Relative $p_T$ (left) and $z_0$ resolution versus pseudorapidity for muons in $t\bar{t}$ events with zero (black dots), 140 (red triangles), and 200 (blue squares) pile-up events per bunch crossing, on average. Results are shown for scenarios in which truncation effects are (markers) or are not (lines) considered in the emulation of L1 track processing. The resolutions correspond to intervals in the track parameter distributions that encompass 68% (filled markers and solid lines) or 90% (open markers and dashed lines) of all tracks with $p_T > 3$ GeV [342].

For $t\bar{t}$ events, the efficiency to identify the primary vertex correctly is $\sim$ 95% at an average pile-up level of 140, and $\sim$ 93% at an average pile-up level of 200. The vertex algorithm used is the same as the one used in Run 2 for a pile-up of about 35, therefore it is not yet optimized for vertex reconstruction at very high pile-up. In Figure 5.5 the resolution of the vertex position in the $x$, $y$, and $z$ coordinates is shown as a function of the number of tracks associated to the vertex. The vertex position resolution is almost independent of the amount of pile-up in the event and the longitudinal resolution is only 50% worse than the transverse one, as expected given the pixel dimensions of the inner tracker modules.

Given that the CMS HLT tracking is based on the offline tracking code, a similar level of performance is expected. Because of HLT time constraints, a parallelization of the algorithms is already under development and will be applied also in the HLT track reconstruction at HL-LHC.

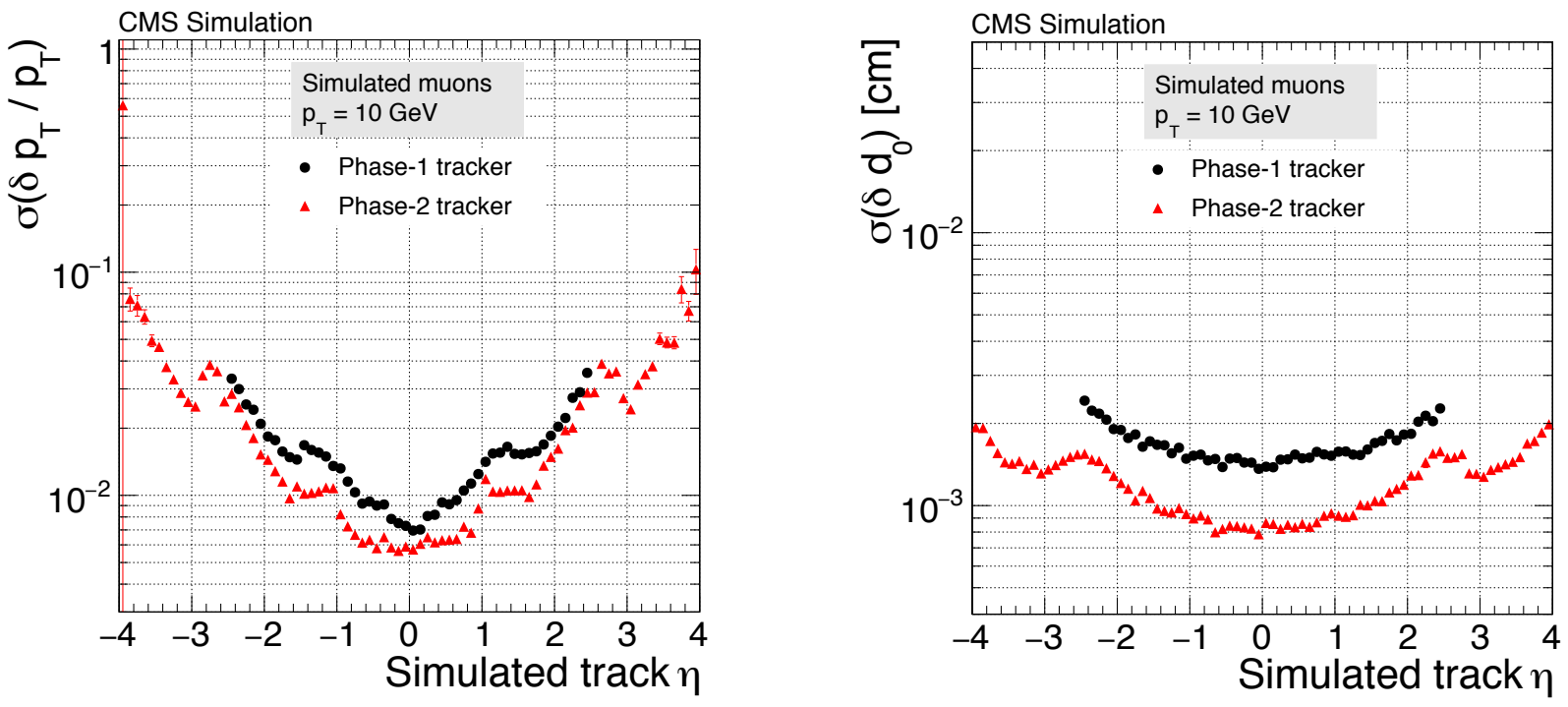

Figure 5.4: Relative resolution of the transverse momentum (left) and transverse impact parameter (right) as a function of the pseudorapidity for the current (black dots) and the upgraded (red triangles) CMS tracker, using single isolated muons with a transverse momentum of 10 GeV [342].

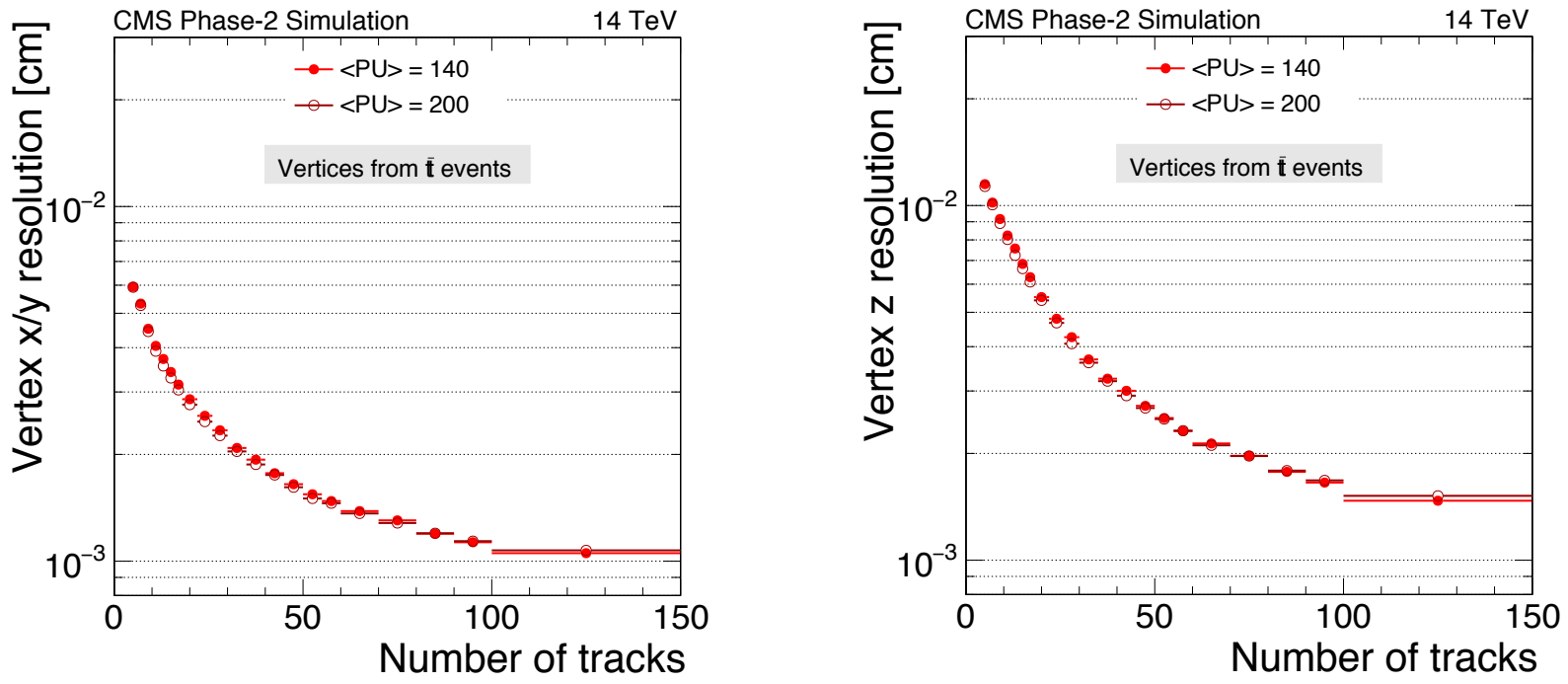

Figure 5.5: Vertex position resolution in $x$ and $y$ (left) and $z$ (right) as a function of the number of tracks associated to the vertex, for $t\bar{t}$ events with an average of 140 (full circles) and 200 (open circles) pile-up interactions per bunch crossing [342].

#### *5.1.2.2 ATLAS Performance*

Excellent tracking performance is also expected with the Inner Tracker (ITk) upgrade of the ATLAS experiment for the HL-LHC era. The left panel of Figure 5.6 shows the track reconstruction efficiency for jets in $Z' \rightarrow \bar{t}t$ events with 200 pile-up for different $\eta$ ranges. The right panel of Figure 5.6 shows the fake rate for reconstructed tracks in $t\bar{t}$ events, and there is clearly substantial improvement over the Run 2 detector performance along with the improved coverage in the forward region.

In Figure 5.7, the resolution of the transverse momentum and the longitudinal impact parameter for single muons with various $p_{\mathrm{T}}$ values is shown as a function of the pseudorapidity both with the current detector and projections for after the HL-LHC upgrade using digital clustering to find the tracks. The improvement is even more marked with analogue clustering: the transverse and longitudinal impact parameter resolutions are shown for different pixel pitches in Figure 5.8.

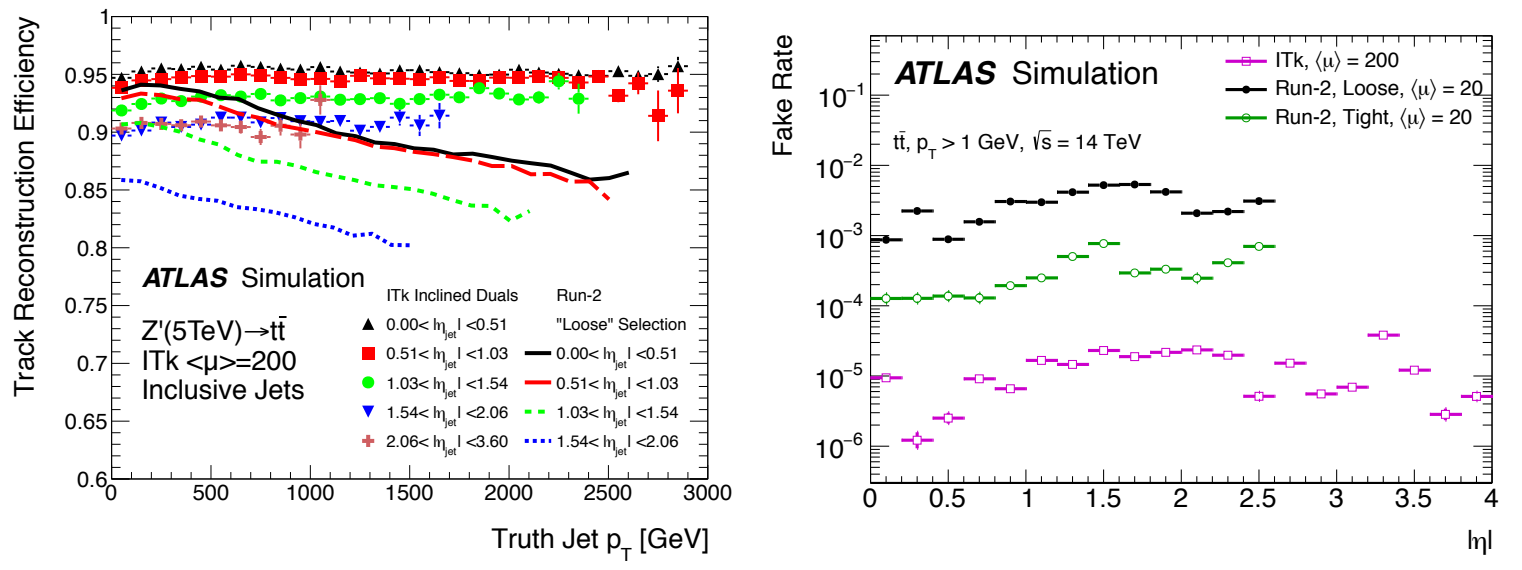

Figure 5.6: **Left:** Track reconstruction efficiency for tracks in jets from $Z' \rightarrow t\bar{t}$ with 200 average pile-up events. The efficiency is shown as a function of jet $p_{\mathrm{T}}$ for different $\eta$ ranges, and $M_{Z'} = 5$ TeV. **Right:** Fake rate for tracks in $t\bar{t}$ events with 200 average pile-up events using ITk; Run 2 detector results are shown for comparison. Both figures are from Ref. [347].

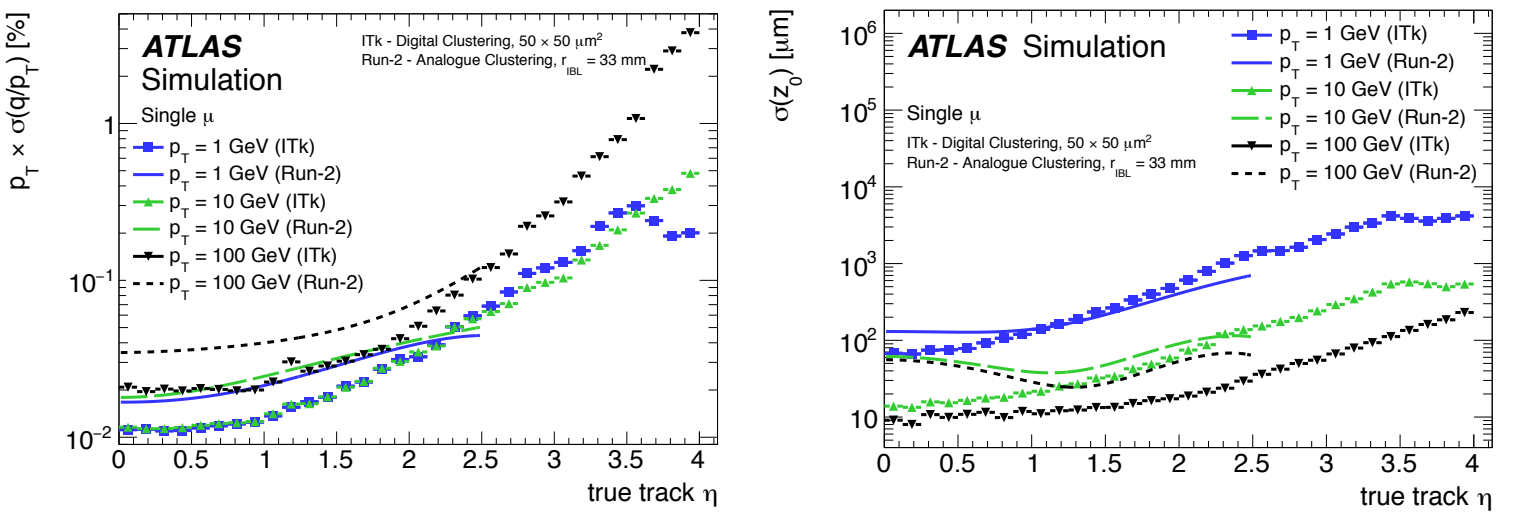

Figure 5.7: Resolution of the transverse momentum (left) and longitudinal impact parameter (right) as a function of the pseudorapidity for the current (solid line) and the upgraded (points) ATLAS tracker [347].

### *5.1.3 Upgrade Projection: LLP Searches*

Searches for long-lived particles are well motivated by various classes of extensions of the Standard Model, discussed at length in Chapter 2.2. Often, the production cross section for such processes is expected to be very small. The HL-LHC will allow for the collection of much larger data sets needed to reach better sensitivity to such BSM scenarios. The prospects are further strengthened with detector and trigger upgrades. This section discusses these potential improvements, and presents sensitivity projections on a number of benchmark LLP search channels with the aforementioned up-grades at the HL-LHC.

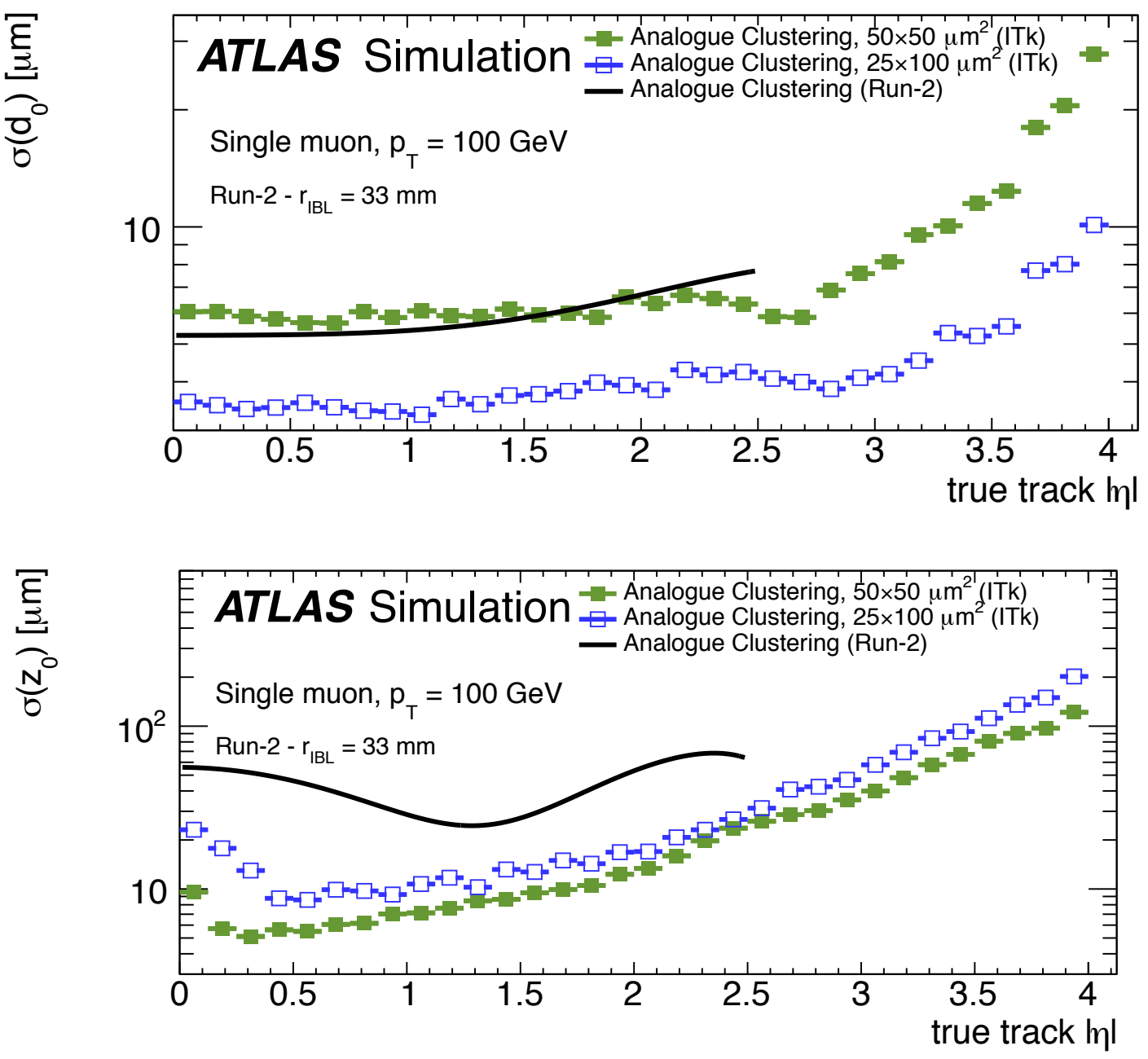

Figure 5.8: Track parameter resolutions using analogue clustering for (left) transverse impact parameter; (right) longitudinal impact parameter. The resolutions are shown for single muons with $p_T = 100$ GeV. The results of ITk are shown for $25 \times 100\ \mu\text{m}^2$ and $50 \times 50\ \mu\text{m}^2$ pixels, along with the current Run 2 detector performance [347].

#### 5.1.3.1 *Heavy Stable Charged Particles in CMS*

A number of new physics scenarios give rise to heavy stable charged particles (HSCPs) with long lifetimes that move with subrelativistical speed through the detector, heavily ionizing the sensor material as they pass through. In split SUSY models, the supersymmetric particles known as the stau ($\tilde{\tau}$) and the gluino ($\tilde{g}$) can have such characteristic signatures. The relevant simplified models are described in Sections 2.4.2–2.4.3, and current searches are described in Section 3.5.1.

*Sensitivity Projection with Tracker Upgrade* Depending on the mass and charge of the new particles, HSCPs experience anomalously high energy losses through ionization ($dE/dx$) in the silicon sensors with respect to the typical energy losses of SM particles, as can be seen in Figure 5.9 (left). At the CMS experiment, the current strip tracker features analog readout. Furthermore, the pixel detector featured analog readout during Phase 0 in 2016 and before, and currently features digital readout during Phase 1, which started at the beginning of the 2017 LHC run. Therefore, these detectors allow for excellent $dE/dx$ measurements.

At the HL-LHC, the upgraded CMS inner pixel detector will continue providing $dE/dx$ measurements, enabled by its time-over-threshold readout, while the outer tracker cannot provide such information, given that the readout is binary. To increase the sensitivity for signatures with anomalously high ionization loss, a second, programmable, threshold has been implemented in the short strip ASICs of the pixel-strip (PS) modules of the outer tracker, and a dedicated readout bit signals if a hit is above this second threshold.

Searches for HSCPs can thus be performed by measuring the energy loss in the inner pixel detector and by discriminating HSCPs from minimum ionizing particles based on the "HIP flag" in the outer tracker. The threshold of the minimum ionization needed to set the HIP flag is an adjustable parameter in each PS module independently. A threshold corresponding to the charge per unit length of 1.4 MIPs, resulting from preliminary optimization studies, is used in the simulation, and the gain in sensitivity obtained by using the HIP flag is studied.

An estimator of the degree of compatibility of the track with the MIP hypothesis is defined to separate candidate HSCPs from tracks from SM background sources. The high resolution $dE/dx$ measurements provided by the inner pixel modules are used for the computation of the $dE/dx$ discriminator. The tracks in background events have a low number of high threshold clusters with HIP flag, compared to those observed for tracks in HSCP signal events and slow-moving protons and kaons in minimum bias events.

In Figure 5.9 (right), the performance of the discriminator is shown by evaluating the signal vs. background efficiency curves to identify tracks from signal events and reject those originating from backgrounds. The performance curves are evaluated for two different strategies for the discriminator: the $dE/dx$ discriminator, which relies solely on the inner pixel modules ($dE/dx$-only, ignoring the HIP flags), and a recomputed discriminator which includes the HIP flags from the outer tracker PS modules ($dE/dx$ + HIP flags). The signal versus background efficiency performance curves demonstrate that for a background efficiency of $10^{-6}$, analogous to the current analysis performance, the $dE/dx$+HIP-based discriminator leads to an expected signal efficiency of 40%, around 4 to 8 times better than the $dE/dx$-only discriminator. In the $dE/dx$-only scenario, the efficiency for the HSCP signal is about 8 times smaller than that obtained in current data. The inclusion of the HIP flag for the PS modules of the Outer Tracker restores much of the efficiency, so that the same sensitivity as in Phase 1 will be realized with about four times the luminosity of Phase 1. The Phase 1 sensitivity will be surpassed with the full expected integrated luminosity of the HL-LHC. This study demonstrates the critical impact of the HIP flag in restoring the sensitivity of the CMS tracker for searches for highly ionizing particles in the HL-LHC era.

Additionally, the current CMS inner pixel detector provides mea-surements of charge deposits in each pixel up to $9,600$ electrons over a range of 4-bits in the digitizer. While it may be difficult to increase the number of bits used to store the charge information due to data rate constraints, it is possible to adopt a dual-slope mapping from charge deposit to ADC counts in the digitizer, which will preserve the granularity for lower charge deposits, while giving more information for highly ionizing particles such as HSCPs. This option is currently being studied to evaluate the potential improvement to $dE/dx$ measurements. Furthermore, tuning of the HIP flag threshold may bring additional improvements.

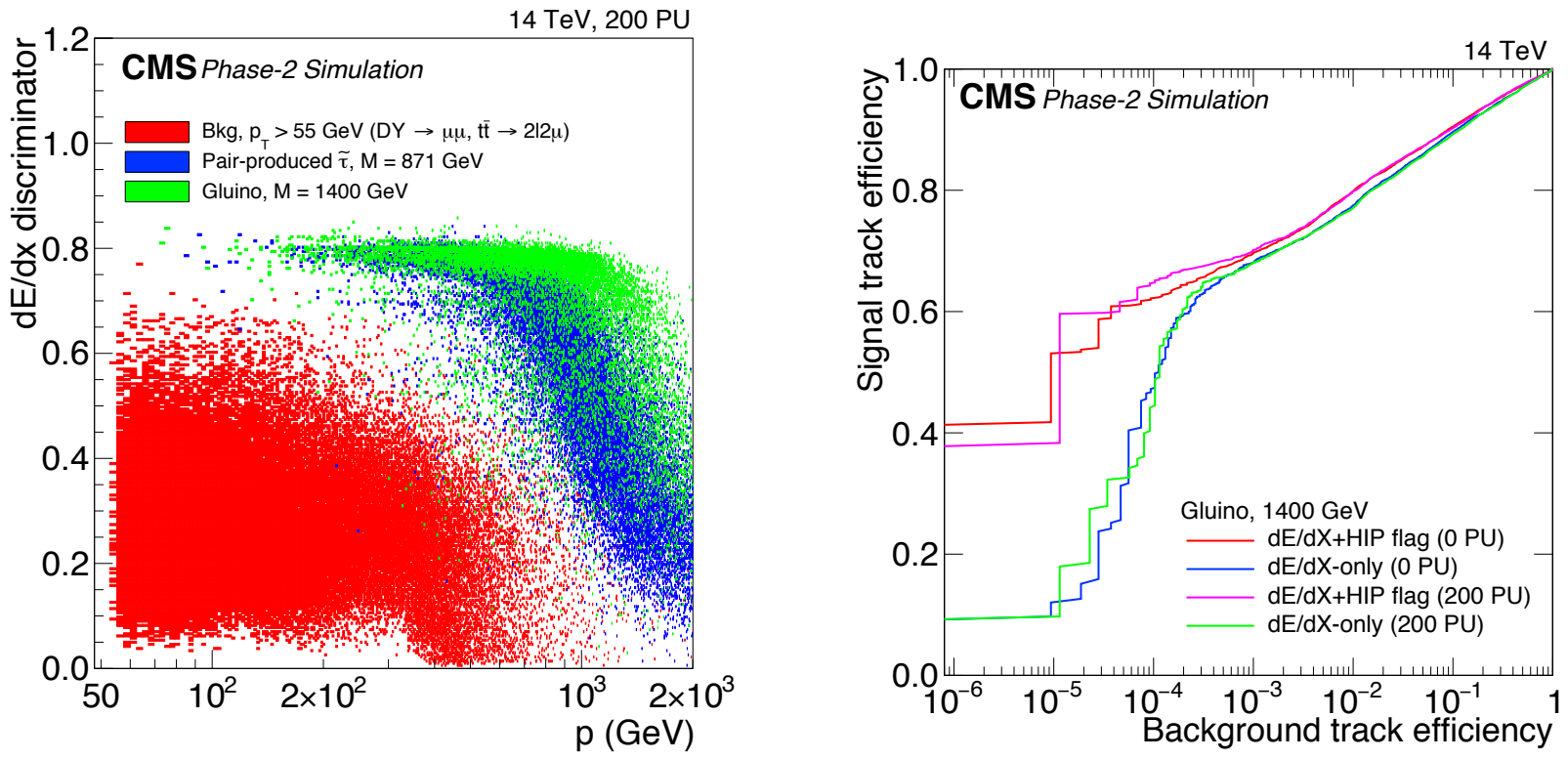

Figure 5.9: **Left:** Distribution in CMS of the $dE/dx$ discriminator versus track momentum ($p$) for tracks with high momentum ($p_T > 55$ GeV) in background events (red) and for candidate signal particles. Pair produced $\tilde{\tau}_S$ with a mass of 871 GeV (blue), and a gluino with a mass of 1400 GeV (green), are shown. **Right:** The performance of the $dE/dx$ discriminator for selecting gluinos in events at rates of 0 pile-up (PU) and 200 PU. The signal vs. background efficiency performance curves for a discriminator making use of both the pixel information and the outer tracker HIP flag (red and magenta) demonstrate a better performance compared to a discriminator trained to exploit only the $dE/dx$ information from the pixel modules (blue and green), for a background rejection of $10^{-6}$ [342].

*HSCP Trigger with Muon Detector Upgrade* The upgrade of the RPC system will allow the trigger and identification of slowly moving particles by measuring their time of flight to each RPC station with a resolution of $\mathcal{O}(1)$ ns. The new RPC detectors have a two-end strip readout, which provides precise measurements of the hit position in the local $y$ or the global $\eta$ coordinate. The speed of muon-like particles and the time (bunch crossing) of their origin will be computed with a fast algorithm to be implemented in the Level-1 trigger at the HL-LHC.

The RPC detectors are synchronized to register muons moving at the speed of light with a local time equal to zero with respect to the collision event that produced the trigger. Slow-moving particles, as HSCPs, will arrive with a delay depending on their speed as shown in Figure 5.10. This time delay, measured by each RPC layer crossed by the HSCP, is exploited in order to trigger on and reconstruct such particles.

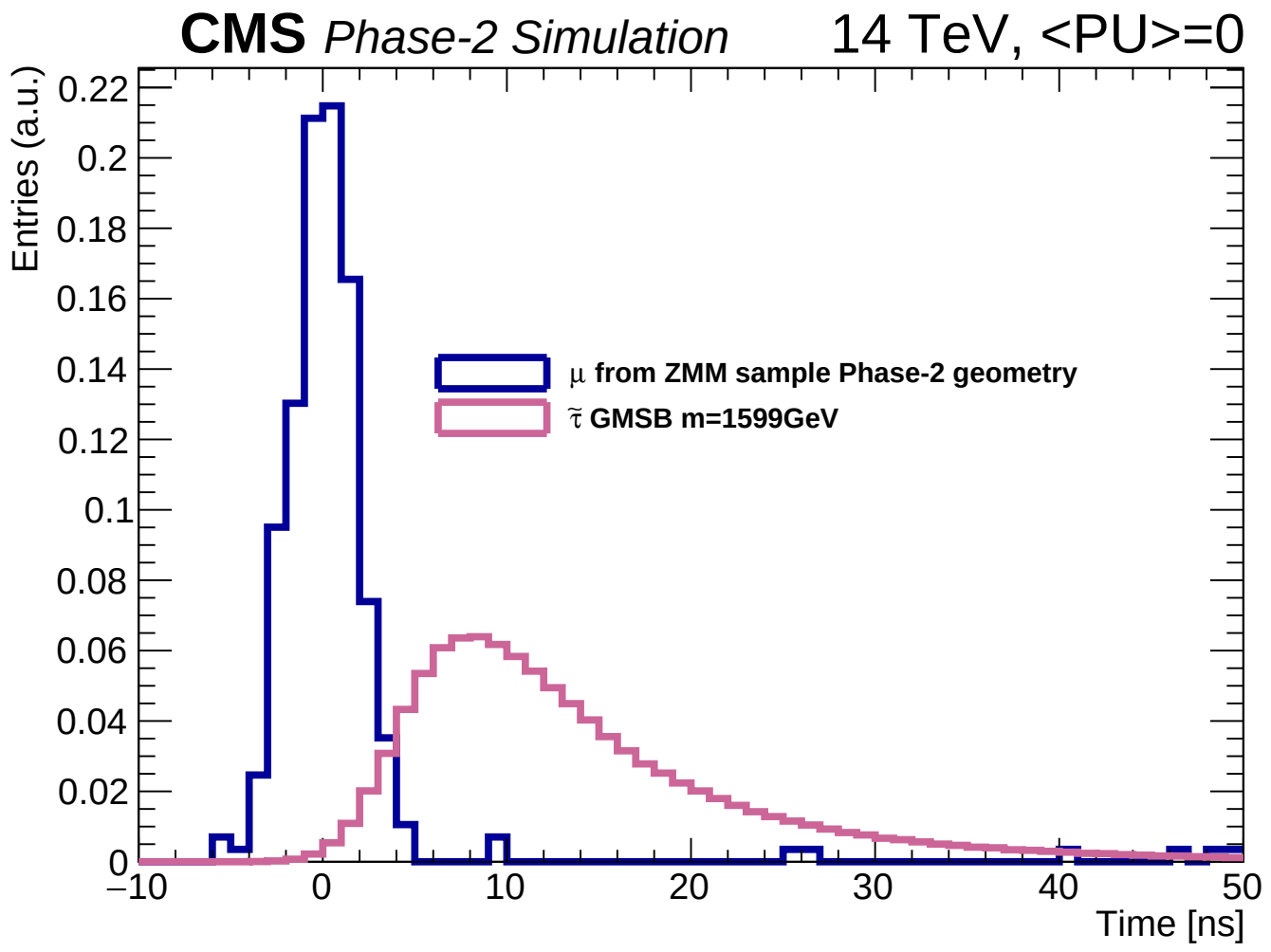

Figure 5.10: RPC hit time measurement distribution in CMS for muons from $Z \rightarrow \mu\mu$ events and for semi-stable $\tilde{\tau}$ particles with m≈1600 GeV, produced in $pp \rightarrow \tilde{\tau}\tilde{\tau}$ processes [344].

The principles of the proposed HSCP trigger algorithm are illustrated in Figure 5.11. In this figure, the vertical axis is the time of signals measured in RPC chambers, as synchronized so that muons moving nearly at the speed of light from a particular collision are measured at the time of the collision. The horizontal axis is the distance from the collision point to the position of the RPC at which the time is measured. The diagram shows three successive bunch crossings, two of which contain muons represented at horizontal lines. The diagram also shows the RPC time measurements from two HSCPs having slopes different from zero due to their traveling significantly slower than the speed of light. The time delay $\Delta t$ is related to the speed $v$ of an HSCP via the following equation:

$$\Delta t = d\left(\frac{1}{v} - \frac{1}{c}\right). \tag{5.1}$$

Here $d$ is the distance between the interaction point (IP) and the point where an HSCP crosses an RPC. For RE4/1 chambers and $\beta = v/c = 0.2$, the delay time is $> 6$ bunch crossings, comparable to 150 ns.

A penetrating charged particle leaves a trail of hits in RPC chambers along its trajectory. The time of flight can be computed in each RPC station with respect to a number of bunch crossing hypotheses. Should there be a common velocity solution, derived from Eq. (5.1), with $\beta < 0.6$, a trigger is formed. For $\beta > 0.6$, the delays are small and can be handled by the Phase 1 trigger. The performance of this algorithm has been studied with the CMS full simulation. All the detector effects (e.g., electronics jitter, signal time propagation along strips) are considered. A particle-speed measurement resolution is shown in Figure 5.12 (right) for the case of 25 ns signal sampling time (Phase 1) and 1.56 ns sampling time provided with the upgraded RPC Link Board System. For both plots, an HSCP signal is shown.

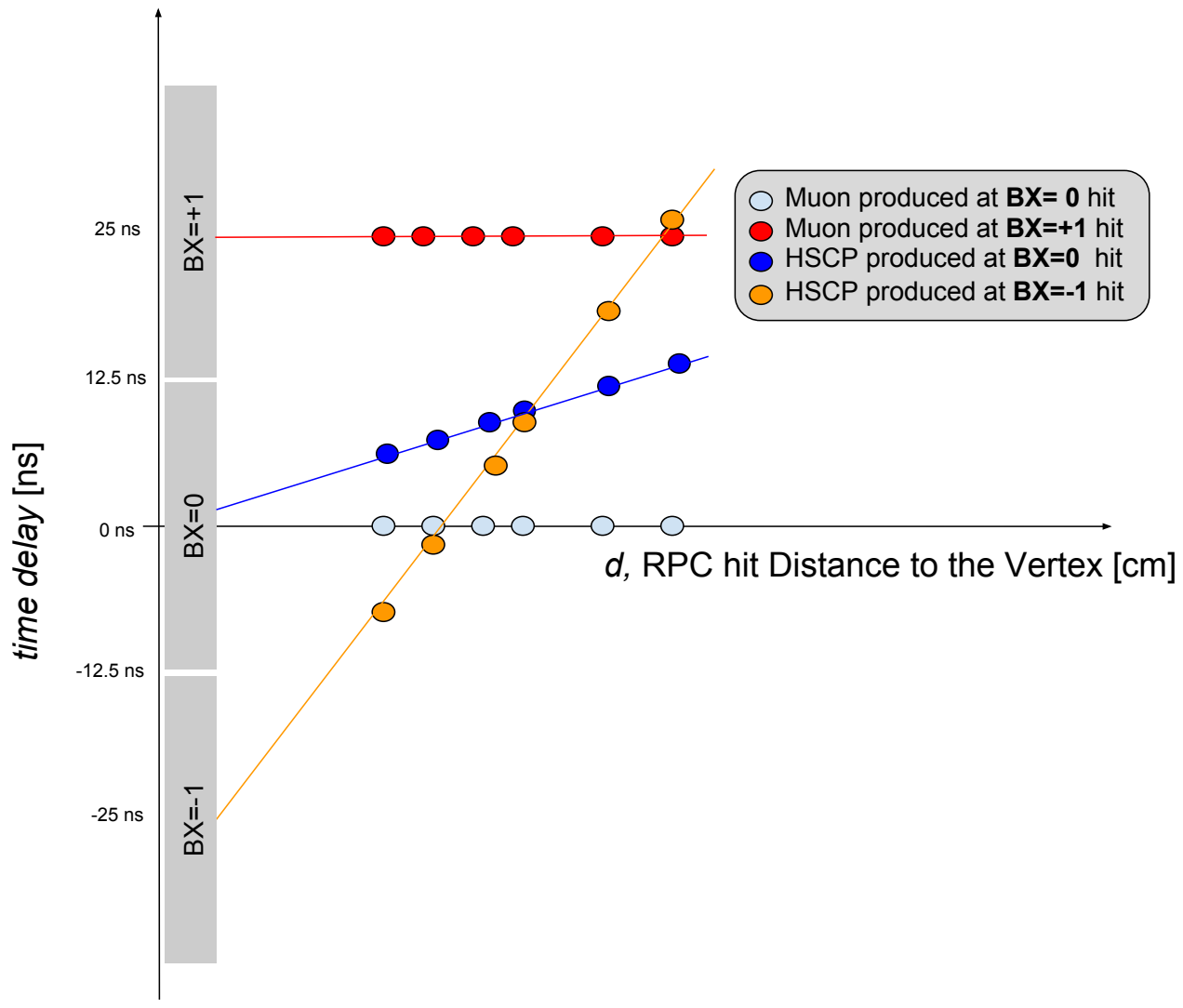

Figure 5.11: Diagram showing times measured at different RPC stations for particles originating at different bunch crossings and with different velocities in CMS. The $x$-axis represents the distances from IP to RPC detectors, while the $y$-axis corresponds to time. The clock at each RPC station is tuned so that particles moving with the speed of light are registered with the exact same "local" times. Hence, relativistic particles are represented by horizontal lines on this diagram [344].

The efficiency of the RPC-HSCP algorithm as a function of $\beta$ is studied and compared with the standard L1 muon trigger. The results are shown in Figure 5.12 (right). The current CMS-HSCP Phase 1 trigger performs well down to $\beta \approx 0.75$. The upgraded RPC Link Board System will allow for the triggering, at the correct bunch crossing, on possible HSCPs with velocities as low as $\beta \sim 0.25$.

Possible improvements for this trigger proposal in the $\beta$ measurement could be achieved by matching tracks in the track trigger to the HSCP muon trigger.

#### *5.1.3.2 Displaced Muons in CMS*

Many BSM theories predict particle decays with displaced muon or muon pairs in its final state, such as dark SUSY and GMSB with smuons. In order to demonstrate the physics potential of displaced muons at the HL-LHC with the CMS detector, a particular SUSY model is selected where the displaced signature consists of a dimuon final state emerging from the decay of heavy sparticles (smuons). Searches for the direct production of heavy sparticles with long lifetimes are difficult in the present LHC runs, owing to small cross sections and limited integrated luminosity, and will only become possible at the HL-LHC.

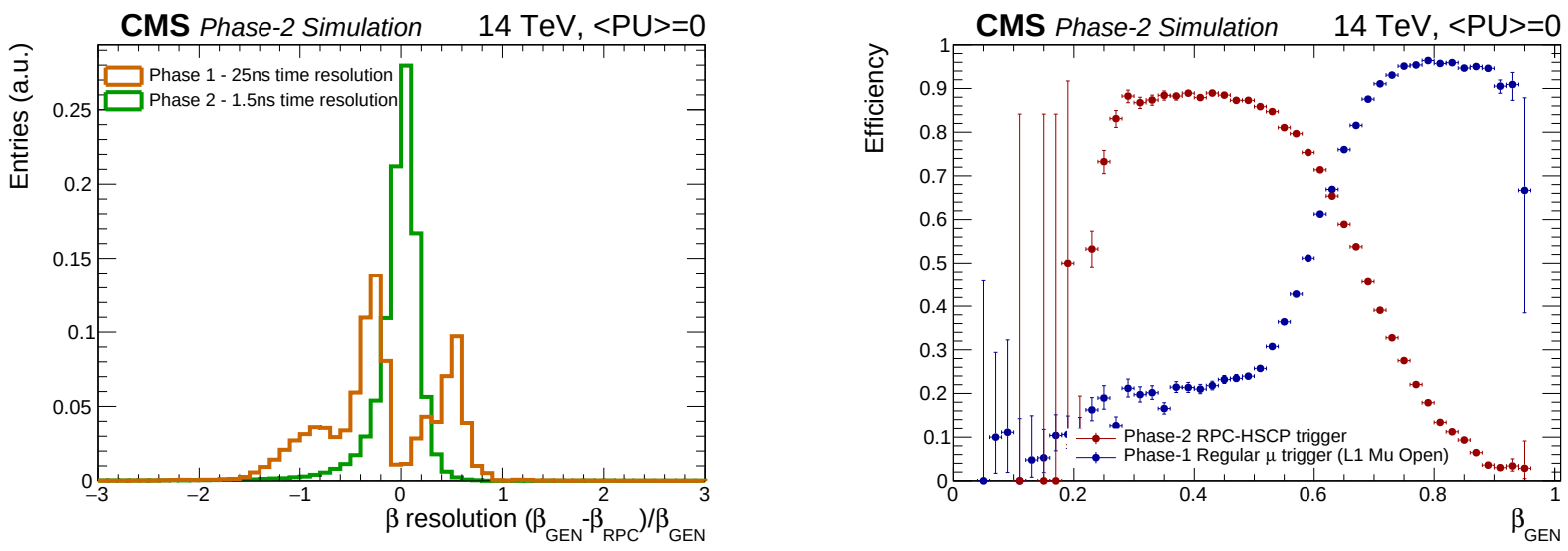

Figure 5.12: **Left:** Resolution of a particle-speed measurement at L1 trigger with Phase 1 and upgraded RPC Link Board System. **Right:** The efficiency as a function of $\beta$ of the standard L1 muon trigger without any $p_T$ threshold, and the RPC-HSCP Phase 2 trigger with 1.56 ns sampling time. For both plots, an HSCP signal is shown [344].

In gauge-mediated SUSY breaking models, smuons can be (co-)NLSPs (next to lightest supersymmetric particles) and decay to a muon and a gravitino [353]. When the slepton is long lived, the final state signature is two displaced oppositely charged muons and significant missing transverse energy. The smuon pair production has the advantage that it can be characterized by a very clean final state topology, and we will therefore focus on the process $q\bar{q} \to \tilde{\mu}\tilde{\mu}$, where the two smuons decay far from the primary interaction vertex. For this process, the muon $|d_0|$ can reach up to approximately one meter (or longer) for sufficiently large lifetimes, as shown in Figure 5.13 (left). Figure 5.13 (right) compares the number of hits created by these displaced muons in the CMS muon system in Phase 2 and the current CMS detector. The hits plotted here are those associated with the displaced stand-alone muon tracks, which is a muon track reconstruction algorithm specifically designed for displaced muons that can only be reconstructed in the muon system [329].

Standard triggers and reconstruction algorithms that use the position of the primary vertex will not efficiently reconstruct tracks with large impact parameters. Consequently, triggering on and reconstructing muons produced far from the IP is challenging and requires dedicated triggers and reconstruction algorithms. The upgrades to the muon system in CMS, as well as the L1 tracking capabilities, significantly improve the experiment's ability to search for displaced muons at the HL-LHC.

*Triggering on Displaced Muons* The momentum resolution of the L1 muon trigger for muons coming from the primary vertex will be greatly improved by adding information from the L1 track trig-ger, discussed previously. The L1 track trigger can also be directly combined with trigger primitives at the first stage of the muon track-finder electronics; this would mirror the offline reconstruction of "Tracker Muons" which improve the efficiency for very low-$p_\mathrm{T}$ muons, especially in the barrel region.

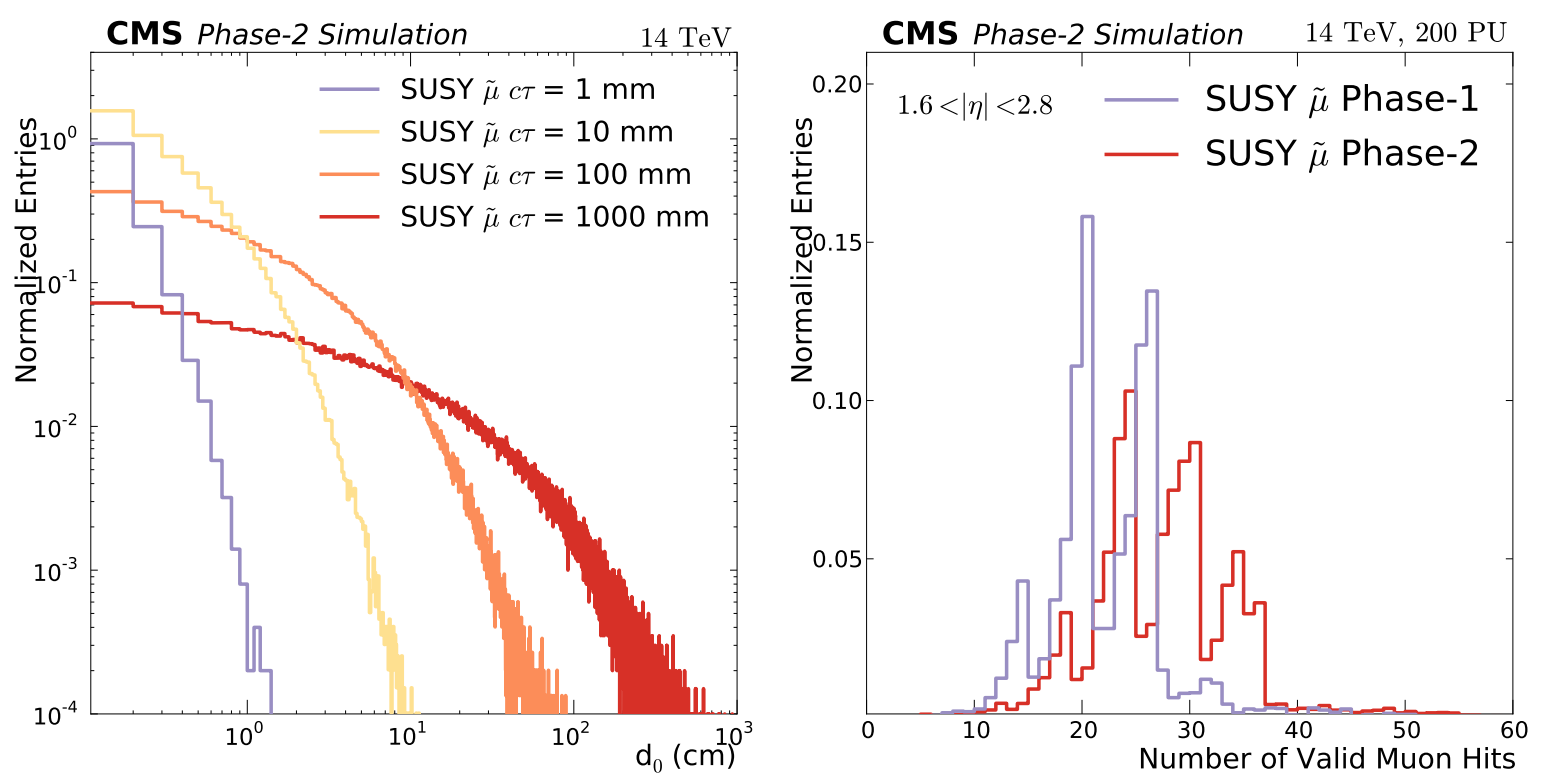

Figure 5.13: **Left:** The muon transverse impact parameter, $|d_0|$, for several simulated smuon decay lengths, $c\tau$, at the generator level. **Right:** Distribution of the minimum number of valid hits in the CMS Phase 2 muon system for a SUSY $\widetilde{\mu}$ with $m_{\widetilde{\mu}}$ = 500 GeV and $\tau$ = 1000 mm for the Run 2 (blue) and Phase 2 (red) detectors [344].

To trigger on both prompt and non-prompt muons effectively at L1, a stand-alone L1 muon generates two $p_\mathrm{T}$ measurements for each muon, prompt and non-prompt, which are matched with L1 tracks. If the track match is successful, the L1 track trigger $p_\mathrm{T}$ is used and a prompt candidate is formed. If the match is unsuccessful and the muon is not vetoed by L1 tracks, the non-prompt L1 muon $p_\mathrm{T}$ is used to form a displaced muon candidate. Figure 5.14 shows good performance for displaced muons with this method, i.e., there is a reasonably high efficiency and a trigger rate for a single muon trigger of around 10 kHz under HL-LHC conditions. Further improvements to the algorithm are underway to accommodate high pile-up conditions. The upgrade of the RPC system will allow slowly-moving particles to pass the trigger and be identified by measuring their time of flight to each RPC station with a resolution of $\mathcal{O}(1)$ ns. The speed of muon-like particles and the time (bunch crossing) of their origin will be computed with a fast algorithm to be implemented in the L1 trigger for the HL-LHC.

*Reconstruction* A dedicated muon reconstruction algorithm was designed for non-prompt muons that leave hits only in the muon system. This displaced stand-alone (DSA) algorithm is seeded by groups of track segments in the muon chambers. For each seed, a muon track is reconstructed with the same Kalman-filter technique as for the standard stand-alone (SA) muon reconstruction algorithm, but without constraining the interaction point. Figure 5.13 (right) shows the distribution of the number of hits in the Run 2 and HL-LHC detectors for displaced muons. The impact of the new stations is clearly visible. The charge mis-identification probability is expected to further decrease with the additional hits.

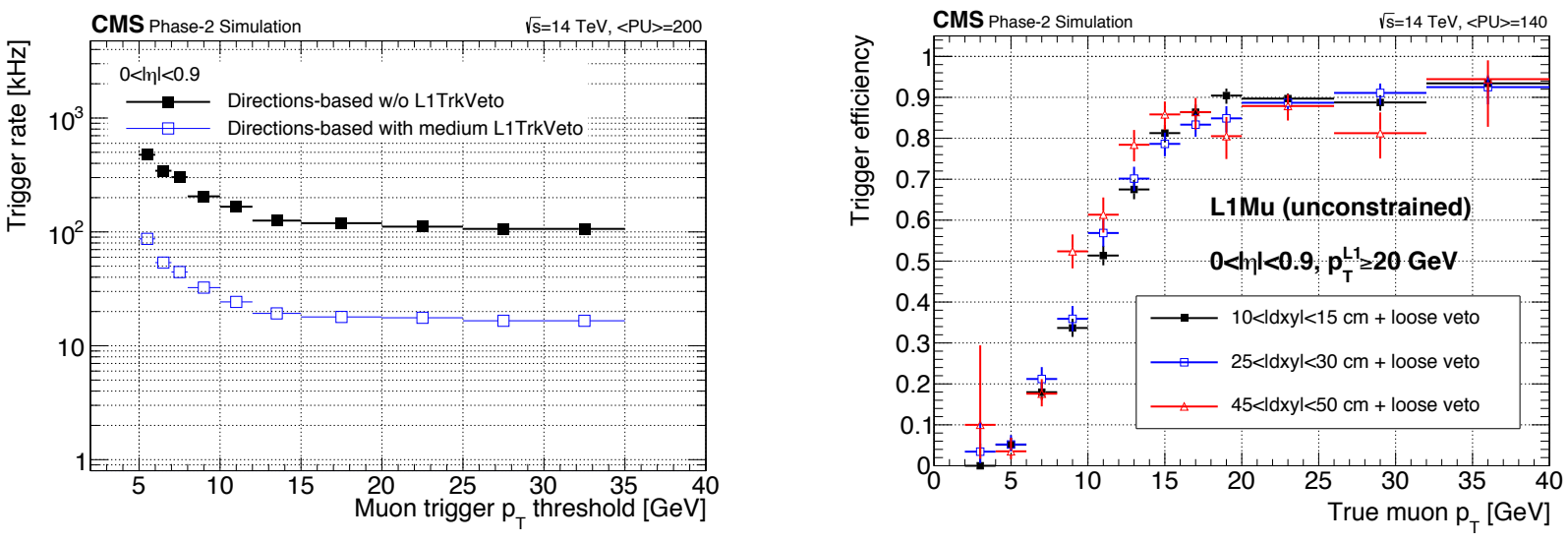

Figure 5.14: L1 Muon trigger rate (left) and efficiency (right) versus muon $p_{\mathrm{T}}$ threshold for the barrel displaced muon algorithm [344].

*Sensitivity Projection with a GMSB Model* To study the impact on physics sensitivity, a particular gauge-mediated SUSY breaking (GMSB) model is selected where the displaced signature consists of a dimuon final state plus gravitinos, emerging from the decay of heavy long-lived sparticles (smuons), where the gravitinos escape detection. This maps to the direct pair production simplified model with neutral LLP decays to muon + invisible in Section 2.4. This signal can serve as a proxy for other models with two LLPs decaying into muons. The final-state signature is then given by two displaced, oppositely-charged muons and significant $\not{E}_{\mathrm{T}}$. Example long-lived particles with $c\tau = 10, 100, 1000$ mm and several mass hypotheses (0.2, 0.5, 1 TeV) are simulated.

The main background for this search comes from multi-jet production (QCD), $t\bar{t}$ production, and Z/DY $\rightarrow \ell\ell$ events where large impact parameters are (mis)reconstructed. Cosmic-ray muons have been studied in Run 2 and these studies can be directly applied to Phase 2 running. In the barrel, they are efficiently rejected by the timing of the hits in the upper leg. Cosmic-ray muons do not originate at the vertex and therefore pass the upper-barrel sectors in reverse direction from outside in. The fraction of cosmic-ray muons in the endcaps is negligible.

Given the very low cross section of the signal process, it is essential to reduce the background efficiently. The best background discriminator is the impact parameter significance $d_0/\sigma(d_0) \geq 5$. Given the signal kinematics, the muons from a signal process are expected to move in roughly opposite directions and $\not{E}_{\mathrm{T}}$ can be expected to be larger than 50 GeV. After this selection the signal efficiency is about 4–5% for $c\tau \approx 1000$ mm, nearly independent of the smuon mass, and $10^{-5}$ – $10^{-4}$ for QCD, $t\bar{t}$, and DY backgrounds.

In Figure 5.15, the expected exclusion limits are shown for the GMSB model in which the smuon is a (co-)next-to-lightest supersymmetric particle (NLSP, where "LSP" indicates the lightest supersymmetric particle), for the predicted cross section as well as for a factor 100 larger cross section. The exclusion limits are shown as functions of smuon mass in Figure 5.15 (left) and decay length in Figure 5.15 (right).

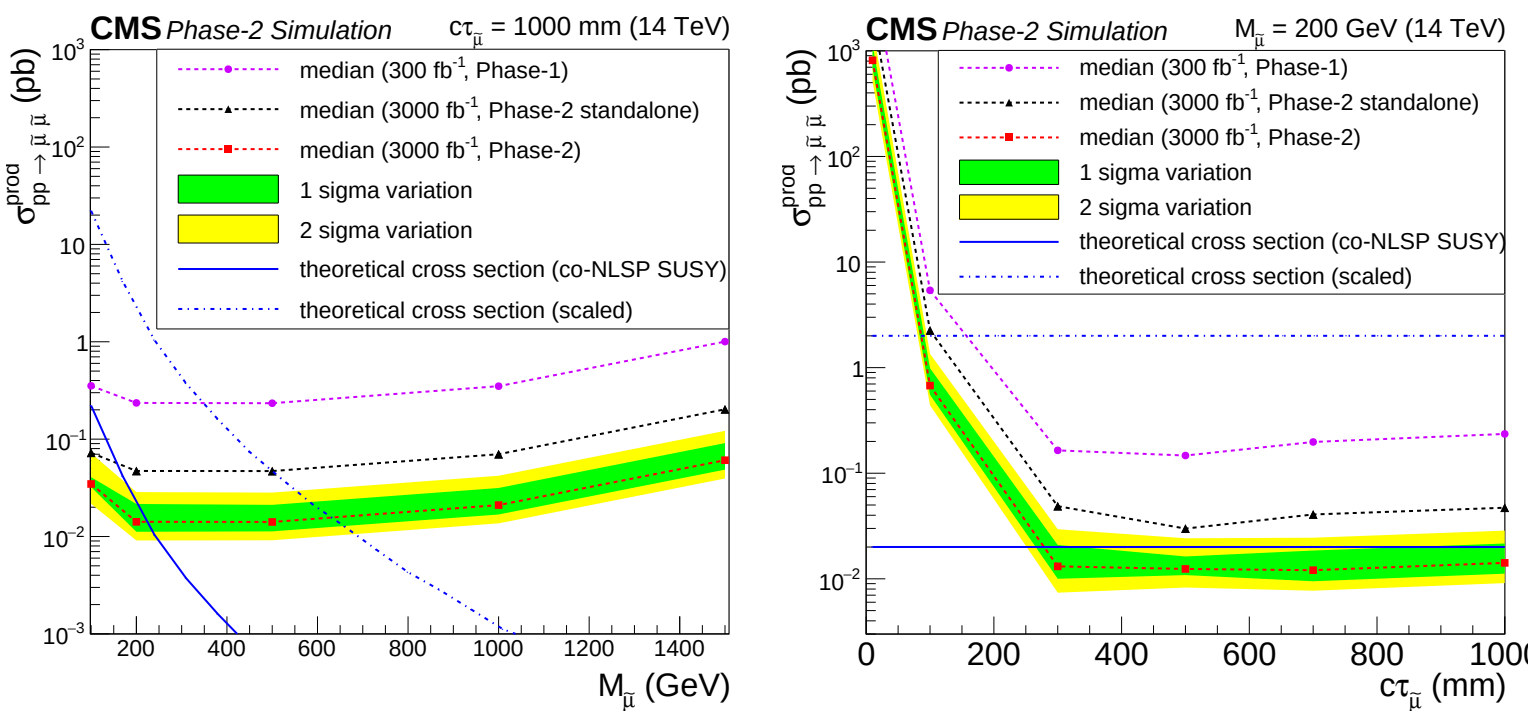

Figure 5.15: The 95% C.L. projected upper limits at CMS for $q\bar{q} \rightarrow \tilde{\mu}\tilde{\mu}, \tilde{\mu} \rightarrow \mu\tilde{G}$ for various mass hypotheses for $c\tau$ = 1 m (left), and as a function of the decay length for $m_{\tilde{\mu}}$ = 200 GeV (right). In both panels, the theoretical cross section for the specific model is represented by the blue solid line. For different SUSY breaking scales, tan $\beta$ or otherwise modified parameters, the cross sections may be 100 times larger, reflected by the blue dash-dotted line. Green (yellow) shaded bands show the one (two) sigma range of variation of the expected 95% C.L. limits. Phase 2 results with an average of 200 pile-up collisions per bunch crossing and an integrated luminosity of 300 fb$^{-1}$are compared to results obtained with 300 fb$^{-1}$. The black line shows the sensitivity without the DSA algorithm, which reduces the reconstruction efficiency by a factor three [344].

The sensitivity depends on $c\tau$ because shorter decay lengths shift the signal closer to background. In Figure 5.15 (right), the resulting physics sensitivities in terms of production cross section for the HL-LHC, normalized to 3000 fb$^{-1}$, are shown for the dedicated reconstruction of displaced muons and for the standard reconstruction. Also shown is the expected sensitivity at the end of Phase 1. Systematic uncertainties for the Phase 1 scenario are taken from current Run 2 analyses and adapted for expected HL-LHC conditions based on the assumptions of reduced systematics described in Ref. [354]. Clearly, only the HL-LHC will allow this process to be studied. The expected exclusion limit is around 200 GeV for $c\tau = 1000$ mm with 3000 fb$^{-1}$. This also illustrates the importance of keeping lepton trigger thresholds at several tens of GeV, even in the environment of 200 pile-up interactions per bunch crossing. Similarly, the discovery sensitivity is assessed assuming that a signal is present in data, and is shown as a function of smuon mass in Figure 5.16 (left) and decay length in Figure 5.16 (right).

*Sensitivity Projection with a Dark SUSY Model* The analysis presented above was reinterpreted using a Dark SUSY model ([30,

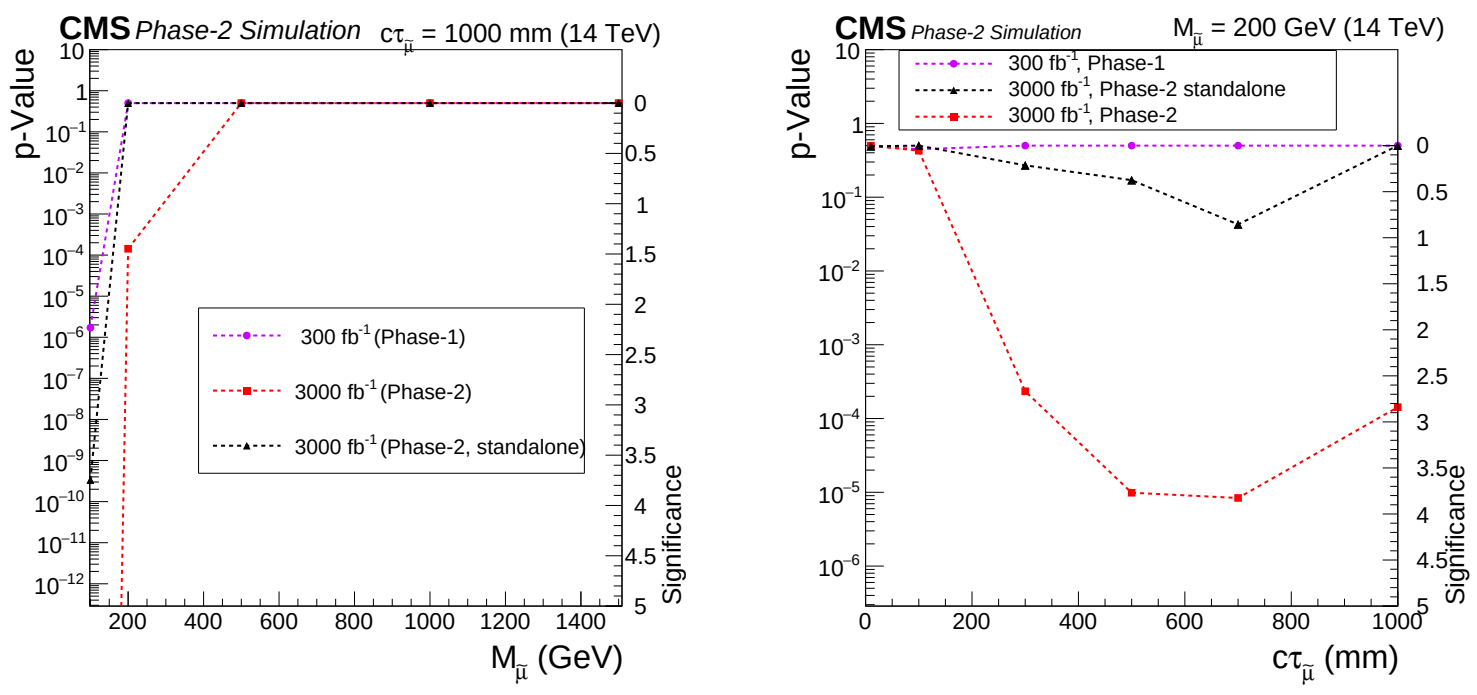

Figure 5.16: The projected discovery sensitivity at CMS for $q\bar{q} \rightarrow \tilde{\mu}\tilde{\mu}, \tilde{\mu} \rightarrow \mu\tilde{G}$ for various mass hypotheses and $c\tau = 1$ m (left) and as a function of the decay length for $m_{\tilde{\mu}} = 200$ GeV (right). Together with the discovery sensitivity the corresponding $p$-value is shown. Phase 2 results with an average of 200 pile-up interactions per bunch crossing and an integrated luminosity of 3000 fb$^{-1}$are compared to results obtained with 300 fb$^{-1}$. The black line shows the sensitivity without the DSA algorithm, which reduces the reconstruction efficiency by a factor three [355].

148]), in which an additional dark $U_D(1)$ symmetry is added as a supersymmetric SM extension. Breaking this symmetry gives rise to an additional massive boson, the so-called dark photon ($\gamma_D$), which couples to SM particles via a small kinetic mixing parameter $\epsilon$. If $\epsilon$ is very weak, the lifetime of the dark photon can range from a few millimeters up to several meters. The lower $\epsilon$, the longer is the dark photon lifetime which then decays displaced from the primary vertex. A golden channel for such searches is the decay to displaced muons.

In the model studied here [356], dark photons are produced in cascade decays of the SM Higgs boson that would first decay to a pair of MSSM-like lightest neutralinos ($n_1$), each of which can decay further to a dark sector neutralino ($n_D$) and the dark photon.

For the branching fraction $\mathrm{BR}(H \rightarrow 2\gamma_D + \mathrm{X})$, where X denotes the particles produced in the decay of the SM Higgs boson apart from the dark photons, 20% is used. Neutralino masses $m(n_1) =$ 50 GeV and $m(n_D) = 1$ GeV are assumed. Final states with two and four muons are included in the analysis. In the former case, one dark photon decays to a pair of muons while the other dark photon decays to some other fermions (2-muon final state). In the latter case, both dark photons decay to muon pairs (4-muon final state).

The main background for this search comes from multi-jet production (QCD), ttbar production, and Z/DY $\rightarrow \ell\ell$ events where large impact parameters are (mis)reconstructed. Cosmic ray muons can travel through the detector far away from the primary vertex and mimic the signature of displaced muons. However, thanks to their striking detector signature, muons from cosmic rays can be suppressed by rejecting back-to-back kinematics.

For each event, at least two DSA muons are required. If more than two, the ones with the highest $p_T$ are chosen. The two muons must have opposite charge ($q_{\mu,1} \cdot q_{\mu,2} = -1$) and must be separated by $\Delta R = \sqrt{\Delta\phi^2 + \Delta\eta^2} > 0.05$. The three-dimensional angle between the two displaced muons is required to be less than $\pi - 0.05$ (not back-to-back) in order to suppress cosmic ray backgrounds. Additionally, $\not{E}_T \geq 50$ GeV is imposed to account for the dark neutralinos escaping the detector without leaving any signal.

In order to discriminate between background and signal, the three-dimensional distance from the primary vertex to the point of closest approach of the extrapolated displaced muon track, called $R_{\text{Muon}}$, is used. The event yield after full event selection of both selected muons as a function of $R_{\text{Muon}-1}$ and $R_{\text{Muon}-2}$ is used to search for the signal. The left panel of Figure 5.17 shows $R_{\text{Muon}-1}$ of the first selected muon for signal and background samples.

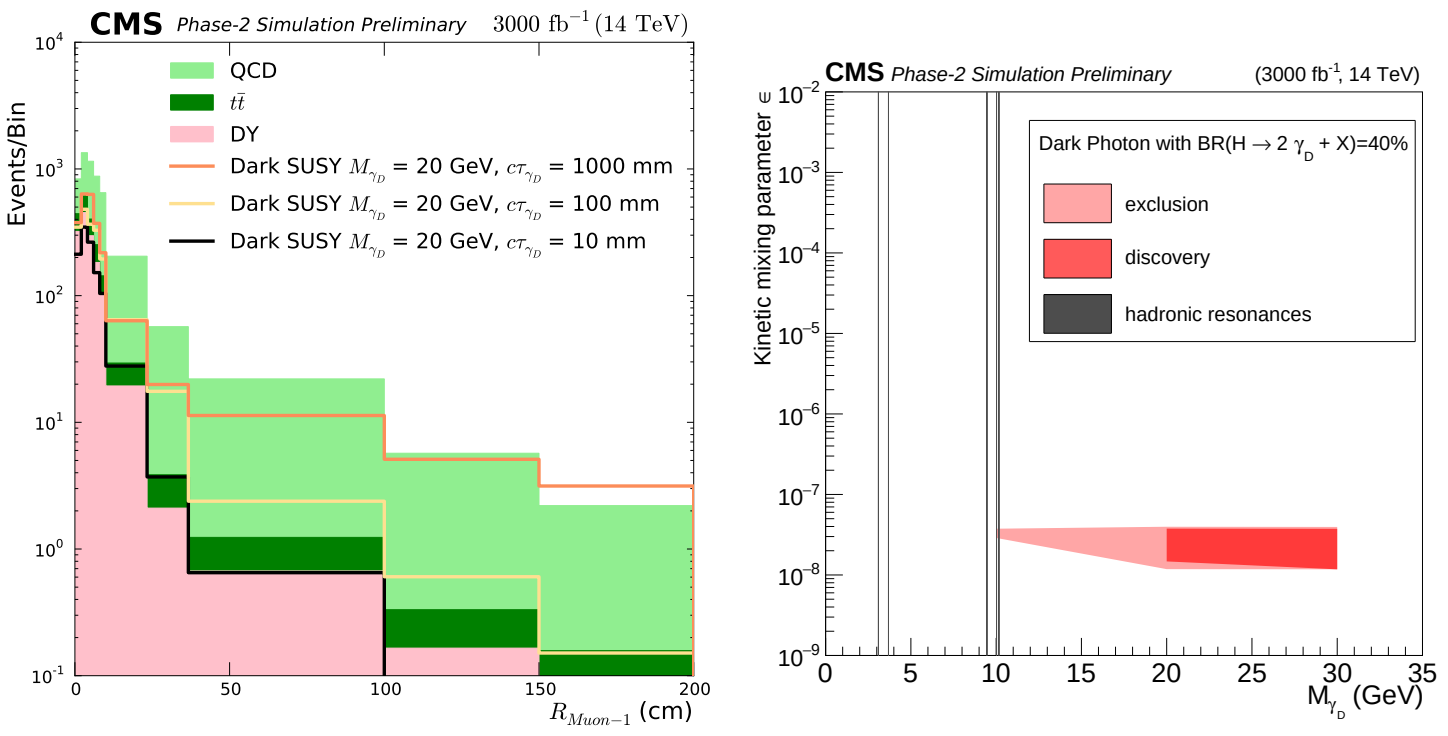

Figure 5.17: **Left:** Distance of the closest approach of the displaced muon track with maximum $p_T$ to the primary interaction vertex, $R_{\text{Muon}-1}$, for signal and background after the final event selection. **Right:** Parameter scan in the $\epsilon - m_{\gamma_D}$ plane. The grey lines indicate the regions of narrow hadronic resonances where the analysis does not claim any sensitivity [356].

The search is performed using a simple counting experiment approach. In presence of the expected signal, significance of the corresponding event excess over the expected background is assessed using the likelihood method. In order to evaluate the discovery sensitivity the same input is used as in the limit calculation, now with the assumption that one would have such a signal in the data. The discovery sensitivity is shown in the two-dimensional $m_{\gamma_D}$-$c\tau$ plane in the right panel of Figure 5.17. This search is sensitive to large decay length of the dark photon.

In absence of a signal, upper limits at 95% C.L. are obtained on a signal event yield with respect to the one expected for the considered model. A Bayesian method with a uniform prior for the signal event rate is used and the nuisance parameters associated with the systematic uncertainties are modeled with log-normal distributions. The resulting limits for the Dark SUSY models are depicted in Figure 5.18. While the results shown in the left panel of Figure 5.18 are for a dark photon with a decay length of 1 m as a function of the dark photon mass, the right panel of Figure 5.18 shows the results for a dark photon mass of 20 GeV as a function of the decay length [356].

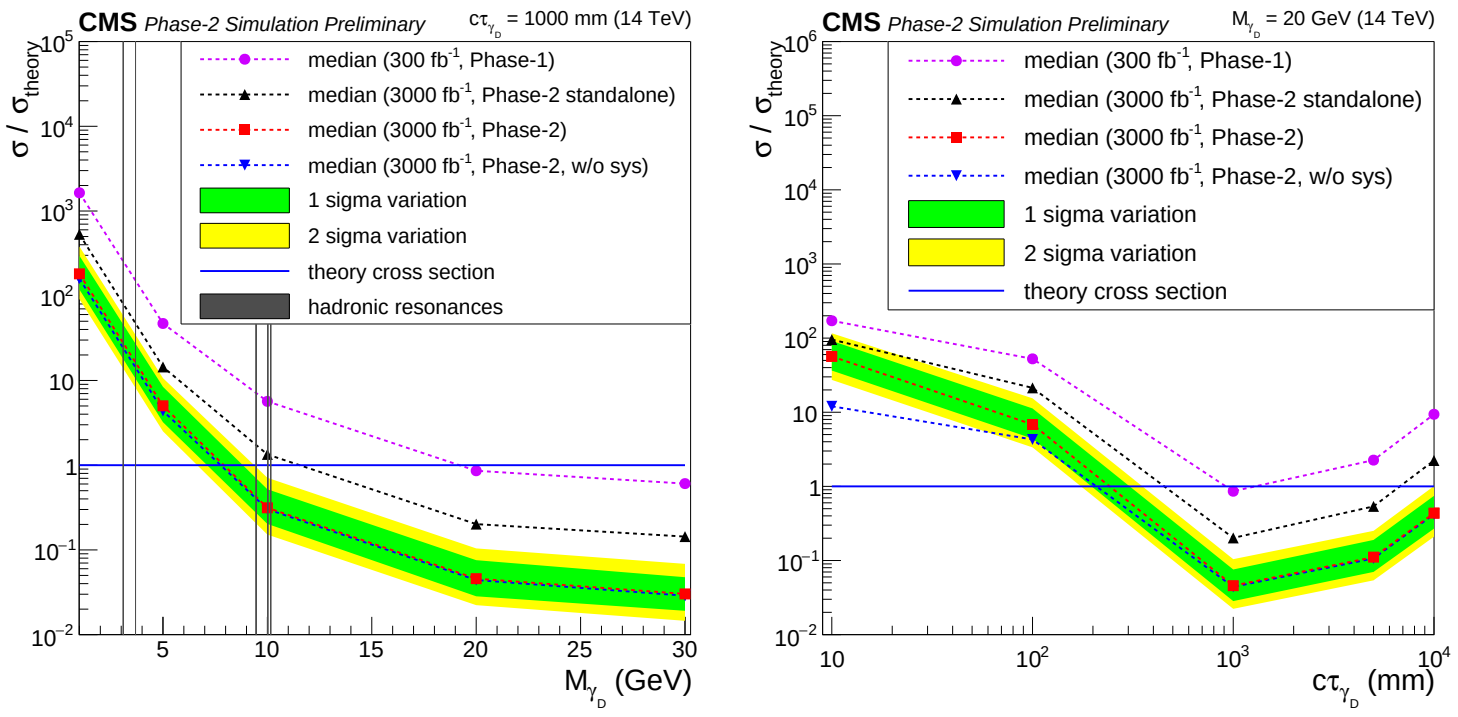

Figure 5.18: Upper limits at 95% C.L. on production cross section $\sigma/\sigma_{\text{theory}}$ for various dark photon mass hypotheses and a fixed decay length of $c\tau = 1000$ mm (left) and a fixed mass of $M_{\gamma_D} = 20$ GeV (right). Green and yellow shaded bands show the one and two sigma range of variation of the expected 95% C.L. limits, respectively. The grey lines indicate the regions of narrow hadronic resonances where the analysis does not claim any sensitivity [356].

### 5.1.3.3 *Displaced Photons at CMS*

A number of new physics scenarios predict new particles that, upon decay, result in displaced photons in the final state (see Section 2.4 and Section 3.4). At the CMS experiment, with the scintillating crystal design of the ECAL that provides excellent resolution but lacks pointing capability, the photon arrival time in the ECAL is the main observable used to distinguish signal from background in displaced photon searches.

One benchmark model for displaced photon searches is the GMSB model where the lightest neutralino ($\tilde{\chi}_0^1$) is the next-to-lightest supersymmetric particle, can be long-lived, and decays to a photon and a gravitino ($\tilde{G}$), which is the LSP, as illustrated in Figure 5.19 (left). For a long-lived neutralino, the photon from the $\tilde{\chi}_0^1 \to \tilde{G} + \gamma$ decay is produced at the $\tilde{\chi}_0^1$ decay vertex, at some distance from the beam line, and reaches the detector at a time later than that of prompt, relativistic particles produced at the interaction point. The time of arrival of the photon at the detector can be used to discriminate the signal from the background. The aforementioned upgrade to the ECAL electronics in the barrel region, and the HGCAL upgrade in the endcaps, will improve photon timing resolution at HL-LHC by an order of magnitude to as little as $\sim 30$ ps for photons with $p_T$ of tens of GeV or above, hence significantly improve the experimental reach of displaced photon searches.

Moreover, the proposed MIP Timing Detector (MTD) will be able to provide another dimension of information to reconstruct LLP decays. The time of flight of the photon inside the detector is the sum of the time of flight of the neutralino before its decay and the time of flight of the photon itself, until it reaches the detector. Since the neutralino is a massive particle the latter is clearly negligible with respect to the former. In order to be sensitive to short neutralino lifetimes of order 1 cm, the performance of the measurement of the photon time of flight is a crucial ingredient of the analysis. Therefore, the excellent resolution of the MTD apparatus can be exploited to determine with high accuracy the time of flight of the neutralino, and similarly the photon, also in case of a short lifetime.

An analysis has been performed at generator level in order to evaluate the sensitivity power of a search for displaced photons at CMS in the scenario where a 30 ps timing resolution is available from the MTD [345]. The events were generated with Pythia 8 [212], exploring neutralino lifetimes ($c\tau$) in the range 0.1–300 cm. The values of the $\Lambda$ scale parameter were considered in the range 100–500 TeV, which is relevant for this model to be consistent with the observation of a 125 GeV Higgs boson. After requiring the neutralino to decay within the CMS ECAL acceptance and the photon energy being above a "trigger-like" threshold, the generator-level photon time of flight was smeared according to the expected experimental resolutions. A cutoff at a photon time greater than $3\sigma$ of the timing resolution is applied and the "signal region" is assumed to be background free to estimate the sensitivity. The signal efficiency of such a requirement is computed and translated, assuming the theoretical cross sections provided in Ref. [59], to an upper limit, at 95% C.L., on the production cross section of the $\tilde{\chi}_0^1 \to \tilde{G} + \gamma$ process.

In Figure 5.19 (right), the analysis sensitivity in terms of the $\Lambda$ scale (and therefore of the neutralino mass) and lifetime is shown for three different assumptions on the timing resolution. The 300 ps resolution is representative of the time-of-flight resolution (TOF) consistent with current CMS detector performance. The 180 ps resolution is representative of the TOF resolution of the upgraded CMS detector without the MTD, in which case the TOF measurement will be dominated by the time spread of the luminous region. The vertex timing provided by the MTD detector will bring the TOF resolution to about 30 ps. As visible in the figure, a full-scope upgrade of the CMS detector with photon and track timing will provide a dramatic increase in sensitivity at short lifetimes and high masses, even with the first 300 fb$^{-1}$ of integrated luminosity.

#### *5.1.3.4 Displaced Jets in ATLAS*

Neutral long-lived particles that can decay into jets displaced from the proton-proton interaction point arise in many BSM theories (see Chapter 2 for an extensive discussion). For example, in hidden sector models a new set of particles and forces is proposed that is weakly coupled to the SM via a communicator particle. The hidden sector is otherwise invisible to the SM sector, but its particles (some of which can be long-lived) may decay to SM particles via the communicator. The lifetimes of these LLPs are typically unconstrained and could be long enough for the LLPs to decay inside the ATLAS detector volume. Such particles produce unique signatures which may have been overlooked by more traditional searches for new physics. One ATLAS search for such displaced jet signatures focuses on LLPs which decay in the ATLAS hadronic calorimeter (HCal) and consequently deposit most of their energy there and very little or none in the electromagnetic calorimeter, and also have few or no charged tracks pointing at the hadronic energy deposits. A signature-driven trigger that optimizes the acceptance for this class of events is required for the online selection.

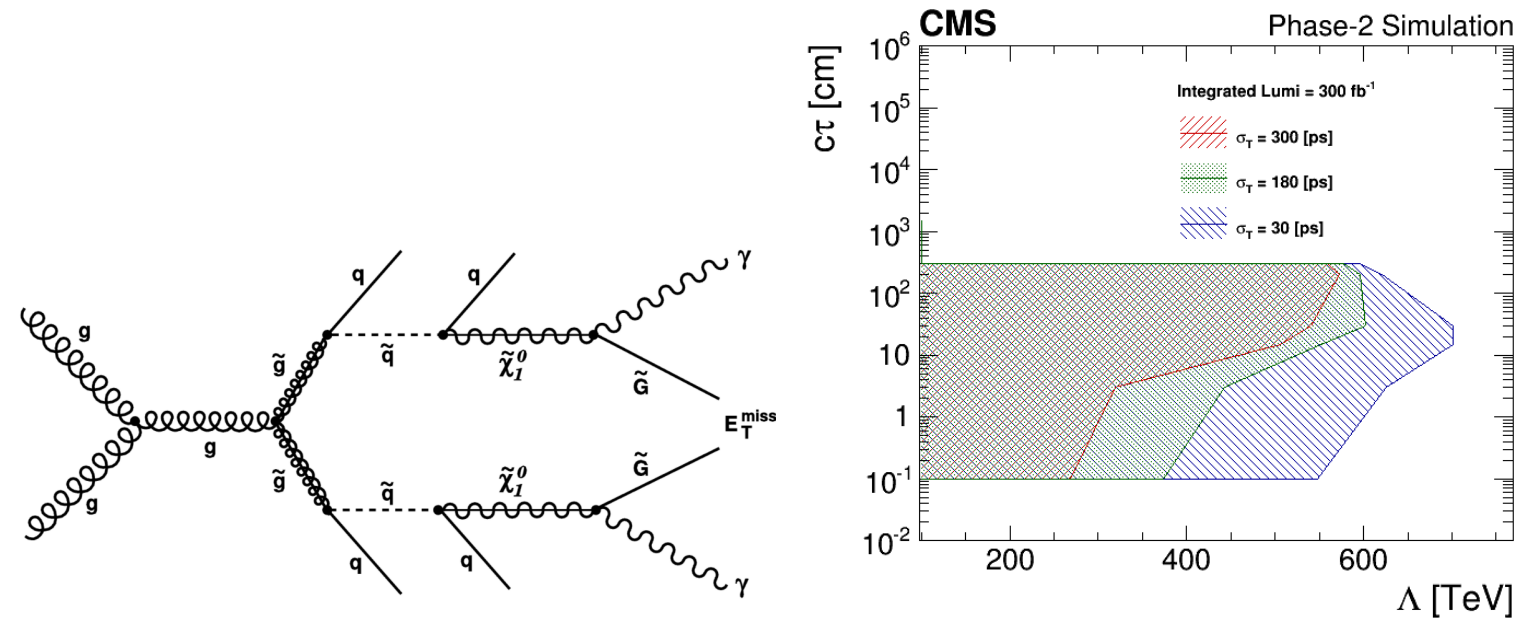

Figure 5.19: **Left:** Diagrams for a SUSY process that results in a diphoton final state through gluino production at the LHC. **Right:** Sensitivity to GMSB $\tilde{\chi}_0^1 \rightarrow \tilde{G} + \gamma$ signals expressed in terms of neutralino lifetimes for 300, 180 and 30 ps resolution, corresponding to the current detector, the HL-LHC detector with photon timing without MTD and with MTD, respectively [345].

The existing ATLAS analysis described in Section 3.1 can likely be improved by planned upgrades by using HCal information splitting the B-C layers in the calorimeter to identify the LLP decay position. The splitting between B-C layers will provide more information on the longitudinal shower profile, see Figure 5.20.

This analysis uses a hidden sector benchmark model and considers the decay of a heavy boson to two long-lived neutral scalars; this maps to the simplified model with Higgs boson production of hadronically decaying LLPs in Section 2.4. The heavy scalar bosons decaying into LLPs have masses ranging from 125 GeV to 1000 GeV, and the LLP scalars have masses ranging between 5 GeV and 400 GeV. Background processes, dominated by QCD dijet production, are suppressed in this analysis by requiring both scalars to decay in the calorimeter. Final results of the sensitivity projections are pending.

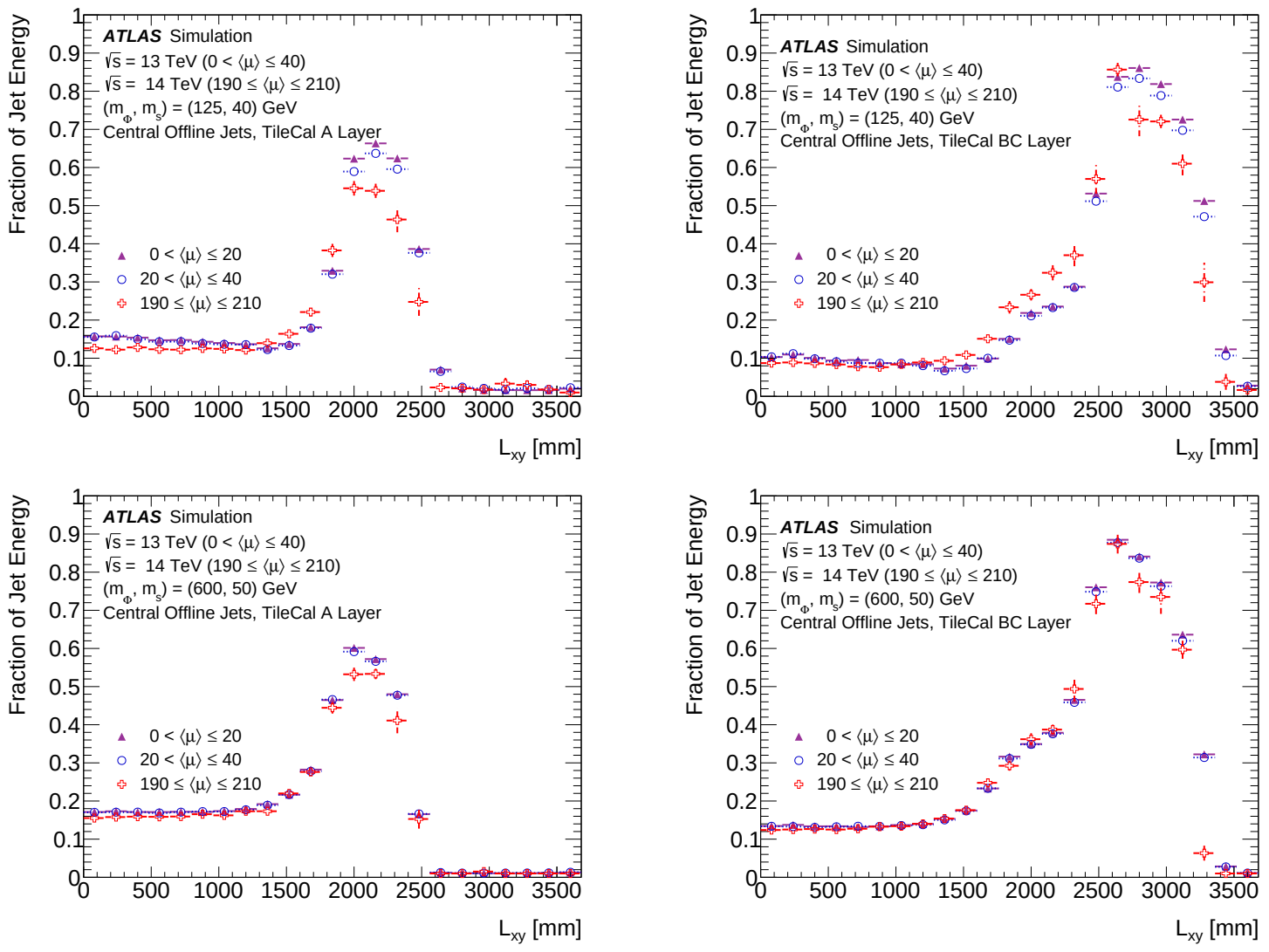

Figure 5.20: Simulation projections of the sensitivity of the ATLAS experiment upgrade plans to neutral LLPs. Top: The fraction of the jet energy deposited in the A-layer (left) and BC-layers (right) of the ATLAS tile calorimeter as a function of the transverse decay position of the LLP in events with a 125 GeV Higgs-like boson decaying to two 40 GeV LLPs. Bottom: The same for events with a 600 GeV Higgs-like boson decaying to two 50 GeV LLPs.

### *5.1.3.5 Disappearing Tracks*

In the ATLAS experiment, the planned ITk upgrade of the tracking volume for the HL-LHC can be exploited to improve the existing searches for BSM particles with a disappearing track [347].

Such particles are predicted by many well-motivated supersymmetric models such as anomaly-mediated SUSY-breaking scenarios, where the supersymmetric partners of the Standard Model $W$ bosons, the wino fermions, are the lightest SUSY state. In such models, the lightest neutralino and chargino can be nearly mass-degenerate, with a mass splitting around 160 MeV, and the chargino is consequently long lived. The combination of long lifetime and boost when produced in a high-energy collider allows the chargino to leave multiple hits in the traversed tracking layers before decaying. When performing searches for such signatures in ATLAS, selected events are typically required to contain at least one short track, hereafter called a tracklet. To study how such searches can be improved with the ITk upgrade, some assumptions have been made for simulating events and projected detector response. The detector response is parametrised by using response functions based on studies performed with Geant4 simulations of the upgraded detector in high-luminosity pile-up conditions.

The tracklet reconstruction efficiency for signal charginos as a function of the decay radius inside the detector is shown in Figure 5.21. A 30% systematic uncertainty on the background yields is assumed, as observed for the Run 2 analysis. The expected background is largely dominated by fake tracklets due to random crossings.

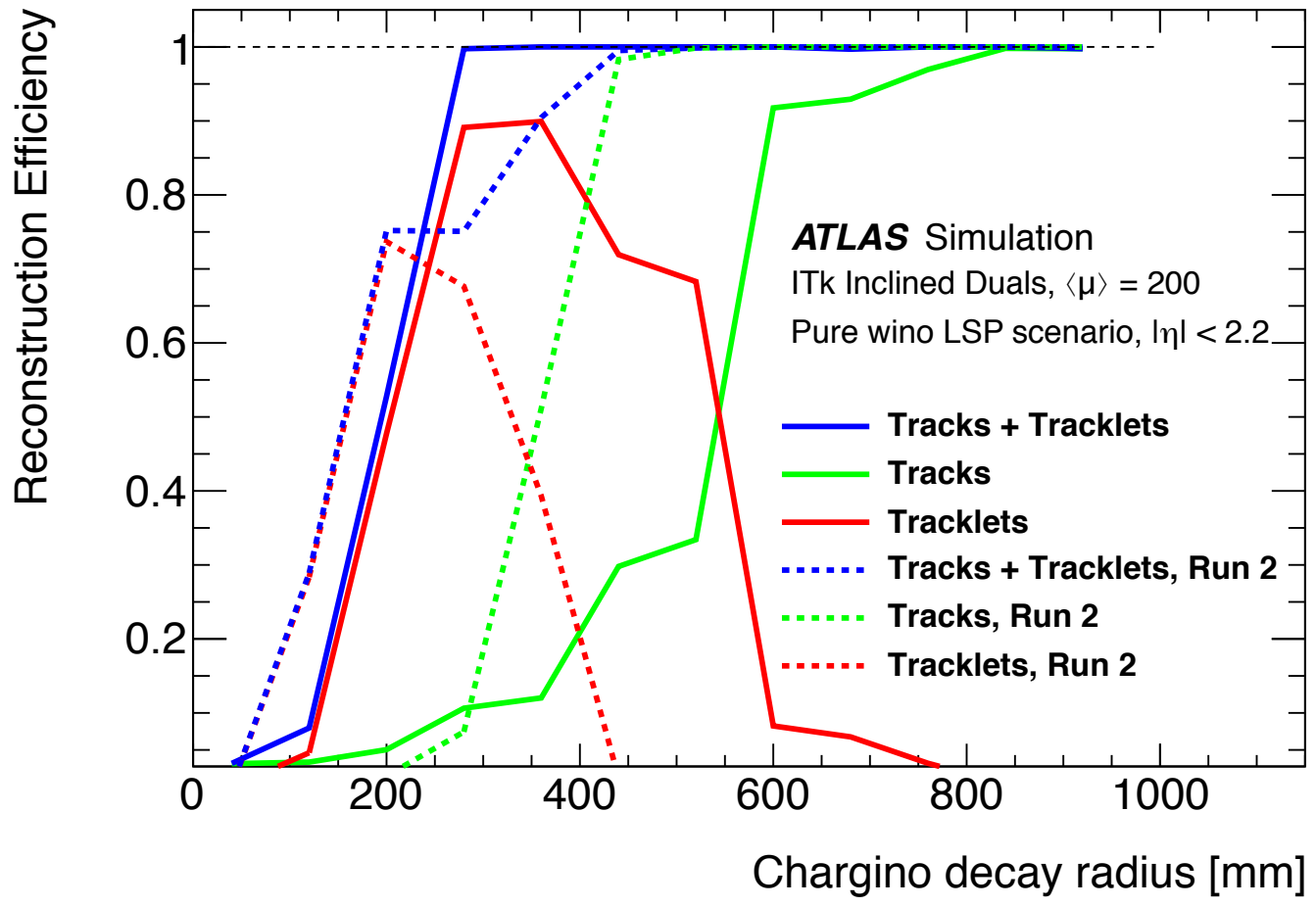

Figure 5.21: Reconstruction efficiency of the disappearing chargino as a function of decay radius using two track reconstruction techniques: "tracklets" refers to the short tracks mentioned in the text, while "tracks" refers to standard track reconstruction. The corresponding reconstruction efficiencies for Run 2 are also shown [347].

Expected limits at 95% C.L. are shown in Figure 5.22 as a function of the chargino mass and lifetime for a pure wino and a pure higgsino LSP scenario. For comparison, the current Run 2 limit for 36.1 $fb^{-1}$ in the wino LSP scenario is also shown. The HL-LHC dataset of 3000 $fb^{-1}$ extends the sensitivity of neutrinos and charginos up to 250 GeV, assuming a pure higgsino scenario.

### *5.1.3.6 Displaced Vertices in ATLAS*

Massive, long-lived particles with lifetimes of the order of O(10) ps to O(10) ns can decay inside the inner tracker into charged and stable particles. The products of these decays are reconstructed as tracks with measurably distant impact parameters with respect to the IP. The reconstruction of such displaced tracks is very challenging compared to the track reconstruction of prompt particles, due to fewer hits along the track and a larger parameter phase space for track finding. In order to identify a displaced vertex, one must first identify the tracks from the decaying LLP.

Several physics models predict the existence of long-lived, massive particles. For example, a standard SUSY scenario can contain a gluino with a mass of 2 TeV and a lifetime of 1 ns. The long-lived gluino hadronises into an $R$-hadron, which then decays into a 100 GeV neutralino and hadrons.

In ATLAS, this topology has been investigated in Run 2 by using a dedicated algorithm for reconstructing displaced vertices [248].

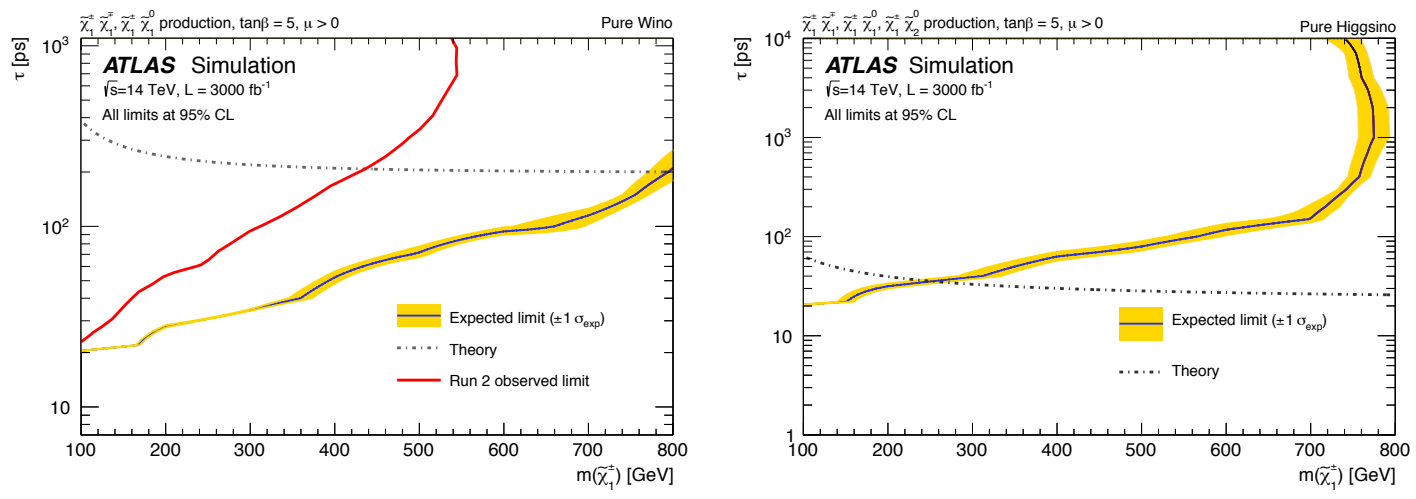

Figure 5.22: Projected 95% C.L. limits on a degenerate chargino-neutralino scenario assuming the chargino is a (left) pure wino; (right) pure neutralino. The limits include both pair production $\tilde{\chi}_1^+\tilde{\chi}_1^-$ and associated production $\tilde{\chi}_1^\pm\tilde{\chi}_1^0$ (the Higgsino model also includes the $\tilde{\chi}_1^\pm\tilde{\chi}_2^0$ mode) [347].

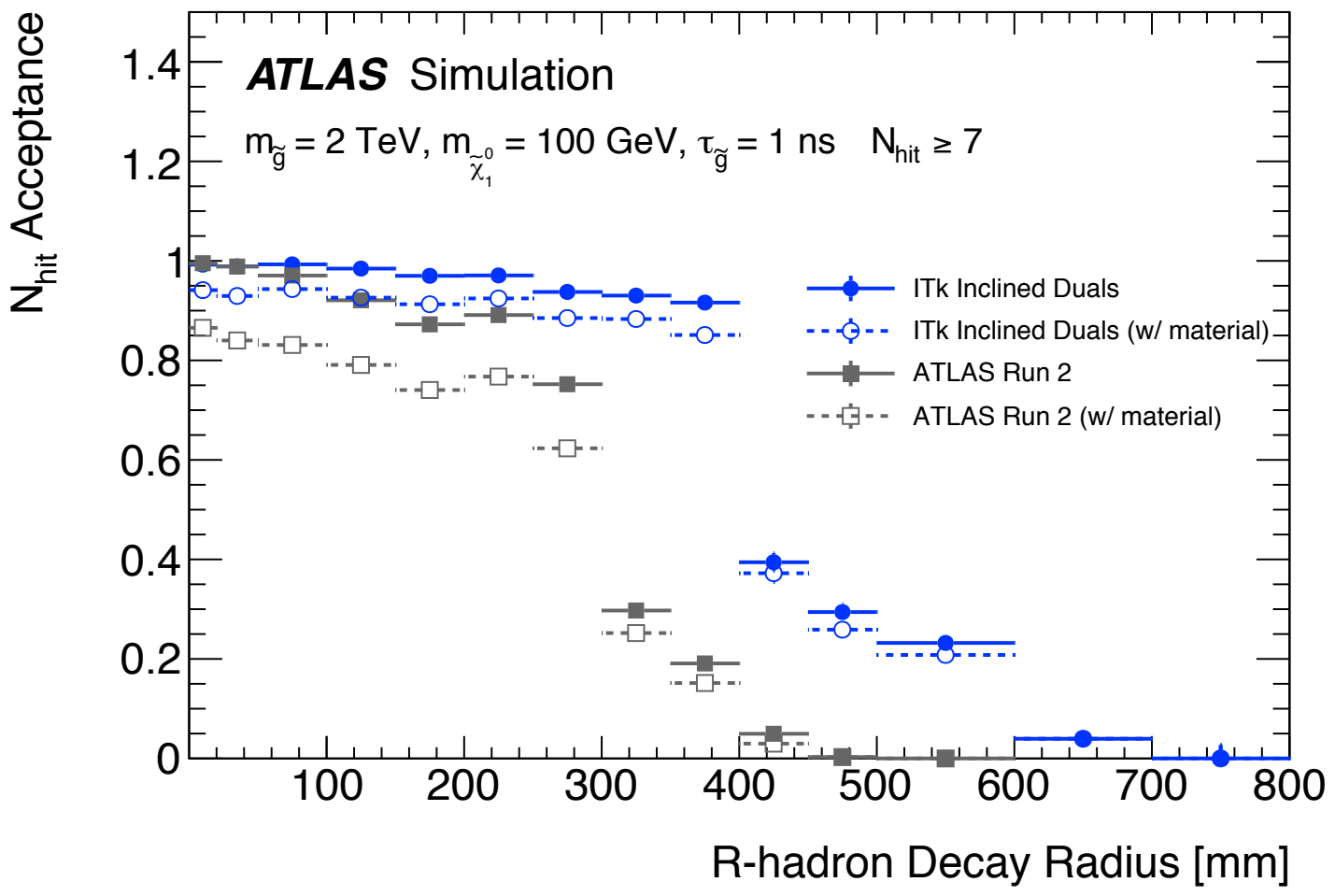

Figure 5.23: Probability that tracks with $p_T > 1$ GeV will traverse at least seven silicon ITk layers if they are produced in the decay of an $R$-hadron with mass 2 TeV and $\tau = 1$ ns. The probability is also shown for at least seven hits with the Run 2 layout [347].

The performance expected to be achieved for the ITk upgrade for this signal has been tested with a simplified simulation which has a description of the ITk active sensors and a modeling of the magnetic field. The kinematics and location of the decay products of the R-hadron are injected into the simulation and their trajectories are extrapolated through the detector model. The probability of producing at least seven silicon hits in the ITk geometry is shown in Figure fig:atlas-tdr-030-fig3-41 for a 2 TeV $R$-hadron with $\tau = 1$ ns. The increased volume and number of ITk layers leads to improved efficiency for the reconstruction of tracks from displaced vertices (and hence the vertices themselves) out to radial distances of up to 550 mm. The increased number of silicon layers also gives more room to veto tracks with missing hits, further suppressing backgrounds.

#### 5.1.3.7 *LLP Searches with Precision Timing at CMS*

The CMS MTD will provide new, powerful information in searches for long-lived particles. In addition to the aforementioned displaced photon search, the additional timing information can be used to provide full kinematic reconstruction of LLP decays, and can be a powerful tool in background suppression.

*Possible Improvements in the Ability to Reconstruct LLP Mass* A precision MIP timing detector allows each reconstructed vertex to be assigned a time and therefore to measure the time of flight of LLPs between primary and secondary vertices. Using the measured displacement between primary and secondary vertices in space and time, the velocity of an LLP in the laboratory frame, $\vec{\beta}^{p}_{LAB}$ (or, equivalently, the boost $\gamma^{p}$), can be measured. In such scenarios, the LLP can decay to fully visible or partially invisible systems. Using the measured energy and momentum of the visible portion of the decay, one can calculate its energy in the LLP rest frame and reconstruct the mass of the LLP, assuming that the mass of the invisible system is known.

This capability can be demonstrated in scenarios where the LLP decay produces a $Z$ boson, which then decays to an electron-positron pair. For example, in the GMSB model where the $\tilde{\chi}^{1}_{0}$ couples to the gravitino $\tilde{G}$ via higher-dimension operators sensitive to the SUSY breaking scale, the $\tilde{\chi}^{1}_{0}$ may have a long lifetime, and can be produced in top-squark pair production with $\tilde{t} \rightarrow t + \tilde{\chi}^{1}_{0}$, $\tilde{\chi}^{1}_{0} \rightarrow Z + \tilde{G}$, $Z \rightarrow ee$.

Studies with simulated event samples have been performed to estimate the possible improved sensitivity of the search with the CMS MTD upgrade. The events are generated with Pythia 8, and the masses of the top-squark and neutralino are set to 1000 GeV and 700 GeV, respectively. Generator-level quantities are smeared according to the expected experimental resolutions. A position resolution of 12 $\mu$m in each of the three spatial directions is assumed for the primary vertex. The secondary vertex position for the electron-positron pair is reconstructed assuming 30 $\mu$m position resolution in the transverse direction. The momentum resolution for electrons is assumed to be 2%. Finally, the time resolution of charged tracks at the displaced vertex is assumed to be 30 ps.

The mass of the LLP is reconstructed assuming that the gravitino is massless. The fraction of events with separation between primary and secondary vertices exceeding $3\sigma$ in both space and time as a function of the MTD resolution is shown in Figure 5.25 (left). The mass resolution, defined as half of the shortest mass interval that contains 68% of events with $3\sigma$ displacement is shown in Figure 5.25 (right), as a function of the MTD resolution.

A similar study has been performed with another SUSY scenario where the two lightest neutralinos and light chargino are Higgsino-like. The light charginos and neutralinos are nearly mass degenerate and may become long-lived as a consequence of the heavy higgsinos. In both studies, the additional timing information from the MTD facilitates the reconstruction of the LLP mass, the resolution and efficiency of which are further improved with the excellent timing resolution of the MTD.

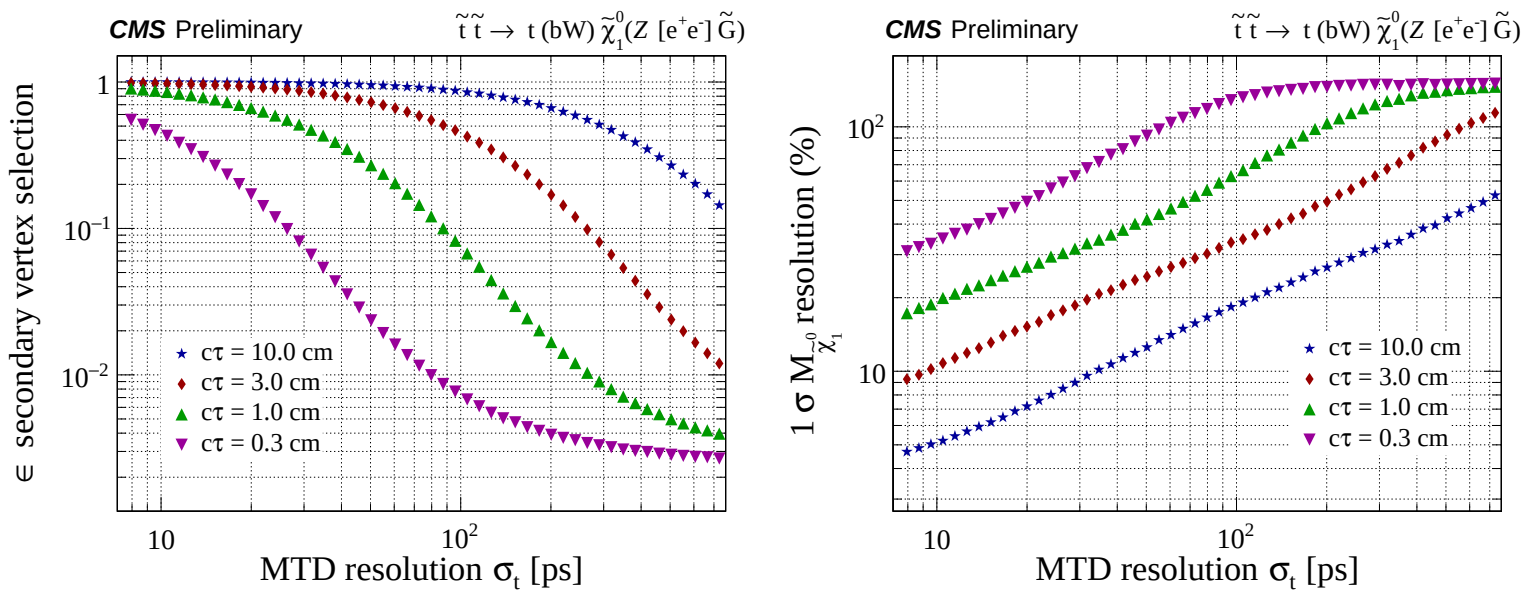

Figure 5.24: Efficiency (left) and mass resolution (right) as a function of the timing resolution of the MTD for reconstruction of the $\tilde{\chi}^1_0$ mass in the SUSY GMSB example of $\tilde{\chi}^1_0 \to \tilde{G}e^+e^-$, with mass of $\tilde{\chi}^1_0 = 700$ GeV, considering events with a separation of primary and secondary vertices by more than $3\sigma$ in both space and time [345].

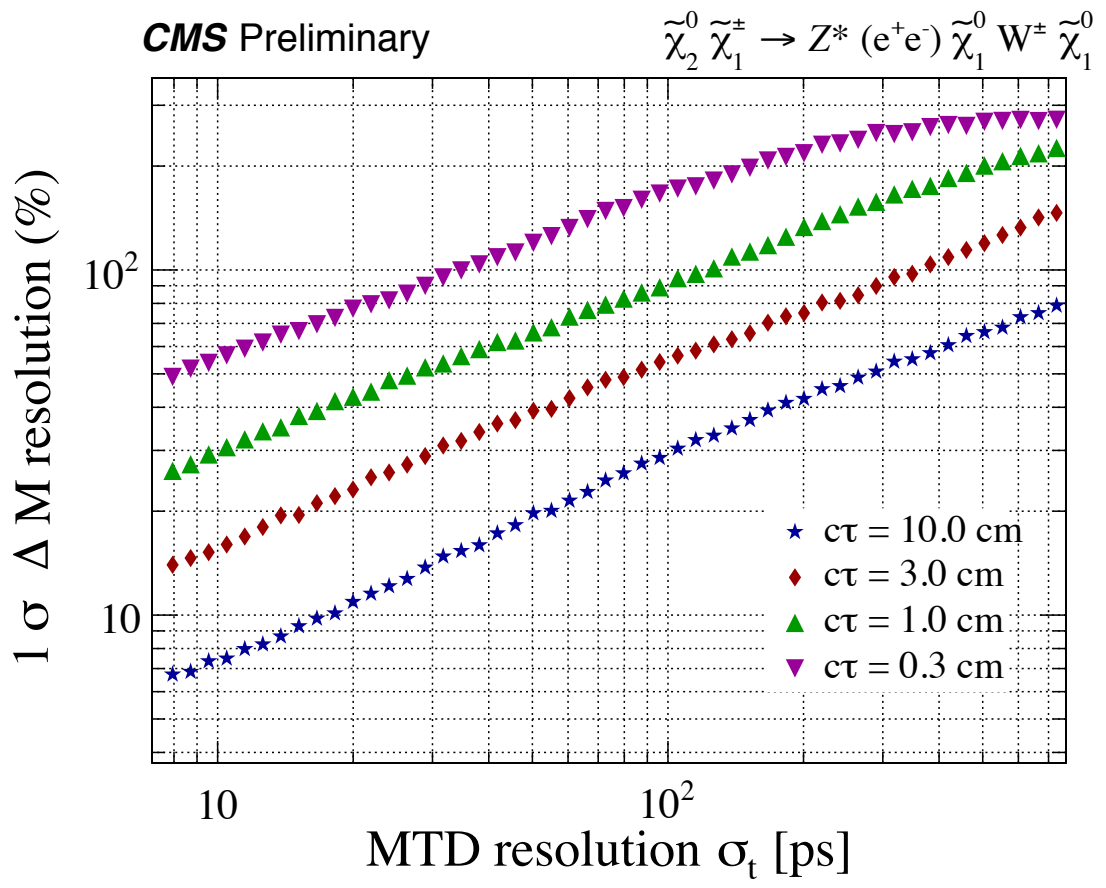

Figure 5.25: Resolution for the reconstruction of the LLP mass in the Higgsino scenario outlined in this section. The LLP mass resolution is shown as a function of lifetime and the MTD resolution [345].

*Possible Improvement to LLPs Searches in General Using Timing Information* Precision timing at CMS will provide a new tool to suppress the background and enhance the reach for LLPs in the HL-LHC era.

A schematic of a typical signal event containing an LLP is shown in Figure 5.26. An LLP, denoted as $X$, travels a distance $\ell_X$ into a detector volume and decays into two light SM particles $a$ and $b$, which then reach a timing layer at a transverse distance $L_{T_2}$ away from the beam axis. In a typical hard collision, the SM particles generally travel very close to the speed of light. Hence, the decay products of $X$ (using particle $a$ as an example) arrives at the timing layer with a time delay of

$$\Delta t = \frac{\ell_X}{\beta_X} + \frac{\ell_a}{\beta_a} - \frac{\ell_{\rm SM}}{\beta_{\rm SM}}, \tag{5.2}$$

with $\beta_a \simeq \beta_{\rm SM} \simeq 1$. An ISR jet could easily be present for all processes, and can be used to "timestamp", i.e., to derive the time of the hard collision at the primary vertex.

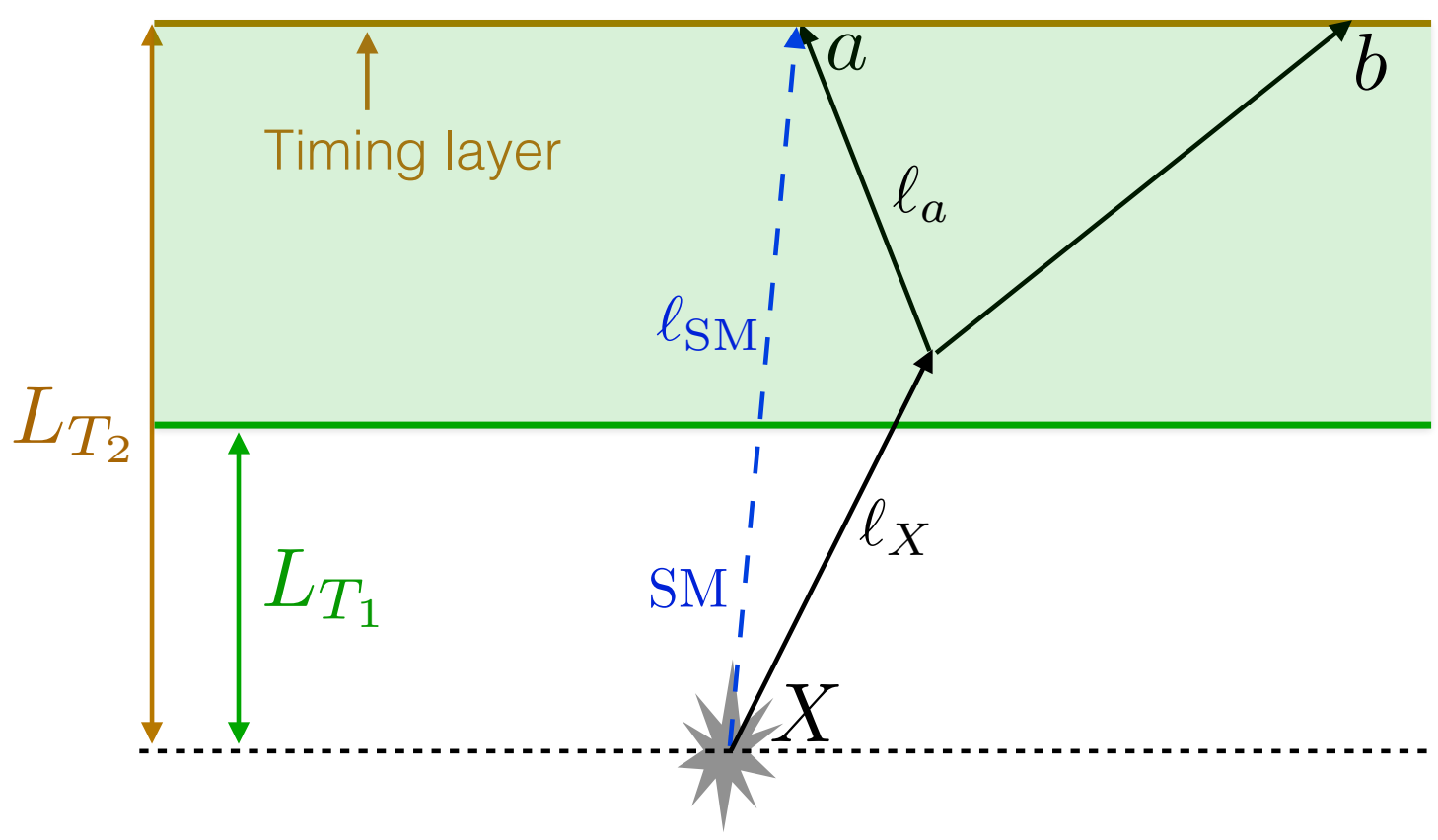

Figure 5.26: An event topology with an LLP $X$ decaying to two light SM particles $a$ and $b$. A timing layer, at a transverse distance $L_{T_2}$ away from the beam axis (horizontal gray dotted line), is placed at the end of the detector volume (shaded region). The trajectory of a potential SM background particle is also shown (blue dashed line). The gray polygon indicates the primary vertex. Taken from Ref. [357].

For the CMS MTD located just outside the tracker volume, $\ell_{\rm SM}/\beta_{\rm SM}$ is about $O(1\text{ ns})$. As a result, with tens of picosecond (ps) timing resolution, a sensitivity to percent-level time delay caused by slow LLP motion, e.g., $1-\beta_X > 0.01$ with boost factor $\gamma < 7$, is expected to be achieved.

A theory study has been done in Ref. [357] examining potential gains in sensitivity for hadronic displaced vertex reconstruction using timing information. A new trigger strategy for a delayed jet is studied by comparing the prompt jet with $p_{\rm T} > 30$ GeV that reconstructs the four-dimensional primary vertex (PV4d) with the arrival time of another jet at the timing layer. The delayed and displaced jet signal, after requiring a minimal decay transverse distance of 0.2 m ($L_{T_1}$), will typically not have good tracks associated with it. Consequently, the major SM background is from trackless jets. The origins of this background can be classified into two categories: hard collision from a same vertex (SV), and pile-up (PU) from different vertices. Other types of background such as cosmic rays, beam halo, material interactions, etc, are out-of-time and will become important after most of the hard collision background is removed using selections based on reconstructing vertices and timing delay. Out-of-time backgrounds may have other distinctive features that would allow their effective suppression. While the study in Ref. [357] can serve as an optimistic inspiration for increasing LLP sensitivity with timing information, detailed experimental studies are needed to determine the actual sensitivity gain.

The jet faking a displaced signal, behaving as a trackless jet, has an intrinsic time delay $\Delta t = 0$. However, due to the limited timing resolution in reconstructing the PV4d, it can have a time spread. The background differential distribution with respect to apparent delay time ($\Delta t$) can be estimated as

$$\frac{\partial N_{\rm bkg}(t)^{\rm SV}}{\partial \Delta t} = N_{\rm bkg}^{\rm SV} \mathcal{P}(\Delta t; \delta_t^{\rm SV}). \tag{5.3}$$

The time delay selection on $\Delta t$ reduces such a background through a very small factor of $\mathcal{P}(\Delta t; \delta_t^{\rm SV})$ for large $\Delta t / \delta_t^{\rm SV}$, where $\delta_t^{\rm SV}$ is the timing resolution for the SV background, dominant by the timing detector resolution. The LLP signal pays a much smaller penalty factor than the background due to its intrinsic delay.

The background from the pile-up requires the coincidence of a triggered hard event and objects from a pile-up (hard) collision whose PV4d fails to be reconstructed and that can mimic a signal. The differential background from pile-up can be estimated as

$$\frac{\partial N_{\rm bkg}^{\rm PU}(\Delta t)}{\partial \Delta t} \simeq N_{\rm bkg}^{\rm PU} \mathcal{P}(\Delta t; \delta_t^{\rm PU}). \tag{5.4}$$

Similar to Eq. 5.3, $\delta_t^{\rm PU}$ is the timing resolution for the PU background, dominated by the beam spread. The key difference between the background from the pile-up and the background from the same hard collision is that the typical time spread is determined by the beam property for the former, and by the timing resolution for the latter. They typically differ by a factor of a few, e.g., 190 ps versus 30 ps for CMS with the current upgrade plan.

Using this estimation, a sensitivity projection is presented with an example signal of a Higgs boson decaying to LLPs with the subsequent decay of the LLPs into $b\bar{b}$ pairs, with only minimal requirements of one low-$p_{\rm T}$ ISR jet, with $p_{\rm T,j} > 30$ GeV and $|\eta_j| < 2.5$, and at least one LLP decay inside the detector. Timing information is used to suppress backgrounds. The 95% C.L. sensitivity is shown in Figure 5.27. The decay branching ratio of the LLP $X \to jj$ is assumed to be 100%. The projection with 30 ps timing resolution of the CMS MTD is plotted with thick dashed lines. Compared to other 13 TeV HL-LHC projections (with 3 ab$^{-1}$ of integrated luminosity) without the timing information (shown in thinner, dotted and dashed lines), it is suggested that the addition of a selection on timing, under the set of assumptions described above, greatly reduces background and improves sensitivity. The possibility of such a significant improvement is enticing; this projection is determined from theoretical studies [357], however, and it remains to be seen whether it is possible to realize these gains in the experiment. For example, the low threshold of the single jet would require timing information at early levels of the trigger, and out-of-time backgrounds could also reduce the gains from timing information. Nevertheless, this high-level theory analysis provides an inspiration to the experimental collaborations to perform more detailed, internal studies that will ultimately determine how realistic the projections are.

In general, the prospect of improvements in LLP searches at the HL-LHC due to precision timing upgrades for CMS (and ATLAS) remains understudied, and deserves more comprehensive experimental and phenomenological studies, including understanding and reducing out-of-time backgrounds.

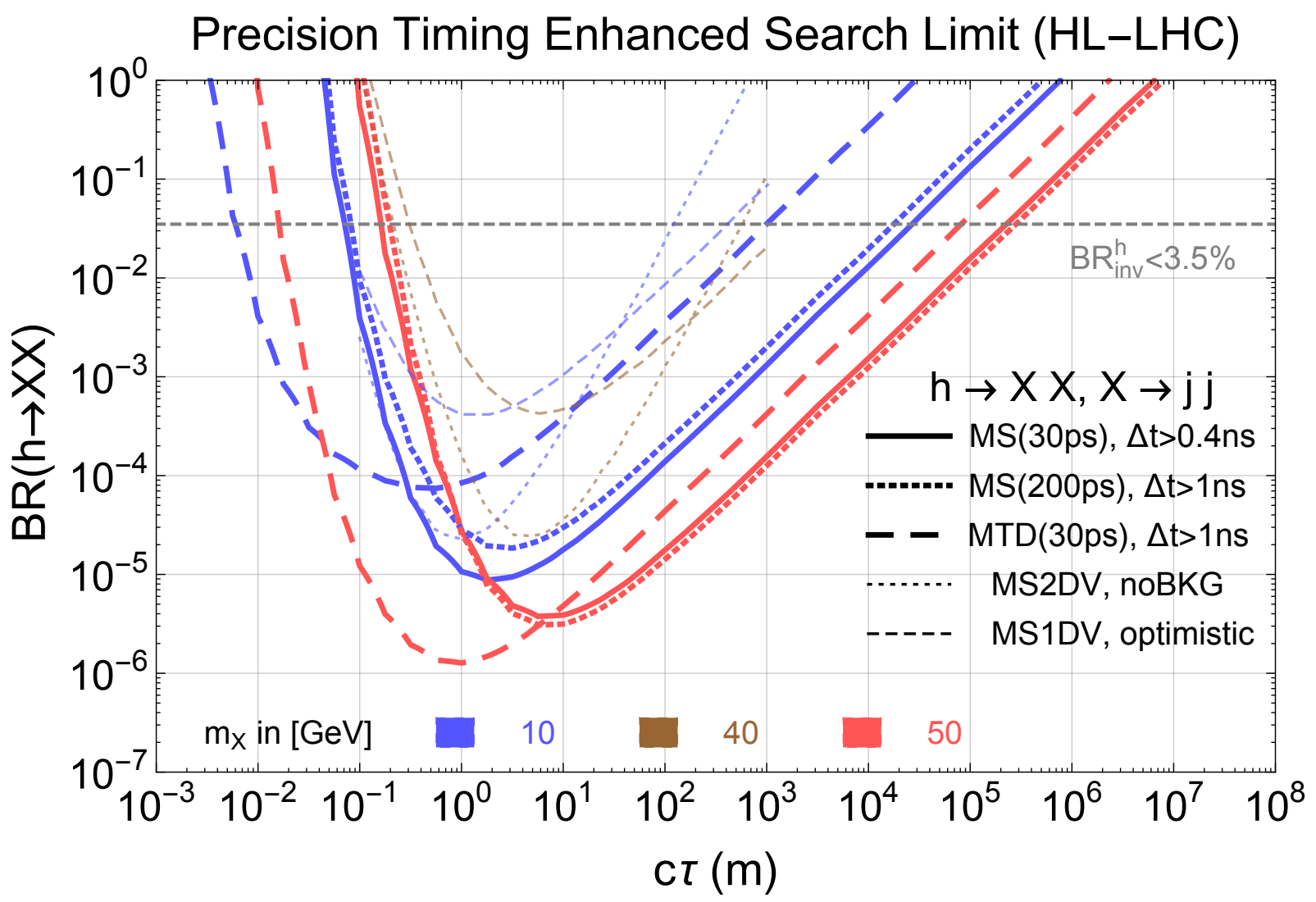

Figure 5.27: Theory projection from Ref. [357] of the 95% C.L. limit on BR$(h \to XX)$, where $X$ is an LLP, for the production of $pp \to jh$ with the subsequent decay of $h \to XX$ and $X \to jj$ subject to assumptions in the text (one ISR jet with $p_{T,j} > 30$ GeV and $|\eta_j| < 2.5$, and at least one LLP inside the detector). Different colors indicate different masses of the particle $X$. The thick, long-dashed lines indicate searches with the CMS MTD plus the timing requirements. The thick solid and dotted lines indicate searches with a hypothetical timing layer outside the ATLAS muon spectrometer plus timing requirements. The numbers in parentheses are the assumed timing resolutions. This provides motivation to see whether these gains can be realized by studies from within the collaborations.

### *5.1.3.8 LLP Searches with a Level-1 Track Trigger in CMS*

As discussed in Section 5.1.1, a central feature of the CMS upgrade at the HL-LHC will be a new silicon outer tracker which allows track reconstruction for every LHC bunch crossing (at a rate of 40 MHz). The $p_T$ selection for stubs (hit pairs in the $p_T$ modules of the outer tracker) to be read out is determined by the bandwidth from the detector to the back end electronics, and is fixed at about 2 GeV. On the other hand, the choice of track finding algorithm and hardware is still being finalized, and there could be significant benefits to extending the L1 track trigger capabilities to trigger on off-pointing tracks.

To illustrate the case, a simple toy simulation to study rare Higgs boson decays into new particles with lifetimes of order of a few mm has been performed [358]. This study considers all-hadronic final states with low $H_T$, taking SM Higgs boson decays into four jets as an example. Theoretical motivation to look for such decays is very strong; see Section 2.2 for a detailed discussion. In particular, there is currently a blind spot for lifetimes of order 1 cm in searches for new long-lived scalars in Higgs decays, i.e. $h \to \phi\phi$. The goal is to probe very small branching fractions of the Higgs boson, so for this study, $BR[h \to \phi\phi \to 4q] = 10^{-5}$ is assumed. For prompt decays, the background is overwhelming, but if the $\phi$ has $c\tau$ of a few mm, the offline analysis has very low backgrounds. The problem is getting such events on tape, in particular through L1. This toy study estimates how an off-pointing track reconstruction at L1 can help. To estimate the efficacy of the approach, the resulting projections are compared with the best alternatives in the absence of an off-pointing track trigger, by using associated Higgs boson production with a $W$ boson that decays leptonically to pass a lepton trigger, or considering L1 calorimeter jets with no associated prompt tracks.

Once these positive results were obtained, a more detailed study using the full Phase 2 simulation of the CMS detector was performed [359]. The more mature exploration found that a plausible extension of the L1 track trigger to tracks with an impact parameter of a few centimeters results in dramatic gains in the trigger efficiency. The gains are even larger for additional heavy SM-like Higgs bosons with the same decay. These results are in agreement with the toy study described above. A few details of the mature study will be described below.

The study focuses on small or moderate decay lengths of the new particles, 1–50 mm, and assumes that the offline selection can remove all SM backgrounds with only a moderate loss of efficiency. While this study focuses on the specific Higgs boson decay to light scalars, the results and the proposed triggers are relevant for a broad spectrum of new physics searches, with or without macroscopic decay lengths.

The authors propose a simple jet clustering algorithm implementable in firmware, and compare it with anti-$k_t$ jets [360] with a size parameter of $R = 0.3$, as produced by FastJet [361]. This simple algorithm produces a similar performance, in both L1 trigger efficiency and rate, to the full jet clustering using the anti-$k_t$ algorithm.

Then, the performance of an algorithm for reconstruction of tracks with non-zero impact parameter is presented. This approach extends the baseline CMS L1 Track Trigger design to handle tracks with non-zero impact parameter and to include the impact parame-ter in the track fit. This enhanced design is feasible without greatly altering the track finding approach, but will require more FPGA computational power than the current proposal, which only considers only prompt tracks. Tracks passing the selection are clustered using the same algorithm as described above, and clusters containing tracks with high impact parameters are flagged as displaced jets. Though the baseline design of the L1 Track Trigger currently is optimized to find prompt tracks, these studies show that an enhanced L1 Track Trigger can extend the L1 trigger acceptance to include new BSM physics signals.

For now, the extended L1 track reconstruction is limited to the barrel region. In order to compare the results with prompt and extended track reconstruction, one needs to make a correction for the rapidity coverage: prompt tracks are found in $|\eta| < 2.4$, while the extended track algorithm currently only reconstructs tracks in $|\eta| < 1.0$. To scale the efficiency for finding track jets to the full $|\eta| < 2.4$ range, a scale factor based on efficiency in the full $\eta$ range and the central $\eta$ range was used. The scale factor was comparable to the increase in L1 rate.

Figure 5.28 shows the expected trigger rate as a function of efficiency for the SM and the heavy SM-like Higgs bosons.

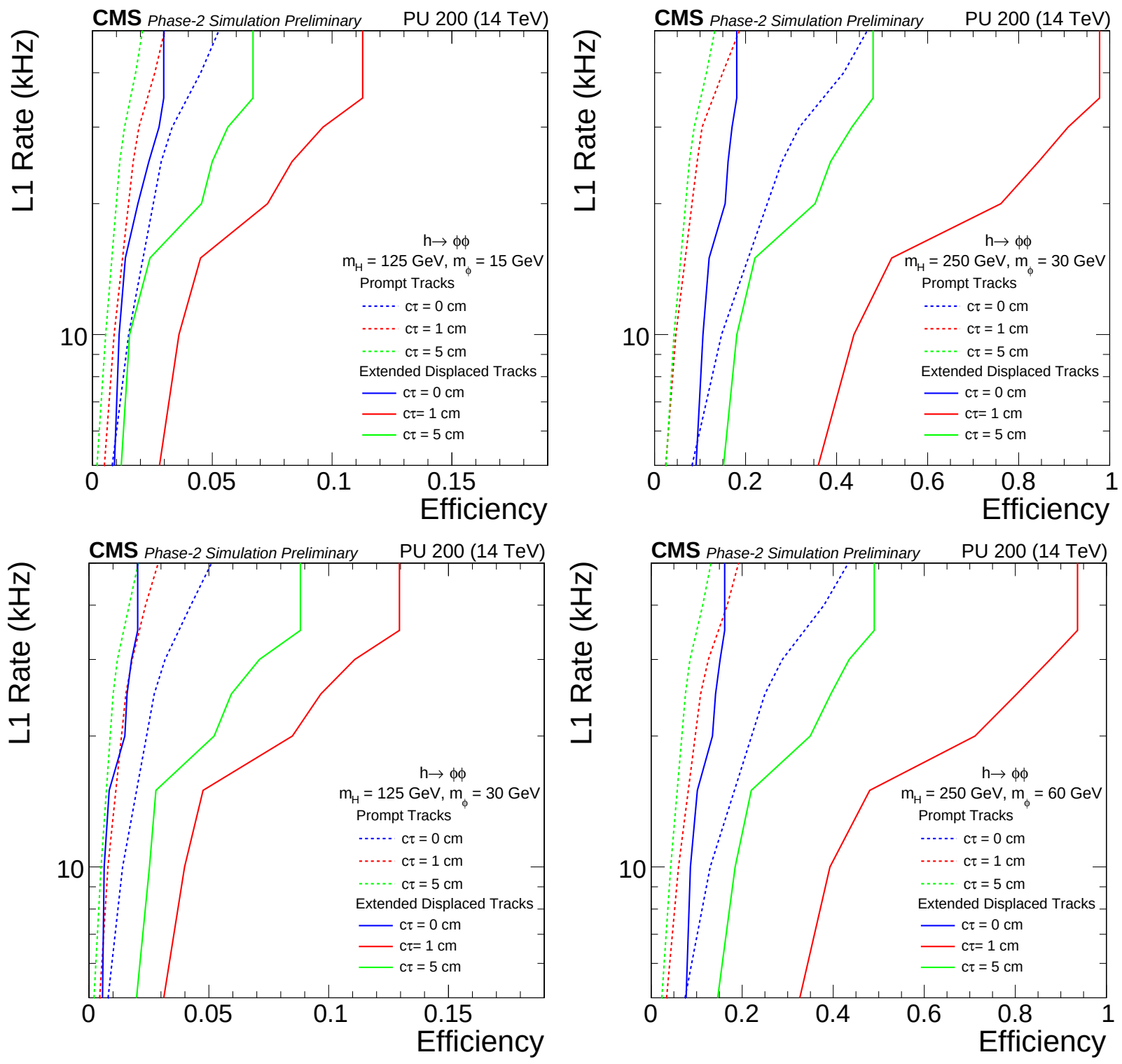

Figure 5.28: The rate of the track jet $H_T$ trigger as a function of signal efficiency using extended track finding for the SM Higgs (left) and the heavy SM-like Higgs (right). The extended track finding performance is extrapolated to the full outer tracker acceptance as described in the text and in Ref. [359].

The available bandwith for the triggers described above, if implemented, will be decided as a part of the full trigger menu optimization. Here, two cases are considered: 5 and 25 kHz. The expected event yield for triggers using extended and prompt tracking are shown in Figure 5.29, assuming branching fraction $\mathcal{B}[h \to \phi\phi] = 10^{-5}$ for the SM Higgs boson. For the heavy Higgs boson, the expected number of produced signal events is set to be the same as for the SM Higgs by requiring $\sigma_{pp\to h(250)}\mathcal{B}[\Phi \to \phi\phi] = 10^{-5}\sigma_{pp\to h(125)}$.

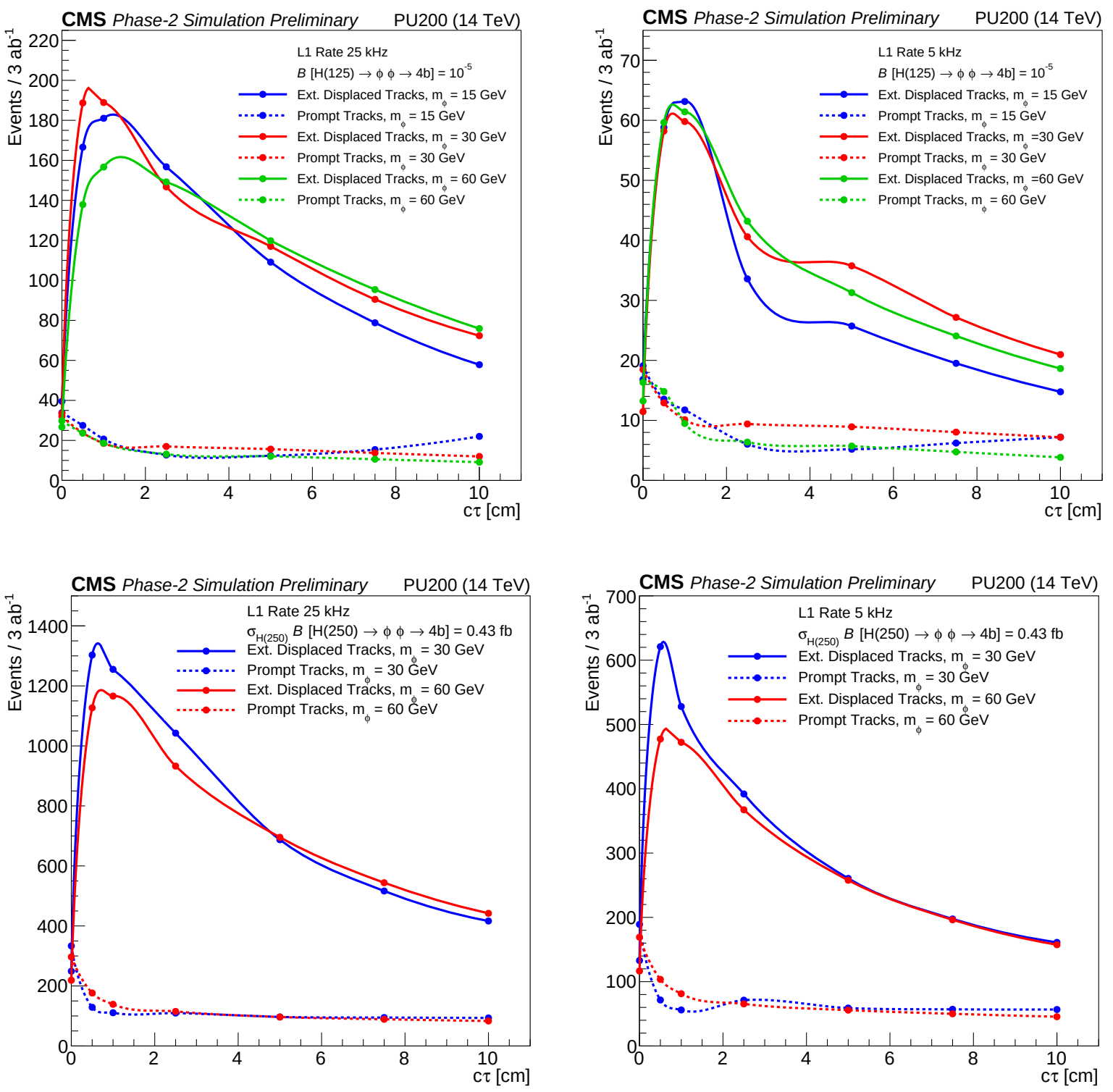

Figure 5.29: This plot shows the number of triggered Higgs events (assuming $\mathcal{B}[h \to \phi\phi] = 10^{-5}$, corresponding to 1700 events) as a function of $c\tau$ for two choices for the trigger rates: 25 kHz (left), 5 kHz (right). Two triggers are compared: one based on prompt track finding (dotted lines) and another that is based on extended track finding with a displaced jet tag (solid lines) [359].

For the exotic Higgs decays considered, given the total Phase 2 dataset of 3 inverse attobarns and branching fraction of $10^{-5}$, CMS would collect $\mathcal{O}(10)$ events, which should be sufficient for discovery. A plausible extension of the L1 track finder was considered, to select tracks with impact parameters of a few cm. That approach improves the yield by more than an order of magnitude. The gains for the extended L1 track finding are even larger for the events with larger $H_T$, as demonstrated by the simulations of heavy Higgs boson decays.

### 5.1.4 Open Questions and New Ideas

The higher data rate and more powerful machinery in the high-luminosity era bring new prospects to LLP searches. Searches that are too challenging for the current machine may become feasible with the HL-LHC upgrades, and it is important to explore such possibilities to the full extent. Existing search methods, triggering, and reconstruction algorithms should also be updated to take full advantage of the new hardware capabilities. As the upgrade scope is being defined and finalized in the near future, it is of particular importance to evaluate the physics cases, which will also motivate the upgrade and inform its designs.

#### 5.1.4.1 New Studies for the HL-LHC

In Section 5.1.3, various analyses with displaced signatures are presented in the context of how detector and trigger upgrades at the ATLAS and CMS experiments for the HL-LHC can improve their sensitivity. As was shown in Section 5.1.1, while the upgrades for both experiments differ in detail, they are similar in concept and scope. It is therefore important to perform the same LLP search projections for both experiments to compare performance and evaluate the complementarity of the two experiments.

For example, both experiments explore the idea of a fast timing detector at HL-LHC. The CMS MTD aims for full-coverage of a barrel layer plus endcap disks, while the ATLAS HGTD plans for multiple layers of endcaps. How the different $\eta$ coverage of the timing layer might impact LLP searches will be an interesting question to answer. In the case of calorimetry, the ATLAS LAr electromagnetic calorimeter is segmented in $z$, providing the additional pointing capability to find displaced photons. On the other hand, the CMS ECAL electronics upgrade in the barrel, and HGCAL upgrade in the endcaps, will considerably improve its timing resolution to identify photons that are out-of-time. A sensitivity comparison of both experiments will provide helpful information in understanding photon reconstruction and the combined reach of both experiments.

Additional opportunities exist for improving LLP sensitivity in Run 3 or at the HL-LHC, such as the ability to perform an analysis using physics objects at the trigger level, rather than their more complex, offline counterparts. This analysis method — sometimes referred to as *trigger-level analysis* or *data-scouting* — has been used in searches for di-jet resonances using data from Runs 1 and 2 by the CMS Collaboration [362, 363], data from Run 2 by the ATLAS Collaboration [364] and by the LHCb Collaboration for the searches for dark photons [262]. Such data-scouting methods can be used, for example, to reduce the $p_{\mathrm{T}}$ thresholds of jets and muons used in dark photon or hidden particle searches. Such possibilities could potentially assist searches for LLPs by, for example, allowing changes in the L1 muon patterns in order to trigger on non-pointing muons.

Moreover, the applications of machine learning techniques to LLP searches is a currently under-studied realm. Such techniques could be used in the context of LLPs in particle identification, reconstruction, or generation, in addition to analysis or reinterpretation. Further studies to supplement those that already exist are of high priority for the community.

Furthermore, as new models are being proposed and new channels open up in the realm of LLP searches, many with final states challenging for the current detector conditions, it is important to evaluate their sensitivities in the high luminosity era. Here are a few such interesting searches that are on the agenda.

- **Inelastic dark matter**

  While stringent limits are currently placed on WIMP-type dark matter from direct detection, indirect detection, and collider searches, dark matter particles can exist in a hidden sector with additional particles and forces. A representative example of a hidden sector consists of a dark matter particle which is charged under a hidden gauge or global symmetry. The DM can have both a symmetry-preserving mass and, if the symmetry is spontaneously broken, also a symmetry-violating mass, which splits the mass eigenstates. This inelastic dark matter (iDM) scenario [125, 126, 128] consists of two DM states that couple only off-diagonally to one another. Such iDM can be probed at colliders with the production of $\mathrm{DM} + \mathrm{DM}^*$ (where $\mathrm{DM}^*$ denotes the heavier DM state) in association with a hard SM object $X$, followed by the subsequent decay of $\mathrm{DM}^* \to \mathrm{DM} + Y$ for some potentially different SM states $Y$. The production is summarized as

$$\begin{aligned} pp &\to X + \mathrm{DM} + \mathrm{DM}^* \\ &\to X + \mathrm{DM} + \Big(\mathrm{DM}^* \to \mathrm{DM} + Y\Big) \equiv X + \not{E}_\mathrm{T} + Y \,, \end{aligned}$$

  where $X$ is any state that can be used to trigger on the event and reconstruct $\not{E}_\mathrm{T}$, such as an ISR jet, and $Y$ depends on the mode by which DM couples inelastically to the SM.

  One concrete, representative version of such models can be realized when the mediator is a dark photon and $Y$ is a pair of leptons. The weak coupling between the SM and the hidden sector suggest the heavy eigenstate is meta-stable, creating a displaced signature. Because the mass splitting is small between the two eigenstates, the lepton pair is also typically softer compared with GMSB models, with $p_\mathrm{T}$ values of a few to tens of GeV for a $O(10\ \mathrm{GeV})$ DM. The displaced muon trigger and reconstruction strategy with the muon system upgrade at CMS can likely improve searches for this scenario. The soft $p_\mathrm{T}$ spectrum in this case particularly motivates the lowering of the $p_\mathrm{T}$ threshold in the displaced muon trigger turn-on. Moreover, the additional timing information from the fast-timing detector opens up the possibility of reconstructing the mass splitting, while taking advantage of the good resolution of the timing detector for even low-$p_T$ particles. Sensitivity studies for iDM with a dark photon mediator are planned for the CMS MTD upgrade. The projections will also be of value to searches for other types of dark sector models, such as self-interacting dark matter [365], that give rise to soft displaced lepton pairs.

- **Dark showers**

  If the hidden sector has a QCD-like structure with dark quarks and hidden forces, a mediator between the SM and the hidden sector, such as a $Z'$ or heavy Higgs boson, can decay into some number of dark quarks that subsequently shower and hadronize into dark mesons, some of which are meta-stable and decay back into SM particles after a macroscopic distance from the proton-proton interaction point. A showering dark sector can yield a particularly rich collider phenomenology that may give rise to a high multiplicity of displaced objects often low in $p_T$. Depending on the final state, searches for a hadronic shower with emerging [325] or semi-visible [366] jets can be improved with the increased acceptance and enhanced resolution due to the tracker upgrades at both ATLAS and CMS, as well as the finer granularity endcap calorimeter (HGCAL) at CMS. A dark shower with displaced photon pairs can benefit from the improved timing resolution with the ECAL electronics upgrade, as well as the new timing detectors. A dark shower with displaced lepton pairs, similar to the aforementioned case of inelastic dark matter, can be probed with better sensitivity with the muon system upgrades, as well as the fast timing information. For more details on dark showers, see Chapter 7.

#### *5.1.4.2 New Detectors at Future Collider*

If we are only limited by our imagination, what new detectors may exist at a future collider that can open up new capabilities for LLP searches? Regardless of practical constraints, studies for such bold new proposals also help us understand our current experimental reach and optimize our search strategies.

- **4-dimensional tracker with timing**

  Extending from the "timing layer" approach for HL-LHC upgrades, it is clear more time measurements on the track are favorable. In particular, the time of flight between layers is a powerful discriminant for choosing hits on a track. A future tracker that provides 4-dimensional information, including precise timing information for every layer, can significantly improve track reconstruction by reducing combinatorics, providing purer track seeds, and remove fake stubs. Assuming high-$p_T$ particles, each layer's time measurement is advanced by 30 ps and so preserves the differences in vertex times. Low $p_T$ particles will have even more discrimination power since their paths leave longer times between hits in consecutive layers of the tracker.

Major technical advances will need to happen to achieve such an implementation, including development of fine-pitch sensors with good timing resolution, improvements in scaling and power consumption, electronics upgrade, and alterations to present pattern recognition methods in track finding.

- **Timing detector outside ATLAS Muon System**

  As discussed in Section 5.1.3, by using a prompt object, such as an ISR jet, to "timestamp" an event, and requiring a timing delay from the LLP, the new timing layer (MTD) at CMS can help significantly reduce the background and improve LLP search sensitivity. The CMS MTD is located between the tracker and the ECAL with a distance to the beamline of about 1.2 m. If one imagines a fast timing layer outside the ATLAS muon system (MS) with a distance of approximately 10 m [357], timing information at such distances could provide additional discriminating power for particles with longer lifetimes. In Figure 5.27, the projection with a 30 ps timing resolution of such a hypothetical detector outside the ATLAS muon system is plotted in a thick solid line. A less-precise timing resolution (150 ps) has been also considered with a selection $\Delta t > 1$ ns to suppress background. While this study is optimistic in assuming that selections on timing can eliminate all backgrounds to the LLP search, it nevertheless serves as important inspiration for more detailed experimental studies to understand the actual extent of the sensitivity gain. The LLP efficiency is largely unaffected by this change, while low-mass LLPs lose sensitivity by a factor of a few.

  The CMS MTD timing upgrade for the HL-LHC already provides significant improvement [357]. The timing detector outside the ATLAS muon system has the notable benefits of lower background, a larger volume for the LLP to decay and more substantial time delay for the LLP signal due to longer travel distance. Moreover, due to the extended time delay of the LLPs in the volume of the muon system, less-precise timing can still achieve similar physics goals. As a result, with the above caveats it can serve as an estimate of the best achievable sensitivity using timing information in LLP searches.

- **New double-sided tracking layer very close to the beamline**

  Inspired by the L1 track trigger design at CMS for the HL-LHC, a scenario may be imagined where an additional tracking layer close to the beamline is added, mechanics and radiation hardness permitting. This would allow a track veto close to the IP to be implemented, and extend the LLP sensitivity to even smaller displacements. Moreover, if such a layer can be designed with double-sided modules, similar to the CMS outer tracker approach for the HL-LHC, hit pairs (stubs) from this layer can be used in a L1 track trigger for displaced tracks or disappearing tracks. The feasibility of such an approach would depend on the ability to store and process data at a rate not allowed by current technology. On the other hand, it is helpful to consider more innovative data-scouting methods at trigger level to store and select information, or introduce additional discriminating factors to reduce the rate for particular data streams, for the potential of exploring such signatures.

## 5.2 *LHCb Upgrade*

The LHCb experiment is designed to detect decays of long-lived particles in the Standard Model, namely bottom and charmed hadrons. As such, it is naturally suited for the search of BSM long-lived particles in a mass and lifetime range comparable to these hadrons. It is the only LHC experiment to be fully instrumented in the forward region $2 < \eta < 5$, where *b*- and *c*-hadrons are abundantly produced and their decay length is enhanced due to the large longitudinal boost. In this region, detector occupancy is extremely high and thus the LHCb experiment has been run at reduced luminosity compared to ATLAS and CMS during Runs 1 and 2. However, an upgrade of the detector is planned to allow running at a luminosity of $2 \times 10^{33}$ cm$^{-2}$s$^{-1}$ in LHC Run 3 (starting in 2020) while maintaining or improving the current physics performance [367]. This is five times larger than the luminosity during Runs 1 and 2. This first upgrade phase (Phase 1a) will entail a novel trigger paradigm where all sub-detectors are fast enough to be read out in real time and the first trigger decisions are done in software. This trigger scheme is flexible and offers a great opportunity for searches of striking signatures like those of BSM long-lived particles. This upgrade comes earlier than the planned ATLAS and CMS upgrades for the HL-LHC phase which are planned to be installed during LHC long shutdown 3 (by 2025). In this shutdown, LHCb plans to consolidate and modestly enhance the Phase 1a upgrade detector (Phase 1b), while a Phase 2 upgrade to run at an even higher luminosity up to $\sim 2 \times 10^{34}$cm$^{-2}$ s$^{-1}$ is planned to be installed later, during long shutdown 4 (by 2030) [368].

Section 5.2.1 gives a brief overview of the Phase 1 upgraded-LHCb detector design and the expected performance of the LHCb sub-detectors. An overview of the upgraded LHCb capabilities in the context of LLP searches is given in Section 5.2.2 with a few example signatures. Finally, an overview of the plans for the Phase 2 upgrade and some thoughts on the opportunities given by putative additional detector features are reported in Section 5.2.3.

### 5.2.1 LHCb Detector and Trigger Upgrade for Run 3 (Phase 1a)

In LHC Run 3, LHCb plans to take data at an instantaneous luminosity of $2 \times 10^{33}$ cm$^{-2}$s$^{-1}$, a factor of five larger than the current luminosity. The LHCb detector needs to be upgraded to cope with the higher radiation dose and, most importantly, to avoid the saturation of the trigger rate and exploit the higher luminosity. The main bottleneck in the current trigger is in the first stage, which reduces the accepted rate from 30 MHz to 1 MHz at hardware level. For the upgrade, the hardware trigger will be removed and the full event will be read out at the bunch crossing rate of the LHC (40 MHz), with a flexible software-based trigger.

| | Current | Upgrade 1a | Upgrade 1b | Upgrade 2 |
|---|---|---|---|---|
| $\mathcal{L}/(\mathrm{cm}^{-2}\mathrm{s}^{-1})$ | $4 \times 10^{32}$ | $2 \times 10^{33}$ | $2 \times 10^{33}$ | $2 \times 10^{34}$ |
| $\int \mathcal{L}$ | 8 fb$^{-1}$ | 23 fb$^{-1}$ | 50 fb$^{-1}$ | 300 fb$^{-1}$ |
| $\sqrt{s}$ | 7, 8, 13 TeV | 14 TeV | 14 TeV | 14 TeV |
| $\mu$ | $\sim 1$ | $\sim 5$ | $\sim 5$ | $\sim 50$ |

Table 5.1: LHCb current and upgraded operating conditions [368]. Instantaneous luminosity $\mathcal{L}$, integrated luminosity $\int \mathcal{L}$ (including previous runs), $pp$ collision energy $\sqrt{s}$ and average number of visible proton interactions $\mu$ are listed.

In Table 5.1 the current and upgraded conditions of the detector are summarised. To cope with the larger occupancy and higher rate of the upgraded detector, the electronics of all of the sub-detectors must be upgraded, and some sub-detectors must be fully replaced. For example, the tracking system, which plays a crucial role in LLP searches, must be replaced.

The upgraded tracking system consists of the VErtex LOcator (VELO), surrounding the interaction point, the Upstream Tracker, a tracking station placed before the magnet, and the Scintillating Fibers tracker (SciFi), three stations after the magnet. In the VELO [369], the current strips will be replaced by pixel detectors, with a custom developed ASIC (VeloPix) able to cope with a maximum hit rate of 900 Mhits/s/ASIC. The Upstream Tracker [370] is composed of four silicon micro-strip planes, with finer granularity and larger acceptance compared to the current tracker. Each station of the SciFi has four planes of 2.5 m long scintillating fibres read out by silicon photo-multipliers.

The upgrade components most important for LLP searches are the VELO, the tracking and the software trigger. In the following subsections a brief description of the design and capabilities is given for each.

#### 5.2.1.1 VELO Upgrade

The VELO plays a fundamental role in LLP searches at LHCb: due to the large boost particles typically experience in the forward direction, a precise measurement of the LLP vertex position allows

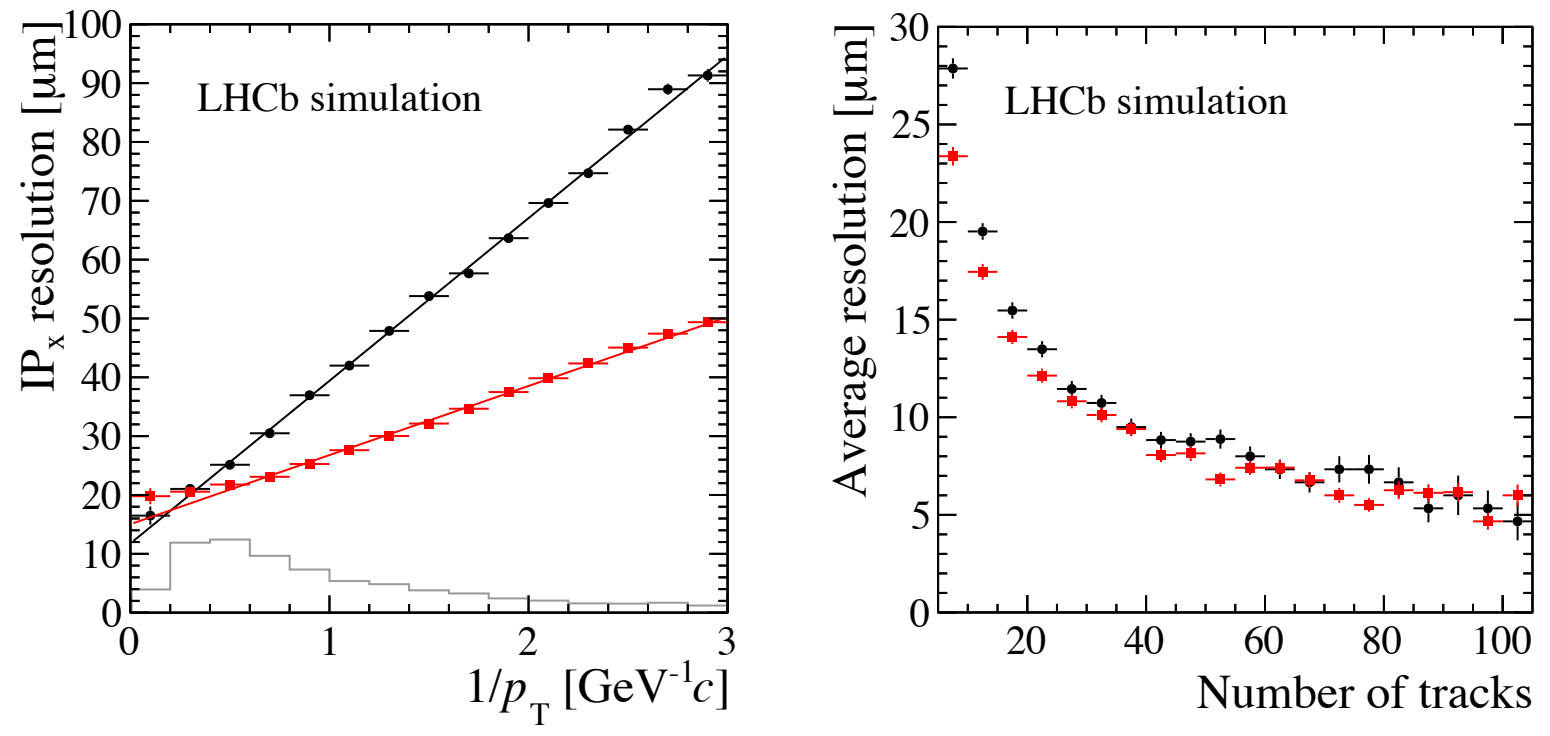

Figure 5.30: **Left:** The impact parameter resolution (in the $x$ direction) as a function of the inverse $p_\mathrm{T}$. **Right:** The resolution on a vertex as a function of number of tracks. The current LHCb VELO performance is shown in black points while the one of the upgraded VELO in red points [369].

LHCb to efficiently reject tracks that belong to a different $b$ or $c$ vertex (see Section 4.2).

In upgrade conditions, the number of tracks and primary vertices will increase by about a factor of five, making it much more difficult to identify displaced vertices close to the beam-line; it is an additional challenge to accomplish this in real time. The VELO was thus completely redesigned [369] to cope with the new expected high-luminosity conditions, maintaining high physics performance and allowing real-time readout for the software trigger. The new VELO has a pixel rather than strip geometry and its distance from the LHC beams is reduced from 8 to 5 mm. This leads to an improvement in the vertex resolution (see Figure 5.30) and reduces the rate of unphysical (ghost) tracks.

The pattern recognition efficiency for track reconstruction is superior to the one of the current VELO when evaluated in high-luminosity conditions. Particularly important for LLP searches is the efficiency of track reconstruction as a function of the displacement from the origin along the beam axis. As shown in Figure 5.31, for the upgraded VELO the efficiency approaches 100% and is uniform in a window of 20 cm around the interaction point, thanks to the new configuration of the modules in the $z$ direction and the shorter distance from the beam. The VELO acceptance degrades quickly after 20 cm in $z$, giving an upper limit for LLP decay lengths with vertices reconstructible in the VELO. A display of the upgraded VELO geometry and its acceptance in both the forward and backward directions is shown in Figure 5.32.

Another important metric of detector performance is impact parameter resolution, especially in LLP searches where it can be exploited to reduce the background due to fake tracks. With the upgrade, the impact parameter resolution significantly improves for low-$p_\mathrm{T}$ tracks. For example, the impact parameter resolution along $x$ for tracks with $p_\mathrm{T}$ of 0.5 GeV is 40 $\mu$m in the upgrade versus

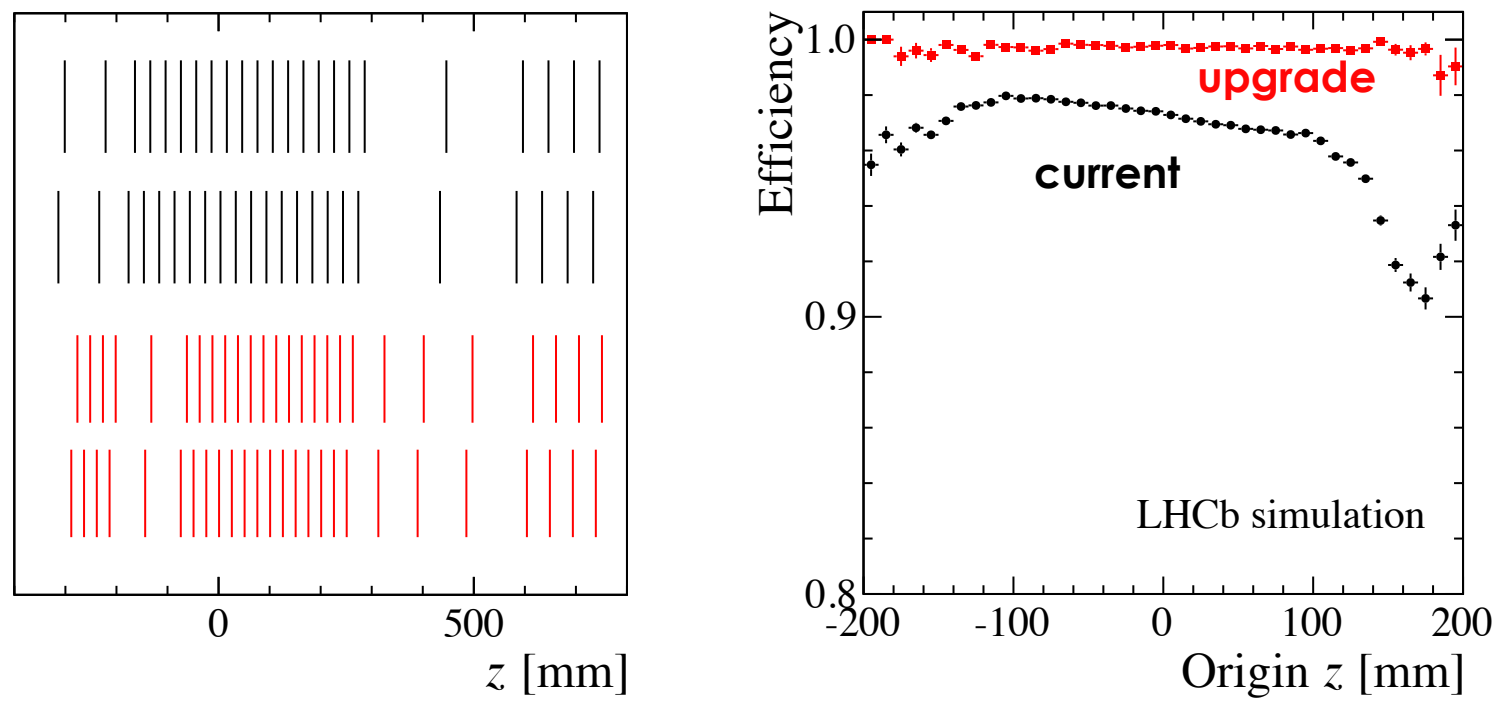

Figure 5.31: **Left:** A comparison of the current and upgrade VELO $z-$layouts for LHCb. The top layout (black) is the current VELO while the bottom layout (red) is the upgrade VELO. **Right:** The track reconstruction efficiency as a function of the origin in $z$ for the current (black) and upgrade (black) VELO in upgrade conditions [369].

70 $\mu$m in the current VELO. The replacement of strips with pixel sensors also makes the pattern recognition faster for the same multiplicity. This can be used in the trigger to find tracks and identify displaced vertices in the trigger, making it possible to soften or remove inefficient $p_T$ requirements.

Real displaced tracks created by interactions in the VELO material can be a significant background to LLP searches (further details are discussed in Chapter 4). The total material budget of the upgraded VELO is similar to that of the current detector (with a radiation length of about 20%) and is dominated by interactions in the Radio-Frequency foil separating the beam vacuum from the vacuum of the sensors. However, the average percentage of radiation length before the first measured point is significantly reduced in the upgrade VELO, passing from $4.6\% X_0$ in the current design to $1.7\% X_0$ in the upgraded VELO (where $X_0$ is the radiation length).

#### 5.2.1.2 *Upgraded Trigger*

The online event selection in the LHCb experiment during the 2010-2018 running period has been performed by a trigger composed of a hardware level (L0), and two software levels: High Level Trigger 1 (HLT1) and High Level Trigger 2 (HLT2). The L0 reduces the rate from 40 MHz (the LHC bunch-crossing rate) to 1 MHz using information from the calorimeter and muon systems. Typical requirements in the L0 are $p_T > 1.4\,\text{GeV}$ for muons and $E > 2.5\,\text{GeV}$ for electrons. The software trigger performs a partial event reconstruction at HLT1, reconstructing tracks and primary vertices for any particle down to $p_T = 500\,\text{MeV}$, followed by a complete event reconstruction at HLT2, reducing further the rate to 12.5 kHz (in LHC Run 2).

In the Phase 1 upgrade, LHCb foresees to run at a luminosity of $2 \times 10^{33}$ cm$^{-2}$s$^{-1}$, about a factor of five larger than that experienced during LHC Run 2. For this reason, a new trigger system, able to fully exploit the LHC potential, has been designed [371, 372]. The upgraded LHCb trigger is based on two paradigms: a triggerless readout and a full software trigger. In addition, as already tested in Run 2, a real-time alignment and calibration will achieve offline-quality reconstruction already in the trigger, allowing a higher signal purity of interesting decay channels. Figure 5.33 shows the current and the upgraded trigger schemes.

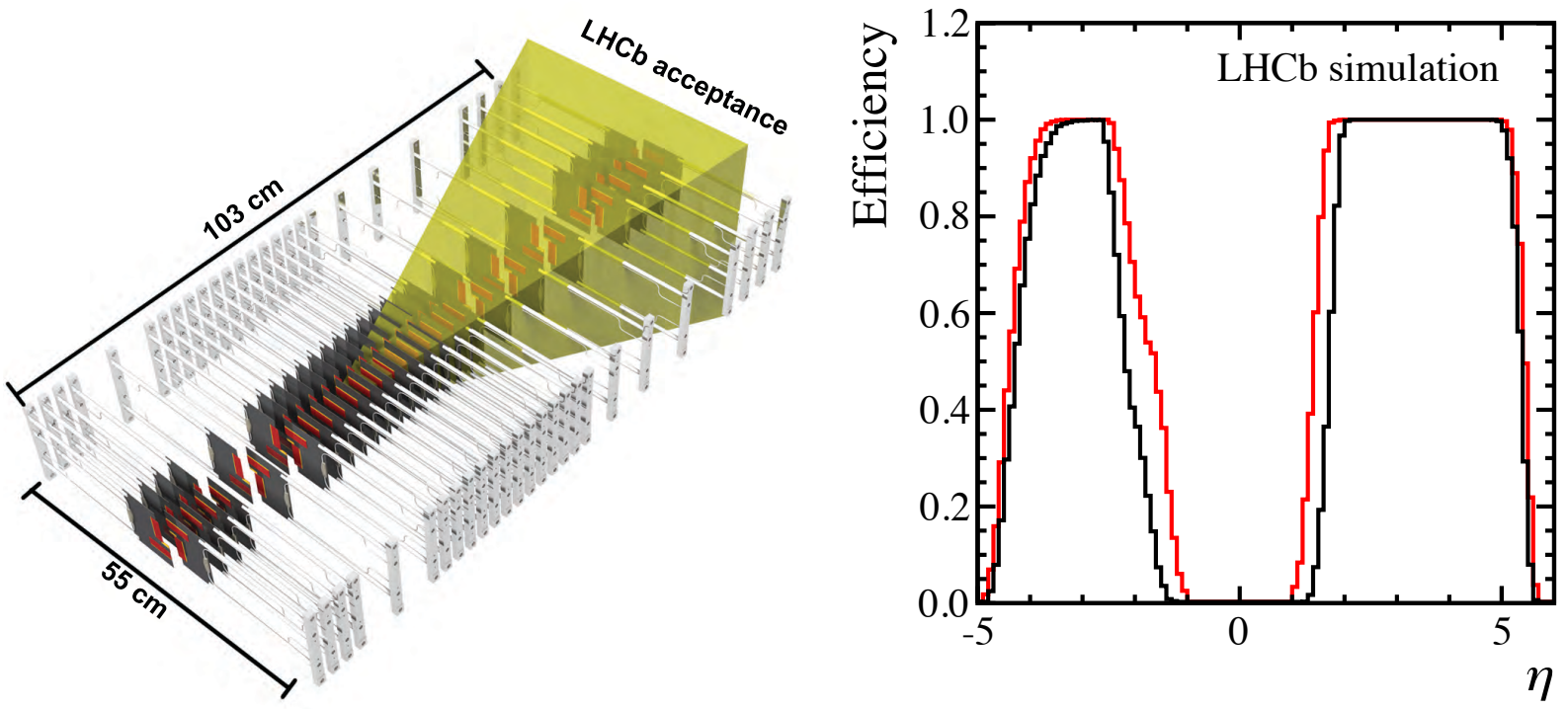

Figure 5.32: **Left:** A display of the upgrade VELO geometry and a comparison with the LHCb spectrometer acceptance which is shaded in yellow. **Right:** The $\eta$ acceptance of the upgrade VELO geometry both for forward and backward tracks. The fraction of tracks crossing three and four modules is given in red and black, respectively [369].

*Triggerless Readout and Full-Software Trigger* With the LHCb trigger upgrade, the 1 MHz readout limitation will be removed, allowing the full event rate to be processed in software. This will increase the efficiency for several channels which otherwise would not benefit from the higher luminosity because they would saturate the L0 trigger rate. Figure 5.34 shows how the rate for non-muonic $B$ decays saturates the L0 trigger with increasing luminosity. Most of these channels saturate the L0 trigger already at the Run 2 luminosity ($4 \times 10^{32}$ cm$^{-2}$s$^{-1}$). Moreover, a purely software trigger will not be subject to the $p_{\mathrm{T}}$ requirements currently applied at the hardware trigger level. Several physics programs involving low-$p_{\mathrm{T}}$ particles, currently prohibited by of the low L0 efficiency, will therefore become possible.

*Turbo Stream* Starting from Run 2, offline-quality alignment and calibration has been applied between the HLT1 and the HLT2 levels. This was made possible by the design of a new dedicated trigger output called Turbo Stream. The event record is written directly from the trigger and processed so that it can be used for physics analysis without the need for offline reconstruction [270]. Several variations of Turbo have been introduced in the last several years. For 2015 data taking, the first version of Turbo allowed only the triggered signal candidate objects to be saved, without keeping the rest of the event and thereby discarding all sub-detector information. While the event size was an order of magnitude smaller than for full stream data, any analysis relying on additional information from the surrounding event could not use Turbo Stream data.

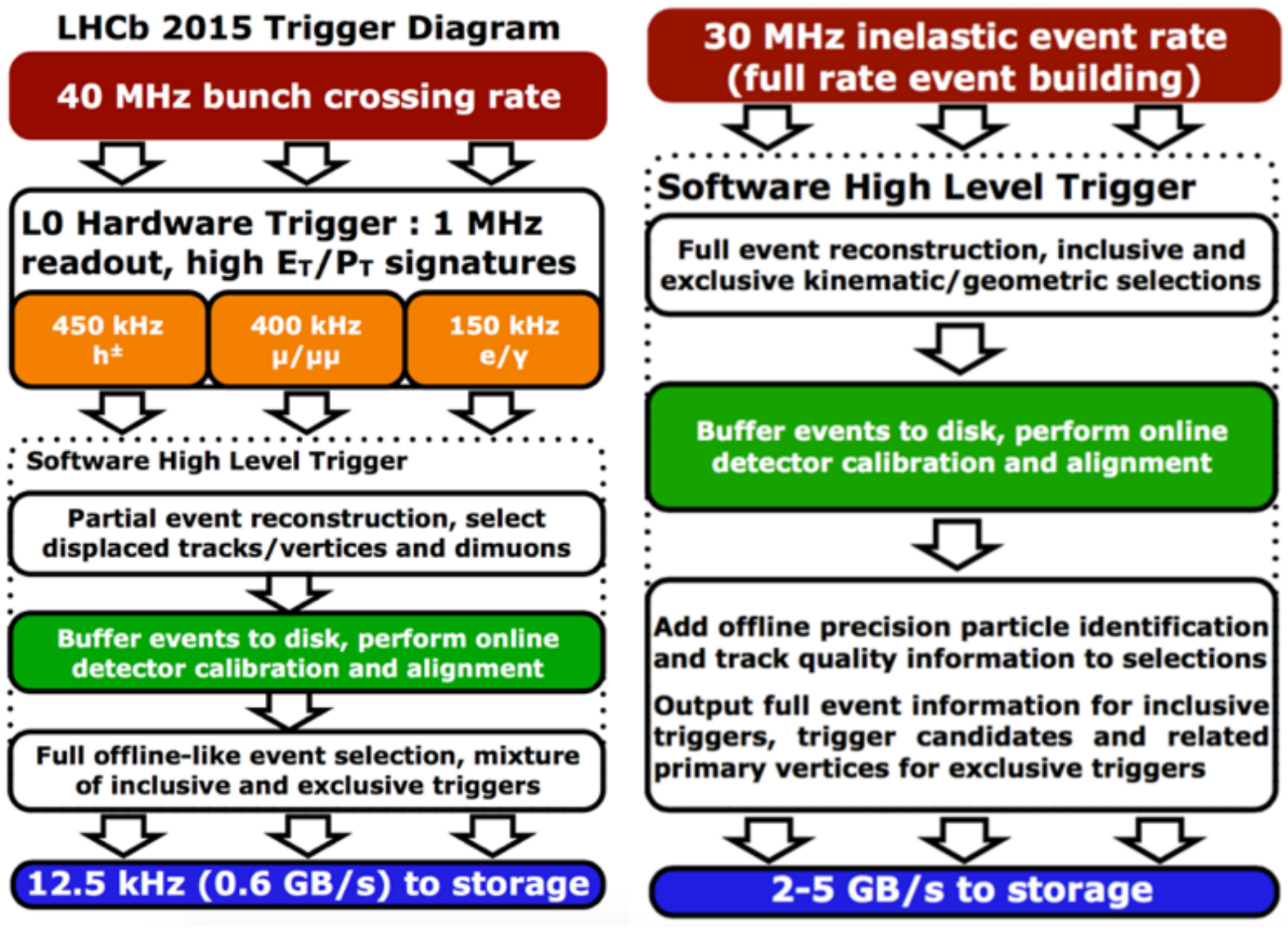

Figure 5.33: Scheme of the current LHCb trigger (left), and of the upgraded trigger (right) [373].

For this reason, full event reconstruction was implemented as Turbo++ in 2016. Finally, a new intermediate solution between Turbo and Turbo++ called Turbo SP (Selective Persistence) was implemented in 2017. With Turbo SP, both the trigger candidate objects and a subset of the other objects in the event are saved. This flexible solution allows the analyzer to choose which objects to save, minimizing the size of the stored event. Once the trigger is upgraded to run purely at the software level, the large majority of LHCb analyses will be moved to Turbo. The reduced event size allows the storage of particle candidates at a high rate, fully exploiting the reduction in $p_T$ thresholds due to the removal of the L0 hardware trigger.

*Triggers on Downstream Tracks* Most of the LLP searches at the LHCb experiment use so-called "long tracks" to reconstruct the candidates, which are tracks where inputs from both the VELO and the tracking stations are considered. These tracks have an excellent spatial and momentum resolution, and result from an LLP decaying within the VELO region. This is the reason why most of the LLP searches done by LHCb correspond to long-lived candidates with displacements typically up to $\mathcal{O}(10)$ cm, with the VELO tracking algorithm optimized for displacements of up to 20 cm. However, for long-lived candidates with displacements larger than 2 cm, a different type of track which considers information only from the tracking stations, "downstream tracks", has to be used instead of the "long track" type. Unfortunately, these tracks have worse vertex and momentum resolution, limiting the capabilities of LHCb for this displacement range. In order to improve the detector sensitivity in the displacement ranges between 20 and 200 cm, a proposal to develop new trigger lines to select "downstream tracks" is being studied [372]. An interesting idea which can help to achieve this task and to significantly reduce the CPU computing time is the implementation of a system of specialized processors used to rapidly find the downstream tracks through look-up tables, and present these tracks to the software trigger in parallel with all the raw detector information in the event. This system, named "retina", is being studied and has its own R&D programme [375].

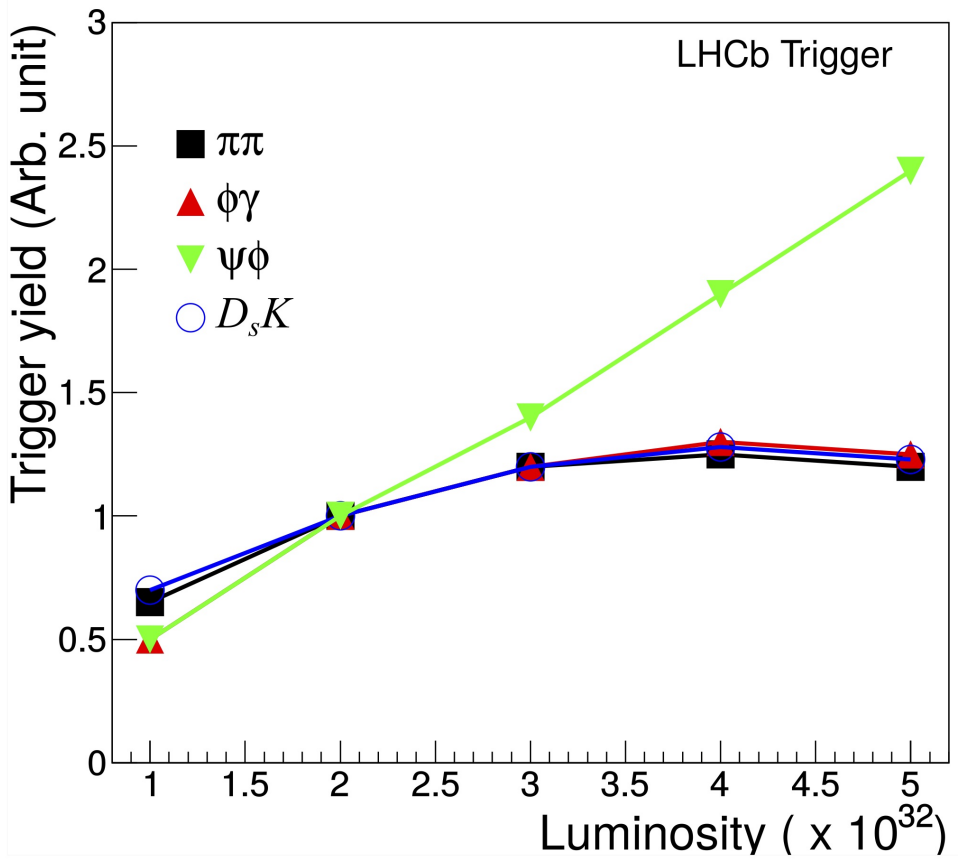

Figure 5.34: Trigger yields for $B \to \pi^+\pi^-$, $B_s \to \phi(K^+K^-)\gamma$, $B_s \to J/\psi(\mu^+\mu^-)\phi(K^+K^-)$ and $B_s \to D_s^-(K^+K^-\pi^-)K^+$ as a function of the luminosity with the current LHCb trigger scheme. For non-muonic channels, the saturation effect due to the L0 trigger can be observed [374].

The much higher luminosity and improved capabilities of the upgraded LHCb detector are expected to significantly improve the capabilities for LLP searches in LHC Run 3. In the following sections, the projected sensitivities to several benchmark LLP signatures are shown to illustrate the potential of the upgraded detector. However, the potential of the upgraded triggerless readout has not been completely explored yet and its great flexibility could be exploited in several ways beyond those shown here.

### 5.2.2 *LHCb Upgrade Phase 1a Projections for LLP Signatures*

#### 5.2.2.1 *Displaced Di-Leptons*

The upgraded LHCb experiment is expected to have exceptional sensitivity to low-mass, displaced dilepton signatures thanks to the mass resolution, excellent vertexing, and the online selection allowed by the triggerless readout.

The upgrade LHCb sensitivity to dileptons has been explored in the literature in the context of dark photon searches. Two com-plementary signatures have been considered: an inclusive search for BSM gauge bosons (dark photons) in $A' \rightarrow \mu^+\mu^-$, and a search for $A'$ using radiative charm decays $D^{*0} \rightarrow D^0 A'$, $A' \rightarrow e^+e^-$. The inclusive search [335] scans a large region from the dimuon threshold $2m_\mu$ all the way up to the $Z$ pole. The second proposed signature [376] exploits a tag of the radiative decay of the $D^{*0}$ using its reconstructed invariant mass and a di-electron dark photon final state to probe a much lower mass range allowed by the decay kinematics, $[2m_e$,142 MeV$]$. For both signatures, searches for both a prompt and a displaced dark photon vertex are carried out. The prompt search is expected to probe mixing parameters $\epsilon^2$ below $10^{-7}$ despite the large irreducible background from Drell-Yan and QCD [2]. Since the dark photons with masses above the $\eta$ mass decay promptly for couplings that are accessible within LHCb, no attempt is made to probe displaced dark photons in that region.

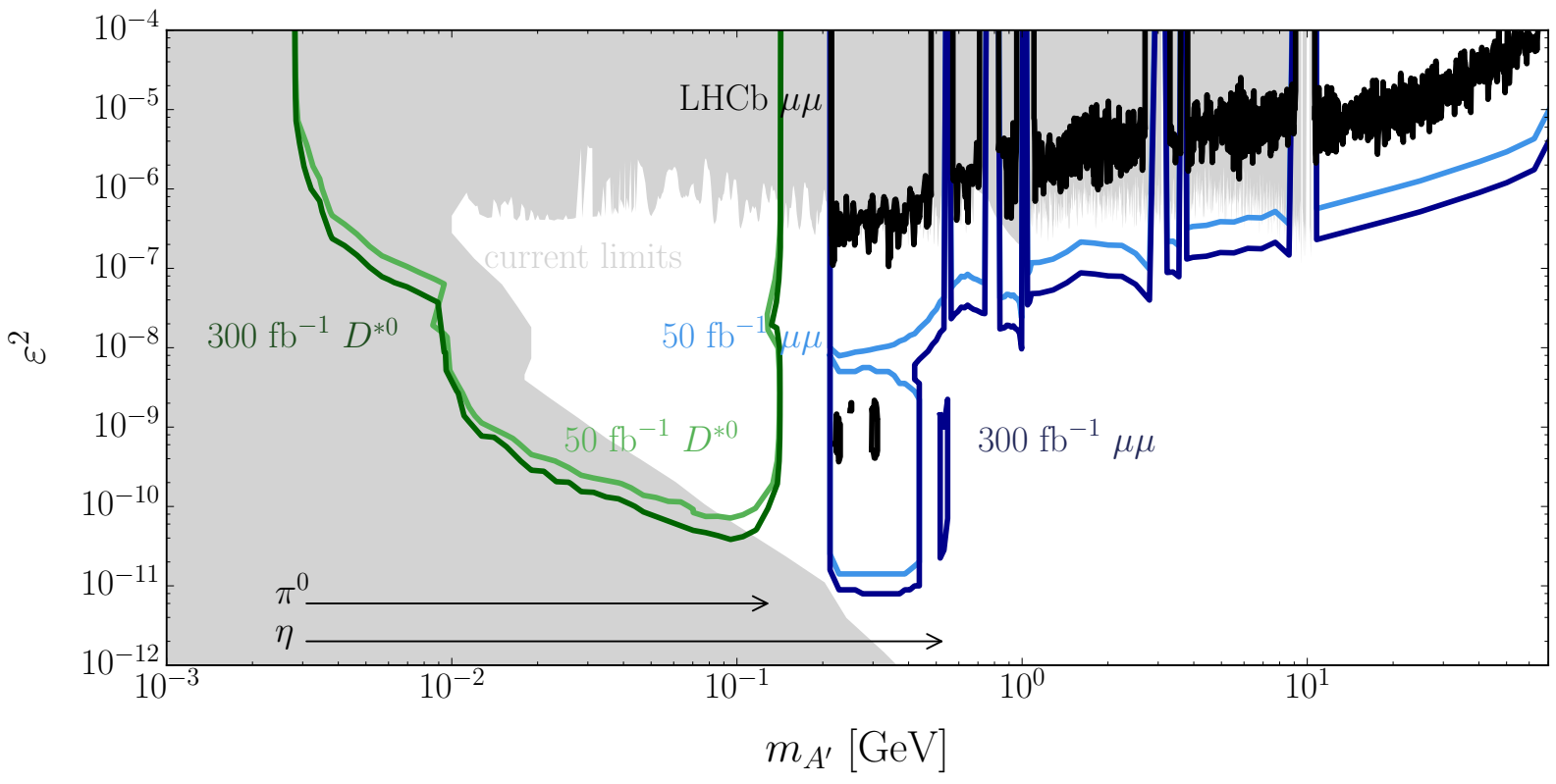

Figure 5.35: Current and expected limits in the dark photon parameter space mixing $\epsilon^2$ versus $A'$ mass [368]. The black line represents current LHCb limits from Ref. [262], while grey-shaded regions are existing limits from other experiments. Expected limits from the proposed inclusive search with 50 and 300 fb$^{-1}$ are shown in shades of blue. The expected sensitivity of $D^{*0} \rightarrow D^0 A'(ee)$ at low mass is shown in shades of green. The arrows indicate the available mass range from light meson decays into $e^+e^-\gamma$.

[2] In the limit $m_{A'} \ll m_Z$, the coupling of the dark photon to SM particles with charge $Q$ is approximately $Qe\epsilon$.

An inclusive search for dark photons has already been performed with LHCb data collected in 2016 [262]. This was possible due to the high reconstruction and identification efficiency of soft di-muons at LHCb. These results demonstrated the unique sensitivity that can be reached at LHCb. The planned increase in luminosity and removal of the hardware-trigger stage in Run 3 should increase the number of expected $A' \rightarrow \mu\mu$ decays in the low-mass region by a factor of $\mathcal{O}(100-1000)$ compared to the 2016 data sample. The limits placed by the current data and the sensitivity expected with future LHC runs is shown in Figure 5.35.

The exclusive search for $D^{*0} \rightarrow D^0 A'$, $A' \rightarrow e^+e^-$ is much more challenging and not feasible prior to the upgrade, since the hardware trigger and the higher material budget degrade the sensitivity.

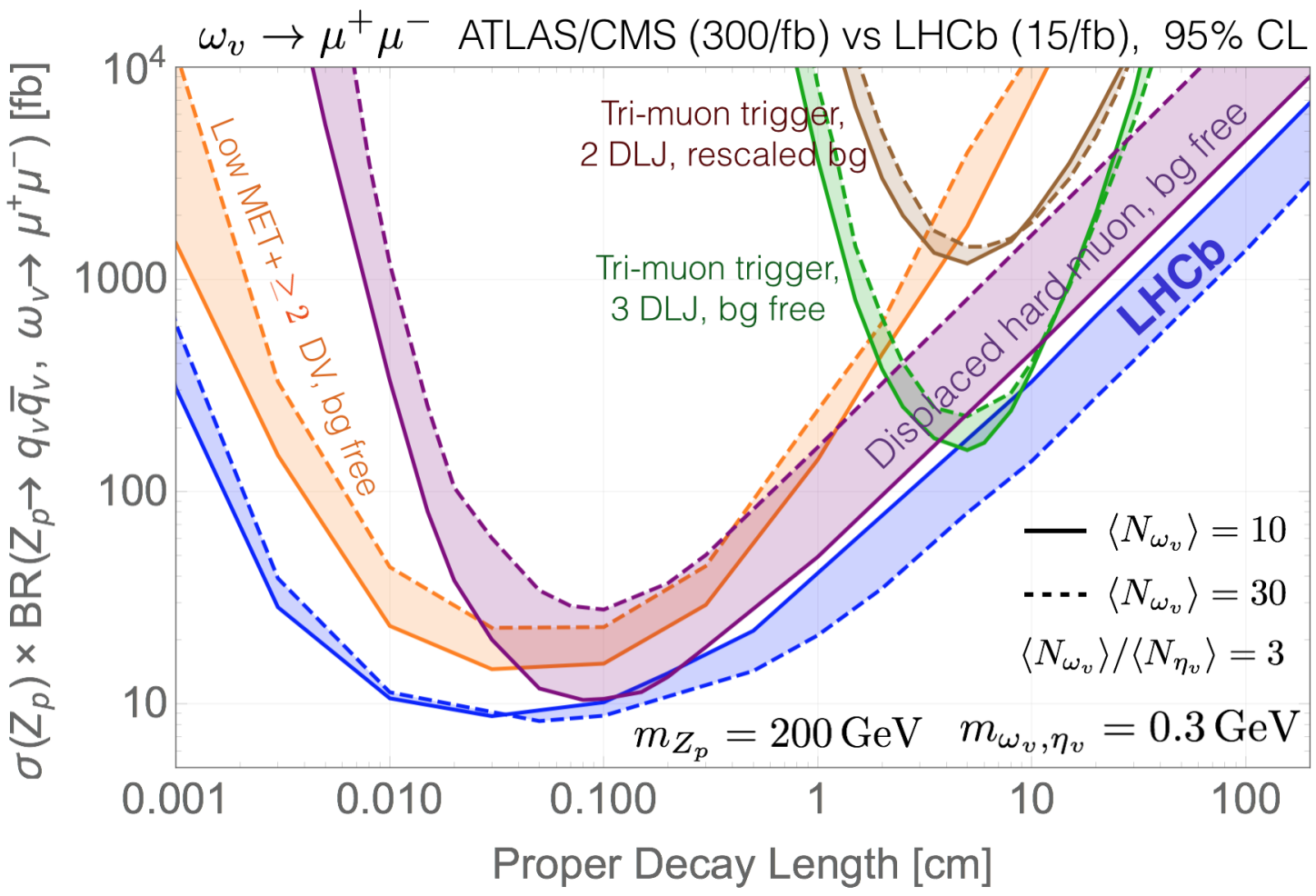

Figure 5.36: Projected bounds from various ATLAS/CMS searches and the LHCb search for Hidden Valley-like models with a $Z_p$ decaying to a $q_{HV}q_{HV}$ pair which then undergo hadronization in the dark sector, eventually leading to dark meson decays to di-muon final states [377].

This search highly relies on the online identification of $e^+e^-$ pairs, since over 5 trillion of these $D^{*0}$ decays are expected in Run 3. The expected sensitivity probes unexplored regions of phase space at low $A'$ mass and mixing $\epsilon^2$ that is usually in the realm of beam-dump experiments.

A displaced di-muon signature also appears in some Hidden Valley (HV) scenarios [59], in which a hidden sector with strong dynamics showers and hadronizes into dark mesons that can have an appreciable decay rate to leptons. (For more information on dark showers, see Chapter 7.) The upgraded LHCb prospects for this type of signature have been explored in Ref. [377] and are very promising. In the scenario explored therein, dark mesons are produced with large multiplicities of between 10 and 30. Selection criteria inspired by the proposed dark photon search [335] in the region after the first VELO module are applied. The expected reach for the proposed searches using Run 3 data from LHCb (15 $\mathrm{fb}^{-1}$) and from ATLAS/CMS (300 $\mathrm{fb}^{-1}$) are shown in Figure 5.36. The model studied involves a 200 GeV $U(1)'$ gauge boson $Z_p$ decaying to a HV quark pair; showering and hadronization in the dark sector leads to a large multiplicity of hidden hadrons $\omega_V$ (with $m_{\omega_V} = 0.3$ GeV/$c^2$) that can decay to di-muons. In this context, the upgraded LHCb detector could have better sensitivity than other proposed searches at ATLAS and CMS.

#### 5.2.2.2 *Displaced Jets*

Signatures with displaced jets are common in the context of LLP searches. As summarised in Section 3.1.3, LHCb has started to explore its sensitivity to displaced jets by using the data collected during Run 1. In LHCb Upgrade Ia, the background rejection for displaced jet searches is expected to improve thanks to the improved VELO resolution. The selection efficiency should also be significantly higher due to the prospects for online displaced vertex identification. The main focus of LHCb will be to probe the region at small lifetime where it has already proven to be the most competitive (see discussion in Section 3). The background from QCD and material interactions is the main limiting factor, and the improved vertex resolution of the upgraded VELO together with the lower material budget and the use of a detailed material map are expected to bring large improvements in the upgraded detector.

In some dark sector models, the number of hadronic displaced vertices can be large, even in the limited LHCb acceptance (see, for example, Ref. [325] and the discussion in Chapter 7). An inclusive displaced vertex search would likely include a requirement on the isolation of the displaced vertex from other tracks in the event (such as the one used for the dark photon analysis [262]). This requirement could be very inefficient in the context of a dark shower and so a dedicated search strategy would be needed. Furthermore, a dedicated software trigger looking for a large number of displaced vertices in the VELO and soft $p_{\mathrm{T}}$ requirements could in principle improve the sensitivity to this kind of models, but studies are needed to fully understand its potential and compare LHCb to other experiments.

#### 5.2.2.3 *Displaced Mesons*

Quark-antiquark pairs from the decay of a low mass particle in the SM often hadronize into SM mesons that subsequently decay with known branching ratios. For example, a dark sector meson with a displaced decay to $c\bar{c}$ often produces two $D$ mesons which in turn have non-negligible lifetimes. In this scenario, the authors of Ref. [377] have investigated the prospects of using more or less inclusive reconstruction of the two $D$ meson decays at the upgraded LHCb. A similar approach could be used to target decays to $b\bar{b}$ that hadronize to $B$ mesons since the latter is likely to produce $D$ mesons in its decay. Since LHCb is designed to reconstruct heavy flavor decays, it can be competitive in this kind of search [377] (see Figure 5.37). Furthermore, searches for displaced mesons will greatly profit from the software trigger, which is mainly designed to improve the efficiency of similar-looking hadronic decays of heavy flavor mesons.

### 5.2.3 *After Phase 1a Upgrade: Phase 1b and Phase 2*

After a consolidation phase (Phase 1b) during Long Shutdown (LS) 3 aiming to run in the same conditions as in Phase 1a, the LHCb detector plans a Phase 2 upgrade during LS 4 (by 2030). The Phase 2 upgrade is needed to face even more challenging conditions than during previous runs. For example, particle multiplicity, pile-up, and radiation damage are expected to be ten times higher than those experienced in Phase 1. The LHCb experiment expects to collect at least 300 fb$^{-1}$ by the end of Upgrade 2 [368].

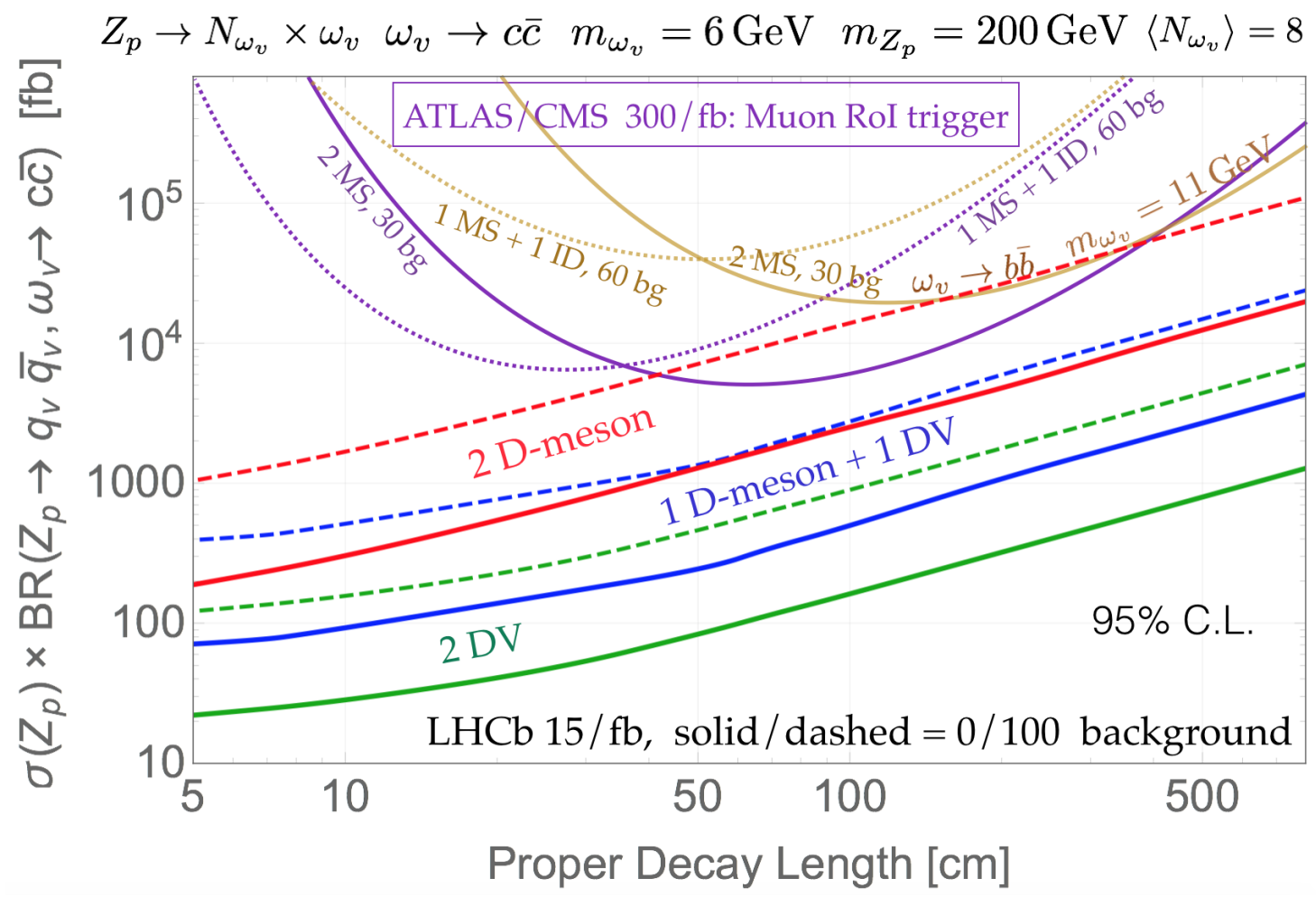

Figure 5.37: Projected bounds from various proposed searches for confining hidden valley models exploiting $c\bar{c}$ decays of hidden valley mesons $\omega_v$ (taken from Ref. [377]). The sensitivities with various signatures at LHCb are shown: two displaced vertices (green), one reconstructed $D$ meson and one DV (blue) and two reconstructed $D$ mesons (red). Searches for this same particular model at ATLAS/CMS are shown not to be competitive (more details can be found in Ref. [377]).

Major improvements to the LHCb detector during Phase 1b and 2 that are relevant for LLP searches are described in the following, as well as some naïve projections of Run 1 results to an an integrated luminosity of 300 fb$^{-1}$.

*Magnet Stations* Along with the “long” and “downstream” track types, “upstream” tracks are also considered useful for LLP searches. These tracks correspond to soft charged particles bending out of the detector acceptance, produced from LLP candidates which decay within the VELO region. Aside from the installation of a new tracker during LS2, the Upstream Tracker, a proposal to add magnet stations inside the LHCb magnet to improve low momentum resolution is considered [378]. These magnet stations have been proposed to be installed for Phase 1b, and are foreseen to highly improve the tracking of low momentum particles produced from certain kind of LLP candidates, such as for example soft pions from “disappearing” chargino tracks.

*Material Interactions* The presence of a VELO envelope at approximately 5 mm from the beam line in order to separate the VELO and the LHC vacuums, named “RF-foil”, strongly affects the background composition of LLP searches in the LHCb experiment. Namely, for LLP candidates decaying below 5 mm from the beam line, the main source of background is due to heavy flavour decays, while material interactions with the RF-foil compose the main background contribution for LLPs decaying above 5 mm from the beam line. While the former is purely due to QCD processes and hence not reducible, the latter is kept under control by the use of a detailed veto map (see Ref. [262]). However, the ideal case would be to completely remove the RF-foil during the Phase 2 upgrade [378], which would result in a large reduction of the background component due to material interactions. The improvements foreseen to the Phase 2 VELO (probably based on an updated version of the Phase 1 VELO), are expected to increase the sensitivity to shorter lifetimes and better primary vertex and impact parameter resolution (see Figure 5.38). Unfortunately, the removal of the RF-foil requires the development of new techniques to isolate the sensors from the beam radio frequency and is not necessarily seen as a viable option. Therefore, improving the material veto maps as much as possible by accurately modeling the material interactions would be desirable as a more realistic option for the HL-LHC era.

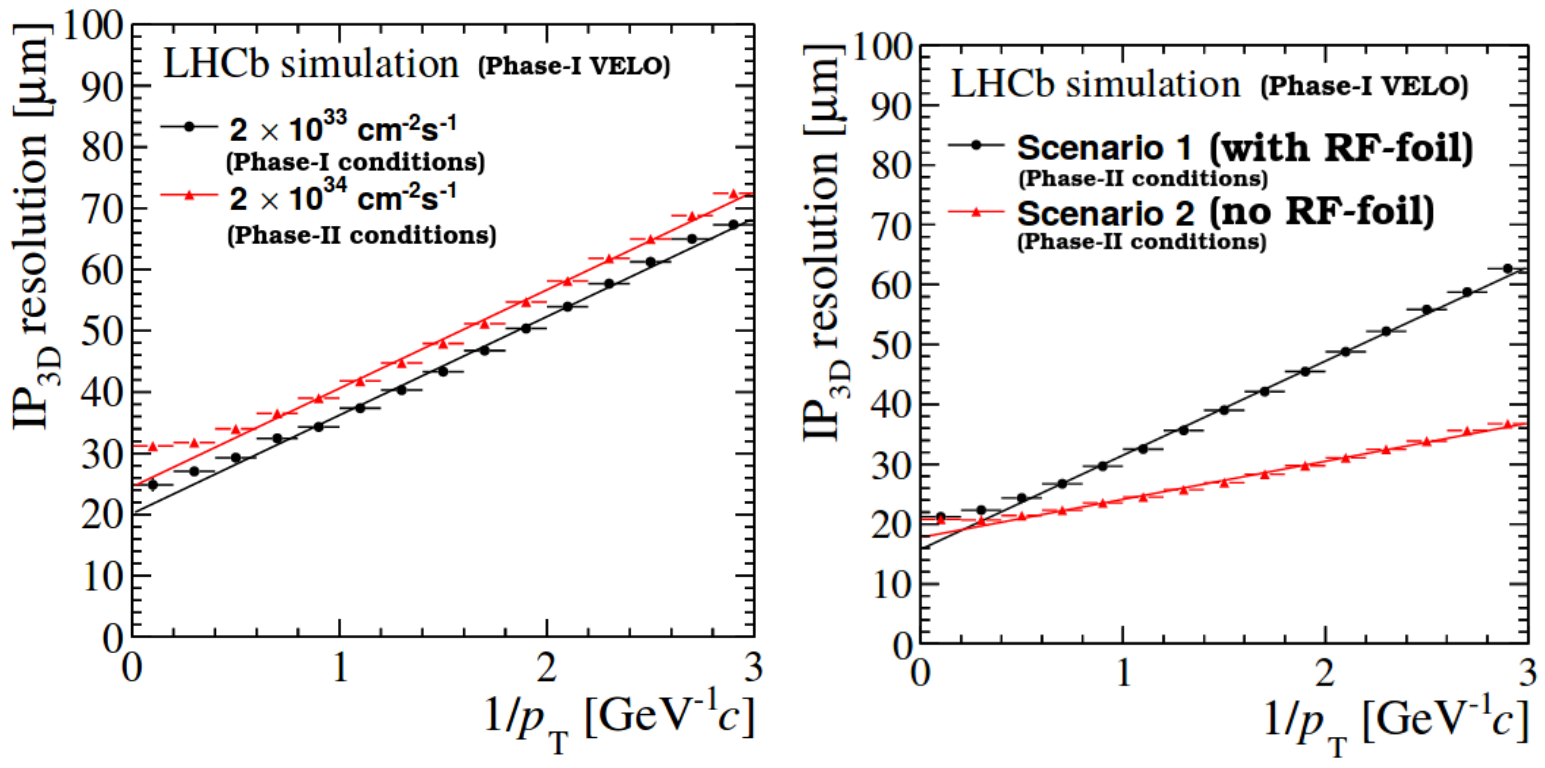

Figure 5.38: Impact parameter resolution estimations from simulation using the Phase 1 VELO model for the LHCb experiment: (left) a comparison if considering Phase 1 or Phase 2 conditions; (right) the effect of the RF-foil removal under Phase 2 conditions [378].

*Naïve Projections of Run 1 Results to the HL-LHC Era* By taking the published Run 1 results from the single displaced dijet search [246] at LHCb, a naïve extrapolation to the integrated luminosity foreseen to be recorded by LHCb during Phase 2, 300 fb$^{-1}$, is presented in Figure 5.39. These numbers have been obtained by simply scaling signal and background with the expected increase in cross section and luminosity, neglecting pile-up effects and expected detector improvements. The removal of neutral objects from jet reconstruction, and the use of machine-learning techniques are expected to assist in the required suppression of pile-up. Furthermore, jet substructure techniques are foreseen to improve the quality of di-jet object with lower mass.

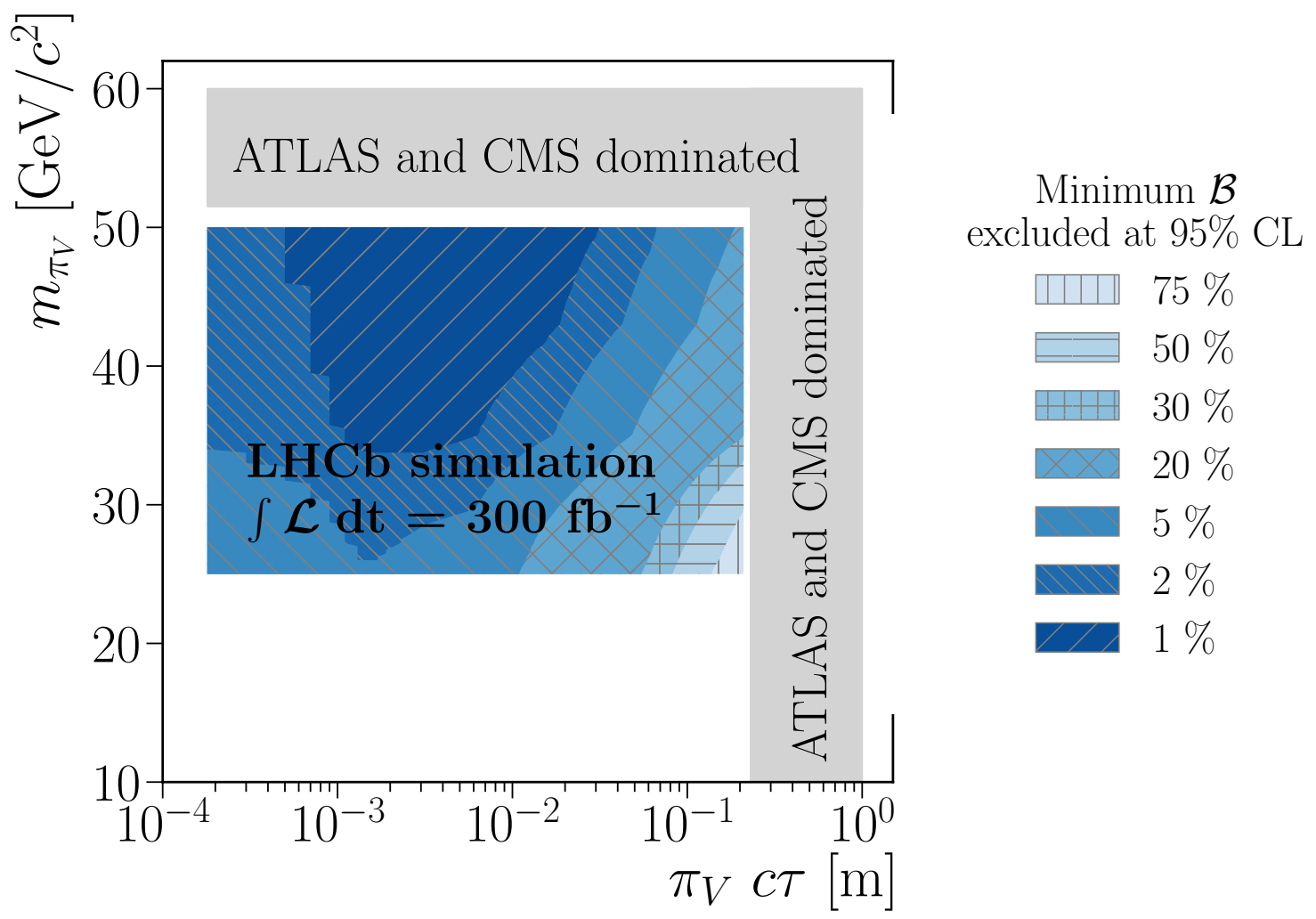

Figure 5.39: Naive sensitivity projections for searches of displaced dijet vertices at LHCb for the expected luminosity to be collected with the Phase 2 upgraded detector (300 fb$^{-1}$) [379]. The projected limit is on Higgs boson decays to dark pions $\pi_V$ with branching fraction $\mathcal{B}$, and the dark pions each in turn decay to a pair of jets.

## 5.3 *Dedicated Detectors for LLPs*

### 5.3.1 *Introduction*

Despite a wide and seemingly comprehensive research program — both existing and, in this document, proposed to be expanded — for LLPs at the LHC, in cases of ultra-low-mass particles, ultra-long lifetimes, or unusual LLP charges, it is hard or impossible to trigger on and/or reconstruct such events in the main ATLAS, CMS, and LHCb detectors. This has led to new proposals for dedicated experiments to look for LLPs in new regimes that are otherwise inaccessible at the LHC. These experiments provide the best sensitivity to new millicharged LLPs, magnetic monopoles, and other LLPs arising from models such as those containing Higgs-portal hidden sectors, dark photons, and Majorana neutrinos.

As discussed in Section 2.2, LLPs beyond the Standard Model are theoretically well motivated and come in wide range of masses and lifetimes. ATLAS and CMS have excellent sensitivity for fairly high mass LLPs, regardless of their lifetime (see, e.g., Refs. [4, 239, 380, 381]). Low mass and/or softer final states are more challenging due to both background and triggering limitations. In the short-lifetime regime, for $c\tau$ of the scale of the VELO, LHCb has sensitivity to somewhat lower masses and can trigger on softer muonic final states, generating complementary reach provided the LLP has a significant branching ratio to muons [246, 247, 272, 302, 382]. Finally, the low-mass/soft final states with rather long lifetimes are challenging for all three experiments. These signatures can be covered partially by NA62 [383] operating in beam dump mode, or by SHiP [384], or by dedicated LHC experiments like CODEX-b [97] (see Section 5.3.5), FASER [98] (see Section 5.3.6), or MATHUSLA [96] (see Section 5.3.4). Each of these dedicated experiments is sensitive to different LLP lifetimes, masses, and production modes based on their position and orientation.

### 5.3.2 *MoEDAL Experiment and Future Developments*

MoEDAL (Monopole and Exotics Detector At the LHC) [94][3] is designed to search for manifestations of new physics through highly-ionising (HI) particles in a manner complementary to ATLAS and CMS [303]. The main motivation for the MoEDAL experiment is to pursue the quest for magnetic monopoles at LHC energies. Nonetheless the detector is also designed to search for any massive, long-lived, slow-moving particle [385, 386] with single or multiple electric charges arising in many scenarios of physics beyond the Standard Model [387].

[3] For general information on the MoEDAL experiment, see: http://moedal.web.cern.ch/.

The MoEDAL detector [304] is deployed around the intersection region at LHC Point 8 (IP8) in the LHCb Vertex Locator (VELO) cavern. A schematic view of the MoEDAL experiment is shown in Figure 5.40. It is a unique and largely passive detector comprising different detector technologies.

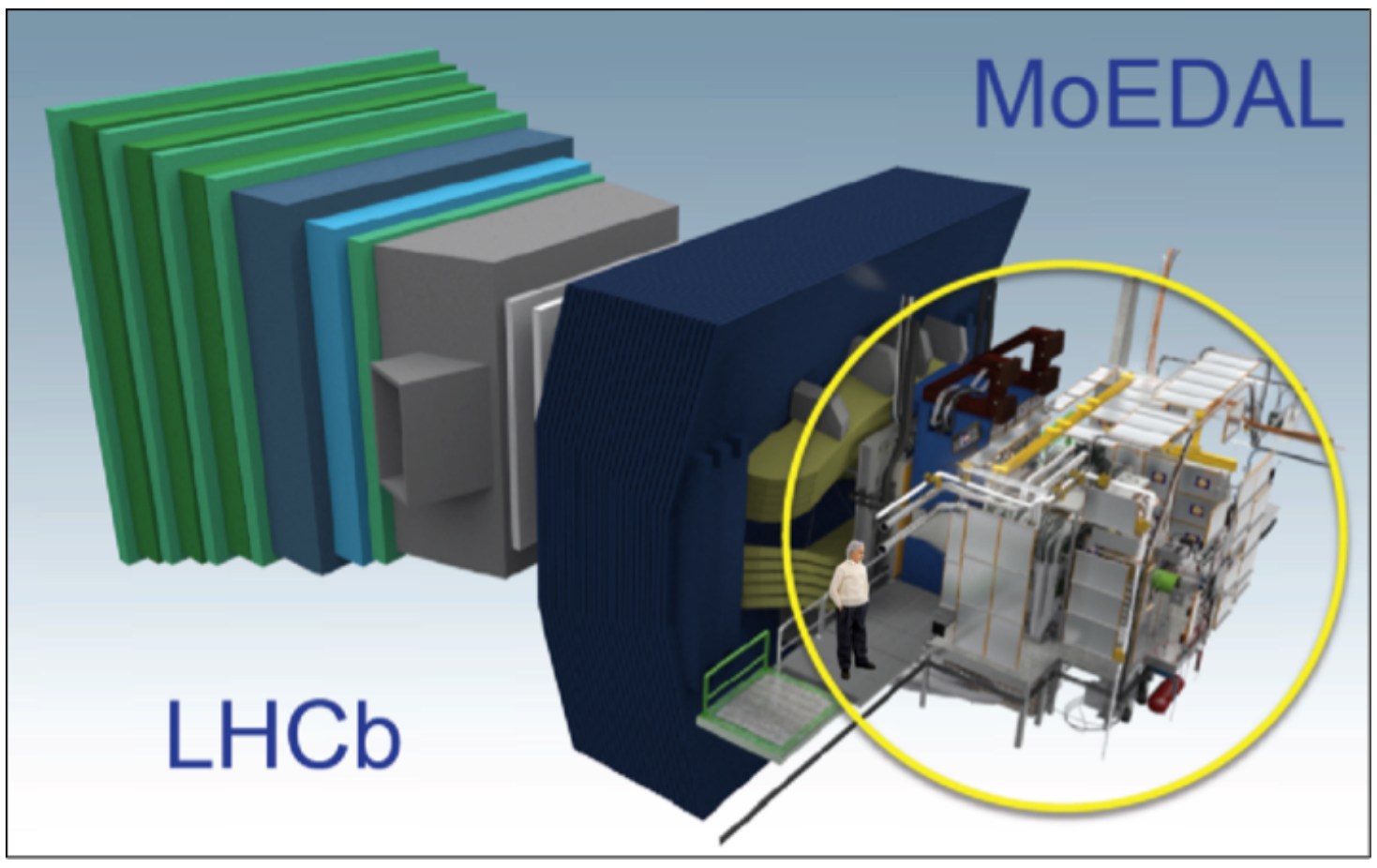

Figure 5.40: A three-dimensional schematic view of the MoEDAL detector (in the yellow circle) around the LHCb VELO region at Point 8 of the LHC.

The main sub-detector system is made of a large array of CR-39, Makrofol® and Lexan™ nuclear track detector (NTD) stacks surrounding the intersection area. The passage of a HI particle through the plastic detector is marked by an invisible damage zone along the trajectory. The damage zone is revealed as a cone-shaped etch-pit when the plastic detector is chemically etched. Then the sheets of plastics are scanned looking for aligned etch pits in multiple sheets. The MoEDAL NTDs have a threshold of $z/\beta \sim 5$, where $z$ is the charge and $\beta = v/c$ the velocity of the incident particle.

Another type of NTD installed is the Very High Charge Catcher

($z/\beta \sim 50$). It consists of two flexible low-mass stacks of Makrofol, deployed in the LHCb acceptance between RICH1 and the Trigger Tracker. It is the only NTD (partly) covering the forward region, adding only $\sim 0.5\%$ to the LHCb material budget while enhancing considerably the overall geometrical coverage of MoEDAL.

A unique feature of the MoEDAL detector is the use of paramagnetic magnetic-monopole trappers (MMTs) to capture magnetically-charged HI particles. The high magnetic charge of a monopole — being at least one Dirac charge $g_{\mathrm{D}} = 68.5e$ — implies a strong magnetic dipole moment, which may result in strong binding of the monopole with the nuclei of the aluminium MMTs. In such a case, the presence of a trapped monopole would de detected through the induction technique by measuring the *persistent current*, defined as the difference between the superconducting magnetometer currents before and after the passage of the MMT bar through the sensing coil [388, 389].

The only non-passive MoEDAL sub-detector is an array of TimePix pixel devices distributed throughout the MoEDAL cavern, forming a real-time radiation monitoring system of HI beam-related backgrounds. The operation in time-over-threshold mode allows a 3D mapping of the charge spreading in the volume of the silicon sensor, thus differentiating between various particles species from mixed radiation fields and measuring their energy deposition.

The MoEDAL detector is designed to fully exploit the energy-loss mechanisms of magnetically charged particles [390–393] in order to optimise its potential to discover these messengers of new physics. Mulitple theoretical scenarios [394] have been proposed over the years in which magnetic charge would be produced at the LHC [387], resulting in such possible new particles as light 't Hooft-Polyakov monopoles [392, 393, 395], electroweak monopoles [396–400], global monopoles [401–406], and monopolium [391, 407–409]. Magnetic monopoles that carry a non-zero magnetic charge and dyons possessing both magnetic and electric charge are predicted by many theories including grand-unified and superstring theories [410–412].

A possible explanation for the non-observation of monopoles so far is Dirac's proposal [390, 391, 407] that monopoles are not seen freely because they form a bound state called *monopolium* [408, 409, 413, 414] being confined by strong magnetic forces. Monopolium is a neutral state, difficult to detect directly at a collider detector, although its decay into two photons would give a rather clear signal for ATLAS and CMS [415]. Nevertheless the LHC radiation detector systems can be used to detect final-state protons $pp \to pXp$ exiting the LHC beam vacuum chamber at locations determined by their fractional momentum losses [416]. Such a technique would be appealing for detecting monopolia.

The MoEDAL detector is also designed to search for any massive, long-lived, slow-moving particles [385, 386] with single or multiple electric charges arising in many scenarios of physics beyond the Standard Model. Supersymmetric long-lived particles [417], quirks, strangelets, Q-balls, and many others fall into this category [387]. A generic search for high-electric-charge objects is currently underway [418].

For the 2016 run at 13 TeV, the MMT included 672 aluminium rods (for a total mass of 222 kg) that were placed 1.62 m from the IP8 LHC interaction point under the beam pipe on the side opposite to the LHCb detector. The MMT bars were analysed and no magnetic charge $> 0.5 g_{\rm D}$ was detected in any of the exposed samples when passed through the ETH Zurich SQUID, which is a DC SQUID long-core magnetometer [419]. Hence cross section limits are obtained for Drell-Yan pair production of spin-1, spin-1/2 and spin-0 monopoles for $1 g_{\rm D} \leq |g| \leq 5 g_{\rm D}$ at 13 TeV [419] improving previous bounds set by MoEDAL at 8 TeV [304] and 13 TeV [420]. Monopole production via photon fusion is also now considered in MoEDAL monopole search analyses [421] following recent studies [422]. However, the large monopole-photon coupling invalidates any perturbative treatment of the cross section calculation and hence any result based on the latter is only indicative. This situation may be resolved if thermal production in heavy-ion collisions — that does not rely on perturbation theory — is considered [423], or by including a magnetic-moment term in monopoles with spin [422].

To recapitulate, under the assumption of Drell-Yan cross sections, MoEDAL has derived mass limits for $1 g_{\rm D} \leq |g| \leq 5 g_{\rm D}$, complementing ATLAS results [208, 424], which placed limits for monopoles with magnetic charge $|g| \leq 1.5 g_{\rm D}$, as shown in Figure 5.41. The ATLAS bounds are better that the MoEDAL ones for $|g| = 1 g_{\rm D}$ due to the higher luminosity delivered in ATLAS and the loss of acceptance in MoEDAL for small magnetic charges. On the other hand, higher charges are difficult to be probed in ATLAS due to the limitations of the level-1 trigger deployed for such searches. Limits on production cross sections of singly-charged magnetic monopoles set by various colliders are presented in Refs. [410, 411], while general limits including searches in cosmic radiation are reviewed in Refs. [394, 425].

MoEDAL is proposing to deploy the MAPP (MoEDAL Apparatus for detecting Penetrating Particles) in a tunnel shielded by some 30 m to 50 m of rock and concrete from the IP8 [427, 428]. A prototype of the MAPP detector was installed in 2017. It is envisaged that the full detector will be installed in LS2 to take data in LHC Run 3. The purpose of the detector is to search for particles with fractional charge as small as one-thousandth the charge of an electron. This detector would also be sensitive to neutral particles from new physics scenarios via their interaction or decay in flight in the $\sim 10$ m decay zone in front of the detector or in the detector itself. The isolation of the detector means that the huge background from SM processes in the main detectors is largely absent. Also, the detector can be placed at various angles to the beam axis (from $5^\circ$ to 25°). The ability to vary depth and angle enhances MoEDAL to be able to distinguish between theoretical scenarios in the event a signal is observed.

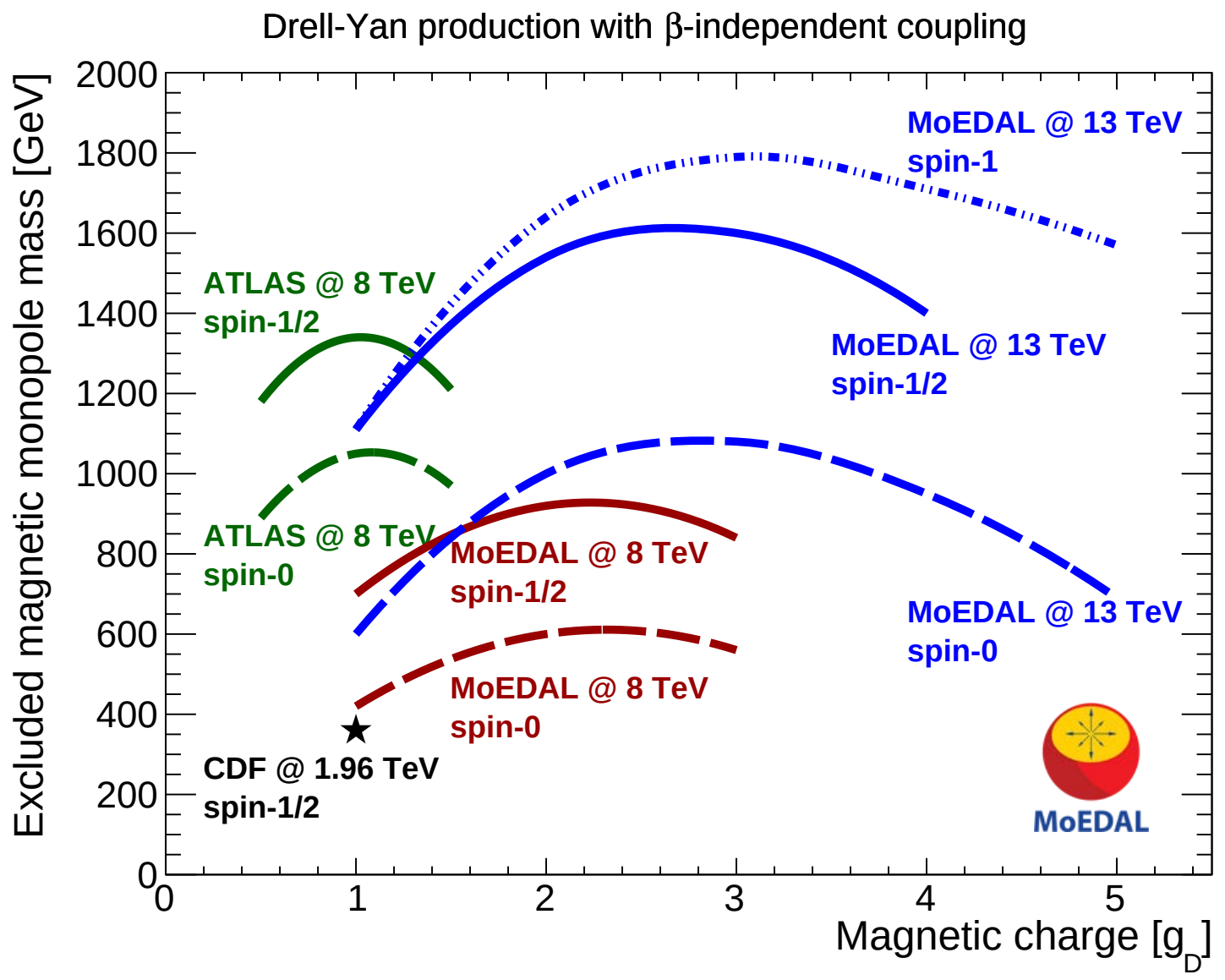

Figure 5.41: Magnetic monopole mass limits from CDF [426], ATLAS [208] and MoEDAL searches [304, 419] as a function of magnetic charge for various spins, assuming a Drell-Yan pair-production mechanism.

The first apparatus specifically designed to detect mini-charged particles was the SLAC (Stanford Linear Accelerator Centre) "beam dump"-type detector, comprising scintillator bars read out by photomultiplier tubes [429]. MoEDAL's new detector, shown in Figure 5.42, and milliQan (discussed above in Section 5.3.3) proposed for deployment near to the CMS detector [95] also designed to search for milli-charged particles, both have a design that harks back to the original SLAC detector. In order to reduce backgrounds from natural radiation the photomultiplier tubes and scintillator detectors of the MoEDAL apparatus will be constructed from materials with low natural backgrounds currently utilised in the astroparticle-physics arena. Its calibration system utilises neutral density filters to reduce the received light of high incident muons that manage to penetrate to the sheltered detector from the interaction point, in order to mimic the much lower light levels expected from particles with fractional charges.

MoEDAL is also planning another new sub-detector called MALL (MoEDAL Apparatus for detecting extremely Long Lived particles) [428]. In this case MoEDAL trapping volumes, after they have been scanned through the ETH Zurich SQUID facility to identify any trapped monopole will be shipped to a remote underground facility to be monitored for the decay of pseudo-stable massive charged particles that may also have become trapped. MALL is the detector that monitors MoEDAL trapping volumes for decays of captured particles. It is envisaged that MALL will be installed deep underground at SNOLAB in Canada where cosmic backgrounds are minimised to one muon per 0.27 $m^2$/day. Background is further reduced by the ability to determine if a detected track originated within the monitored volume and also by energy cuts on deposited signals. A schematic view of the detector is shown in Figure 5.43. Initial estimates indicate that lifetimes up to around 10 years can be probed. The MALL detector is designed to be sensitive to charged particles and to photons, with energy as small as 1 GeV. It is envisaged that construction of the detector will begin after the MAPP detector is full deployed.

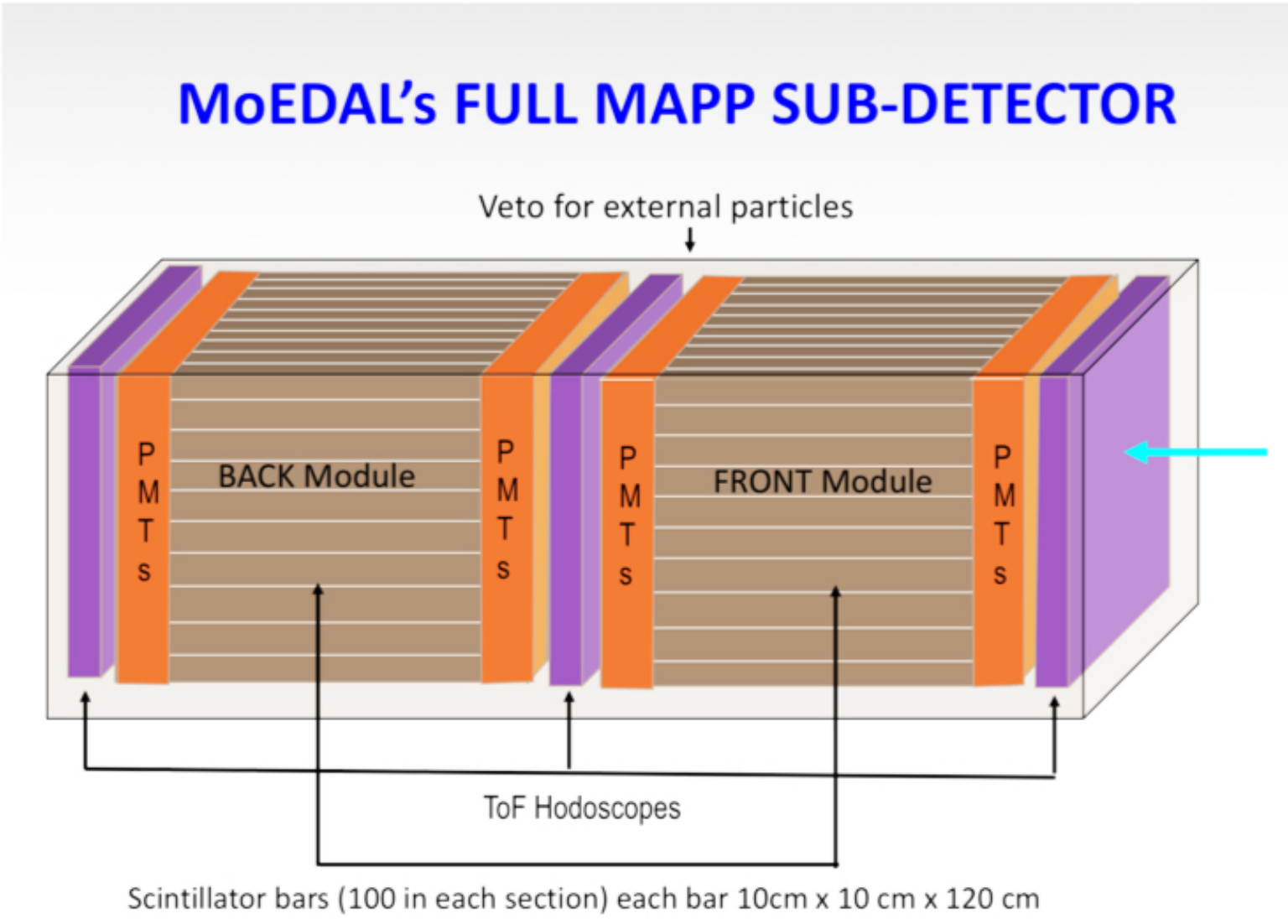

Figure 5.42: A depiction of the MoEDAL's MAPP sub-detector [428].

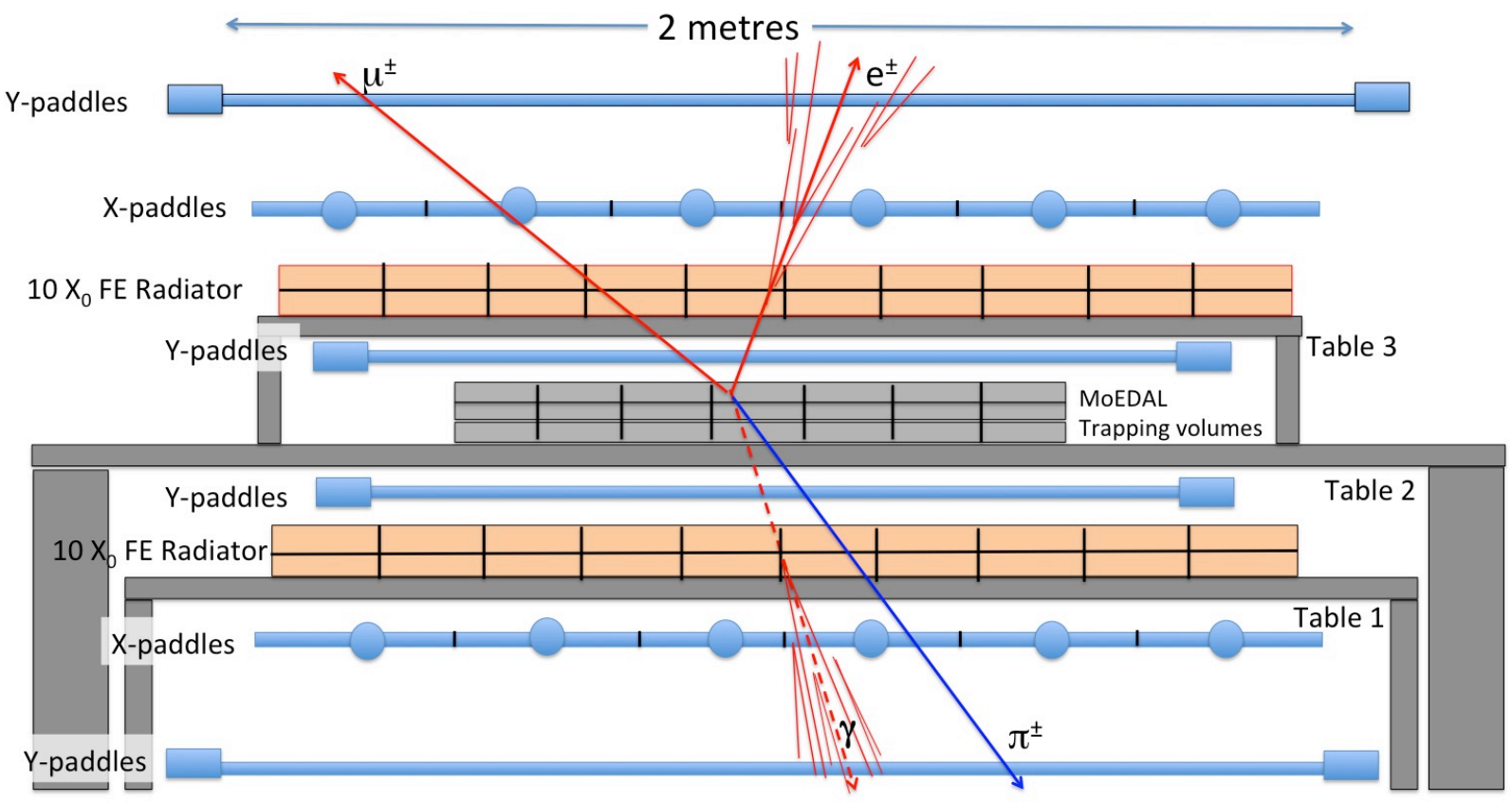

Figure 5.43: The MALL sub-detector designed to monitor MoEDAL trapping volumes for the decays of trapped electrically charged particles with lifetimes as long as 10 years [428].

The possibility of analysing decommissioned parts of the LHC

beam-pipe system at the ATLAS, CMS and LHCb/MoEDAL sites with a SQUID to search for trapped magnetic monopoles has been proposed [430]. In this context the MoEDAL experiment may serve as a formal platform for coordinating machining, scanning and analysis work, in collaboration with interested ATLAS, CMS and LHCb members.

The induction technique has been successfully employed at the LHC with the dedicated MoEDAL trapping detector. Additional searches for trapped monopoles in beam pipe material would access wide windows of magnetic charges and production cross sections to which other LHC experiments are insensitive. The decommissioned central beryllium beam pipe sections of ATLAS and CMS, with a $4\pi$ coverage and exposure to the highest rates of 7 and 8 TeV $pp$ collisions, are by far the most attractive samples to be analysed. The analysis on the CMS beam pipe is expected to be carried out in 2019.

### 5.3.3 *The MilliQan Experiment*

MilliQan is a dedicated experiment at the LHC to search for millicharged particles (mCP) [95, 431]. The milliQan experiment is part of a general program to search for hidden sectors [432] and other BSM scenarios [433]. As an illustrative example, we show the sensitivity of milliQan to an extra Abelian gauge field coupled to a massive Dirac fermion ("dark QED") that mixes with hypercharge through the kinetic term [115]. The result is that the new matter field is charged under hypercharge with a fractional electric charge of $\epsilon$, where $\epsilon \ll 1$. The milliQan experiment targets an unexplored part of the parameter space, namely mCP masses $0.1 \lesssim M_{\text{mCP}} \lesssim 100\,\text{GeV}$, for charges $Q$ at the $10^{-3}\ e - 10^{-1}\ e$ level.

The experimental apparatus consists of three scintillator detector layers of roughly 1 $\text{m}^3$ each, positioned near one of the high-luminosity interaction points of the LHC. The experimental signature consists of a few photo-electrons (PE) arising from the small ionization produced by the mCPs that travel unimpeded through material after escaping the LHC detectors.

The milliQan experiment is planned to be sited in the PX56 Observation and Drainage gallery above the CMS underground experimental cavern. The proposed gallery is limited in space. The detector will be located in this tunnel at an optimized location that is 33 m from the CMS interaction point (IP), behind 17 m of rock, and at an angle of 43.1 degrees from the horizontal plane. The selected location in a 3D model is shown in Figure 5.44.

The milliQan detector is a 1 m $\times$ 1 m $\times$ 3 m plastic scintillator array. The array will be oriented such that the long axis points at the nominal CMS IP. The array is subdivided into 3 sections each containing 400 5 cm $\times$ 5 cm $\times$ 80 cm scintillator bars optically coupled to high-gain photomultiplier tubes (PMTs). The detector will be shielded from other background sources such as activity in the scintillator and environmental radiation. With an estimated detection efficiency of about 10%, milliQan expects an average of $\mathcal{O}(1)$ PE from each attached phototube for each mCP with $Q = \mathcal{O}(10^{-3})\ e$ that traverses 80 cm of plastic scintillator. The signal is a longitudinal triple-incidence of hits with one or more PEs; a triple-incidence within a 15 ns time window along longitudinally contiguous bars in each of the three sections is required to reduce the background from dark-current noise. Requiring triple-incidence is expected to control background to $\mathcal{O}(10)$ events per year with $N_{\text{PE}} \geq 1$. The milliQan detector will self-trigger to a dedicated readout with no dead time from readout, up to rates of $\sim 1$ kHz. Energy calibration will be done in situ using an $^{241}$Am source.

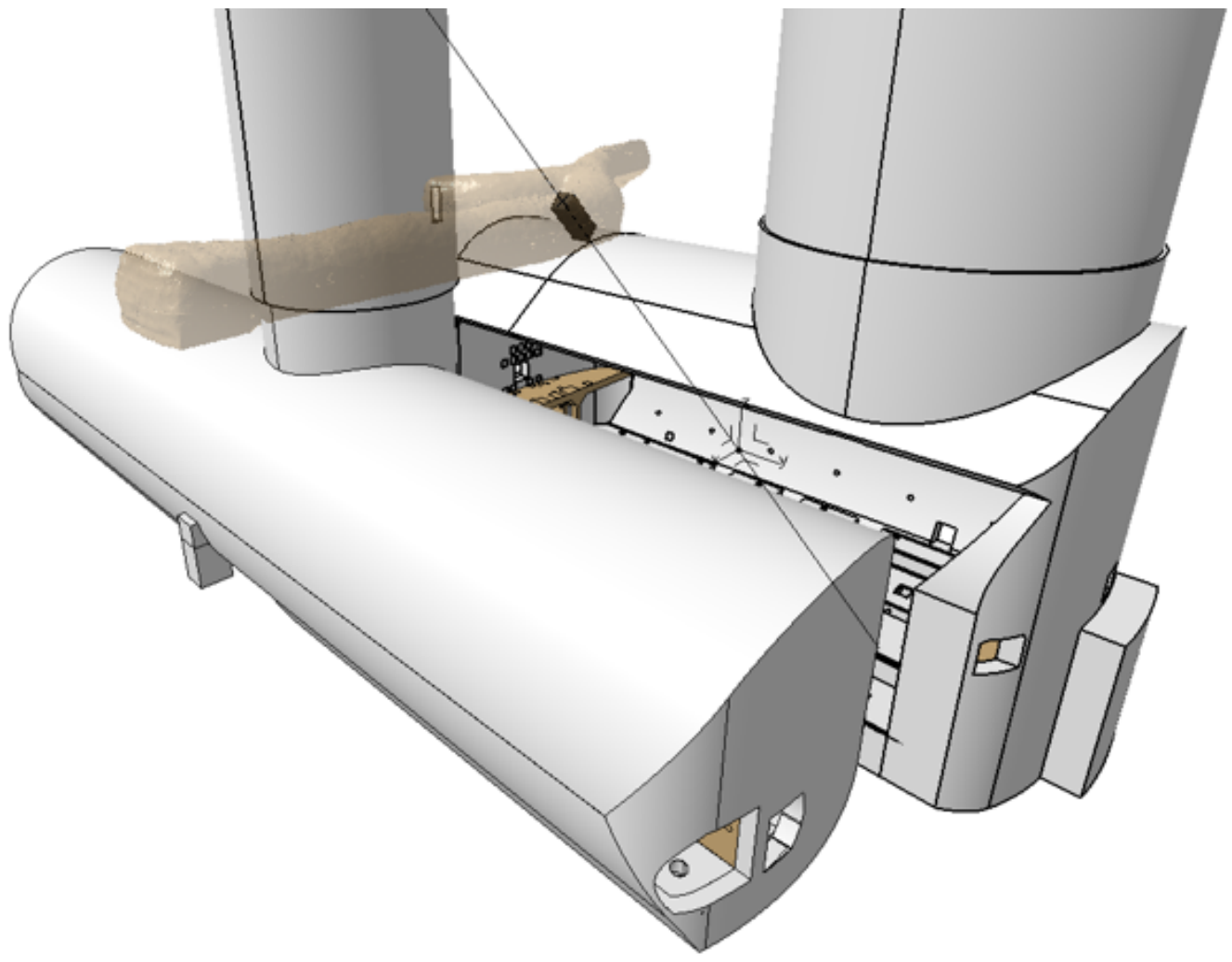

Figure 5.44: A 3D model showing the optimal position of milliQan within the PX56 Drainage and Observation gallery located above CMS UXC.

The dominant background is expected to come from dark-current pulses in the PMTs. Pulses from background radiation, including cosmic muons, will consist of 1000s of PEs that can be easily vetoed offline. Assuming a total background rate per PMT of $\nu_B$ =500 Hz, with a time window of $(\Delta t)_{\text{online}} = 100$ ns, milliQan expects a double coincident trigger rate per board of 1.5 Hz. The entire detector will be read out if one board triggers and there will be 50 such boards in total. Therefore, the full background trigger rate is expected to be 75 Hz. Offline, the time window will be tightened to $(\Delta t)_{\text{offline}} = 15$ ns, yielding an offline background rate for a triple coincidence of $2.8 \times 10^{-8}$ Hz. Since there are 400 such sets, the total offline background rate is estimated to be $1.1 \times 10^{-5}$ Hz. With these background rates, milliQan estimates a total of 165 (330) background events in 300 (3000) fb$^{-1}$ of integrated luminosity.

The milliQan collaboration has performed a full simulation of the experiment to evaluate the projected sensitivity to various mCP electric charges and masses. The simulation is performed in two stages. In the first, the production of mCP particles via Drell-Yan,

J/Ψ, Υ(1S), Υ(2S), and Υ(3S) channels at a center-of-mass energy of 14 TeV is performed. Particles produced at the IP are propagated to the proposed experimental site described above using a map of the CMS magnetic field. The effects of multiple scattering and energy loss are included using a simplified model of the CMS detector material budget and a region of rock spanning 17 m between the CMS experimental cavern and the proposed experimental site. The number of expected mCP particles per $fb^{-1}$ of integrated luminosity incident at the detector is computed as a function of the mass of the milli-charged particle. The signal efficiency is then estimated after processing the calculated particles as they would emerge from the interaction point through a full GEANT4 simulation of the detector; this is necessary because the small charge regime is sensitive to details such as the reflectivity, the light attenuation length, and the shape of the scintillator. These details, as well as the quantum efficiency, light-emission spectrum and the fast-time constants were modeled in GEANT4 using the specifications provided by the manufacturers for the scintillator and PMTs. Combining the estimated background rates discussed above with the cross sections, acceptances and efficiencies calculated for all masses and electric charges, the sensitivity projections of the milliQan experiment for LHC and HL-LHC are shown in Figure 5.45.

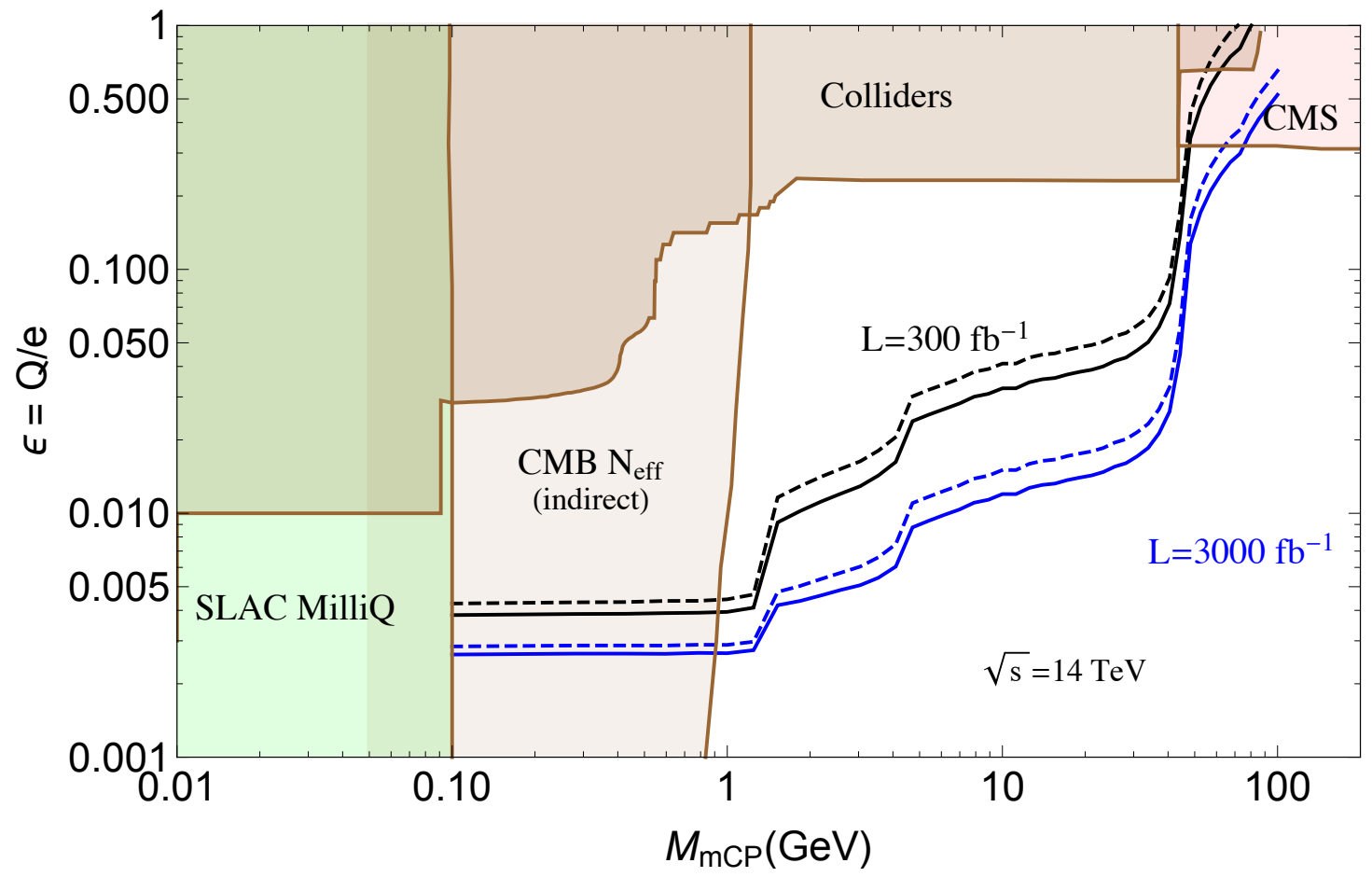

Figure 5.45: The expected sensitivity of the milliQan experiment for different LHC luminosity scenarios. The black line shows the expected 95% C.L. exclusion (solid) and 3$\sigma$ sensitivity (dashed), assuming 300 $fb^{-1}$ of integrated luminosity. In blue is shown the corresponding expectations for 3000 $fb^{-1}$.

A 1/100th scale "demonstrator" of milliQan to validate the detector concept was installed in the PX56 location at CERN during Technical Stop 2 of 2017 and was upgraded during the 2017–2018 year-end technical stop. This demonstrator, shown in Figure 5.46, has been recording data since its installation, and expects to have first results later this year. If funding is secured, construction of the full milliQan apparatus is planned for 2019, with installation in the

tunnel in 2020.

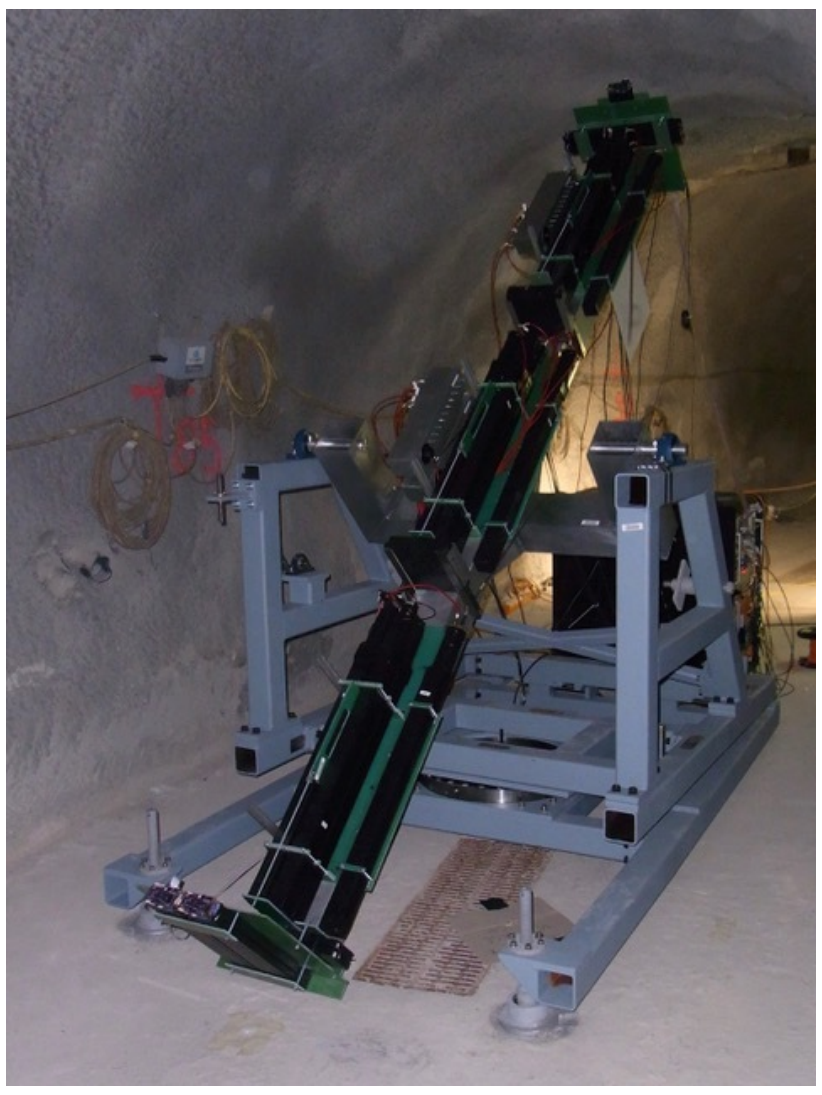

Figure 5.46: The 1/100th scale "demonstrator" of milliQan installed in the PX56 drainage gallery 33 m from the CMS IP.

### 5.3.4 The MATHUSLA Experiment

The basic motivation for the MATHUSLA (MAssive Timing Hodoscope for Ultra-Stable neutraL pArticles) detector [96] is the search for LLPs with lifetimes much greater than the size of the LHC main detectors, $c\tau \gg 100$ m. Any detector that can be reasonably constructed could only catch a small fraction of such LLPs decaying inside of its volume. Even with potentially large LLP production rates at the LHC, suppression of backgrounds is therefore crucial for discovery.

For LLP searches with high-energy or lepton-containing final states, the spectacular nature of displaced-vertex (DV) signals leads to very low backgrounds in searches at LHC detectors such as ATLAS or CMS. Any other class of LLP signature suffers from backgrounds and triggering limitations that can be significant. This greatly curtails the main detectors' ability to discover LLPs with very long lifetimes.

To address this broad blind spot of existing detectors, MATHUSLA is proposed to be a large, relatively simple surface detector that can robustly reconstruct DVs with good timing resolution. This gives MATHUSLA a similar geometric acceptance to LLP decays in the long-lifetime limit as the main detectors, while providing shielding from QCD backgrounds and sufficient tracking to reject ubiquitous cosmic rays (CRs). As a result, MATHUSLA is able to detect LLPs produced with $\sim 1$ pb cross sections at the HL-LHC with lifetimes near lifetimes of $\sim 0.1$ s, which is generally the limit imposed by Big-Bang Nucleosynthesis (BBN).

The simplified detector design for MATHUSLA is shown in Figure 5.47. The main component of the detector is a tracker ar-ray situated above an air-filled decay volume that is 20 m tall and $200\ \text{m} \times 200\ \text{m}$ in area. The tracker should have on the order of five planes to provide robust tracking with $\sim$ ns timing and $\sim$ cm spatial resolution. The current MATHUSLA design employs proven and relatively cheap technologies to allow for MATHUSLA's construction in time for the HL-LHC upgrade. Therefore, the trackers are envisioned to be implemented with Resistive Plate Chambers (RPCs), which have been used for very large area experiments in the past [434, 435], while plastic scintillators provide the surrounding veto.

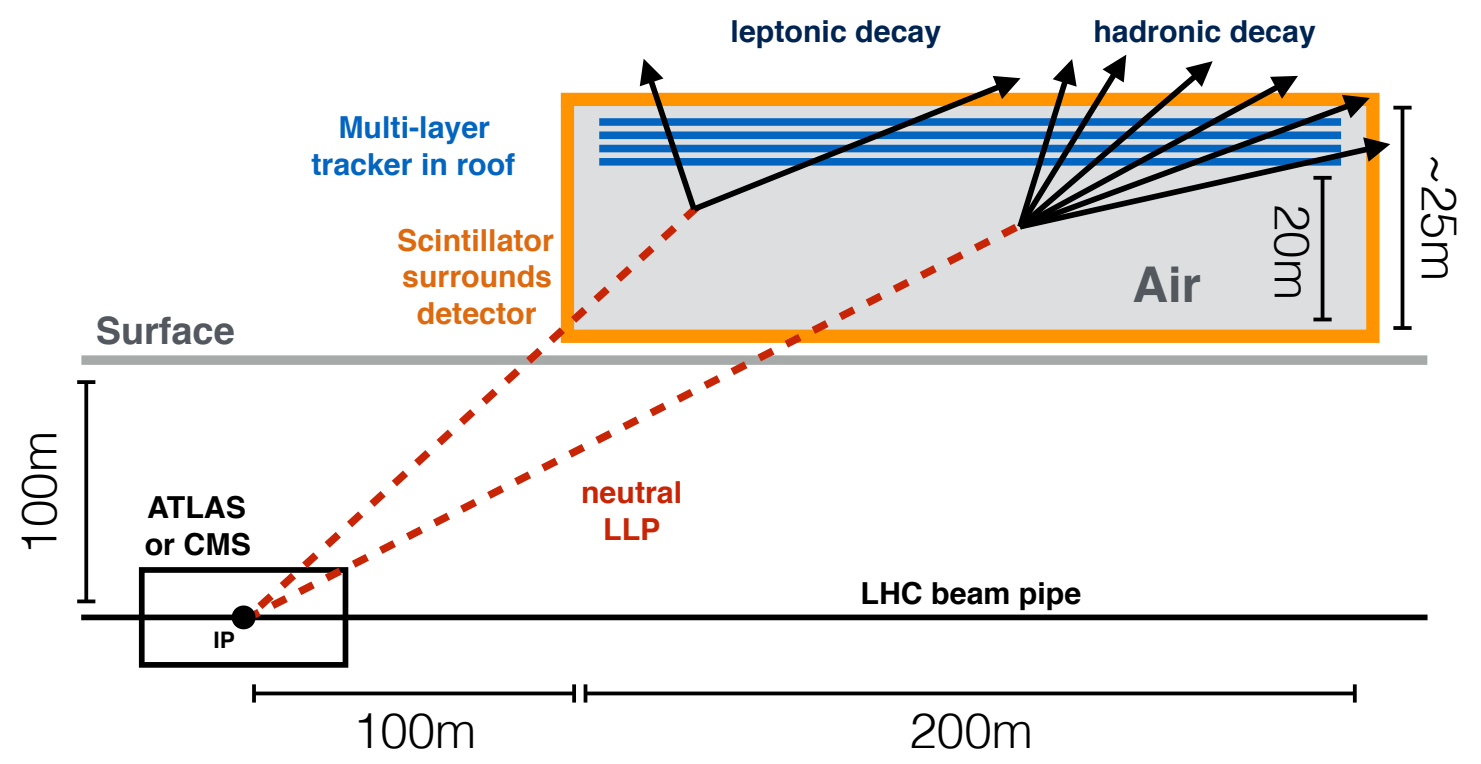

Figure 5.47: Simplified MATHUSLA detector layout showing the position of the $200\ \text{m} \times 200\ \text{m} \times 20\ \text{m}$ LLP decay volume.

This minimal design has been shown to be capable of measuring the LLP boost on an event-by-event basis [436] using the geometry of the LLP visible decay products. Final-state multiplicity provides a straightforward discriminant between hadronic and electromagnetic decays. Additional particle identification capability, as well as detection of final-state photons, might be possible by inserting an additional material layer between tracking layers to induce an electromagnetic shower that can be used to distinguish electrons, photons and muons.

As argued in Ref. [96], MATHUSLA could search for LLPs decaying into charged particles with little or no backgrounds. In Figure 5.47 is shown, schematically, the two main MATHUSLA signals, LLPs decaying into at least two charged leptons, or into jets that contain $\mathcal{O}(10)$ charged hadrons [436]. In 50-90% of leptonic decays and practically 100% of hadronic decays [436], two or more charged partices hit the ceiling due to the LLP boost and are recorded by the tracker. The charged particle trajectories can be fitted to reconstruct a DV. Unlike most analyses in the main detectors, these DVs must satisfy the additional stringent requirement that all trajectories coincide in time at the DV. The scintillator is used as a veto to ensure that the charged particles originate at the DV: there can be no hits along the line between the vertex and the LHC main interaction point (IP), nor along the lines obtained by extrapolating the individual charged particle trajectories backwards. These exhaustive geometric and timing requirements make it extremely difficult for backgrounds to fake the LLP signal. Cosmic rays can be rejected since they travel in the wrong direction (as well as occurring mostly in cosmic ray showers with extremely highly correlated particle multiplicity in the detector). Muons from the LHC collision do not satisfy the DV signal requirement. The rare occasions in which muons scatter inside the decay volume can be vetoed with the scintillators. Finally, neutrinos from cosmic rays and LHC collisions can scatter in the decay volume and produce displaced vertices, but those can also be rejected with geometric and timing requirements. We refer the reader to Ref. [96] for more details. More comprehensive studies of these backgrounds and their rejection strategies, including full simulations, are currently in progress.

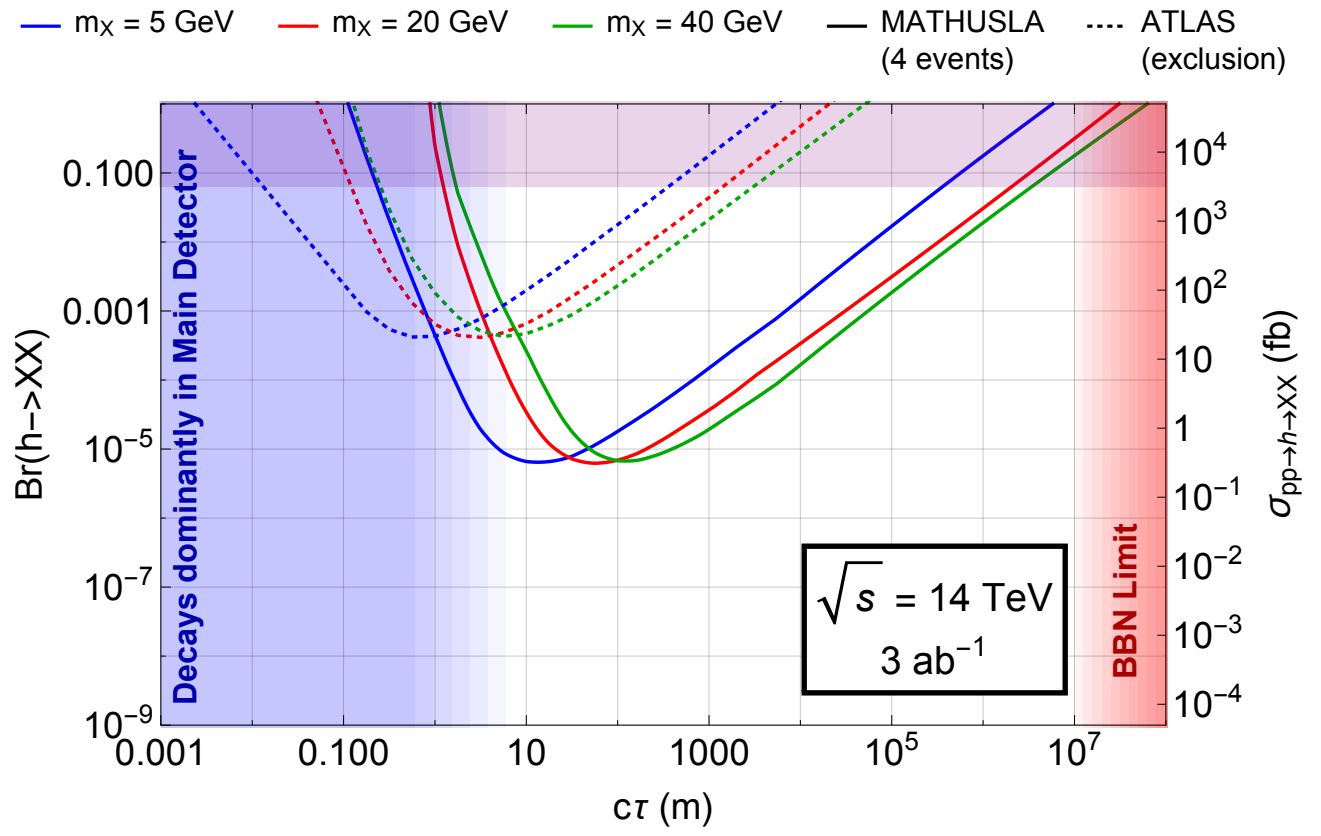

Figure 5.48: Sensitivity of MATHUSLA to hadronically decaying LLPs produced in exotic Higgs decays, where the solid lines correspond to four decays in the detector [96]. The dotted lines are the most optimistic ATLAS projection, using a very inclusive search for a single DV in the muon chamber [255].

An important general class of signals are LLPs with masses $\lesssim$ 100 GeV that decay hadronically and are produced without other highly visible signals like high MET or high-energy leptons. MATHULSLA can improve the sensitivity to these LLP production cross sections by a factor of $\sim 10^3$ compared to searches with the main detectors alone. This is illustrated in Figure 5.48, which shows MATHUSLA's sensitivity to hadronically decaying LLPs produced in exotic Higgs decays [96] compared to an optimistic projection for searches for a single DV in the ATLAS muon spectrometer [255]. For branching ratios of $\sim$ 10%, the BBN lifetime limit [437] can be probed.

Apart from reaching the BBN upper limit or ceiling of LLP parameter space, MATHUSLA extends the power of other LHC measurements. If a $\sim$ 10% invisible branching ratio for the Higgs was detected via coupling measurements at the HL-LHC, the absence of a MATHUSLA signal would lend strong support to the interpretation that the Higgs decayed to a stable component of dark matter.

Beyond LLPs produced in rare Higgs decays, MATHUSLA is a general-purpose LLP detector that is sensitive to a wide range of other models. Its reach for other BSM scenarios has been explored in [165, 167, 191, 438–444]. The general physics case for MATHUSLA's construction is made systematically in the recently released whitepaper [2].

It is important to point out that MATHUSLA is a very flexible detector concept that is completely scalable. The sensitivity of MATHUSLA is roughly proportional to the decay volume (which scales with surface area) and roughly independent of the precise geometry or location of the detector, as long as it is $\mathcal{O}(100 \text{ m})$ horizontally displaced from the IP. Therefore, a detector with smaller or larger volume than the benchmark in Figure 5.47 may be implemented, depending on available space and budget. Furthermore, the detector volume can be divided into smaller, independent modules (which cooperate for triggering purposes). These can be mass-produced economically and arranged according to the requirements of the experimental site. Such a modular construction also allows for a natural way to improve or extend the physics program of the experiment in an efficient way, by eventually upgrading, for example, one or a few modules with additional capabilities, such as higher tracking resolution for reconstruction of very low-mass LLPs below 10 MeV, or an additional material layer between the trackers for particle ID.

The MATHUSLA collaboration is currently studying such a modular design with the aim of producing a Letter-of-Intent in 2018. Crucial to this endeavor is the data from the MATHUSLA test stand, a $\sim 3 \times 3 \times 5$ m MATHUSLA-type detector that took a few days of data in the ATLAS instrument hall in 2017 and will take data for a few months in 2018. This allows local cosmic ray backgrounds to be measured, background rejection and signal reconstruction strategies to be tested, and simulation frameworks to be calibrated.

In conclusion, the MATHUSLA detector concept calls for a dedicated LLP surface detector above ATLAS or CMS. A detector volume of $\sim 10^6$ m$^3$ gives sensitivity to LLPs near the BBN lifetime limit if they are produced with $\sim$ pb cross section. This improves LLP sensitivity, compared to the main detectors alone, by several orders of magnitude for many LLP scenarios. The detector is simple and relies on proven technology, making its construction in time for the HL-LHC upgrade feasible. Once constructed, MATHUSLA could function without modification as a detector for the HE-LHC (with increased sensitivity). A small-scale test stand detector is already taking data at CERN, and studies are underway to finalize a detailed design for the full-scale detector.

### 5.3.5 A COmpact Detector for EXotics at LHCb (CODEX-b)

The CODEX-b proposal involves housing a small detector in the LHCb cavern — external to the LHCb detector itself — in a space approximately 25 m from the interaction point (IP8), behind the 3-m thick concrete UXA shield wall. This space is presently occupied by the LHCb data acquisition (DAQ) system, but will become available before Run 3 once the DAQ system is relocated to the surface. The layout of the cavern is shown in Figure 5.49, with the location of CODEX-b overlaid. The nominal CODEX-b configuration features a $10\times10\times10$-m volume instrumented with RPC tracking layers or other off-the-shelf tracking technology, as well as roughly 25 interaction lengths of shielding near IP8 — e.g., 4.5 m of Pb — to suppress primary and secondary $K_L$, neutron, and other hadronic backgrounds. This shield requires an active muon veto with an efficiency of $\mathcal{O}(10^{-5})$, in order to reject backgrounds induced by muons or other charged particles in the downstream parts of the shield. The veto is located several metres within the shield such that backgrounds induced by neutral particles remain suppressed. See Ref. [97] for a study of a proof-of-concept example detector layout and corresponding tracking efficiency, as well as a detailed study of the backgrounds. More ambitious technologies, including calorimetry, precision time-of-flight measurements, or integration into the LHCb readout may also be feasible.

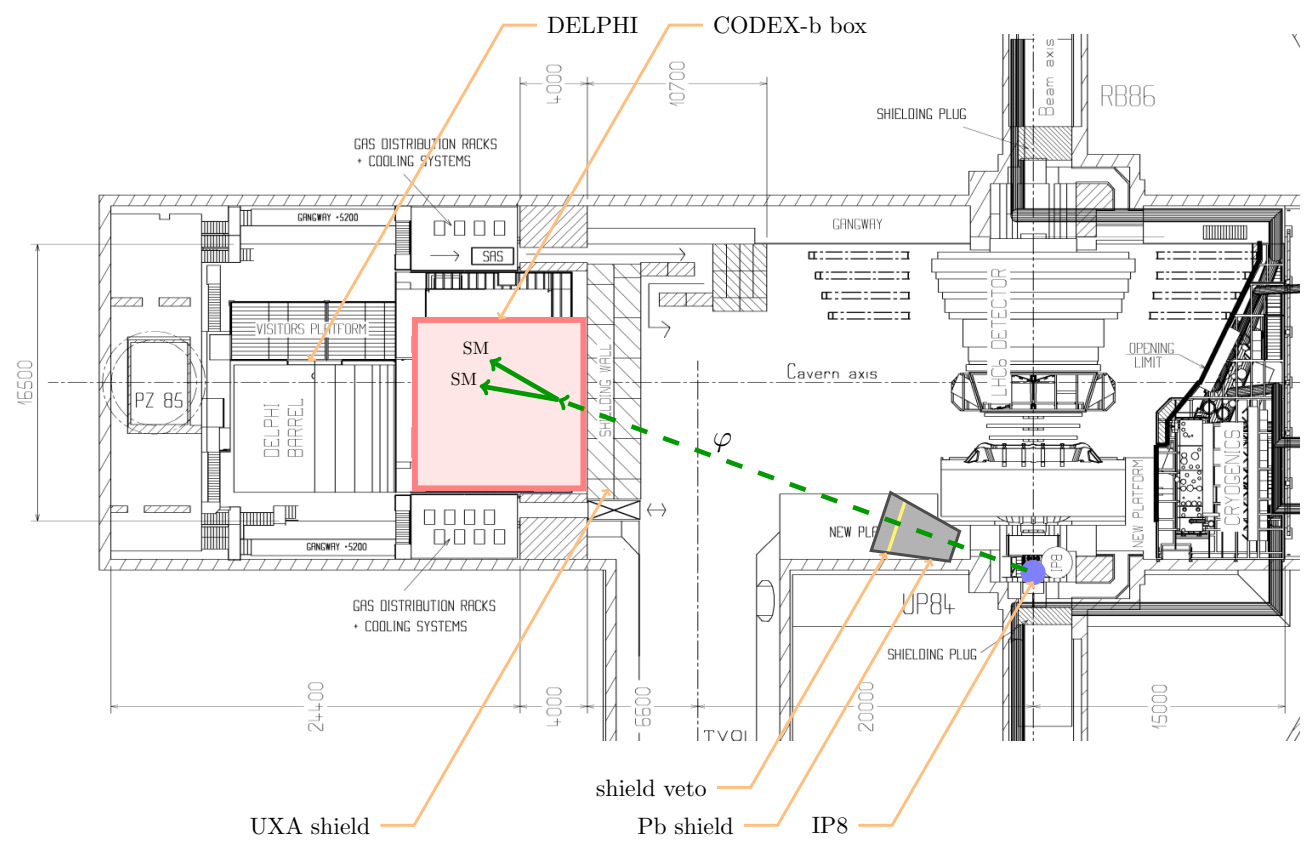

Figure 5.49: Layout of the LHCb experimental cavern UX85 at point 8 of the LHC [445], overlaid with the proposed CODEX-b location.

For the current discussion, the reach of CODEX-b for two benchmark models is quantified: First, a light scalar field, $\varphi$, that mixes with the SM Higgs boson is considered. If $m_\varphi \lesssim 5$ GeV, the production mode is primarily through inclusive $b \to s\varphi$ decays [446–448]. The LLP $\varphi$ subsequently decays back to SM fermions through the same Higgs portal. The reach in terms of $m_\varphi$ and the mixing angle $s_\theta$ is shown in the left-hand panel of Figure 5.50. CODEX-b significantly extends the projected reach of LHCb using only VELO-based displaced vertex reconstruction, and covers part of the parameter space to which SHiP [449] and MATHUSLA [440] are projected to be sensitive. Studies of the potential LHCb reach to longer lifetimes using downstream tracking are ongoing [372, 450]. The right-hand panel of Figure 5.50 indicates the reach for more general models, where the lifetime and production rate of $\varphi$ are unrelated.

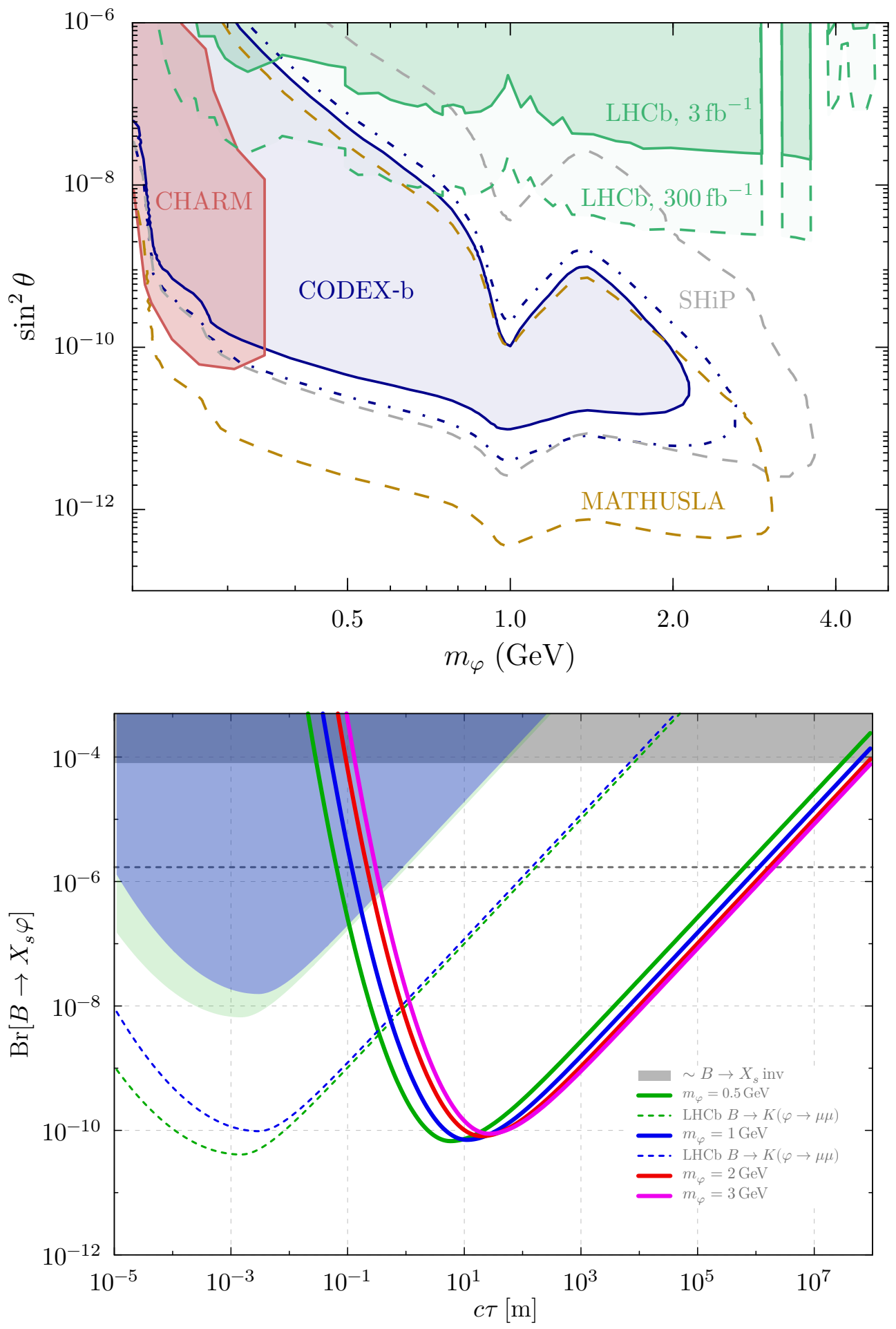

Figure 5.50: **Top:** CODEX-b reach for $B \to X_s\varphi$ in the $s^2_\theta$–$m_\varphi$ plane. Solid (dot-dashed) line assumes $\mathcal{L} = 300\,\mathrm{fb}^{-1}$ ($\mathcal{L} = 1\,\mathrm{ab}^{-1}$). **Bottom:** Inclusive CODEX-b $B \to X_s\varphi$ reach (solid lines). The shaded regions (dashed lines) indicate current LHCb limits (300 fb$^{-1}$ projection) from $B \to K(\varphi \to \mu\mu)$, rescaled to the inclusive process and assuming $\mathrm{Br}[\varphi \to \mu\mu] \simeq 30\%$ and 10% for $m_\varphi = 0.5$ GeV and 1 GeV, respectively. The gray shaded area and the dashed line indicate the approximate current [451] and projected [452] limits, respectively, for Belle II, from $B \to K^{(*)}\nu\bar{\nu}$ precision measurements.

For the second benchmark, a dark boson, $\gamma_\mathrm{d}$, produced through the exotic Higgs decay $h \to \gamma_\mathrm{d}\gamma_\mathrm{d}$ is considered. For concreteness the $\gamma_\mathrm{d}$ is taken to be a spin-1 field which can decay through mixing with the SM photon [63, 223, 453, 454]. In this benchmark, the production and decay are therefore controlled by different portals. The projected reach is shown in Figure 5.51, overlaid with the reach of ATLAS [255, 266] and MATHUSLA [96]. In particular, at low $\gamma_d$ masses, CODEX-b complements and significantly extends the reach of ATLAS and CMS.

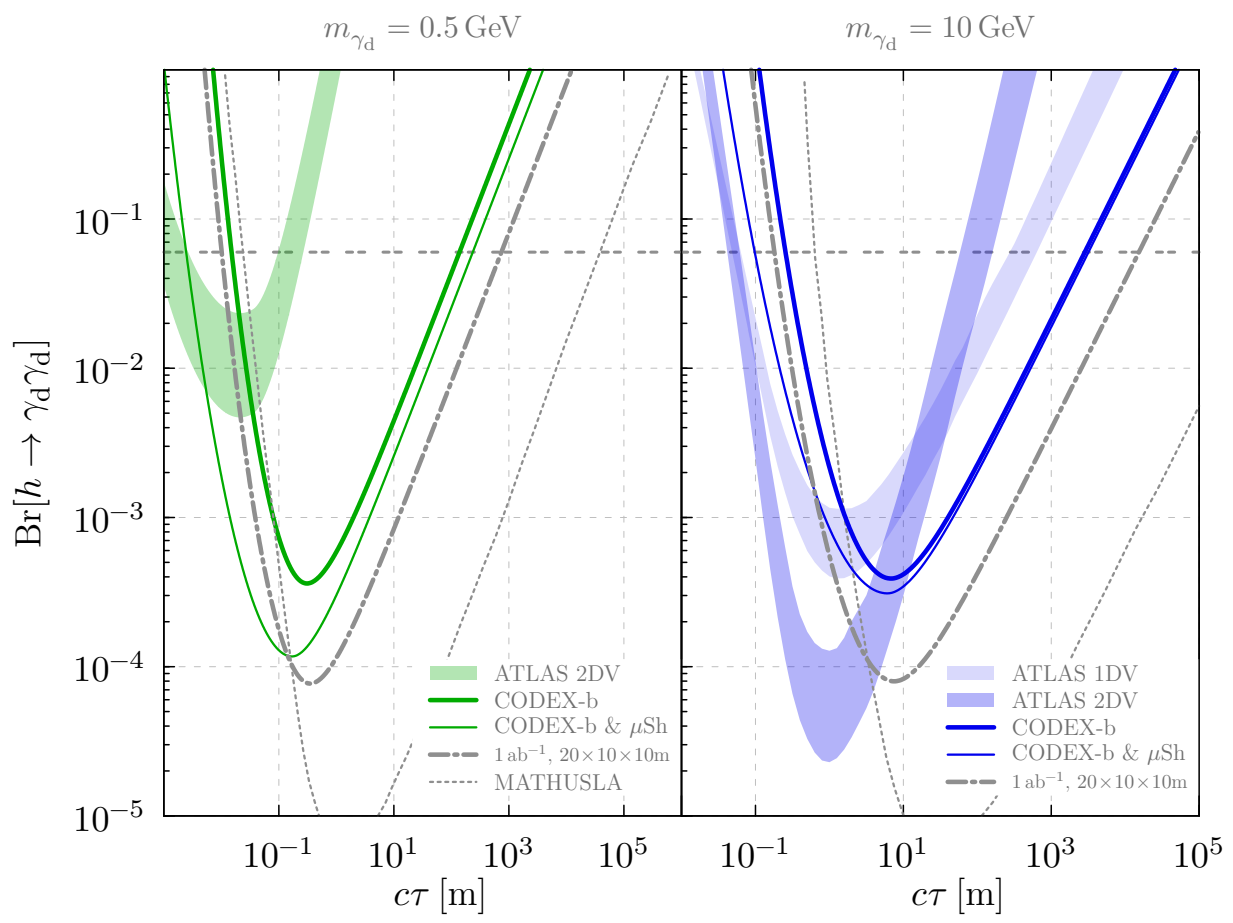

Figure 5.51: Comparison of experimental sensitivity to the BSM decay of an SM-like Higgs to dark photons with and without the muon shadow ('$\mu$Sh') for a few different existing and proposed detectors. The $\gamma_d \to \mu\mu$ branching ratio is taken from $e^+e^-$ data [455]. The CODEX-b sensitivity here is the so-called "optimistic" reach, with $\mathcal{L} = 1\,\text{ab}^{-1}$ and a larger volume, assuming DELPHI is removed.

Finally, it might be possible to install a larger version of the CODEX-b detector at IP2, after the ALICE collaboration concludes its heavy ion program. This option was named "A Laboratory for Long-Lived eXotics" (AL3X) [456], and can make use of the existing ALICE TPC, supplemented with a thick hadron absorber. The B-field from the L3 magnet would provide a good momentum measurement of the tracks, something which is absent in MATHUSLA and CODEX-b. The feasibility of this proposal is however contingent upon a luminosity upgrade of IP2, as well as the possible continuation of the ALICE physics during Run 5.

### *5.3.6 The ForwArd Search ExpeRiment (FASER)*

If new long-lived particles are light compared to the weak scale and very weakly coupled, the focus at the LHC on searches for new particles at high $p_T$ may be completely misguided. In contrast to TeV-scale particles, which are produced more or less isotropically, light particles with masses in the MeV-GeV range are dominantly produced at low $p_T$. In addition, because the new particles are extremely weakly coupled, very large standard model event rates are required to discover the rare new physics events. These rates are not available at high $p_T$, but they are available at low $p_T$: at the 13 TeV LHC, the total inelastic $pp$ scattering cross section is $\sigma_{\text{inel}}(13\text{ TeV}) \approx 75$ mb [457, 458], with most of it in the very forward direction. This implies

$$N_{\text{inel}} \approx 2.3 \times 10^{16}\ (2.3 \times 10^{17}) \tag{5.5}$$

inelastic $pp$ scattering events for an integrated luminosity of 300 fb$^{-1}$ at the LHC (3 ab$^{-1}$ at the HL-LHC). Even extremely weakly-coupled new particles may therefore be produced in sufficient numbers in the very forward region. Due to their weak coupling to the SM, such particles are typically long lived and travel a macroscopic distance before decaying back into SM particles. Moreover, such particles may be highly collimated. For example, new particles that are produced in pion ($B$-meson) decays are typically produced within angles of $\theta \sim \Lambda_{\text{QCD}}/E$ ($m_B/E$) off the beam-collision axis, where $E$ is the energy of the particle. For $E \sim$ TeV, this implies that even $\sim$ 500 m downstream, such particles will have only spread out $\sim$ 10 cm $-$ 1 m in the transverse plane. A small and inexpensive detector placed in the very forward region may therefore be capable of extremely sensitive searches to LLPs, provided a suitable location can be found and the signal can be differentiated from the SM background.

FASER [98, 459–462], the ForwArd Search ExpeRiment, is an experiment designed to take advantage of this opportunity. It is a small detector, with volume $\sim$ 1 m$^3$, that is proposed to be placed along the beam-collision axis, several hundreds of meters downstream from the ATLAS or CMS interaction point (IP). In the following, we present a promising location of FASER, discuss the properties of the signal and the required detector design, and present the new physics reach for representative models.

As shown in Figure 5.52, FASER will be placed along the beam collision axis, several hundreds of meters downstream from the ATLAS or CMS IP after the LHC tunnel starts to curve. A particularly promising location is a few meters outside the main LHC tunnel, 480 m downstream from the ATLAS IP, in service tunnel TI12, as shown in the bottom panels of Figure 5.52. (A symmetric location on the other side of ATLAS in tunnel TI18 is also possible.) This tunnel was formerly used to connect the SPS to the LEP tunnel, but is currently empty and unused. As shown on the tunnel map in the lower left panel of Figure 5.52, the beam collision axis passes through TI12 close to where it merges with the main LHC tunnel. A more detailed study of the intersection between the beam collision axis and TI12 verifies that there exists space for FASER in the tunnel, as shown in the lower-right panel of Figure 5.52.

In this location, FASER harnesses the enormous, previously completely unused and unexamined cross section for very forward physics ($\sigma \sim$ 100 mb). This cross section implies that even very weakly coupled new particles can be produced in large numbers at the LHC. In addition, the production of LLPs at high center-of-mass energy results in long average propagation distances ($\bar{d} \sim \mathcal{O}(100)$ m) and decays that are far beyond the main LHC infrastructure in regions where the backgrounds are expected to be negligible.

FASER will search for LLPs that are produced at or close to the ATLAS IP in the very-forward direction, travel approximately 480 m, and then decay via LLP $\to$ charged tracks $+ X$. When LLPs are produced in the very-forward region of the beam collision axis, they typically have very high energies $E \sim$ TeV. Although the identity of the LLP decay products depends on the mass of the LLP and the concrete new-physics model, some of the standard, characteristic LLP decay signatures are generically expected, such as two or more stable charged particles, such as electrons, muons or pions. This leads to a striking signature at FASER: two oppositely charged tracks with very high energy that emanate from a vertex inside the detector and which have a combined momentum that points back to the IP. A measurement of individual tracks with sufficient resolution and an identification of their charges is therefore imperative if the apparatus is to make use of kinematic features to distinguish signal from background. A tracking-based technology, supplemented by a magnet and a calorimeter to allow for energy measurements, will be the key components of FASER. Details of the detector design can be found in the Letter of Intent [459] and Technical Proposal [461].

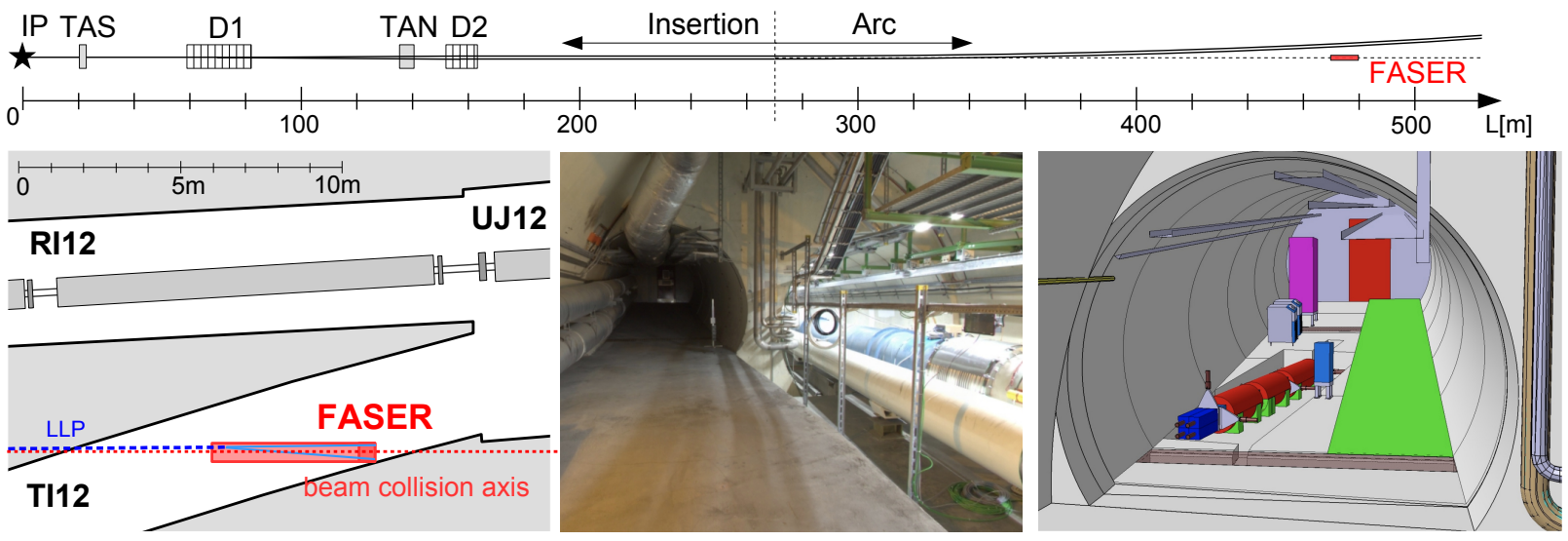

Figure 5.52: Proposed location for FASER in TI12. Top panel: a schematic drawing of the LHC and the very forward infrastructure downstream from the ATLAS and CMS interaction points; FASER is to be located 480 m from the IP, after the LHC ring starts to curve. Bottom panels: a map of the tunnel TI12 including the beam collision axis (left), a photo of this location (center), and a model of the experiment integrated in the TI12 tunnel (right).

The FASER signals consist of two extremely energetic ($\sim$TeV) coincident tracks or photons starting at a common vertex and pointing back to the ATLAS IP. Muons and neutrinos are the only known particles that can transport such energies through 100 m of rock and concrete between the IP and FASER. The CERN Sources, Targets, and Interactions (STI) group has computed the fluxes of muons and neutrinos at the FASER location using a FLUKA simulation [463–465]. These muon fluxes then allow one to estimate the rate and energy spectrum of muon-associated radiative processes near the detector. Preliminary estimates show that muon-associated radiative processes and neutrino-induced backgrounds may be reduced to negligible levels [461].

Emulsion detectors and battery-operated radiation monitors were installed in both TI12 and TI18 during Technical Stops in 2018. The results from these *in situ* measurements have validated the estimates of the FLUKA simulation, confirming that the high-energy particle background is highly suppressed and radiation levels are also very low and not expected to be problematic for detector electronics. Additional work is ongoing to refine background estimates, evaluate signal efficiencies, and optimize the detector.

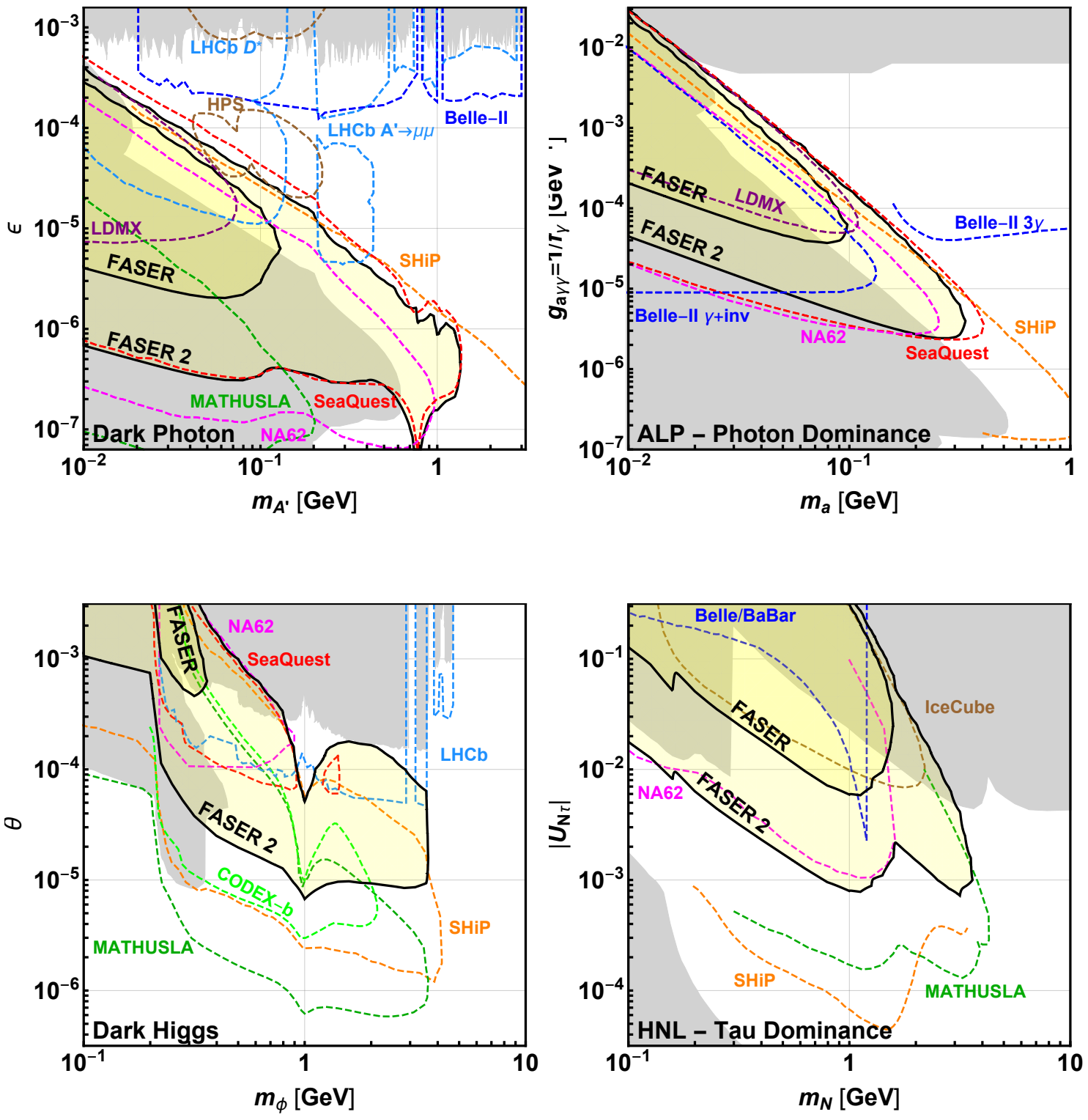

Figure 5.53: Projected FASER exclusion reach for benchmark new-physics scenarios containing dark photons (top left), ALPs with dominantly photon couplings (top right), dark Higgs bosons (bottom left), and HNLs with dominantly $\tau$-mixing (bottom right) in the corresponding coupling vs. mass planes. The gray shaded regions are excluded by current experimental bounds, and the colored contours represent projected future sensitivities of other proposed experiments that search for LLPs. See Ref. [460] for details.

In its first stage, FASER is an extremely compact detector, sensitive to decays in a cylindrical region of radius $R = 10$ cm and length $L = 1.5$ m. FASER is planned to be constructed and installed in Long Shutdown 2 and will collect data during Run 3 of the 14 TeV LHC (2021-23). After FASER's successful operation, FASER 2, a much larger successor with roughly $R \sim 1$ m and $L \sim 5$ m, could be constructed in Long Shutdown 3 and collect data during the HL-LHC era (2026-35). More details on the FASER timeline can be found in the Letter of Intent [459] and Technical Proposal [461].

The sensitivity of FASER to detect LLPs has been studied in a plethora of new physics models, including dark photons [98], dark Higgs bosons [466], heavy neutral leptons (HNLs) [290], axion-like particles (ALPs) [467], inelastic dark matter [468], flavor-specific scalar mediators [469], $U(1)_{B-L}$-gauge bosons [470], R-parity violating supersymmetry [167, 471] and strongly interacting massive particles (SIMPs) [472]. A summary on FASER's physics reach for LLP can be found in Ref. [460]. Combined, these studies establish FASER's significant impact on the LHC's discovery reach for LLPs. The physics reach at FASER and FASER 2 for these models is shown in Figure 5.53. Here we assume that backgrounds can be reduced to negligible levels. The gray-shaded regions of parameter space have already been excluded by previous experiments. For comparison we also show the projected reaches of other proposed experiments that search for long-lived particles.

Dark photons and ALPs (upper panels) are mainly produced in the decay of light mesons, via dark bremsstrahlung, or through the Primakoff process, and they are therefore very collimated around the beam collision axis. Already a very small detector is able to probe large and unconstrained regions of parameter space, making dark photons an ideal short-term goal for FASER. In contrast, dark Higgs bosons and HNLs (lower panels) define good long-term physics goals. They are both mainly produced in heavy meson decays, leading to a larger spread around the beam collision axis. A larger, but still relatively small, detector with $R \sim 1$ m is then required to exploit the full potential of FASER.

# 6

# *Reinterpretation and Recommendations for the Presentation of Search Results*

**Contents**

**Chapter editors:** Giovanna Cottin, Nishita Desai, Sabine Kraml, Andre Lessa, Zhen Liu

**Contributors:** Juliette Alimena, Will Buttinger, Eric Conte, Yanou Cui, Jared Evans, Benjamin Fuks, Lukas Heinrich, Jan Heisig, Gavin Hesketh, David Michael Morse, Michael Ramsey-Musolf, Ennio Salvioni, Michele Selvaggi, Brian Shuve, Yuhsin Tsai

## 6.1 Introduction

Models and scenarios with LLPs have seen an enormous rise in interest in recent years. They include supersymmetric scenarios with almost mass-degenerate lightest states [56, 58], highly split spectra [106, 473], very weakly interacting lightest supersymmetric particles (LSPs) like gravitinos or axinos [474, 475], or *R*-parity violation [52, 103, 104], as well as equivalent scenarios in other SM extensions (e.g., extra-dimensional models) with new SM gauge-charged particles. More recent ideas include models with feebly interacting dark matter [139] (supersymmetric or not), asymmetric dark matter [476], Hidden Valley models [59], and other dark-sector models; for a comprehensive discussion, see the classification of existing well-motivated theories with LLPs in Chapter 2.

All of these models can feature a large variety of possible LLP signatures. In Hidden Valley models [59], for instance, new particles can either decay into invisible dark particles or back to the SM, thus possibly leading to a mix of long-lived and prompt signatures, with or without missing transverse energy ($\not{E}_T$). Furthermore, new theoretical frameworks are constantly emerging, often motivated by new approaches to the hierarchy problem or dark matter. It is therefore of great interest to our community to be able to reinterpret the LLP experimental results for new models which may be developed in the future [1].

The reinterpretation of experimental results can generically be done in two ways: by applying appropriate simplified-model results to more complete models, or by reproducing the experimental analysis in a Monte Carlo (MC) simulation. Clearly, the former is easier and faster, while the latter is more generally applicable but also more difficult and much more time consuming [2].

In the context of searches for prompt signatures with $\not{E}_T$, the use of simplified models has been shown to be a fruitful approach for both the experimental and theoretical communities [67, 69, 71, 80, 81, 478, 479]. Dedicated tools, notably SModelS [480–482] and Fastlim [483], are publicly available and allow the user to reinterpret SUSY simplified-model results within the context of a full model. The coverage of a full model can, however, be severely limited by the kind of simplified-model results available, as discussed recently in Ref. [99] for the case of the phenomenological MSSM. Indeed, for some models the large number of relevant simplified-model topologies and their complexity can make the simplified-model approach inexpedient. In this case, a more complete and robust recasting procedure is necessary.

Again, for prompt signatures, a general recasting approach is available through several public tools, notably CheckMATE [484, 485], MadAnalysis5 [486, 487], Rivet [488] (v2.5 onwards) and Gambit's ColliderBit [489]. These tools allow the user to reproduce experimental analyses by means of MC event simulation coupled to an approximate emulation of detector effects. [3] For the latter, CheckMATE and MadAnalysis5 rely on Delphes [336], in some cases supplemented with appropriate tuning, while Rivet and ColliderBit employ object smearing and analysis-specific reconstruction efficiencies. The ATLAS and CMS collaborations are helping these recasting efforts by providing more and more detailed information about their analyses and results. As an example, covariance matrices for the background correlations in the framework of simplified likelihoods were provided for the analysis of Ref. [490].

[1] In this context we refer the reader also to the activities of the "Forum on the Interpretation of the LHC Results for BSM studies" [477].

[2] When no new backgrounds need to be considered and the hypothesized signal does not affect control regions, one can simply determine the event counts in the signal regions and compare them to the 95% C.L. observed limits, or take the numbers of observed events and expected backgrounds to compute a likelihood.

[3] For completeness it should be noted that, while all of these tools include a more-or-less extensive set of SUSY searches, many of the searches for other, "exotic" types of new physics cannot yet be reproduced outside the experimental collaborations. This concerns in particular searches relying on multivariate techniques such as Boosted Decision Trees.

The situation is, so far, quite different for LLP searches. First, the presentation of results in terms of simplified models is still limited to a few topologies and does not always include all of the parameters required for a general purpose reinterpretation. It is worth noting that, compared with simplified models for prompt searches, simplified models for LLP searches always have at least one additional free model parameter: the lifetime of the LLP. Second, recasting LLP searches outside of the experimental collaborations is a difficult task since the searches are very sensitive to the detector response, which in most cases cannot be easily emulated by a fast detector simulation (and many backgrounds are also challenging to simulate due to their non-standard nature; see Chapter 4 for a full discussion of backgrounds to LLP searches). As a result, while first steps towards the treatment of LLPs were achieved in SModelS [491] and MadAnalysis5 [492], the implementation of LLP searches in public recasting tools is still in its infancy, thus limiting the applicability of the experimental results.

In order to allow for a more extensive reinterpretation or recasting of the experimental analyses, detailed information concerning the detector performance and object reconstruction is needed. These can in principle be provided in the format of efficiencies [4] for selection and reconstruction of relevant objects, as demonstrated already by some pioneering experimental publications [230, 493]. One difficulty in this respect is that the information needed for recasting LLP searches is clearly analysis dependent, which means an additional workload for the analysis groups to provide this information on a case-by-case basis.

[4] We employ the term "efficiency" in a broad sense. It can refer to reconstruction efficiencies, selection efficiencies, overall signal efficiencies, etc., which will be further specified below.

The objective of this chapter is to discuss the presentation of LLP search results with the aim that they can be re-used for interpretations beyond the models considered in the experimental publications. To this end, we first discuss in Section 6.2 the various options for presenting the LLP results (and for a summary, see Figure 6.1), and compare their advantages and shortcomings. In Section 6.3, we discuss in more details how the simplified models defined in Section 2 can be used to reinterpret LLP searches. In Section 6.4, we present several attempts at recasting LLP searches, according to the LLP signature, focusing on heavy stable charged particles and various displaced objects. For each case, we elaborate on the lessons learned from the reinterpretation effort. Section 6.5 presents a first attempt to extend the public detector simulator DELPHES to deal with LLP searches, while Section 6.6 focuses on reinterpretations performed within the experiments themselves, including the RECAST framework [494]. In Section 6.7, we discuss complementary constraints on LLPs from reinterpreting prompt searches. We conclude in Section 6.8 with our recommendations for the presentation of LLP results.

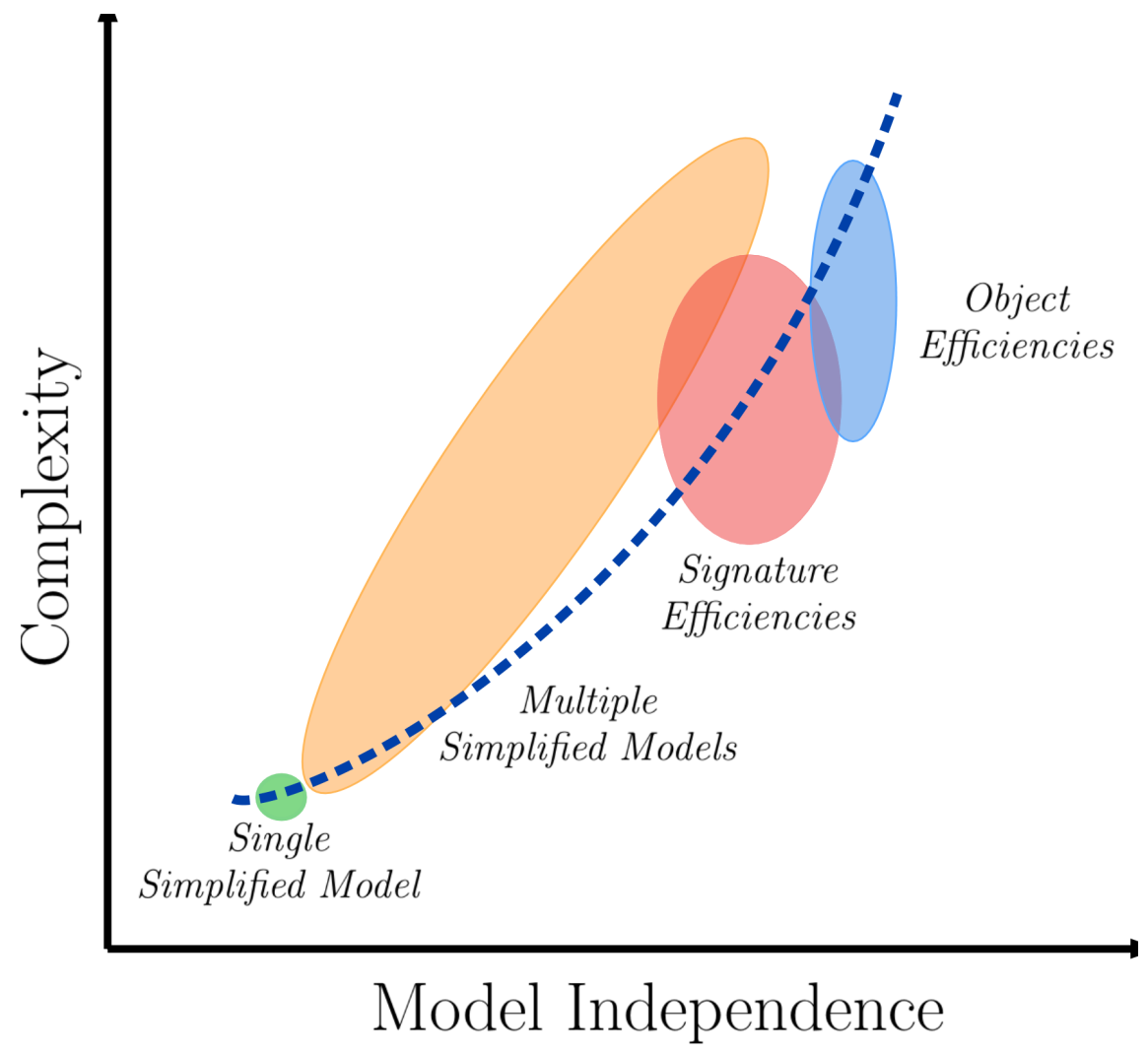

Figure 6.1: A qualitative overview of the possibilities for the presentation of results discussed in this chapter. The axes represent the complexity of information required by each format and the corresponding level of model independence.

## 6.2 *Options for Presenting Experimental Results*

A qualitative view of the various possibilities for presentation of search results is illustrated in Figure 6.1. We broadly classify these possibilities according to the type of information provided for reinterpretation (in other words, the type of efficiency). Each type refers to distinct signal objects, as illustrated in Figure 6.2. As we can see, each possibility relies on distinct assumptions about the signal, resulting in different levels of model dependence. Below, we provide a brief discussion of the advantages and limitations of the various possibilities of presentation of LLP search results. We gradually progress from the simplest case towards more complexity but also better re-usability.

*Simplified Models* In most cases, the simplified-model topology corresponds to single or pair production of the LLPs, though in principle simplified models for production through cascade decay of heavier states can also be envisaged. In any case, the simplified model incorporates important assumptions on the LLP production mode, decay mode and quantum numbers. Simplified-model results can be presented at different levels of sophistication and re-usability:

- exclusion curves in, e.g., a mass-vs-mass or mass-vs-lifetime plane, are highly model dependent and can rarely be used for reinterpretation without significant effort;
- cross section upper limits [5] can be applied to a larger variety of models in which the same LLP production and decay mode occurs through a rescaling of the cross section times branching ratio factor;
- simplified-model efficiencies go one step further: they make it possible to combine different topology contributions to the same signal region and compute an approximate likelihood using the number of expected and observed events.

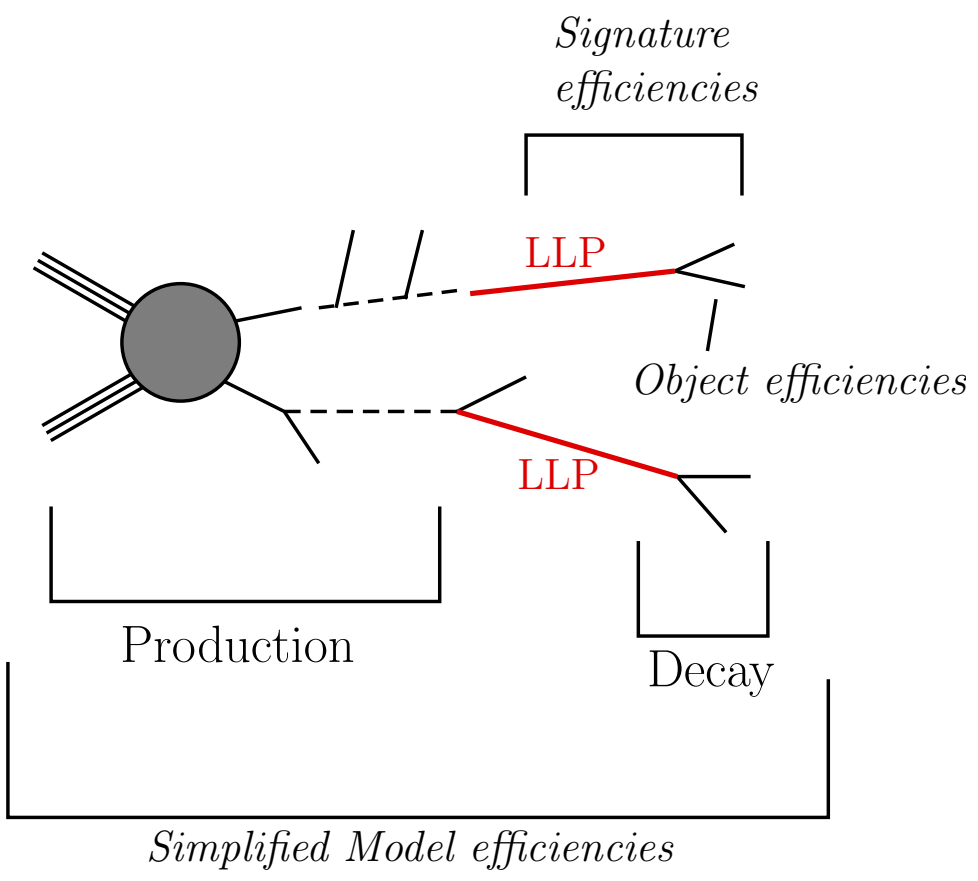

Figure 6.2: Possibilities for the presentation of results: simplified-model efficiencies assuming a specific topology of LLP production and decay, signature efficiencies assuming only a specific LLP decay, and final-state object efficiencies which are independent of the specific decay mode.

[5] For the sake of re-usability, cross section upper limits in absolute terms are much preferred over limits on the signal strength.

The main advantages of simplified models are a parametrization of the signal sensitivity in terms of few physical parameters and a unified language and format applicable to a wide range of searches. Also, when re-using simplified-model results one avoids detector simulation uncertainties. The disadvantages are that the simplified-model result cannot be applied to other LLP production or decay modes, resulting in overly conservative limits if the LLP signal is composed from multiple topologies. This can be important if the LLP is a color singlet, but there are several heavier color-charged states which can be produced and decay to the LLP. In principle this can be overcome if efficiencies are provided for a sufficiently large number of simplified models (including cascade decays), as a function of the simplified-model parameters, which should include the LLP lifetime. These efficiencies can then be combined in order to compute the corresponding constraints to complex models, where multiple topologies are present [6]. We stress, however, that in order for this combination to be possible, signal efficiencies and not cross section upper limits must be provided. The major drawback of this approach is that in order for the results to be applicable to a broad class of models, the number of required simplified models and their complexity can easily become very large. For achieving a high level of model independence, it is therefore desirable that the experimental analysis can be recast with MC event simulation. Two ways of presenting results are useful to this end: *signature efficiencies* and *object efficiencies*.

[6] These points are illustrated by the reinterpretation of the CMS search for heavy stable charged particles [493] discussed in Section 6.3: for the specific model considered in Ref. [90], constraints obtained using only a single simplified model (direct production of $\tilde{\tau}$s in this case) can underestimate the bounds on the LLP mass by almost a factor of two.

*Signature Efficiencies* Signature efficiencies are efficiencies for the reconstruction of the main LLP signature treated as a single "object" (single charged track, displaced vertex, disappearing track, etc.) as a function of the LLP kinematic parameters and the lifetime. Signal efficiencies require the assumption of a specific LLP decay mode, but are otherwise highly model independent since they make no assumption on the LLP production mode. In addition, they are fully model independent for stable particles (within the detector volume), since in this case no assumptions about the LLP decay mode are required at all. In many cases, however, the reconstruction efficiencies depend on multiple kinematic variables, such as the LLP $p_T$, its transverse decay position, impact parameter, etc. As illustrated in Section 6.4.6, these efficiencies can be very useful for recasting LLP searches, but they are not often provided by the experimental collaborations. Many recasting efforts consist of extracting these efficiencies from the provided information, but this procedure can result in large uncertainties.

*Object Efficiencies* Object efficiencies are efficiencies for the reconstruction of the physics objects relevant for building the LLP signature; for example, if an analysis reconstructs displaced vertices out of tracks, the associated object efficiencies would be the efficiency of reconstructing each displaced track that is later combined to form the vertex. They can clearly be applied to a wide range of LLP decay and production modes, since no specific assumptions about either is required. For instance, a displaced lepton reconstruction efficiency can be provided as a function of the lepton $p_T$ and its transverse impact parameter $d_0$. As discussed in Section 6.4.2, these efficiencies can be used to recast LLP searches to an acceptable accuracy ($\sim 20\%$). Furthermore, as illustrated in Section 6.4.6, knowledge of object efficiencies are essential for a general purpose recasting of the search. Within this approach the model dependence is minimal and can be restricted to a few general assumptions about the nature of the LLP. Also, object efficiencies could be included in fast-detector simulators, thus providing a way of recasting LLP searches on a similar footing as prompt searches. The main difficulty with providing such object efficiencies is the potentially large number of parameters required for their parametrization.

## 6.3 Reinterpretation using Simplified Models

One of the possibilities for extending the experimental results from LLP searches to a large variety of scenarios is through the use of simplified-model topologies. Simplified models (or simplified-model spectra, SMS) have been widely used for the interpretation of prompt and LLP searches. As discussed in Chapter 2, a large number of SMS topologies are possible for the distinct LLP signatures, which can be grouped by the LLP production mode, decay and lifetime. These SMS topologies aim to capture the main physical properties of the LLP signal and can then be used to constraint other scenarios containing similar topologies. The use of simplified-model results to constrain full models has been shown to be possible [86, 99, 480, 483, 495, 496], even though it has its shortcomings [99]. Also within the context of LLP searches, the use of simplified model results for reinterpretation can be a good alternative, e.g., when a recasting based on a MC simulation is difficult or is too computationally expensive. In this section, we briefly review how SMS results can be used to reinterpret searches for full models as well as the particular challenges presented by LLP searches. A concrete example of reinterpretation using simplified models is given in Section 6.3.2, based on the results of Ref. [90].

### 6.3.1 From Simplified to Full Models

The interpretation of experimental results using simplified models typically corresponds to producing upper limits on the production cross section or signal efficiencies for a specific SMS topology (production and decay channel). These results are provided as a function of the simplified model parameters, which have been largely taken to be the masses of the BSM particles appearing in the topology. For LLP topologies, however, a new parameter must be considered: the LLP lifetime (see Chapter 2). With the exception of searches for stable particles, the lifetime is one of the main parameters affecting the topology efficiency and upper limit.

Once signal efficiencies [7] ($\epsilon$) are provided for one or more SMS topologies, these can be used, under some approximations, to quickly compute the number of expected signal events ($S$) for a full model:

$$S = \mathcal{L} \times \left( \sum_{\text{SMS}} \sigma_{\text{SMS}} \times BR_{\text{SMS}} \times \epsilon_{\text{SMS}} \right) , \tag{6.1}$$

where $\mathcal{L}$ is the luminosity for the respective search and the sum runs over simplified model topologies. Since the production cross section ($\sigma_{\text{SMS}}$) and branching ratios ($BR_{\text{SMS}}$) for each topology can be quickly computed for any full model, the simplified model signal efficiencies ($\epsilon_{\text{SMS}}$) can be directly used to obtain the signal yield. This procedure does not rely on any Monte Carlo simulation or recasting of LLP searches and can be easily applied to a wide variety of models, provided $\epsilon_{\text{SMS}}$ is known. The main limitation of this approach comes from the limited (although growing) number of SMS results available. Since $\epsilon_{\text{SMS}}$ is typically known only for very few simplified models, the sum in Eq. (6.1) is limited to the number of available topologies, resulting in an under-estimation of $S$.

[7] For simplicity we will refer to the signal acceptance times efficiency as "signal efficiency". This efficiency is a function of the simplified model parameters, including the LLP lifetime.

For prompt SUSY searches, a systematic approach for reinterpreting simplified model results based on the procedure outlined above has been developed in Refs. [480, 483]. Furthermore, using the large number of available SUSY SMS results, public tools are available for constraining full models using these results [481, 483]. The same procedure can also be applied to LLP SMS results, as shown in Ref. [90] for the case of heavy stable charged particles (HSCPs) and implemented recently in SModelS [491]. In the next section we review some of the results found in Ref. [90]. Although these have been obtained within the context of HSCPs, the main results can be generalized to other LLP signatures, and demonstrate some of the advantages and shortcomings of reinterpretations using LLP simplified models.

### *6.3.2 Reinterpretation using HSCP Simplified Models*

The CMS search for HSCPs in Ref. [493] provided signal efficiencies for the simplified model topology $pp \to \tilde{\tau}\tilde{\tau}$ as a function of the stau mass. In the language of the simplified models of Section 2.4.2, this is the direct pair production mode of a charged LLP. The stau is assumed to be stable (at detector scales), thus producing a highly ionizing track, which can be used to search for this scenario (see Section 3.5). Since the stau lifetime ($\tau$) is assumed to be $\gg$ 10 ns, the signal efficiencies do not depend on $\tau$, thus simplifying the SMS parameter space, which reduces to the stau mass. [8] The relevant selection efficiencies required for a general purpose MC recasting of the HSCP search have also been provided by the CMS analysis; see Section 6.4.1 for details.

[8] We point out that it is still possible to apply these simplified model results to models with smaller LLP lifetimes if we include the suppression factor from the LLP decay length distribution, as discussed in Section 6.4.1.2.

The efficiencies for the stau simplified model can be used to constrain a full BSM scenario which contains HSCPs. In Ref. [90], the region of the CMSSM parameter space with $m_{\tilde{\tau}} - m_{\tilde{\chi}_1^0} < m_\tau$ has been considered, since it provides a possible solution to the Lithium problem [497, 498]. Due to the small mass difference, the stau is long-lived and decays outside the detector, thus generating a HSCP signal. In Figure 6.3, we show the constraint on the CMSSM parameter space obtained using only the simplified model provided by CMS (direct stau production). Since the simplified model only contains one parameter, it translates to a limit on the stau mass ($m_{\tilde{\tau}} < 260$ GeV), as shown by the blue region in Figure 6.3. In this CMSSM scenario, however, direct production of staus only con-tributes to a small fraction of the total HSCP signal, since staus are typically produced from cascade decays of heavier SUSY states, such as charginos, squarks and gluinos. (Note that these correspond to the heavy-parent modes of Section 2.4.2.) Furthermore, there are several possible topologies which contain a stau and the LSP ($\tilde{\chi}_1^0$) in the final state, thus resulting in a mixed missing energy-HSCP signature. Therefore using only the CMS constraints for the direct stau production simplified model largely underestimates the sensitivity of the CMS search.

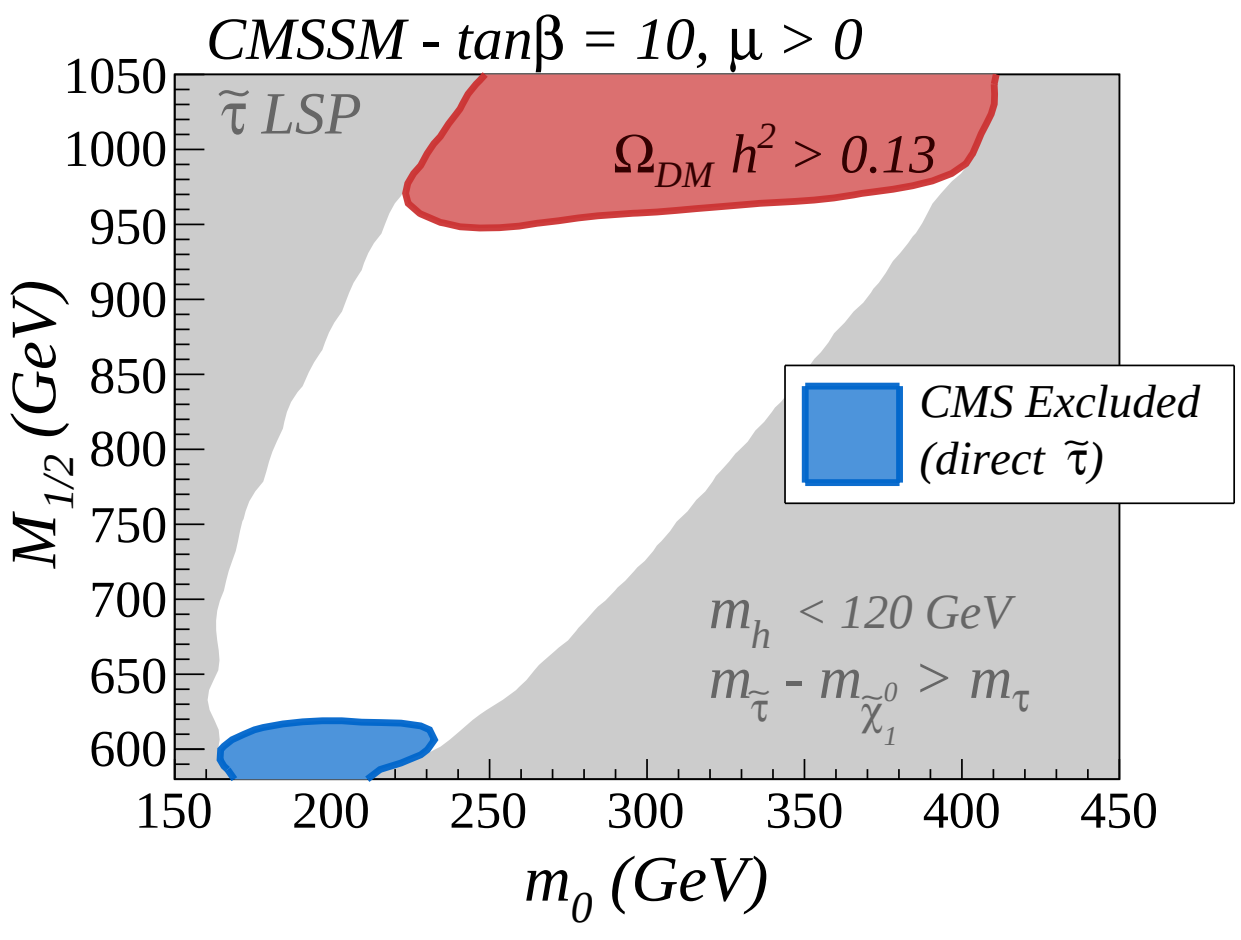

Figure 6.3: Region of the CMSSM parameter space with long-lived NLSP staus. The light gray regions are excluded by the requirements $m_{\tilde{\tau}} \gtrsim m_{\tilde{\chi}_1^0}$ and 120 GeV $\leq m_h \leq$ 130 GeV. The top red region is excluded by the upper limit on the neutralino relic density, while the lower blue region is excluded by the CMS constraints on direct production of long-lived staus. For more details see Ref. [90].

In order to improve the constraints shown in Figure 6.3, one must have efficiencies for several SMS topologies. Fortunately, a Monte Carlo recasting of the 8 TeV CMS search is possible (see Section 6.4.1 for details) and can be used to compute simplified model efficiencies. In Ref. [90], seven additional simplified models containing cascade decays were considered and their efficiencies computed as a function of the masses appearing in the topology. A summary of the topologies considered are shown in Table 6.1. It is important to point out that it is not necessary to specify the Standard Model final states appearing in the simplified models, since the HSCP search is inclusive and the efficiencies do not depend on the additional event activity. Using this extended database of simplified model efficiencies and Eq. (6.1), we can compute a more inclusive signal yield for each point of the CMSSM parameter space and improve the constraints on the model. The results are shown in Figure 6.4, where we see a drastic improvement in the region excluded by the constraints on HSCPs, as expected. For this specific scenario (with $\tan\beta = 10$), all the parameter space is excluded either by the CMS or dark matter constraints [90].

| SMS topology | Notation in Chapter 2 |
|---|---|
| $pp \to X\ X$ | DPP |
| $pp \to Y_1\ Y_1, Y_1\ \to SM\ X$ | HP |
| $pp \to Y_1\ Y_1, Y_1\ \to SM\ Y_2, Y_2\ \to SM\ X$ | - |
| $pp \to Y_1\ Y_2, Y_1\ \to SM\ Y_2, Y_2\ \to SM\ X$ | - |
| $pp \to Y\ Y, Y\ \to SM\ SM\ X$ | HP |
| $pp \to inv\ X$ | CC |
| $pp \to Y_1\ Y_2, Y_1\ \to SM\ inv, Y_2\ \to SM\ X$ | - |
| $pp \to Y_1\ Y_2, Y_1\ \to SM\ inv, Y_2\ \to Y_3\ SM, Y_3\ \to SM\ X$ | - |

Table 6.1: Definitions of the HSCP simplified models considered in this section. The symbol $X$ represents the HSCP, $Y_i$ represent intermediate BSM particles, $SM$ represents any Standard Model particle and $inv$ represents an invisible final state, such as the neutralino. The correspondance with the simplified models language of Chapter 2 (Section 2.3.1) is also given.

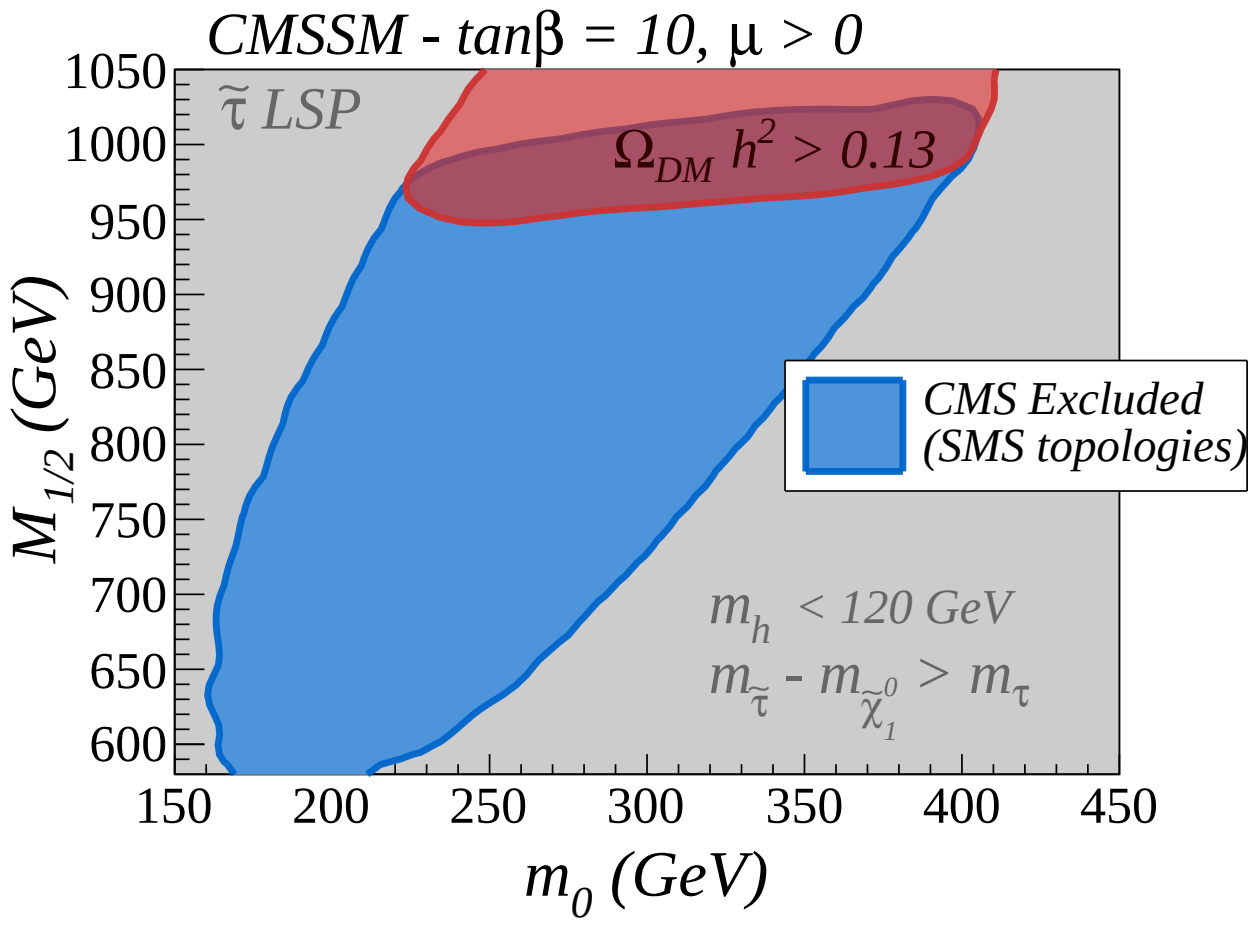

Figure 6.4: Same as Figure 6.4, but using all the simplified models listed in Table 6.1.

Figure 6.4 illustrates the feasibility of using simplified model efficiencies to constrain full models. This approach has the advantage of being computationally inexpensive (once the efficiencies are known) and can be used to quickly test a large number of model points. However the approach relies on a few approximations and is never fully inclusive, since the number of SMS topologies with published efficiencies is always limited. Hence, it is important to verify how close the simplified model reinterpretation comes to the full recasting using a Monte Carlo simulation. In Ref. [90] it was shown that, within the CMSSM scenario discussed above and using eight simplified model topologies, the SMS results reproduce the full simulation within 20% or better. Since this error is of the order of the uncertainties in recasting, the use of simplified models becomes a viable alternative to full recasting. The SMS reinterpreta-tion is even more relevant for the cases where a straightforward MC recasting is not possible.

**Lessons Learned**

The example discussed in Section 6.3.2 illustrates how simplified models for LLPs can be used as a reinterpretation tool. One important point which becomes clear once we compare Figures 6.3 and 6.4 is the importance of a sufficiently inclusive database of simplified model efficiencies. In particular, depending on the full BSM scenario considered, the minimal set of simplified models proposed in Chapter 2 and Appendix A may not be sufficient to allow for a reinterpretation based on simplified model results alone. The reason is that this set was derived as the minimal set to generate a collection of relatively inclusive signatures, but may not correctly model the efficiency of every UV theory leading to that signature. In these cases, results for additional simplified model topologies are necessary and can be provided by the experimental collaborations or generated by theory groups if a full recasting is available. Furthermore, since LLP searches can be very inclusive, the simplified models considered can also be defined inclusively, as discussed above. In this way a limited number of simplified models can cover a large number of event topologies, thus increasing the SMS coverage of full models.

## 6.4 *Recasting Examples for Specific Searches*

Here we provide examples of recasting specific experimental searches for several LLP signatures: searches for heavy stable charge particles (HSCPs), displaced leptons, displaced jets, displaced lepton-jets (LJs), non-pointing photons, and displaced vertices (DV). These recasting attempts have been made outside the experimental collaborations, making use of the public information provided by the experimental note or publication. The aim here is to highlight the challenges faced when recasting LLP searches and also to highlight the cases where the experimental information provided is straightforward and useful for recasting.

### 6.4.1 *Heavy Stable Charged Particles (HSCPs)*

Searches for HSCPs are based on the signature of highly ionizing tracks and/or an anomalous time of flight between the particle's production at the interaction point and its arrival in the muon detector [385] (see Section 3.5.1 for more details). Both signatures are sensitive to the particle's velocity and exploit the production of HSCPs outside of the ultra-relativistic regime, allowing for a powerful discrimination against the highly boosted Standard Model backgrounds. HSCP searches assume particles are sufficiently long-lived to traverse the entire detector. They have been interpreted for HSCPs that are purely electrically charged or carry color charge, the latter of which hadronize to form $R$-hadrons as they propagate through the detector [10]. Typically, the HSCP signature yields high sensitivities providing a very strong background rejection while still allowing for large signal efficiencies. As a consequence, search strategies for new physics models with HSCPs typically do not benefit from more model-dependent selection criteria, like requiring additional particles in the event [88]. The corresponding searches can, hence, be performed in a mostly inclusive manner concentrating on the HSCP candidate itself. This fact opens up the possibility of providing a widely applicable recasting based on signature efficiencies. This approach has been followed by the CMS Collaboration [493], which has provided probabilities for HSCP candidates to pass the on- and off-line selection criteria for Run 1 as a function of the relevant kinematical parameters.

In this section we describe the recasting of the 8 TeV CMS search for HSCPs and discuss its validation and applicability. Furthermore, we comment on the attempt to extrapolate the 8 TeV signature efficiencies to the corresponding 13 TeV analysis, for which the corresponding efficiencies have not been provided by CMS.

#### 6.4.1.1 *Recasting Using Signature Efficiencies*

Ref. [493] provides efficiencies for the reconstruction and selection of HSCP candidates with $|Q| = 1$ in the form of on- and off-line probabilities, $P_{\rm on}(\boldsymbol{k})$ and $P_{\rm off}(\boldsymbol{k})$. These are given as a function of the truth-level kinematic variables velocity ($\beta$), pseudo-rapidity ($\eta$) and transverse momentum ($p_{\rm T}$) of isolated HSCP candidates, so the vector $\boldsymbol{k}$ is defined as: $\boldsymbol{k} = (\beta, \eta, p_{\rm T})$. The on- and off-line probabilities must be applied to isolated HSCP candidates, which are required to fulfill

$$\left( \sum_{i}^{\substack{\text{charged} \\ \Delta R<0.3}} p_{\rm T}^{i} \right) < 50\ \text{GeV}\,, \qquad \left( \sum_{i}^{\substack{\text{visible} \\ \Delta R<0.3}} \frac{E^i}{|\boldsymbol{p}|} \right) < 0.3\,, \tag{6.2}$$

where the sums include all charged and visible particles, respectively, within a radius of $\Delta R = \sqrt{\Delta\eta^2 + \Delta\phi^2} < 0.3$ around the HSCP candidate track, $p_{\rm T}^i$ denotes their transverse momenta and $E^i$ their energy. Muons are not counted as visible particles and the HSCP itself is not included in either sum.

If an event contains one or more HSCPs satisfying the above isolation criteria, the efficiency for the event to pass the analysis selection is given by:

$$\epsilon = \epsilon_{\rm on} \times \epsilon_{\rm off}\,. \tag{6.3}$$

For an event with one HSCP candidate $\epsilon_{\rm on/off}$ is directly given by the signature efficiencies $P_{\rm on/off}(\boldsymbol{k})$. For an event with two candidates, the efficiency reads [493]

$$\epsilon_{\rm on/off} = P_{\rm on/off}(\boldsymbol{k}^1) + P_{\rm on/off}(\boldsymbol{k}^2) - P_{\rm on/off}(\boldsymbol{k}^1) P_{\rm on/off}(\boldsymbol{k}^2)\,, \tag{6.4}$$

where $\boldsymbol{k}^{1,2}$ are the kinematical vectors of the two HSCPs in the given event. Therefore the on- and off-line probabilities combined with the isolation criteria allow for the complete recasting of the HSCP search using only truth-level events generated in MC.

The recasting of the 8 TeV search was performed in Ref. [90] using the procedure described above. Events were simulated using Pythia 6 [499] and the total signal efficiency for a given model was then computed using:

$$\epsilon = \frac{1}{N} \sum_{i=1}^{N} \epsilon_i \ .$$

where the sum runs over all the ($N$) generated events and $\epsilon_i$ is the efficiency for each event computed using Eq. (6.3). Since the probabilities $P_{\text{on/off}}(\boldsymbol{k})$ are given for four distinct cuts on the reconstructed HSCP mass ($m_{\text{rec}}$), these were considered as four different signal regions. The number of observed events and the expected background for which of these cuts are reported in Ref. [493].

### 6.4.1.2 *Validation and Applicability*

A validation of the method described above was done in Ref. [90] using the same gauge-mediated supersymmetry breaking (GMSB) model considered by CMS [493]. This supersymmetric model features a gravitino and a long-lived stau as the lightest and next-to-lightest supersymmetric particle, respectively. Since the stau only decays outside the detector volume, all cascade decays of the produced sparticles terminate in the lightest stau, which provides the HSCP signature. The left pane of Figure 6.5 compares the resulting signal efficiency obtained by the recasting and the full CMS detector simulation. The signal efficiencies agree within 3%, demonstrating that the recast is an excellent approximation to the full CMS simulation. The differences are of the order of the statistical uncertainties from the MC simulation of the signal. In the right pane of Figure 6.5, we show the 95% C.L. upper limits on the inclusive production cross sections, which, again, agree (within $\sim$ 3%) with the ones obtained by the full simulation in Ref. [493]. Note that both limits are based on the discrete mass cuts on $m_{\text{rec}}$ mentioned above. In the full CMS analysis [299], an event-based mass cut is used, resulting in slightly stronger constraints for some HSCP masses.

Due to the inclusive nature of the search, the above recasting provides a widely applicable and highly reliable way to reinterpret the HSCP search for arbitrary models containing detector-stable HSCPs. Accordingly it has been used in a variety of phenomenological studies. For instance, it has been used for reinterpretations of supersymmetric models [89, 198, 500, 501] and non-supersymmetric models of very weakly interacting dark matter [135, 141]. In Refs. [89, 135], the recasting has been used to reinterpret the HSCP search for finite lifetimes by convolving the signature efficiency with the fraction of HSCPs that decay after traversing the relevant parts of the detector. The recasting has also been used for a reinterpretation in terms of simplified models, as discussed in Section 6.3.2.

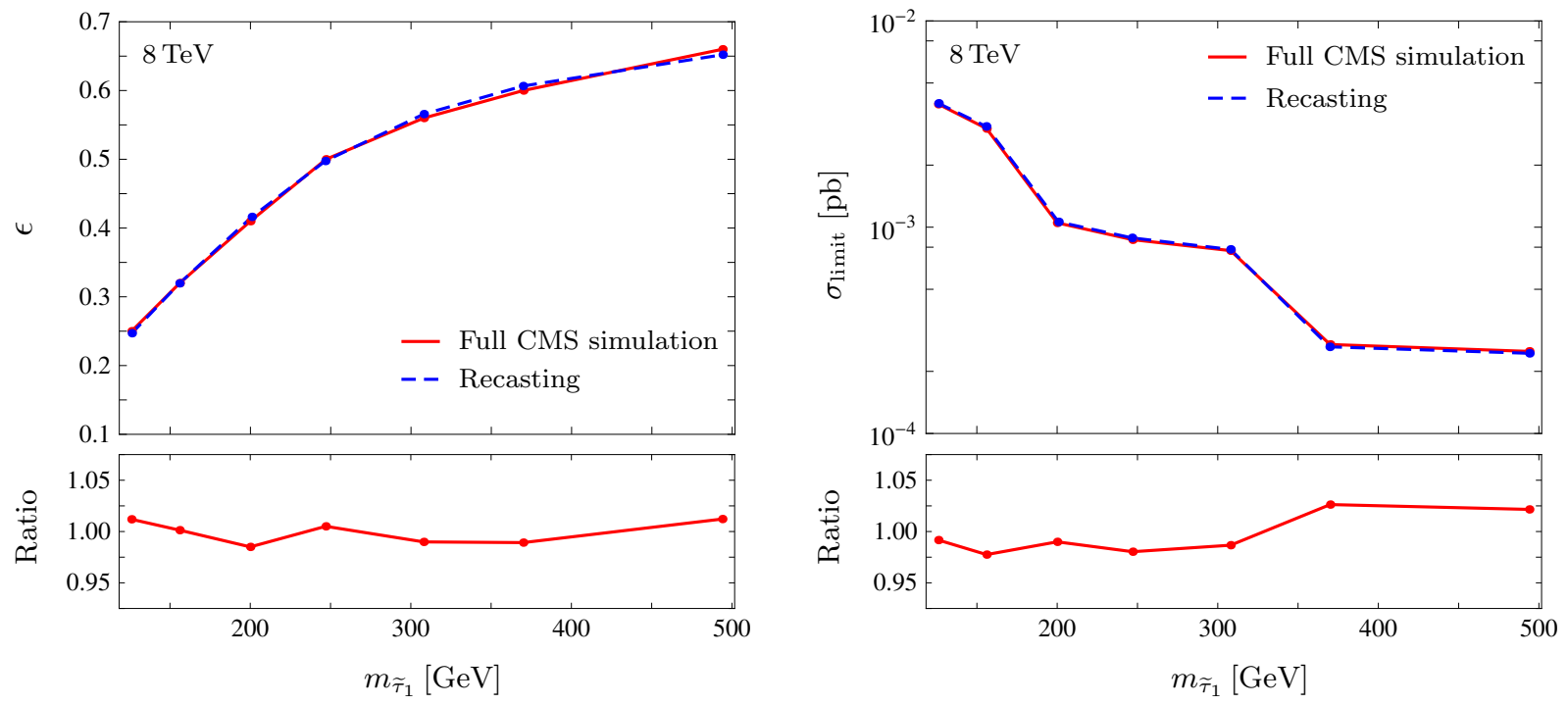

Figure 6.5: **Left:** Signal efficiency $\epsilon$ of the HSCP search. **Right:** 95% C.L. cross section upper limit for the GMSB model as the function of the stau mass. We compare the CMS analysis [493] from the full detector simulation (red solid lines) with the recasting using signature efficiencies (blue dashed lines). In the lower frames we show the respective ratios $\epsilon^{\text{Full}}/\epsilon^{\text{Recast}}$, $\sigma_{\text{limit}}^{\text{Full}}/\sigma_{\text{limit}}^{\text{Recast}}$. Taken from Ref. [90].

#### 6.4.1.3 *Extrapolation to 13 TeV*

While the CMS search for HSCPs at 8 TeV has provided the signature efficiencies discussed above, the same is not true for the 13 TeV analysis [4]. Therefore a straightforward recasting of the Run 2 search is not possible. Nonetheless, since the 8 TeV CMS search has proven to be extremely useful in constraining models with long-lived charged particles, it would be desirable to recast the 13 TeV analysis as well. In the following we discuss an attempt [502] to obtain a similar recasting for the corresponding HSCP search at 13 TeV. Our aim is to extrapolate the public 8 TeV efficiencies to Run 2 by introducing a correction function $F$ that accounts for the differences between both runs:

$$P_{\text{off}}^{13\text{ TeV}}(\boldsymbol{k}) = F(\beta) \times P_{\text{off}}^{8\text{ TeV}}(\boldsymbol{k}) \,, \tag{6.5}$$

where we have assumed that the correction function is mainly dependent on the HSCP velocity. If $F(\beta)$ is sufficient to account for the difference between both runs and can be computed, we can directly obtain $P_{\text{off}}^{13\text{ TeV}}$ and, using the procedure described in Section 6.4.1.1, recast the 13 TeV analysis.

In order to compute the correction function $F(\beta)$, we use the total signal efficiencies reported by the 13 TeV CMS analysis [4] for direct production of long-lived staus. Since the signal efficiencies have been provided for six distinct values of the stau mass, we perform a fit of $F$ to the efficiencies reported. We chose to parametrize the correction function $F(\beta)$ by eight parameters ($C_i$). Using MadGraph5_aMC@NLO [219] and Pythia 6 [499], we obtain generator-level events for each of the stau mass points at 13 TeV. Then, comparing the total signal efficiencies obtained for a given set $C_i$ to the efficiencies reported in Ref. [4], we can determine the best-fit values for the $C_i$ parameters and consequently the best-fit for the correction function ($F_{\text{best-fit}}$). The result of the best-fit function and its $1\sigma$ uncertatinty is shown in Figure 6.6. The deviation of $F$ from 1 implies a decrease or increase of the respective detector and signal efficiency between the 8 and 13 TeV analyses. Figure 6.6 also shows that the function is loosely constrained for low values of $\beta$.

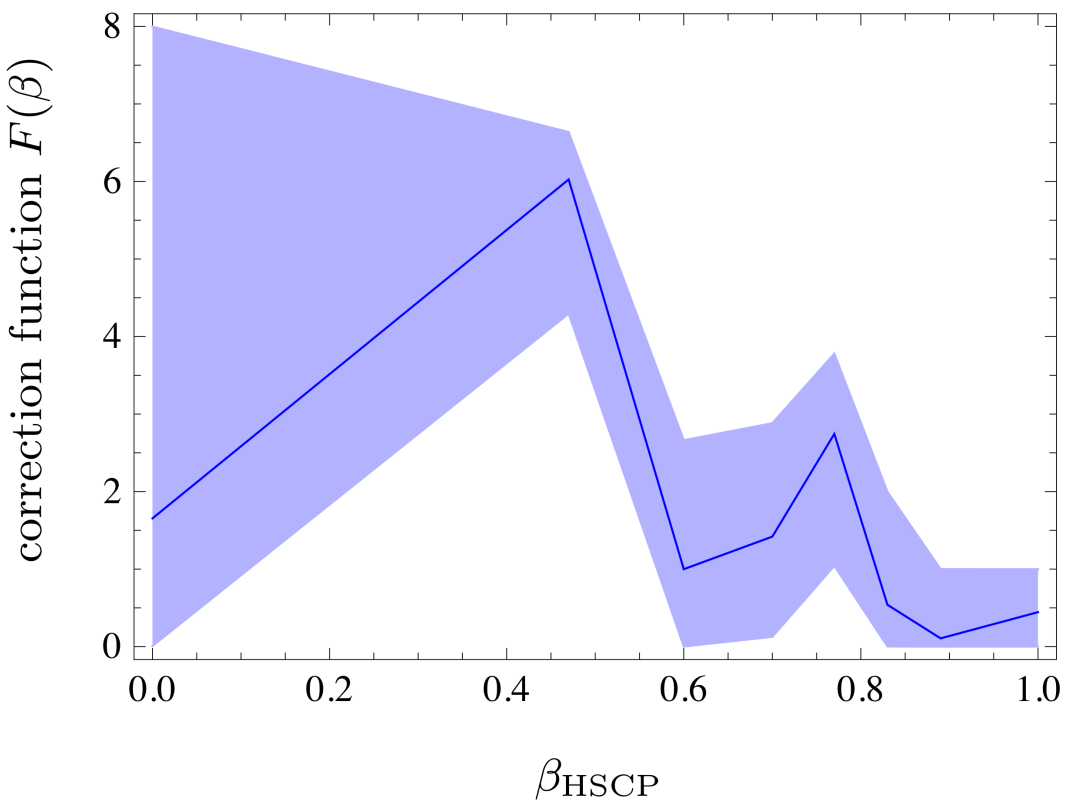

Figure 6.6: Best-fit correction function $F(\beta)$ and its $\pm 1\sigma$ band.

| $m_{\rm HSCP}$ [GeV] | direct production $\epsilon$(CMS) | $\epsilon(F_{\text{best-fit}})$ | $\epsilon(F=1)$ |
|---|---|---|---|
| 200 | 0.235 | 0.232 | 0.259 |
| 308 | 0.294 | 0.298 | 0.346 |
| 494 | 0.387 | 0.384 | 0.452 |
| 651 | 0.450 | 0.448 | 0.503 |
| 1029 | 0.497 | 0.501 | 0.466 |
| 1599 | 0.428 | 0.429 | 0.225 |

Table 6.2: Efficiencies for the 13 TeV LHC for the six benchmark masses in the direct stau production scenario reported by CMS (second column) and obtained through recasting using $F_{\text{best-fit}}$ and without the inclusion of the correction function ($F = 1$).

In order to verify the validity of the extrapolation to 13 TeV, we use $F_{\text{best-fit}}$ and Eq. (6.5) to compute the total signal efficiencies for the same six benchmark points used in the fit. A comparison between the results obtained through recasting and the efficiencies reported by CMS is shown by the second and third columns in Table 6.2. The results reproduce the CMS values well within the expected uncertainties, thus validating the fitting procedure. Furthermore, the inclusion of the correction function significantly improves the agreement with respect to the direct extrapolation of the 8 TeV efficiencies ($F = 1$), as shown by the fourth column in Table 6.2.

The high level of agreement obtained with $F_{\text{best-fit}}$ for the direct stau benchmark points is expected, since these points were used in order to fit the correction function. Therefore an independent test of the above fit must be performed in order to properly validate the recasting of the 13 TeV analysis. Fortunately CMS has also reported efficiencies for a second scenario, the GMSB model with long lived staus. This scenario not only contains direct stau production, but also includes production through cascade decays of heavier sparticles. Since the GMSB model produces distinct event topologies, it provides a good test for the validity of the recasting procedure.

| | GMSB | | |
|---|---|---|---|
| $m_{\text{HSCP}}$ [GeV] | $\epsilon$(CMS) | $\epsilon(F_{\text{best-fit}})$ | $\epsilon(F = 1)$ |
| 200 | 0.276 | 0.297 | 0.279 |
| 308 | 0.429 | 0.401 | 0.423 |
| 494 | 0.569 | 0.494 | 0.556 |
| 651 | 0.628 | 0.524 | 0.580 |
| 1029 | 0.665 | 0.538 | 0.493 |
| 1599 | 0.481 | 0.442 | 0.228 |

Table 6.3: Efficiencies for the 13 TeV LHC for the six GMSB model benchmark points reported by CMS (second column) and obtained through recasting using $F_{\text{best-fit}}$ and without the inclusion of the correction function ($F = 1$).

The results for the six GMSB bechmark points considered in Ref. [4] are shown in Table 6.3. As we can see, they deviate from the CMS values by up to 20% for large stau masses, where our estimate undershoots the CMS efficiencies. Although the overall agreement is improved by the correction function, the result is not entirely satisfactory, given that the uncertainties for the 8 TeV recasting were under 5% (see Figure 6.5). The observed difference might arise from several shortcomings in our description.

In particular, we assume $F$ to only depend on $\beta$ whereas the full probability maps are parametrized in the three kinematic variables $\beta, \eta$ and $p_{\text{T}}$. However, assuming a dependence of all three kinematic variables is clearly not feasible given the very limited amount of information provided by the 13 TeV CMS analysis. Therefore we conclude that it is not possible to extrapolate the 8 TeV efficiencies in a straighforward way without additional information from the experimental collaboration.

**Lessons Learned**

The prominent signature of HSCPs allows for a mostly inclusive search strategy concentrating on the HSCP track itself. Hence, searches for HSCPs can be reinterpreted using signature efficiencies in a widely applicable and highly reliable way. This possibility has been followed by the CMS Collaboration providing signature efficiency maps for the 8 TeV LHC. The validation reveals an excellent performance. The recast has been successfully used in the literature.

The signature efficiencies for 8 TeV can also be used to estimate the ones for Run 2 by applying a multiplicative correction function. While such an extrapolation introduces some level of approximation, a better knowledge of the underlying changes between both runs might reduce the uncertainties.

| **Cut Summary of CMS displaced $e\mu$** |
|---|
| **Preselection** |
| 1 OS $e^{\pm}\mu^{\mp}$ pair |
| $\lvert d_0\rvert_\ell > 100\,\mu$m |
| $p_{T,\ell} > 25$ GeV, $\lvert\eta_\ell\rvert < 2.5$ |
| Reject $1.44 < \lvert\eta_e\rvert < 1.56$ |
| $I^{calo,e}_{\Delta R<0.3} < 0.10$, $I^{calo,\mu}_{\Delta R<0.4} < 0.12$ |
| $\Delta R_{\ell j} > 0.5\ \forall$ jets with $p_T > 10$ GeV |
| $\Delta R_{e\mu} > 0.5$ |
| $v_{T,\tilde{\ell}} < 4$ cm, $v_{Z,\tilde{\ell}} < 30$ cm |
| Veto additional leptons |

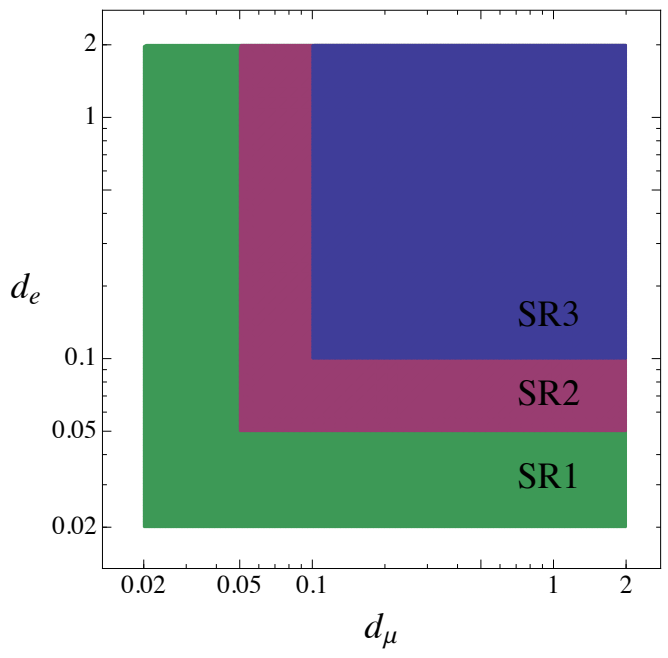

Table 6.4: **Left:** Preselection cuts in Ref. [268] (see also [503, 504]). **Right:** The transverse impact parameter bins that define the exclusive signal regions. Table and figure taken from Ref. [198].

### 6.4.2 *Displaced Leptons*

Searching for displaced leptons by requiring the leptons to have large impact parameters with respect to the primary vertex is a very clean strategy due to the low backgrounds, and such searches are usually very straightforward to recast. The CMS displaced $e\mu$ search [268] demands two oppositely charged, different flavour $(e, \mu)$ leptons with large impact parameters (see also Section 3.2). The recast is fairly straightforward to do, and the biggest difficulty in doing so is locating all of the relevant information as it is not all provided within the main document. The "standard" isolation requirements used in the search can be found in an earlier version of the search [503]. The necessary cuts on the displaced decay position $(v_T, v_Z)$ as well as the selection (as a function of $p_T$), reconstruction (as a function of impact parameter, $d_0$) and trigger efficiencies can be found on an additional website [504] containing auxiliary information for recasting. Although all of this information is excellent and greatly facilitates recasting the search, it is a challenge to collect the relevant information due to the fact that the additional material is not referenced in the document.

The benchmark model used in this search is the direct pair production of stops that decay through small lepton-flavor-universal RPV $\lambda'_{ijk}L_iQ_jD^c_k$ couplings ($\lambda'_{133} = \lambda'_{233} = \lambda'_{333}$) to yield displaced $\tilde{t} \to eb$, $\mu b$, and $\tau b$ decays. The signal is simple to generate, where the only challenge is in handling the displacement properly. The most identifying pre-selection requirement of this search is that the transverse impact parameter, $|d_0|$, is required to be larger than 100 $\mu$m for both the electron and muon. The impact parameter is not the point where the parent object (e.g., $\tau$ or $b$) decays, i.e., the $v$ mentioned above, but rather the distance to the point of closest approach of the lepton's track relative to the center of the beampipe in the transverse plane. Backgrounds in this search from $Z \to \tau\tau$ or heavy flavor tend to result in leptons that are nearly collinear with the parent due to the small mass-to-momentum ratio, and yield a small impact parameter even for decays well on the lifetime tail of the parent. Events are binned across three exclusive signal regions: SR3, where both leptons have transverse impact parameters $|d_0|$ between 0.1 and 2.0 cm; SR2, with $|d_0|$ of both leptons between 0.05 and 2.0 cm, but not satisfying the requirement of SR3; and SR1, with $|d_0|$ between 0.02 and 2.0 cm, but not within SR2 or SR3. All selection requirements are summarized in Table 6.4.

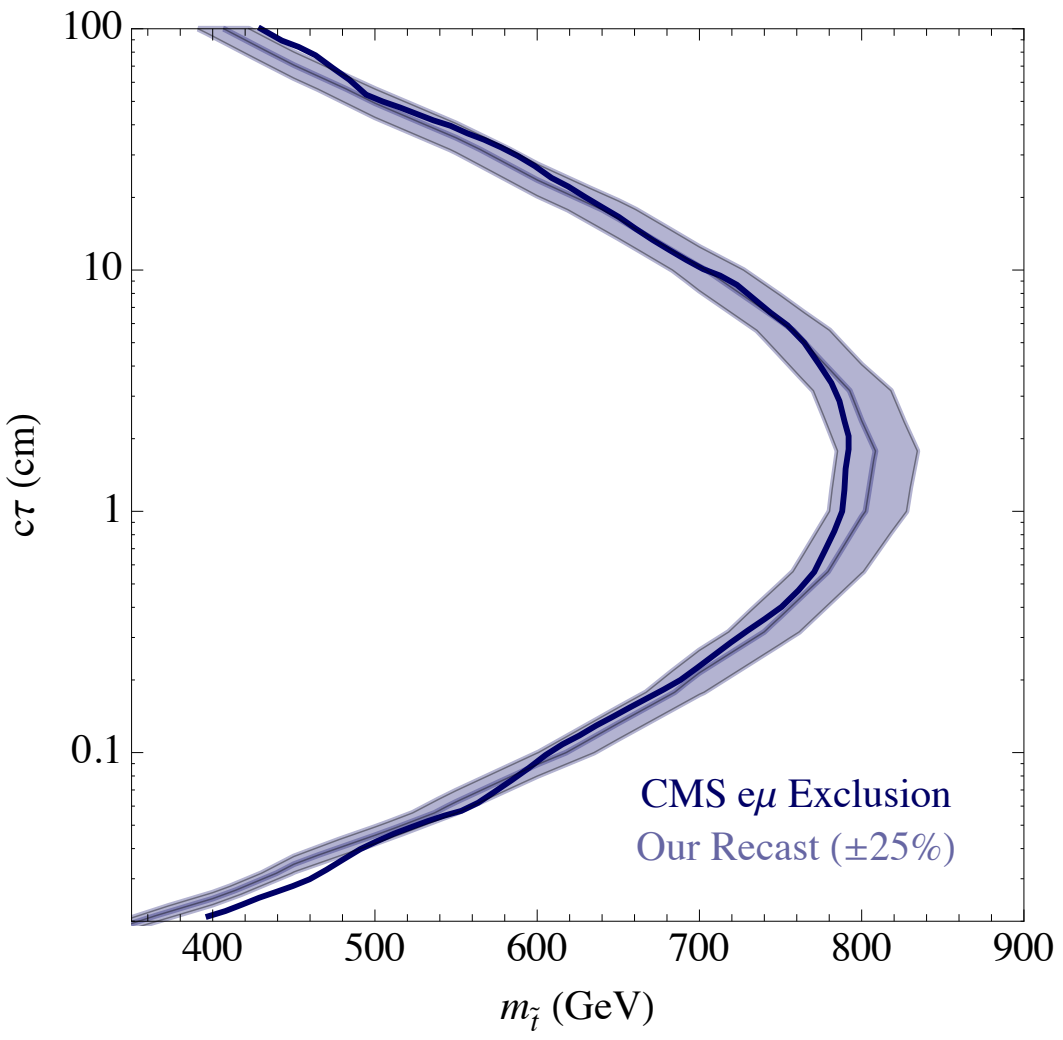

Figure 6.7: Validation of the CMS displaced $e\mu$ search [268] for the displaced supersymmetry benchmark model [505]. Figure taken from Ref.[198].

In Figure 6.7, we present the validation of the CMS displaced $e\mu$ search [268] from the study performed in Ref. [198]. For this search, we show the recommended 25% modeling uncertainty [9]. The recast agrees very well with the results from the CMS displaced $e\mu$ search in the region of highest sensitivity, 300 $\mu$m $\lesssim c\tau \lesssim$ 50 cm, but exhibits a moderate deviation on the tails. As this extremely low efficiency region is overly sensitive to the tails of kinematic distributions, it may be the case that the sensitivity is slightly underestimated for lifetimes near 1 m or 100 $\mu$m, but this discrepancy typically has no qualitative impact on any application of the results.

[9] A similar validation was done in Ref. [89] with details provided in its Appendix D. After applying a flat 80% efficiency, a 20% modeling uncertainty is found to be appropriate, as shown in Figure 14 of the reference

#### 6.4.2.1 *Extrapolation to 13 TeV*

We now show another reinterpretation example of the CMS displaced $e\mu$ in order to highlight the comparison between 8 TeV [268] and 13 TeV [190] analyses. We compare in Figure 6.8 our reproduction of expected signal events with the published validation material for the 8 TeV version, and the partially-available validation material for the 13 TeV search. Information on efficiency maps from the 8 TeV analysis was needed to obtain an extrapolation to 13 TeV, as the 13 TeV efficiency maps are not yet public. As we can see, the 8 TeV recast for the CMS displaced lepton search [268] agrees very well in the region of highest sensitivity. The 13 TeV recast-ing, however, under-estimates the CMS values by a factor of two or more. This is likely due to the fact that the lepton efficiencies can not be directly extrapolated from 8 to 13 TeV, as assumed in making Figure 6.8. Also, with the absence of a cut-flow table, it is impossible to verify where the mis-match arises, whether it is due to mis-modeling of the signal region cuts or due to genuine changes in the efficiencies.

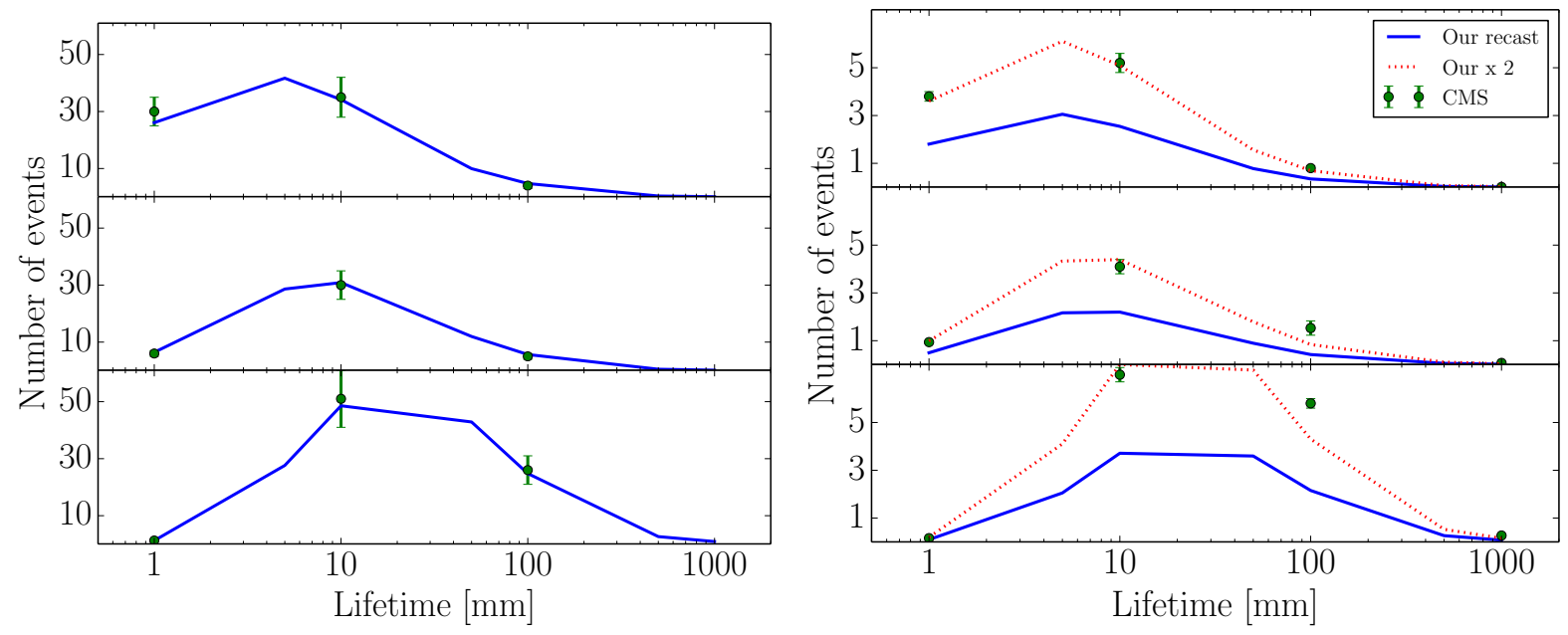

Figure 6.8: Number of expected events in the signal regions defined based on $|d_0|$ for the CMS displaced $e\mu$ search. The green points refer to the expected signal published by the analysis. **Left:** Validation for 8 TeV analysis. Production cross section assumed NLO+NLL value 85.6 fb for $M_{\tilde{t}_1} = 500$ GeV, BR = 0.33 in each $\ell$-channel. **Right:** Validation for 13 TeV analysis. Production cross section assumed NLO+NLL value 67.0 fb for $M_{\tilde{t}_1} = 700$ GeV. The 13 TeV numbers are made using efficiency maps published for the 8 TeV search, as the 13 TeV maps are not yet public. Figures taken from Ref. [502].

**Lessons Learned**

The selection and trigger efficiencies provided by CMS are very useful for recasting the 8 TeV CMS search for displaced leptons [268] and allow for a very good level of agreement. The main challenge, however, consisted in collecting all the available information, which was not provided by the main document in Ref. [268]. Furthermore, the corresponding information for the 13 TeV search is not publicly available and an extrapolation of the 8 TeV efficiencies was shown to be inadequate.

### *6.4.3 Displaced Jets*

Searches for displaced jets are less straightforward to reinterpret than displaced leptons. Interest in accurate reinterpretation is increasing, as many new physics models give rise to this particular signature. The CMS search for displaced di-jets [506] was reinterpreted in Ref. [26] to explore long-lived particle signatures for certain weak-scale models of baryogenesis [24, 26, 507], as well as a study [89] to understand current limits on long-lived signatures in supersymmetry.

The CMS search [26] uses a multivariate discriminant composed of observables that are challenging to model in Monte Carlo, such as the track-multiplicity of the vertex and the root-mean-square of a cluster track-multiplicity variable. The reinterpretation approach in Refs. [26, 89] was to construct track information at truth level based on the output of a parton shower program (such as PYTHIA 8), and then use the truth-level information to construct the various vertex, cluster, and track-level observables for each event. As it is difficult to adequately account for inefficiencies of track and vertex reconstruction, the efficiency of passing the cuts with truth-level observables was considered and then it was normalized to the results from CMS. To do so, the authors of Ref. [26] simulated identical signal models to those with efficiencies reported by the CMS collaboration, assumed that the MC truth-level reconstruction gave an adequate description of kinematics but *not* track and vertex reconstruction, and so computed a ratio of truth-level efficiencies to those reported by CMS. The resulting efficiency ratios were used to re-scale the truth-level results of other models, leading to a reinterpretation of the CMS search for different models beyond the ones they considered. The details can be found in Ref. [26]; a more sophisticated approach in which object efficiencies were estimated and applied to tracks and displaced vertices in Ref. [89].

To validate this approach, truth-level quantities for the models constrained in Ref. [506] were computed and compared to the numbers and distributions reported. For example, comparisons of the distributions of the observables going into the multivariate reinterpretation discriminant, as well as the output of the multivariate discriminant itself, could be performed. While the truth-level distributions disagreed with those of CMS for individual observables, the actual multivariate discriminant output agreed with that of CMS at better than 25%. The ratio of truth-level efficiencies to CMS efficiencies are also compared for different LLP masses and kinematics, and these typically agree with one another at the factor-of-two level [26]. This suggests that this reinterpretation of the CMS results in terms of cross-section limits is likely accurate to the factor-of-two level.

**Lessons Learned**

We find that a rather naïve truth-level reconstruction of the event could give a reinterpretation of cross-section limits to agree within a factor of two, provided the efficiencies were normalized to the experimental values using an overlapping set of signal models. One of the major obstructions to improving the accuracy of the estimate was the model dependence observed among the ratios of efficiencies between truth-level information and CMS. For instance, it was found that highly boosted models showed a much lower relative reconstruction rate in data vs. truth-level MC than less-boosted LLPs. Since pair production of LLPs was considered near threshold in Ref. [26], this degradation in the performance of highly boosted LLPs does not greatly affect the confidence in the final result. However, it does suggest that characterizing the effects of the particle boost are important for reinterpretation.

In addition, with a larger and more diverse set of signal benchmarks available, the prospects for the reinterpretation of search results are better. The reasons are twofold:

- Increasing the number of presented signal models by the collaboration allows for more cross-checks between MC and the results in data. This allows for more sophisticated tuning of efficiencies as applied to truth-level events;
- Having a more diverse set of benchmark signal models means that it is easier to disentangle various kinematic effects on the efficiency (such as the LLP mass, boost, etc.) and find a signal benchmark that most closely matches the model for which one wants to derive sensitivity.

### *6.4.4 Displaced Lepton-Jets*

A variety of scenarios predict the existence of LLPs decaying to a pair of highly collimated leptons, also known as lepton-jets (LJs) [120]. Models giving rise to LJ signatures include heavy right-handed neutrinos, exotic Higgs decays, and dark gauge bosons. The relevant signature is one or more LJs emerging from a DV. In many cases, there can also be associated prompt objects, such as a prompt lepton produced in conjunction with the right-handed neutrino (corresponding to charged-current production in the simplified model framework of Section 2.4).

*Existing Searches* Current search results are outlined in Section 3.2. The search in Ref. [266] was interpreted in the framework of the Falkowski-Ruderman-Volanksy-Zupan (FRVZ) model [148] for the Higgs boson interacting with a hidden sector containing a massive dark photon ($\gamma_d$), massive neutralinos, and three hidden scalars. Displaced LJs are produced at the end of Higgs cascade decay that also yields two hidden light stable particles (HLSPs). Depending on the hidden-sector spectrum, the cascade decay may yield two or more $\gamma_d$ that each decay to pairs of Standard Model charged particles via kinetic mixing with the hypercharge gauge boson. The $\gamma_d$ decay products are highly collimated. For $m_{\gamma_d} < 500$ MeV, LJs are the dominant decay mode, while for larger dark photon masses, displaced hadronic jets can also be significant.

Results are presented as limits on $\sigma \times \text{BR}(H \to n\gamma_d + X)$ ($n = 2$, 4) as a function of the $\gamma_d$ $c\tau$. The strongest limits from the 8 TeV dataset arise from events that require at least one LJ with muons. For $n = 2$, a $\sigma \times \text{BR}(H \to 2\gamma_d + X)$ of $\gtrsim 1$ pb is excluded for $c\tau \sim 50$ mm.

*Recast: Dark Photon with Non-Abelian Kinetic Mixing* The FRVZ model on which the ATLAS analysis was based implies the pres-

ence of at least two displaced LJs in the final state as well as two additional unobserved HLSPs that corresponde to missing energy. The ATLAS analysis did not impose any $\not{E}_T$ cuts. The presence of the two HLSPs affects the kinematics and topology of the signal event but does not explicitly enter the event selection or reconstruction. Thus, one should be able to reinterpret the ATLAS analysis in terms of any other model containing two or more displaced LJ, along with potentially additional, unobserved objects.

We consider a scenario for a dark photon $X_\mu$ that mixes with the neutral Standard Model SU(2)$_L$ gauge boson $W^3_\mu$ via a higher dimensional operator. Two possibilities have recently been considered: a dimension six operator involving the SM Higgs doublet, the SU(2)$_L$ field strength $W^a_{\mu\nu}$ and the corresponding U(1)$'$ field strength $X_{\mu\nu}$ [508]; and a dimension five operator involving $W^a_{\mu\nu}$, $X_{\mu\nu}$, and a real scalar triplet $\Sigma \sim (1, 3, 0)$ [509]. We consider the latter since it can yield an event topology similar to that of the FRVZ model and since it is the leading operator that may generate non-abelian kinetic mixing (NAKM) of the U(1)$'$ gauge boson with the SM. We will henceforth refer to this model as the NAKM5 scenario. The corresponding operator is

$$\mathcal{O}^{(5)}_{WX} = -\frac{\beta}{\Lambda}\,\mathrm{Tr}\left(W_{\mu\nu}\Sigma\right) X^{\mu\nu} \tag{6.6}$$

where $\Lambda$ is the associated effective field theory mass scale and $\beta$ is a dimensionless coupling that is nominally $\mathcal{O}(1)$. When the neutral component of the $\Sigma$ obtains a vacuum expectation value (vev) $v_\Sigma$, one has the ratio of the dark photon and SM photon couplings

$$\epsilon = \beta \sin\theta_W \left(\frac{v_\Sigma}{\Lambda}\right) \quad . \tag{6.7}$$

Note that the $\rho$-parameter constrains $v_\Sigma$ to be smaller than about 4 GeV. Consequently, for $\Lambda$ of order one TeV, $\epsilon$ is small (and consistent with present experimental constraints) without requiring the presence of a suppressed coupling in the Lagrangian. This feature distinguishes the dimension-five non-abelian kinetic mixing from the dimension-four kinetic mixing between $X_\mu$ and the SM hypercharge gauge boson.

The final state with two $\gamma_d$ (resulting from the $X_\mu$-$W^3_\mu$ mixing) can be produced in one of two ways: (a) electroweak Drell-Yan pair production of triplet scalars that subsequently decay to a $\gamma_d$ and SM gauge boson via the operator in Eq. (6.6); (b) production of the $\gamma_d$ and a triplet scalar via $\mathcal{O}^{(5)}_{WX}$ with a subsequent decay of the triplet to a second $\gamma_d$ plus a SM gauge boson via the same operator. In each case, one would expect the presence of two $\gamma_d$ plus one or more unobserved massive prompt bosons in the final state. The ATLAS DV plus LJ analysis can be recast in a straightforward way for these event topologies, as no information about the unobserved prompt object has been used. It is worth emphasizing that both the production processes as well as the $\Sigma$ decay rate are independent of the triplet vev, $v_\Sigma$. The latter only enters $\epsilon$ and, thus, only affects the dark photon $c\tau$.

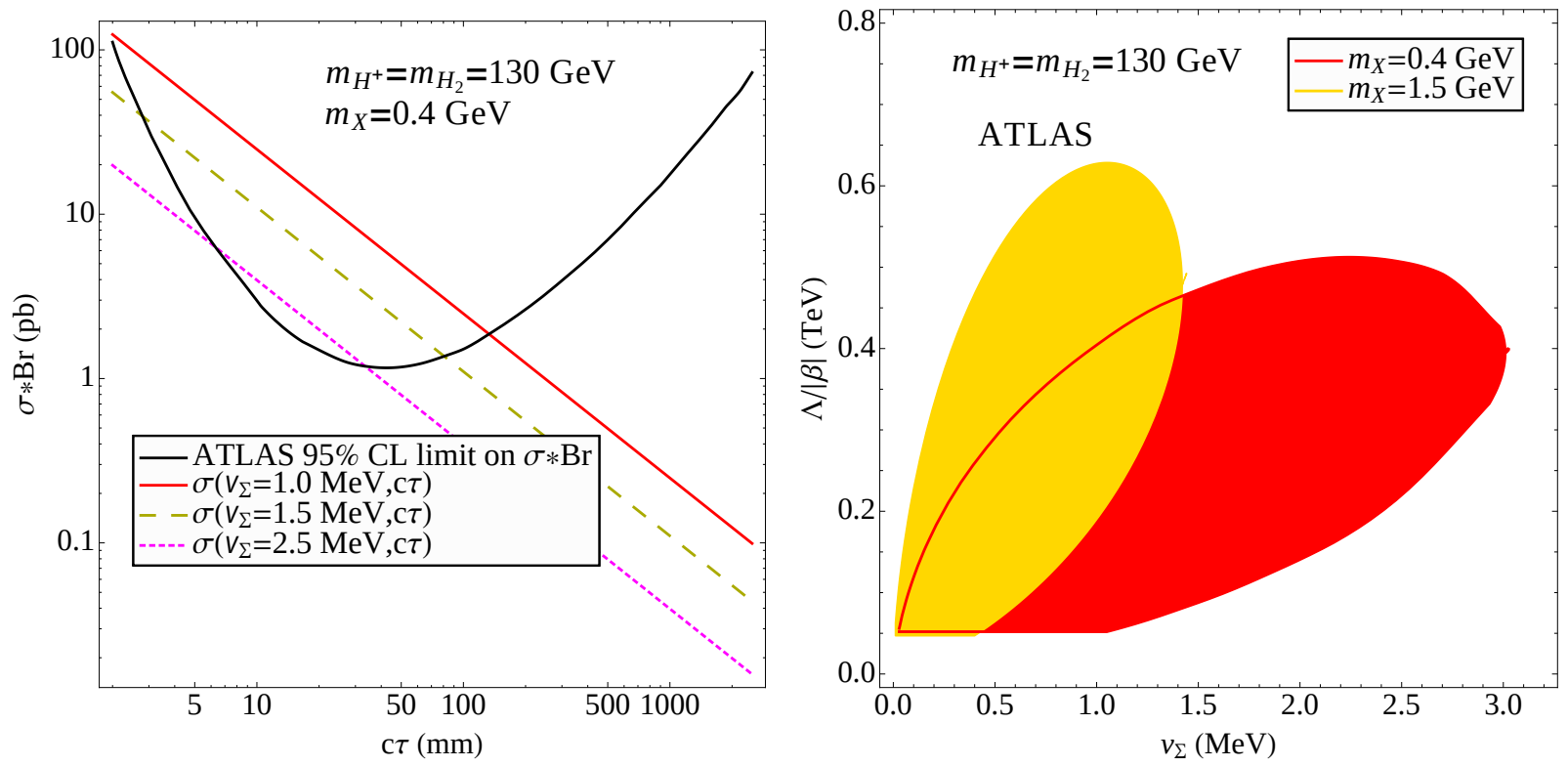

Figure 6.9: Constraints on the NAKM5 scenario, recast from the ATLAS search reported in Ref. [265]. The left panel gives the exclusion in the ($c\tau$, $\sigma \times$ BR) plane, where the region above the parabola is excluded. The diagonal lines indicate the dependence of $\sigma \times$ BR on $c\tau$ for different representative choices of $v_\Sigma$. The right panel gives the exclusion region in the($v_\Sigma$, $\Lambda/\beta$) plane for $m_X = 0.4$ GeV (red region) and $m_X = 1.5$ GeV (yellow region).

The corresponding implications of the ATLAS 8 TeV results are indicated in Figure 6.9. The first panel shows the limits on the $\sigma \times$ Br as a function of $c\tau$. The corresponding model sensitivity is shown for different choices of $v_\Sigma$ by the diagonal lines. Note that for fixed $v_\Sigma$ both the $\sigma \times$ Br and $c\tau$ depend on the operator coefficient $\beta/\Lambda$, leading to a non-trivial relationship between the two experimental quantities. This situation differs from the FRVZ model, where $\sigma \times$ Br is independent of $c\tau$ since the mixing parameter $\epsilon$ does not depend on the parameters governing production of the hidden sector particles or their cascade decays. The intersections of the model lines with the ATLAS limits can then be translated into bounds on $\Lambda/\beta$ as a function of $v_\Sigma$ for different choices of the dark photon mass (denoted here as $m_X$), as shown in the second panel of Figure 6.9. For a 1.5 GeV dark photon, the excluded region reaches 600 GeV for $v_\Sigma = 1$ MeV.

The aforementioned recast does not require detailed information on event topology, other than the dark photon decay length. Consequently, the 13 TeV limits [266] translate rather straightforwardly into stronger bounds on the model parameter space, reaching to somewhat larger $\Lambda/\beta$.

**Lessons Learned**

The triggering requirements used in the ATLAS analysis thus far limit the reach of displaced LJ searches. For models having a signal with only the displaced LJ and no other observable objects, such as the FRVZ model, triggering solely on MS tracks not associated with ID tracks is appropriate, though even here the 3mu6 trigger may not be sufficiently inclusive, as it requires at least three ROIs in the MS (see Section 3.2.1). Events with LJ pairs for which neither LJ can be resolved into two separate ROIs will be missed.

It is clear that triggering on associated prompt objects, such as the lepton from one of the final state vector bosons in the NAKM5 scenario or from the $W$ boson in heavy right-handed neutrino models (charged-current LLP production), could significantly enhance the triggering efficiency and extend the reach of displaced LJ searches to a wider class of models and to a broader range of model parameter space. Inclusion of an associated prompt object in triggering may also enhance background rejection and relax the requirement on $\Delta\phi$ between LJs.

In addition, the implications of the mass scale of the intermediate BSM particles and the number of final state prompt objects remains to be investigated. The ATLAS 8 TeV search assumed the hidden sector particles are light compared to the mass of the SM Higgs boson, whose decay chain leads to the final states involving multiple $\gamma_d$ and HLSPs. For the NAKM5 scenario and models with heavy right-handed neutrinos, these assumptions about mass hierarchy may not apply. It is also not clear what impact the DV LJ isolation requirements have when there is an associated prompt lepton in the signal event.

### *6.4.5 Non-Pointing Photons*

The search for non-pointing photons produced in association with missing transverse energy ($\not{E}_T$) [294] plays an important role in probing BSM particles that decay to a SM photon and an invisible particle through a highly suppressed coupling. Besides the gauge-mediated supersymmetry breaking (GMSB) models [510], which were the main motivation for the non-pointing photon search, this type of signal can also appear in many hidden-sector models. For example, in the dipole-mediated DM model (the "Dark Penguin") [149], the production of two heavier dark fermions $pp \to Z^*/\gamma^* \to \chi_h\bar{\chi}_h$ is followed by the decays $\chi_h \to \chi_l + \gamma$. If the flavor structures of the DM mass and coupling are aligned, $\chi_h$ can be long-lived and give rise to non-pointing photons. Another example is provided by the dark-shower scenario [511, 512] that explains the galactic center gamma-ray excess. In this model, many hidden pions can be produced in the same LHC event. Some of these have displaced decays to a pair of SM photons, while others decay outside of the detector, yielding $\not{E}_T$. Notice that in this case the topology of the events is different from the previous examples, as the non-pointing photons and $\not{E}_T$ originate from separate particles (and for more discussion on dark showers, see Chapter 7).

Here, we describe a method used to recast the bounds of Ref. [294] to a BSM scenario that is different from the GMSB model, using the Dark Penguin signal [149] as an example.

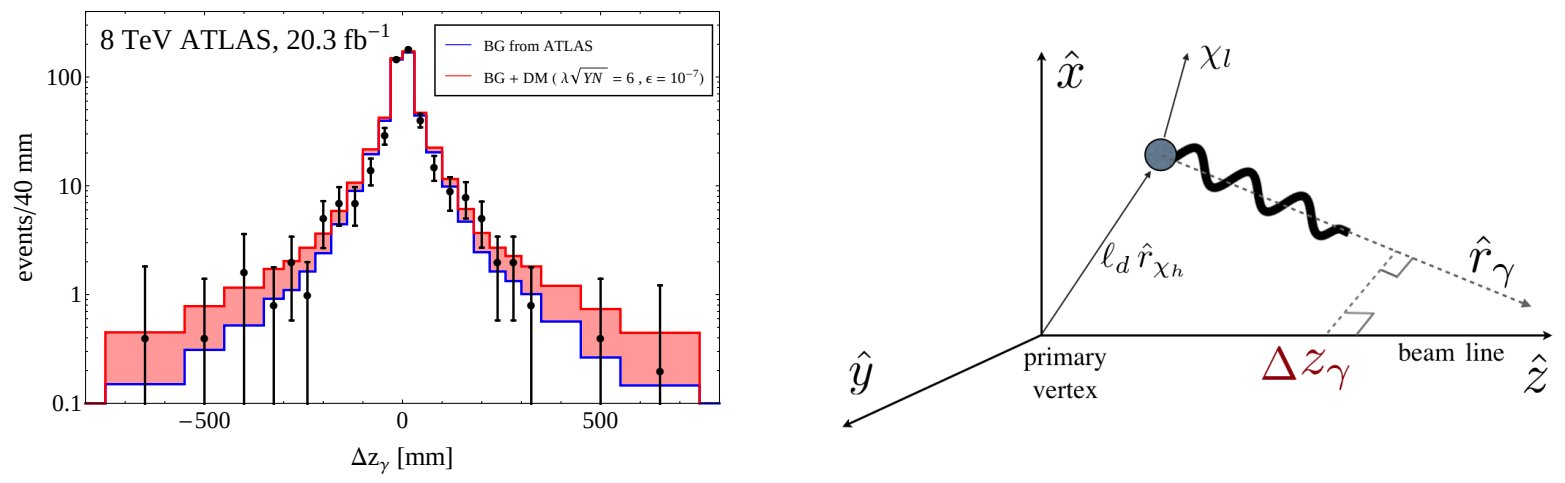

Figure 6.10: **Left:** The $\Delta z_\gamma$ distribution of the non-pointing photon signals measured by ATLAS. The background reported by ATLAS (blue histogram) was obtained from a data-driven analysis, using diphoton events with $\not{E}_T < 20$ GeV. Also shown, stacked on top of the background (red histogram), is the signal distribution from the dipole-mediated DM scenario with $(m_{\chi_h}, m_{\chi_l}, M) = (300, 10, 300)$ GeV, $\lambda\sqrt{NY} = 6$, and $\varepsilon = 10^{-7}$. See Ref. [149] for more details. **Right:** The geometry of the displaced signals.

#### 6.4.5.1 *Calculation of the Signal Efficiency for the Non-Pointing Photon Search*

We follow the non-pointing photon analysis in Ref. [294], performed by the ATLAS collaboration on about 20 fb$^{-1}$ of 8 TeV data. In Ref. [294] delayed photons were also considered, but here we focus only on the measurement of the $\Delta z_\gamma$ of non-pointing photons (see Figure 6.10). For DM signals given by the long-lived $\chi_h \to \chi_l \gamma$ decay, $\Delta z_\gamma$ can be related to the $\chi_h$ decay length $\ell_d$ in the lab frame:

$$\begin{aligned}\Delta z_\gamma &= \ell_d \left( \hat{r}_{\chi_h,z} - \frac{\hat{r}_{\chi_h,T} \cdot \hat{r}_{\gamma,T}}{1-(\hat{r}_{\gamma,z})^2} \hat{r}_{\gamma,z} \right) \\ &= \ell_d \Big[ \cos\theta_{\chi_h} - \cos\big(\phi_{\chi_h} - \phi_\gamma\big) \cot\theta_\gamma \sin\theta_{\chi_h} \Big] \end{aligned} \tag{6.8}$$

where $\hat{r}_{T,z}$ represent the transverse and longitudinal components of the unit vector $\hat{r}$, respectively, as shown in the right pane of Figure 6.10. To obtain the $\Delta z_\gamma$ distribution of the DM decay, we first simulate the prompt process, $p\,p \to \chi_h \bar{\chi}_h, \chi_h \to \chi_l \gamma, \bar{\chi}_h \to \bar{\chi}_l \gamma$ in MadGraph5, then we apply the cuts performed in the ATLAS analysis, and finally reweight the events using the dark penguin form factors. Then we calculate the proper lifetime of $\chi_h$ and boost it to the lab frame using the momenta of each parton-level event. The angular information of the photon and $\chi_h$ allow us to calculate $\Delta z_\gamma$ in Eq. (6.8) as a function of the decay length. Using this, each simulated MC event contributes to the differential cross section in $\Delta z_\gamma$ as

$$\frac{d\sigma_{\text{displaced}}}{d\Delta z_\gamma} = \sigma_{\text{prompt}} \frac{dP}{d\Delta z_\gamma} = \sigma_{\text{prompt}} \frac{|\mu|}{2} e^{-\mu \Delta z_\gamma}, \tag{6.9}$$

where the $\mu$ characterizing the probability distribution $dP/d\Delta z_\gamma$ of the decay is defined as

$$\mu \equiv \frac{\Gamma_{\chi_h} m_{\chi_h}}{p_{\chi_h}} \left( \hat{r}_{\chi_h,z} - \frac{\hat{r}_{\chi_h,T} \cdot \hat{r}_{\gamma,T}}{1-(\hat{r}_{\gamma,z})^2} \hat{r}_{\gamma,z} \right)^{-1}. \tag{6.10}$$

Summing the distributions derived from all the simulated events we obtain the differential cross section in $\Delta z_\gamma$ shown in Figure 6.10.

The ATLAS search requires at least two loose photons with $|\eta| < 2.37$ and $E_T > 50$ GeV. At least one photon is required to be in the barrel region $|\eta| < 1.37$. To avoid collisions due to satellite bunches, both photons are required to have an arrival time at the ECAL $t_\gamma$ smaller than 4 ns, with zero defined as the expected time of arrival for a prompt photon from the primary vertex. We approximate $t_\gamma$ with the time of flight of the $\chi_h$, requiring it to be smaller than 4 ns. In our sensitivity estimate, we do not include the detailed isolation cuts on the photon. We also neglect the effect of the displaced decay on the angular acceptance of the photons, simply imposing the requirements on $|\eta|$ at the level of the prompt event. The signal region also requires $\not{E}_T > 75$ GeV.

Finally, to simplify the discussion we assume that every event has a reconstructed primary vertex in the geometrical center of the detector.

For events where only one photon satisfies $|\eta| < 1.37$ (i.e., it is in the barrel calorimeter), this photon is used for the measurement of $\Delta z_\gamma$. For events where both photons are in the barrel, the photon with larger $t_\gamma$ is used. We approximate this timing condition by taking the photon emitted by the more boosted $\chi_h$, in which case the average decay is more delayed. In Figure 6.10 the generated $\Delta z_\gamma$ signal distribution is shown on top of the expected background. The latter is taken from Figure 4 of the ATLAS paper [294]. Because we are focusing on the non-pointing photon signals, to set constraints on the DM couplings in Ref. [149] we remove events with $|\Delta z_\gamma| < 30$ mm. In our exploratory analysis we only consider the statistical uncertainty on the background, neglecting the effect of systematics.

**Lessons Learned**

The ATLAS paper gives detailed descriptions of the cuts and background analysis, which makes an approximate estimation of the signal efficiency quite straightforward.

The background analysis in Ref. [294] is based on a data-driven study, for which events passing the diphoton selection with $\not{E}_T < 20$ GeV are used as control region sample. It is challenging for theorists to simulate the background for different energy cuts. This is particularly true for the $\not{E}_T$ cut that plays a vital role in the DM and Hidden Valley searches.

To obtain a more precise result, it would be useful if the ATLAS collaboration could provide the reconstruction efficiency of non-pointing photons as function of $\Delta z_\gamma$, or of the angle between the photon and the surface of the ECAL, a variable that may be relevant to the efficiency. The paper does provide a table of signal acceptance times efficiency for the GMSB model used therein. However, the numbers depend on details of the particular model used, and it is hard to extract the efficiency that is associated to the non-pointing photon reconstruction. Therefore, when estimating the signal efficiency, we only consider efficiency from the selection cuts and do not include possible suppressions from photon reconstruction.

It would also be very useful to have a table of background events for different cuts on $\not{E}_T$. For instance, the Dark Penguin signature has $\not{E}_T$ from the decay of $\chi_h\bar{\chi}_h$ with electroweak-scale $m_{\chi_h}$, and $\not{E}_T$ can easily be higher than 100 GeV. By contrast, in the dark shower scenario where soft hidden pions decay to two SM photons, the $\not{E}_T$ originating from additional late-decaying pions can be much lower than the 75 GeV cut used in the ATLAS analysis. Knowing the background and systematic uncertainty for different $\not{E}_T$ cuts would be very important to constrain different models with potentially very different kinematics.

### 6.4.6 *Displaced Vertices*

Displaced-vertex searches differ from displaced jets and displaced leptons due to the requirement of an actual secondary vertex from the displaced objects. These searches are sensitive to LLP lifetimes that allow it to decay in the inner trackers or muon spectrometer of the LHC detectors, where vertexing is possible [189, 230, 241, 244, 246, 258, 272]. These searches have extremely low backgrounds as there are no irreducible contributions from the SM, making them sensitive to very small signals of new physics (for more information, see Chapter 4). Moreover, the identification of displaced decays can be used to extract kinematical information in a decay, such as (invisible) particle masses [513, 514].

In this section we review some reinterpretations of displaced-vertex searches, differentiating between reinterpretations making use of only truth-level information to identify displaced decays and reinterpretations in which an attempt is made to reconstruct displaced vertices from displaced tracks (with an approximate detector response).

#### 6.4.6.1 *Truth Level Displaced Vertices*

The work in Ref. [26] reinterpreted the 8 TeV ATLAS search for a displaced muon and a multi-track vertex (DV+$\mu$) [515], where long-lived particle signatures for certain weak-scale models of baryogenesis [24, 26, 507] were explored. For reinterpreting this search, a similar procedure described in Section 6.4.3, on displaced jets, of constructing ratios of truth-level vs. ATLAS efficiencies for the ATLAS multi-track vertex analysis [515] was performed, with similar results for the validation being correct within approximately a factor of two.

This DV+$\mu$ analysis has since been superseded by Ref. [189], in which a displaced vertex is searched for at 8 TeV in association with either a muon, electrons, jets, or missing transverse momenta. Recently, an updated ATLAS analysis [230], which looks for multi-track displaced vertices at 13 TeV in association with large $\not{E}_T$, was made public. This search now includes a prescription using parametrized efficiencies as a function of vertex radial distance, number of tracks and mass. Their prescription can be applied to vertices and events passing certain particle level acceptance requirements using the truth MC event record.

Here we validate the prescription with parametrized selection efficiencies in Ref. [230] [10]. The results of this search are interpreted by ATLAS in a split-SUSY simplified model with a long-lived gluino that hadronizes, forming an $R$-hadron before decaying as $\tilde{g} \rightarrow q\bar{q}\tilde{\chi}_1^0$. Event samples are generated with Pythia 8.2 [213]. We use truth-level $\not{E}_T$ and we identify the truth $R$-hadron decay position and decay products, as the ATLAS collaboration provides selection efficiencies that can be directly applied to these truth-level quantities. These efficiencies can be found in the auxiliary material in Ref. [516], and are given at the event-level as a function of the truth $\not{E}_T$ and displaced-vertex radial distance, and at the vertex level parametrized as functions of vertex invariant mass and number of tracks. The efficiencies are given for different detector regions, encapsulating also the effect of the material veto cut.

[10] This prescription is also validated in Ref. [502].

The selection of events used for the signal region requires:

- truth level $\not{E}_T > 200$ GeV;
- one trackless jet with $p_T > 70$ GeV, or two trackless jets with $p_T > 25$ GeV. A trackless jet is defined as a jet with $\sum_{tracks} p_T < 5$ GeV.

In addition, events must have at least one displaced vertex with:

- transverse distance between the IP and the decay position > 4 mm;
- the decay position must lie in the fiducial region $r_{DV} < 300$ mm and $|z_{DV}| < 300$ mm;
- the number of selected decay products must be at least 5, where selected decay products are charged and stable, with $p_T > 1$ GeV and $|d_0| > 2$ mm;
- the invariant mass of the truth vertex must be larger than 10 GeV, and is constructed assuming all decay products have the mass of the pion.

Applying these cuts and efficiencies, we get event-level efficiencies for two of the benchmarks (with gluino masses of 1400 GeV or 2000 TeV, and the neutralino mass is fixed to 100 GeV). Based on the efficiencies obtained and the estimated number of background vertices, we can extract 95% C.L. upper limits on the total visible cross section for the two gluino masses. For reference, assuming 100% efficiency, we get an upper limit of 0.091 fb. The curves in Figure 6.11 show our recasting results compared to ATLAS. The level of agreement is very good, within 20% for most of the lifetime values. We also point out that the recasting limits agree well even for regions where the efficiency is very low ($\tau > 10$ ns and $\tau < 10^{-2}$ ns). This 13 TeV ATLAS multitrack analysis [230] has also been reinterpreted in the context of long-lived sterile neutrinos [47, 257].

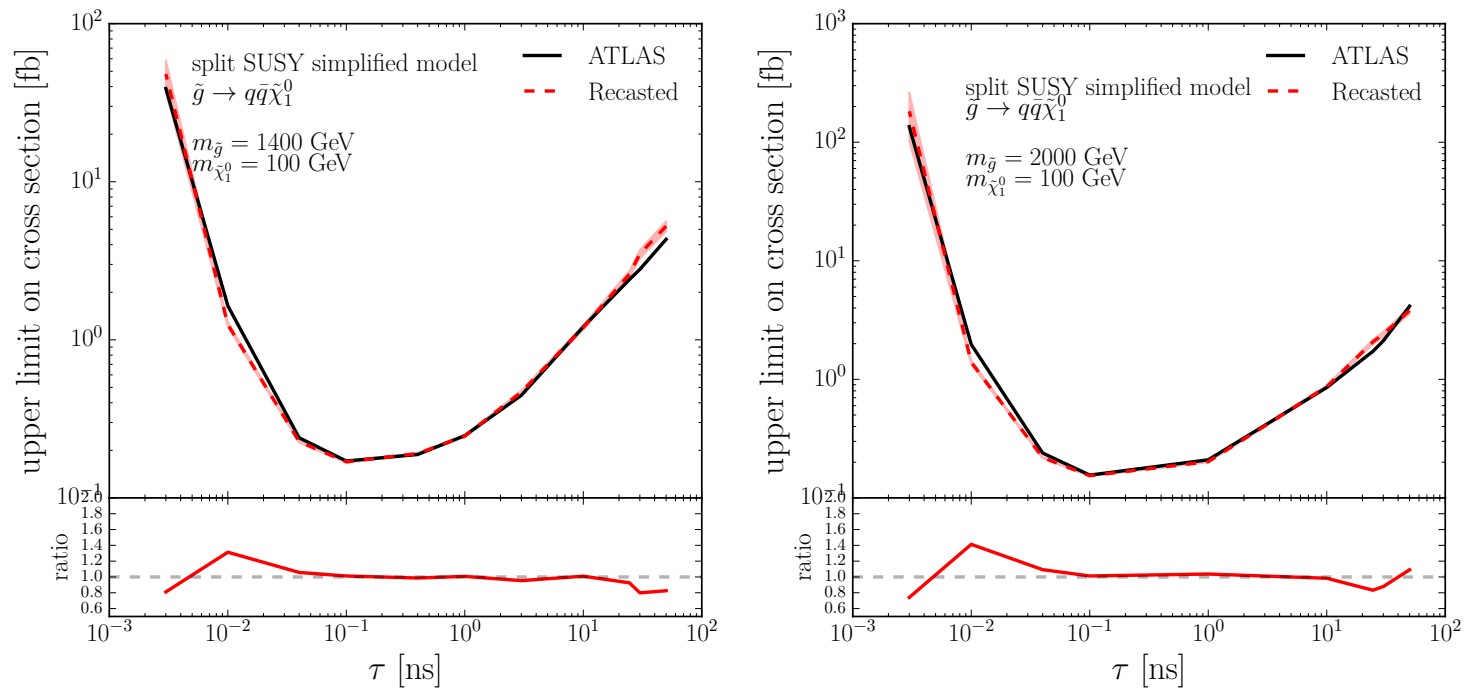

Figure 6.11: Comparison of our recast and ATLAS [230] on the upper limit on the gluino production cross section vs. its proper lifetime [502].

### 6.4.6.2 *Displaced-Vertex Reconstruction*

Before parametrized efficiencies applicable for truth-level displaced vertices were made public, attempts to recast displaced-vertex searches were made by performing their reconstruction from charged tracks only, with an approximate detector response. In Ref. [199] the ATLAS DV+jets multitrack analysis [189] was recast. Reinterpretation was performed using generator-level events and the detector fiducial region was reproduced as well as possible. The jets are clustered according to the anti-$k_T$ prescription [360] in the analysis with momentum smearing, validated by reproducing the jets+$\not{E}_T$ exclusion curve of prompt searches. The selection of events used for the signal region (and the approximations to real detector simulation) were as follows: A tracking efficiency of the form

$$\begin{aligned}\varepsilon_{\text{trk}} =& 0.5 \times \left(1 - \exp\left(\frac{-p_{\text{T}}}{4.0\ \text{GeV}}\right)\right) \times \exp\left(\frac{-z_{\text{DV}}}{270\ \text{mm}}\right) \\ & \times \max\left(-0.0022 \times \frac{r_{\text{DV}}}{1\ \text{mm}} + 0.8, 0\right),\end{aligned} \tag{6.11}$$

is used, where $r_{\text{DV}}$ and $z_{\text{DV}}$ are the transverse and longitudinal distance of the track's production vertex (which is the same as the displaced vertex origin when using truth-level generator information). This functional form is designed to take into account the size of the detector (i.e., it imposes a linear dependence on $r_{\text{DV}}$, and an exponential dependence on $z_{\text{DV}}$), as well as a turn-on like feature dependent on the $p_{\text{T}}$ of the track. It reproduces the overall behavior of efficiency falling off with vertex displacement. The parameters were determined by fitting the efficiency curve (with lifetime dependence), for three benchmarks in the analyses. We find that fitting only one benchmark does not correctly reproduce the efficiency curve for any of the others. This is most likely due to insufficient dimensionality of the efficiency map. We expect that a full tracking-efficiency parametrization depends not only on $r_{\text{DV}}$, $z_{\text{DV}}$ and $p_T$, but also on transverse and longitudinal impact parameters ($d_0$,$z_0$), and on the charge and pseudorapidity of the track. Furthermore, we expect a vertex efficiency that depends on the topology of the event and the nature of the particles forming the vertex. The fit for the event efficiencies from this tracking function can be seen in Figure 6.12.

| | |
|---|---|
| DV jets | 4 or 5 or 6 jets with $\|\eta\| < 2.8$ and $p_T > 90, 65, 55$ GeV, each |
| DV reconstruction* | DV made from tracks with $p_T > 1$ GeV, $\|\eta\| < 2.5$ and $\|d_0\| > 2$ mm. Vertices within 1 mm are merged. Note: a tracking efficiency needed here; we assume a functional form given by equation 6.11 |
| DV fiducial | DV within 4 mm $< r_{DV} < 300$ mm and $\|z_{DV}\| < 300$ mm |
| DV material* | No DV in regions near beampipe or within pixel layers. Discard tracks with $r_{DV}/\text{mm} \in \{[25, 38], [45, 60], [85, 95], [120, 130]\}$. |
| $N_{\text{trk}}$ | DV track multiplicity $\geq 5$ |
| $m_{DV}$ | DV mass $> 10$ GeV |

Table 6.5: Implementation of cuts applied in the ATLAS multitrack DV + jets search, from Ref. [189]. Cuts denoted by an asterisk (*) are approximations to the experimental analysis in the absence of the full detector simulation.

**Lessons Learned**

With a larger and more diverse set of signal benchmarks, the prospects for reinterpretation are better. For example, the ATLAS analysis examined in Ref. [515] only showed limits for three signal model benchmark points for which efficiencies were shown, making it challenging to find a benchmark whose kinematics matched the desired signal models for the reinterpretations of other models (for example, in Ref. [26]). Because the efficiencies and limits were shown for either a high-mass, low-boost LLP or a low-mass, high-boost LLP, this made it more challenging to reinterpret the results for other types of kinematics.

The new parametrized efficiencies presented by ATLAS in Ref. [516] are extremely useful. They constitute an optimal efficiency map for recasting these type of analyses, as they can by applied in a straightforward way to truth-level quantities. Before this information was made public, efficiency tables (for vertex-level efficiency) in terms of $r_{\text{DV}}$ were only available for few channels and for a single benchmark. It was not clear how to translate this infor-mation to other channels, or to parent particles of a different mass. In this case, a functional parametrization for track efficiency was needed (as derived in Ref. [199]) to be able to reproduce the experimental results. Finding this kind of parametrization is not easy, as it needs to be validated across different benchmarks. For example, Ref. [199] found that fitting only one benchmark did not correctly reproduce the event-level efficiency curve for any of the others.

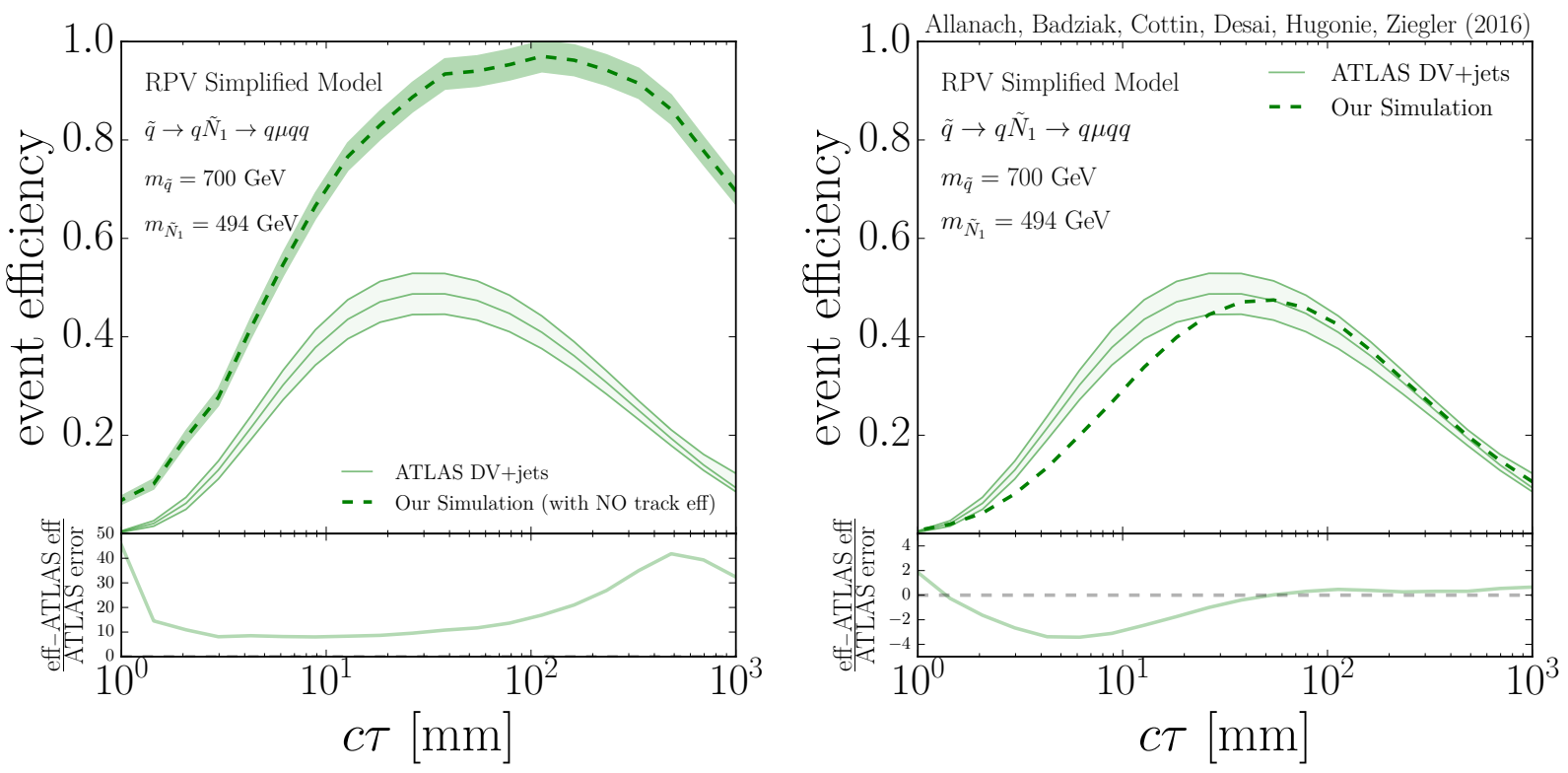

Figure 6.12: Validation of the DV + jets search for the ATLAS benchmark of a simplified RPV model with a 700 GeV squark decaying to a neutralino, $\tilde{q} \to q(\tilde{N}_1 \to \mu\bar{u}d)$ . **Left:** Without any tracking efficiency. **Right:** With a tracking efficiency function given by equation 6.11, taken from Ref. [199].

## 6.5 *Handling Long-Lived Particles in* Delphes*-Based Detector Simulations*

### 6.5.1 *Long-Lived Particle Simulation in* Delphes 3.4.1

The Delphes package [336] allows for the generic simulation of the response of a typical detector used in high-energy physics experiments. It is widely used for simulating the effects of the ATLAS and CMS detectors or the hypothetical detectors that could be used for the future FCC and CLIC projects. The architecture of Delphes is composed of distinct and specialized modules that interact with each other. The detector is described by the user through an input card, where the modules to be used in the simulation are sequentially enumerated and their input parameters are specified.

The detector simulation relies on a mix of parametric and algorithmic modules. More precisely, tracking is simulated through efficiency and smearing functions that are applied to the properties of the electrically-charged stable particles. The particles are then propagated to the calorimeters and dedicated modules simulate the energy deposits in the electromagnetic and hadronic calorimeters. The output of such a step consists of a list of calorimetric towers. Moreover, Delphes includes a particle-flow-like algorithm that combines tracking and calorimetric data in order to improve the identification of the final-state objects and the resolution on their reconstructed momenta.

Jet clustering is performed via an internal call to the FastJet package [361], which takes as input the list of calorimeter towers or, alternatively, the particle-flow candidates, and outputs jet objects. Lepton and photon isolation is then handled through a specific isolation module. Finally, Delphes takes care of removing the double-counting of objects that could be simultaneously identified as elements of different collections. The final output is stored in a Root file.

In addition, Delphes allows for the simulation of pile-up effects by superimposing minimum-bias events attached to displaced interaction vertices along the beam direction to the hard scattering event. Procedures mimicking pile-up removal can then be configured in the input card. The subtraction of the charged particles belonging to pile-up vertices is performed at the tracker level. Neutral particles are removed by applying the jet area method [517] supplied within the FastJet package. More advanced methods are also available in Delphes, such as the Puppi [518] or SoftKiller [519] techniques, and they can easily be added to the input card. The loss of performance originating from pile-up, in particular relative to the isolation, is automatically accounted for.

The official Delphes package (version 3.4.1) with the default detector cards needs to be adapted for the proper handling of LLPs. By default, the decay products of a long-lived particle enter the simulation as if the corresponding decay would have occurred within the tracker volume. However, the user has the possibility to define a volume in which the particle can decay and still be detected outside the tracker volume. This is achieved in practice by making use of the `RadiusMax` parameter of the `ParticlePropagator` module, that is by default set to the tracker radius stored in the `Radius` parameter. When setting the `Radius` and `RadiusMax` parameters to different values, the particles decaying outside the tracker volume, but inside the "decay volume" of radius `RadiusMax`, are included in the collection of stable particles stored in the output Root file. They can in this way be used for an offline, more correct, treatment.

Moreover, several modules that are not used in the default ATLAS and CMS cards could serve for a better simulation of the long-lived particles in Delphes. For instance, the `TrackSmearing` module allows the user to smear the track momentum according to the impact parameter in the transverse plane (i.e., the $d_0$ parameter) and in the longitudinal plane (i.e., the $d_z$ parameter).

By default, the detector simulation in Delphes totally ignores the presence of any LLP. While this is convenient for neutral particles like a neutralino which could be considered as invisible from the detector standpoint, a charged particle leaves tracks in the tracker and would interact with the calorimeters if its lifetime is large enough. In this case, if the long-lived particle decays inside the tracker, its trajectory is properly propagated to the calorimeters and the displaced vertex is correctly accounted for. However, if the long-lived particle decays outside the tracker, its decay products are ignored in DELPHES, unless the `RadiusMax` parameter has been specified to be larger than the tracker `Radius` parameter. In this case the decay products can be found in the `Delphes/stableParticles` output collection and treated with adequate smearing functions and efficiency directly from the DELPHES output.

Finally, disappearing tracks are simply treated as missing energy in DELPHES. Emerging tracks or tracks containing kinks are not treated appropriately, in the sense that the parameterizations required for a proper description of such signatures has not been implemented yet. Also, DELPHES does not include any trigger simulation, and the latter is in general complex in the case of LLPs.

### 6.5.2 *Displaced Tracks With the* MADANALYSIS 5 *Tune of* DELPHES

The DELPHES-LLP package can be installed from the version v1.6 of MADANALYSIS 5 [486, 520] and contains improvements of DELPHES specific to LLPs. It leads to new possibilities for phenomenological investigations of long-lived particles and the recasting of related LHC analyses [11].

This new package was designed to handle neutral LLPs that decay into leptons within the tracker volume. Realistic efficiencies are applied to the displaced tracks and several parameters specific to this kind of analysis have been made available within MADANALYSIS 5. An extension to the case of neutral LLPs decaying into muons outside the tracker volume can be easily implemented, the muons being thus reconstructed only through their hits in the muon chambers. The simulation of the displaced leptons is performed through efficiencies and resolution functions to be specified by the user. Furthermore, another extension allowing the user to handle long-lived charged particles that decay into leptons could be implemented. A similar dedicated tune of DELPHES and MADANALYSIS was devised for the studies of neutral particles decaying to displaced leptons and jets in Refs. [44, 48, 162].[12]

The DELPHES-LLP package contains a new module called `MA5EfficiencyD0` that allows for the definition of a track reconstruction efficiency parameterized as a function of the $|d_0|$ and $d_z$ parameters (named `d0` and `dz` in the DELPHES input card). The default efficiency function, specified via a `DelphesFormula` is taken from the 8 TeV tracking performance of CMS [268] [13],

[11] More information is available at the following website: https://madanalysis.irmp.ucl.ac.be/wiki/MA5LongLivedParticle

[12] The code is available to download at https://sites.google.com/site/leftrighthep/

[13] https://twiki.cern.ch/twiki/bin/view/CMSPublic/DisplacedSusyParametrisationStudyForUser

```
set EfficiencyFormula {
  (d0<=20) * (-5.06107e-7 * d0**6 + 0.0000272756 * d0**5 - 0.00049321 * d0**4
     + 0.00287189 * d0**3 + 0.00522007 * d0**2 - 0.0917957 * d0 +  0.924921) +
  (d0> 20) * (0.00)
}
```

In addition, the data-format of DELPHES has been extended so that the `Muon` and `Electron` classes now include the transverse ($|d_0|$) and longitudinal ($d_z$) impact parameters relative to the closest approach point (encoded in the `d0` and `dz` variables), the coordinates of the closest approach point $(x_d, y_d, z_d)$ (encoded in the `xd`, `yd` and `zd` variables), and the four-vector of the vertex from which the lepton is originating from $(t_p, x_p, y_p, z_p)$ (encoded within the `tp`, `xp`, `yp` and `zp` variables), the latter quantity being evaluated from Monte Carlo information.

| Region | $c\tau_{\tilde{t}}$ [cm] | MA5 | CMS | Difference [%] |
|---|---|---|---|---|
| **SR-1** | 0.1 | 3.89 | 3.8 | 2 |
| | 1 | 4.44 | 5.2 | 15 |
| | 10 | 0.697 | 0.8 | 15 |
| | 100 | 0.0610 | 0.009 | $> 100\%$ |
| **SR-2** | 0.1 | 0.924 | 0.94 | 2 |
| | 1 | 3.87 | 4.1 | 5 |
| | 10 | 0.854 | 1.0 | 15 |
| | 100 | 0.0662 | 0.03 | $\sim 100\%$ |
| **SR-3** | 0.1 | 0.139 | 0.16 | 15 |
| | 1 | 6.19 | 7.0 | 10 |
| | 10 | 4.45 | 5.8 | 25 |
| | 100 | 0.497 | 0.27 | 85 |

Table 6.6: Number of events populating the three signal regions (SR-1, SR-2 and SR-3) of Ref. [190] for different stop decay lengths ($c\tau_{\tilde{t}}$). We compare the CMS and MadAnalysis 5 (MA5) results in the second and third column of the table, respectively, and the difference taken relatively to CMS is shown in the last column.

Consequently, the data-format of MadAnalysis 5 has been extended, so that the `Muon` and `Electron` classes now contain the `d0()` and `dz()` methods allowing to access the value of the $|d_0|$ and $d_z$ parameters, the `closestPoint()` method that returns the coordinates of the closest approach point (through the `X()`, `Y()` and `Z()` daughter methods) and the `vertexProd()` method that returns coordinates of the displaced vertex from which the lepton originates (through the `X()`, `Y()` and `Z()` daughter methods). An analysis example [492] can be found in the public analysis database of MadAnalysis 5 [14], where information about the re-implementation in MadAnalysis 5 of Ref. [190], an analysis of 2.6 fb$^{-1}$ of 13 TeV LHC data, is available. This is a search for long-lived particles decaying into electrons and muons, where signal events are selected by requiring the presence of either an electron or a muon whose transverse impact parameter lies between 200 $\mu$m and 10 cm. For a given benchmark signal where a pair of long-lived stops is produced through QCD interactions and where each stop further decays into a displaced $b$-jet and a displaced lepton. In Table 6.6, we present the number of events surviving the selection of the three different signal regions of the CMS analysis. Our event generation has been performed with Pythia 8 [212] and for the benchmark scenario used. We observe that a good agreement is obtained, except in the case of long stop lifetimes, $c\tau \gtrsim 1$ m. This is however the region in which no public information on the CMS reconstruction efficiencies is available.

[14] `http://madanalysis.irmp.ucl.ac.be/wiki/PublicAnalysisDatabase`

The MadAnalysis 5 tune of Delphes has neither been designed for disappearing (or appearing) tracks nor for track kinks. Concerning the disappearing (or appearing) tracks, the only missing experimental ingredients are the track reconstruction efficiency and resolution as a function of the number of missing hits in the (inner) outer layers of the tracker. There is to this date no public material on the tracking performance description related to track kinks.

### 6.5.3 *What About Other LLP Signatures?*

In this section, we briefly discuss how Delphes could be improved for a better handling of LLP signatures.

#### 6.5.3.1 *Displaced Jets*

Displaced jets are jets that are reconstructed either from standalone calorimeter information, or from the particle-flow input with a minimum requirement on the multiplicity of tracks with high transverse displacement (see Ref. [244]). Conceptually, such jets can be handled in Delphes provided that the displaced tracks are properly parametrized. As described above, a module designed to smear the full set of track properties, including their transverse and longitudinal displacement, exists (i.e., the `TrackSmearing` module). In addition, efficiencies based on displacement parameters have already been implemented in MadAnalysis 5 (see above) and a module that performs the matching of an existing jet collection with a track collection based on track displacements is very similar to the already existing `TrackCountingBTagging` module. Minor modifications to this module are hence needed to be able to select tracks based on an absolute displacement instead of the impact parameter significance. Finally, in order to be able to perform a displaced jet selection, one would need a (not yet existing) module that performs jet clustering on the basis of the secondary vertices and the displaced tracks matched with these jets. Alternatively, a module that includes a vertex reconstruction efficiency and a module including a vertex position smearing could be implemented.

#### 6.5.3.2 *Displaced Vertices*

The missing modules described in Section 6.5.3.1 could perfectly serve the purpose of a displaced vertex analysis. Provided that tracking efficiencies and resolutions are available as a function of the full set of tracking parameters ($d_0$, $d_z$, $p_T$, $\phi$, and $\theta$ ) and, eventually, of the Monte Carlo truth vertex position, a simple vertexing algorithm can be implemented in Delphes.

#### 6.5.3.3 *Discussion:* Delphes *Versus a Specific Parametric Simulation*

In a fully parametric simulation of the detector, the detector effects are encoded in terms of efficiencies and resolution functions. Such a simulation is typically very fast, but is likely to suffer from a lack of accuracy in the modeling of complex observables such as jet properties and missing energy. The Delphes simulation is an admixture of such a parametric simulation and of an algorithmic one. This is slower, but has the clear advantage of correctly treating in particular the reconstruction of the jets and the missing energy.

In order to be able to answer whether Delphes should be used in place of a fully parametric simulation in recasting LLP analyses, further studies are needed. In the meantime, the following guidelines could be used. If the signal selection is based only on displaced tracks, a simple parametric simulation should in principle be sufficient. This simulation could encapsulate the track reconstruction efficiency and resolution, including pile-up effects. Delphes could then optionally be used to mix the resulting "reconstructed" tracks with the additional tracks originating from the pile-up vertices. On the other hand, if the analysis under consideration additionally uses calorimetric information (i.e., jets or missing energy), Delphes should be preferred to a fully parametric framework. However, a precise quantification of these effects cannot be assessed without detailed comparative studies between the two approaches. Finally, it should be pointed out that neither of these techniques can be used to correctly simulate the instrumental background, which is challenging to simulate and is discussed in Chapter 4.

## 6.6 *Recasting Inside the Experimental Collaborations*

Reinterpretations performed within the experiments themselves present unique advantages and disadvantages. They allow for thorough and consistent treatment of detector effects and geometry, object reconstruction, and systematic uncertainties in a way which is impossible through external recasting. Groups can share resources and easily communicate all necessary details. On the other hand, they are of course limited to the model(s) chosen for reinterpretation. In the ideal situation, reinterpretations which provide meaningful results can be performed with minimal overhead to a given analysis.

As the LHC enters an era of deccelerating luminosity growth and analysis techniques are becoming more sophisticated, the LHC analyses become harder to re-implement with sufficient accuracy outside of the experiments compared to cut-based analyses. Analyses increasingly utilize machine-learning algorithms that transform a large number of event-level and particle-level observables into higher-level discriminants which are not easily characterized by low-dimensional efficiency tables and may require inputs that third-party detector simulations are not able to reproduce. In particular non-prompt searches may depend on non-traditional reconstruction objects and details of the detector simulation and geometry in ways that require a more detailed simulation than is achievable by, e.g., third-party simulators. Hence, experiments are investigating approaches that enable internal reinterpretation using the full set of available information.

Full-fidelity reinterpretations are especially relevant for long-lived particles, since the signal simulation may depend more heavily on details not well captured by third-party simulation tools. For example, for sufficiently high lifetime, the decays must be handled by a full detector simulation such as GEANT (or some complex interaction between GEANT and MC packages such as Pythia). Such decays are not well-covered by tools like Delphes as the response of such in-detector decays may require access to a more detailed geometry description.

### *6.6.1 The RECAST Framework*

The RECAST Framework [494] is a developing platform for experiments as well as researchers external to the collaborations who wish to reinterpret LHC analyses. RECAST enables cloud-based analysis execution and common presentation of reinterpretation results. The framework consists of two components:

*The RECAST Front End* This is a web-based service in which reinterpretation of analyses can be suggested by interested authors that provide necessary inputs such as UFO model files [218], process and parameter card templates, or suggested scan grids. Responses to such requests, possibly by more than one analysis implementation, can then be uploaded. Such a web service, interfaced with services such as HepData, may then serve as a resource for the LHC community to organize and share reinterpretation results obtained by the various analysis implementations.

*The RECAST Back End* An important objective of the framework is to enable a full-fidelity reinterpretation of an LHC analysis using the original analysis code developed within the experiment that can be approved by the collaboration and be placed on an equal footing with the original publication. In contrast to third-party recasting tools, in which multiple analyses are implemented using a single, common framework that is executed on a single computing element, such an exact analysis re-execution often necessitates a distributed data analysis using a number of different frameworks in use within the experiments. Therefore, RECAST has developed a flexible graph-based analysis description and execution back end [521] that enables a faithful re-execution of nearly arbitrary analysis code on cloud platforms such as those offered by CERN [15]. The back-end provides experiments with an access-controlled interface to view reinterpretation requests, retrieve the necessary analysis description from repositories such as the CERN Analysis Preservation Portal (CAP) [522], execute the analysis on datasets for the new model and — if approved — upload the results to the public front-end. Figure 6.14 shows a screenshot of the current prototype user interface, giving collaboration members an overview over requested points as well as controls to steer processing and submission.

[15] Thanks to this flexibility, the popular third-party recasting tools can be easily integrated into this back-end as well, with current integrations being available for CheckMate and Rivet.

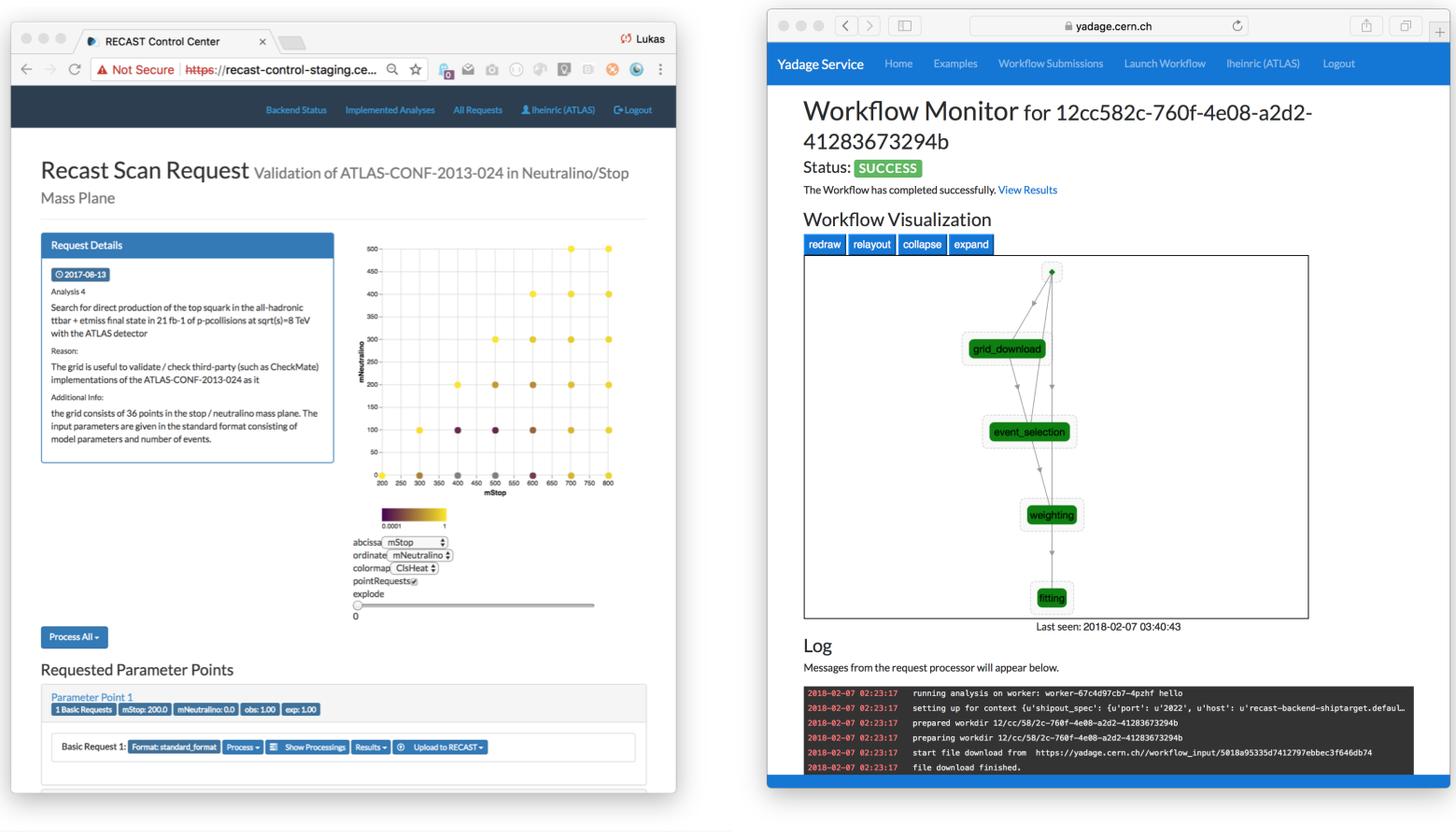

Figure 6.13: **Left:** Web-based interface for the RECAST back end for and experiment that presents the requested parameter points and color-coded results. **Right:** An experiment-internal reinterpretation executing on the distributed infrastructure built by CAP and RECAST.

These services are being developed in close collaboration with the CERN Analysis Preservation project, which is a common project supported by the four major LHC experiments. While this integration work is on-going, the computing back end for RECAST has been successfully used for a number of Run 1 and Run 2 reinterpretations published by the ATLAS Experiment.

### 6.6.2 *Analysis Preservation as a Driver for Reinterpretation*

Within the LHC experiments, the ability to reinterpret analyses is, perhaps unintuitively, mostly limited by the internal availability of the analysis routines to the wider collaboration as opposed to, e.g., availability of computing resource constraints. The large number of measurements and searches, the heterogeneity and complexity of the analysis software, as well the size of the collaboration, all lead to a situation in which very often only a small number of analyzers of the original analysis team is able to execute any given analysis. Furthermore, due to the collaborative development model, analyzers are typically responsible for only a subset of the analysis, which results in knowledge fragmentation. Therefore, both ATLAS and CMS are now designing an interface to store analysis-relevant information (the CERN Analysis Preservation Portal, CAP [522]) to mitigate this problem. In the context of RECAST, the software and analysis-workflow preservation aspects of this effort are most relevant. The former is mainly implemented through archiving of source-code repositories and the archival Linux Containers, which now enjoy wide-spread industry support, while internal structure of the analysis workflow is archived in CAP in the form of declarative workflow specifications such as yadage [523], which has been developed for RECAST, and the Common Workflow Language (CWL) [524]. It is planned that CAP and RECAST will utilize a common computing back-end in order to re-execute the analyses that have been preserved in the portal. As the preservation is enabled by recent technological advances and the process of archival is increasingly streamlined, it is expected that a higher number of experiment-internal analysis codes will be available for RECAST.

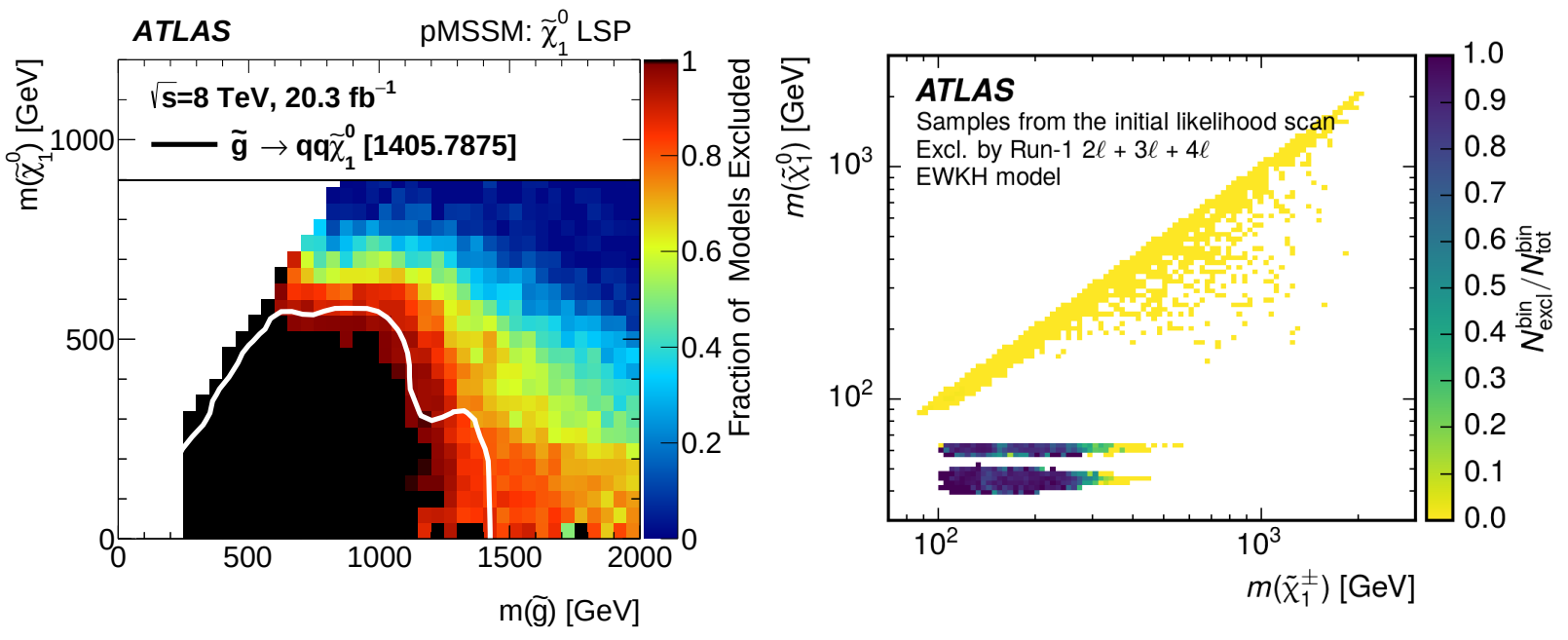

Figure 6.14: **Left:** pMSSM exclusion in the gluino-neutralino mass-plane. Results partially provided by RECAST. **Right:** Follow-up dark-matter reinterpretation. Exclusions presented in chargino-neutralino mass-plane.

### 6.6.3 *RECAST Examples*

*ATLAS-Internal Analysis Examples and Results:* A number of reinterpretation publications have been supported by the back end underpinning RECAST. After Run 1, ATLAS has conducted a thorough reinterpretation of the SUSY landscape in the context of the phenomenological pMSSM [525], a study involving 20 SUSY analyses and 50,000 fully simulated pMSSM parameter points. While at that time, most analyses had to be reinterpreted manually, the 2L electroweak analysis [526] included in that paper served as a prototype analysis and provided results using the highly automated RECAST back-ends.

The analysis was then later re-used with minimal additional effort in two further publications that focused on more domain-specific SUSY realizations: a five-dimensional dark-matter reinterpretation of electroweak seaches [527], as well as a reinterpretations in the context of general gauge-mediated models [528].

Recently ATLAS has reinterpreted ten analyses [236] in terms of models of supersymmetry with non-vanishing baryon number-violating coupling strength $\lambda''$ partly through the use of the new analysis preservation infrastructure. In such models the lighest neutralino is unstable with a decay length of

$$L(cm) = \frac{0.9\beta\gamma}{\lambda''^2}\left(\frac{m(\tilde{q})}{100\text{GeV}}\right)^4\left(\frac{1\text{GeV}}{m(\tilde{\chi}_1^0)}\right)^5 . \qquad (6.12)$$

This reinterpretation required a joint re-execution of a mix of analyses originally designed for R-parity-conserving and R-parity-violating models and special systematics have been added into the statistical analyses to account for the detector response and flavor-tagging rate of displaced jets. Results were presented in a two-dimensional parameter space of gluino mass and neutralino lifetime (or analogously the $\lambda''$ coupling) as shown in Figure 6.15. Such reinterpretations are difficult to perform outside of the experiments, as the publicly available information lack details of the detector and analayses.

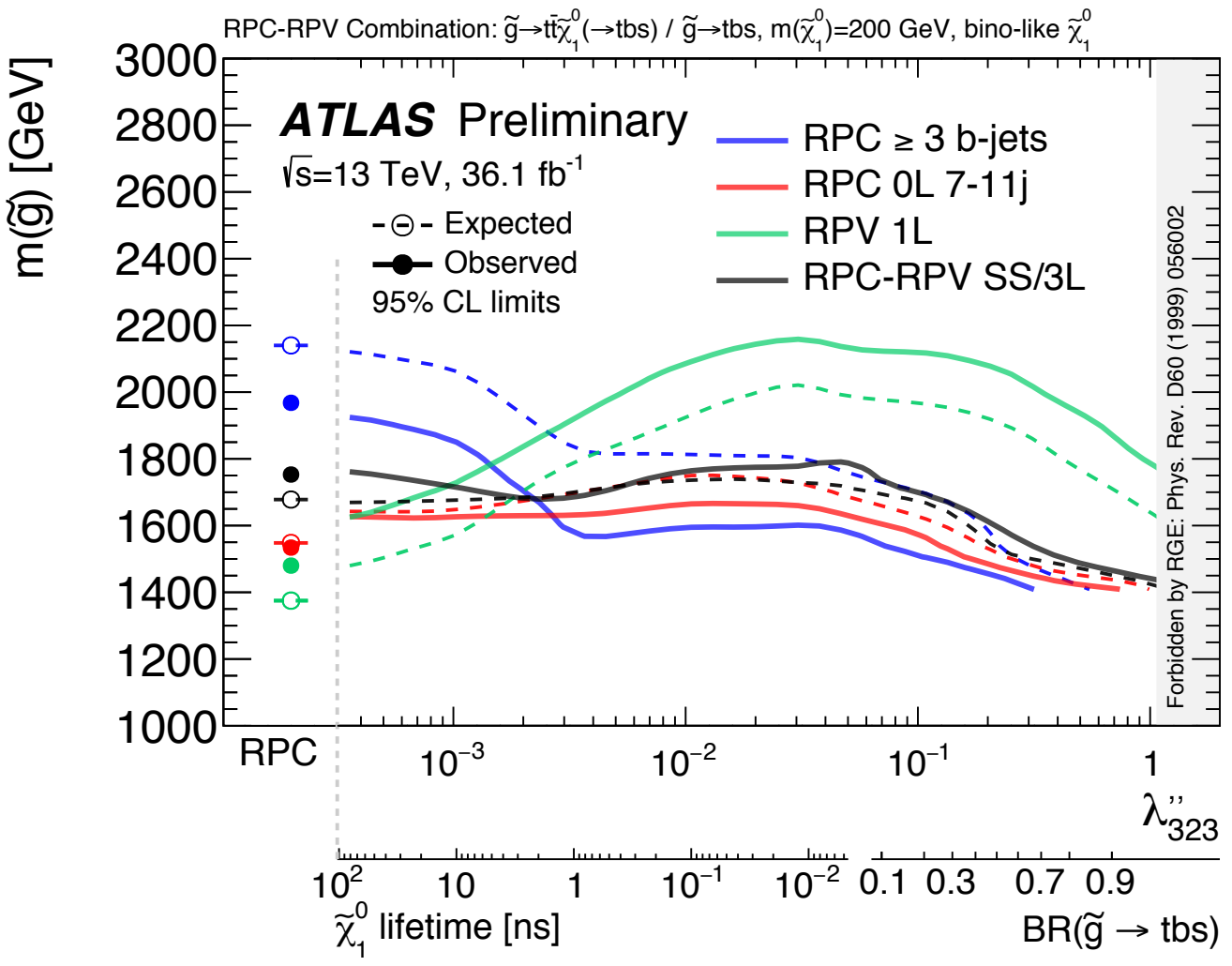

Figure 6.15: Exclusion limits as a function of $\lambda''_{323}$ and $m(\tilde{g})$ as a result of an internal reinterpretation of several ATLAS SUSY results. Expected limits are shown with dashed lines, and observed as solid. Taken from Ref. [236].

*Third-Party Tool Integration* Both the CheckMate [484, 485] and the Rivet [488] analysis catalogues have been implemented in the analysis execution framework. Both are configured to analyze events that are provided in the HepMC format [529]. Due to the modular approach of the analysis back end, a number of MC generation workflows, such as Herwig [530], SHERPA [531] or MadGraph [231], can be used depending on their ability to correctly model the desired signal.

For analyses where multiple implementations exists, e.g., from multiple third-party tools such as Rivet BSM, CheckMate or Mad-

Analysis [486, 487] as well as from multiple experiment-internal configurations (fast simulation, full simulation), RECAST will allow the community to compare and contrast reinterpretation results.

### 6.6.4 Outlook

Thanks to industry-backed technological advances, a realistic technical solution to the problem of analysis preservation for the LHC experiments and the original RECAST proposal has come into view. The initial use of such infrastructure for the reinterpretation for prompt SUSY searches is generalized easily for long-lived particle searches as the tools used for signal simulation, analysis preservation, and execution do not make simplifying assumptions on the nature of the BSM signal or analysis structure. As such, RECAST may cover reinterpretation use cases, where either third-party reinterpretations are impossible due to missing public information or limitations of third-party tools or accurate, experiment-approved results are desired.

## 6.7 Reinterpretation with Prompt Analyses

Since the decay time probability of an unstable particle follows an exponential decay law (dependent upon the mean lifetime), some percentage of the decays of the LLP will occure outside the detector, leading to an $\not{E}_T$ signature if the LLP is electrically and color neutral. Likewise, some part of the LLP decays can appear “promptly". Prompt searches with and without $\not{E}_T$ can therefore provide additional, corroborating constraints on models with LLPs, especially for short lifetimes. Therefore, it is important to understand the sensitivity of prompt searches to displaced objects.

Reinterpreting prompt searches in the context of LLPs is, however, quite nontrivial, because *a)* prompt searches may or may not make explicit requirements on the primary vertex and *b)* it is currently not documented how reconstruction efficiencies drop as a function of small displacement. Thus, the reinterpretation of prompt searches in the context of LLPs is currently best done within the collaborations themselves.

An example of such an experiment-internal reinterpretation can be found in a CMS search for an RPV SUSY model where pair-production of stops each proceed through an R-parity-violating decay to a $b$ quark and a lepton. A dedicated long-lived search for this model exists in the $e\mu$ channel [190]. This search includes selection criteria which require the transverse impact parameter to the interaction point be greater than 10 $\mu$m. This maximizes sensitivity to the long-lived model and greatly reduces standard model backgrounds. It also necessarily highly reduces the sensitivity of the search at low stop lifetime. The exclusion curve in the stop lifetime ($c\tau$) vs. top squark mass is shown in the left frame of Figure 6.16.

A reinterpretation was performed of a search for pair-production of second generation leptoquarks (LQs) [532, 533]. In this model, massive leptoquarks are pair-produced. Each of these bosons then decays to a muon and $c$ quark, leading to a final state with two muons and two jets from $c$ quarks. In the prompt limit of the RPV SUSY model with a final state of two muons and two jets from $b$ quarks, the kinematics of the LQs are nearly identical. In the LQ search, no selection is made on jet flavor. The LQ search uses final selections which are optimized to the event kinematics for each LQ mass hypothesis, but the search in general strives to remain as model-independent as possible. In this case, the reinterpretation was simply performed using the original LQ analysis, and replacing the signal samples with the long-lived RPV samples, only taking into account the reduced branching fraction to the final state with two muons and two jets. The expected and observed exclusion curves of the reinterpretation are shown in the right frame of Figure 6.16.

The reinterpretation gives large improvements for lifetimes $\leq$ 1mm, and as expected, contributes little in the large lifetime limit. This type of reinterpretation is valuable not only because it extends coverage of a given model, but also because it helps guide the analysts performing the dedicated search to focus their efforts in areas which are truly uncovered.

Reinterpretations like this one, which provide meaningful results without placing a large burden on analysis teams, should be highly encouraged. Other reinterpretations along these lines can be found in Refs. [534–536]. Another relatively simple reinterpretation is Ref. [236], although in this case, it should be noted that the original analysis was modified in a simple way, that is, the reliance on tracking information to identify jets and suppress non-collision background was removed, in order to be sensitive to long-lived gluinos.

## 6.8 *Our Proposals for the Presentation of Results*

Here we summarize the recommendations for the presentation of searches involving long-lived particles. These recommendations follow from the detailed examples presented in Sections 6.3 and 6.4.

Our primary recommendation is that the experiments provide as detailed information as possible to make a general recasting feasible. We therefore encourage the experiments to:

*A.1.* Provide LLP reconstruction and selection efficiencies at the signature or object level. Although the parametrization of efficiencies is strongly analysis dependent, it is advantageous if they are given as a function of model-independent variables (such as functions of displaced vertex $d_0$, $p_\mathrm{T}$, $\eta$, etc.), so they do not rely on a specific LLP decay or production mode;

*A.2.* Present results for at least two distinct benchmark models, with different event topologies, since it greatly helps to validate the recasting. For clarity, the input cards for the benchmark

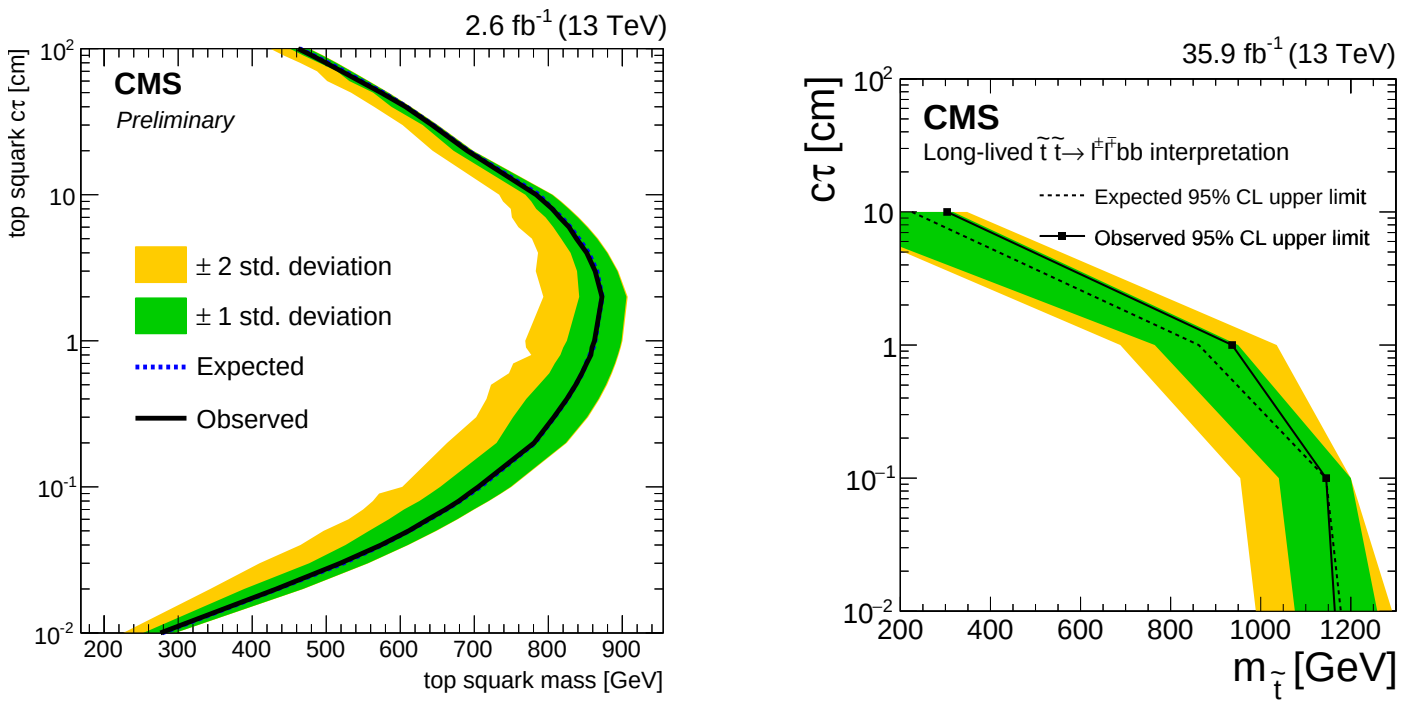

Figure 6.16: **Left:** Expected and observed 95% confidence level limits in the $c\tau$-$M_{\tilde{t}}$ plane from the displaced lepton search [190] for pair-production of long-lived stops decaying to $b$ quarks and leptons. **Right:** Exclusion on the pair-production of long-lived stops decaying to $b$ quarks and leptons in the $c\tau$-$M_{\tilde{t}}$ plane, from the prompt reinterpretation of the second generation LQ search [533].

points should also be provided;

*A.3.* Present cut-flow tables, for both the signal benchmarks and the background, since these are very useful for validating the recasting;

*A.4.* When an analysis is superseded, differences and commonalities with previous versions of the same analysis should be made clear, especially if the amount of information presented in both analyses differs. The understanding as the extent to which the information presented in an old version can be used directly in a later version greatly helps the recasting procedure, and also highlights ways in which the new search gains or loses sensitivity relative to the superseded analysis;

*A.5.* Provide all this material in numerical form, preferably on HEPdata, or on the collaboration wiki page. A very useful resource we also highly encourage is a truth-code snippet illustrating the event and object selections, such as the one from the ATLAS disappearing-track search [211] provided in HEPdata under "Common Resources".

We realize that implementing the above recommendations requires an enormous amount of time and effort by the collaborations and may not always be feasible to the full extent. However, good examples of presentations are already available, such as the parametrized efficiencies provided by the ATLAS 13 TeV displaced vertex [230] (see the auxiliary material to Ref. [516]) and the CMS 8 TeV heavy stable charged particle [493] analyses.

When the object- or signature-level efficiency maps are not feasible, providing efficiencies for an extensive, diverse array of simplified models can be useful for reinterpretation. Concerning simplified-model results, we recommend that the experiments:

*B.1.* Provide signal efficiencies (acceptance times efficiency) for simplified models and not only upper limits or exclusion curves. Note that efficiencies for *all signal regions*, not only the best one, are necessary for reinterpretation;

*B.2.* Present efficiency maps as a function of the relevant simplified-model parameters, such as the LLP mass and lifetime, with sufficient coverage of the simplified-model parameter space. While for direct production of LLPs the parameter space is 2-dimensional (LLP mass and lifetime), simplified models with cascade decays have a higher-dimensional parameter space. In these cases we strongly recommend efficiencies to be provided for a significant range of *all* the parameters;

*B.3.* Release the efficiencies in digital format (on HEPdata or the collaboration wiki page), going beyond the 2-dimensional parameterization suitable for paper plots whenever necessary. In particular, for auxiliary material, we recommend multidimensional data tables instead of a proliferation of 2-dimensional projections of the parameter space;

*B.4.* Consider in each analysis a range of simplified models which aim to encompass:

   *(a)* Different decay modes, including distinct final-state particles and multiplicities;

   *(b)* Different LLP boosts (for example, provide efficiencies and limits for distinct parent particle masses, which decay to the LLP).

Although extensive, the above recommendations for the choice of simplified models allow for a thorough comparison between the range of validity of the LLP analyses and a detailed test of recasting methods. Furthermore, when an MC-based recasting is not available, one can use the "nearest" simplified model or a combination of them to estimate the constraints on a theory of interest. Finally, if a sufficiently broad spectrum of simplified models is covered, this can be useful for quickly testing complex models which feature a large variety of signatures, and rapidly finding the interesting region in a model scan before going to more precise but computationally more expensive MC simulation.

We hope that our recommendations, in particular, points A.1–A.5, will serve as a guide for best practices and help establish a reliable and robust reinterpretation of LLP searches. The added value for the experiments and the whole HEP community will be the immediate and more precise feedback on the implications of the LLP results for a broad range of theoretical scenarios, including gaps in coverage.

# 7
# New Frontiers: Dark Showers

**Contents**

**Chapter editors:** Simon Knapen, Jessie Shelton

**Contributors:** Michael Adersberger, James Beacham, Malte Buschmann, Cari Cesarotti, Marat Freytsis, Gregor Kasieczka, Dylan Linthorne, Sascha Mehlhase, Matt Reece, Sophie Renner, Jakub Scholtz, Pedro Schwaller, Daniel Stolarski, Yuhsin Tsai

## 7.1 Introduction: The Anatomy of a Dark Shower

Hidden sectors are increasingly common features in many models that address mysteries of particle physics such as the hierarchy problem, the origins of dark matter, baryogenesis, and neutrino masses, in addition to being a generic possibility for physics beyond the Standard Model; "hidden valleys" are one such broad class of hidden-sector scenarios [59, 62]. Given the complexity of the Standard Model, such hidden sectors may very well have rich dynamics of their own, with numerous far-reaching implications for their phenomenology [60, 61, 63, 64, 537]. The main LHC signatures predicted by such hidden valley models are characterized by an injection of a large amount of energy into a hidden sector, which is then shared among a large number of relatively soft particles [63]. We will refer to this class of signatures, where rich dynamics internal to the hidden sector yields a high multiplicity of dark states, as "dark showers". Given that the particles emerging at the end of this process are necessarily both comparatively light and secluded from the SM, their lifetimes can easily become long, thus giving rise to displaced signatures at the LHC.

Long-lived particles are especially generic predictions of hidden valleys with confining gauge groups, similar to QCD in the SM [59]. It is worth noting that QCD already provides many familiar examples of long-lived particles, realizing macroscopic lifetimes through a hierarchy of scales ($\Lambda_{QCD}/m_W$) combined with approximately preserved symmetries ($K_L$, $B$ and $D$ hadrons) or restricted phase space ($n$). Also provided by QCD are numerous examples of particles that have a hierarchy of lifetimes. For instance, charged pions experience a slow decay through a very off-shell $W$ boson, while neutral pions can decay much faster through their anomalous coupling to photons. The neutral pion lifetime is thus orders of magnitude shorter than the lifetime for charged pions. Both long-lived species and a hierarchy of lifetimes between species are generic predictions and nearly unavoidable consequences of confining theories which produce dark showers. However, LLPs with a hierarchy of lifetimes also arise naturally in non-confining hidden sectors, especially in theories with multiple species. A familiar example from the literature is theories with dark photons. Here, a small kinetic mixing can make the dark photon long-lived, but a simple, well-motivated extension of the model by a dark Higgs boson introduces additional dark species and production mechanisms for dark states that are independent of the small coupling controlling the macroscopic lifetime. This naturally yields high-multiplicity events featuring particles with a hierarchy of lifetimes [32, 223, 453].

More generally, a dark shower topology can be broken down in three components (see Figure 7.1), each of which allows for a large degree of variation between models:

1. **Production.** A dark shower event begins with the production of one or more heavy states which decay into the dark sector. These heavy states could be part of the SM, most notably the Higgs boson, or could be a new particle from the menu of BSM states we have become accustomed to ($Z'$s, color triplets/octets, electroweak doublets/triplets, etc.). In some cases the produc-tion mechanism provides an important trigger handle (e.g., $VH$ production for the Higgs), but this is not universal (e.g., $Z'$ production). The options for production are laid out in Section 7.2.

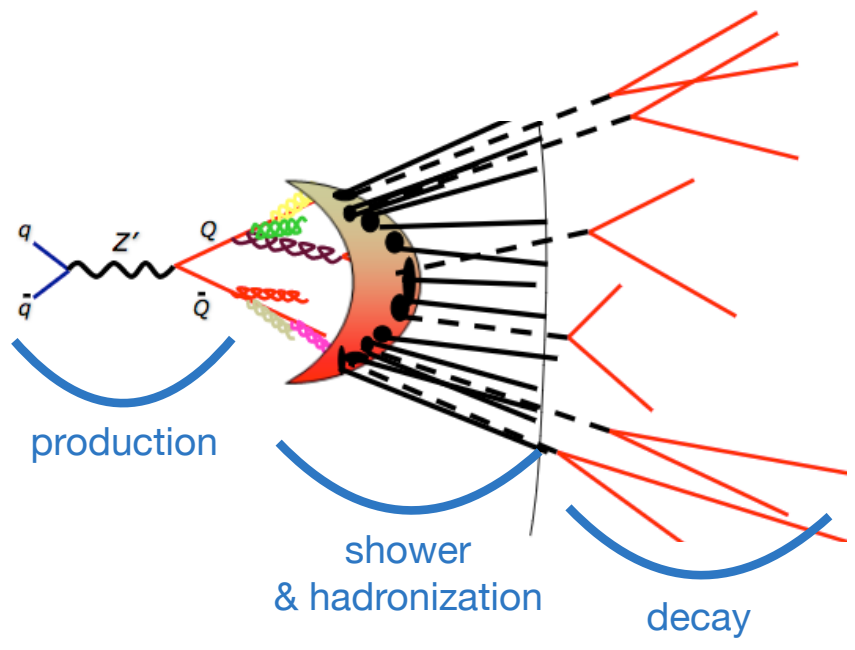

Figure 7.1: Schematic representation of a dark shower event, in this case from hidden valley model with a $Z'$ production portal. Figure adapted from [64].

2. **Showering and hadronization.** If the dark sector contains an asymptotically free gauge group, the originally produced particles will shower and possibly hadronize within the dark sector. This yields a final state with a potentially large number of dark states, similarly to how quarks and gluons undergo showering and hadronization to yield a jet of hadrons. The shape of the shower and the $p_{\mathrm{T}}$ spectrum depends on the coupling of the dark gauge group: the shower may be pencil-like, as in QCD, completely spherically symmetric, or something in between. Alternatively, it is possible that the hidden sector does not contain a gauge sector but instead features a perturbative cascade decay over a large number of states. Indeed, in certain cases the perturbative picture is dual to the strongly coupled showers. In general, showering and hadronization are the sources of greatest uncertainty from a theory perspective; the current status and some new results are discussed in Section 7.3.

3. **Decay.** After the dark degrees of freedom hadronize (or reach the bottom of the cascade in perturbative models), they can decay back to the SM. The decay may occur through the same (off-shell) portal as the production, but this is not essential, and one may expect multiple species with a range of lifetimes. The specifics of the decay step (e.g., muon multiplicity) are particularly important if there is no good trigger handle from the production topology. Decays are frequently interconnected with showering and especially hadronization, however, and it is not tractable to enumerate all possibilities without making simplifying assumptions and/or inserting additional theory prejudice. For this reason it is often useful to focus on the species with the largest multiplicity and/or shortest (macroscopic) lifetime; this frequently provides a reasonable guide to the overall signatures, just as one may obtain a reasonable $O(1)$ picture of QCD jets by considering only their pions. We survey a (non-exhaustive) list of popular decay portals in Section 7.4.

A priori it is typically possible to construct a model by choosing an ingredient from the menu of options for each of the three components outlined above. This is an enormous model space, and it may appear daunting to construct searches capable of capturing all possibilities. On the other hand, the signatures of these models are often so striking that they enable powerful, inclusive searches, sensitive to a very large portion of this overall model space, provided that triggers allow the event to be recorded. Toward this end, it is useful to observe that dark shower events have the following generic features:

1. Events have a *variable and potentially large multiplicity of LLPs*. The number of produced particles of various dark species depends on the details of the parton shower and/or hadronization, as in QCD, and varies from event to event. Typically there are more than two LLPs per event.

2. The BSM species produced in dark showers exhibit a *hierarchy of proper lifetimes*. This could result in production of, e.g., mostly prompt particles with a few displaced decays; mostly invisible detector-stable particles with a few displaced decays within the detector; or anything in between.

3. *LLPs are not generically isolated*, i.e., they often appear within $\Delta R \lesssim 0.4$ of other LLPs and/or prompt objects (such as the decay products of short-lived species originating from the same shower).

4. The *energy flow in the event reflects the evolution of the BSM parton shower and hadronization*, and thus looks non-SM-like. For instance, hidden sector jets may be either narrower or broader than QCD jets, depending on the hidden sector gauge group, gauge coupling, and particle spectrum. Additionally, SM particle multiplicity — i.e., the relative fractions of pions, kaons, etc., produced in a jet — differs from QCD. While energy flow can be a powerful discriminant to separate dark showers from the SM at any stage from the trigger level forward, it is model dependent and must be assessed on a case-by-case basis. Note that this also implies that traditional (displaced) jet triggers do not suffice to cover the general model space, and alternative trigger strategies (e.g., track or muon multiplicity) are needed as well.

The existence of $\gtrsim 2$ LLPs per event, and indeed frequently $N_{LLP} \gg 2$, generally ensures that these events can be easily distinguished from background if they can recorded on tape and subsequently reconstructed. However, the unique features of events with dark showers require new strategies to ensure that this class of theories is actually captured by the trigger. In particular, non-isolation and the hierarchy of proper lifetimes can result in qualitatively novel collider signatures that require new triggering approaches. Meanwhile, event-level observables such as non-SM-like energy flow or particle multiplicity are potentially powerful but highly model-dependent discriminants, and must be considered on a case-by-case basis. We discuss the strengths and shortcomings of existing triggers and off-line strategies in Sections 7.5 and 7.6, as well as a number of new ideas. An executive summary of our main points and recommendations is provided in Section 7.7. The chapter concludes with a collection of example models for which Monte Carlo event samples are currently available (see Section 7.8).

## 7.2 *Production*

Events with dark showers generically begin with the pair production of dark partons $Q_D$. In most cases the two produced partons are of the same species, but this does not necessarily need to be true. For clarity and simplicity, we confine ourselves to the case where $Q_D$ is a SM singlet. Then the production modes for $Q_D$ can be simply related to the production modes discussed for neutral LLPs in Section 2.4, in the chapter on Simplified Models. In particular, the most relevant production modes are:

- **Heavy Parent (HP):** Pair production of a SM-charged heavy parent $X_D$, which subsequently decays via $X_D \to Q_D + (\text{SM})$. The SM quantum numbers of the parent $X_D$, together with its mass, control both the overall production cross section and the typical prompt accompanying objects in the event. Depending on the lifetime of $X_D$, showering can begin before or after its decay. The model of Ref. [325] features this production mechanism, where the parent $X_D$ is a heavy scalar carrying both color and dark color charges. After QCD pair-production, the mediators each decay into a visible jet and a dark shower.

- **Higgs (HIG):** Production in exotic Higgs decays, $h \to Q_D\bar{Q}_D$. As the Higgs boson provides an especially sensitive window into low-mass dark sectors, this production mechanism is one of the best-motivated at the LHC [60, 110]. In particular, Higgs portal production is the dominant production mechanism in many Twin Higgs and related models of neutral naturalness [111, 256, 538]. The Higgs boson determines the overall mass scale of the event, often awkwardly low for LHC triggers, while the branching fraction remains a free parameter. As discussed in Section 2.3.1, the SM Higgs has a characteristic set of accompanying prompt objects, which can extend trigger options. As also emphasized in Section 2.3.1, this category additionally encompasses production through parent Higgs-portal scalars, which may be either heavier or lighter than the SM Higgs.

- $Z'$ **(ZP):** Here a new $Z'$ boson couples to both SM and hidden sector states, allowing for production through $q\bar{q} \to Z' \to Q_D\bar{Q}_D$.

This scenario was developed in the original hidden valley models [59, 62], and considered again recently in Ref [366]. The $Z'$ mass determines the overall mass scale, while production cross sections depend on its couplings to both SM and hidden sector states. The only typical accompanying objects are ISR radiation.

- **Charged current (CC):** Here the parent dark states couple to the SM through the neutrino portal, $\mathcal{O}_{int} = HL\mathcal{O}_{\text{dark}}$. When the dark states carry charge under a dark gauge group, this coupling requires (at least) two dark states, one fermionic and one bosonic. For instance, if the dark sector contains both scalar and fermionic fundamentals, $\phi$ and $\psi$ respectively, then one can construct the dimension five interaction $\mathcal{O}_{int} = HL\phi^*\psi$. This production mode has not been well studied in the literature. It is worth noting that given this dark field content, it is generically possible to also construct Higgs portal couplings, $\mathcal{O}_H = |H|^2|\phi|^2, |H|^2\bar{\psi}\psi$, which can generally include lower dimensional operators.

As we have taken $Q_D$ to be SM singlets, the first process discussed in Section 2.4, direct pair production, is typically negligible. Of course, if BSM mediators connecting the dark sector to the SM, like the $X_D$ and $Z'$ in the examples discussed above, are heavy enough to be integrated out at LHC scales, then the resulting higher-dimension operators can mediate direct pair production of $Q_D$. Single production of $Q_D$ is generally suppressed, and in many cases impossible, in particular when $Q_D$ transforms nontrivially under an unbroken gauge group. In Abelian theories, single production of a dark gauge boson is possible, e.g., through a loop of heavy bi-charged matter; this gauge boson can then subsequently shower. In perturbative cascades, a single BSM state $Q_D$ may certainly be produced, but whether $Q_D$ goes on to produce one or more "showers" is highly dependent on the detailed kinematics of the event.

In contrast with most of the simplified models in Section 2.4, after a parton $Q_D$ is produced it undergoes extensive evolution in the hidden sector, so that there is no one-to-one connection between the initial $Q_D$ and relevant detector objects. Thus, an event that begins with two partons may result in a final state that contains two pencil-like jets, each containing more than one displaced object; a spherical distribution of displaced objects; or anything in between.

The most important consequences of production modes for our purposes are twofold. First and most importantly, the production mode informs the types of prompt accompanying objects in an event, as well as determining overall event rate and energy scale. These accompanying objects can be useful levers to distinguish signals from SM background, beginning at the trigger level. Secondarily, production modes rely on a mediator that couples the dark sector to the SM, which may be a BSM particle like a bifundamental $X_D$ or a $Z'$, or a SM particle like the Higgs boson. In many models, this mediator-SM coupling also ultimately governs the decay of the dark sector states back into the SM. This is the case for the examples of Refs. [59, 60, 62, 111, 256, 325, 366, 538, 539], and is certainly the minimal possibility. It is worth mentioning, however, that decays may be governed by a different interaction than production. As a simple perturbative example, consider the hidden abelian Higgs model. This theory can realize dark showers in a variety of ways; consider for concreteness the perturbative cascade decay chain $h \to ss, s \to Z_D Z_D$, which can yield collimated pairs of dark photons when $m_s \ll m_h$ (see, e.g., [540]). In this case a Higgs portal interaction governs production, but the long-lived $Z_D$ decays back to the SM through a separate vector portal interaction.

## 7.3 Shower

### 7.3.1 Motivation

A familiar feature of QCD is the formation of jets, sprays of approximately collinear hadrons arising from a parton emitted in a hard scattering process. The physics of jet formation is independent of hadronization or confinement, originating in the singularities of the weakly coupled theory at short distances [541], where the 't Hooft coupling is $\lambda \equiv g_s^2 N_c \ll 1$. In perturbation theory, the differential probability for a quark to radiate a gluon carrying a small energy fraction $z$ at small angle $\theta$ is

$$P(z, \theta)\, dz\, d\theta \sim \frac{\lambda}{4\pi^2} \frac{dz}{z} \frac{d\theta}{\theta}, \tag{7.1}$$

independent of the underlying hard process. The logarithmic divergences at $z, \theta \to 0$ indicate that perturbative theories favor radiation that is *soft* (low energy) and *collinear*. This enhanced emission of collinear radiation is the source of jets, even in theories that do not confine at all, such as the perturbative, conformal Banks–Zaks gauge theories [542] [1]. The large logarithms appearing in these calculations can be numerically resummed through Markov Chain algorithms, leading to the *parton showers* widely used to model jets in Monte Carlo simulations.

In the limit of strong 't Hooft coupling[2] ($\lambda \gg 1$) perturbation theory breaks down; soft and collinear radiation are no longer enhanced over more general radiation at wide angles. In QCD, the coupling gradually runs strong in the infrared, but other gauge theories (which could be realized in nature as hidden valleys [59]) exist that have an intrinsically large 't Hooft coupling persisting over a wide energy range. Such theories can be understood with the use of gauge/gravity duality. Examples may be conformal, e.g., strongly coupled $\mathcal{N} = 4$ super-Yang–Mills, or confining, e.g., those detailed in Refs. [544, 545]. Such large-$\lambda$ theories were conjectured to lead to spherical event shapes [63], a result that has been directly proven for strongly-coupled large-$N_c$ CFTs [546]. At colliders, these spherical events would lead to characteristic soft unclustered energy patterns (SUEPs) [175, 235, 547]. An illustration of the range of event shapes from jets to SUEPs is shown in Figure 7.2.

[1] See Ref. [543] for a brief and clear recent introduction to jet physics emphasizing the role of scale invariance.

[2] When making this and similar statements below, we always implicitly mean the coupling at the energy scales being probed by a given measurement. For example, while QCD does become strongly coupled near its confinement scale, treating the theory as weakly coupled is justified provided one only ever talks about jet masses and energies at much larger scales.

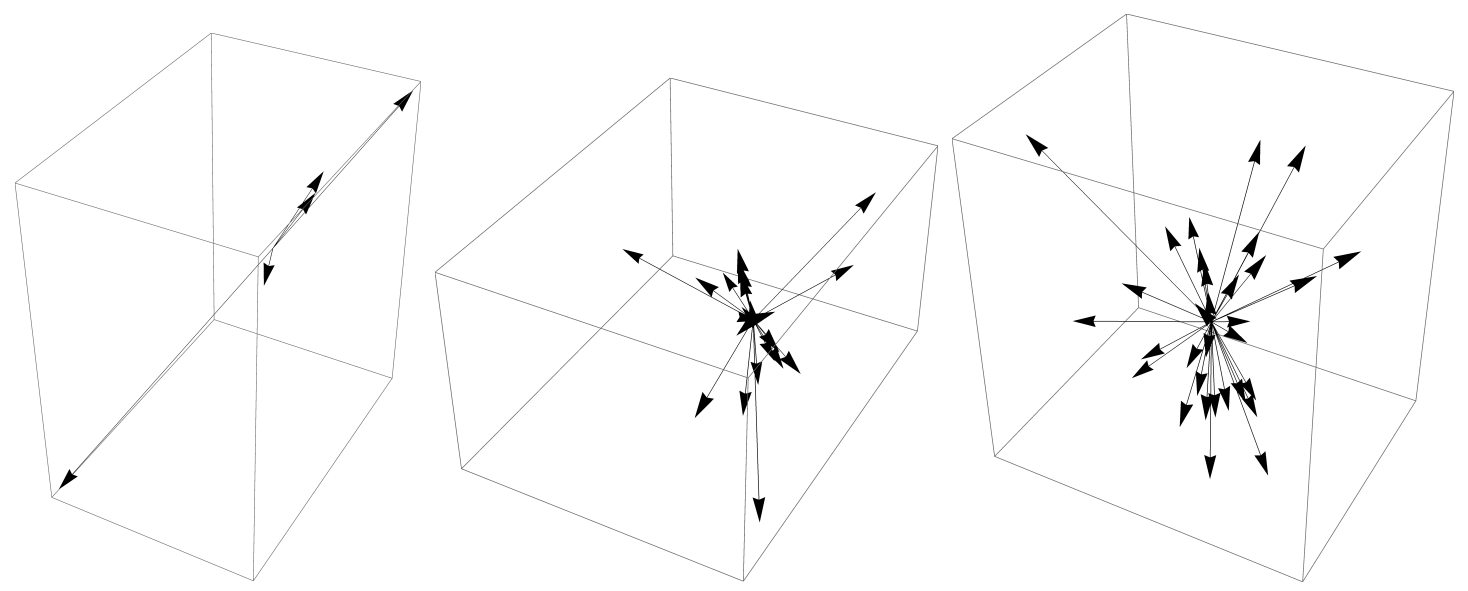

Figure 7.2: Illustrations of event shapes, from jetty (left) to spherical (right) with an intermediate event in the middle. The values of Sphericity and Thrust are: (left) Sphericity = 0.00636, Thrust = 0.991; (center) Sphericity = 0.530, Thrust = 0.706; (right) Sphericity = 0.940, Thrust = 0.521.

Thus we have two well-understood regimes, jets at $\lambda \ll 1$ and SUEPs at $\lambda \gg 1$. There is a gap in the middle when $\lambda \sim 1$ over a wide range of energies. Both perturbative QCD and gravitational duals approach strong coupling in this regime (from different sides) and cannot be trusted to give accurate predictions; for some questions, the predictions may not even be qualitatively accurate.

This regime is of interest because we want to ensure that LHC experiments do not miss a hidden valley signal simply because it looks different than expected. We should aim to be able to trigger on and analyze these events. While sufficiently long lifetimes may provide useful trigger handles, theories at intermediate 't Hooft coupling could also occur with prompt decays while still failing to provide the types of trigger handles typically associated with prompt hard production and decay. Given our inability to reliably calculate the predictions of this scenario, in this section we will take a pragmatic approach: we push the two tools we have, perturbation theory and gauge/gravity duality, into a regime where we do not fully trust them. To the extent that their predictions overlap, one could gain confidence in the qualitative picture we obtain. Where they differ, one would hope that experimental strategies broad enough to encompass the range of possibilities would also be sensitive to poorly modeled scenarios. In the following we present some initial results, which we aim to improve and expand upon in future published work [548].

### *7.3.2 Phenomenological Models*

#### *7.3.2.1 Parton Showers and Their Limits*

One can approach the intermediate regime from the direction of weak 't Hooft coupling using parton shower methods. The coupling is then a direct parameter of the model whose value is easy to vary. However, naïvely setting the coupling to large values would quickly lead to unphysical results. The reason is that the simplest derivation of parton shower evolution equations assume that the final state can be organized as a series of splittings where $z, \theta \ll 1$ at each iteration. Since the full matrix element for a given final state has no uniquely-defined splitting history associated with it, this soft, collinear condition needs to hold as a function of the final-state kinematics alone. As the coupling is increased, the increased showering probability means that there is an increased chance of populating the phase space while violating these conditions. Regions of phase space without nominal soft-collinear enhancements (but in truth populated with comparable probability) will end up under-populated, leading to more jet-like events than the underlying theory actually predicts.

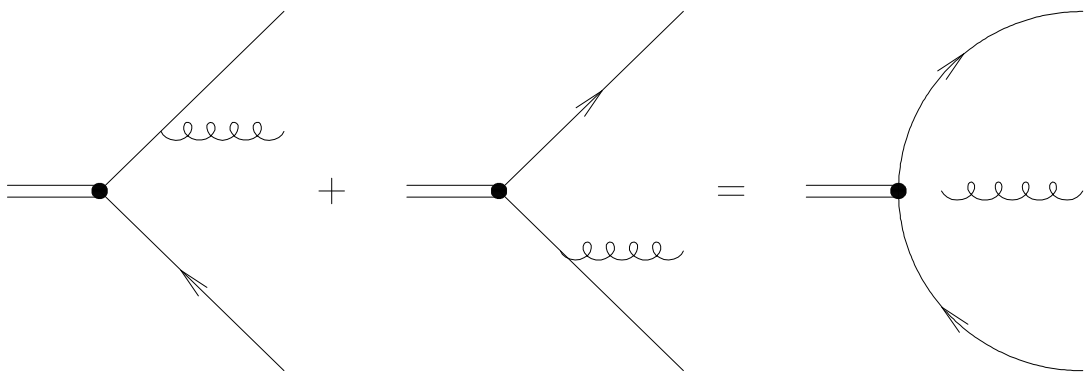

Figure 7.3: Schematic view of a $2 \rightarrow 3$ branching process, with the initial two-particle state approximated as a localized dipole. (Figure courtesy of Ref. [549].)

Things can be somewhat improved by a more careful consideration of parton-shower methods. It is currently known how to implement iterated corrections allowing one to relax one, but not both, of the inequalities given above. Implementations which allow for the correct inclusion of finite-$z$ (but small-$\theta$) effects can be accomplished by extending the approximation of Eq. (7.1) with terms non-singular in $z$.

If we have reason to believe that the wide-angle structure of dark showers will be more critical to detecting models in the transition region, sacrificing finite-$z$ corrections for a better understanding of the finite-$\theta$ region is preferable. This requires going beyond the $1 \rightarrow 2$ splitting picture of Eq. (7.1) as information about multiple particles in the event must be encoded. A known solution is to phrase the shower in the language of $2 \rightarrow 3$ evolution kernels [550], where radiation is treated as coming from dipoles rather than individual charges, as in Figure 7.3. Necessarily more complicated, in this approach the splitting function describing the emission of a gluon with momentum $p^\mu$ is then expressed as

$$P(p_r^\mu)\, d\Phi_r \sim \frac{\lambda}{s_{ij}} \left[ \left(1 - \frac{s_{ir}}{s_{ij}} - \frac{s_{rj}}{s_{ij}}\right) \left(\frac{2s_{ij}^2}{s_{ir}s_{rj}} + \frac{s_{ir}}{s_{rj}} + \frac{s_{rj}}{s_{ir}}\right) \right. \quad (7.2)$$

$$\left. + \text{non-singular} \right] d\Phi_r, \quad (7.3)$$

$$, \quad s_{ij} = (p_i + p_j)^2. \quad (7.4)$$

for the process $a + b \rightarrow a + r + b$. Features related to the expected broadening of jets at larger coupling will be well-modeled deeper into the transition region with a dipole shower. Angular structure will be correctly reproduced even at larger couplings while final-state energy sharing of collinear particles becomes increasingly untrustworthy. When some sort of angular localization of energy flow can be expected (i.e., soft-collinear enhancements are still present) the corrections at finite splitting angle may still provide a good approximation.

#### 7.3.2.2 *Gauge/Gravity Duality: Spheres to Jets*

Gauge/gravity duality allows calculation in $\lambda \gg 1$ theories with spherical events, from which we extrapolate toward the $\lambda \sim 1$ regime. The simplest case is AdS/CFT duality, where events are perfectly spherical [546]. Anti-de Sitter (AdS) space is a warped product of a $(3+1)$D Minkowski space with an infinite fifth dimension. To obtain a theory with a discrete spectrum, we cut off the fifth dimension of AdS space at a hard wall or *IR brane* as in Randall–Sundrum (RS) models [551]. 5D fields then decompose into a tower of 4D Kaluza–Klein (KK) modes (dual to hadrons in the gauge theory), each with an associated wave function in the fifth dimension. This wave function, as well as the 4D mass, is calculated by solving the equations of motion up to quadratic order. The couplings between 4D mass eigenstates are proportional to the overlap of their wave functions in the fifth dimension.

Heavy KK modes decay to lighter modes, which decay to still lighter modes, populating a cascade of particles. The case of a flat (unwarped) extra dimension yields sinusoidal wave functions and modes with linearly spaced masses. In this case, the *KK number* is a conserved discrete momentum in the fifth dimension, and KK modes decay only to daughters at threshold. The RS model breaks translation invariance in the fifth dimension, so KK number is no longer exactly conserved and a variety of decays are possible. Nonetheless, for the simple model of a bulk cubic coupling, KK number is still approximately conserved, as illustrated in Figure 7.4, top panel, and the sum of the two daughter modes is approximately equal to the KK mode of the parent. At each step of the cascade, daughter particles have small momentum. This results in spherical event shapes, with no highly boosted daughters [552]. Here we have studied a scalar field with a $\Phi^3$ interaction for simplicity, with the expectation that qualitative features of the event shapes persist in more realistic models.

To push our toy model into the jetty regime, we introduce interactions that explicitly break KK number themselves. We continue to study the simple case of a scalar field, but now, to move toward jettier events, we include the $\Phi^3$ interaction as a boundary term, restricted to the IR brane. As shown in Figure 7.4, right panel, this opens up a much wider range of possible decays with greater phase space, which lead to more variety in the resulting event shapes. We emphasize that moving the $\Phi^3$ term to the boundary is a simple toy model for the purpose of this section that accomplishes the goal of altering event shapes. However, it is not expected to be a good

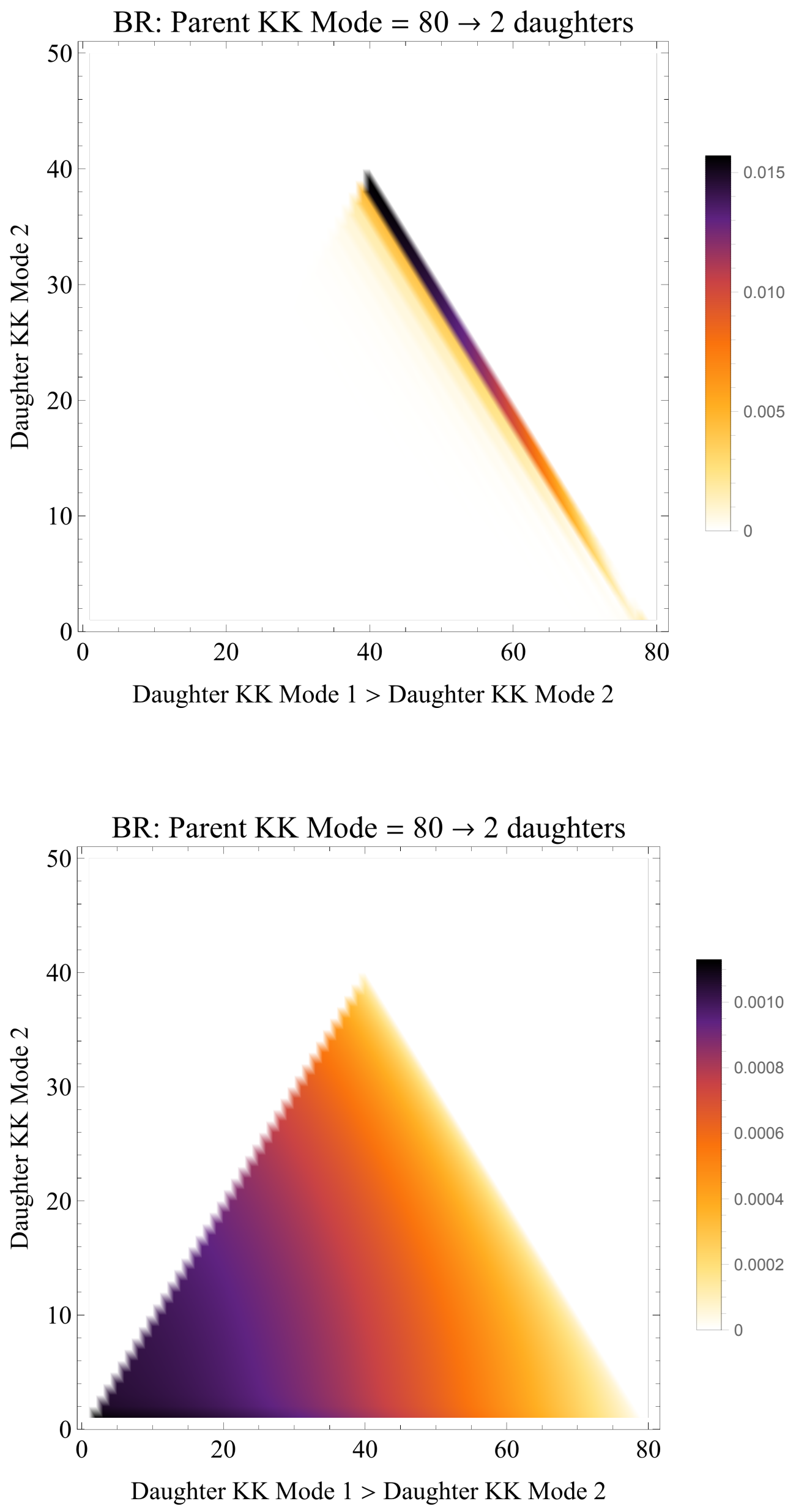

Figure 7.4: The branching ratio from a parent with KK-number 80 into two daughters for a cubic interaction on the bulk (top) or on the boundary (bottom). The nonzero probabilities occur along the line where KK-number is conserved for the bulk interaction, whereas the boundary interaction has non-zero probability to decay with large jumps in phase space.

approximation of the dual of an actual confining gauge theory with smaller $\lambda$.

Although the IR brane interaction is used here as an ad hoc tool for generating less spherical events, we expect that similar results can be achieved in a more principled way. As $\lambda$ decreases toward $O(1)$ values, the expectation from gauge/gravity duality is that more bulk fields, dual to the many single-trace operators in the gauge theory, become light. This suggests that models with several interacting bulk fields are a more faithful toy model of the physics of the dual gauge theories at intermediate $\lambda$. Such models could be of interest for a wider set of questions than event shapes, as they move the AdS/QCD toolkit—always a toy model at best—somewhat qualitatively closer to QCD. Such models and their consequences will be studied in a forthcoming publication [548].

### 7.3.3 *Results*

We wish to characterize to what extent our available methods allow us to model the range of behaviors we might expect from new showering sectors. Since we do not expect collimated sprays of final-state particles to be a sensible way of organizing information in the event in all cases, for our purposes here we focus on observables that can be defined globally. In particular, we study a pair of event-shape variables that have proven useful to both establish and provide precision data on the non-abelian nature of QCD.

Sphericity is defined as the scalar sum of the two smaller eigenvalues of the sphericity tensor

$$S^{ab} = \frac{\sum_i p_i^a p_i^b}{\sum_i |\boldsymbol{p}_i|^2}, \tag{7.5}$$

where the sum is over all final-state particles in the event [553]. With the eigenvalues $\lambda_i$ defined in decreasing order, we have $S = \frac{3}{2}(\lambda_2 + \lambda_3)$, which can take on the values $0 \le S \le 1$. Thrust is instead defined via a maximization procedure with respect to all possible axes in the event [554] ,

$$T = \max_{|\boldsymbol{n}|=1} \frac{\sum_i |\boldsymbol{n} \cdot \boldsymbol{p}_i|}{\sum_i |\boldsymbol{p}_i|}. \tag{7.6}$$

Both observables essentially measure the divergence of an event from the pencil-like final-state structure of a $2 \to 2$ scattering process without making any direct reference to jets.

Historically, thrust had the advantage of being infrared and collinear safe. In a given event, the change in the thrust due to an additional radiated parton vanishes as the parton becomes soft or collinear to the thrust axis. Singular regions of phase space thus do not contribute to finite values of the thrust, and its measured distribution in QCD is well described by a perturbative calculation up to corrections that scale as $O(\Lambda_{\text{QCD}}/Q)^2$, where $Q$ is a high-energy scale associated with the total system being probed by the thrust. This is not the case for sphericity—specifically a perfectly collinear splitting still changes the value of the sphericity tensor. A perturbative calculation of the sphericity is then divergent for finite values of sphericity, a divergence which can be tamed by either an explicit cutoff at the hadronization scale supplemented by a phenomenological hadronization model (the approach taken by Monte Carlo generators) or by absorbing it into a form factor (as done in analytic calculations). In either case $O(1)$ sensitivity is induced to the region of phase space dominated by non-perturbative hadronization effects. For our concerns, this difference can be turned into an advantage, as any apparent difference between the two observables can act as a diagnostic of the sensitivity of our predictions to the non-perturbative parameters in the parton shower.

We generated events from models expected to yield a range of behaviors. We considered extra-dimensional models with both bulk and boundary interactions, with the former expected to yield very isotropic events. For the parton shower, we used a modified version of the VINCIA dipole-antenna parton shower [555] in which an $SU(N)$ gauge theory with only light quarks showers and hadronizes into light mesons with masses $m_{\pi_v}/\Lambda_v \sim m_\pi/\Lambda_{\text{QCD}}$. We then varied the Coupling boundary conditions and one-loop running while adjusting shower cutoffs to ensure that couplings remain perturbative throughout the parton shower.

We summarize the results in Figure 7.5, with the uncertainty in the parton shower distributions coming from considering both transverse-momentum and dipole-virtuality shower ordering. The similar behavior of the two distributions indicates that sensitivity to non-perturbative effects from hadronization are not large. The lowest sphericity/highest thrust distribution is fairly close to that expected from QCD, but a wide range of non-QCD-like behaviors is observable. A significant fraction of the allowed range for these observables is populated by a combination of the extra-dimensional and the parton shower approaches, and the boundary interaction KK model and parton shower models give similar results.

Examining the behavior of observables that are perturbatively incalculable in the weakly coupled theory indicates that a degree of caution is warranted, however, with the two approaches giving qualitatively different results to certain questions. As an example, we look at the correlation between sphericity and total event multiplicity for the two closest parton shower and extra-dimensional models. Displayed in Figure 7.6, we see that multiplicity is broadly correlated with sphericity in the parton shower, while being nearly sphericity independent for the KK model. Such qualitative differences between the parton shower and the extra-dimensional models warrant more detailed study of event behavior in the transition region, while both approaches would benefit from further consideration of how jet-level observables, whether physically well-justified or not, vary over their accessible ranges.

## 7.4 *Decay*

In general, the mass spectrum of confining hidden sectors is poorly known, with the exception of a few special cases, like pure glue confining theories [556] and the SM QCD sector. The dynamical process of hadronization is even less well understood, and one must typically resort to uncontrolled extrapolations from the measured fragmentation functions in the SM. Moreover, states with different $CP$ and spin quantum numbers can have greatly different lifetimes even in well-understood examples [59, 537, 557], a problem that is further exacerbated if the hidden sector contains one or more approximate flavor symmetries. For concreteness we here primarily focus on the case where only one dark species has a detector-scale lifetime; other dark states may either decay promptly to the SM or appear as MET. A useful special case realizing this scenario is one in which all internal hidden sector decays are sufficiently rapid for heavier hidden states to promptly decay down to the lightest state in the spectrum, which in turn decays to the SM. The discussion below is focused on the possible properties of this lightest hidden state which decays to the SM. We refer to it as "the" long-lived particle (LLP), keeping in mind that in complete models there may very well be multiple species of LLPs. It is therefore important to emphasize that *experimental strategies should not be overly optimized* towards the naïve assumption of a single LLP giving rise to all visible decays, as it may not be generically true. In particular, an inclusive experimental strategy should avoid making detailed assumptions about the distributions of proper lifetimes of long-lived states in the event.

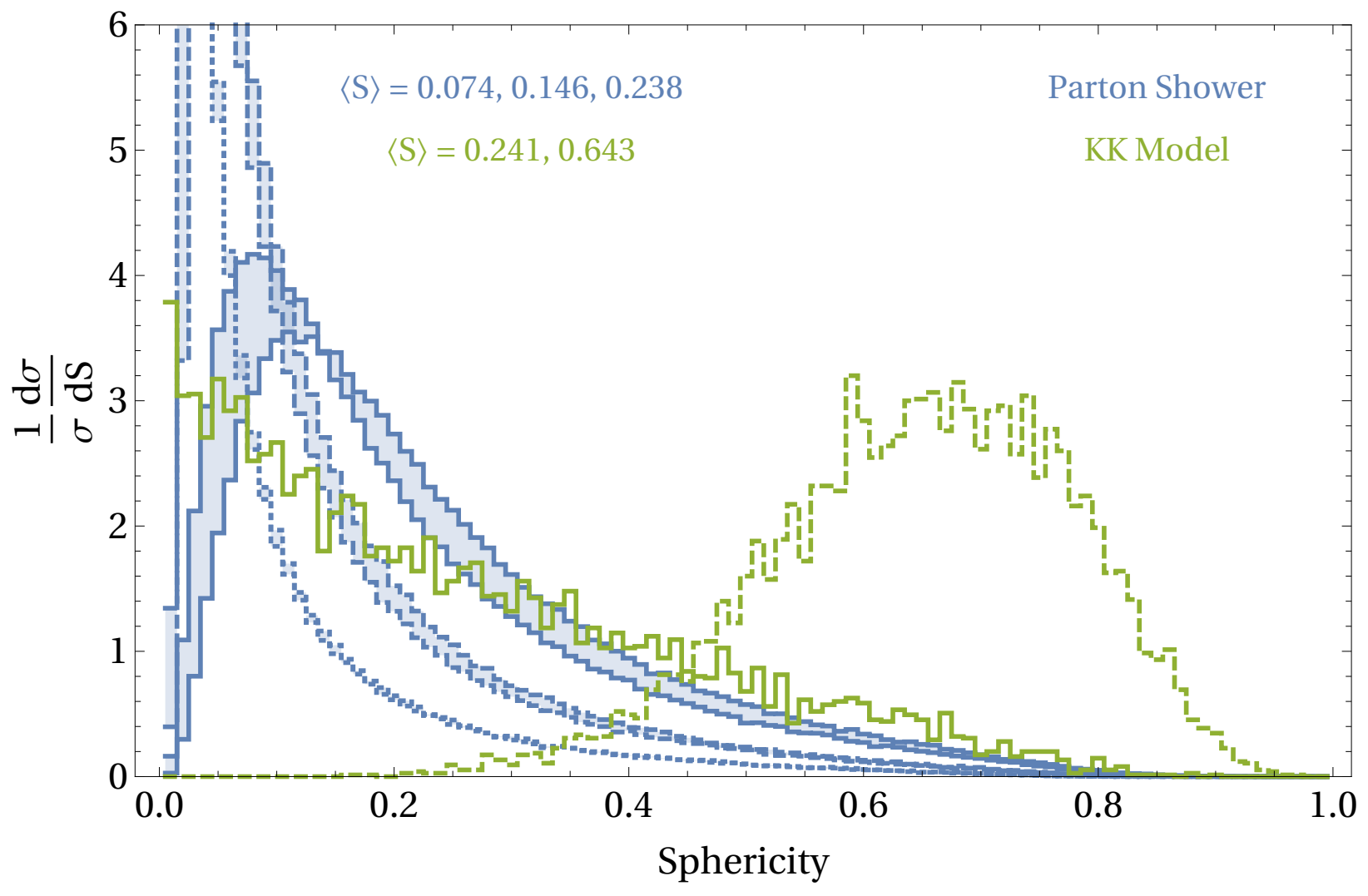

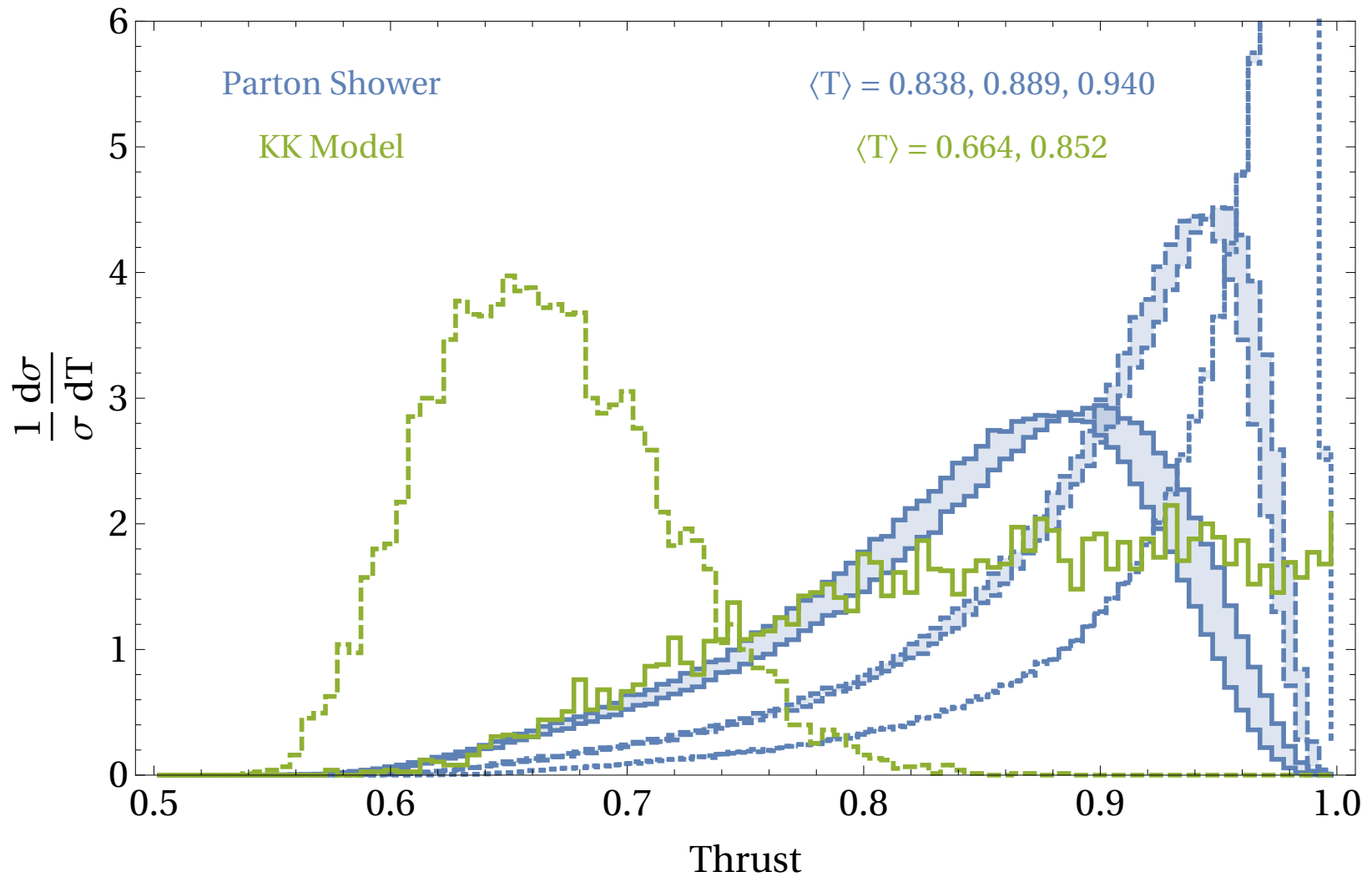

Figure 7.5: Comparison between accessible ranges of parton shower and AdS/CFT-inspired models (labeled "KK model") for sphericity and thrust. In the parton shower, the curves correspond to events produced at $\sqrt{s} = 100\Lambda$, with confinement scale $\Lambda$; the shading indicates the uncertainty resulting from comparing transverse momentum and dipole virtuality shower ordering. The $\beta$ functions of the theories are tuned such that $g^2(100\Lambda)/4\pi$ is $0.06, 0.12, 0.24$ for the 3 distributions. In the KK model case, the dashed curve corresponds to a bulk interaction while the solid curve corresponds to a boundary interaction. We also show expectation values for all distributions.

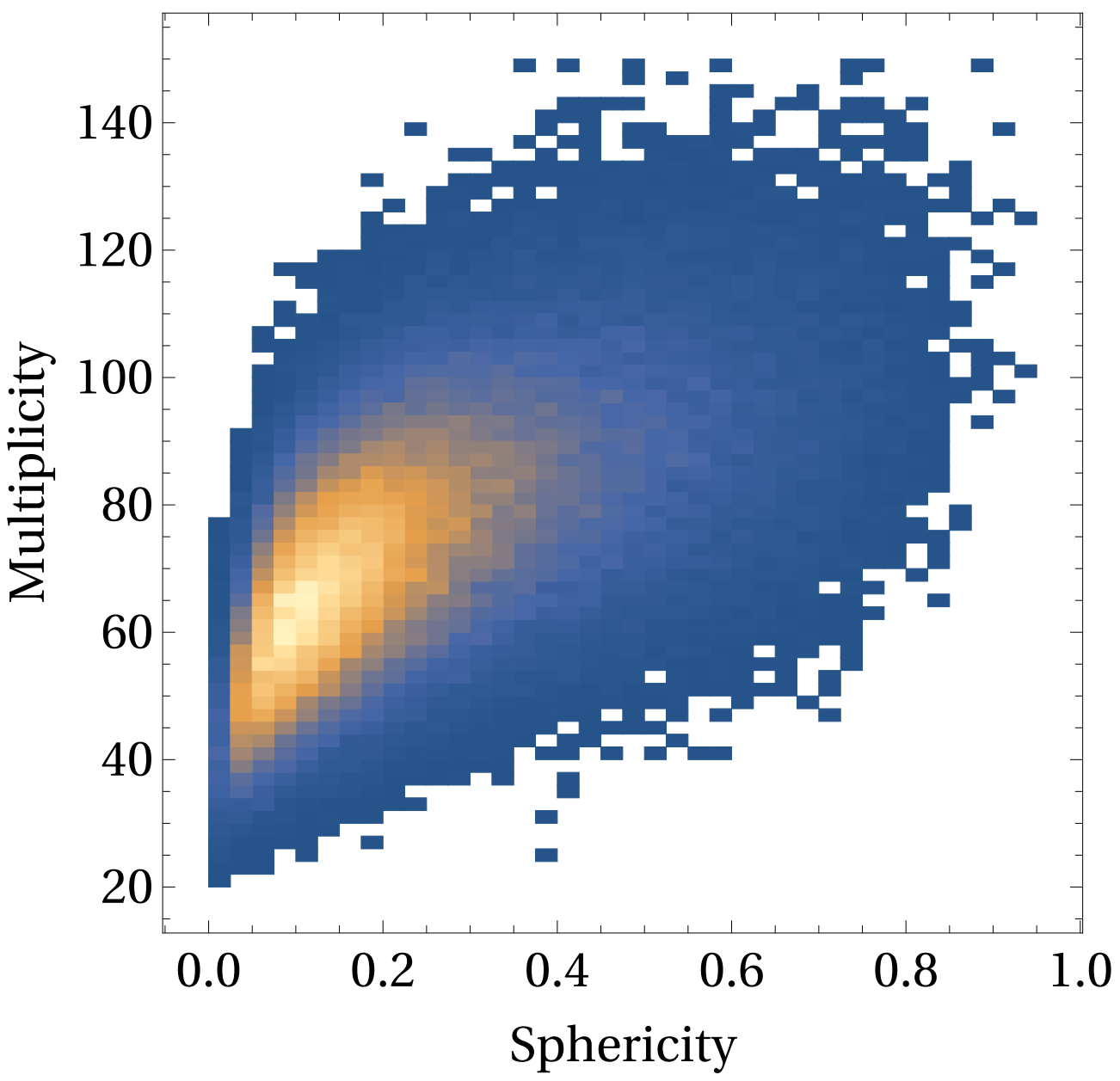

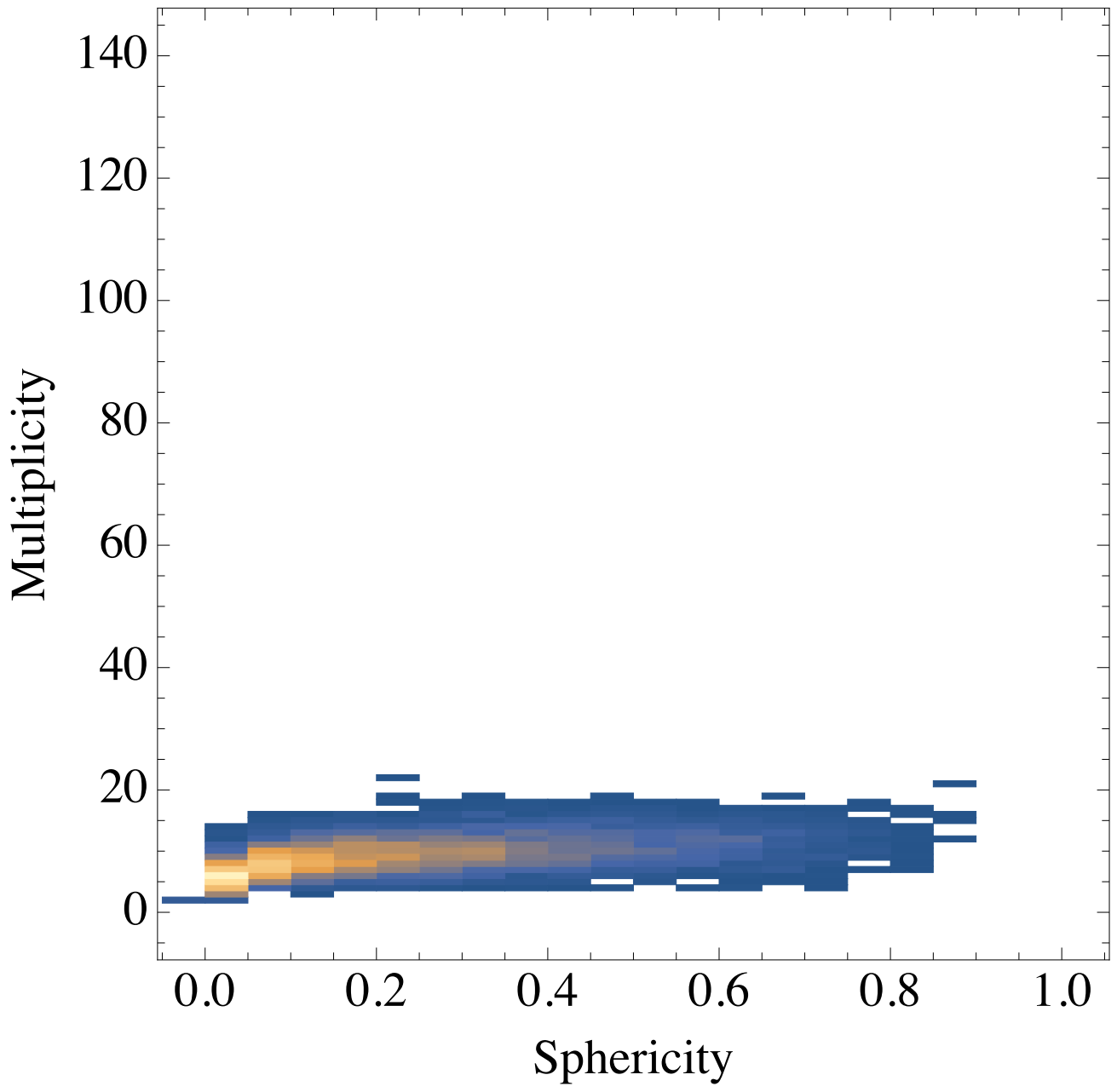

Figure 7.6: Correlation between sphericity and hadron multiplicity for events generated with parton shower (top) and AdS/CFT-inspired models with boundary interaction (bottom).

### *7.4.1 Portals and Branching Ratios*

The available decay channels and branching ratios of the LLP are critical both for the trigger strategy and the off-line analysis. In particular, various (multi-)muon triggers can be very effective for models where the LLP branching ratio to muons is not too small. At the same time, displaced hadronic decays tend to give more discriminating power in the off-line analysis, since they produce a larger number of tracks as compared to the leptonic modes. Since each event contains multiple LLPs, a single event can contain *both* leptonic and hadronic vertices, where one uses the former for the trigger and the latter for off-line background rejection. For this reason it is not straightforward to interpolate sensitivity between lepton-rich and hadron-rich hidden showers, since the optimal search strategy for the intermediate cases is qualitatively different from the strategies for the two extremes.

As a starting point for our exploration, we therefore recommend a small number of theoretically motivated decay portals, which cover the fully leptonic and fully hadronic cases, as well as two other intermediate scenarios. We focus our attention on operators of dimension four or five which do not induce additional flavor violation in the quark sector:

- **Neutrino portal:** If the LLP is a neutral, possibly composite, fermion $X$, it may decay through a small mixing with the SM neutrinos. This state predominantly decays through the $X \to \ell^+ W^{-(*)}$ and $X \to \nu Z^{(*)}$ channels and its decays tend to be rich in leptons. The muon fraction in the final states depends on the mixing angle of the $X$ with the muon neutrino, which is model dependent. However, even if this angle is accidentally small as compared to the mixing angles with electron and tau neutrinos, one still expects a muon in roughly $\sim 10\%$ of all decays through the $W^{(*)}$ channel. In $\tau$-rich scenarios, muons also originate from leptonic $\tau$ decays.

- **Hypercharge portal:** It is plausible for a dark sector to contain a vector particle, whether it is an elementary $U(1)$ gauge boson or a composite (the analogue of the $\rho$ in the SM). An elementary vector boson can be copiously produced through decays of a hidden sector meson through a chiral anomaly, analogous to $\pi^0$ decay in the SM. Whether elementary or composite, such a vector state generically mixes with SM hypercharge through the kinetic mixing operator $\epsilon B^{\mu\nu} F'_{\mu\nu}$, where $F'_{\mu\nu}$ is the dark vector boson field strength. The branching ratios of such a state then depend on its mass and can be extracted from data, as shown in Table 7.1 (see, e.g., Refs. [223, 455]).

- **Higgs portal:** In the same spirit, it is possible that a hidden sector scalar $S$ mixes with the Standard Model Higgs boson through the $SH^\dagger H$ operator. In this case, its branching ratios to SM fermions are proportional to $m_f^2$, with the caveat that non-perturbative effects modify the story substantially for $m_S \lesssim 5$ GeV. For $m_S \lesssim 1$ GeV, hadronic branching ratios can be obtained through chiral perturbation theory; however, in the intermediate range $1\,\text{GeV} \lesssim m_S \lesssim 5$ GeV, the theory uncertainties are substantial and we do not attempt to make any quantitative statements in this regime (see Ref. [558] and references therein). As shown in Table 7.1, the muon branching fraction predicted by Higgs-portal decays is smaller than for the previous two portals, but it can still be relevant if the (non-isolated) muons from $B$-meson decays are taken into account.

- **Gluon portal:** The hidden scalar ($S$) or pseudoscalar ($a$) could also decay to the SM through a coupling to gluons of the form $S\text{Tr}G^{\mu\nu}G_{\mu\nu}$ or $a\text{Tr}G^{\mu\nu}\tilde{G}_{\mu\nu}$. In this case the direct leptonic branching ratio is zero, although a small number of muons may still be produced in the hadronization process of the gluons.

- **Photon portal:** Similar to the gluon portal, the LLP could decay to two photons through the $SF^{\mu\nu}F_{\mu\nu}$ or $aF^{\mu\nu}\tilde{F}_{\mu\nu}$ operators. The signature for this case is qualitatively different from the previous four, since there are many fewer tracks. In particular, tracks only originate from photon conversions in the detector and suppressed Dalitz decays to $e^+e^-\gamma$. The signal is therefore a trackless jet with a high rate of energy depositions in the ECAL relative to the HCAL.

| mass (GeV) | photon | Higgs |
|---|---|---|
| 0.5 | 0/0.4 | 0/0.09 |
| 1.2 | 0/0.35 | / |
| 8 | 0.08/0.16 | 0.25/0.02 |
| 15 | 0.1/0.15 | 0.3/0.05 |

Table 7.1: Probability of finding exactly one muon / two muons in one LLP decay through the photon and Higgs portals. For the lowest mass points, the branching ratios for the photon and Higgs portals were taken from Refs. [455] and [437] respectively. For the 8 and 15 GeV benchmark the hadronization was performed with Pythia 8 [212, 499].

If the lifetime of the LLP is $\lesssim \mathcal{O}(50)$ m, dark showers generically give rise to multiple decays within the detector volume. In Table 7.2 we show the probability for the Higgs and hypercharge portals to produce an event which contains at least three or four muons, as a function of the number of decaying LLPs in the event. Assuming the event has at least four LLP decays, a multi-muon trigger has good efficiency for the Higgs portal above the $b\bar{b}$ threshold [3] and for all masses for the hypercharge portal[4]. This trigger strategy should also have good efficiency for neutrino-portal decays, although this scenario is more model dependent. For the gluon and photon portals a different set of triggers is needed, as discussed below. The lesson is that multi-muon triggers do not suffice to cover all possible options, but could provide a reasonably generic trigger path for an important subset of the relevant models.

[3] The importance of muons in decays which are rich in heavy flavor was emphasized in Ref. [64].

[4] Again, this statement is LLP-lifetime dependent, and for sufficiently long-lived LLPs this strategy will break down as the probability of getting sufficiently large numbers of LLP decays within the detector becomes small.

| | mass (GeV) | 3 decays | 4 decays | 5 decays | 6 decays | 7 decays |
|---|---|---|---|---|---|---|
| photon | 0.5 | 0.36/0.36 | 0.53/0.53 | 0.67/0.67 | 0.77/0.77 | 0.85/0.85 |
| | 1.2 | 0.29/0.29 | 0.44/0.44 | 0.57/0.57 | 0.68/0.68 | 0.77/0.77 |
| | 8 | 0.13/0.07 | 0.23/0.13 | 0.33/0.21 | 0.42/0.28 | 0.5 /0.35 |
| | 15 | 0.14/0.07 | 0.23/0.13 | 0.33/0.2 | 0.42/0.27 | 0.51/0.35 |
| Higgs | 0.5 | 0.02/0.02 | 0.04/0.04 | 0.07/0.07 | 0.1/0.1 | 0.13/0.13 |
| | 8 | 0.04/0.0 | 0.09/0.02 | 0.15/0.04 | 0.23/0.07 | 0.31/0.12 |
| | 15 | 0.11/0.02 | 0.21/0.07 | 0.33/0.13 | 0.44/0.21 | 0.55/0.31 |

Table 7.2: Probability of finding at least 3 muons / at least 4 muons in an event for the photon and Higgs portals, as a function of the number of LLP decays. Branching ratios and hadronization were determined as in Table 7.1.

### 7.4.2 *Lifetime*

Without additional model assumptions, the theory prior on the lifetime of the LLP is rather weak. We can, however, extract some insight from the generic scaling of the width of the LLP as a function of its mass ($m$): for the Higgs and hypercharge portals, $\Gamma \sim m$; for the photon and gluon portals, $\Gamma \sim m^3/f^2$ with $f$ the decay constant of the LLP (in this case, the LLP behaves similar to an axion-like particle, or ALP); for the neutrino portal, $\Gamma \sim m^5/m_W^4$. The obvious trend in all cases is that the lifetime rises as the mass decreases, steeply so for the case of the neutrino portal. It is furthermore important to note that the above scalings are lower bounds, and in many models the leading decay portal involves a higher-dimensional operator, leading to a stronger scaling with mass. This is especially relevant in confining models where the hidden states are composite particles. For example, in a pure-glue hidden valley coupled though the Higgs portal [557], the lightest glueball can decay through its mixing with the Higgs, but as the portal coupling contains the dimension-four combination of dark gluon field strengths $G^a_{\mu\nu}G^{a\mu\nu}$, its width scales as $\Gamma \sim m^7 \times (v_h/M^2 m_h^2)^2$, where we have taken $m \sim \Lambda_D$, the scale of dark QCD, and $M$ is the suppression scale of the portal operator.

An additional consideration is that the LLP can be discovered directly in (low energy) collider or beam-dump experiments. Since the LLP has an irreducible production cross section through the same coupling that governs its decay, collider experiments effectively impose a lower bound on the lifetime of these states. Beam-dump and supernova constraints on the other hand rely on a displaced signal and constrain the lifetime from above. These bounds are, however, not typically applicable for masses above a few hundreds of MeV (and for a summary of some of the latest constraints on the hypercharge, photon and gluon portals see, e.g., Refs. [559] and [560]). However, particles that can be produced in sizable numbers at the LHC are in general coupled sufficiently strongly to the SM that they face upper limits on their lifetimes from Big Bang Nucleosynthesis (BBN). In most cases, this upper bound is essentially mass independent and requires $\tau \lesssim 1$ second, but this upper bound can be much weaker for lighter particles and/or particles that decay to hadron-poor final states. For the photon, gluon and hypercharge portals, prompt decays are currently allowed for masses all the way down to the electron threshold, though for the hypercharge portal LHCb is expected to completely close this window below $\sim 100$ MeV [335, 376]. For the Higgs portal, current LHCb results from $b$ to $s$ transitions [260, 261] already require $c\tau \gtrsim 0.5$ mm for masses below 4 GeV. For the neutrino portal, lifetimes $c\tau \lesssim 10$ m are excluded if the mass is below 1 GeV and the mixing is predominantly with the electron and muon neutrino [561].

As argued above, if the mass of the LLP is larger than a few hundred MeV, their lifetime is only constrained from above by BBN. In the context of a dark-shower topology, there is, however, a much more stringent upper bound to lifetimes that can be probed at the LHC. In particular, to observe at least a handful of decays in the detector volume, the lifetime in the lab frame should satisfy

$$c\tau_{\text{lab}} \lesssim 10\,\text{m} \times \frac{N_{\text{LLP}}}{N_{\text{decays}}} \tag{7.7}$$

with $N_{\text{LLP}}$ the typical LLP multiplicity and $N_{\text{decays}}$ the number of observed decays in the detector[5]. For instance, for $N_{\text{decays}} \sim$ few and $N_{\text{LLP}} \sim \mathcal{O}(10)$ this implies $c\tau_{\text{lab}} \lesssim 50$ m.

[5] If only one or two LLPs are expected in the detector, then the dark shower maps onto the topologies with one or two displaced vertices discussed earlier in this document.

Near this heuristic upper limit in the LHC's sensitivity to lifetime, the shower effectively gets stretched out over the detector elements, and it is useful to study how the decays are distributed over the different detector elements. We can use a toy MC to make some simple estimates of this effect by making the following simplifying assumptions: (1) the number of LLPs is Poisson distributed; (2) their decay lengths follow the usual exponential distribution with uniform average lifetime in the lab frame $c\tau_{\text{lab}}$; (3) the angular distribution of the vertices is approximately spherical in the lab frame. Figure 7.7 shows the distribution of the number of decays occurring in the tracker, calorimeter and muon chamber for two benchmark points. We hereby assume for concreteness the approximate CMS geometry with the "tracker", "calorimeter" and "muon chambers" defined by $r < 100\,\text{cm}$, $100\,\text{cm} < r < 200\,\text{cm}$ and $400\,\text{cm} < r < 700\,\text{cm}$ respectively. We find that the number of decays in all detector components is roughly equal, with slightly more decays in the muon chamber for a higher number of LLP decays in the shower. The latter result is easily understood from the larger size of the muon system as compared to the tracker and the calorimeter. The qualitative conclusion is that tracker-based searches for multiple displaced vertices are likely to have reasonable sensitivity over the whole relevant lifetime range. For the longer lifetimes, the sensitivity necessarily degrades although it can be partially recovered by incorporating the muon chamber in the analysis. On the other hand, searches relying exclusively on the muon chamber or the calorimeter are less sensitive to a wide range of lifetimes. To quantify this effect more accurately, detailed simulations for specific benchmark models accounting for more realistic boost and $\eta$ distributions are needed, although the relative importance of the tracker is likely to continue to hold. The displaced SUEP scenario is an interesting exception, as discussed in Section 7.6.2.2.

## 7.5 *Trigger Strategies*

### 7.5.1 *General Considerations*

Due to the immense diversity of hidden sector models, there is no trigger strategy which can comprehensively cover all options, nor is it straightforward to compile a list of such strategies. This problem is further compounded by the lack of available MC simulations, except in a handful of special cases. Many of the considerations informing the trigger strategies discussed for singly and doubly produced LLPs in Chapters 3 and 5 apply to dark shower events as well. However, the four characteristic features of dark showers laid out in Section 7.1 can pose both additional opportunities and challenges:

1. *The particle multiplicity can range from high to very high.* This in itself can provide a very powerful trigger strategy, especially when the muon fraction in final states is appreciable, as discussed in Section 7.4.1. For models where this is not the case or where the muons are too soft to fire the trigger, triggers relying on a large multiplicity of tracks or an over-density of pixel hits could provide an alternative strategy. We will return to this discussion later in this section.

2. *The hidden sector may contain states with vastly different lifetimes.* While this makes it challenging to get a comprehensive grasp on the full space of possible event topologies, this feature provides interesting opportunities from the point of view of the trigger. For instance, often the most striking off-line signature is a set of displaced vertices in the tracker, which are notoriously challenging targets for the trigger. It is completely plausible that these vertices would originate from the decay of one species of hidden sector particles, while a second species decays either promptly or very displaced. This means that the event may very well come with prompt leptons, $H_T$, an appreciable amount of $\not{E}_{\mathrm{T}}$, and/or activity in the muon chamber.

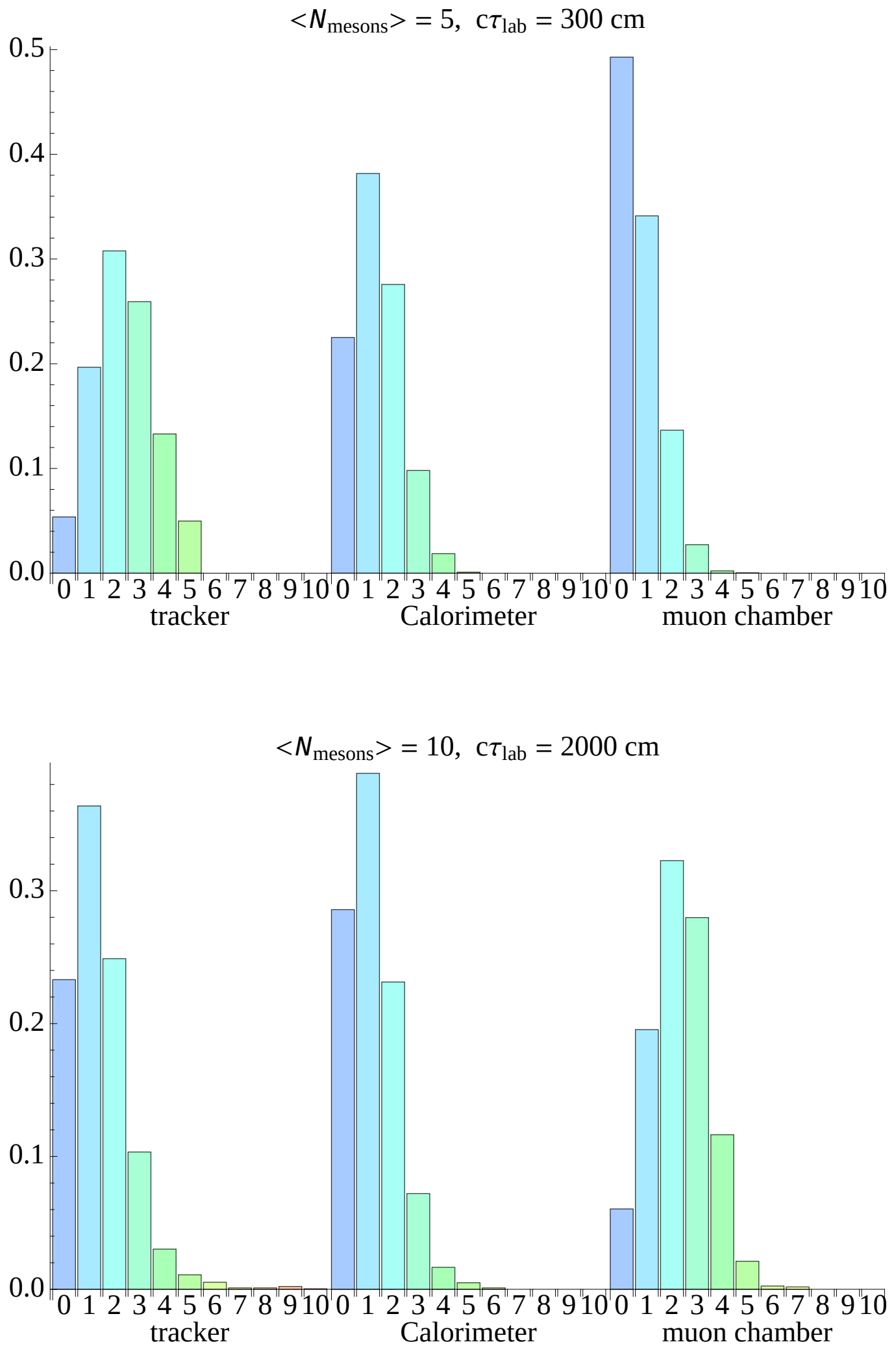

Figure 7.7: Distribution of decays occurring in the various detector elements, for those events with more than three decays occurring in the fiducial volume.

3. *The decay products may not be isolated.* Combining the first two features, we should prepare for the possibility that the objects we aim to use as trigger handles may not be isolated. Especially for triggers on muons and (displaced) tracks, it is advisable to relax isolation criteria as much as possible and instead rely on the presence of one or more additional objects to reduce rates as necessary. For example, one would expect that a trigger on two isolated muons would be less effective than a trigger on three or four non-isolated muons or a trigger with two non-isolated muons and a moderate amount of $\not{E}_{\mathrm{T}}$. Non-isolation is likely to be a limiting factor for the acceptance of dark shower events in specialized triggers designed for singly- or doubly-produced LLPs.

4. *The energy flow in the dark shower may be non-standard.* This implies that a dark shower may be broader or narrower than SM QCD jets, and the momentum distribution of the particles in the shower may also differ substantially from QCD predictions. This means that triggers for (trackless) jets could be effective for some models, but would by no means capture the full range of possibilities. Since such $p_{\mathrm{T}}$ distributions are very model dependent, it is moreover sensible to keep object $p_{\mathrm{T}}$ thresholds as low as possible, and instead rely on the high multiplicity by requiring multiple low $p_{\mathrm{T}}$ objects, rather than a few high $p_{\mathrm{T}}$ objects.

In summary, while some dark shower models will be very challenging to trigger on, this is not true for the entire class of models. For example, some (although certainly not all) models can readily be captured on the traditional $\not{E}_{\mathrm{T}}$, $H_T$ or lepton triggers. This means that these traditional triggers *are an excellent place to start, but a bad place to stop.* Indeed, in the context of limited resources, it makes sense to first pursue the scenarios where the trigger is not a major challenge and use these scenarios to develop off-line reconstruction techniques. It is, however, crucial to follow up with more innovative trigger strategies. While this is largely still a topic of study, we can identify two categories of non-traditional triggers:

1. *Triggers on displaced objects:* This type of trigger is notoriously challenging due to bandwidth and online computing limitations, but nevertheless a lot of progress has been made in the last few years. This is discussed in Chapters 3 and 5. For our purposes here, it suffices to note that high multiplicity of LLPs in dark shower events can help provide excellent acceptance in such triggers, provided that any isolation criteria in the trigger are not spoiled by shorter-lived objects in the event.

2. *Specialized Triggers for dark showers:* There are a number of (speculative) ideas for triggers which exploit the high multiplicity and/or energy features of a dark shower, as described above, although this area is much less developed than the triggers on displaced objects. We will discuss these ideas for specialized triggers briefly in the remainder of this section, and comment on the unique capabilities of LHCb when it comes to triggering on dark showers.

### 7.5.2 Specialized Trigger Ideas

In this subsection we briefly highlight some proposals for specialized trigger strategies that inherently depend on the unique features of dark shower events.

#### 7.5.2.1 Multiplicity Triggers

As emphasized throughout this chapter, dark shower models can produce a high multiplicity of dark particles, ranging from several tens up to thousands of particles in the most extreme case. This motivates triggers that aim to minimize $p_T$ thresholds and instead exploit high particle multiplicity. The most obvious option is a suitable multi-muon trigger (three, four, or more muons), where the priority should be to avoid tight isolation criteria, if needed at the expense of increasing the muon multiplicity and/or $p_T$ thresholds. Prompt multi-muon triggers would, however, only capture (nearly) prompt decays, and it is critical to pursue the design and implementation of a trigger with good efficiency for displaced muons; see Section 5.1.3 for more on planned displaced muon triggers.

Another way of exploiting the particle multiplicity is to trigger on an anomalously large number of tracks originating from the primary vertex. This would likely only work for prompt or nearly prompt decays, in models which produce a rather low muon fraction and/or a very soft $p_T$ spectrum. For this strategy to work, one must first pass the L1 trigger with MET, which can be provided by the shower itself or by initial state radiation (ISR). At the HLT it may be possible to count the number of tracks with the planned ATLAS hardware track trigger (HTT) system or the future CMS hardware track triggers. Alternatively, if the tracks are too soft and the multiplicity high enough, it is possible to pass the HLT by looking for an over-density of hits in the innermost tracking layer, centered around the $z$-coordinate of the primary vertex [235].

#### 7.5.2.2 Substructure Triggers

Substructure triggers may open up the possibility of using substructure techniques to distinguish between QCD jets and other types of showers already at the trigger level. CMS currently uses jet trimming at HLT level to enhance acceptance for jets with hard splittings, optimized for boosted electroweak gauge bosons, (see, e.g., Refs. [562, 563]), and we are aware of an effort within ATLAS to investigate the inclusion of substructure routines at the HLT level. This strategy could be helpful in cases such as photon-jets (dark showers that decay into SM photons), which can otherwise pose significant trigger challenges: substructure variables such as $n$-subjettiness [564] and energy-energy correlations can be very effective at separating QCD jets, photons and photon-jets [565, 566]. In practical terms, this means that one may be able to use the relatively low threshold jet/tau trigger for the L1 trigger. This could be followed by a HLT trigger that uses, for instance, the ratio of the energy deposited in the ECAL/HCAL or substructure variables.

An advantage of substructure methods is that they are fairly robust with respect to the lifetime of the dark shower final states, provided the lifetime is not too long compared to the size of the relevant detector sub-system. As long as the showering process itself happens quickly (and provided that hadronization represents a small correction to the typical particle momenta), the separation between energy depositions in the calorimeter is already pre-determined. In particular, under these assumptions it does not matter if the SM decay products of the lightest dark states are themselves collimated. A clear difficulty with substructure triggers is that the actual dark showering and hadronization process, and therefore the substructure, is theoretically only understood in a handful of example models as discussed in Section 7.3. While such a trigger is clearly valuable for a number of models, it is at the moment difficult to evaluate how broadly applicable it could be to different shower shapes.

#### 7.5.2.3 *Non-Isolated Displaced Vertex Trigger in the ATLAS Muon Spectrometer*

For sufficiently long-lived particles, fallback trigger strategies can be provided by existing ATLAS triggers that require a cluster of hits in the muon spectrometer (MS) [234, 250]. As these triggers look directly for the presence of displaced objects and do not rely on any features of the rest of the event, they offer potential sensitivity to signals dominantly produced at low mass scales. The existing MS displaced vertex (DV) trigger requires at least three (barrel) or four (endcap) tracks in the MS [250]; a dedicated vertex-finding algorithm is run on these events at the analysis level. Thus this trigger option offers good sensitivity to hadronic decays of particles with mass $\gtrsim$ 5 GeV, but cannot provide sensitivity to individual di-leptonic or photophilic decays, nor to decays of LLPs with GeV-scale masses.

In Run 1, LLP searches with displaced vertices in the MS relied on a trigger that recorded events where the displaced vertices additionally passed isolation requirements. Specifically, events were required to have no tracks with $p_T > 5$ GeV in the inner detector within $\Delta R < 0.4$ of the displaced vertex, and the distance to the nearest jet (with $E_T > 30$ GeV) was required to be $\Delta R_j > 0.7$. However, this isolation cut was not imposed for jets with anomalously large energy deposition in the hadronic calorimeter, $\log_{10}(E_{\text{had}}/E_{\text{EM}}) > 0.5$ in order to improve acceptance for LLPs decaying in the outer edges of the hadronic calorimeter [234]. This isolation requirement can limit acceptance for models with dark showers where multiple BSM species can spoil each other's isolation, especially in cases where the shower contains both short-lived as well as long-lived dark particles. It is an open question whether tracks from sufficiently displaced vertices in the inner detector would contribute to the track veto, but particles that decay within the inner detector would be likely to spoil the isolation requirements through calorimeter deposits alone, unless their decays involved only muons.

New in Run 2 is a trigger on *non-isolated* clusters of hits in the MS [251], which uses the same triggering algorithm for the muon system hits but does not impose isolation criteria. The primary aim of this trigger is to enable background estimation for single DV searches [255], but it would also efficiently record dark shower signal events, regardless of event shape or event energy scale, provided at least one particle species in the shower has $m \gtrsim 5$ GeV, decays to final states with at least three charged particles, and has reasonable probability to decay in the MS. This trigger may be especially useful for high-multiplicity signals, where the probability for getting one particle to decay in the MS can become appreciable even for shorter-lived LLPs.

#### 7.5.2.4 *Capabilities at LHCb*

The LHCb detector is designed for studying displaced decays of SM mesons. This provides a good environment to search for dark shower events containing long-lived particles with lifetimes below the meter scale [61]. Although LHCb does not yet have a dedicated trigger for dark shower events, many of the trigger strategies used for SM meson searches may be applied to dark shower signals. The LHCb detector is a single-arm forward spectrometer covering the pseudorapidity range $2 < \eta < 5$, which means the signals for which it is optimal have typical $p_{\text{T}}$ that is much lower than the signals targeted by ATLAS and CMS. Hence, the LHCb signal trigger usually has much weaker $p_{\text{T}}$ requirements compared to an ATLAS or CMS search. This allows for better observation of dark shower events that come from the decay of light parent particles boosted in the forward direction, such as dark mesons from the decay of a light $Z'$.

For example, Ref. [382] searches for a pair of long-lived particles decaying into jet pairs. Although the signal is different from a dark shower, the trigger strategy in the search can be useful for dark shower events. The hardware trigger (L0) requires a single hadron, electron, muon, or photon with object-dependent $p_{\text{T}}$ thresholds. For muons (hadrons), the thresholds are $p_{\text{T}} > 1.48$ (3.5) GeV, and given the rapidity range of the signal, this corresponds to momenta $p > 6$ (13) GeV for muons (hadrons). These low $p_{\text{T}}$ requirements can thus keep soft and low-mass dark meson decays that can be hard to trigger on at ATLAS and CMS. The software trigger contains algorithms that run a simplified version of track reconstruction and an identification of displaced tracks and vertices, and the first software stage (HLT1) requires a high-quality displaced track satisfying the above $p_T$ requirement. In Ref. [382], the final trigger stage (HLT2) further requires a displaced vertex with $\geq$ 4 charged tracks, and either a $>$ 2 mm decay length in the transverse direction or the reconstructed vertex mass $>$ 10 GeV. Since the search focuses on $b$-quark final states, the HLT2 trigger also contains a multivariate algorithm to identify $b$-hadron decays. Although the trigger is designed for long-lived particles heavier than 10 GeV scale, it is important for LHCb to determine if the same strategy can be applied to even lighter particles or different decay final states.

For specific decay channels of dark mesons, we can also adapt triggers from existing SM hadron searches that look for the same final states [377]. For example, if dark mesons decay into muon pairs or $c$-quarks, we can use similar triggers as in the $K_s \to \mu^+\mu^-$ [567] or $B^0 \to D^+D^-$ [568] searches to study the signal. A more recent muon trigger is used for the 13 TeV dark photon $A' \to \mu^+\mu^-$ search [262], where a muon with $p_T > 1.8$ GeV is required at the hardware trigger level, and further quality cuts on the displaced vertex and muon identification are required. This search can be useful for dark mesons that decay through a kinetic mixing, in which muon final states have a sizable branching ratio.

Planned upgrades at LHCb are further discussed in Section 5.2. These upgrades will make LHCb an even more powerful facility for studying LLPs, particularly in the low-mass and short-lifetime regime, and further study of LHCb's capabilities for dark showers is well warranted.

#### 7.5.2.5 *Low Pile-Up Data*

While Run 2 of the LHC has brought unprecedented opportunities for discovery of new physics with a large $\sim$ 150 fb$^{-1}$ dataset of 13 TeV pp collisions, the ability to explore the energy frontier comes at a cost: high trigger thresholds and challenging experimental conditions may limit the sensitivity of LHC experiments to BSM models such as SUEPs with soft and diffuse signatures. During Run 2 data-taking, the typical $\langle\mu\rangle$ value has already reached $\sim$ 60 interactions per bunch crossing.

The low pile-up datasets provided by the LHC during Run 2 therefore present an interesting opportunity for dark showers. There are two such data sets: one with 0.5 fb$^{-1}$ at 13 TeV and $\langle\mu\rangle =$ 2 and one with with 0.3 fb$^{-1}$ at 4 TeV and $\langle\mu\rangle = 5$. While the size of these datasets is much lower than the 13 TeV high-$\mu$ pp dataset, the change in data-taking conditions is amenable to searches for BSM scenarios which would normally be difficult to distinguish from pile-up noise, and low-background searches may be performed under circumstances that would otherwise be impossible.

In low pile-up data, object-multiplicity (track- or cluster-based) triggers (Section 7.5.2.1) are typically run with lower thresholds and higher rates than otherwise may be possible. The ability to cleanly reconstruct low-$p_{\mathrm{T}}$ tracks and vertices or soft calorimeter clusters could provide the only way to experimentally access some low-mass LLP scenarios. Simple analyses in this modest dataset may provide the first limits on some models, and these results could help direct the more advanced developments discussed throughout this report.

#### 7.5.2.6 *Zero Bias Strategy*

If all other strategies fail to capture the signal, the zero bias strategy is an ultimate fallback [569]. In this case one would simply rely on passing the L1 trigger due to an object in an unrelated pile-up event. The effective dataset one can ultimately obtain this way is only $\sim 0.5\,\mathrm{fb}^{-1}$, though it might increase to $\sim 50\,\mathrm{fb}^{-1}$ if a specialized selection could be made at the high level trigger.

## 7.6 *Off-Line Analysis*

In this section we discuss aspects of the off-line analyses needed to discover dark showers. As has been a primary theme throughout this chapter, hidden sector states with displaced decays are generic, but prompt decays are possible as well. Naturally cases with and without promptly decaying species require different strategies. Though the main focus of this document is on long-lived phenomena, showers where *all* species are short-lived are closely related to the signals of our main interest and provide valuable illustrations of tools and techniques that can ultimately shed light on the underlying hidden sector dynamics itself, and we accordingly provide a discussion of prompt showers as well.

### 7.6.1 *Prompt Decays*

We begin by discussing dark showers with promptly decaying BSM states, i.e., with no reconstructable displaced decays in the event. Promptly-decaying dark showers are substantially more challenging to separate from backgrounds than showers containing LLPs, as the presence of multiple displaced objects is a very powerful background-suppression tool. However, the techniques that have been proposed and/or used for prompt showers are important for several reasons. First, they provide a useful illustration of how the unique properties of showering events have been approached in analyses to date without introducing the separate complication of displacement. Second, dark showers that produce prompt SM particles may very well also produce LLPs, thanks to the hierarchies of lifetimes that are generic in confining theories. Such events can thereby produce *semi-visible jets* [366], which contain detector-stable invisible states as well as promptly-decaying states. These semi-visible jets pose some specific challenges in analysis and reconstruction, as we briefly review below.

If SM particles resulting from the dark shower are produced promptly, there are two possible experimental handles: the portal through which the hidden sector couples to the SM, and the structure of the dark shower itself. The former case depends on the operators mediating the production and decay of showering states, as described in Sections 7.2 and 7.4. For example, for decays governed by the Higgs, photon, and neutrino portals, one may expect an unusually muon-rich jet (Section 7.4). Similarly, the jet may be semi-visible (Section 7.6.1.1) if some of the states do not decay in the detector volume. On the other hand, if the SM final states are almost purely hadronic, as when decays are governed by the gluon portal, no such obvious handles are available and one must look at the substructure of the jets themselves to find differences from those in typical QCD events. The discussion below is organized according to the typical size of the dark shower, going from narrow QCD-like jets to large-radius jets and finally to fully spherical topologies (SUEPs).

#### 7.6.1.1 QCD-Like Jets ($R \sim 0.4$)

The case of *lepton jets*, originally motivated by dark matter considerations, provides an example of showering with noticeably different particle content than a QCD shower. Here, visible decays of the dark states are primarily to leptons [148, 570], which can occur either due to other decays being kinematically inaccessible or due to selection rules. The most striking signature of the model is the presence of collimated sprays of leptons rather than hadrons in the final state. Reconstructing these objects requires the elimination of lepton isolation criteria typically imposed on leptonic final states, together with variables such as layer-specific cuts on energy deposited in the ECAL, and a fraction of high-threshold TRT hits and/or activity in the muon chambers, selecting for electromagnetically dominated radiation. Searches for prompt lepton jets have been performed by the ATLAS and CMS collaborations [267, 571].

A more elaborate prompt scenario is the case of *semi-visible jets* [366]. Here, the visible final states are hadrons, as in typical QCD jets, but missing energy is interleaved with the visible final states as only some fraction of the states produced in the dark shower decay back to SM particles, while others escape the detector. Semi-visible jets fail to provide the dramatic handles of the lepton jet scenario. However, by considering combinations of large amounts of $\not{E}_T$ and differing cuts on $\Delta\phi$, the angular separation between the $\not{E}_T$ and the closest jet, control of both SM backgrounds and discrimination from more typical SUSY-like high $\not{E}_T$ scenarios can be achieved. Bounds on showers with a fraction of between 0.2 and 0.9 of produced dark hadrons decaying invisibly should allow for bounds comparable to conventional resonance searches [572].

If event-level handles like lepton (or heavy flavor) multiplicity or $\not{E}_{\mathrm{T}}$ are not present, sensitivity might still be achieved by looking toward the internal structure of the jets themselves. While this topic is presently relatively uncovered, with only a few exploratory results available [573], similar techniques to those that have proven useful for quark/gluon discrimination seem promising. In both cases the lack of a perturbatively-generated hard scale means that parametric separation of signal and background is challenging. Infrared and collinear (IRC) safe observables related to jet mass, such as girth and two-point energy correlations are reasonably well studied in QCD, with results readily generalizable to other gauge theories [574–576]. For quark/gluon discrimination, observables characterized by Poisson-like distributions such as particle/track multiplicities or production ratios of particular SM particles tend to yield better discrimination, but suffer from being IRC unsafe and thus subject to large non-perturbative modeling uncertainties [577], a significant point of concern when extending their use beyond QCD. Here, some recently developed IRC-safe generalizations of multiplicity [578] may prove useful. Another approach might be to explore machine-learning techniques that do not require fully labelled data for training (so-called weak supervision) [579–581]. These can be trained directly on data, here at least in the case of the QCD background, so that modeling concerns about the shower can be partially alleviated. While achieving a sizable signal-to-background ratio is likely to be the major challenge in using such jet observables to separate dark showers from QCD, it is vital to note that if signal/background separation can be achieved through other means, e.g., by reconstructing displaced vertices, these same tools offer an enormously powerful window onto the underlying dynamics of the dark shower.

#### 7.6.1.2 *Large-R Jets ($R \sim 1.0$)*

As discussed in Section 7.3, dark showers at larger couplings are currently poorly understood. If couplings controlling the shower can no longer be treated as small at any scale, the QCD-inspired picture of pencil-like jets starts to break down, and we might expect that showers radiate more copiously at ever larger angles, so that large-$R$ jets can become necessary to adequately capture the underlying hard process. This is first and foremost a triggering challenge, since it is no longer clear that triggers designed for local hard depositions can maintain their sensitivity. It is likely that a viable trigger path either needs to rely on a prompt associated object from the production mechanism (ISR jet, lepton, etc.) or on a high multiplicity of leptons in the shower itself. Of course, $H_T$ triggers continue to be useful for those models with sufficiently high event energy scales, provided that a sufficient amount of the radiation in the event clusters into trigger objects of sufficiently high $p_{\mathrm{T}}$ ($\sim$ few $\times$ 10 GeV) to be counted in the $H_T$ trigger. As long as the event can be triggered on, observables similar to those at the end of Section 7.6.1.1 should still prove useful to separate signal from background, possibly even more so, since the radiation pattern in the dark shower can now be expected to be substantially different from QCD.

#### 7.6.1.3 *Spherical Showers*

In the large-'t Hooft coupling regime, in which we expect much more showering at large angles than in QCD (see Section 7.3), the final states are soft and spherically distributed in the rest frame of the shower. This leads to so-called soft unclustered energy patterns (SUEPs) in the detector. In the rest frame of the event, the momenta of the final-state SM particles are on the order of the hadronization scale of the dark strong dynamics. If this scale is much lower than 100 MeV, the decay products are too soft to be reconstructed as tracks and the entire shower is effectively invisible. In this case existing jets+MET searches can apply. For a hadronization scale around $\mathcal{O}(100\,\text{MeV})$, some amount of ISR can be needed to boost the particles enough to render a fraction of them reconstructable. Finally, if the decay products are on average harder than $\mathcal{O}(100\,\text{MeV})$, the tracks associated to the SUEP vertex can typically be reconstructed off-line, subject to momentum-dependent reconstruction efficiencies. The main parameters of interest are accordingly the number of charged particles produced and the corresponding $p_\text{T}$ spectrum. Also important are the fraction of invisible particles and the composition of SM particles in the final state. In particular, since the momenta of the invisible particles balance on average, they are are not likely to give rise to a substantial $\not{E}_\text{T}$ signal unless the dark shower is recoiling against a relatively hard ISR jet. A large fraction of invisible particles will instead degrade the sensitivity of a strategy relying on track multiplicity. The muon fraction, on the other hand, can be a powerful handle: even though the average muon $p_\text{T}$ may be very low, due to the high particle multiplicity a handful of muons may still be hard enough to pass the trigger and reconstruction thresholds of the muon systems.

A major background to SUEP-like signals comes from pile-up, which also yields a large number of isotropically distributed soft tracks, though in contrast to SUEP signatures, pile-up tracks arise from multiple vertices. Studies on the multiplicity of charged tracks from minimum-bias interactions like single, double or non-diffractive collisions are described in Ref. [582]. As shown in Figure 7.8, the fraction of 13 TeV $pp$ collisions having 80 or more associated charged tracks is O($10^{-3}$), as measured in a pile-up-free environment. Some benchmark SUEP models are described in Ref. [235], one with a low-mass Higgs-mediator and two higher-mass scalar models, and charged particle multiplicities for these benchmarks are shown in Figure 7.9; for these models, it is assumed that SUEP particles decay only to electrons and muons. Ref. [235] demonstrates that counting tracks associated to the PV should provide a very powerful discriminant against pile-up background during off-line reconstruction for high-mass mediators, while discrimination for the Higgs-mediator model is extremely challenging. The crucial discriminating factor is the number of particle tracks that can actually be reconstructed, and both ATLAS and CMS can reconstruct tracks with a minimum momentum of roughly 400 MeV [583]. The multiplicity of tracks fulfilling this minimum-$p_T$ requirement is shown in Figure 7.9c for the three SUEP benchmark models. It was shown in Ref. [235] that, even for the highest-mass mediator, no significant losses of the tracking efficiency are expected in spite of the hundreds of tracks present per event. It is moreover worth noting that, while the track multiplicity is large compared to what is generated by SM proton-proton collisions, it is still relatively small compared to the multiplicities that can be reconstructed in heavy ion collisions.

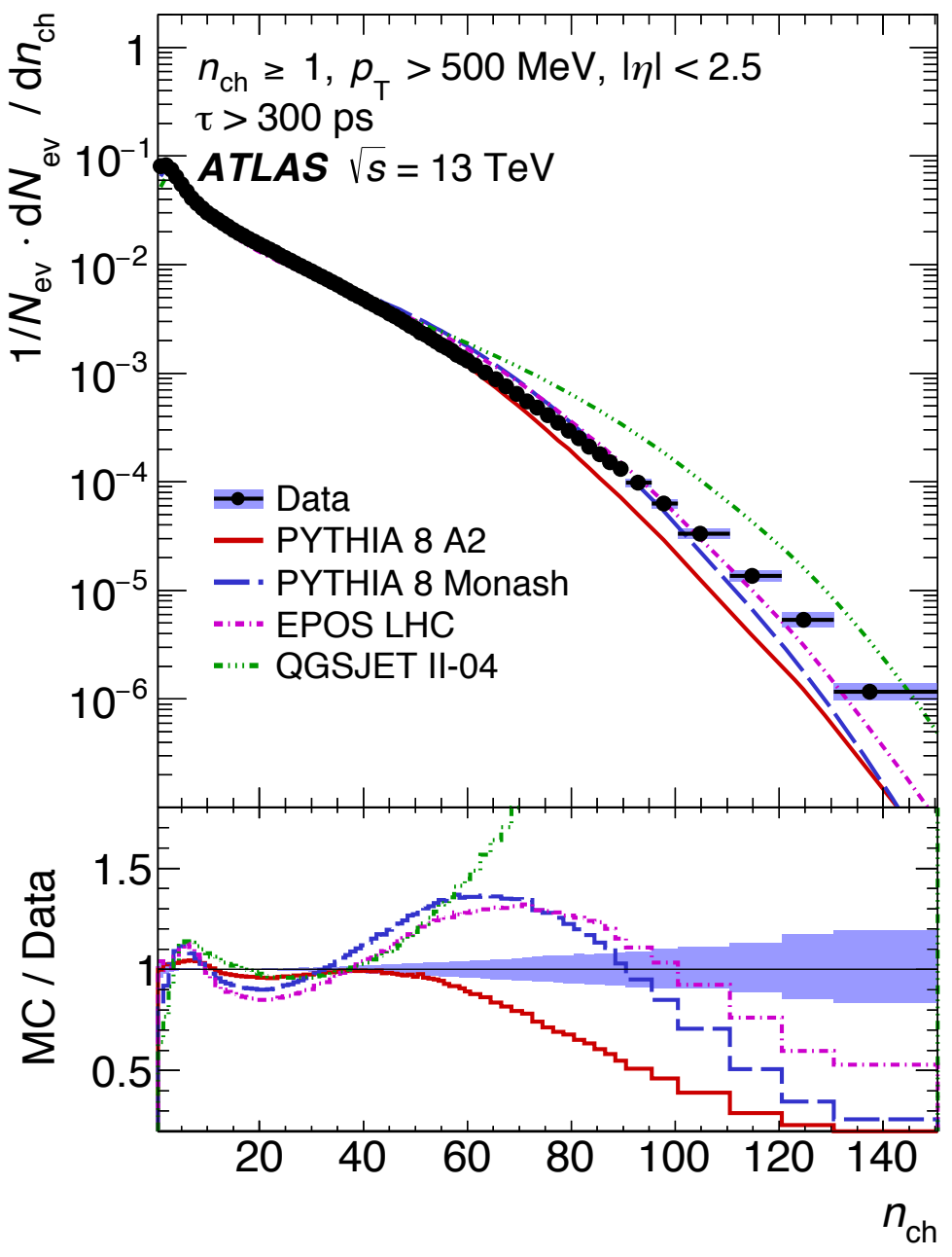

Figure 7.8: Charged particle multiplicities measured in minimum bias events with the ATLAS detector for events with at least one track with a minimum $p_T$ of 500 MeV and $|\eta| < 2.5$ [582].

A large fraction of electrically neutral hadrons produced in the SUEP can further lower the number of associated tracks, which makes signal/background discrimination very difficult for the lowest-mass-mediator signals. A possible additional discriminant for those cases can be the hemisphere mass. This mass is estimated by dividing the event into a hemisphere associated to the ISR jet and one associated to the SUEP, while calculating the "jet" mass, as in Section 7.6.1.1, from the tracks in the SUEP hemisphere. This variable should be significantly higher for the SUEP events, where the heavy mediators are decaying, than for pile-up events.

Another possible background to SUEP signals is QCD multi-jet events. As described in Section 7.5.2.1,, triggering SUEP events based on $\not{E}_T$ relies on the emission of an ISR jet recoiling against the system. QCD scenarios similar to SUEP signatures could arise when one hard jet is recoiling against a system of multiple soft jets. The mean number of charged tracks per jet is less than ten for jets $O(100\text{ GeV})$ [584]. To get to $O(100)$ charged tracks, as expected from a SUEP signature, a very high number of jets is needed, hence the perturbative cross section is heavily suppressed by many orders in $\alpha_s$. This background can nevertheless play a significant role for the low-mass mediator models, where significantly fewer tracks are expected. A veto against large calorimeter deposits can be a powerful handle to reject those events, as rather hard jets are needed to get many associated charged tracks.

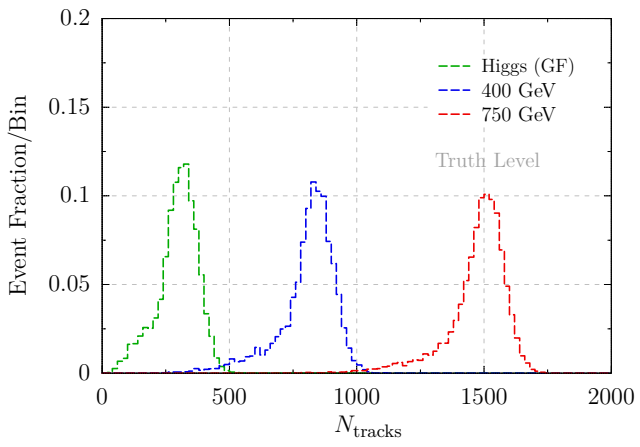

(a) Inclusive

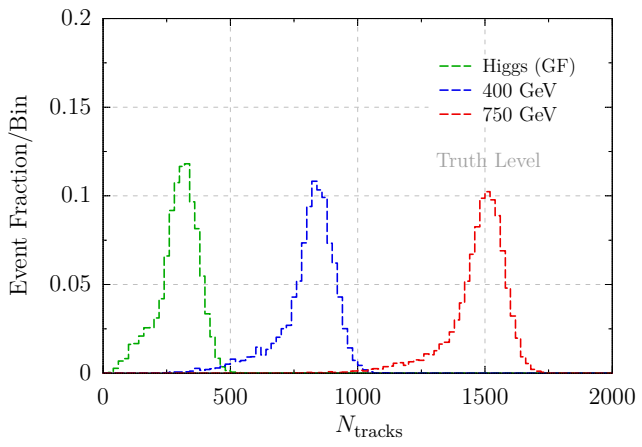

(b) $p_T > 10$ MeV

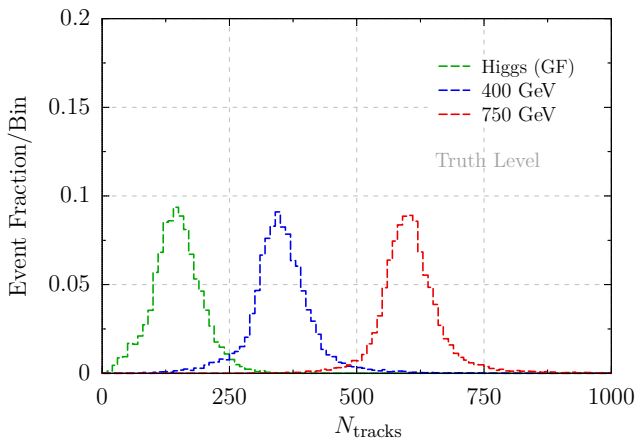

(c) $p_T > 400$ MeV

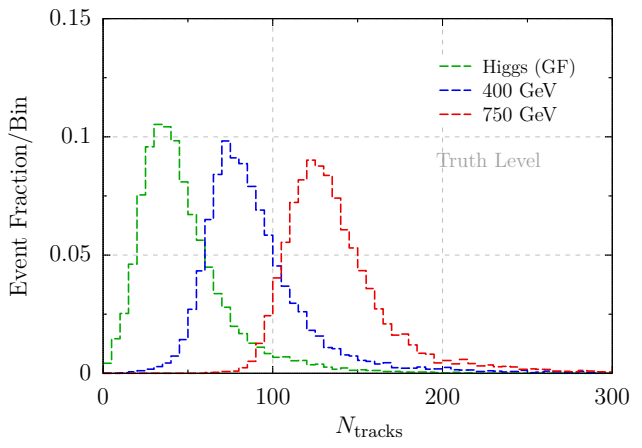

(d) $p_T > 1$ GeV

Figure 7.9: Event multiplicities of charged particle tracks fulfilling the respective indicated $p_{\rm T}$ requirements for three benchmark SUEP models. Figure from Ref. [235].

### 7.6.2 *Displaced Decays*

Outside the prompt regime, the lifetimes of the various dark states and the composition of SM final states produced in their decays drive the phenomenology. LLPs result when there are one or more species of dark states in the hidden sector spectrum that would be stable in the absence of couplings to the SM. In particular, different species naturally come with vastly different lifetimes, as is the case for the SM $\pi^0$ and $\pi^\pm$ mesons. In what follows, we consider the single-species case in most detail, and treat the case of primarily leptonic decays (lepton-jets) separately from the cases with substantial branching fractions to hadrons (emerging jets). We finally comment on the multi-species case, and relate it to the semi-visible jet scenario mentioned in the previous section.

Even in high multiplicity events, the isolation of displaced decays from each other is not likely to pose difficulties in reconstructing

DVs in the inner tracker. Two nearby DVs can be separately resolved down to separations of $\sim$ mm. Even in the case where two DVs are closer than 1 mm, all of the tracks associated to both vertices will simply be reconstructed as a single DV with larger track multiplicity.

#### 7.6.2.1 *Single Species Leptonic Dark Showers (Displaced Lepton-jets)*

One signature predicted by dark shower models is lepton-jets [148, 570], whose prompt decays are discussed in Section 7.6.1. Depending on the lifetime of the decaying state, these leptons can easily be produced with a measurable displacement. Explicit searches for this signature have been performed by the ATLAS Collaboration at both 8 and 13 TeV [265, 266], as detailed in Section 3.2. It is worth remarking that the experimental signature of a displaced decay to electrons can be very similar to the signature of a displaced decay to a pair of charged pions, and thus searches for lepton-jets frequently cover pionic final states as well. Searches to date have targeted lepton-jets containing up to two lepton (pion) pairs.

To further extend coverage into low-mass regimes, it may be of interest here to investigate the samples stored through data scouting for, e.g., di-muon resonances, although the limited information retained in scouted events likely makes this a promising avenue only for sufficiently short lifetimes that dedicated displaced track reconstruction is not necessary.

#### 7.6.2.2 *Single Species Hadronic Dark Showers (Emerging Jets)*

The ATLAS, CMS, and LHCb collaborations have developed powerful searches for pairs of DVs in the trackers, typically in association with missing energy or large $H_T$, as described in Section 3.1. These searches are nearly background-free and inclusive in the number of vertices, and thus have good sensitivity for any model of dark showers that has a large enough signal acceptance in such searches, regardless of the detailed features of the shower shape.

While these searches demonstrate the power and flexibility of inclusive low-background searches, it is unfortunately very easy for signal acceptances in existing displaced vertex searches to be prohibitively small. Primary drivers for the loss of acceptance are the requirement of associated objects (leptons, MET, sizable visible $H_T$, etc.), and/or cuts on the invariant mass of or number of tracks belonging to the displaced vertex. Given that a hidden shower tends to produce a multitude of LLPs, it should be possible to maintain very low background levels—and therefore the power and inclusivity of the search—by relaxing many of these requirements and demanding a larger number of displaced vertices or even displaced tracks instead.

When the lightest decaying dark state has mass $\lesssim$ 10 GeV, backgrounds to DV searches do become more important, since the background DV rate rises rapidly as the number of tracks associated to the vertex falls; also, irreducible heavy-flavor backgrounds are important in this mass range. The number of associated tracks is also crucial for the vertex reconstruction in both the inner detector and the muon spectrometer. Hence, the composition of hadronic and leptonic particles in the final states may have a significant impact on the ability to reconstruct the associated vertices and/or the size of the expected backgrounds. Moreover, as the track momentum and multiplicity in the vertices drop, the odds increase that one or more tracks belonging to a particular vertex are not reconstructed. The efficiency of reconstructing displaced tracks in the inner tracker falls off substantially with displacement, as particles traverse fewer layers of the tracker: for particles produced at $r = 30$ cm from the interaction point, this efficiency is $\sim 0.35$ at ATLAS [248] and $\sim 0.25$ at CMS [585]. Explicitly requiring vertex reconstruction can thus come at a large cost in signal acceptance, especially for LLP decays that produce only two tracks per vertex. However, a large number of unassociated tracks with a large impact parameter is still a striking signature, despite the larger backgrounds that come with relaxing requirements on vertex reconstruction. This is the idea behind a recent CMS search [310], based on the model of Ref. [325] (see below). Whether or not explicit DV reconstruction is helpful depends on relative signal and background rates, and can be model dependent.

Backgrounds also increase as the lifetime of the decaying dark state becomes shorter; as track reconstruction efficiencies are very good for particles at small production radii, explicit vertex reconstruction is likely to be useful here to help keep backgrounds under control. Even in the short lifetime regime, a dark shower offers many additional handles for signal/background discrimination beyond the number of vertices, such as a common mass scale for reconstructed vertices and non-SM-like particle multiplicity distributions.

Given the striking nature of these high multiplicity signal events, it is typically not challenging to separate them from backgrounds once the event is on tape, provided that sufficiently many displaced tracks (and possibly vertices) can be reconstructed. At ATLAS, reconstructing displaced tracks can require running dedicated re-tracking algorithms in order to identify highly displaced tracks. This re-tracking can be computationally expensive, and necessitates the preselection of at most $\sim 5\%$ of the total event sample on tape. In this case, the preselection criteria are likely to be the limiting factor in signal acceptance at the analysis level. At CMS, the standard tracking algorithm is iterative and automatically reconstructs highly-displaced tracks, so preselection is not necessary.

Dark showers produced through mediators carrying SM charges, such as the scenario of Ref. [325], generally provide ample preselection criteria through associated objects and relatively high overall event $H_T$ scales. For instance, typical events for the model in Ref. [325] contain two emerging jets and two QCD jets, though the additional two QCD jets can be absent in other models. The emerging jets can be reconstructed using default anti-$k_t$ [360] $R = 0.4$ jets. A baseline preselection requires each jet to have $p_T > 200$ GeV, $|\eta| < 2.5$ and $H_T$ exceeding 1000 GeV. These criteria assume that jets can be reconstructed using the calorimeters, which should be very efficient for the considered lifetimes in the model. CMS has recently published a search for this model [310], finding excellent sensitivity for the benchmark model developed in Ref. [325]; this analysis is an excellent demonstration that searches of this type are feasible despite the challenging nature of the signal.

When dark showers originate from a SM singlet, such as the Higgs boson or a $Z'$, the problem of preselection becomes more acute. The overall mass scale of signal events can easily be small, making event $H_T$ useless for signal separation. Here perhaps one of the most robust avenues for preselection is muon multiplicity: as discussed in Section 7.4, many of the operators governing LLP decays tend to predict muon-rich final states. With $\gg 2$ LLPs in an event, muon number becomes a useful and inclusive preselection criterion that places no demands on the possible presence of associated objects, event $H_T$, or detailed shower shape.

For muon-poor, low-mass dark shower events, pre-selection criteria become more model dependent. When such dark showers originate from an exotic Higgs decay, SM Higgs production provides a suite of associated prompt objects that offer pre-selection handles at some acceptance cost; see Section 2.3.1. However it is also important to study to what extent features reflecting the presence of (multiple) LLPs in the event itself could be used as pre-selection criteria, e.g., anomalously track-poor jets, unusual ECAL/HCAL ratios (as realized by, for instance, LLPs with $c\tau \lesssim$ m decaying dominantly to electrons and/or photons), anomalously large numbers of poorly reconstructed or high impact parameter tracks, etc.

Off-line reconstruction for events with low-mass and/or very soft vertices, as occurs in SUEP-style models, poses additional challenges. For the reconstruction of secondary vertices in the inner detector in ATLAS and CMS, currently a minimum track $p_T$ of 1 GeV is required [230, 586]. Having vertices with several tracks fulfilling this minimum $p_T$ requirement might be rare if the hidden sector hadronization scale is low, and this is particularly true for low mediator masses. In this case, dedicated search strategies might be needed (particularly for short lifetimes). For example, it may be possible to look for increased multiplicity of hits in the outer layers of the inner detector compared to the inner parts, though unassociated hits from secondaries may be a significant background. A subtraction of hits from tracks stemming from vertices close to the beam pipe might be helpful, but dedicated studies are needed to assess the viability of this approach. In addition, the calorimeter may be another handle on this type of event: a large collection of soft particles could collectively contribute a non-standard energy pattern in the calorimeters provided the LLP lifetime is suffi-ciently long and if the calorimeter segmentation in radial direction is exploited. However, if the LLPs can reach the calorimeter, their decays typically take place over a distance longer than or comparable to the size of the calorimeter itself. This effect could wash out the signal to the extent that it may be difficult to observe above background, though a full truth-level simulation of this effect is certainly warranted.

Finally, one of the great advantages of an inclusive, low-background search strategy is that it enables a single search to apply robustly to a vast class of models. To fully realize the power and impact of such searches, it will be critical to make it possible to reinterpretable searches with publicly available material. In searches for DVs in a particular detector element, it may be advantageous to bin signal and background in terms of the number of reconstructed vertices with very loose $\eta$ and $\phi$ requirements. While more difficult experimentally, it would also be highly valuable to eventually supply the transverse distance of each vertex from the interaction point on an event-by-event basis, combining the different detector elements. This would allow for straightforward reinterpretation of searches to models with different shower shapes, lifetimes and $p_T$ spectra, allowing a single search to transparently apply to a very broad model space. If possible, it would also be useful to provide the distribution of background vertices as a function of the number of tracks, as this would help theorists estimate the sensitivity to models with different masses and LLP decay modes.

For searches that rely directly on displaced tracks, it is especially important to publish information on displaced track reconstruction that allows for reliable recasting since this is challenging to model accurately using only public data. For instance, the analysis of Ref. [310] relies on the transverse impact parameter significance of a track to construct discriminating variables, but theorists cannot reconstruct impact parameter significance without clear examples of the detector response. For these searches, it would again be useful to bin signal and background in terms of the number of reconstructed objects and their geometric location within the detector, as well as their $p_T$.

We conclude by reiterating that,. while inclusive searches are naturally desirable, there are substantial practical obstacles for such a program, some of which have been discussed above. In reality, it is likely that a number of different semi-inclusive searches with partially overlapping acceptance are necessary. To that end, further theoretical studies are needed to map out which selection cuts are most robust against the varying of the specifics of particular models.

#### *7.6.2.3 Multi-Species Dark Showers*

If there are multiple species in the dark shower that decay to SM final states, the lifetimes of these species will in general be very different. The intermediate case where most LLP species decay with a macroscopic lifetime produces similar phenomenology to the single-species models discussed above. Another regime occurs when at least one species decays promptly, while the others are either stable or decay outside of the detector, which gives rise to a semi-visible jet [366] (see Section 7.6.1.1 above). The intermediate scenario, where some dark states decay promptly while others have macroscopic detector-scale lifetimes, is equally generic. A combination of semi-visible techniques with DV reconstruction can be used to boost sensitivity in this case. It is especially important to pay attention to potential isolation criteria on DVs, as the presence of potentially large numbers of promptly decaying dark particles may necessitate their relaxation.

## 7.7 *Executive Summary*

Dark showers are a common prediction of a wide range of hidden sector theories. Although the model space is dauntingly enormous, it is possible to make some very general statements about the *signature* space of interest: dark shower events can be characterized by (1) a variable and frequently large multiplicity of particles per event; (2) BSM states with a hierarchy of proper lifetimes, making the existence of at least one LLP species generic; (3) frequently, non-isolation of LLPs from other objects in the event; and (4) non-SM-like energy flow and particle multiplicity. These features typically ensure that dark shower events are very distinctive signatures in the generic regime where at least one species has a detector-scale lifetime. Thus. it should be possible to design powerful and inclusive searches which would be sensitive to a very large portion of the vast model space without needing to rely on poorly predicted (and model-dependent) properties such as shower shape. Toward this end we have identified several promising directions for future study for both theory and experiment, and provide some recommendations here.

As always, one of the primary challenges in searches for dark showers is ensuring the events are recorded on tape. In some models, dark showers may be accompanied by $\not{E}_T$, $H_T$, or a number of associated leptons that can be triggered on, but there are many scenarios where this is not the case, or where these handles come at the expense of a large reduction in signal rate. The foremost example of this is when the dark shower is initiated by an exotic Higgs decay. It is thus critical to pursue dedicated trigger strategies, and we present a few ideas here. We expect that displaced triggers designed for singly- or doubly-produced LLPs will typically have reasonable acceptance for dark shower events, though we caution that in some cases the non-isolation of LLPs in dark shower events may limit their acceptance. Another promising avenue for triggering on dark showers exploits the high particle multiplicity typical of such events, and we recommend study of triggers on (displaced) multi-muons in particular.

The off-line analysis of dark shower events poses several challenges as well. At this time, only the limiting cases of pencil-like or fully spherical showers are under good theoretical control, and hadronization can introduce significant additional uncertainties, especially for models with a spectrum substantially different from QCD. New studies of showering in the intermediate regime were performed for the purpose of this document (presented in Section 7.3), which revealed that different approaches can yield qualitatively different phenomenology. On the one hand, this provides interesting opportunities for searches to make use of event shape and/or jet substructure variables, in particular if the decays of hidden sector states occur promptly or with small displacements. On the other hand, for larger displacements it implies that one should be careful not to heavily bias the selection choices of a search towards a particular shower shape, and rely instead on the displaced LLP decays to separate signal from background. We expect that the most inclusive, most broadly applicable, and most readily reinterpreted searches will be those in which the data is binned in terms of the number of reconstructed displaced objects and their locations within the detector. For recasting purposes, and to assist with unraveling the underlying physics in the event of a discovery, it would moreover be important to supply information concerning distributions of the number of tracks per vertex and/or the vertex mass whenever possible.

The presence of $N_{DV} \gg 2$ displaced vertices in an event will be enormously powerful for background suppression, provided those vertices can be reconstructed. To identify highly displaced tracks, it may be necessary to run dedicated and computationally expensive re-tracking algorithms, requiring the imposition of some pre-selection criteria to identify events of interest. As these pre-selection criteria are likely to be the limiting factor controlling post-trigger signal acceptance in many models, we recommend developing criteria for dark showers that rely on particle multiplicity and, if possible, the presence of multiple LLPs in the event, while keeping $p_{\rm T}$ thresholds as low as possible.

**Acknowledgments**

We thank Laura Jeanty, Yuri Gershtein, Ted Kolberg, Simone Pagan Griso, Brian Shuve and Matt Strassler for useful discussions.

## *7.8 Appendix: Example Models*

In this appendix we survey the models for which Monte Carlo simulations are currently available. Please also see Appendix A as some of the models are included in simplified model library.

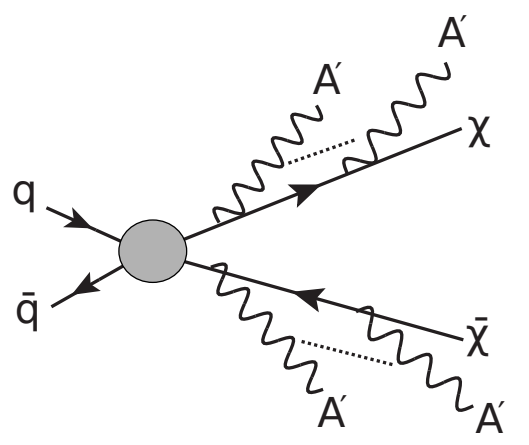

Figure 7.10: The process that gives radiating dark matter its name: production of two DM particles $\chi$, followed by the emission of several soft or collinear dark photons $A'$ [587].

### 7.8.1 Lepton Jets from Radiating Dark Matter

One of the simplest and most widely discussed types of low-mass hidden sector particle is a dark photon $A'$, i.e., a new gauge boson associated with a local $U(1)$ symmetry in the dark sector. By kinetically mixing with the SM hypercharge gauge boson, a dark photon can act as the mediator of dark matter–SM interactions, in addition to being responsible for DM self-interactions. The range of dark photon masses of relevance to dark showers is between MeV and GeV. The dark sector Lagrangian in such a scenario reads

$$\mathcal{L}_{\text{dark}} \equiv \bar{\chi}(i\partial\!\!\!/ - m_\chi + i g_{A'} A\!\!\!/')\chi - \frac{1}{4}F'_{\mu\nu}F'^{\mu\nu} + \frac{1}{2}m^2_{A'}A'_\mu A'^\mu - \frac{\epsilon}{2}F'_{\mu\nu}F^{\mu\nu} \,. \tag{7.8}$$

Here, $\chi$ is the fermionic DM particle with mass $m_\chi$, and $g_{A'} = \sqrt{4\pi\alpha'}$ is the $U(1)'$ gauge coupling. From the point of view of dark showers, interesting values for the coupling strength are $\alpha' \gtrsim 0.01$. If $\alpha'$ is much smaller, there is too little radiation to form a dark shower. The dark photon mass is denoted by $m_{A'}$, and the kinetic mixing parameter by $\epsilon$. Typically, $\epsilon$ is constrained by current limits to be $\lesssim 10^{-3}$. The particular mechanism by which the dark photon mass arises is not important in this context—it could originate from a dark sector Higgs mechanism or from the Stückelberg mechanism.

If $m_\chi \ll 100$ GeV, DM particles may be produced at the LHC with a large boost. This entails a large probability for radiating additional collinear $A'$ bosons (see Figure 7.10). A detailed analytical and numerical description of such dark photon showers is presented in Ref. [587]. The $A'$ bosons eventually decay to observable SM particles through the kinetic mixing term in Eq. (7.8). Depending on the value of $\epsilon$, the decays can be either prompt or displaced. Phenomenologically, the final state of the process $pp \to \bar{\chi}\chi + nA'$ thus consists of two "jets" of collimated $A'$ decay products, plus missing energy.

The $A'$ branching ratios into different SM final states depend sensitively on $m_{A'}$ (see Figure 7.11). At $m_{A'} \lesssim 400$ MeV, the dominant decay modes are $A' \to e^+e^-$ and $A' \to \mu^+\mu^-$. The decay rate

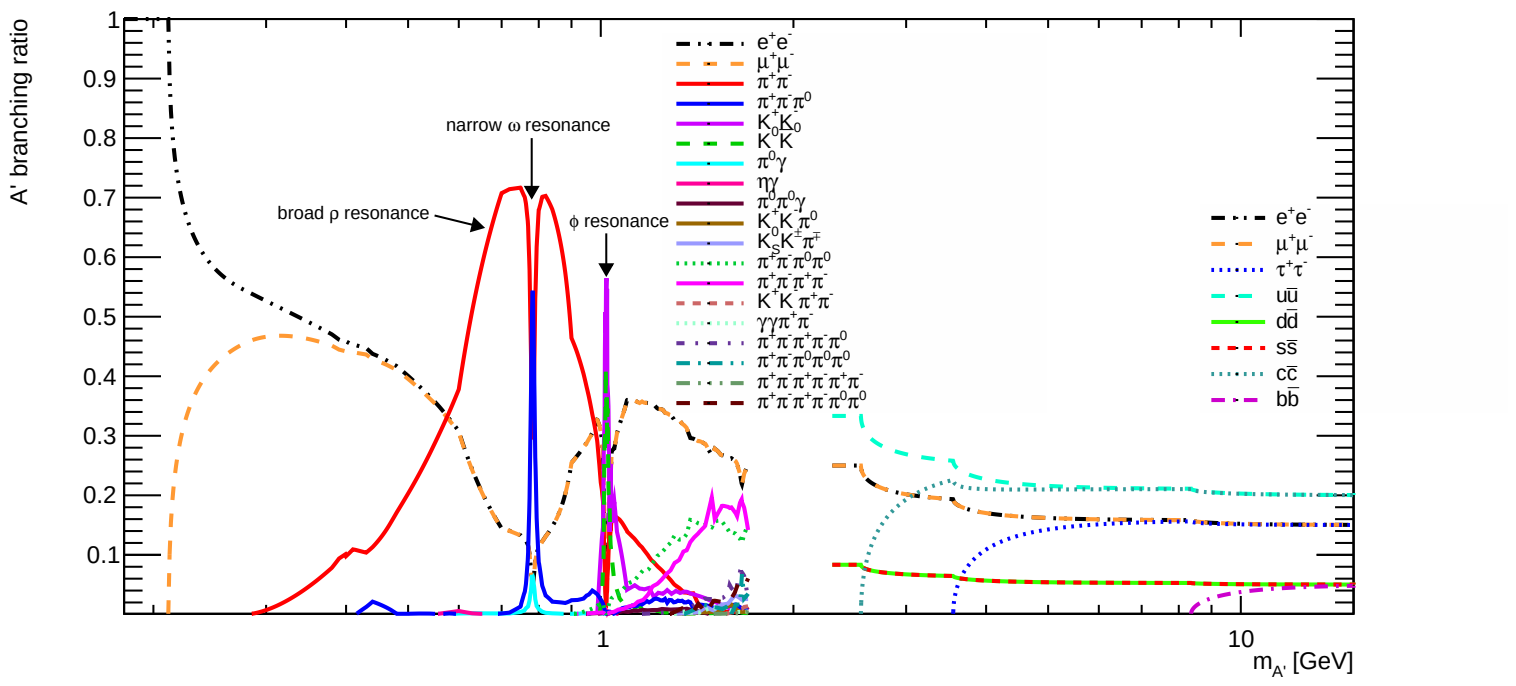

[t]
Figure 7.11: Branching ratios of a kinetically mixed dark photon $A'$as a function of $m_{A'}$ [587].

into each lepton flavor $\ell$ is

$$\Gamma(A' \to \ell^+\ell^-) = \frac{1}{3}\alpha\epsilon^2 m_{A'} = \frac{1}{8\times 10^{-6}\,\text{cm}}\left(\frac{\epsilon}{10^{-3}}\right)^2\left(\frac{m_{A'}}{\text{GeV}}\right), \quad (7.9)$$

where $m_\ell \ll m_{A'}$ has been assumed for simplicity. For purely leptonic decays, the $A'$ shower thus corresponds to a "lepton-jet", i.e., a set of collimated leptons. This signature has been previously discussed, for instance, in Refs. [30, 32, 110, 120, 177, 570, 588–591]. Experimental searches for lepton-jets have been presented in Refs. [263, 265–267], where Refs. [265, 266] focus on lepton-jets with displaced vertices.

Lepton-jet searches may also be sensitive to dark photons with masses $\gtrsim$ 400 MeV, even though the leptonic branching ratio is reduced to between 20% and 70% in this regime. The mix of leptons and hadrons expected from $A'$ decays at $m_{A'} \gtrsim$ 400 MeV implies, however, that most leptons will not be isolated, but occur in conjunction with hadronic activity in the same detector region. In view of this, dedicated trigger and analysis strategies may significantly boost the sensitivity (see Section 7.6.2.1).

In certain parameter regions, the decays of radiated $A'$ bosons may closely resemble a purely hadronic QCD jet. This will happen in particular when $m_{A'}$ is close to a QCD resonance, or when the average $A'$ multiplicity in each shower is low, and the hadronic branching ratio is sizeable. In this case, separation of the signal from the QCD background will most likely be possible only if $\epsilon$ is so small that $A'$ decays are displaced.

Let us summarize several important considerations to take into account when devising a search for dark sector radiation:

- **There may or may not be a signal in the tracking detector.** Prompt $A'$ decays will typically lead to such signals, except for specific $A'$ decay modes to neutral particles, for instance $A' \to K^0\bar{K}^0$ and $A' \to \pi^0\gamma$. Displaced $A'$ decays can also leave a signal in the tracker, however if the lifetime is sufficiently long more decays will occur in the muon chamber (see Section 7.4.2).
- **There may or may not be a signal in the calorimeters.** Most $A'$

decay modes will be visible to the calorimeters. However, it is important to realize that a signal in the hadronic calorimeter can arise not only from hadronic activity, but also from displaced decays to leptons occurring inside the calorimeter.

- **There will be missing energy contained within the lepton jet.** As the decaying $A'$ bosons are aligned with the DM particle from which they were radiated, the corresponding missing momentum vector points in the same direction. However, unless there is significant initial state radiation, the missing momentum will typically be balanced between the two showers shown in Figure 7.10, and may therefore be small.

In Figure 7.12, we show for illustration the exclusion limits that past LHC searches place on the dark photon parameters.

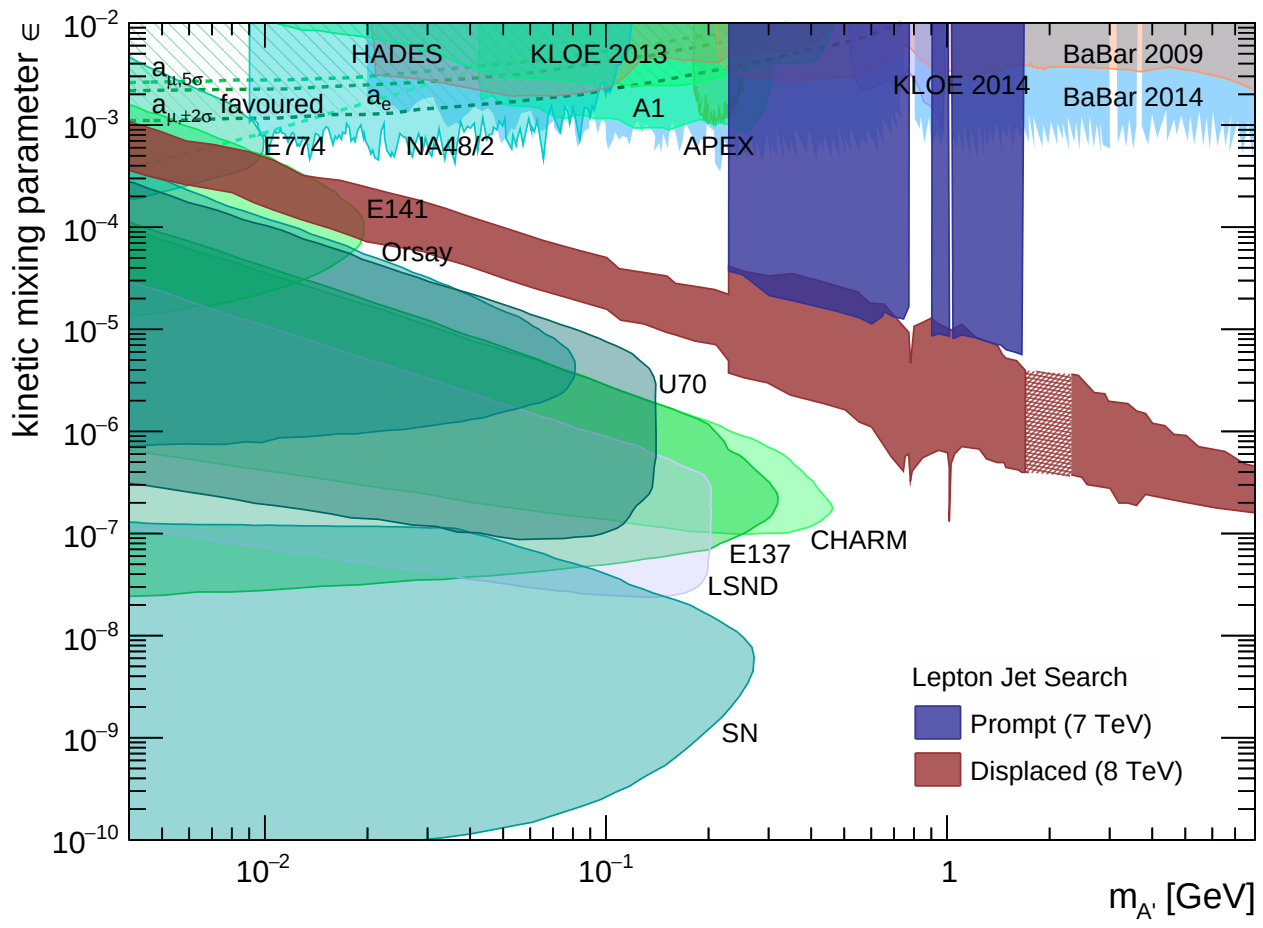

Figure 7.12: Limits, at the 95% C.L., on the dark photon mass $m_{A'}$ and the kinetic mixing parameter $\epsilon$, for $m_\chi = 4$ GeV, $\alpha_{A'} = 0.2$. We have assumed $\chi$ production through a $Z'$ portal with a mass of 1 TeV, with a production cross section of 0.58 pb at $\sqrt{s} = 7$ TeV and 0.85 pb at 8 TeV. We show exclusion limits from the ATLAS search for prompt lepton-jets in 5 fb$^{-1}$ of 7 TeV data [592] (blue shaded region) and from the ATLAS displaced lepton-jet search in 20.3 fb$^{-1}$ of 8 TeV data [265] (red shaded region). The lighter colored region around $m_{A'} = 2$ GeV corresponds to the transition region between an analysis in terms of hadron final states and an analysis in terms of quark final states and is based on interpolation. The computation was carried out in Pythia 8 [213, 593, 594]; see Ref. [587] for details. We also show the existing 90% C.L. exclusion limits from the electron and muon anomalous magnetic moment [19, 595, 596], HADES [597], KLOE 2013 [598] and 2014 [599], the test run results from APEX [600], BaBar 2009 [601] and 2014 [602], beam dump experiments E137, E141, and E774 [603–605], A1 [606], Orsay [607], U70 [608], CHARM [609], LSND [610], as well as constraints from astrophysical observations [611, 612] and $\pi^0$ decays [613]. Figure based on Ref. [587].

### 7.8.2 Emerging and Semi-Visible Jets

Hidden valley models [59] with QCD-like hidden sectors allow for interesting collider signatures. Thus we consider a confined dark gauge group SU($N_d$), where $N_d \geq 2$, with confinement scale $\Lambda_d$ which sets the mass of the dark hadrons. There are also $n_f$ flavors of dark quarks whose bare masses are lighter than $\Lambda_d$. The hidden valley comes equipped with a portal that couples the dark sector to the SM, and the mass is usually taken to be $M \gg \Lambda_d$. The portal can be an $s$-channel vector mediator, $Z_d$, which couples to SM quarks and dark quarks:

$$\mathcal{L} \supset -Z_{d,\mu} \sum_{i,a} \left( g_q \, \bar{q}_i \gamma^\mu q_i + g_{q_d} \, \bar{q}_{d,a} \gamma^\mu q_{d,a} \right) , \qquad (7.10)$$

Here $g_{q/q_d}$ are coupling constants and $i, a$ are flavor indices. One can also have a $t$-channel scalar bifundamental mediator, $X$, which carries color and dark color and can decay to a quark and a dark quark. In the case of the scalar mediator, the only allowable coupling is of the form

$$\mathcal{L} \supset \kappa_{ij} \bar{q}_i q_{d,j} X + \text{h.c.} , \qquad (7.11)$$

where $\kappa_{ij}$ is a $3 \times n_f$ matrix of Yukawa couplings. One could also add multiple flavors of $X$ mediators, something that has also been implemented [539].

When dark quarks are produced, they shower and hadronize and the same tools that are familiar for QCD can be used to simulate these processes. Because of the large gap between the mediator mass and the confining scale, there will be large particle multiplicity and the dark hadrons will typically form into jet-like structures. In the large $N_d$ limit, the fraction of dark baryons produced is suppressed, and in the case of QCD this fraction is $\mathcal{O}(0.1)$. Therefore the hadronization in these simulations is typically dominated by dark mesons. The lightest hadronic states are the dark pions $\pi_d$, acting as goldstone bosons of the $U(n_f) \times U(n_f)$ dark flavor symmetry. When a heavier mesonic state is produced it will promptly decay into dark pions if kinematically allowed, making the dark pions the dominant component of the dark showering process.

These events can be simulated with the Hidden Valley [593] module / package that appears in `Pythia8` [213]. `Pythia8` hosts a hidden valley class which incorporates the $SU(N_d)$ model, allowing the user to vary the masses of the spectrum ($\pi_d, \rho_d, etc$), number of flavors $n_f$, and parameters of the running coupling (Version $\geq$ 8.226). Pair production of $X$ and resonant production of $Z_d$ are implemented at tree level, and the decay of dark mesons to SM states is also present.

In Refs. [572, 614], production of the heavy mediators with ISR/FSR was considered. This was done by interfacing with

`Madgraph5_aMC@NLO` [219] using a modified version of the spin-1 `DMsimp` model[6] implemented through `FeynRules` [615]. The models are located in the repository[7] folder `DMsimp_s_spin1`. The genera-tion files for the $t$-channel exchange of the scalar $X$ are located in the folder `DMsimp_tchannel`. The bi-fundamentals are denoted with `su11, su12, su21, su22...`, where `u` explicitly specifies the QCD flavor index and the numbers are the explicit dark non-Abelian group indices. Similarly, the dark quarks are labeled as `qv11, qv12, qv21, qv22`. A `FeynRules` model file (`DMsimp_tchannel.fr`) as well as the Mathematica notebook (`DMsimp_tchannel.nb`) used to generated the UFO output are also provided. The showering and hadronization in the dark sector can still be performed in `Pythia8`.

[6] `http://feynrules.irmp.ucl.ac.be/wiki/DMsimp`

#### *7.8.2.1 Semi-Visible Jets*

Generically, some of the dark pions could decay promptly while some could be long-lived or even collider-stable, analogous to the appearance of neutral vs. charged pions in the SM. If both stable and unstable hadrons are produced in a collision, the missing energy could be aligned along one of the jets, resulting in low acceptance for traditional monojet-style searches. This is the *semi-visible jet* scenario, and Refs. [366, 572] introduce a simplified model-like parameterization to map the complicated dynamics of the underlying dark sector onto a limited number of physically-motivated variables—in particular, the fraction of stable vs. decaying pions, the characteristic mass scale of dark pions and the dark coupling strength. The Monte Carlo production described here along with the `Pythia8` Hidden Valley module allows the user to vary these parameters for the $s$- and $t$-channel UV completions. See Refs. [366, 572] for further details.

#### *7.8.2.2 Emerging Jets*

By contrast, the dark pions can be taken to have detector-scale lifetimes in the *emerging jets* scenario [325]. The expected lifetime of the dark pion can be quantified using Eq. 7.11. Under the assumption of universal couplings $\kappa_{ij} = \kappa$ and $m_q > \Lambda_d$, we can calculate the proper lifetime of the dark pions:

$$c\tau_0 \approx 80\text{mm} \times \frac{1}{\kappa} \times \left(\frac{2\ \text{GeV}}{f_{\pi_d}}\right)^2 \left(\frac{100\ \text{MeV}}{m_q}\right) \left(\frac{2\ \text{GeV}}{m_{\pi_d}}\right) \left(\frac{M_X}{1\ \text{TeV}}\right)^4 . \tag{7.12}$$

Here $f_{\pi_d}$ is the dark pion decay constant, and $m_q$ is the mass of the SM quark in the final state. A similar formula applies for the $Z_d$ mediator. Therefore a jet of dark hadrons will be created as dominantly invisible particles, but at long distance, the dark pions will decay back to Standard Model particles and appear in shape like an ordinary jet, although it contains a large number of displaced vertices. The jet emerges as it travels through the detectors.

In the case of purely $t$-channel interactions, having a non-trivial $\kappa_{ij}$ in Eq. 7.11 will break the $U(n_f) \times U(n_f)$ dark flavor symmetry. With an appropriate $n_f$, the dark quark flavors can be exactly aligned with the SM down quark flavors. It is immediately clear that such alignment leads to dark pions with lifetimes that differ significantly from one another. In return, the flavor composition of the emerging jets will vary throughout the detector volume. This differs from the $s$-channel interaction, in which no breaking occurs, and $n_f$-stable vs. $n_f(n_f - 1)$-unstable dark pions exist.

### 7.8.3 SUEPs

In the strongly coupled regime, dark showers can be become spherical and almost arbitrarily soft, leading to soft, unclustered energy patterns, or SUEPs, discussed earlier in this chapter. In this case the sheer multiplicity of the final states becomes the most important experimental handle. It has been shown through the AdS/CFT correspondence that their spectrum should follow an approximately thermal distribution [616], which is what was assumed in the phenomenological study in [235]. The production mechanism for these studies was the Higgs or a heavy Higgs-like scalar. The Monte Carlo code that was written for this study is currently not available publicly, though events in hepmc format can be obtained by contacting the authors. At this time only leptonic decays of the hidden mesons are implemented.

# 8 Conclusions

The research program carried out over the first nine years of the Large Hadron Collider (LHC) at CERN has been an unqualified success. The discovery, in 2012, at a center-of-mass energy of 7 and 8 TeV, of a new particle thus far consistent with the Standard Model (SM) Higgs boson has opened numerous new research directions and has begun to shed light upon the source of electroweak symmetry breaking, vector boson scattering amplitudes, and the origin of particle masses. And the establishment of a wide range of searches for new physics at 7, 8, and 13 TeV with the ATLAS, CMS, and LHCb detectors — searches thus far consistent with SM expectations — has inspired new ideas and thinking about the most prominent open issues of physics, such as the nature of dark matter, the hierarchy problem, neutrino masses, and the possible existence of supersymmetry.

The overwhelming majority of searches for new physics have been performed under the assumption that the new particles decay promptly, i.e., very close to the proton-proton interaction point (IP), leading to well-defined objects such as jets, leptons, photons, and missing transverse momentum. Such objects are constructed requiring information from all parts of the detector including hits close to the IP, calorimeter deposits known to be signatures of particles originating from the IP, and muons with tracks that traverse the entirety of the detector, moving out from the IP. However, given the large range of particle lifetimes in the SM —– resulting from general concepts such as approximately preserved symmetries, scale hierarchies, or phase space restrictions —– and the lack of clear, objective motivation related to any particular model or theory beyond the SM, the lifetime of hypothetical new particles is best treated as a free parameter. This leads to a wide variety of spectacular signatures in the LHC detectors that would evade prompt searches, and which have received modest attention compared to searches for promptly decaying new particles. Because such signatures require significantly customized analysis techniques and are usually performed by a smaller number of physicists working on the experimental collaborations, a comprehensive overview and critical review of beyond-SM (BSM) LLPs at the LHC has been performed by a community of experimentalists, theorists, and phenomenolo-gists. This effort ensures that such avenues of the possible discovery of new physics at the LHC are not overlooked. The results of this initiative have been presented in the current document.

We developed a set of simplified models and tools, in Chapter 2, that can be used to parametrize the space of LLP signatures. The simplified models were organized around generic ways that various BSM LLPs can be produced and decay to displaced or non-standard objects in the LHC detectors, rather than emphasizing any one particular theory or physics motivation. These can serve as a useful grammar by which to compare coverage of LLP signature space and model classes among current and future experiments.

To that end, in Chapter 3, we utilized these models and tools to assess the coverage of current LLP searches and we identified multiple avenues for improving and extending the existing LLP search program. Opportunities for new and improved triggering strategies, searches, and open questions for the experimental collaborations to explore centrally were presented as a list at the end of the chapter.

Moreover, due to the non-standard nature of LLP searches, many of them are performed under very low-background conditions. As a result, sources of backgrounds largely irrelevant to searches for promptly decaying BSM particles are important for LLP analyses and can be surprising and unexpected. In Chapter 4 we discussed several sources of backgrounds for LLP searches, collecting the knowledge gained, often by trial-and-error, by experimentalists over many years of searches.

Also with an eye to the future, in Chapter 5 we explored the potential for the expanded capabilities of proposed detector upgrades at ATLAS, CMS, LHCb, and related dedicated detectors, highlighting areas where new technologies can have a large impact on sensitivity to LLP signatures and suggesting several studies to be performed by the collaborations to ensure new physics potential is not missed for the upcoming era of the High-Luminosity LHC.

Additionally, to ensure that current and future searches can be maximally useful in the future, in Chapter 6 we explored how current searches can apply to new models and performed a comprehensive overview of some of the challenges and pitfalls inherent in attempting to recast existing LLP analyses, leading to recommendations for the presentation of search results in the future.

Finally, in Chapter 7, we looked toward the newest frontiers of LLP searches, namely high-multiplicity or "dark shower" LLP signals that can, for example, be signatures of complex hidden sectors with strong dynamics and internal hadronization. In this chapter, we elaborated on the theoretical and experimental challenges and opportunities in expanding the LHC reach of these signals. Such dark shower signatures have a high potential for being overlooked with existing triggering strategies and analysis techniques. Moreover, the dark-QCD-like theoretical models from which they can arise are currently being explored in depth and the resulting

LHC phenomenology is in the process of being understood. We discussed the current state of this work and we anticipate exciting independent developments in the near future.

This document is incomplete by design, since it is a record of the critical thinking and examination of the state of LLP signatures by a large number of independently organized members of the LHC LLP Community as it has evolved from 2016 to 2019, and a major component of the work has been the identification of several open questions and opportunities for discovery in such signatures. As these questions are addressed and new searches emerge from the experimental collaborations, so, too, will new ideas emerge and evolve from the community. We expect this document to be followed by future papers to record, review, and summarize the evolution of LLP signatures and searches, always with the intention of more effectively facilitating the discovery of new particles at the LHC and beyond.

**Acknowledgments:**

We are very grateful to Andreas Albert, Tao Huang, Laura Jeanty, Joachim Kopp, Matt LeBlanc, Larry Lee, Haolin Li, Siddharth Mishra-Sharma, Simone Pagan Griso, Michele Papucci, Dean Robinson, Matt Strassler, and Tien-Tien Yu for helpful conversations and comments on the draft.

*Editors:* J. Beacham acknowledges support from the U.S. Department of Energy (DOE) and the U.S. National Science Foundation (NSF). G. Cottin acknowledges support from the Ministry of Science and Technology of Taiwan (MOST) under Grant No. MOST-107-2811-M-002-3120. N. Desai was supported in part by the OCEVU Labex (ANR-11-LABX-0060) and the A*MIDEX project (ANR-11-IDEX-0001-02) funded by the "Investissements d'Avenir" French government program managed by the ANR. J. A. Evans is supported by DOE grant DE-SC0011784. S. Knapen is supported by DOE grant DE-SC0009988. A. Lessa is supported by the Sao Paulo Research Foundation (FAPESP), project 2015/20570-1. Z. Liu is supported in part by the NSF under Grant No. PHY-1620074, and by the Maryland Center for Fundamental Physics. S. Mehlhase is supported by the BMBF, Germany. M. J. Ramsey-Musolf is supported by DOE grant DE-SC0011095. The work of P. Schwaller has been supported by the Cluster of Excellence "Precision Physics, Fundamental Interactions, and Structure of Matter" (PRISMA+ EXC 2118/1) funded by the German Research Foundation (DFG) within the German Excellence Strategy (Project ID 39083149). The work of J. Shelton is supported in part by DOE under grant DE-SC0017840. The work of B. Shuve is supported by NSF under grant PHY-1820770. X.C. Vidal is supported by MINECO through the Ramón y Cajal program RYC-2016-20073 and by XuntaGal under the ED431F 2018/01 project.

*Contributors:* M. Adersberger is supported by the BMBF, Germany. The work of C. Alpigiani, A. Kvam, E Torro-Pastor, M. Profit, and G. Watts is supported in part by the NSF. Y. Cui is supported in part by DOE Grant DE-SC0008541. J. L. Feng is supported in part by Simons Investigator Award #376204 and by NSF Grant No. PHY-1620638. I. Galon is supported by DOE Grant DE-SC0010008. K. Hahn is supported by DOE Grant DE-SC0015973. J. Heisig acknowledges support from the F.R.S.-FNRS, of which he is a postdoctoral researcher. The work of F. Kling was supported by NSF under Grant No. PHY-1620638. H. Lubatti thanks the NSF for support. P. Mermod was supported by Grant PP00P2_150583 of the Swiss National Science Foundation. V. Mitsou acknowledges support by the Generalitat Valenciana (GV) through MoEDAL-supporting agreements and the GV Excellence Project PROMETEO-II/2017/033, by the Spanish MINECO under the project FPA2015-65652-C4-1-R, by the Severo Ochoa Excellence Centre Project SEV-2014-0398 and by a 2017 Leonardo Grant for Researchers and Cultural Creators, BBVA Foundation. J. Prisciandaro is supported by funding from FNRS. M. Reece is supported by DOE Grant DE-SC0013607. D. Stolarski is supported in part by the Natural Sciences and Engineering Research Council of Canada (NSERC). S. Trojanowski is supported by Lancaster-Manchester-Sheffield Consortium for Fundamental Physics under STFC grant ST/L000520/1. S. Xie is supported by the California Institute of Technology High Energy Physics Contract DE-SC0011925 with the DOE. This manuscript has been partially authored by Fermi Research Alliance, LLC under Contract No. DE-AC02-07CH11359 with the U.S. Department of Energy, Office of Science, Office of High Energy Physics.

# A
# Appendix: Simplified Model Library

**Contents**

**Chapter Editor:** Brian Shuve

**Contributors:** Eshwen Bhal, David Curtin, Alessandro Davoli, Andrea De Simone, Jared Evans, Thomas Jacques, Zhen Liu, Alessandro Morandini, Michael Ramsey-Musolf, Jessie Shelton, Jiang-Hao Yu

## A.1 Instructions for the Simplified Model Library

The simplified model library is available at the LHC LLP Community website [1], hosted at CERN. We refer here extensively to the simplified models in Tables 2.1-2.3. Because it is already quite an extensive task to come up with simplified models for so many (production)×(decay) modes, we for now restrict ourselves predominantly to the "filled" entries in Tables 2.1-2.3. If you are interested in performing an experimental search or developing a simplified model library entry for one of the "unfilled" entries, please contact the chapter editor.

[1] `http://cern.ch/longlivedparticles`

There are essentially two possible pipelines to simulate LLP events with the library:

1. *LLP decay as part of matrix-element calculation:* Using UFO models in the library, it is possible to generate the production and decay of LLPs at the parton level using calculations of the matrix element for production and decay. As a concrete example, we provide cards that allow the production and decay of LLPs using `MadGraph5_aMC@NLO` [219] and the accompanying `MadSpin` [220] package. This employs the narrow-width approximation, but otherwise gives rise to the correct angular distribution of LLP decay products. The downside is that if a particular decay is not allowed from the interactions in the UFO model file, the UFO must be modified to include the new coupling. The output of `MadGraph5_aMC@NLO` is then fed into programs such as `Pythia 8` [212, 213] for showering, implementation of underlying event and other particle-level processes.

2. *Phenomenological LLP decays:* Using UFO models in the library, it is possible to generate the production of LLPs, leaving them stable as outputs of the matrix-element-level calculation recorded in LHE format [217]. The LLP can then be subsequently decayed in programs such as `Pythia 8`, which allows a particle to decay into any final state, albeit without correctly modeling the angular distribution. This could be convenient for models where the interactions leading to LLP decay are not included in the UFO, or where computational time is a concern and the angular distribution of LLP decays is irrelevant. We provide detailed instructions in the library files for how to implement decays of LHE files via Pythia.

In the final version of the library, we aim to provide example cards to direct the production and decay of LLPs in both pipelines. Note that in all of the simplified model proposals below, any particles *not* present in the production or decay chain should have their masses set to a very large value ($M \gtrsim 5$ TeV) to ensure they are sufficiently decoupled from direct production at the LHC.

Currently, we only provide simplified model libraries for neutral LLPs. The simplified models for LLPs with electric or color charges are equally compelling, but their simulation is more subtle. In particular, the simulation of the propagation and decay of the LLPs are more challenging if the LLP carries a SM gauge charge. Such effects can be included in detector simulations using GEANT4 [224], but decay processes typically need to be hard-coded into GEANT or otherwise interfaced with other MC programs. This is an important issue for the community to address, as discussed in Sections 2.4.2-2.4.3. Without implementing the decays, it is straightforward to use the SUSY model to simulate the production of any of the electrically or color charged LLPs.

Finally, we note that there is not currently a set of minimal simplified models to cover dark showers. However, we are collecting models used in studies of dark showers that could potentially be helpful for experimentalists and theorists alike. The existing models are included in the library, along with a very brief description in Section A.1.2.

### *A.1.1 Neutral LLPs*

The instructions for simulating the simplified model channels for neutral LLPs are given below. Note that it is often true that the same simplified model production and decay channel can be simulated using several simplified models. However, the philosophy of the simplified model approach is that the UV model used to simulate the process is not important when sensitivity is expressed in terms of physical masses and cross sections. As a result, we typically provide only one set of instructions for simulating each simplified model channel.

We begin by presenting the simplified model library instructions for Double Pair Production (DPP) in Table A.1, Heavy Parent (HP, QCD-charged parent) in Table A.2. We then proceed to the Higgs (HIG) production modes in Tables A.4-A.6. For the $Z'$ (ZP) production modes, we use a set of simplified models described in Tables A.7-A.9. A relatively simple model file is provided for each table. In addition, a more adjustable 'advanced' model file is provided which includes all ZP production and decay modes, and allows for features such as individual couplings to each generation of quarks. This comes at the cost of a greatly increased set of parameters, and the possibility of including unwanted diagrams if the process is not carefully specified. A simple python script is provided to generate the processes and set unwanted parameters to zero for those users wishing to use the advanced model files.

Finally, we provide instructions for the charged-current (CC) production modes in Table A.10. This production mode is most easily simulated using a left-right symmetric model or other right-handed-neutrino model.

| Decay Mode | Simplified Model Library Process |
|---|---|
| $X \to \gamma$+inv. | MSSM+GMSB. LLP is a bino ($\tilde{\chi}^0$) produced due to $pp \to \tilde{\chi}^0\tilde{\chi}^0$ via $t$-channel squark exchange ($M_{\tilde{q}} > 5$ TeV). Bino decays to photon + gravitino, $\tilde{\chi}^0 \to \gamma + \tilde{G}$. |
| $X \to jj$ | MSSM+RPV. LLP is sneutrino LSP ($\tilde{\nu}$) that is pair-produced via weak gauge interactions. $\tilde{\nu} \to q\bar{q}$ via the $QLd^c$ operator. |
| $X \to jj$+inv. | MSSM. LLP is second neutralino (wino) LSP $\tilde{\chi}^0_2$ that is pair-produced via weak gauge interactions. $\tilde{\chi}^0_2 \to q\bar{q}\tilde{\chi}^0_1$ via an off-shell sfermion, and the $\tilde{\chi}^0_1$ is invisible with arbitrary mass. |
| $X \to jjj$ | MSSM+RPV. While this is partially covered by $jj +$ inv. in the case where the additional quark is not reconstructed, we include it here for completeness. LLP is wino LSP ($\tilde{\chi}^0$) that is pair-produced via weak interactions. $\tilde{\chi}^0 \to q_\alpha q_\alpha q_\beta$ via an off-shell sfermion and the $u^c_\alpha d^c_\alpha d^c_\beta$ operator. |
| $X \to jj\ell_\alpha$ | MSSM+RPV. LLP is wino LSP ($\tilde{\chi}^0$) that is pair-produced via weak interactions. $\tilde{\chi}^0 \to \ell_\alpha q\bar{q}$ via an off-shell sfermion and $L_\alpha Q d^c$ operator. |
| $X \to \ell^+_\alpha \ell^-_\alpha$ | MSSM+RPV. LLP is sneutrino $\tilde{\nu}_\beta$ of flavor $\beta$ that is pair-produced via weak interactions. $\tilde{\nu}_\beta \to \ell^+_\alpha \ell^-_\alpha$ via the $L_\alpha L_\beta E^c_\alpha$ operator. |
| $X \to \ell^+_\alpha \ell^-_\alpha$(+inv.) | MSSM. LLP is second neutralino $\tilde{\chi}^0_2$ that is pair-produced via weak interactions. $\tilde{\chi}^0_2 \to \tilde{\chi}^0_1 \ell^+_\alpha \ell^-_\alpha$ via an off-shell slepton. |
| $X \to \ell^+_\alpha \ell^-_\beta$(+inv.) | MSSM+RPV. LLP is sneutrino $\tilde{\nu}_\alpha$ of flavor $\alpha$ that is pair-produced via weak interactions. $\tilde{\nu}_\alpha \to \ell^+_\alpha \ell^-_\beta$ via the $L_\alpha L_\beta E^c_\alpha$ operator. An additional massless invisible final state can be obtained with a wino LLP decaying into $\ell^+_\alpha \ell^-_\beta \nu_\alpha$ through the same operator and an off-shell slepton. The massive invisible case is less motivated for $\alpha \neq \beta$. |

Table A.1: Simplified model library process proposals for Double Pair Production (DPP) production mode. Where a "wino" LSP is specified, an admixture of Higgsino is required to lead to direct pair production of the neutral wino component. As an alternative, one could have $pp \to \tilde{\chi}^\pm\tilde{\chi}^0$, $\tilde{\chi}^\pm \to W^{\pm *}\tilde{\chi}^0$ promptly, and take the $\tilde{\chi}^\pm$ to be degenerate with $\tilde{\chi}^0$ such that the additional charged decay products are essentially unobservable.

| Decay Mode | Simplified Model Library Process |
|---|---|
| $X \to \gamma$+inv. | MSSM+GMSB. LLP is a bino $(\tilde{\chi})$ produced via $pp \to \tilde{q}\tilde{q}^*$, $\tilde{q} \to \tilde{\chi} + q$. Bino decays to photon+ gravitino, $\tilde{\chi} \to \gamma + \tilde{G}$. |
| $X \to jj$+inv. | MSSM. LLP is wino LSP $\tilde{\chi}^0_2$ that is produced via $pp \to \tilde{q}\tilde{q}^*$, $\tilde{q} \to q\tilde{\chi}^0_2$. Then, $\tilde{\chi}^0_2 \to q\bar{q}\tilde{\chi}^0_1$ via an off-shell quark. |
| $X \to jjj$ | MSSM+RPV. LLP is bino LSP $\tilde{\chi}$ that is produced via $pp \to \tilde{q}\tilde{q}^*$, $\tilde{q} \to q\tilde{\chi}$. Then, $\tilde{\chi} \to q_\alpha q_\alpha q_\beta$ via the $u^c_\alpha d^c_\alpha d^c_\beta$ operator. |
| $X \to jj\ell_\alpha$ | MSSM+RPV. LLP is bino LSP $(\tilde{\chi})$ that is produced via $pp \to \tilde{q}\tilde{q}^*$, $\tilde{q} \to q\tilde{\chi}$. $\tilde{\chi} \to \ell_\alpha q\bar{q}$ via an off-shell sfermion and $L_\alpha Q d^c$ operator. |
| $X \to \ell^+_\alpha \ell^-_\alpha$ or $\ell^+_\alpha \ell^-_\beta$ | MSSM+RPV. LLP is sneutrino $(\tilde{\nu})$ that is produced via $pp \to \tilde{g}\tilde{g}$, $\tilde{g} \to jj\tilde{\chi}$, $\tilde{\chi} \to \tilde{\nu}\bar{\nu}$. Then, $\tilde{\nu}_\alpha \to \ell^+_\alpha \ell^-_\beta$ or $\tilde{\nu}_\beta \to \ell^+_\alpha \ell^-_\alpha$ via the $L_\alpha L_\beta E^c_\alpha$ operator. |
| $X \to \ell^+_\alpha \ell^-_\alpha$+inv. | MSSM. LLP is second neutralino $(\tilde{\chi}^0_2)$ that is produced via $pp \to \tilde{q}\tilde{q}^*$, $\tilde{q} \to q\tilde{\chi}^0_2$. Then, $\tilde{\chi}^0_2 \to \ell^+_\alpha \ell^-_\alpha \tilde{\chi}^0_1$. |
| $X \to \ell^+_\alpha \ell^-_\beta$+inv. | MSSM+RPV. LLP is bino $(\tilde{\chi}^0)$ that is produced via $pp \to \tilde{q}\tilde{q}^*$, $\tilde{q} \to q\tilde{\chi}^0$. Then, $\tilde{\chi}^0 \to \ell^+_\alpha \ell^-_\beta \nu_\alpha$ via $L_\alpha L_\beta E^c_\alpha$ operator and off-shell slepton (massless invisible only). |

Table A.2: Simplified model library process proposals for Heavy Parent (HP) production mode where the parent particle carries a QCD charge. In most of the above cases, a squark parent can be replaced by a gluino parent with an additional jet in its decay.

| Decay Mode | Simplified Model Library Process |
|---|---|
| $X \to \gamma$+inv. | MSSM+GMSB. LLP is a bino $(\tilde{\chi}^0)$ produced via $pp \to \tilde{\chi}^+\tilde{\chi}^-$, $\tilde{\chi}^+ \to W^+\tilde{\chi}^0$ ($\tilde{\chi}^+$ is a wino). Bino decays to photon+ gravitino, $\tilde{\chi} \to \gamma + \tilde{G}$. |
| $X \to jj$ | MSSM+RPV. LLP is sneutrino LSP $(\tilde{\nu})$ produced via $pp \to \tilde{\chi}^+\tilde{\chi}^-$, $\tilde{\chi}^+ \to \tilde{\nu}\ell^+$. The sneutrino decays via $\tilde{\nu} \to q\bar{q}$ via the $u^c_\alpha d^c_\alpha d^c_\beta$ operator. |
| $X \to jj$+inv. | MSSM. LLP is wino $\tilde{\chi}^0_2$ that is produced via $pp \to \tilde{\chi}^+\tilde{\chi}^-$, $\tilde{\chi}^+_2 \to W^+\tilde{\chi}^0_2$. Then, $\tilde{\chi}^0_2 \to q\bar{q}\tilde{\chi}^0_1$ via an off-shell squark. |
| $X \to jjj$ | MSSM+RPV. LLP is bino LSP $(\tilde{\chi}^0)$ that is produced via $pp \to \tilde{\chi}^+\tilde{\chi}^-$, $\tilde{\chi}^+ \to W^+\tilde{\chi}^0$. Then, $\tilde{\chi}^0 \to qqq$ via an off-shell sfermion and the $u^c d^c d^c$ operator. |
| $X \to jj\ell_\alpha$ | MSSM+RPV. LLP is bino LSP $(\tilde{\chi}^0)$ that is produced via $pp \to \tilde{\chi}^+\tilde{\chi}^-$, $\tilde{\chi}^+ \to W^+\tilde{\chi}^0$. Then, $\tilde{\chi}^0 \to qq'\ell_\alpha$ via an off-shell sfermion and the $Qd^cL_\alpha$ operator. |
| $X \to \ell^+_\alpha \ell^-_\alpha$ or $\ell^+_\alpha \ell^-_\beta$ | MSSM+RPV. LLP is sneutrino $(\tilde{\nu})$ that is produced via $pp \to \tilde{\chi}^+\tilde{\chi}^-$, $\tilde{\chi}^+ \to \ell^+\tilde{\nu}$. Then, $\tilde{\nu}_\alpha \to \ell^+_\alpha \ell^-_\beta$ or $\tilde{\nu}_\beta \to \ell^+_\alpha \ell^-_\alpha$ via the $L_\alpha L_\beta E^c_\alpha$ operator. |
| $X \to \ell^+_\alpha \ell^-_\alpha$+inv. | MSSM. LLP is second neutralino $(\tilde{\chi}^0_2)$ that is produced via $pp \to \tilde{\chi}^+\tilde{\chi}^-$, $\tilde{\chi}^+ \to W^+\tilde{\chi}^0_2$. Then, $\tilde{\chi}^0_2 \to \ell^+_\alpha \ell^-_\alpha \tilde{\chi}^0_1$ via an off-shell slepton and the $L_\alpha L_\beta E^c_\alpha$ operator. |
| $X \to \ell^+_\alpha \ell^-_\beta$+inv. | MSSM+RPV. LLP is bino $(\tilde{\chi}^0)$ that is produced via $pp \to \tilde{\chi}^+\tilde{\chi}^-$, $\tilde{\chi}^+ \to W^+\tilde{\chi}^0$. Then, $\tilde{\chi}^0 \to \ell^+_\alpha \ell^-_\alpha \nu_\beta$ or $\tilde{\chi}^0 \to \ell^+_\alpha \ell^-_\beta \nu_\alpha$ via an off-shell slepton and the $L_\alpha L_\beta E^c_\alpha$ operator. |

Table A.3: Simplified model library process proposals for Heavy Parent (HP) production mode where the parent particle carries electroweak charge.

| Decay Mode | Simplified Model Library Process |
|---|---|
| $X \to \gamma\gamma$ | (N)MSSM. LLP is the lightest pseudoscalar Higgs ($a$) produced via $pp \to h, h \to aa$. Then, $a \to \gamma\gamma$. |
| $X \to \gamma\gamma$+inv. | MSSM. LLP is the second neutralino ($\tilde{\chi}^0_2$) produced via $pp \to h, h \to \tilde{\chi}^0_2\tilde{\chi}^0_2$. Then, $\tilde{\chi}^0_2 \to \tilde{\chi}^0_1\gamma\gamma$ via an off-shell SM Higgs. |
| $X \to jj$ | (N)MSSM. LLP is the lightest pseudoscalar Higgs ($a$) produced via $pp \to h, h \to aa$. Then, $a \to jj$. |
| $X \to jj$+inv. | MSSM. LLP is the second neutralino ($\tilde{\chi}^0_2$) that is produced via $pp \to h, h \to \tilde{\chi}^0_2\tilde{\chi}^0_2$. Then, $\tilde{\chi}^0_2 \to jj\tilde{\chi}^0_1$ via an off-shell squark. |
| $X \to \ell^+_\alpha \ell^-_\alpha$ | (N)MSSM. LLP is the lightest pseudoscalar Higgs ($a$) produced via $pp \to h, h \to aa$. Then, $a \to \ell^+_\alpha \ell^-_\alpha$. |
| $X \to \ell^+_\alpha \ell^-_\alpha$+inv. | MSSM. LLP is the second neutralino ($\tilde{\chi}^0_2$) produced via $pp \to h, h \to \tilde{\chi}^0_2\tilde{\chi}^0_2$. Then, $\tilde{\chi}^0_2 \to \ell^+_\alpha \ell^-_\alpha \tilde{\chi}^0_1$ via an off-shell slepton. |
| $X \to \ell^+_\alpha \ell^-_\beta$+inv. | MSSM+RPV. LLP is the second neutralino ($\tilde{\chi}^0_2$) produced via $pp \to h, h \to \tilde{\chi}^0_2\tilde{\chi}^0_2$. Then, $\tilde{\chi}^0_2 \to \ell^+_\alpha \ell^-_\beta \nu$ via an off-shell slepton and RPV couplings. |
| $X \to \ell^+_\alpha jj$ | MSSM+RPV. LLP is the second neutralino ($\tilde{\chi}^0_2$) produced via $pp \to h, h \to \tilde{\chi}^0_2\tilde{\chi}^0_2$. Then, $\tilde{\chi}^0_2 \to \ell^+_\alpha jj$ via an off-shell slepton and RPV couplings. |

Table A.4: Simplified model library process proposals for Higgs (HIG) production mode where the Higgs decays to two LLPs. These modes are particularly important because they can come in association with forward jets (VBF) or leptons and $\not{E}_T$ (VH). Note that, in cases of $M_X > M_h/2$, the same production modes could still occur if the Higgs is taken to be off-shell.

| Decay Mode | Simplified Model Library Process |
|---|---|
| $X \to \gamma\gamma$ | (N)MSSM. LLP is a pseudoscalar or singlino ($a$) produced via $pp \to h, h \to \tilde{\chi}^0_2\tilde{\chi}^0_2$, $\tilde{\chi}^0_2 \to \tilde{\chi}^0_1 a$. Finally, $a \to \gamma\gamma$. |
| $X \to \gamma\gamma$+inv. | MSSM. LLP is the second neutralino ($\tilde{\chi}^0_2$) produced via $pp \to h, h \to \tilde{\nu}\tilde{\nu}^*$, $\tilde{\nu} \to \tilde{\chi}^0_2\nu$. Then, $\tilde{\chi}^0_2 \to \tilde{\chi}^0_1\gamma\gamma$ via an off-shell SM Higgs. |
| $X \to jj$ | MSSM+RPV. LLP is a sneutrino ($\tilde{\nu}$) produced via $pp \to h, h \to \tilde{\chi}^0_1\tilde{\chi}^0_1$, $\tilde{\chi}^0_1 \to \tilde{\nu}\bar{\nu}$. Then, $\tilde{\nu} \to jj$ via the RPV operator $LQd^c$. |
| $X \to jj$+inv. | MSSM. LLP is the second neutralino ($\tilde{\chi}^0_2$) that is produced via $pp \to h$, $h \to \tilde{\nu}\tilde{\nu}^*$, $\tilde{\nu} \to \nu\tilde{\chi}^0_2$. Then, $\tilde{\chi}^0_2 \to jj\tilde{\chi}^0_1$ via an off-shell squark. |
| $X \to \ell^+_\alpha \ell^-_\alpha$ | MSSM+RPV. LLP is a sneutrino ($\tilde{\nu}_\beta$) produced via $pp \to h, h \to \tilde{\chi}^0_1\tilde{\chi}^0_1$, $\tilde{\chi}^0_1 \to \tilde{\nu}_\beta\bar{\nu}_\beta$. Then, $\tilde{\nu}_\beta \to \ell^+_\alpha \ell^-_\alpha$ via the RPV operator $L_\alpha L_\beta E^c_\alpha$. |
| $X \to \ell^+_\alpha \ell^-_\alpha$+inv. | MSSM. LLP is the second neutralino ($\tilde{\chi}^0_2$) that is produced via $pp \to h$, $h \to \tilde{\nu}\tilde{\nu}^*$, $\tilde{\nu} \to \nu\tilde{\chi}^0_2$. Then, $\tilde{\chi}^0_2 \to \ell^+_\alpha \ell^-_\alpha \tilde{\chi}^0_1$ via an off-shell slepton. |

Table A.5: Simplified model library process proposals for Higgs (HIG) production mode where the Higgs decays to two LLPs plus invisible. These modes are particularly important because they can come in association with forward jets (VBF) or leptons and $\not{E}_T$ (VH). Note that, in cases of $M_X > M_h/2$, the same production modes could still occur if the Higgs is taken to be off-shell.

| Decay Mode | Simplified Model Library Process |
|---|---|
| $X \to \gamma\gamma$+inv. | MSSM. LLP is the second neutralino ($\tilde{\chi}_2^0$) produced via $pp \to h$, $h \to \tilde{\chi}_2^0\tilde{\chi}_1^0$. Then, $\tilde{\chi}_2^0 \to \tilde{\chi}_1^0\gamma\gamma$ via an off-shell SM Higgs. |
| $X \to jj$+inv. | MSSM. LLP is the second neutralino ($\tilde{\chi}_2^0$) that is produced via $pp \to h$, $h \to \tilde{\chi}_2^0\tilde{\chi}_1^0$. Then, $\tilde{\chi}_2^0 \to jj\tilde{\chi}_1^0$ via an off-shell squark. |
| $X \to \ell_\alpha^+\ell_\alpha^-$+inv. | MSSM. LLP is the second neutralino ($\tilde{\chi}_2^0$) that is produced via $pp \to h$, $h \to \tilde{\chi}_2^0\tilde{\chi}_1^0$. Then, $\tilde{\chi}_2^0 \to \ell_\alpha^+\ell_\alpha^-\tilde{\chi}_1^0$ via an off-shell slepton. |
| $X \to \ell_\alpha^+ jj$ | MSSM+RPV. LLP is the second neutralino ($\tilde{\chi}_2^0$) that is produced via $pp \to h$, $h \to \tilde{\chi}_2^0\tilde{\chi}_1^0$. Then, $\tilde{\chi}_2^0\ell_\alpha^+ jj$ via the RPV operator $QLd^c$. |
| $X \to \ell_\alpha^+\ell_\beta^-$+inv. | MSSM+RPV. LLP is the second neutralino ($\tilde{\chi}_2^0$) that is produced via $pp \to h$, $h \to \tilde{\chi}_2^0\tilde{\chi}_1^0$. Then, $\tilde{\chi}_2^0 \to \ell_\alpha^+\ell_\beta^-\nu_\sigma$ via the RPV operator $L_\beta L_\sigma E_\alpha^c$. |

Table A.6: Simplified model library process proposals for Higgs (HIG) production mode where the Higgs decays to single LLP plus invisible. These modes are particularly important because they can come in association with forward jets (VBF) or leptons and $\not{E}_T$ (VH). Note that, in cases of $M_X > M_h/2$, the same production modes could still occur if the Higgs is taken to be off-shell.

| Decay Mode | Simplified Model Library Process |
|---|---|
| $X \to \gamma\gamma$ | DMSM. LLP is scalar $s_2$, produced via $pp \to Z' \to s_2 s_2$, then $s_2 \to \gamma\gamma$. |
| $X \to \gamma\gamma$+inv. | DMSM. LLP is fermion $x_2$, produced via $pp \to Z' \to x_2 x_2$, then $x_2 \to \gamma\gamma x_1$. |
| $X \to jj$ | DMSM. LLP is scalar $s_2$, produced via $pp \to Z' \to s_2 s_2$, then $s_2 \to q\bar{q}$. $s_2 \to \ell_\alpha^+\ell_\alpha^-$ couplings are proportional to SM Yukawa couplings. |
| $X \to jj$+inv. | DMSM. LLP is fermion $x_2$, produced via $pp \to Z' \to x_2 x_2$, then $x_2 \to q\bar{q}x_1$. |
| $X \to \ell_\alpha^+\ell_\alpha^-$ | DMSM. LLP is scalar $s_2$, produced via $pp \to Z' \to s_2 s_2$, then $s_2 \to \ell_\alpha^+\ell_\alpha^-$. $s_2 \to \ell_\alpha^+\ell_\alpha^-$ couplings are proportional to SM Yukawa couplings. |
| $X \to \ell_\alpha^+\ell_\alpha^-$+inv. | DMSM. LLP is fermion $x_2$, produced via $pp \to Z' \to x_2 x_2$, then $x_2 \to \ell_\alpha^+\ell_\alpha^- x_1$. |
| $X \to \ell_\alpha^+ jj$. | LRSM. LLP is fermion $\nu_R$, produced via $pp \to Z/Z'/h \to \nu_R\nu_R$, then $\nu_R \to \ell_\alpha^+ jj$ via off-shell $W/W'$. |

Table A.7: Simplified model library process proposals for $Z/Z'$ (ZP) production mode where the $Z'$ decays to two LLPs. For this section, we mostly use a DM simplified model, where fermion $x_2$ is the LLP for $X \to SM+$ inv modes and scalar $s_2$ is the LLP for $X \to SM$ modes. The same models can also be used for off-shell $Z/Z'$ where $M_X > M_{Z'}/2$. The model file includes all processes in the table; the undesired couplings can be set to zero and energy scales to be very large. The final entry uses a left-right symmetric model; the mass of the intermediate $Z/Z'/h$ can be changed to simulate the desired decay kinematics.

| Decay Mode | Simplified Model Library Process |
|---|---|
| $X \to \gamma\gamma$ | DMSM. LLP is scalar $s_2$, produced via $pp \to Z' \to x_3 x_3, x_3 \to s_2 x_1$, then $s_2 \to \gamma\gamma$. |
| $X \to \gamma\gamma$+inv. | DMSM. LLP is fermion $x_2$, produced via $pp \to Z' \to x_3 x_3, x_3 \to x_2 s_1$, then $x_2 \to \gamma\gamma x_1$. |
| $X \to jj$ | DMSM. LLP is scalar $s_2$, produced via $pp \to Z' \to x_3 x_3, x_3 \to s_2 x_1$, then $s_2 \to q\bar{q}$. $s_2 \to \ell_\alpha^+ \ell_\alpha^-$ couplings are proportional to SM Yukawa couplings. |
| $X \to jj$+inv. | DMSM. LLP is fermion $x_2$, produced via $pp \to Z' \to x_3 x_3, x_3 \to x_2 s_1$, then $x_2 \to q\bar{q} x_1$. |
| $X \to \ell_\alpha^+ \ell_\alpha^-$ | DMSM. LLP is scalar $s_2$, produced via $pp \to Z' \to x_3 x_3, x_3 \to s_2 x_1$, then $s_2 \to \ell_\alpha^+ \ell_\alpha^-$. $s_2 \to \ell_\alpha^+ \ell_\alpha^-$ couplings are proportional to SM Yukawa couplings. |
| $X \to \ell_\alpha^+ \ell_\alpha^-$+inv. | DMSM. LLP is fermion $x_2$, produced via $pp \to Z' \to x_3 x_3, x_3 \to x_2 s_1$, then $x_2 \to \ell_\alpha^+ \ell_\alpha^- x_1$. |

Table A.8: Simplified model library process proposals for $Z/Z'$ (ZP) production mode where the $Z/Z'$ decays to two LLPs plus invisible. For this section, we use a DM simplified model, where the $Z'$ decays into $x_3 x_3$, and $x_3$ then decays into the LLP (fermion $x_2$ for SM+inv decay mode or scalar $s_2$ for SM decay mode), plus invisible (scalar $s_1$ or fermion $x_1$ respectively). The same models can also be used for off-shell $Z/Z'$ where $M_X > M_{Z'}/2$. One model file (DMSM) includes all processes in the table.

| Decay Mode | Simplified Model Library Process |
|---|---|
| $X \to \gamma\gamma$+inv. | DMSM. LLP is fermion $x_2$, produced via $pp \to Z' \to x_1 x_2$, then $x_2 \to \gamma\gamma x_1$; or a scalar $s_2$, produced via $pp \to Z' \to s_1 s_2$, then $s_2 \to \gamma\gamma$. |
| $X \to jj$+inv. | DMSM. LLP is fermion $x_2$, produced via $pp \to Z' \to x_1 x_2$, then $x_2 \to jj x_1$; or a scalar $s_2$, produced via $pp \to Z' \to s_1 s_2$, then $s_2 \to jj$. |
| $X \to \ell_\alpha^+ \ell_\alpha^-$+inv. | DMSM. LLP is fermion $x_2$, produced via $pp \to Z' \to x_1 x_2$, then $x_2 \to \ell^+ \ell^- x_1$; or a scalar $s_2$, produced via $pp \to Z' \to s_1 s_2$, then $s_2 \to \ell^+ \ell^-$. |
| $X \to \ell_\alpha^+ jj$ | LRSM. LLP is fermion $\nu_R$, produced via $pp \to Z/Z'/h \to \bar{\nu}_L \nu_R$, then $\nu_R \to \ell^+ jj$. |

Table A.9: Simplified model library process proposals for $Z/Z'$ (ZP) production mode where the $Z/Z'$ decays to single LLP plus invisible. For this section, we use a DM simplified model, where the $Z'$ couples to an $x_1\, x_2$ or $s_1\, s_2$ pair. $x_1$ and $s_1$ behave as DM, the LLP $x_2$ decays into $x_1 + SM$, and the LLP $s_2$ decays into $SM$. The DMSM model file again includes all processes in the table except for the last, which is LRSM.

| Decay Mode | Simplified Model Library Process |
|---|---|
| $X \to jj$+inv. | LRSM. LLP is the right-handed neutrino ($\nu_R$) produced via $pp \to W^\pm$, $W^\pm \to \ell^\pm \nu_R$. Then, $\nu_R \to q\bar{q}\nu$ via an off-shell $Z$. For massive invisible state, it may be possible to use a cascade $\nu_{R2} \to q\bar{q}\nu_{R1}$ treating the lightest right-handed neutrino as stable. |
| $X \to jj\ell^\pm$ | LRSM. LLP is the right-handed neutrino ($\nu_R$) produced via $pp \to W^\pm$, $W^\pm \to \ell^\pm \nu_R$. Then, $\nu_R \to q\bar{q}'\ell^\pm$ via an off-shell $W$. Alternately, production and decay can be mediated by $W_R$. |
| $X \to \ell_\alpha^+ \ell_\alpha^-$+inv. or $X \to \ell_\alpha^+ \ell_\beta^-$+inv. | LRSM. LLP is the right-handed neutrino ($\nu_R$) produced via $pp \to W^\pm$, $W^\pm \to \ell^\pm \nu_R$. Then, $\nu_R \to \ell_\alpha^+ \ell_\alpha^- \nu_\beta$ or $\nu_R \to \ell_\alpha^+ \ell_\beta^- \nu_\alpha$ via an off-shell $W/Z$. |

Table A.10: Simplified model library process proposals for charged current (CC) production mode, $W_{SM}^\pm / W'^\pm \to X + \ell^\pm$; these can be simulated using left-right symmetric models using either the $W$ or $W'$ (for simplicity, in the table above we only state explicitly $W$). Right-handed neutrino lifetimes are most naturally long for sub-weak-scale masses.

### A.1.2 Dark Shower Models

In the repository, we also include two models of dark showers. These are not meant to be representative in the same way that the simplified models are; however, it does allow theorists and experimentalists to simulate these particular dark showers. The currently-included models are:

1. *Radiating lepton jets:* This is a model where dark matter is charged under a dark $U(1)'$ gauge interaction. The dark photon can be produced via final-state radiation from dark matter produced in a collider. Due to the collinear enhancement radiation of the dark photon, this can lead to a perturbative dark shower [587].

2. *Semi-visible jets:* This is a model in which hidden-sector states are charged under a new confining gauge group, leading to QCD-like showers. The showers produce both stable, invisible dark matter, as well as unstable states that decay back to the SM. This produces "semi-visible" jets [366, 572].